\documentclass[a4paper,UKenglish,cleveref, autoref, thm-restate]{lipics-v2021}

\usepackage{teamwork}
\ANNnoAnnotations      
\input macrosAgdaAdaption.tex
\usepackage{latex-before-sed/agda}
\usepackage{hyperref}
\usepackage{multicol}
\usepackage{amssymb} 
\usepackage{mathabx}

\newcommand{\semanticopen}{\mathopen{\hbox{$[\kern-1.5pt[$}}}
\newcommand{\semanticclose}{\mathclose{\hbox{$]\kern-1.5pt]$}}}
\newcommand{\semantic}[1]{\semanticopen\,#1\,\semanticclose}

\newcommand{\all}{\forall}
\newcommand{\ar}{\to}

\renewcommand{\iff}{\leftrightarrow}

\newcommand{\cross}{\times}

\newcommand{\Script}{$\textsc{script}$}  %

\newcommand{\AgdaMaybe}{\AgdaDatatype{Maybe}}
\newcommand{\AgdaReturn}{\AgdaFunction{return}}
\newcommand{\AgdaJust}{\AgdaInductiveConstructor{just}}
\newcommand{\AgdaNothing}{\AgdaInductiveConstructor{nothing}}
\newcommand{\AgdaTrue}{\AgdaInductiveConstructor{true}}

\newcommand{\AgdaBind}{\AgdaOperator{\AgdaFunction{\ensuremath{\mathbin{{>}\mkern-8.5mu{>}\mkern-2mu{=}}}}}}
\newcommand{\AgdaSemantic}[1]{\AgdaFunction{⟦}\; #1\;\AgdaFunction{⟧}}
\newcommand{\AgdaAr}{\AgdaSymbol{→}}
\newcommand{\AgdaInstructionBasic}{\AgdaDatatype{InstructionBasic}}
\newcommand{\AgdaBitcoinScriptBasic}{\AgdaDatatype{BitcoinScriptBasic}}
\newcommand{\AgdaN}{\AgdaDatatype{ℕ}}
\newcommand{\AgdaMsg}{\AgdaDatatype{Msg}}
\newcommand{\AgdaSemanticsStackOp}[1]{\AgdaFunction{⟦}\; #1\; \AgdaFunction{⟧ₛ}}
\newcommand{\AgdaSemanticsStackOpGeneric}{\AgdaFunction{⟦\AgdaUnderscore{}⟧ₛ}}
\newcommand{\AgdaSemanticsStackOpssGeneric}{\AgdaOperator{\AgdaFunction{⟦\AgdaUnderscore{}⟧ₛˢ}}}
\newcommand{\AgdaColon}{\AgdaSymbol{:}}

\newcommand{\AgdaDoubleAdd}{\AgdaOperator{\AgdaFunction{\ensuremath{+\!\!+}}}}
\newcommand{\AgdaOpEqual}{\AgdaInductiveConstructor{opEqual}}

\newcommand{\AgdaOpPush}{\AgdaInductiveConstructor{opPush}}
\newcommand{\AgdaOpCheckSig}{\AgdaInductiveConstructor{opCheckSig}}

\newcommand{\AgdaExecuteStackEquality}{\AgdaFunction{executeStackEquality}}

\newcommand{\AgdaExecuteStackDup}{\AgdaFunction{executeStackDup}}
\newcommand{\AgdaExecuteOpHash}{\AgdaFunction{executeOpHash}}
\newcommand{\AgdaexecuteStackVerify}{\AgdaFunction{executeStackVerify}}
\newcommand{\AgdaExecuteStackCheckSig}{\AgdaFunction{executeStackCheckSig}}

\newcommand{\AgdaaceeptCond}{\AgdaFunction{acceptState}}

\newcommand{\AgdaAcceptStateSt}{\AgdaFunction{acceptStateˢ}}
\newcommand{\AgdaAcceptStone}{\AgdaFunction{accept\ensuremath{{}^{\mathsf{s}}_1}}} 
\newcommand{\AgdaAcceptSttwo}{\AgdaFunction{accept\ensuremath{{}^{\mathsf{s}}_2}}} 
\newcommand{\AgdaAcceptStthree}{\AgdaFunction{accept\ensuremath{{}^{\mathsf{s}}_3}}} 
\newcommand{\AgdaAcceptStfour}{\AgdaFunction{accept\ensuremath{{}^{\mathsf{s}}_4}}}  
\newcommand{\AgdaAcceptStfive}{\AgdaFunction{accept\ensuremath{{}^{\mathsf{s}}_5}}}  
\newcommand{\AgdaAcceptone}{\AgdaFunction{accept₁}}
\newcommand{\AgdaAccepttwo}{\AgdaFunction{accept₂}}

\newcommand{\AgdaWPreCondPtoPKHst}{\AgdaFunction{wPreCondP2PKHˢ}}
\newcommand{\AgdaWPreCondPtoPKH}{\AgdaFunction{wPreCondP2PKH}}
\newcommand{\AgdaAcceptState}{\AgdaFunction{acceptState}}
\newcommand{\AgdaCorrectone}{\AgdaFunction{correct-1}}
\newcommand{\AgdaCorrecttwo}{\AgdaFunction{correct-2}}

\newcommand{\AgdaCorrectopchecksig}{\AgdaFunction{correct-opCheckSig}}
\newcommand{\AgdaCorrectopverify}{\AgdaFunction{correct-opVerify}}

\newcommand{\AgdaopPushList}{\AgdaFunction{opPushList}}

\newcommand{\AgdastackPredtwoSPred}{\AgdaFunction{stackPred2SPred}}
\newcommand{\AgdaStackPredicate}{\AgdaFunction{StackPredicate}}
\newcommand{\AgdaStackStatePred}{\AgdaFunction{StackStatePred}}

\newcommand{\AgdaHoareTriple}[3]{\AgdaOperator{\AgdaRecord{<}}\AgdaSpace{}%
\AgdaBound{#1}\AgdaSpace{}%
\AgdaOperator{\AgdaRecord{>iff}}\AgdaSpace{}%
\AgdaBound{#2}\AgdaSpace{}%
\AgdaOperator{\AgdaRecord{<}}\AgdaSpace{}%
\AgdaBound{#3}\AgdaSpace{}%
\AgdaOperator{\AgdaRecord{>}}}

\newcommand{\AgdaDoubleColon}{\AgdaOperator{\AgdaInductiveConstructor{∷}}}
\newcommand{\AgdaIsSigned}{\AgdaFunction{IsSigned}}

\newcommand{\StackStateAgda}{\AgdaRecord{StackState}}
\newcommand{\TimeAgda}{\AgdaFunction{Time}}
\newcommand{\MsgAgda}{\AgdaDatatype{Msg}}
\newcommand{\Maybe}{\AgdaDatatype{Maybe}}
\newcommand{\AgdaStack}{\AgdaFunction{Stack}}
\newcommand{\StackAgda}{\AgdaFunction{Stack}} 
\newcommand{\Agdaleft}{\AgdaRecord{<}}
\newcommand{\Agdaright}{\AgdaRecord{>}}
\newcommand{\Agdarightiff}{\AgdaRecord{>ⁱᶠᶠ}}

\newcommand{\Agdastepl}{\AgdaRecord{⟩ᵉ}}
\newcommand{\Agdasteplnoe}{\AgdaRecord{⟩}}

\newcommand{\AgdaSemanticPlus}[1]{\AgdaOperator{\AgdaFunction{⟦}\;#1\;\AgdaFunction{⟧⁺}}}
\newcommand{\AgdaLiftPlus}{\AgdaOperator{\AgdaFunction{⁺}}}
\newcommand{\AgdaBot}{\AgdaDatatype{⊥}}
\newcommand{\AgdaHoareIff}[3]{\AgdaOperator{\AgdaRecord{<}}\AgdaSpace{}#1%
\AgdaSpace{}\AgdaOperator{\AgdaRecord{>ⁱᶠᶠ}}\AgdaSpace{}#2\AgdaSpace{}%
\AgdaOperator{\AgdaRecord{<}}\AgdaSpace{}#3\AgdaSpace{}\AgdaOperator{\AgdaRecord{>}}}

\newcommand{\AgdaMsgOnePostulate}{\AgdaPostulate{msg₁}}
\newcommand{\AgdaStackOnePostulate}{\AgdaPostulate{stack₁}}
\newcommand{\AgdaStackxOnePostulate}{\AgdaPostulate{x₁}}
\newcommand{\AgdaSet}{\AgdaPrimitive{Set}}
\newcommand{\AgdaEquivPred}{\AgdaOperator{\AgdaRecord{<=>ᵖ}}}

\newcommand{\AgdaEmptyList}{\AgdaInductiveConstructor{[]}}

\newcommand{\AgdaAbstrFun}{\AgdaFunction{abstrFun}}

\newcommand{\AgdaCompareNaturals}{\AgdaFunction{compareNaturals}}
\newcommand{\AgdaStackOne}{\AgdaPostulate{stack₁}}
\newcommand{\AgdaStackTwo}{\AgdaPostulate{stack₂}}
\newcommand{\AgdaSuc}{\AgdaInductiveConstructor{suc}}
\newcommand{\AgdaXonePostulate}{\AgdaPostulate{x₁}}
\newcommand{\pbkhone}{\ensuremath{pbkh₁}}
\newcommand{\pbkhtwo}{\ensuremath{pbkh₂}}
\newcommand{\pbkone}{\ensuremath{pbk₁}}

\newcommand{\AgdaPbkPostulate}{\AgdaPostulate{pbk}}
\newcommand{\AgdaPbkhPostulate}{\AgdaPostulate{pbkh}}
\newcommand{\AgdaHashFun}{\AgdaFunction{hashFun}}
\newcommand{\AgdaSigOne}{\AgdaPostulate{sig₁}}
\newcommand{\ptopkhFunctionDecoded}{\AgdaFunction{p2pkhFunctionDecoded}}
\newcommand{\AgdaPhi}{\AgdaBound{φ}}

\newcommand{\AgdaExecuteMultiSig}{\AgdaFunction{executeMultiSig}}
\newcommand{\AgdaExecuteMultiSigThree}{\AgdaFunction{executeMultiSig3}}
\newcommand{\AgdacmpSigs}{\AgdaFunction{cmpMultiSigs}}

\newcommand{\AgdaliftStackFunToStackState}{\AgdaFunction{liftStackFun2StackState}}

\newcommand{\PredicateAgda}{\AgdaFunction{Predicate}}
\newcommand{\AgdaEquiv}{\AgdaOperator{\AgdaDatatype{≡}}}
\newcommand{\AgdaPrivate}{\AgdaKeyword{private}}

\newcommand{\AgdaHoareGens}[3]{\AgdaOperator{\AgdaRecord{<}}#1%
\AgdaOperator{\AgdaRecord{>gˢ}}#2%
\AgdaOperator{\AgdaRecord{<}}#3\AgdaOperator{\AgdaRecord{>}}}

\newcommand{\AgdaHoareGen}[3]{\AgdaOperator{\AgdaRecord{<}}#1%
\AgdaOperator{\AgdaRecord{>g}}#2%
\AgdaOperator{\AgdaRecord{<}}#3\AgdaOperator{\AgdaRecord{>}}}

\newcommand{\AgdaHoareTripleIffNoSpace}[3]{\AgdaOperator{\AgdaRecord{<}}%
\AgdaBound{#1}%
\AgdaOperator{\AgdaRecord{>ⁱᶠᶠ}}%
\AgdaBound{#2}%
\AgdaOperator{\AgdaRecord{<}}%
\AgdaBound{#3}%
\AgdaOperator{\AgdaRecord{>}}}

\newcommand{\AgdaTimeOne}{\ensuremath{time₁}}

\newcommand{\myparagraph}[1]{\subparagraph*{#1}}
\newcommand{\vspaceBeforeAgdaCode}{}
\newcommand{\vspaceAfterAgdaCode}{}
\newcommand{\vspacetwocm}{}
\newcommand{\vspacethreecm}{}
\newcommand{\vspacezeroonecm}{}
\newcommand{\vspacezerofourfivecm}{}
 
\newcommand{\Agadaleftarrow}{\AgdaOperator{\AgdaFunction{←}}}
\newcommand{\AgdaAccepti}{\AgdaFunction{accept$_i$}}

\newcommand{\AgdaAndsp}{\AgdaOperator{\AgdaFunction{∧sp}}}
\newcommand{\Agdaland}{\AgdaFunction{$\land$}}

\supplementdetails[subcategory={Agda Source Code}]{Software}{https://github.com/fahad1985lab/Smart--Contracts--Verification--With--Agda}

\title{Verification of Bitcoin Script in Agda using Weakest Preconditions for Access Control}
\titlerunning{Verification of Bitcoin Script in Agda}
\author{Fahad F. Alhabardi}{Swansea University, Dept.~of Computer Science \and \url{}}{fahadalhabardi@gmail.com}{https://orcid.org/0000-0003-0992-2709}{}
\author{Arnold Beckmann}{Swansea University,  Dept.~of Computer Science \and \url{https://www.beckmann.pro/} }{a.beckmann@swansea.ac.uk}{https://orcid.org/0000-0001-7958-5790}{}
\author{Bogdan Lazar}{University of Bath}{lazarbogdan90@yahoo.com}{}{}
\author{Anton Setzer}{Swansea University,  Dept.~of Computer Science \and \url{http://www.cs.swan.ac.uk/~csetzer/} }{a.g.setzer@swansea.ac.uk}{https://orcid.org/0000-0001-5322-6060}{}
\authorrunning{F. Alhabardi, A. Beckmann, B. Lazar, and  A. Setzer}
\keywords{Blockchain; 
Cryptocurrency; 
Bitcoin; 
Agda; 
Verification; 
Hoare logic; 
Bitcoin script; 
P2PKH; 
P2MS; 
Access control; 
Weakest precondition;
Predicate transformer semantics;
Provable correctness; 
Symbolic execution;
Smart contracts}
\Copyright{Fahad F. Alhabardi, Arnold Beckmann, Bogdan Lazar, Anton Setzer}
\acknowledgements{We would like to thank the anonymous referees for valuable comments and suggestions.}

\ccsdesc[500]{Theory of computation~Hoare logic}
\ccsdesc[500]{Theory of computation~Type theory}
\ccsdesc[300]{Theory of computation~Programming logic}
\ccsdesc[500]{Theory of computation~Interactive proof systems}
\ccsdesc[300]{Theory of computation~Operational semantics}
\ccsdesc[300]{Theory of computation~Denotational semantics}
\ccsdesc[500]{Security and privacy~Access control}
\ccsdesc[500]{Security and privacy~Logic and verification}
\ccsdesc[500]{Applied computing~Digital cash}

\EventEditors{Henning Basold, Jesper Cockx, and Silvia Ghilezan}
\EventNoEds{3}
\EventLongTitle{27th International Conference on Types for Proofs and Programs (TYPES 2021)}
\EventShortTitle{TYPES 2021}
\EventAcronym{TYPES}
\EventYear{2021}
\EventDate{June 14--18, 2021}
\EventLocation{Leiden, The Netherlands (Virtual Conference)}
\EventLogo{}
\SeriesVolume{239}
\ArticleNo{1}

\ANNannotationDone{section to thanks all referees.}
\nolinenumbers 
\hideLIPIcs

\begin{document}
%

\AgdaHide{
\begin{code}%
\>[0]\AgdaKeyword{open}\AgdaSpace{}%
\AgdaKeyword{import}\AgdaSpace{}%
\AgdaModule{basicBitcoinDataType}\<%
\\
\\[\AgdaEmptyExtraSkip]%
\>[0]\AgdaKeyword{module}\AgdaSpace{}%
\AgdaModule{exampleGeneratedWeakPreCond}\AgdaSpace{}%
\AgdaSymbol{(}\AgdaBound{param}\AgdaSpace{}%
\AgdaSymbol{:}\AgdaSpace{}%
\AgdaRecord{GlobalParameters}\AgdaSymbol{)}\AgdaSpace{}%
\AgdaKeyword{where}\<%
\\
\\[\AgdaEmptyExtraSkip]%
\>[0]\AgdaKeyword{open}\AgdaSpace{}%
\AgdaKeyword{import}\AgdaSpace{}%
\AgdaModule{libraries.listLib}\<%
\\
\>[0]\AgdaKeyword{open}\AgdaSpace{}%
\AgdaKeyword{import}\AgdaSpace{}%
\AgdaModule{Data.List.Base}\<%
\\
\>[0]\AgdaKeyword{open}\AgdaSpace{}%
\AgdaKeyword{import}\AgdaSpace{}%
\AgdaModule{libraries.natLib}\<%
\\
\>[0]\AgdaKeyword{open}\AgdaSpace{}%
\AgdaKeyword{import}\AgdaSpace{}%
\AgdaModule{Data.Nat}%
\>[22]\AgdaKeyword{renaming}\AgdaSpace{}%
\AgdaSymbol{(}\AgdaOperator{\AgdaDatatype{\AgdaUnderscore{}≤\AgdaUnderscore{}}}\AgdaSpace{}%
\AgdaSymbol{to}\AgdaSpace{}%
\AgdaOperator{\AgdaDatatype{\AgdaUnderscore{}≤'\AgdaUnderscore{}}}\AgdaSpace{}%
\AgdaSymbol{;}\AgdaSpace{}%
\AgdaOperator{\AgdaFunction{\AgdaUnderscore{}<\AgdaUnderscore{}}}\AgdaSpace{}%
\AgdaSymbol{to}\AgdaSpace{}%
\AgdaOperator{\AgdaFunction{\AgdaUnderscore{}<'\AgdaUnderscore{}}}\AgdaSymbol{)}\<%
\\
\>[0]\AgdaKeyword{open}\AgdaSpace{}%
\AgdaKeyword{import}\AgdaSpace{}%
\AgdaModule{Data.List}\AgdaSpace{}%
\AgdaKeyword{hiding}\AgdaSpace{}%
\AgdaSymbol{(}\AgdaOperator{\AgdaFunction{\AgdaUnderscore{}\ensuremath{+\!\!+}\AgdaUnderscore{}}}\AgdaSymbol{)}\<%
\\
\>[0]\AgdaKeyword{open}\AgdaSpace{}%
\AgdaKeyword{import}\AgdaSpace{}%
\AgdaModule{Data.Unit}%
\>[23]\AgdaKeyword{hiding}\AgdaSpace{}%
\AgdaSymbol{(}\AgdaOperator{\AgdaRecord{\AgdaUnderscore{}≤\AgdaUnderscore{}}}\AgdaSymbol{)}\<%
\\
\>[0]\AgdaKeyword{open}\AgdaSpace{}%
\AgdaKeyword{import}\AgdaSpace{}%
\AgdaModule{Data.Empty}\<%
\\
\>[0]\AgdaKeyword{open}\AgdaSpace{}%
\AgdaKeyword{import}\AgdaSpace{}%
\AgdaModule{Data.Maybe}\<%
\\
\>[0]\AgdaKeyword{open}\AgdaSpace{}%
\AgdaKeyword{import}\AgdaSpace{}%
\AgdaModule{Data.Bool}%
\>[23]\AgdaKeyword{hiding}\AgdaSpace{}%
\AgdaSymbol{(}\AgdaOperator{\AgdaDatatype{\AgdaUnderscore{}≤\AgdaUnderscore{}}}\AgdaSpace{}%
\AgdaSymbol{;}\AgdaSpace{}%
\AgdaOperator{\AgdaDatatype{\AgdaUnderscore{}<\AgdaUnderscore{}}}\AgdaSpace{}%
\AgdaSymbol{;}\AgdaSpace{}%
\AgdaOperator{\AgdaFunction{if\AgdaUnderscore{}then\AgdaUnderscore{}else\AgdaUnderscore{}}}\AgdaSpace{}%
\AgdaSymbol{)}\AgdaSpace{}%
\AgdaKeyword{renaming}\AgdaSpace{}%
\AgdaSymbol{(}\AgdaOperator{\AgdaFunction{\AgdaUnderscore{}∧\AgdaUnderscore{}}}\AgdaSpace{}%
\AgdaSymbol{to}\AgdaSpace{}%
\AgdaOperator{\AgdaFunction{\AgdaUnderscore{}∧b\AgdaUnderscore{}}}\AgdaSpace{}%
\AgdaSymbol{;}\AgdaSpace{}%
\AgdaOperator{\AgdaFunction{\AgdaUnderscore{}∨\AgdaUnderscore{}}}\AgdaSpace{}%
\AgdaSymbol{to}\AgdaSpace{}%
\AgdaOperator{\AgdaFunction{\AgdaUnderscore{}∨b\AgdaUnderscore{}}}\AgdaSpace{}%
\AgdaSymbol{;}\AgdaSpace{}%
\AgdaFunction{T}\AgdaSpace{}%
\AgdaSymbol{to}\AgdaSpace{}%
\AgdaFunction{True}\AgdaSymbol{)}\<%
\\
\>[0]\AgdaKeyword{open}\AgdaSpace{}%
\AgdaKeyword{import}\AgdaSpace{}%
\AgdaModule{Data.Product}\AgdaSpace{}%
\AgdaKeyword{renaming}\AgdaSpace{}%
\AgdaSymbol{(}\AgdaOperator{\AgdaInductiveConstructor{\AgdaUnderscore{},\AgdaUnderscore{}}}\AgdaSpace{}%
\AgdaSymbol{to}\AgdaSpace{}%
\AgdaOperator{\AgdaInductiveConstructor{\AgdaUnderscore{},,\AgdaUnderscore{}}}\AgdaSpace{}%
\AgdaSymbol{)}\<%
\\
\>[0]\AgdaKeyword{open}\AgdaSpace{}%
\AgdaKeyword{import}\AgdaSpace{}%
\AgdaModule{Data.Nat.Base}\AgdaSpace{}%
\AgdaKeyword{hiding}\AgdaSpace{}%
\AgdaSymbol{(}\AgdaOperator{\AgdaDatatype{\AgdaUnderscore{}≤\AgdaUnderscore{}}}\AgdaSpace{}%
\AgdaSymbol{;}\AgdaSpace{}%
\AgdaOperator{\AgdaFunction{\AgdaUnderscore{}<\AgdaUnderscore{}}}\AgdaSymbol{)}\<%
\\
\>[0]\AgdaKeyword{open}\AgdaSpace{}%
\AgdaKeyword{import}\AgdaSpace{}%
\AgdaModule{Data.List.NonEmpty}\AgdaSpace{}%
\AgdaKeyword{hiding}\AgdaSpace{}%
\AgdaSymbol{(}\AgdaField{head}\AgdaSymbol{;}\AgdaSpace{}%
\AgdaOperator{\AgdaFunction{[\AgdaUnderscore{}]}}\AgdaSymbol{)}\<%
\\
\>[0]\AgdaKeyword{open}\AgdaSpace{}%
\AgdaKeyword{import}\AgdaSpace{}%
\AgdaModule{Data.Nat}\AgdaSpace{}%
\AgdaKeyword{using}\AgdaSpace{}%
\AgdaSymbol{(}\AgdaDatatype{ℕ}\AgdaSymbol{;}\AgdaSpace{}%
\AgdaOperator{\AgdaPrimitive{\AgdaUnderscore{}+\AgdaUnderscore{}}}\AgdaSymbol{;}\AgdaSpace{}%
\AgdaOperator{\AgdaFunction{\AgdaUnderscore{}≥\AgdaUnderscore{}}}\AgdaSymbol{;}\AgdaSpace{}%
\AgdaOperator{\AgdaFunction{\AgdaUnderscore{}>\AgdaUnderscore{}}}\AgdaSymbol{;}\AgdaSpace{}%
\AgdaInductiveConstructor{zero}\AgdaSymbol{;}\AgdaSpace{}%
\AgdaInductiveConstructor{suc}\AgdaSymbol{;}\AgdaSpace{}%
\AgdaInductiveConstructor{s≤s}\AgdaSymbol{;}\AgdaSpace{}%
\AgdaInductiveConstructor{z≤n}\AgdaSymbol{)}\<%
\\
\>[0]\AgdaKeyword{import}\AgdaSpace{}%
\AgdaModule{Relation.Binary.PropositionalEquality}\AgdaSpace{}%
\AgdaSymbol{as}\AgdaSpace{}%
\AgdaModule{Eq}\<%
\\
\>[0]\AgdaKeyword{open}\AgdaSpace{}%
\AgdaModule{Eq}\AgdaSpace{}%
\AgdaKeyword{using}\AgdaSpace{}%
\AgdaSymbol{(}\AgdaOperator{\AgdaDatatype{\AgdaUnderscore{}≡\AgdaUnderscore{}}}\AgdaSymbol{;}\AgdaSpace{}%
\AgdaInductiveConstructor{refl}\AgdaSymbol{;}\AgdaSpace{}%
\AgdaFunction{cong}\AgdaSymbol{;}\AgdaSpace{}%
\AgdaKeyword{module}\AgdaSpace{}%
\AgdaModule{≡-Reasoning}\AgdaSymbol{;}\AgdaSpace{}%
\AgdaFunction{sym}\AgdaSymbol{)}\<%
\\
\>[0]\AgdaKeyword{open}\AgdaSpace{}%
\AgdaModule{≡-Reasoning}\<%
\\
\>[0]\AgdaKeyword{open}\AgdaSpace{}%
\AgdaKeyword{import}\AgdaSpace{}%
\AgdaModule{Agda.Builtin.Equality}\<%
\\
\\[\AgdaEmptyExtraSkip]%
\\[\AgdaEmptyExtraSkip]%
\>[0]\AgdaComment{--our\ libraries}\<%
\\
\>[0]\AgdaKeyword{open}\AgdaSpace{}%
\AgdaKeyword{import}\AgdaSpace{}%
\AgdaModule{libraries.andLib}\<%
\\
\>[0]\AgdaKeyword{open}\AgdaSpace{}%
\AgdaKeyword{import}\AgdaSpace{}%
\AgdaModule{libraries.miscLib}\<%
\\
\>[0]\AgdaKeyword{open}\AgdaSpace{}%
\AgdaKeyword{import}\AgdaSpace{}%
\AgdaModule{libraries.maybeLib}\<%
\\
\>[0]\AgdaKeyword{open}\AgdaSpace{}%
\AgdaKeyword{import}\AgdaSpace{}%
\AgdaModule{libraries.boolLib}\<%
\\
\\[\AgdaEmptyExtraSkip]%
\>[0]\AgdaKeyword{open}\AgdaSpace{}%
\AgdaKeyword{import}\AgdaSpace{}%
\AgdaModule{stack}\<%
\\
\>[0]\AgdaKeyword{open}\AgdaSpace{}%
\AgdaKeyword{import}\AgdaSpace{}%
\AgdaModule{stackPredicate}\<%
\\
\>[0]\AgdaKeyword{open}\AgdaSpace{}%
\AgdaKeyword{import}\AgdaSpace{}%
\AgdaModule{semanticBasicOperations}\AgdaSpace{}%
\AgdaBound{param}\<%
\\
\>[0]\AgdaKeyword{open}\AgdaSpace{}%
\AgdaKeyword{import}\AgdaSpace{}%
\AgdaModule{instructionBasic}\<%
\\
\>[0]\AgdaKeyword{open}\AgdaSpace{}%
\AgdaKeyword{import}\AgdaSpace{}%
\AgdaModule{verificationP2PKHbasic}\AgdaSpace{}%
\AgdaBound{param}\<%
\\
\\[\AgdaEmptyExtraSkip]%
\>[0]\AgdaKeyword{open}\AgdaSpace{}%
\AgdaKeyword{import}\AgdaSpace{}%
\AgdaModule{verificationStackScripts.stackHoareTriple}\AgdaSpace{}%
\AgdaBound{param}\<%
\\
\>[0]\AgdaKeyword{open}\AgdaSpace{}%
\AgdaKeyword{import}\AgdaSpace{}%
\AgdaModule{verificationStackScripts.stackState}\<%
\\
\>[0]\AgdaKeyword{open}\AgdaSpace{}%
\AgdaKeyword{import}\AgdaSpace{}%
\AgdaModule{verificationStackScripts.sPredicate}\<%
\\
\>[0]\AgdaKeyword{open}\AgdaSpace{}%
\AgdaKeyword{import}\AgdaSpace{}%
\AgdaModule{verificationStackScripts.semanticsStackInstructions}\AgdaSpace{}%
\AgdaBound{param}\<%
\\
\>[0]\AgdaKeyword{open}\AgdaSpace{}%
\AgdaKeyword{import}\AgdaSpace{}%
\AgdaModule{verificationStackScripts.stackVerificationLemmas}\AgdaSpace{}%
\AgdaBound{param}\<%
\\
\\[\AgdaEmptyExtraSkip]%
\>[0]\AgdaComment{--\textbackslash{}exampleGeneratedWeakPreCondweakestPreCond}\<%
\end{code}
} 

\newcommand{\exampleGeneratedWeakPreCondweakestPreCond}{
\begin{code}%
\>[0]\AgdaFunction{weakestPreCondˢ}\AgdaSpace{}%
\AgdaSymbol{:}\AgdaSpace{}%
\AgdaFunction{BitcoinScriptBasic}\AgdaSpace{}%
\AgdaSymbol{→}\AgdaSpace{}%
\AgdaFunction{StackStatePred}\AgdaSpace{}%
\AgdaSymbol{→}\AgdaSpace{}%
\AgdaFunction{StackStatePred}\<%
\\
\>[0]\AgdaFunction{weakestPreCondˢ}\AgdaSpace{}%
\AgdaBound{p}\AgdaSpace{}%
\AgdaBound{φ}\AgdaSpace{}%
\AgdaBound{s}\AgdaSpace{}%
\AgdaSymbol{=}\AgdaSpace{}%
\AgdaSymbol{(}\AgdaBound{φ}\AgdaSpace{}%
\AgdaOperator{\AgdaFunction{⁺}}\AgdaSymbol{)}\AgdaSpace{}%
\AgdaSymbol{(}\AgdaOperator{\AgdaFunction{⟦}}\AgdaSpace{}%
\AgdaBound{p}\AgdaSpace{}%
\AgdaOperator{\AgdaFunction{⟧}}\AgdaSpace{}%
\AgdaBound{s}\AgdaSymbol{)}\<%
\end{code}
} 

\AgdaHide{
\begin{code}%
\>[0]\<%
\\
\\[\AgdaEmptyExtraSkip]%
\>[0]\AgdaFunction{testprog}\AgdaSpace{}%
\AgdaSymbol{:}\AgdaSpace{}%
\AgdaFunction{BitcoinScriptBasic}\<%
\\
\>[0]\AgdaComment{--\textbackslash{}exampleGeneratedWeakPreCondtestprog}\<%
\end{code}
} 

\newcommand{\exampleGeneratedWeakPreCondtestprog}{
\begin{code}[inline]%
\>[0]\AgdaFunction{testprog}\AgdaSpace{}%
\AgdaSymbol{=}\AgdaSpace{}%
\AgdaInductiveConstructor{opDrop}\AgdaSpace{}%
\AgdaOperator{\AgdaInductiveConstructor{∷}}\AgdaSpace{}%
\AgdaInductiveConstructor{opDrop}\AgdaSpace{}%
\AgdaOperator{\AgdaInductiveConstructor{∷}}\AgdaSpace{}%
\AgdaOperator{\AgdaFunction{[}}\AgdaSpace{}%
\AgdaInductiveConstructor{opDrop}\AgdaSpace{}%
\AgdaOperator{\AgdaFunction{]}}\<%
\end{code}
} 

\AgdaHide{
\begin{code}%
\>[0]\<%
\\
\>[0]\AgdaFunction{weakestPreCondTestProg}\AgdaSpace{}%
\AgdaSymbol{:}\AgdaSpace{}%
\AgdaFunction{StackStatePred}\<%
\\
\>[0]\AgdaComment{--\textbackslash{}exampleGeneratedWeakPreCondweakestPreCondTestProg}\<%
\end{code}
} 

\newcommand{\exampleGeneratedWeakPreCondweakestPreCondTestProg}{
\begin{code}[inline]%
\>[0]\AgdaFunction{weakestPreCondTestProg}\AgdaSpace{}%
\AgdaSymbol{=}\AgdaSpace{}%
\AgdaFunction{weakestPreCondˢ}\AgdaSpace{}%
\AgdaFunction{testprog}\AgdaSpace{}%
\AgdaFunction{acceptState}\<%
\end{code}
} 

\AgdaHide{
\begin{code}%
\>[0]\<%
\\
\\[\AgdaEmptyExtraSkip]%
\>[0]\AgdaFunction{weakestPreCondTestProgNormalised}\AgdaSpace{}%
\AgdaSymbol{:}\AgdaSpace{}%
\AgdaFunction{StackStatePred}\<%
\\
\>[0]\AgdaComment{--\textbackslash{}exampleGeneratedWeakPreCondweakestPreCondTestProgNormalised}\<%
\end{code}
} 

\newcommand{\exampleGeneratedWeakPreCondweakestPreCondTestProgNormalised}{
\begin{code}%
\>[0]\AgdaFunction{weakestPreCondTestProgNormalised}\AgdaSpace{}%
\AgdaBound{s}\AgdaSpace{}%
\AgdaSymbol{=}\<%
\\
\>[0][@{}l@{\AgdaIndent{0}}]%
\>[2]\AgdaSymbol{(}\AgdaFunction{stackPred2SPred}\AgdaSpace{}%
\AgdaFunction{acceptStateˢ}\AgdaSpace{}%
\AgdaOperator{\AgdaFunction{⁺}}\AgdaSymbol{)}\<%
\\
\>[2][@{}l@{\AgdaIndent{0}}]%
\>[3]\AgdaSymbol{(}\AgdaFunction{stackState2WithMaybe}\AgdaSpace{}%
\AgdaOperator{\AgdaInductiveConstructor{⟨}}\AgdaSpace{}%
\AgdaField{currentTime}\AgdaSpace{}%
\AgdaBound{s}\AgdaSpace{}%
\AgdaOperator{\AgdaInductiveConstructor{,}}\AgdaSpace{}%
\AgdaField{msg}\AgdaSpace{}%
\AgdaBound{s}\AgdaSpace{}%
\AgdaOperator{\AgdaInductiveConstructor{,}}\AgdaSpace{}%
\AgdaFunction{executeStackDrop}\AgdaSpace{}%
\AgdaSymbol{(}\AgdaField{stack}\AgdaSpace{}%
\AgdaBound{s}\AgdaSymbol{)}\AgdaSpace{}%
\AgdaOperator{\AgdaInductiveConstructor{⟩}}\<%
\\
\>[3]\AgdaOperator{\AgdaFunction{\ensuremath{\mathbin{{>}\mkern-8.5mu{>}\mkern-2mu{=}}}}}%
\>[8]\AgdaSymbol{(λ}\AgdaSpace{}%
\AgdaBound{s₁}\AgdaSpace{}%
\AgdaSymbol{→}%
\>[180I]\AgdaFunction{stackState2WithMaybe}%
\>[38]\AgdaOperator{\AgdaInductiveConstructor{⟨}}\AgdaSpace{}%
\AgdaField{currentTime}\AgdaSpace{}%
\AgdaBound{s₁}\AgdaSpace{}%
\AgdaOperator{\AgdaInductiveConstructor{,}}\AgdaSpace{}%
\AgdaField{msg}\AgdaSpace{}%
\AgdaBound{s₁}\AgdaSpace{}%
\AgdaOperator{\AgdaInductiveConstructor{,}}\AgdaSpace{}%
\AgdaFunction{executeStackDrop}\AgdaSpace{}%
\AgdaSymbol{(}\AgdaField{stack}\AgdaSpace{}%
\AgdaBound{s₁}\AgdaSymbol{)}\AgdaSpace{}%
\AgdaOperator{\AgdaInductiveConstructor{⟩}}\<%
\\
\>[180I][@{}l@{\AgdaIndent{0}}]%
\>[17]\AgdaOperator{\AgdaFunction{\ensuremath{\mathbin{{>}\mkern-8.5mu{>}\mkern-2mu{=}}}}}%
\>[22]\AgdaFunction{liftStackFun2StackState}%
\>[47]\AgdaSymbol{(λ}\AgdaSpace{}%
\AgdaBound{time₁}\AgdaSpace{}%
\AgdaBound{msg₁}\AgdaSpace{}%
\AgdaSymbol{→}\AgdaSpace{}%
\AgdaFunction{executeStackDrop}\AgdaSymbol{)))}\<%
\end{code}
} 

\AgdaHide{
\begin{code}%
\>[0]\<%
\end{code}
} 


\AgdaHide{
\begin{code}%
\>[0]\AgdaKeyword{open}\AgdaSpace{}%
\AgdaKeyword{import}\AgdaSpace{}%
\AgdaModule{basicBitcoinDataType}\<%
\\
\\[\AgdaEmptyExtraSkip]%
\>[0]\AgdaKeyword{module}\AgdaSpace{}%
\AgdaModule{verificationStackScripts.sPredicate}\AgdaSpace{}%
\AgdaKeyword{where}\<%
\\
\\[\AgdaEmptyExtraSkip]%
\>[0]\AgdaKeyword{open}\AgdaSpace{}%
\AgdaKeyword{import}\AgdaSpace{}%
\AgdaModule{Data.Nat}%
\>[22]\AgdaKeyword{hiding}\AgdaSpace{}%
\AgdaSymbol{(}\AgdaOperator{\AgdaDatatype{\AgdaUnderscore{}≤\AgdaUnderscore{}}}\AgdaSymbol{)}\<%
\\
\>[0]\AgdaKeyword{open}\AgdaSpace{}%
\AgdaKeyword{import}\AgdaSpace{}%
\AgdaModule{Data.List}\AgdaSpace{}%
\AgdaKeyword{hiding}\AgdaSpace{}%
\AgdaSymbol{(}\AgdaOperator{\AgdaFunction{\AgdaUnderscore{}\ensuremath{+\!\!+}\AgdaUnderscore{}}}\AgdaSymbol{)}\<%
\\
\>[0]\AgdaKeyword{open}\AgdaSpace{}%
\AgdaKeyword{import}\AgdaSpace{}%
\AgdaModule{Data.Unit}%
\>[23]\AgdaKeyword{hiding}\AgdaSpace{}%
\AgdaSymbol{(}\AgdaOperator{\AgdaRecord{\AgdaUnderscore{}≤\AgdaUnderscore{}}}\AgdaSymbol{)}\<%
\\
\>[0]\AgdaKeyword{open}\AgdaSpace{}%
\AgdaKeyword{import}\AgdaSpace{}%
\AgdaModule{Data.Empty}\<%
\\
\>[0]\AgdaKeyword{open}\AgdaSpace{}%
\AgdaKeyword{import}\AgdaSpace{}%
\AgdaModule{Data.Sum}\<%
\\
\>[0]\AgdaKeyword{open}\AgdaSpace{}%
\AgdaKeyword{import}\AgdaSpace{}%
\AgdaModule{Data.Maybe}\<%
\\
\>[0]\AgdaKeyword{open}\AgdaSpace{}%
\AgdaKeyword{import}\AgdaSpace{}%
\AgdaModule{Data.Bool}%
\>[23]\AgdaKeyword{hiding}\AgdaSpace{}%
\AgdaSymbol{(}\AgdaOperator{\AgdaDatatype{\AgdaUnderscore{}≤\AgdaUnderscore{}}}\AgdaSpace{}%
\AgdaSymbol{;}\AgdaSpace{}%
\AgdaOperator{\AgdaFunction{if\AgdaUnderscore{}then\AgdaUnderscore{}else\AgdaUnderscore{}}}\AgdaSpace{}%
\AgdaSymbol{)}\AgdaSpace{}%
\AgdaKeyword{renaming}\AgdaSpace{}%
\AgdaSymbol{(}\AgdaOperator{\AgdaFunction{\AgdaUnderscore{}∧\AgdaUnderscore{}}}\AgdaSpace{}%
\AgdaSymbol{to}\AgdaSpace{}%
\AgdaOperator{\AgdaFunction{\AgdaUnderscore{}∧b\AgdaUnderscore{}}}\AgdaSpace{}%
\AgdaSymbol{;}\AgdaSpace{}%
\AgdaOperator{\AgdaFunction{\AgdaUnderscore{}∨\AgdaUnderscore{}}}\AgdaSpace{}%
\AgdaSymbol{to}\AgdaSpace{}%
\AgdaOperator{\AgdaFunction{\AgdaUnderscore{}∨b\AgdaUnderscore{}}}\AgdaSpace{}%
\AgdaSymbol{;}\AgdaSpace{}%
\AgdaFunction{T}\AgdaSpace{}%
\AgdaSymbol{to}\AgdaSpace{}%
\AgdaFunction{True}\AgdaSymbol{)}\<%
\\
\>[0]\AgdaKeyword{open}\AgdaSpace{}%
\AgdaKeyword{import}\AgdaSpace{}%
\AgdaModule{Data.Bool.Base}\AgdaSpace{}%
\AgdaKeyword{hiding}\AgdaSpace{}%
\AgdaSymbol{(}\AgdaOperator{\AgdaDatatype{\AgdaUnderscore{}≤\AgdaUnderscore{}}}\AgdaSpace{}%
\AgdaSymbol{;}\AgdaSpace{}%
\AgdaOperator{\AgdaFunction{if\AgdaUnderscore{}then\AgdaUnderscore{}else\AgdaUnderscore{}}}\AgdaSpace{}%
\AgdaSymbol{)}\AgdaSpace{}%
\AgdaKeyword{renaming}\AgdaSpace{}%
\AgdaSymbol{(}\AgdaOperator{\AgdaFunction{\AgdaUnderscore{}∧\AgdaUnderscore{}}}\AgdaSpace{}%
\AgdaSymbol{to}\AgdaSpace{}%
\AgdaOperator{\AgdaFunction{\AgdaUnderscore{}∧b\AgdaUnderscore{}}}\AgdaSpace{}%
\AgdaSymbol{;}\AgdaSpace{}%
\AgdaOperator{\AgdaFunction{\AgdaUnderscore{}∨\AgdaUnderscore{}}}\AgdaSpace{}%
\AgdaSymbol{to}\AgdaSpace{}%
\AgdaOperator{\AgdaFunction{\AgdaUnderscore{}∨b\AgdaUnderscore{}}}\AgdaSpace{}%
\AgdaSymbol{;}\AgdaSpace{}%
\AgdaFunction{T}\AgdaSpace{}%
\AgdaSymbol{to}\AgdaSpace{}%
\AgdaFunction{True}\AgdaSymbol{)}\<%
\\
\>[0]\AgdaKeyword{open}\AgdaSpace{}%
\AgdaKeyword{import}\AgdaSpace{}%
\AgdaModule{Data.Product}\AgdaSpace{}%
\AgdaKeyword{renaming}\AgdaSpace{}%
\AgdaSymbol{(}\AgdaOperator{\AgdaInductiveConstructor{\AgdaUnderscore{},\AgdaUnderscore{}}}\AgdaSpace{}%
\AgdaSymbol{to}\AgdaSpace{}%
\AgdaOperator{\AgdaInductiveConstructor{\AgdaUnderscore{},,\AgdaUnderscore{}}}\AgdaSpace{}%
\AgdaSymbol{)}\<%
\\
\>[0]\AgdaKeyword{open}\AgdaSpace{}%
\AgdaKeyword{import}\AgdaSpace{}%
\AgdaModule{Data.Nat.Base}\AgdaSpace{}%
\AgdaKeyword{hiding}\AgdaSpace{}%
\AgdaSymbol{(}\AgdaOperator{\AgdaDatatype{\AgdaUnderscore{}≤\AgdaUnderscore{}}}\AgdaSymbol{)}\<%
\\
\>[0]\AgdaKeyword{open}\AgdaSpace{}%
\AgdaKeyword{import}\AgdaSpace{}%
\AgdaModule{Data.List.NonEmpty}\AgdaSpace{}%
\AgdaKeyword{hiding}\AgdaSpace{}%
\AgdaSymbol{(}\AgdaField{head}\AgdaSymbol{)}\<%
\\
\>[0]\AgdaKeyword{import}\AgdaSpace{}%
\AgdaModule{Relation.Binary.PropositionalEquality}\AgdaSpace{}%
\AgdaSymbol{as}\AgdaSpace{}%
\AgdaModule{Eq}\<%
\\
\>[0]\AgdaKeyword{open}\AgdaSpace{}%
\AgdaModule{Eq}\AgdaSpace{}%
\AgdaKeyword{using}\AgdaSpace{}%
\AgdaSymbol{(}\AgdaOperator{\AgdaDatatype{\AgdaUnderscore{}≡\AgdaUnderscore{}}}\AgdaSymbol{;}\AgdaSpace{}%
\AgdaInductiveConstructor{refl}\AgdaSymbol{;}\AgdaSpace{}%
\AgdaFunction{cong}\AgdaSymbol{;}\AgdaSpace{}%
\AgdaKeyword{module}\AgdaSpace{}%
\AgdaModule{≡-Reasoning}\AgdaSymbol{;}\AgdaSpace{}%
\AgdaFunction{sym}\AgdaSymbol{)}\<%
\\
\>[0]\AgdaKeyword{open}\AgdaSpace{}%
\AgdaModule{≡-Reasoning}\<%
\\
\>[0]\AgdaKeyword{open}\AgdaSpace{}%
\AgdaKeyword{import}\AgdaSpace{}%
\AgdaModule{Agda.Builtin.Equality}\<%
\\
\\[\AgdaEmptyExtraSkip]%
\\[\AgdaEmptyExtraSkip]%
\\[\AgdaEmptyExtraSkip]%
\>[0]\AgdaComment{--our\ libraries}\<%
\\
\\[\AgdaEmptyExtraSkip]%
\>[0]\AgdaKeyword{open}\AgdaSpace{}%
\AgdaKeyword{import}\AgdaSpace{}%
\AgdaModule{libraries.listLib}\<%
\\
\>[0]\AgdaKeyword{open}\AgdaSpace{}%
\AgdaKeyword{import}\AgdaSpace{}%
\AgdaModule{libraries.natLib}\<%
\\
\>[0]\AgdaKeyword{open}\AgdaSpace{}%
\AgdaKeyword{import}\AgdaSpace{}%
\AgdaModule{libraries.boolLib}\<%
\\
\>[0]\AgdaKeyword{open}\AgdaSpace{}%
\AgdaKeyword{import}\AgdaSpace{}%
\AgdaModule{libraries.andLib}\<%
\\
\>[0]\AgdaKeyword{open}\AgdaSpace{}%
\AgdaKeyword{import}\AgdaSpace{}%
\AgdaModule{libraries.miscLib}\<%
\\
\>[0]\AgdaKeyword{open}\AgdaSpace{}%
\AgdaKeyword{import}\AgdaSpace{}%
\AgdaModule{libraries.maybeLib}\<%
\\
\\[\AgdaEmptyExtraSkip]%
\\[\AgdaEmptyExtraSkip]%
\>[0]\AgdaKeyword{open}\AgdaSpace{}%
\AgdaKeyword{import}\AgdaSpace{}%
\AgdaModule{stack}\<%
\\
\>[0]\AgdaKeyword{open}\AgdaSpace{}%
\AgdaKeyword{import}\AgdaSpace{}%
\AgdaModule{stackPredicate}\<%
\\
\>[0]\AgdaKeyword{open}\AgdaSpace{}%
\AgdaKeyword{import}\AgdaSpace{}%
\AgdaModule{verificationStackScripts.stackState}\<%
\\
\\[\AgdaEmptyExtraSkip]%
\\[\AgdaEmptyExtraSkip]%
\\[\AgdaEmptyExtraSkip]%
\>[0]\AgdaComment{--\ Boolean\ valued\ Stack\ Predicate}\<%
\end{code}
} 

\newcommand{\sPredicatettt}{
\begin{code}%
\>[0]\AgdaFunction{BStackStatePred}%
\>[17]\AgdaSymbol{:}\AgdaSpace{}%
\AgdaPrimitive{Set}\<%
\\
\>[0]\AgdaFunction{BStackStatePred}\AgdaSpace{}%
\AgdaSymbol{=}%
\>[19]\AgdaRecord{StackState}\AgdaSpace{}%
\AgdaSymbol{→}%
\>[33]\AgdaDatatype{Bool}\<%
\end{code}
} 

\AgdaHide{
\begin{code}%
\>[0]\<%
\\
\\[\AgdaEmptyExtraSkip]%
\>[0]\AgdaFunction{MaybeBStackStatePred}\AgdaSpace{}%
\AgdaSymbol{:}\AgdaSpace{}%
\AgdaPrimitive{Set}\<%
\\
\>[0]\AgdaFunction{MaybeBStackStatePred}\AgdaSpace{}%
\AgdaSymbol{=}\AgdaSpace{}%
\AgdaDatatype{Maybe}\AgdaSpace{}%
\AgdaRecord{StackState}\AgdaSpace{}%
\AgdaSymbol{→}%
\>[43]\AgdaDatatype{Bool}\<%
\\
\\[\AgdaEmptyExtraSkip]%
\\[\AgdaEmptyExtraSkip]%
\>[0]\AgdaComment{--\ Stack\ Predicate}\<%
\\
\>[0]\AgdaFunction{StackStatePred}\AgdaSpace{}%
\AgdaSymbol{:}\AgdaSpace{}%
\AgdaPrimitive{Set₁}\<%
\\
\>[0]\AgdaComment{--\textbackslash{}sPredicatestackstatepred}\<%
\end{code}
} 

\newcommand{\sPredicatestackstatepred}{
\begin{code}[inline]%
\>[0]\AgdaFunction{StackStatePred}\AgdaSpace{}%
\AgdaSymbol{=}\AgdaSpace{}%
\AgdaRecord{StackState}\AgdaSpace{}%
\AgdaSymbol{→}\AgdaSpace{}%
\AgdaPrimitive{Set}\<%
\end{code}
} 

\AgdaHide{
\begin{code}%
\>[0]\<%
\\
\\[\AgdaEmptyExtraSkip]%
\\[\AgdaEmptyExtraSkip]%
\>[0]\AgdaFunction{predicateAfterPushingx}\AgdaSpace{}%
\AgdaSymbol{:}\AgdaSpace{}%
\AgdaSymbol{(}\AgdaBound{n}\AgdaSpace{}%
\AgdaSymbol{:}\AgdaSpace{}%
\AgdaDatatype{ℕ}\AgdaSymbol{)(}\AgdaBound{φ}\AgdaSpace{}%
\AgdaSymbol{:}\AgdaSpace{}%
\AgdaFunction{StackStatePred}\AgdaSymbol{)}\AgdaSpace{}%
\AgdaSymbol{→}\AgdaSpace{}%
\AgdaFunction{StackStatePred}\<%
\\
\>[0]\AgdaFunction{predicateAfterPushingx}\AgdaSpace{}%
\AgdaBound{n}\AgdaSpace{}%
\AgdaBound{φ}\AgdaSpace{}%
\AgdaOperator{\AgdaInductiveConstructor{⟨}}\AgdaSpace{}%
\AgdaBound{time}\AgdaSpace{}%
\AgdaOperator{\AgdaInductiveConstructor{,}}\AgdaSpace{}%
\AgdaBound{msg₁}\AgdaSpace{}%
\AgdaOperator{\AgdaInductiveConstructor{,}}\AgdaSpace{}%
\AgdaBound{stack₁}\AgdaSpace{}%
\AgdaOperator{\AgdaInductiveConstructor{⟩}}\AgdaSpace{}%
\AgdaSymbol{=}\AgdaSpace{}%
\AgdaBound{φ}\AgdaSpace{}%
\AgdaOperator{\AgdaInductiveConstructor{⟨}}\AgdaSpace{}%
\AgdaBound{time}\AgdaSpace{}%
\AgdaOperator{\AgdaInductiveConstructor{,}}\AgdaSpace{}%
\AgdaBound{msg₁}\AgdaSpace{}%
\AgdaOperator{\AgdaInductiveConstructor{,}}\AgdaSpace{}%
\AgdaBound{n}\AgdaSpace{}%
\AgdaOperator{\AgdaInductiveConstructor{∷}}\AgdaSpace{}%
\AgdaBound{stack₁}\AgdaSpace{}%
\AgdaOperator{\AgdaInductiveConstructor{⟩}}\<%
\\
\\[\AgdaEmptyExtraSkip]%
\\[\AgdaEmptyExtraSkip]%
\>[0]\AgdaFunction{predicateForTopElOfStack}\AgdaSpace{}%
\AgdaSymbol{:}\AgdaSpace{}%
\AgdaSymbol{(}\AgdaBound{n}\AgdaSpace{}%
\AgdaSymbol{:}\AgdaSpace{}%
\AgdaDatatype{ℕ}\AgdaSymbol{)}\AgdaSpace{}%
\AgdaSymbol{→}\AgdaSpace{}%
\AgdaFunction{StackStatePred}\<%
\\
\>[0]\AgdaFunction{predicateForTopElOfStack}\AgdaSpace{}%
\AgdaBound{n}\AgdaSpace{}%
\AgdaOperator{\AgdaInductiveConstructor{⟨}}\AgdaSpace{}%
\AgdaBound{time}\AgdaSpace{}%
\AgdaOperator{\AgdaInductiveConstructor{,}}\AgdaSpace{}%
\AgdaBound{msg₁}\AgdaSpace{}%
\AgdaOperator{\AgdaInductiveConstructor{,}}\AgdaSpace{}%
\AgdaInductiveConstructor{[]}\AgdaSpace{}%
\AgdaOperator{\AgdaInductiveConstructor{⟩}}\AgdaSpace{}%
\AgdaSymbol{=}\AgdaSpace{}%
\AgdaDatatype{⊥}\<%
\\
\>[0]\AgdaFunction{predicateForTopElOfStack}\AgdaSpace{}%
\AgdaBound{n}\AgdaSpace{}%
\AgdaOperator{\AgdaInductiveConstructor{⟨}}\AgdaSpace{}%
\AgdaBound{time}\AgdaSpace{}%
\AgdaOperator{\AgdaInductiveConstructor{,}}\AgdaSpace{}%
\AgdaBound{msg₁}\AgdaSpace{}%
\AgdaOperator{\AgdaInductiveConstructor{,}}\AgdaSpace{}%
\AgdaBound{x}\AgdaSpace{}%
\AgdaOperator{\AgdaInductiveConstructor{∷}}\AgdaSpace{}%
\AgdaBound{stack₁}\AgdaSpace{}%
\AgdaOperator{\AgdaInductiveConstructor{⟩}}\AgdaSpace{}%
\AgdaSymbol{=}\AgdaSpace{}%
\AgdaBound{x}\AgdaSpace{}%
\AgdaOperator{\AgdaDatatype{≡}}\AgdaSpace{}%
\AgdaBound{n}\<%
\\
\\[\AgdaEmptyExtraSkip]%
\\[\AgdaEmptyExtraSkip]%
\\[\AgdaEmptyExtraSkip]%
\>[0]\AgdaOperator{\AgdaFunction{\AgdaUnderscore{}∧p\AgdaUnderscore{}}}\AgdaSpace{}%
\AgdaSymbol{:}\AgdaSpace{}%
\AgdaSymbol{(}\AgdaSpace{}%
\AgdaBound{φ}\AgdaSpace{}%
\AgdaBound{ψ}%
\>[14]\AgdaSymbol{:}\AgdaSpace{}%
\AgdaFunction{StackStatePred}\AgdaSpace{}%
\AgdaSymbol{)}\AgdaSpace{}%
\AgdaSymbol{→}\AgdaSpace{}%
\AgdaFunction{StackStatePred}\<%
\\
\>[0]\AgdaSymbol{(}\AgdaBound{φ}\AgdaSpace{}%
\AgdaOperator{\AgdaFunction{∧p}}\AgdaSpace{}%
\AgdaBound{ψ}\AgdaSpace{}%
\AgdaSymbol{)}\AgdaSpace{}%
\AgdaBound{s}%
\>[13]\AgdaSymbol{=}%
\>[16]\AgdaBound{φ}\AgdaSpace{}%
\AgdaBound{s}%
\>[21]\AgdaOperator{\AgdaRecord{∧}}%
\>[24]\AgdaBound{ψ}\AgdaSpace{}%
\AgdaBound{s}\<%
\\
\\[\AgdaEmptyExtraSkip]%
\\[\AgdaEmptyExtraSkip]%
\\[\AgdaEmptyExtraSkip]%
\>[0]\AgdaFunction{⊥p}\AgdaSpace{}%
\AgdaSymbol{:}\AgdaSpace{}%
\AgdaFunction{StackStatePred}\<%
\\
\>[0]\AgdaFunction{⊥p}\AgdaSpace{}%
\AgdaBound{s}\AgdaSpace{}%
\AgdaSymbol{=}\AgdaSpace{}%
\AgdaDatatype{⊥}\<%
\\
\\[\AgdaEmptyExtraSkip]%
\>[0]\AgdaKeyword{infixl}\AgdaSpace{}%
\AgdaNumber{4}\AgdaSpace{}%
\AgdaOperator{\AgdaFunction{\AgdaUnderscore{}⊎p\AgdaUnderscore{}}}\<%
\\
\\[\AgdaEmptyExtraSkip]%
\>[0]\AgdaOperator{\AgdaFunction{\AgdaUnderscore{}⊎p\AgdaUnderscore{}}}%
\>[7]\AgdaSymbol{:}\AgdaSpace{}%
\AgdaSymbol{(}\AgdaBound{φ}\AgdaSpace{}%
\AgdaBound{ψ}\AgdaSpace{}%
\AgdaSymbol{:}\AgdaSpace{}%
\AgdaFunction{StackStatePred}\AgdaSymbol{)}\AgdaSpace{}%
\AgdaSymbol{→}\AgdaSpace{}%
\AgdaFunction{StackStatePred}\<%
\\
\>[0]\AgdaSymbol{(}\AgdaBound{φ}\AgdaSpace{}%
\AgdaOperator{\AgdaFunction{⊎p}}\AgdaSpace{}%
\AgdaBound{ψ}\AgdaSymbol{)}\AgdaSpace{}%
\AgdaBound{s}\AgdaSpace{}%
\AgdaSymbol{=}\AgdaSpace{}%
\AgdaBound{φ}\AgdaSpace{}%
\AgdaBound{s}\AgdaSpace{}%
\AgdaOperator{\AgdaDatatype{⊎}}\AgdaSpace{}%
\AgdaBound{ψ}\AgdaSpace{}%
\AgdaBound{s}\<%
\\
\\[\AgdaEmptyExtraSkip]%
\\[\AgdaEmptyExtraSkip]%
\\[\AgdaEmptyExtraSkip]%
\>[0]\AgdaFunction{lemma⊎pleft}\AgdaSpace{}%
\AgdaSymbol{:}%
\>[214I]\AgdaSymbol{(}\AgdaBound{ψ}\AgdaSpace{}%
\AgdaBound{ψ'}\AgdaSpace{}%
\AgdaSymbol{:}\AgdaSpace{}%
\AgdaFunction{StackStatePred}\AgdaSymbol{)(}\AgdaBound{s}\AgdaSpace{}%
\AgdaSymbol{:}\AgdaSpace{}%
\AgdaDatatype{Maybe}\AgdaSpace{}%
\AgdaRecord{StackState}\AgdaSymbol{)}\<%
\\
\>[.][@{}l@{}]\<[214I]%
\>[14]\AgdaSymbol{→}\AgdaSpace{}%
\AgdaSymbol{(}\AgdaBound{ψ}\AgdaSpace{}%
\AgdaOperator{\AgdaFunction{⁺}}\AgdaSymbol{)}\AgdaSpace{}%
\AgdaBound{s}\AgdaSpace{}%
\AgdaSymbol{→}\AgdaSpace{}%
\AgdaSymbol{(}\AgdaSpace{}%
\AgdaSymbol{(}\AgdaBound{ψ}\AgdaSpace{}%
\AgdaOperator{\AgdaFunction{⊎p}}\AgdaSpace{}%
\AgdaBound{ψ'}\AgdaSymbol{)}\AgdaSpace{}%
\AgdaOperator{\AgdaFunction{⁺}}\AgdaSymbol{)}\AgdaSpace{}%
\AgdaBound{s}\<%
\\
\>[0]\AgdaFunction{lemma⊎pleft}\AgdaSpace{}%
\AgdaBound{ψ}\AgdaSpace{}%
\AgdaBound{ψ'}\AgdaSpace{}%
\AgdaSymbol{(}\AgdaInductiveConstructor{just}\AgdaSpace{}%
\AgdaBound{x}\AgdaSymbol{)}\AgdaSpace{}%
\AgdaBound{p}\AgdaSpace{}%
\AgdaSymbol{=}\AgdaSpace{}%
\AgdaInductiveConstructor{inj₁}\AgdaSpace{}%
\AgdaBound{p}\<%
\\
\\[\AgdaEmptyExtraSkip]%
\>[0]\AgdaFunction{lemma⊎pright}\AgdaSpace{}%
\AgdaSymbol{:}%
\>[240I]\AgdaSymbol{(}\AgdaBound{ψ}\AgdaSpace{}%
\AgdaBound{ψ'}\AgdaSpace{}%
\AgdaSymbol{:}\AgdaSpace{}%
\AgdaFunction{StackStatePred}\AgdaSymbol{)}\AgdaSpace{}%
\AgdaSymbol{(}\AgdaBound{s}\AgdaSpace{}%
\AgdaSymbol{:}\AgdaSpace{}%
\AgdaDatatype{Maybe}\AgdaSpace{}%
\AgdaRecord{StackState}\AgdaSymbol{)}\<%
\\
\>[.][@{}l@{}]\<[240I]%
\>[15]\AgdaSymbol{→}\AgdaSpace{}%
\AgdaSymbol{(}\AgdaBound{ψ'}\AgdaSpace{}%
\AgdaOperator{\AgdaFunction{⁺}}\AgdaSymbol{)}\AgdaSpace{}%
\AgdaBound{s}\AgdaSpace{}%
\AgdaSymbol{→}\AgdaSpace{}%
\AgdaSymbol{((}\AgdaSpace{}%
\AgdaBound{ψ}\AgdaSpace{}%
\AgdaOperator{\AgdaFunction{⊎p}}\AgdaSpace{}%
\AgdaBound{ψ'}\AgdaSpace{}%
\AgdaSymbol{)}\AgdaSpace{}%
\AgdaOperator{\AgdaFunction{⁺}}\AgdaSpace{}%
\AgdaSymbol{)}\AgdaSpace{}%
\AgdaBound{s}\<%
\\
\>[0]\AgdaFunction{lemma⊎pright}%
\>[14]\AgdaBound{ψ}\AgdaSpace{}%
\AgdaBound{ψ'}\AgdaSpace{}%
\AgdaSymbol{(}\AgdaInductiveConstructor{just}\AgdaSpace{}%
\AgdaBound{x}\AgdaSymbol{)}\AgdaSpace{}%
\AgdaBound{p}\AgdaSpace{}%
\AgdaSymbol{=}\AgdaSpace{}%
\AgdaInductiveConstructor{inj₂}\AgdaSpace{}%
\AgdaBound{p}\<%
\\
\\[\AgdaEmptyExtraSkip]%
\>[0]\AgdaFunction{lemma⊎pinv}\AgdaSpace{}%
\AgdaSymbol{:}%
\>[268I]\AgdaSymbol{(}\AgdaBound{ψ}\AgdaSpace{}%
\AgdaBound{ψ'}\AgdaSpace{}%
\AgdaSymbol{:}\AgdaSpace{}%
\AgdaFunction{StackStatePred}\AgdaSymbol{)(}\AgdaBound{A}\AgdaSpace{}%
\AgdaSymbol{:}\AgdaSpace{}%
\AgdaPrimitive{Set}\AgdaSymbol{)}\AgdaSpace{}%
\AgdaSymbol{(}\AgdaBound{s}\AgdaSpace{}%
\AgdaSymbol{:}\AgdaSpace{}%
\AgdaDatatype{Maybe}\AgdaSpace{}%
\AgdaRecord{StackState}\AgdaSymbol{)}\<%
\\
\>[.][@{}l@{}]\<[268I]%
\>[13]\AgdaSymbol{→}\AgdaSpace{}%
\AgdaSymbol{((}\AgdaBound{ψ}\AgdaSpace{}%
\AgdaOperator{\AgdaFunction{⁺}}\AgdaSymbol{)}\AgdaSpace{}%
\AgdaBound{s}%
\>[25]\AgdaSymbol{→}\AgdaSpace{}%
\AgdaBound{A}\AgdaSymbol{)}\<%
\\
\>[13]\AgdaSymbol{→}\AgdaSpace{}%
\AgdaSymbol{((}\AgdaBound{ψ'}\AgdaSpace{}%
\AgdaOperator{\AgdaFunction{⁺}}\AgdaSymbol{)}\AgdaSpace{}%
\AgdaBound{s}%
\>[26]\AgdaSymbol{→}\AgdaSpace{}%
\AgdaBound{A}\AgdaSymbol{)}\<%
\\
\>[13]\AgdaSymbol{→}%
\>[16]\AgdaSymbol{((}\AgdaBound{ψ}\AgdaSpace{}%
\AgdaOperator{\AgdaFunction{⊎p}}\AgdaSpace{}%
\AgdaBound{ψ'}\AgdaSymbol{)}\AgdaSpace{}%
\AgdaOperator{\AgdaFunction{⁺}}\AgdaSymbol{)}\AgdaSpace{}%
\AgdaBound{s}\AgdaSpace{}%
\AgdaSymbol{→}\AgdaSpace{}%
\AgdaBound{A}\<%
\\
\>[0]\AgdaFunction{lemma⊎pinv}\AgdaSpace{}%
\AgdaBound{ψ}\AgdaSpace{}%
\AgdaBound{ψ'}\AgdaSpace{}%
\AgdaBound{A}\AgdaSpace{}%
\AgdaSymbol{(}\AgdaInductiveConstructor{just}\AgdaSpace{}%
\AgdaBound{x}\AgdaSymbol{)}\AgdaSpace{}%
\AgdaBound{p}\AgdaSpace{}%
\AgdaBound{q}\AgdaSpace{}%
\AgdaSymbol{(}\AgdaInductiveConstructor{inj₁}\AgdaSpace{}%
\AgdaBound{x₁}\AgdaSymbol{)}\AgdaSpace{}%
\AgdaSymbol{=}\AgdaSpace{}%
\AgdaBound{p}\AgdaSpace{}%
\AgdaBound{x₁}\<%
\\
\>[0]\AgdaFunction{lemma⊎pinv}\AgdaSpace{}%
\AgdaBound{ψ}\AgdaSpace{}%
\AgdaBound{ψ'}\AgdaSpace{}%
\AgdaBound{A}\AgdaSpace{}%
\AgdaSymbol{(}\AgdaInductiveConstructor{just}\AgdaSpace{}%
\AgdaBound{x}\AgdaSymbol{)}\AgdaSpace{}%
\AgdaBound{p}\AgdaSpace{}%
\AgdaBound{q}\AgdaSpace{}%
\AgdaSymbol{(}\AgdaInductiveConstructor{inj₂}\AgdaSpace{}%
\AgdaBound{y}\AgdaSymbol{)}\AgdaSpace{}%
\AgdaSymbol{=}\AgdaSpace{}%
\AgdaBound{q}\AgdaSpace{}%
\AgdaBound{y}\<%
\\
\\[\AgdaEmptyExtraSkip]%
\\[\AgdaEmptyExtraSkip]%
\\[\AgdaEmptyExtraSkip]%
\\[\AgdaEmptyExtraSkip]%
\>[0]\AgdaFunction{stackPred2SPred}\AgdaSpace{}%
\AgdaSymbol{:}\AgdaSpace{}%
\AgdaFunction{StackPredicate}\AgdaSpace{}%
\AgdaSymbol{→}\AgdaSpace{}%
\AgdaFunction{StackStatePred}\<%
\\
\>[0]\AgdaFunction{stackPred2SPred}\AgdaSpace{}%
\AgdaBound{\,f\,}\AgdaSpace{}%
\AgdaOperator{\AgdaInductiveConstructor{⟨}}\AgdaSpace{}%
\AgdaBound{time}\AgdaSpace{}%
\AgdaOperator{\AgdaInductiveConstructor{,}}\AgdaSpace{}%
\AgdaBound{msg₁}\AgdaSpace{}%
\AgdaOperator{\AgdaInductiveConstructor{,}}\AgdaSpace{}%
\AgdaBound{stack₁}\AgdaSpace{}%
\AgdaOperator{\AgdaInductiveConstructor{⟩}}\AgdaSpace{}%
\AgdaSymbol{=}\AgdaSpace{}%
\AgdaBound{\,f\,}\AgdaSpace{}%
\AgdaBound{time}\AgdaSpace{}%
\AgdaBound{msg₁}\AgdaSpace{}%
\AgdaBound{stack₁}\<%
\\
\\[\AgdaEmptyExtraSkip]%
\\[\AgdaEmptyExtraSkip]%
\\[\AgdaEmptyExtraSkip]%
\>[0]\AgdaFunction{stackPred2SPredBool}\AgdaSpace{}%
\AgdaSymbol{:}\AgdaSpace{}%
\AgdaSymbol{(}\AgdaSpace{}%
\AgdaFunction{Time}\AgdaSpace{}%
\AgdaSymbol{→}\AgdaSpace{}%
\AgdaDatatype{Msg}\AgdaSpace{}%
\AgdaSymbol{→}\AgdaSpace{}%
\AgdaFunction{Stack}\AgdaSpace{}%
\AgdaSymbol{→}\AgdaSpace{}%
\AgdaDatatype{Bool}\AgdaSpace{}%
\AgdaSymbol{)}\AgdaSpace{}%
\AgdaSymbol{→}\AgdaSpace{}%
\AgdaSymbol{(}%
\>[57]\AgdaRecord{StackState}\AgdaSpace{}%
\AgdaSymbol{→}%
\>[71]\AgdaDatatype{Bool}\AgdaSpace{}%
\AgdaSymbol{)}\<%
\\
\>[0]\AgdaFunction{stackPred2SPredBool}\AgdaSpace{}%
\AgdaBound{\,f\,}\AgdaSpace{}%
\AgdaOperator{\AgdaInductiveConstructor{⟨}}\AgdaSpace{}%
\AgdaBound{currentTime₁}\AgdaSpace{}%
\AgdaOperator{\AgdaInductiveConstructor{,}}\AgdaSpace{}%
\AgdaBound{msg₁}\AgdaSpace{}%
\AgdaOperator{\AgdaInductiveConstructor{,}}\AgdaSpace{}%
\AgdaBound{stack₁}\AgdaSpace{}%
\AgdaOperator{\AgdaInductiveConstructor{⟩}}\<%
\\
\>[0][@{}l@{\AgdaIndent{0}}]%
\>[14]\AgdaSymbol{=}\AgdaSpace{}%
\AgdaBound{\,f\,}\AgdaSpace{}%
\AgdaBound{currentTime₁}\AgdaSpace{}%
\AgdaBound{msg₁}\AgdaSpace{}%
\AgdaBound{stack₁}\<%
\\
\\[\AgdaEmptyExtraSkip]%
\\[\AgdaEmptyExtraSkip]%
\\[\AgdaEmptyExtraSkip]%
\>[0]\AgdaFunction{topElStack=0}\AgdaSpace{}%
\AgdaSymbol{:}\AgdaSpace{}%
\AgdaFunction{StackStatePred}\<%
\\
\>[0]\AgdaFunction{topElStack=0}\AgdaSpace{}%
\AgdaOperator{\AgdaInductiveConstructor{⟨}}\AgdaSpace{}%
\AgdaBound{time}\AgdaSpace{}%
\AgdaOperator{\AgdaInductiveConstructor{,}}\AgdaSpace{}%
\AgdaBound{msg₁}\AgdaSpace{}%
\AgdaOperator{\AgdaInductiveConstructor{,}}\AgdaSpace{}%
\AgdaInductiveConstructor{[]}\AgdaSpace{}%
\AgdaOperator{\AgdaInductiveConstructor{⟩}}\AgdaSpace{}%
\AgdaSymbol{=}\AgdaSpace{}%
\AgdaDatatype{⊥}\<%
\\
\>[0]\AgdaFunction{topElStack=0}\AgdaSpace{}%
\AgdaOperator{\AgdaInductiveConstructor{⟨}}\AgdaSpace{}%
\AgdaBound{time}\AgdaSpace{}%
\AgdaOperator{\AgdaInductiveConstructor{,}}\AgdaSpace{}%
\AgdaBound{msg₁}\AgdaSpace{}%
\AgdaOperator{\AgdaInductiveConstructor{,}}\AgdaSpace{}%
\AgdaInductiveConstructor{zero}\AgdaSpace{}%
\AgdaOperator{\AgdaInductiveConstructor{∷}}\AgdaSpace{}%
\AgdaBound{stack₁}\AgdaSpace{}%
\AgdaOperator{\AgdaInductiveConstructor{⟩}}\AgdaSpace{}%
\AgdaSymbol{=}\AgdaSpace{}%
\AgdaRecord{⊤}\<%
\\
\>[0]\AgdaFunction{topElStack=0}\AgdaSpace{}%
\AgdaOperator{\AgdaInductiveConstructor{⟨}}\AgdaSpace{}%
\AgdaBound{time}\AgdaSpace{}%
\AgdaOperator{\AgdaInductiveConstructor{,}}\AgdaSpace{}%
\AgdaBound{msg₁}\AgdaSpace{}%
\AgdaOperator{\AgdaInductiveConstructor{,}}\AgdaSpace{}%
\AgdaInductiveConstructor{suc}\AgdaSpace{}%
\AgdaBound{x}\AgdaSpace{}%
\AgdaOperator{\AgdaInductiveConstructor{∷}}\AgdaSpace{}%
\AgdaBound{stack₁}\AgdaSpace{}%
\AgdaOperator{\AgdaInductiveConstructor{⟩}}\AgdaSpace{}%
\AgdaSymbol{=}\AgdaSpace{}%
\AgdaDatatype{⊥}\<%
\\
\\[\AgdaEmptyExtraSkip]%
\\[\AgdaEmptyExtraSkip]%
\\[\AgdaEmptyExtraSkip]%
\>[0]\AgdaFunction{truePred}\AgdaSpace{}%
\AgdaSymbol{:}\AgdaSpace{}%
\AgdaFunction{StackPredicate}\AgdaSpace{}%
\AgdaSymbol{→}\AgdaSpace{}%
\AgdaFunction{StackStatePred}\<%
\\
\>[0]\AgdaFunction{truePred}\AgdaSpace{}%
\AgdaBound{φ}\AgdaSpace{}%
\AgdaSymbol{=}\AgdaSpace{}%
\AgdaFunction{stackPred2SPred}\AgdaSpace{}%
\AgdaSymbol{(}\AgdaFunction{truePredaux}%
\>[43]\AgdaBound{φ}\AgdaSymbol{)}\<%
\\
\\[\AgdaEmptyExtraSkip]%
\\[\AgdaEmptyExtraSkip]%
\\[\AgdaEmptyExtraSkip]%
\>[0]\AgdaFunction{falsePredaux}\AgdaSpace{}%
\AgdaSymbol{:}\AgdaSpace{}%
\AgdaFunction{StackPredicate}\AgdaSpace{}%
\AgdaSymbol{→}\AgdaSpace{}%
\AgdaFunction{StackPredicate}\<%
\\
\>[0]\AgdaFunction{falsePredaux}\AgdaSpace{}%
\AgdaBound{φ}\AgdaSpace{}%
\AgdaBound{time}\AgdaSpace{}%
\AgdaBound{msg}\AgdaSpace{}%
\AgdaInductiveConstructor{[]}\AgdaSpace{}%
\AgdaSymbol{=}\AgdaSpace{}%
\AgdaDatatype{⊥}\<%
\\
\>[0]\AgdaFunction{falsePredaux}\AgdaSpace{}%
\AgdaBound{φ}\AgdaSpace{}%
\AgdaBound{time}\AgdaSpace{}%
\AgdaBound{msg}\AgdaSpace{}%
\AgdaSymbol{(}\AgdaInductiveConstructor{zero}\AgdaSpace{}%
\AgdaOperator{\AgdaInductiveConstructor{∷}}\AgdaSpace{}%
\AgdaBound{st}\AgdaSymbol{)}\AgdaSpace{}%
\AgdaSymbol{=}\AgdaSpace{}%
\AgdaBound{φ}\AgdaSpace{}%
\AgdaBound{time}\AgdaSpace{}%
\AgdaBound{msg}%
\>[50]\AgdaBound{st}\<%
\\
\>[0]\AgdaFunction{falsePredaux}\AgdaSpace{}%
\AgdaBound{φ}\AgdaSpace{}%
\AgdaBound{time}\AgdaSpace{}%
\AgdaBound{msg}\AgdaSpace{}%
\AgdaSymbol{(}\AgdaInductiveConstructor{suc}\AgdaSpace{}%
\AgdaBound{x}\AgdaSpace{}%
\AgdaOperator{\AgdaInductiveConstructor{∷}}\AgdaSpace{}%
\AgdaBound{st}\AgdaSymbol{)}\AgdaSpace{}%
\AgdaSymbol{=}\AgdaSpace{}%
\AgdaDatatype{⊥}\<%
\\
\\[\AgdaEmptyExtraSkip]%
\>[0]\AgdaFunction{falsePred}\AgdaSpace{}%
\AgdaSymbol{:}\AgdaSpace{}%
\AgdaFunction{StackPredicate}\AgdaSpace{}%
\AgdaSymbol{→}\AgdaSpace{}%
\AgdaFunction{StackStatePred}\<%
\\
\>[0]\AgdaFunction{falsePred}\AgdaSpace{}%
\AgdaBound{φ}\AgdaSpace{}%
\AgdaSymbol{=}\AgdaSpace{}%
\AgdaFunction{stackPred2SPred}\AgdaSpace{}%
\AgdaSymbol{(}\AgdaFunction{falsePredaux}%
\>[45]\AgdaBound{φ}\AgdaSymbol{)}\<%
\\
\\[\AgdaEmptyExtraSkip]%
\\[\AgdaEmptyExtraSkip]%
\>[0]\AgdaFunction{liftAddingx}\AgdaSpace{}%
\AgdaSymbol{:}\AgdaSpace{}%
\AgdaSymbol{(}\AgdaBound{n}\AgdaSpace{}%
\AgdaSymbol{:}\AgdaSpace{}%
\AgdaDatatype{ℕ}\AgdaSymbol{)(}\AgdaSpace{}%
\AgdaBound{φ}\AgdaSpace{}%
\AgdaSymbol{:}\AgdaSpace{}%
\AgdaFunction{StackPredicate}\AgdaSpace{}%
\AgdaSymbol{)}\AgdaSpace{}%
\AgdaSymbol{→}%
\>[47]\AgdaFunction{StackStatePred}\<%
\\
\>[0]\AgdaFunction{liftAddingx}\AgdaSpace{}%
\AgdaBound{n}\AgdaSpace{}%
\AgdaBound{φ}\AgdaSpace{}%
\AgdaSymbol{=}\AgdaSpace{}%
\AgdaFunction{predicateAfterPushingx}\AgdaSpace{}%
\AgdaBound{n}\AgdaSpace{}%
\AgdaSymbol{(}\AgdaFunction{stackPred2SPred}\AgdaSpace{}%
\AgdaBound{φ}\AgdaSymbol{)}\<%
\\
\\[\AgdaEmptyExtraSkip]%
\\[\AgdaEmptyExtraSkip]%
\\[\AgdaEmptyExtraSkip]%
\>[0]\AgdaComment{--\ acceptingState\ \ :\ BStackStatePred}\<%
\\
\>[0]\AgdaComment{--\ acceptingState\ \ =\ stackPred2SPredBool\ acceptingBasic}\<%
\\
\\[\AgdaEmptyExtraSkip]%
\\[\AgdaEmptyExtraSkip]%
\>[0]\AgdaComment{--\ acceptingMaybeState\ \ :\ MaybeBStackStatePred}\<%
\\
\>[0]\AgdaComment{--\ acceptingMaybeState\ \ =\ acceptingState\ ⁺ᵇ}\<%
\\
\\[\AgdaEmptyExtraSkip]%
\\[\AgdaEmptyExtraSkip]%
\>[0]\AgdaFunction{acceptState}\AgdaSpace{}%
\AgdaSymbol{:}\AgdaSpace{}%
\AgdaFunction{StackStatePred}\<%
\\
\>[0]\AgdaFunction{acceptState}\AgdaSpace{}%
\AgdaSymbol{=}\AgdaSpace{}%
\AgdaFunction{stackPred2SPred}\AgdaSpace{}%
\AgdaFunction{acceptStateˢ}\<%
\\
\\[\AgdaEmptyExtraSkip]%
\>[0]\AgdaComment{--\ AcceptingMaybeState\ :\ Maybe\ StackState\ →\ Set}\<%
\\
\>[0]\AgdaComment{--\ AcceptingMaybeState\ s\ =\ True\ (acceptingMaybeState\ s)}\<%
\end{code}
} 


\AgdaHide{
\begin{code}%
\>[0]\AgdaKeyword{open}\AgdaSpace{}%
\AgdaKeyword{import}\AgdaSpace{}%
\AgdaModule{basicBitcoinDataType}\<%
\\
\\[\AgdaEmptyExtraSkip]%
\>[0]\AgdaKeyword{module}\AgdaSpace{}%
\AgdaModule{verificationStackScripts.stackVerificationP2PKHextractedProgram}\AgdaSpace{}%
\AgdaSymbol{(}\AgdaBound{param}\AgdaSpace{}%
\AgdaSymbol{:}\AgdaSpace{}%
\AgdaRecord{GlobalParameters}\AgdaSymbol{)}%
\>[99]\AgdaKeyword{where}\<%
\\
\\[\AgdaEmptyExtraSkip]%
\\[\AgdaEmptyExtraSkip]%
\\[\AgdaEmptyExtraSkip]%
\>[0]\AgdaKeyword{open}\AgdaSpace{}%
\AgdaKeyword{import}\AgdaSpace{}%
\AgdaModule{Data.Nat}%
\>[22]\AgdaKeyword{hiding}\AgdaSpace{}%
\AgdaSymbol{(}\AgdaOperator{\AgdaDatatype{\AgdaUnderscore{}≤\AgdaUnderscore{}}}\AgdaSymbol{)}\<%
\\
\>[0]\AgdaKeyword{open}\AgdaSpace{}%
\AgdaKeyword{import}\AgdaSpace{}%
\AgdaModule{Data.List}\AgdaSpace{}%
\AgdaKeyword{hiding}\AgdaSpace{}%
\AgdaSymbol{(}\AgdaOperator{\AgdaFunction{\AgdaUnderscore{}\ensuremath{+\!\!+}\AgdaUnderscore{}}}\AgdaSymbol{)}\<%
\\
\>[0]\AgdaKeyword{open}\AgdaSpace{}%
\AgdaKeyword{import}\AgdaSpace{}%
\AgdaModule{Data.Unit}%
\>[23]\AgdaKeyword{hiding}\AgdaSpace{}%
\AgdaSymbol{(}\AgdaOperator{\AgdaRecord{\AgdaUnderscore{}≤\AgdaUnderscore{}}}\AgdaSymbol{)}\<%
\\
\>[0]\AgdaKeyword{open}\AgdaSpace{}%
\AgdaKeyword{import}\AgdaSpace{}%
\AgdaModule{Data.Empty}\<%
\\
\>[0]\AgdaKeyword{open}\AgdaSpace{}%
\AgdaKeyword{import}\AgdaSpace{}%
\AgdaModule{Data.Sum}\<%
\\
\>[0]\AgdaKeyword{open}\AgdaSpace{}%
\AgdaKeyword{import}\AgdaSpace{}%
\AgdaModule{Data.Bool}%
\>[23]\AgdaKeyword{hiding}\AgdaSpace{}%
\AgdaSymbol{(}\AgdaOperator{\AgdaDatatype{\AgdaUnderscore{}≤\AgdaUnderscore{}}}\AgdaSpace{}%
\AgdaSymbol{;}\AgdaSpace{}%
\AgdaOperator{\AgdaFunction{if\AgdaUnderscore{}then\AgdaUnderscore{}else\AgdaUnderscore{}}}\AgdaSpace{}%
\AgdaSymbol{)}\AgdaSpace{}%
\AgdaKeyword{renaming}\AgdaSpace{}%
\AgdaSymbol{(}\AgdaOperator{\AgdaFunction{\AgdaUnderscore{}∧\AgdaUnderscore{}}}\AgdaSpace{}%
\AgdaSymbol{to}\AgdaSpace{}%
\AgdaOperator{\AgdaFunction{\AgdaUnderscore{}∧b\AgdaUnderscore{}}}\AgdaSpace{}%
\AgdaSymbol{;}\AgdaSpace{}%
\AgdaOperator{\AgdaFunction{\AgdaUnderscore{}∨\AgdaUnderscore{}}}\AgdaSpace{}%
\AgdaSymbol{to}\AgdaSpace{}%
\AgdaOperator{\AgdaFunction{\AgdaUnderscore{}∨b\AgdaUnderscore{}}}\AgdaSpace{}%
\AgdaSymbol{;}\AgdaSpace{}%
\AgdaFunction{T}\AgdaSpace{}%
\AgdaSymbol{to}\AgdaSpace{}%
\AgdaFunction{True}\AgdaSymbol{)}\<%
\\
\>[0]\AgdaKeyword{open}\AgdaSpace{}%
\AgdaKeyword{import}\AgdaSpace{}%
\AgdaModule{Data.Product}\AgdaSpace{}%
\AgdaKeyword{renaming}\AgdaSpace{}%
\AgdaSymbol{(}\AgdaOperator{\AgdaInductiveConstructor{\AgdaUnderscore{},\AgdaUnderscore{}}}\AgdaSpace{}%
\AgdaSymbol{to}\AgdaSpace{}%
\AgdaOperator{\AgdaInductiveConstructor{\AgdaUnderscore{},,\AgdaUnderscore{}}}\AgdaSpace{}%
\AgdaSymbol{)}\<%
\\
\>[0]\AgdaKeyword{open}\AgdaSpace{}%
\AgdaKeyword{import}\AgdaSpace{}%
\AgdaModule{Data.Nat.Base}\AgdaSpace{}%
\AgdaKeyword{hiding}\AgdaSpace{}%
\AgdaSymbol{(}\AgdaOperator{\AgdaDatatype{\AgdaUnderscore{}≤\AgdaUnderscore{}}}\AgdaSymbol{)}\<%
\\
\>[0]\AgdaKeyword{open}\AgdaSpace{}%
\AgdaKeyword{import}\AgdaSpace{}%
\AgdaModule{Data.List.NonEmpty}\AgdaSpace{}%
\AgdaKeyword{hiding}\AgdaSpace{}%
\AgdaSymbol{(}\AgdaField{head}\AgdaSpace{}%
\AgdaSymbol{;}\AgdaSpace{}%
\AgdaOperator{\AgdaFunction{[\AgdaUnderscore{}]}}\AgdaSpace{}%
\AgdaSymbol{)}\<%
\\
\>[0]\AgdaKeyword{open}\AgdaSpace{}%
\AgdaKeyword{import}\AgdaSpace{}%
\AgdaModule{Data.Maybe}\<%
\\
\>[0]\AgdaKeyword{open}\AgdaSpace{}%
\AgdaKeyword{import}\AgdaSpace{}%
\AgdaModule{Relation.Nullary}\<%
\\
\\[\AgdaEmptyExtraSkip]%
\>[0]\AgdaKeyword{import}\AgdaSpace{}%
\AgdaModule{Relation.Binary.PropositionalEquality}\AgdaSpace{}%
\AgdaSymbol{as}\AgdaSpace{}%
\AgdaModule{Eq}\<%
\\
\>[0]\AgdaKeyword{open}\AgdaSpace{}%
\AgdaModule{Eq}\AgdaSpace{}%
\AgdaKeyword{using}\AgdaSpace{}%
\AgdaSymbol{(}\AgdaOperator{\AgdaDatatype{\AgdaUnderscore{}≡\AgdaUnderscore{}}}\AgdaSymbol{;}\AgdaSpace{}%
\AgdaInductiveConstructor{refl}\AgdaSymbol{;}\AgdaSpace{}%
\AgdaFunction{cong}\AgdaSymbol{;}\AgdaSpace{}%
\AgdaKeyword{module}\AgdaSpace{}%
\AgdaModule{≡-Reasoning}\AgdaSymbol{;}\AgdaSpace{}%
\AgdaFunction{sym}\AgdaSymbol{)}\<%
\\
\>[0]\AgdaKeyword{open}\AgdaSpace{}%
\AgdaModule{≡-Reasoning}\<%
\\
\>[0]\AgdaKeyword{open}\AgdaSpace{}%
\AgdaKeyword{import}\AgdaSpace{}%
\AgdaModule{Agda.Builtin.Equality}\<%
\\
\\[\AgdaEmptyExtraSkip]%
\\[\AgdaEmptyExtraSkip]%
\>[0]\AgdaComment{--our\ libraries}\<%
\\
\>[0]\AgdaKeyword{open}\AgdaSpace{}%
\AgdaKeyword{import}\AgdaSpace{}%
\AgdaModule{libraries.listLib}\<%
\\
\>[0]\AgdaKeyword{open}\AgdaSpace{}%
\AgdaKeyword{import}\AgdaSpace{}%
\AgdaModule{libraries.equalityLib}\<%
\\
\>[0]\AgdaKeyword{open}\AgdaSpace{}%
\AgdaKeyword{import}\AgdaSpace{}%
\AgdaModule{libraries.natLib}\<%
\\
\>[0]\AgdaKeyword{open}\AgdaSpace{}%
\AgdaKeyword{import}\AgdaSpace{}%
\AgdaModule{libraries.boolLib}\<%
\\
\>[0]\AgdaKeyword{open}\AgdaSpace{}%
\AgdaKeyword{import}\AgdaSpace{}%
\AgdaModule{libraries.emptyLib}\<%
\\
\>[0]\AgdaKeyword{open}\AgdaSpace{}%
\AgdaKeyword{import}\AgdaSpace{}%
\AgdaModule{libraries.andLib}\<%
\\
\>[0]\AgdaKeyword{open}\AgdaSpace{}%
\AgdaKeyword{import}\AgdaSpace{}%
\AgdaModule{libraries.miscLib}\<%
\\
\>[0]\AgdaKeyword{open}\AgdaSpace{}%
\AgdaKeyword{import}\AgdaSpace{}%
\AgdaModule{libraries.maybeLib}\<%
\\
\\[\AgdaEmptyExtraSkip]%
\>[0]\AgdaKeyword{open}\AgdaSpace{}%
\AgdaKeyword{import}\AgdaSpace{}%
\AgdaModule{stack}\<%
\\
\>[0]\AgdaKeyword{open}\AgdaSpace{}%
\AgdaKeyword{import}\AgdaSpace{}%
\AgdaModule{stackPredicate}\<%
\\
\>[0]\AgdaKeyword{open}\AgdaSpace{}%
\AgdaKeyword{import}\AgdaSpace{}%
\AgdaModule{semanticBasicOperations}\AgdaSpace{}%
\AgdaBound{param}\<%
\\
\>[0]\AgdaKeyword{open}\AgdaSpace{}%
\AgdaKeyword{import}\AgdaSpace{}%
\AgdaModule{hoareTripleStack}\AgdaSpace{}%
\AgdaBound{param}\<%
\\
\>[0]\AgdaKeyword{open}\AgdaSpace{}%
\AgdaKeyword{import}\AgdaSpace{}%
\AgdaModule{instruction}\<%
\\
\>[0]\AgdaKeyword{open}\AgdaSpace{}%
\AgdaKeyword{import}\AgdaSpace{}%
\AgdaModule{stackSemanticsInstructions}\AgdaSpace{}%
\AgdaBound{param}\<%
\\
\>[0]\AgdaKeyword{open}\AgdaSpace{}%
\AgdaKeyword{import}\AgdaSpace{}%
\AgdaModule{verificationP2PKHbasic}\AgdaSpace{}%
\AgdaBound{param}\<%
\\
\\[\AgdaEmptyExtraSkip]%
\>[0]\AgdaKeyword{open}\AgdaSpace{}%
\AgdaKeyword{import}\AgdaSpace{}%
\AgdaModule{verificationStackScripts.stackState}\<%
\\
\>[0]\AgdaKeyword{open}\AgdaSpace{}%
\AgdaKeyword{import}\AgdaSpace{}%
\AgdaModule{verificationStackScripts.sPredicate}\<%
\\
\>[0]\AgdaKeyword{open}\AgdaSpace{}%
\AgdaKeyword{import}\AgdaSpace{}%
\AgdaModule{verificationWithIfStack.hoareTripleStackScript}\AgdaSpace{}%
\AgdaBound{param}\<%
\\
\>[0]\AgdaKeyword{open}\AgdaSpace{}%
\AgdaKeyword{import}\AgdaSpace{}%
\AgdaModule{verificationStackScripts.stackHoareTriple}\AgdaSpace{}%
\AgdaBound{param}\<%
\\
\>[0]\AgdaKeyword{open}\AgdaSpace{}%
\AgdaKeyword{import}\AgdaSpace{}%
\AgdaModule{verificationStackScripts.stackVerificationLemmas}\AgdaSpace{}%
\AgdaBound{param}\<%
\\
\>[0]\AgdaKeyword{open}\AgdaSpace{}%
\AgdaKeyword{import}\AgdaSpace{}%
\AgdaModule{verificationStackScripts.stackSemanticsInstructionsBasic}\AgdaSpace{}%
\AgdaBound{param}\<%
\\
\>[0]\AgdaKeyword{open}\AgdaSpace{}%
\AgdaKeyword{import}\AgdaSpace{}%
\AgdaModule{verificationStackScripts.semanticsStackInstructions}\AgdaSpace{}%
\AgdaBound{param}\<%
\\
\>[0]\AgdaKeyword{open}\AgdaSpace{}%
\AgdaKeyword{import}\AgdaSpace{}%
\AgdaModule{verificationStackScripts.stackVerificationP2PKH}\AgdaSpace{}%
\AgdaBound{param}\<%
\\
\>[0]\AgdaKeyword{open}\AgdaSpace{}%
\AgdaKeyword{import}\AgdaSpace{}%
\AgdaModule{verificationStackScripts.stackVerificationP2PKHindexed}\AgdaSpace{}%
\AgdaBound{param}\<%
\\
\>[0]\AgdaKeyword{open}\AgdaSpace{}%
\AgdaKeyword{import}\AgdaSpace{}%
\AgdaModule{verificationStackScripts.hoareTripleStackBasic}\AgdaSpace{}%
\AgdaBound{param}\<%
\\
\>[0]\AgdaKeyword{open}\AgdaSpace{}%
\AgdaKeyword{import}\AgdaSpace{}%
\AgdaModule{verificationStackScripts.stackVerificationLemmasPart2}\AgdaSpace{}%
\AgdaBound{param}\<%
\\
\\[\AgdaEmptyExtraSkip]%
\\[\AgdaEmptyExtraSkip]%
\>[0]\AgdaComment{\{-\ Now\ we\ have\ decoded\ the\ function\ -\}}\<%
\\
\\[\AgdaEmptyExtraSkip]%
\>[0]\AgdaComment{--\ p2pkhFunctionDecodedAux1\ should\ be\ equal\ to\ p2PKHNonEmptyStackAbstr\ (ignorning\ time)}\<%
\\
\>[0]\AgdaComment{--\ p2pkhFunctionDecoded\ should\ be\ equal\ to\ \ \ ⟦\ scriptP2PKHᵇ\ pubKeyHash\ ⟧ˢ\ time₁\ msg₁\ stack₁\ \ \ (where\ time\ is\ irrelevant)}\<%
\\
\\[\AgdaEmptyExtraSkip]%
\>[0]\AgdaKeyword{mutual}\<%
\end{code}
} 

\newcommand{\stackVerificationPtoPKHextractedProgramptopkhFunctionDecoded}{
\begin{code}%
\>[0][@{}l@{\AgdaIndent{1}}]%
\>[2]\AgdaFunction{p2pkhFunctionDecoded}\AgdaSpace{}%
\AgdaSymbol{:}\AgdaSpace{}%
\AgdaSymbol{(}\AgdaBound{pbkh}\AgdaSpace{}%
\AgdaSymbol{:}\AgdaSpace{}%
\AgdaDatatype{ℕ}\AgdaSymbol{)(}\AgdaBound{msg₁}\AgdaSpace{}%
\AgdaSymbol{:}\AgdaSpace{}%
\AgdaDatatype{Msg}\AgdaSymbol{)(}\AgdaBound{stack₁}\AgdaSpace{}%
\AgdaSymbol{:}\AgdaSpace{}%
\AgdaFunction{Stack}\AgdaSymbol{)}\AgdaSpace{}%
\AgdaSymbol{→}\AgdaSpace{}%
\AgdaDatatype{Maybe}\AgdaSpace{}%
\AgdaFunction{Stack}\<%
\\
\>[2]\AgdaFunction{p2pkhFunctionDecoded}%
\>[24]\AgdaBound{pbkh}%
\>[30]\AgdaBound{msg₁}%
\>[36]\AgdaInductiveConstructor{[]}%
\>[52]\AgdaSymbol{=}%
\>[55]\AgdaInductiveConstructor{nothing}\<%
\\
\>[2]\AgdaFunction{p2pkhFunctionDecoded}%
\>[24]\AgdaBound{pbkh}%
\>[30]\AgdaBound{msg₁}%
\>[36]\AgdaSymbol{(}\AgdaBound{pbk}\AgdaSpace{}%
\AgdaOperator{\AgdaInductiveConstructor{∷}}\AgdaSpace{}%
\AgdaBound{stack₁}\AgdaSymbol{)}%
\>[52]\AgdaSymbol{=}%
\>[55]\AgdaFunction{p2pkhFunctionDecodedAux1}\AgdaSpace{}%
\AgdaBound{pbk}\AgdaSpace{}%
\AgdaBound{msg₁}\AgdaSpace{}%
\AgdaBound{stack₁}\<%
\\
\>[55]\AgdaSymbol{(}\AgdaFunction{compareNaturals}\AgdaSpace{}%
\AgdaBound{pbkh}\AgdaSpace{}%
\AgdaSymbol{(}\AgdaFunction{hashFun}\AgdaSpace{}%
\AgdaBound{pbk}\AgdaSymbol{))}\<%
\\
\\[\AgdaEmptyExtraSkip]%
\>[2]\AgdaFunction{p2pkhFunctionDecodedAux1}\AgdaSpace{}%
\AgdaSymbol{:}\AgdaSpace{}%
\AgdaSymbol{(}\AgdaBound{pbk}\AgdaSpace{}%
\AgdaSymbol{:}\AgdaSpace{}%
\AgdaDatatype{ℕ}\AgdaSymbol{)(}\AgdaBound{msg₁}\AgdaSpace{}%
\AgdaSymbol{:}\AgdaSpace{}%
\AgdaDatatype{Msg}\AgdaSymbol{)(}\AgdaBound{stack₁}\AgdaSpace{}%
\AgdaSymbol{:}\AgdaSpace{}%
\AgdaFunction{Stack}\AgdaSymbol{)(}\AgdaBound{cpRes}\AgdaSpace{}%
\AgdaSymbol{:}\AgdaSpace{}%
\AgdaDatatype{ℕ}\AgdaSymbol{)}\AgdaSpace{}%
\AgdaSymbol{→}\AgdaSpace{}%
\AgdaDatatype{Maybe}\AgdaSpace{}%
\AgdaFunction{Stack}\<%
\\
\>[2]\AgdaFunction{p2pkhFunctionDecodedAux1}%
\>[28]\AgdaBound{pbk}%
\>[33]\AgdaBound{msg₁}%
\>[39]\AgdaInductiveConstructor{[]}%
\>[56]\AgdaBound{cpRes}%
\>[69]\AgdaSymbol{=}%
\>[72]\AgdaInductiveConstructor{nothing}\<%
\\
\>[2]\AgdaFunction{p2pkhFunctionDecodedAux1}%
\>[28]\AgdaBound{pbk}%
\>[33]\AgdaBound{msg₁}%
\>[39]\AgdaSymbol{(}\AgdaBound{sig₁}\AgdaSpace{}%
\AgdaOperator{\AgdaInductiveConstructor{∷}}\AgdaSpace{}%
\AgdaBound{stack₁}\AgdaSymbol{)}%
\>[56]\AgdaInductiveConstructor{zero}%
\>[69]\AgdaSymbol{=}%
\>[72]\AgdaInductiveConstructor{nothing}\<%
\\
\>[2]\AgdaFunction{p2pkhFunctionDecodedAux1}%
\>[28]\AgdaBound{pbk}%
\>[33]\AgdaBound{msg₁}%
\>[39]\AgdaSymbol{(}\AgdaBound{sig₁}\AgdaSpace{}%
\AgdaOperator{\AgdaInductiveConstructor{∷}}\AgdaSpace{}%
\AgdaBound{stack₁}\AgdaSymbol{)}%
\>[56]\AgdaSymbol{(}\AgdaInductiveConstructor{suc}\AgdaSpace{}%
\AgdaBound{cpRes}\AgdaSymbol{)}%
\>[69]\AgdaSymbol{=}\<%
\\
\>[28]\AgdaInductiveConstructor{just}\AgdaSpace{}%
\AgdaSymbol{(}\AgdaFunction{boolToNat}\AgdaSpace{}%
\AgdaSymbol{(}\AgdaFunction{isSigned}%
\>[55]\AgdaBound{msg₁}\AgdaSpace{}%
\AgdaBound{sig₁}\AgdaSpace{}%
\AgdaBound{pbk}\AgdaSymbol{)}\AgdaSpace{}%
\AgdaOperator{\AgdaInductiveConstructor{∷}}\AgdaSpace{}%
\AgdaBound{stack₁}\AgdaSymbol{)}\<%
\end{code}
} 

\AgdaHide{
\begin{code}%
\>[0]\<%
\end{code}
} 


\AgdaHide{
\begin{code}%
\>[0]\AgdaComment{--\ \textbackslash{}stackVerificationPtoPKHsymbolicExecution}\<%
\\
\>[0]\AgdaKeyword{open}\AgdaSpace{}%
\AgdaKeyword{import}\AgdaSpace{}%
\AgdaModule{basicBitcoinDataType}\<%
\\
\\[\AgdaEmptyExtraSkip]%
\>[0]\AgdaKeyword{module}\AgdaSpace{}%
\AgdaModule{paperTypes2021PostProceed.stackVerificationP2PKHsymbolicExecutionPaperVersion}\AgdaSpace{}%
\AgdaSymbol{(}\AgdaBound{param}\AgdaSpace{}%
\AgdaSymbol{:}\AgdaSpace{}%
\AgdaRecord{GlobalParameters}\AgdaSymbol{)}%
\>[113]\AgdaKeyword{where}\<%
\\
\\[\AgdaEmptyExtraSkip]%
\>[0]\AgdaKeyword{open}\AgdaSpace{}%
\AgdaKeyword{import}\AgdaSpace{}%
\AgdaModule{Data.Nat}%
\>[22]\AgdaKeyword{hiding}\AgdaSpace{}%
\AgdaSymbol{(}\AgdaOperator{\AgdaDatatype{\AgdaUnderscore{}≤\AgdaUnderscore{}}}\AgdaSymbol{)}\<%
\\
\>[0]\AgdaKeyword{open}\AgdaSpace{}%
\AgdaKeyword{import}\AgdaSpace{}%
\AgdaModule{Data.List}\AgdaSpace{}%
\AgdaKeyword{hiding}\AgdaSpace{}%
\AgdaSymbol{(}\AgdaOperator{\AgdaFunction{\AgdaUnderscore{}\ensuremath{+\!\!+}\AgdaUnderscore{}}}\AgdaSymbol{)}\<%
\\
\>[0]\AgdaKeyword{open}\AgdaSpace{}%
\AgdaKeyword{import}\AgdaSpace{}%
\AgdaModule{Data.Unit}%
\>[23]\AgdaKeyword{hiding}\AgdaSpace{}%
\AgdaSymbol{(}\AgdaOperator{\AgdaRecord{\AgdaUnderscore{}≤\AgdaUnderscore{}}}\AgdaSymbol{)}\<%
\\
\>[0]\AgdaKeyword{open}\AgdaSpace{}%
\AgdaKeyword{import}\AgdaSpace{}%
\AgdaModule{Data.Empty}\<%
\\
\>[0]\AgdaKeyword{open}\AgdaSpace{}%
\AgdaKeyword{import}\AgdaSpace{}%
\AgdaModule{Data.Sum}\<%
\\
\>[0]\AgdaKeyword{open}\AgdaSpace{}%
\AgdaKeyword{import}\AgdaSpace{}%
\AgdaModule{Data.Bool}%
\>[23]\AgdaKeyword{hiding}\AgdaSpace{}%
\AgdaSymbol{(}\AgdaOperator{\AgdaDatatype{\AgdaUnderscore{}≤\AgdaUnderscore{}}}\AgdaSpace{}%
\AgdaSymbol{;}\AgdaSpace{}%
\AgdaOperator{\AgdaFunction{if\AgdaUnderscore{}then\AgdaUnderscore{}else\AgdaUnderscore{}}}\AgdaSpace{}%
\AgdaSymbol{)}\AgdaSpace{}%
\AgdaKeyword{renaming}\AgdaSpace{}%
\AgdaSymbol{(}\AgdaOperator{\AgdaFunction{\AgdaUnderscore{}∧\AgdaUnderscore{}}}\AgdaSpace{}%
\AgdaSymbol{to}\AgdaSpace{}%
\AgdaOperator{\AgdaFunction{\AgdaUnderscore{}∧b\AgdaUnderscore{}}}\AgdaSpace{}%
\AgdaSymbol{;}\AgdaSpace{}%
\AgdaOperator{\AgdaFunction{\AgdaUnderscore{}∨\AgdaUnderscore{}}}\AgdaSpace{}%
\AgdaSymbol{to}\AgdaSpace{}%
\AgdaOperator{\AgdaFunction{\AgdaUnderscore{}∨b\AgdaUnderscore{}}}\AgdaSpace{}%
\AgdaSymbol{;}\AgdaSpace{}%
\AgdaFunction{T}\AgdaSpace{}%
\AgdaSymbol{to}\AgdaSpace{}%
\AgdaFunction{True}\AgdaSymbol{)}\<%
\\
\>[0]\AgdaKeyword{open}\AgdaSpace{}%
\AgdaKeyword{import}\AgdaSpace{}%
\AgdaModule{Data.Product}\AgdaSpace{}%
\AgdaKeyword{renaming}\AgdaSpace{}%
\AgdaSymbol{(}\AgdaOperator{\AgdaInductiveConstructor{\AgdaUnderscore{},\AgdaUnderscore{}}}\AgdaSpace{}%
\AgdaSymbol{to}\AgdaSpace{}%
\AgdaOperator{\AgdaInductiveConstructor{\AgdaUnderscore{},,\AgdaUnderscore{}}}\AgdaSpace{}%
\AgdaSymbol{)}\<%
\\
\>[0]\AgdaKeyword{open}\AgdaSpace{}%
\AgdaKeyword{import}\AgdaSpace{}%
\AgdaModule{Data.Nat.Base}\AgdaSpace{}%
\AgdaKeyword{hiding}\AgdaSpace{}%
\AgdaSymbol{(}\AgdaOperator{\AgdaDatatype{\AgdaUnderscore{}≤\AgdaUnderscore{}}}\AgdaSymbol{)}\<%
\\
\>[0]\AgdaKeyword{open}\AgdaSpace{}%
\AgdaKeyword{import}\AgdaSpace{}%
\AgdaModule{Data.List.NonEmpty}\AgdaSpace{}%
\AgdaKeyword{hiding}\AgdaSpace{}%
\AgdaSymbol{(}\AgdaField{head}\AgdaSpace{}%
\AgdaSymbol{;}\AgdaSpace{}%
\AgdaOperator{\AgdaFunction{[\AgdaUnderscore{}]}}\AgdaSpace{}%
\AgdaSymbol{)}\<%
\\
\>[0]\AgdaKeyword{open}\AgdaSpace{}%
\AgdaKeyword{import}\AgdaSpace{}%
\AgdaModule{Data.Maybe}\<%
\\
\>[0]\AgdaKeyword{open}\AgdaSpace{}%
\AgdaKeyword{import}\AgdaSpace{}%
\AgdaModule{Relation.Nullary}\<%
\\
\\[\AgdaEmptyExtraSkip]%
\>[0]\AgdaKeyword{import}\AgdaSpace{}%
\AgdaModule{Relation.Binary.PropositionalEquality}\AgdaSpace{}%
\AgdaSymbol{as}\AgdaSpace{}%
\AgdaModule{Eq}\<%
\\
\>[0]\AgdaKeyword{open}\AgdaSpace{}%
\AgdaModule{Eq}\AgdaSpace{}%
\AgdaKeyword{using}\AgdaSpace{}%
\AgdaSymbol{(}\AgdaOperator{\AgdaDatatype{\AgdaUnderscore{}≡\AgdaUnderscore{}}}\AgdaSymbol{;}\AgdaSpace{}%
\AgdaInductiveConstructor{refl}\AgdaSymbol{;}\AgdaSpace{}%
\AgdaFunction{cong}\AgdaSymbol{;}\AgdaSpace{}%
\AgdaKeyword{module}\AgdaSpace{}%
\AgdaModule{≡-Reasoning}\AgdaSymbol{;}\AgdaSpace{}%
\AgdaFunction{sym}\AgdaSymbol{)}\<%
\\
\>[0]\AgdaKeyword{open}\AgdaSpace{}%
\AgdaModule{≡-Reasoning}\<%
\\
\>[0]\AgdaKeyword{open}\AgdaSpace{}%
\AgdaKeyword{import}\AgdaSpace{}%
\AgdaModule{Agda.Builtin.Equality}\<%
\\
\\[\AgdaEmptyExtraSkip]%
\\[\AgdaEmptyExtraSkip]%
\>[0]\AgdaComment{--our\ libraries}\<%
\\
\>[0]\AgdaKeyword{open}\AgdaSpace{}%
\AgdaKeyword{import}\AgdaSpace{}%
\AgdaModule{libraries.listLib}\<%
\\
\>[0]\AgdaKeyword{open}\AgdaSpace{}%
\AgdaKeyword{import}\AgdaSpace{}%
\AgdaModule{libraries.equalityLib}\<%
\\
\>[0]\AgdaKeyword{open}\AgdaSpace{}%
\AgdaKeyword{import}\AgdaSpace{}%
\AgdaModule{libraries.natLib}\<%
\\
\>[0]\AgdaKeyword{open}\AgdaSpace{}%
\AgdaKeyword{import}\AgdaSpace{}%
\AgdaModule{libraries.boolLib}\<%
\\
\>[0]\AgdaKeyword{open}\AgdaSpace{}%
\AgdaKeyword{import}\AgdaSpace{}%
\AgdaModule{libraries.emptyLib}\<%
\\
\>[0]\AgdaKeyword{open}\AgdaSpace{}%
\AgdaKeyword{import}\AgdaSpace{}%
\AgdaModule{libraries.andLib}\<%
\\
\>[0]\AgdaKeyword{open}\AgdaSpace{}%
\AgdaKeyword{import}\AgdaSpace{}%
\AgdaModule{libraries.miscLib}\<%
\\
\>[0]\AgdaKeyword{open}\AgdaSpace{}%
\AgdaKeyword{import}\AgdaSpace{}%
\AgdaModule{libraries.maybeLib}\<%
\\
\\[\AgdaEmptyExtraSkip]%
\>[0]\AgdaKeyword{open}\AgdaSpace{}%
\AgdaKeyword{import}\AgdaSpace{}%
\AgdaModule{stack}\<%
\\
\>[0]\AgdaKeyword{open}\AgdaSpace{}%
\AgdaKeyword{import}\AgdaSpace{}%
\AgdaModule{stackPredicate}\<%
\\
\>[0]\AgdaKeyword{open}\AgdaSpace{}%
\AgdaKeyword{import}\AgdaSpace{}%
\AgdaModule{semanticBasicOperations}\AgdaSpace{}%
\AgdaBound{param}\<%
\\
\>[0]\AgdaKeyword{open}\AgdaSpace{}%
\AgdaKeyword{import}\AgdaSpace{}%
\AgdaModule{hoareTripleStack}\AgdaSpace{}%
\AgdaBound{param}\<%
\\
\>[0]\AgdaKeyword{open}\AgdaSpace{}%
\AgdaKeyword{import}\AgdaSpace{}%
\AgdaModule{instruction}\<%
\\
\>[0]\AgdaKeyword{open}\AgdaSpace{}%
\AgdaKeyword{import}\AgdaSpace{}%
\AgdaModule{verificationStackScripts.stackState}\<%
\\
\>[0]\AgdaKeyword{open}\AgdaSpace{}%
\AgdaKeyword{import}\AgdaSpace{}%
\AgdaModule{verificationStackScripts.sPredicate}\<%
\\
\>[0]\AgdaKeyword{open}\AgdaSpace{}%
\AgdaKeyword{import}\AgdaSpace{}%
\AgdaModule{verificationStackScripts.stackHoareTriple}\AgdaSpace{}%
\AgdaBound{param}\<%
\\
\>[0]\AgdaKeyword{open}\AgdaSpace{}%
\AgdaKeyword{import}\AgdaSpace{}%
\AgdaModule{verificationStackScripts.stackVerificationLemmas}\AgdaSpace{}%
\AgdaBound{param}\<%
\\
\>[0]\AgdaKeyword{open}\AgdaSpace{}%
\AgdaKeyword{import}\AgdaSpace{}%
\AgdaModule{verificationStackScripts.stackSemanticsInstructionsBasic}\AgdaSpace{}%
\AgdaBound{param}\<%
\\
\>[0]\AgdaKeyword{open}\AgdaSpace{}%
\AgdaKeyword{import}\AgdaSpace{}%
\AgdaModule{verificationStackScripts.semanticsStackInstructions}\AgdaSpace{}%
\AgdaBound{param}\<%
\\
\>[0]\AgdaKeyword{open}\AgdaSpace{}%
\AgdaKeyword{import}\AgdaSpace{}%
\AgdaModule{verificationStackScripts.stackVerificationP2PKH}\AgdaSpace{}%
\AgdaBound{param}\<%
\\
\>[0]\AgdaKeyword{open}\AgdaSpace{}%
\AgdaKeyword{import}\AgdaSpace{}%
\AgdaModule{verificationStackScripts.stackVerificationP2PKHindexed}\AgdaSpace{}%
\AgdaBound{param}\<%
\\
\\[\AgdaEmptyExtraSkip]%
\>[0]\AgdaComment{-------------------------------------------------------------}\<%
\\
\>[0]\AgdaComment{--\ \ \ This\ file\ explores\ the\ symoblic\ execution\ of\ the\ P2PKH\ program}\<%
\\
\>[0]\AgdaComment{--\ \ \ \ \ in\ order\ to\ determine\ the\ case\ distinction}\<%
\\
\>[0]\AgdaComment{--\ \ \ \ \ and\ extract\ a\ program\ from\ it}\<%
\\
\>[0]\AgdaComment{--}\<%
\\
\>[0]\AgdaComment{--\ \ \ This\ is\ done\ by\ postulating\ parameters\ and\ applying\ ⟦\ scriptP2PKHᵇ\ pbkh\ ⟧ˢ}\<%
\\
\>[0]\AgdaComment{--\ \ \ \ \ to\ parameters}\<%
\\
\>[0]\AgdaComment{--------------------------------------------------------------------------------}\<%
\\
\\[\AgdaEmptyExtraSkip]%
\>[0]\AgdaKeyword{private}\<%
\\
\>[0][@{}l@{\AgdaIndent{0}}]%
\>[2]\AgdaKeyword{postulate}\AgdaSpace{}%
\AgdaPostulate{time₁}\AgdaSpace{}%
\AgdaSymbol{:}\AgdaSpace{}%
\AgdaFunction{Time}\<%
\\
\>[2]\AgdaKeyword{postulate}\AgdaSpace{}%
\AgdaPostulate{msg₁}\AgdaSpace{}%
\AgdaSymbol{:}\AgdaSpace{}%
\AgdaDatatype{Msg}\<%
\\
\>[2]\AgdaKeyword{postulate}\AgdaSpace{}%
\AgdaPostulate{stack₁}\AgdaSpace{}%
\AgdaSymbol{:}\AgdaSpace{}%
\AgdaFunction{Stack}\<%
\\
\>[2]\AgdaKeyword{postulate}\AgdaSpace{}%
\AgdaPostulate{pbkh}\AgdaSpace{}%
\AgdaSymbol{:}\AgdaSpace{}%
\AgdaDatatype{ℕ}\<%
\\
\>[2]\AgdaKeyword{postulate}\AgdaSpace{}%
\AgdaPostulate{pbk}\AgdaSpace{}%
\AgdaSymbol{:}\AgdaSpace{}%
\AgdaDatatype{ℕ}\<%
\\
\>[2]\AgdaKeyword{postulate}\AgdaSpace{}%
\AgdaPostulate{x₁}\AgdaSpace{}%
\AgdaSymbol{:}\AgdaSpace{}%
\AgdaDatatype{ℕ}\<%
\\
\>[2]\AgdaKeyword{postulate}\AgdaSpace{}%
\AgdaPostulate{sig₁}\AgdaSpace{}%
\AgdaSymbol{:}\AgdaSpace{}%
\AgdaDatatype{ℕ}\<%
\\
\\[\AgdaEmptyExtraSkip]%
\\[\AgdaEmptyExtraSkip]%
\>[0]\AgdaComment{\{-\ We\ first\ create\ a\ symbolic\ execution\ of\ the\ scriptP2PKHᵇ\ pbkh\ to\ see\ what\ kind}\<%
\\
\>[0]\AgdaComment{\ \ \ of\ case\ distinction\ happens\ -\}}\<%
\\
\\[\AgdaEmptyExtraSkip]%
\>[0]\AgdaFunction{check}\AgdaSpace{}%
\AgdaSymbol{=}\AgdaSpace{}%
\AgdaFunction{scriptP2PKHᵇ}\<%
\\
\\[\AgdaEmptyExtraSkip]%
\>[0]\AgdaFunction{testP2PKHscript}\AgdaSpace{}%
\AgdaSymbol{:}\AgdaSpace{}%
\AgdaDatatype{Maybe}\AgdaSpace{}%
\AgdaFunction{Stack}\<%
\\
\>[0]\AgdaFunction{testP2PKHscript}\AgdaSpace{}%
\AgdaSymbol{=}\<%
\\
\>[0]\AgdaComment{--\ \textbackslash{}stackVerificationPtoPKHsymbolicExecutiontestPtoP}\<%
\end{code}
} 

\newcommand{\stackVerificationPtoPKHsymbolicExecutiontestPtoP}{
\begin{code}[inline]%
\>[0][@{}l@{\AgdaIndent{1}}]%
\>[1]\AgdaOperator{\AgdaFunction{⟦}}\AgdaSpace{}%
\AgdaFunction{scriptP2PKHᵇ}\AgdaSpace{}%
\AgdaPostulate{pbkh}\AgdaSpace{}%
\AgdaOperator{\AgdaFunction{⟧ˢ}}\AgdaSpace{}%
\AgdaPostulate{time₁}\AgdaSpace{}%
\AgdaPostulate{msg₁}\AgdaSpace{}%
\AgdaPostulate{stack₁}\<%
\end{code}
} 

\AgdaHide{
\begin{code}%
\>[0]\<%
\\
\\[\AgdaEmptyExtraSkip]%
\\[\AgdaEmptyExtraSkip]%
\>[0]\AgdaComment{--⟦\ scriptP2PKHᵇ\ pbkh\ ⟧ˢ\ time₁\ msg₁\ stack}\<%
\\
\\[\AgdaEmptyExtraSkip]%
\>[0]\AgdaComment{\{-\ evaluation\ gives}\<%
\\
\>[0]\<%
\\
\>[0]\AgdaComment{executeStackDup\ stack₁\ \ensuremath{\mathbin{{>}\mkern-8.5mu{>}\mkern-2mu{=}}}}\<%
\\
\>[0]\AgdaComment{(λ\ stack₂\ →}\<%
\\
\>[0]\AgdaComment{\ \ \ executeOpHash\ stack₂\ \ensuremath{\mathbin{{>}\mkern-8.5mu{>}\mkern-2mu{=}}}}\<%
\\
\>[0]\AgdaComment{\ \ \ (λ\ stack₃\ →}\<%
\\
\>[0]\AgdaComment{\ \ \ \ \ \ executeStackEquality\ (pbkh\ ∷\ stack₃)\ \ensuremath{\mathbin{{>}\mkern-8.5mu{>}\mkern-2mu{=}}}}\<%
\\
\>[0]\AgdaComment{\ \ \ \ \ \ (λ\ stack₄\ →}\<%
\\
\>[0]\AgdaComment{\ \ \ \ \ \ \ \ \ executeStackVerify\ stack₄\ \ensuremath{\mathbin{{>}\mkern-8.5mu{>}\mkern-2mu{=}}}}\<%
\\
\>[0]\AgdaComment{\ \ \ \ \ \ \ \ \ (λ\ stack₅\ →\ executeStackCheckSig\ msg₁\ stack₅))))}\<%
\\
\>[0]\<%
\\
\>[0]\AgdaComment{Improved\ layout}\<%
\\
\>[0]\AgdaComment{executeStackDup\ stack₁\ \ensuremath{\mathbin{{>}\mkern-8.5mu{>}\mkern-2mu{=}}}\ (λ\ stack₂\ →\ \ executeOpHash\ stack₂\ \ensuremath{\mathbin{{>}\mkern-8.5mu{>}\mkern-2mu{=}}}}\<%
\\
\>[0]\AgdaComment{\ \ \ (λ\ stack₃\ →\ \ executeStackEquality\ (pbkh\ ∷\ stack₃)\ \ensuremath{\mathbin{{>}\mkern-8.5mu{>}\mkern-2mu{=}}}}\<%
\\
\>[0]\AgdaComment{\ \ \ (λ\ stack₄\ →\ \ executeStackVerify\ stack₄\ \ensuremath{\mathbin{{>}\mkern-8.5mu{>}\mkern-2mu{=}}}}\<%
\\
\>[0]\AgdaComment{\ \ \ (λ\ stack₅\ →\ executeStackCheckSig\ msg₁\ stack₅))))}\<%
\\
\>[0]\<%
\\
\>[0]\<%
\\
\>[0]\<%
\\
\>[0]\AgdaComment{--\ old\ version\ was}\<%
\\
\>[0]\<%
\\
\>[0]\AgdaComment{lift2Maybe}\<%
\\
\>[0]\AgdaComment{(λ\ stack₂\ →}\<%
\\
\>[0]\AgdaComment{\ \ \ lift2Maybe}\<%
\\
\>[0]\AgdaComment{\ \ \ (λ\ stack₃\ →}\<%
\\
\>[0]\AgdaComment{\ \ \ \ \ \ lift2Maybe}\<%
\\
\>[0]\AgdaComment{\ \ \ \ \ \ (λ\ stack₄\ →}\<%
\\
\>[0]\AgdaComment{\ \ \ \ \ \ \ \ \ lift2Maybe\ (λ\ stack₅\ →\ executeStackCheckSig\ msg₁\ stack₅)}\<%
\\
\>[0]\AgdaComment{\ \ \ \ \ \ \ \ \ (executeStackVerify\ stack₄))}\<%
\\
\>[0]\AgdaComment{\ \ \ \ \ \ (executeStackEquality\ (pbkh\ ∷\ stack₃)))}\<%
\\
\>[0]\AgdaComment{\ \ \ (executeOpHash\ stack₂))}\<%
\\
\>[0]\AgdaComment{(executeStackDup\ stack₁)}\<%
\\
\>[0]\<%
\\
\>[0]\AgdaComment{-\}}\<%
\\
\\[\AgdaEmptyExtraSkip]%
\>[0]\AgdaComment{--\ We\ define\ a\ term\ giving\ the\ result\ of\ the\ evaluation}\<%
\\
\\[\AgdaEmptyExtraSkip]%
\>[0]\AgdaFunction{testP2PKHscript2}\AgdaSpace{}%
\AgdaSymbol{:}\AgdaSpace{}%
\AgdaDatatype{Maybe}\AgdaSpace{}%
\AgdaFunction{Stack}\<%
\\
\>[0]\AgdaFunction{testP2PKHscript2}\AgdaSpace{}%
\AgdaSymbol{=}\<%
\\
\>[0]\AgdaComment{--\ \textbackslash{}stackVerificationPtoPKHsymbolicExecutiontestPtoPKHscripttwo}\<%
\end{code}
} 

\newcommand{\stackVerificationPtoPKHsymbolicExecutiontestPtoPKHscripttwo}{
\begin{code}%
\>[0][@{}l@{\AgdaIndent{1}}]%
\>[1]\AgdaFunction{executeStackDup}\AgdaSpace{}%
\AgdaPostulate{stack₁}%
\>[39]\AgdaOperator{\AgdaFunction{\ensuremath{\mathbin{{>}\mkern-8.5mu{>}\mkern-2mu{=}}}}}%
\>[44]\AgdaSymbol{λ}\AgdaSpace{}%
\AgdaBound{stack₂}\AgdaSpace{}%
\AgdaSymbol{→}%
\>[56]\AgdaFunction{executeOpHash}\AgdaSpace{}%
\AgdaBound{stack₂}%
\>[83]\AgdaOperator{\AgdaFunction{\ensuremath{\mathbin{{>}\mkern-8.5mu{>}\mkern-2mu{=}}}}}\AgdaSpace{}%
\AgdaSymbol{λ}\AgdaSpace{}%
\AgdaBound{stack₃}\AgdaSpace{}%
\AgdaSymbol{→}\<%
\\
\>[1]\AgdaFunction{executeStackEquality}\AgdaSpace{}%
\AgdaSymbol{(}\AgdaPostulate{pbkh}\AgdaSpace{}%
\AgdaOperator{\AgdaInductiveConstructor{∷}}\AgdaSpace{}%
\AgdaBound{stack₃}\AgdaSymbol{)}%
\>[39]\AgdaOperator{\AgdaFunction{\ensuremath{\mathbin{{>}\mkern-8.5mu{>}\mkern-2mu{=}}}}}%
\>[44]\AgdaSymbol{λ}\AgdaSpace{}%
\AgdaBound{stack₄}\AgdaSpace{}%
\AgdaSymbol{→}%
\>[56]\AgdaFunction{executeStackVerify}\AgdaSpace{}%
\AgdaBound{stack₄}%
\>[83]\AgdaOperator{\AgdaFunction{\ensuremath{\mathbin{{>}\mkern-8.5mu{>}\mkern-2mu{=}}}}}\AgdaSpace{}%
\AgdaSymbol{λ}\AgdaSpace{}%
\AgdaBound{stack₅}\AgdaSpace{}%
\AgdaSymbol{→}\<%
\\
\>[1]\AgdaFunction{executeStackCheckSig}\AgdaSpace{}%
\AgdaPostulate{msg₁}\AgdaSpace{}%
\AgdaBound{stack₅}\<%
\end{code}
} 

\AgdaHide{
\begin{code}%
\>[0]\<%
\\
\>[0]\AgdaFunction{testP2PKHscript2UsingMoreSpace}\AgdaSpace{}%
\AgdaSymbol{=}\<%
\\
\>[0]\AgdaComment{--\ \textbackslash{}stackVerificationPtoPKHsymbolicExecutiontestPtoPKHscripttwoUsingMoreSpace}\<%
\end{code}
} 

\newcommand{\stackVerificationPtoPKHsymbolicExecutiontestPtoPKHscripttwoUsingMoreSpace}{
\begin{code}%
\>[0][@{}l@{\AgdaIndent{1}}]%
\>[1]\AgdaFunction{executeStackDup}\AgdaSpace{}%
\AgdaPostulate{stack₁}%
\>[52]\AgdaOperator{\AgdaFunction{\ensuremath{\mathbin{{>}\mkern-8.5mu{>}\mkern-2mu{=}}}}}\<%
\\
\>[1]\AgdaSymbol{λ}\AgdaSpace{}%
\AgdaBound{stack₂}\AgdaSpace{}%
\AgdaSymbol{→}%
\>[13]\AgdaFunction{executeOpHash}\AgdaSpace{}%
\AgdaBound{stack₂}%
\>[52]\AgdaOperator{\AgdaFunction{\ensuremath{\mathbin{{>}\mkern-8.5mu{>}\mkern-2mu{=}}}}}\<%
\\
\>[1]\AgdaSymbol{λ}\AgdaSpace{}%
\AgdaBound{stack₃}\AgdaSpace{}%
\AgdaSymbol{→}%
\>[13]\AgdaFunction{executeStackEquality}\AgdaSpace{}%
\AgdaSymbol{(}\AgdaPostulate{pbkh}\AgdaSpace{}%
\AgdaOperator{\AgdaInductiveConstructor{∷}}\AgdaSpace{}%
\AgdaBound{stack₃}\AgdaSymbol{)}%
\>[52]\AgdaOperator{\AgdaFunction{\ensuremath{\mathbin{{>}\mkern-8.5mu{>}\mkern-2mu{=}}}}}\<%
\\
\>[1]\AgdaSymbol{λ}\AgdaSpace{}%
\AgdaBound{stack₄}\AgdaSpace{}%
\AgdaSymbol{→}%
\>[13]\AgdaFunction{executeStackVerify}\AgdaSpace{}%
\AgdaBound{stack₄}%
\>[52]\AgdaOperator{\AgdaFunction{\ensuremath{\mathbin{{>}\mkern-8.5mu{>}\mkern-2mu{=}}}}}\<%
\\
\>[1]\AgdaSymbol{λ}\AgdaSpace{}%
\AgdaBound{stack₅}\AgdaSpace{}%
\AgdaSymbol{→}%
\>[13]\AgdaFunction{executeStackCheckSig}\AgdaSpace{}%
\AgdaPostulate{msg₁}\AgdaSpace{}%
\AgdaBound{stack₅}\<%
\end{code}
} 

\AgdaHide{
\begin{code}%
\>[0]\<%
\\
\\[\AgdaEmptyExtraSkip]%
\>[0]\AgdaFunction{testP2PKHscript2UsingMoreSpaceUsingDo}\AgdaSpace{}%
\AgdaSymbol{=}\<%
\\
\>[0]\AgdaComment{--\ \textbackslash{}stackVerificationPtoPKHsymbolicExecutiontestPtoPKHscripttwoUsingDo}\<%
\end{code}
} 

\newcommand{\stackVerificationPtoPKHsymbolicExecutiontestPtoPKHscripttwoUsingDo}{
\begin{code}%
\>[0][@{}l@{\AgdaIndent{1}}]%
\>[1]\AgdaKeyword{do}%
\>[5]\AgdaBound{stack₂}\AgdaSpace{}%
\AgdaOperator{\AgdaFunction{←}}\AgdaSpace{}%
\AgdaFunction{executeStackDup}\AgdaSpace{}%
\AgdaPostulate{stack₁}\<%
\\
\>[5]\AgdaBound{stack₃}\AgdaSpace{}%
\AgdaOperator{\AgdaFunction{←}}\AgdaSpace{}%
\AgdaFunction{executeOpHash}\AgdaSpace{}%
\AgdaBound{stack₂}\<%
\\
\>[5]\AgdaBound{stack₄}\AgdaSpace{}%
\AgdaOperator{\AgdaFunction{←}}\AgdaSpace{}%
\AgdaFunction{executeStackEquality}\AgdaSpace{}%
\AgdaSymbol{(}\AgdaPostulate{pbkh}\AgdaSpace{}%
\AgdaOperator{\AgdaInductiveConstructor{∷}}\AgdaSpace{}%
\AgdaBound{stack₃}\AgdaSymbol{)}\<%
\\
\>[5]\AgdaBound{stack₅}\AgdaSpace{}%
\AgdaOperator{\AgdaFunction{←}}\AgdaSpace{}%
\AgdaFunction{executeStackVerify}\AgdaSpace{}%
\AgdaBound{stack₄}\<%
\\
\>[5]\AgdaFunction{executeStackCheckSig}\AgdaSpace{}%
\AgdaPostulate{msg₁}\AgdaSpace{}%
\AgdaBound{stack₅}\<%
\end{code}
} 

\AgdaHide{
\begin{code}%
\>[0]\<%
\\
\\[\AgdaEmptyExtraSkip]%
\>[0]\AgdaComment{\{-\ in\ imperative\ programming\ we\ would\ write}\<%
\\
\>[0]\<%
\\
\>[0]\AgdaComment{\ \ \ stack₂\ :=\ executeStackDup\ stack₁;}\<%
\\
\>[0]\AgdaComment{\ \ \ stack₃\ :=\ executeOpHash\ stack₂;}\<%
\\
\>[0]\AgdaComment{\ \ \ stack₄\ :=\ executeStackEquality\ (pbkh\ ∷\ stack₃);}\<%
\\
\>[0]\AgdaComment{\ \ \ stack₅\ :=\ executeStackVerify\ stack₄;}\<%
\\
\>[0]\AgdaComment{\ \ \ executeStackCheckSig\ msg₁\ stack₅;}\<%
\\
\>[0]\<%
\\
\>[0]\AgdaComment{-\}}\<%
\\
\\[\AgdaEmptyExtraSkip]%
\\[\AgdaEmptyExtraSkip]%
\\[\AgdaEmptyExtraSkip]%
\\[\AgdaEmptyExtraSkip]%
\\[\AgdaEmptyExtraSkip]%
\>[0]\AgdaComment{\{-}\<%
\\
\>[0]\AgdaComment{If\ we\ execute\ the\ first\ line}\<%
\\
\>[0]\AgdaComment{(executeStackDup\ stack₁)}\<%
\\
\>[0]\<%
\\
\>[0]\AgdaComment{we\ see\ it\ will\ give}\<%
\\
\>[0]\AgdaComment{nothing\ if\ stack₁\ =\ []}\<%
\\
\>[0]\AgdaComment{just\ something\ if\ stack₁\ is\ nonempty}\<%
\\
\>[0]\<%
\\
\>[0]\AgdaComment{So\ let's\ check\ what\ happens\ if\ stack₁\ =\ []}\<%
\\
\>[0]\AgdaComment{-\}}\<%
\\
\\[\AgdaEmptyExtraSkip]%
\\[\AgdaEmptyExtraSkip]%
\>[0]\AgdaFunction{testP2PKHscriptEmpty}\AgdaSpace{}%
\AgdaSymbol{:}\AgdaSpace{}%
\AgdaDatatype{Maybe}\AgdaSpace{}%
\AgdaFunction{Stack}\<%
\\
\>[0]\AgdaFunction{testP2PKHscriptEmpty}\AgdaSpace{}%
\AgdaSymbol{=}\<%
\\
\>[0]\AgdaComment{--\ \textbackslash{}stackVerificationPtoPKHsymbolicExecutionEmpty}\<%
\end{code}
} 

\newcommand{\stackVerificationPtoPKHsymbolicExecutionEmpty}{
\begin{code}[inline]%
\>[0][@{}l@{\AgdaIndent{1}}]%
\>[1]\AgdaOperator{\AgdaFunction{⟦}}\AgdaSpace{}%
\AgdaFunction{scriptP2PKHᵇ}\AgdaSpace{}%
\AgdaPostulate{pbkh}\AgdaSpace{}%
\AgdaOperator{\AgdaFunction{⟧ˢ}}\AgdaSpace{}%
\AgdaPostulate{time₁}\AgdaSpace{}%
\AgdaPostulate{msg₁}\AgdaSpace{}%
\AgdaInductiveConstructor{[]}\<%
\end{code}
} 

\AgdaHide{
\begin{code}%
\>[0]\<%
\\
\\[\AgdaEmptyExtraSkip]%
\>[0]\AgdaComment{--⟦\ scriptP2PKHᵇ\ pbkh\ ⟧ˢ\ time₁\ msg₁\ []}\<%
\\
\\[\AgdaEmptyExtraSkip]%
\\[\AgdaEmptyExtraSkip]%
\\[\AgdaEmptyExtraSkip]%
\>[0]\AgdaComment{\{-\ if\ we\ evaluate\ it\ we\ get:}\<%
\\
\>[0]\<%
\\
\>[0]\AgdaComment{nothing}\<%
\\
\>[0]\<%
\\
\>[0]\AgdaComment{So\ now\ get\ the\ first\ (trivial)\ theorem}\<%
\\
\>[0]\AgdaComment{(without\ the\ postulated\ parameters)}\<%
\\
\>[0]\AgdaComment{-\}}\<%
\\
\\[\AgdaEmptyExtraSkip]%
\>[0]\<%
\end{code}
} 

\newcommand{\stackVerificationPtoPKHsymbolicExecutionnothing}{
\begin{code}%
\>[0]\AgdaFunction{stackfunP2PKHemptyIsNothing}\AgdaSpace{}%
\AgdaSymbol{:}%
\>[224I]\AgdaSymbol{(}\AgdaBound{pubKeyHash}\AgdaSpace{}%
\AgdaSymbol{:}\AgdaSpace{}%
\AgdaDatatype{ℕ}\AgdaSymbol{)(}\AgdaBound{time₁}\AgdaSpace{}%
\AgdaSymbol{:}\AgdaSpace{}%
\AgdaFunction{Time}\AgdaSymbol{)(}\AgdaBound{msg₁}\AgdaSpace{}%
\AgdaSymbol{:}\AgdaSpace{}%
\AgdaDatatype{Msg}\AgdaSymbol{)}\<%
\\
\>[.][@{}l@{}]\<[224I]%
\>[30]\AgdaSymbol{→}\AgdaSpace{}%
\AgdaOperator{\AgdaFunction{⟦}}\AgdaSpace{}%
\AgdaFunction{scriptP2PKHᵇ}\AgdaSpace{}%
\AgdaBound{pubKeyHash}\AgdaSpace{}%
\AgdaOperator{\AgdaFunction{⟧ˢ}}\AgdaSpace{}%
\AgdaBound{time₁}\AgdaSpace{}%
\AgdaBound{msg₁}\AgdaSpace{}%
\AgdaInductiveConstructor{[]}\AgdaSpace{}%
\AgdaOperator{\AgdaDatatype{≡}}\AgdaSpace{}%
\AgdaInductiveConstructor{nothing}\<%
\\
\>[0]\AgdaFunction{stackfunP2PKHemptyIsNothing}\AgdaSpace{}%
\AgdaBound{pubKeyHash}\AgdaSpace{}%
\AgdaBound{time₁}\AgdaSpace{}%
\AgdaBound{msg₁}\AgdaSpace{}%
\AgdaSymbol{=}\AgdaSpace{}%
\AgdaInductiveConstructor{refl}\<%
\end{code}
} 

\AgdaHide{
\begin{code}%
\>[0]\<%
\\
\\[\AgdaEmptyExtraSkip]%
\\[\AgdaEmptyExtraSkip]%
\>[0]\AgdaComment{\{-\ Now\ we\ look\ at\ what\ happens\ if\ the\ stack\ is\ non\ empty}\<%
\\
\>[0]\<%
\\
\>[0]\AgdaComment{lets\ a\ test\ for\ symbolic\ execution\ -\}}\<%
\\
\\[\AgdaEmptyExtraSkip]%
\\[\AgdaEmptyExtraSkip]%
\>[0]\AgdaFunction{teststackfunP2PKHNonEmptyStack}\AgdaSpace{}%
\AgdaSymbol{:}%
\>[34]\AgdaDatatype{Maybe}\AgdaSpace{}%
\AgdaFunction{Stack}\<%
\\
\>[0]\AgdaFunction{teststackfunP2PKHNonEmptyStack}\AgdaSpace{}%
\AgdaSymbol{=}\<%
\\
\>[0]\AgdaComment{--\ \textbackslash{}stackVerificationPtoPKHsymbolicExecutionnonestack}\<%
\end{code}
} 

\newcommand{\stackVerificationPtoPKHsymbolicExecutionnonestack}{
\begin{code}[inline]%
\>[0][@{}l@{\AgdaIndent{1}}]%
\>[2]\AgdaOperator{\AgdaFunction{⟦}}\AgdaSpace{}%
\AgdaFunction{scriptP2PKHᵇ}\AgdaSpace{}%
\AgdaPostulate{pbkh}\AgdaSpace{}%
\AgdaOperator{\AgdaFunction{⟧ˢ}}\AgdaSpace{}%
\AgdaPostulate{time₁}\AgdaSpace{}%
\AgdaPostulate{msg₁}\AgdaSpace{}%
\AgdaSymbol{(}\AgdaPostulate{pbk}\AgdaSpace{}%
\AgdaOperator{\AgdaInductiveConstructor{∷}}\AgdaSpace{}%
\AgdaPostulate{stack₁}\AgdaSymbol{)}\<%
\end{code}
} 

\AgdaHide{
\begin{code}%
\>[0]\<%
\\
\>[0]\AgdaComment{\{-\ If\ we\ compute\ it\ we\ get}\<%
\\
\>[0]\AgdaComment{executeStackVerify\ (compareNaturals\ pbkh\ (param\ .hash\ pbk)\ ∷\ pbk\ ∷\ stack₁)}\<%
\\
\>[0]\AgdaComment{\ensuremath{\mathbin{{>}\mkern-8.5mu{>}\mkern-2mu{=}}}\ (λ\ stack₂\ →\ executeStackCheckSig\ msg₁\ stack₂)}\<%
\\
\>[0]\<%
\\
\>[0]\AgdaComment{--\ older\ version\ was}\<%
\\
\>[0]\AgdaComment{lift2Maybe\ (λ\ stack₂\ →\ executeStackCheckSig\ msg₁\ stack₂)}\<%
\\
\>[0]\AgdaComment{(executeStackVerify}\<%
\\
\>[0]\AgdaComment{\ (compareNaturals\ pbkh\ (hashFun\ pbk)\ ∷\ pbk\ ∷\ stack₁))}\<%
\\
\>[0]\AgdaComment{-\}}\<%
\\
\\[\AgdaEmptyExtraSkip]%
\>[0]\AgdaFunction{stackfunP2PKHNonEmptyStackNormalForm}\AgdaSpace{}%
\AgdaSymbol{:}%
\>[40]\AgdaDatatype{Maybe}\AgdaSpace{}%
\AgdaFunction{Stack}\<%
\\
\>[0]\AgdaFunction{stackfunP2PKHNonEmptyStackNormalForm}\AgdaSpace{}%
\AgdaSymbol{=}\<%
\\
\>[0]\AgdaComment{--\ \textbackslash{}stackVerificationPtoPKHsymbolicExecutionNonEmptyStackNormalForm}\<%
\end{code}
} 

\newcommand{\stackVerificationPtoPKHsymbolicExecutionNonEmptyStackNormalForm}{
\begin{code}%
\>[0][@{}l@{\AgdaIndent{1}}]%
\>[1]\AgdaFunction{executeStackVerify}\AgdaSpace{}%
\AgdaSymbol{(}\AgdaFunction{compareNaturals}\AgdaSpace{}%
\AgdaPostulate{pbkh}\AgdaSpace{}%
\AgdaSymbol{(}\AgdaFunction{hashFun}\AgdaSpace{}%
\AgdaPostulate{pbk}\AgdaSymbol{)}\AgdaSpace{}%
\AgdaOperator{\AgdaInductiveConstructor{∷}}\AgdaSpace{}%
\AgdaPostulate{pbk}\AgdaSpace{}%
\AgdaOperator{\AgdaInductiveConstructor{∷}}\AgdaSpace{}%
\AgdaPostulate{stack₁}\AgdaSymbol{)}\AgdaSpace{}%
\AgdaOperator{\AgdaFunction{\ensuremath{\mathbin{{>}\mkern-8.5mu{>}\mkern-2mu{=}}}}}\<%
\\
\>[1]\AgdaFunction{executeStackCheckSig}\AgdaSpace{}%
\AgdaPostulate{msg₁}\<%
\end{code}
} 

\AgdaHide{
\begin{code}%
\>[0]\<%
\\
\>[0]\AgdaFunction{stackfunP2PKHNonEmptyStackNormalFormDo}\AgdaSpace{}%
\AgdaSymbol{:}%
\>[42]\AgdaDatatype{Maybe}\AgdaSpace{}%
\AgdaFunction{Stack}\<%
\\
\>[0]\AgdaFunction{stackfunP2PKHNonEmptyStackNormalFormDo}\AgdaSpace{}%
\AgdaSymbol{=}\<%
\\
\>[0]\AgdaComment{--\ \textbackslash{}stackVerificationPtoPKHsymbolicExecutionNonEmptyStackNormalForm}\<%
\end{code}
} 

\newcommand{\stackVerificationPtoPKHsymbolicExecutionNonEmptyStackNormalFormUsingDo}{
\begin{code}%
\>[0][@{}l@{\AgdaIndent{1}}]%
\>[1]\AgdaKeyword{do}%
\>[6]\AgdaBound{stack₅}\AgdaSpace{}%
\AgdaOperator{\AgdaFunction{←}}\AgdaSpace{}%
\AgdaFunction{executeStackVerify}\AgdaSpace{}%
\AgdaSymbol{(}\AgdaFunction{compareNaturals}\AgdaSpace{}%
\AgdaPostulate{pbkh}\AgdaSpace{}%
\AgdaSymbol{(}\AgdaFunction{hashFun}\AgdaSpace{}%
\AgdaPostulate{pbk}\AgdaSymbol{)}\AgdaSpace{}%
\AgdaOperator{\AgdaInductiveConstructor{∷}}\AgdaSpace{}%
\AgdaPostulate{pbk}\AgdaSpace{}%
\AgdaOperator{\AgdaInductiveConstructor{∷}}\AgdaSpace{}%
\AgdaPostulate{stack₁}\AgdaSymbol{)}\<%
\\
\>[6]\AgdaFunction{executeStackCheckSig}\AgdaSpace{}%
\AgdaPostulate{msg₁}\AgdaSpace{}%
\AgdaBound{stack₅}\<%
\end{code}
} 

\AgdaHide{
\begin{code}%
\>[0]\<%
\\
\\[\AgdaEmptyExtraSkip]%
\\[\AgdaEmptyExtraSkip]%
\\[\AgdaEmptyExtraSkip]%
\>[0]\AgdaFunction{stackfunP2PKHNonEmptyStackNormalFormFirstPart}\AgdaSpace{}%
\AgdaSymbol{:}%
\>[49]\AgdaDatatype{Maybe}\AgdaSpace{}%
\AgdaFunction{Stack}\<%
\\
\>[0]\AgdaFunction{stackfunP2PKHNonEmptyStackNormalFormFirstPart}\AgdaSpace{}%
\AgdaSymbol{=}\<%
\\
\>[0]\AgdaComment{--\ \textbackslash{}stackVerificationPtoPKHsymbolicExecutionNonEmptyStackNormalFormFirstPart}\<%
\end{code}
} 

\newcommand{\stackVerificationPtoPKHsymbolicExecutionNonEmptyStackNormalFormFirstPart}{
\begin{code}[inline]%
\>[0][@{}l@{\AgdaIndent{1}}]%
\>[1]\AgdaFunction{executeStackVerify}\AgdaSpace{}%
\AgdaSymbol{(}\AgdaFunction{compareNaturals}\AgdaSpace{}%
\AgdaPostulate{pbkh}\AgdaSpace{}%
\AgdaSymbol{(}\AgdaFunction{hashFun}\AgdaSpace{}%
\AgdaPostulate{pbk}\AgdaSymbol{)}\AgdaSpace{}%
\AgdaOperator{\AgdaInductiveConstructor{∷}}\AgdaSpace{}%
\AgdaPostulate{pbk}\AgdaSpace{}%
\AgdaOperator{\AgdaInductiveConstructor{∷}}\AgdaSpace{}%
\AgdaPostulate{stack₁}\AgdaSymbol{)}\<%
\end{code}
} 

\AgdaHide{
\begin{code}%
\>[0]\<%
\\
\>[0]\AgdaFunction{stackfunP2PKHNonEmptyStackNormalFormFirstPartZoomedIn}\AgdaSpace{}%
\AgdaSymbol{:}%
\>[57]\AgdaDatatype{ℕ}\<%
\\
\>[0]\AgdaFunction{stackfunP2PKHNonEmptyStackNormalFormFirstPartZoomedIn}\AgdaSpace{}%
\AgdaSymbol{=}\<%
\\
\>[0]\AgdaComment{--\ \textbackslash{}stackVerificationPtoPKHsymbolicExecutionNonEmptyStackNormalFormFirstPartZoomedIn}\<%
\end{code}
} 

\newcommand{\stackVerificationPtoPKHsymbolicExecutionNonEmptyStackNormalFormFirstPartZoomedIn}{
\begin{code}[inline]%
\>[0][@{}l@{\AgdaIndent{1}}]%
\>[1]\AgdaFunction{compareNaturals}\AgdaSpace{}%
\AgdaPostulate{pbkh}\AgdaSpace{}%
\AgdaSymbol{(}\AgdaFunction{hashFun}\AgdaSpace{}%
\AgdaPostulate{pbk}\AgdaSymbol{)}\<%
\end{code}
} 

\AgdaHide{
\begin{code}%
\>[0]\<%
\\
\\[\AgdaEmptyExtraSkip]%
\\[\AgdaEmptyExtraSkip]%
\\[\AgdaEmptyExtraSkip]%
\>[0]\AgdaComment{\{-}\<%
\\
\>[0]\AgdaComment{We\ see\ that}\<%
\\
\>[0]\<%
\\
\>[0]\AgdaComment{(λ\ stack₂\ →\ executeStackCheckSig\ msg₁\ stack₂)\ =\ executeStackCheckSig\ msg₁}\<%
\\
\>[0]\<%
\\
\>[0]\AgdaComment{and\ can\ therefore\ use}\<%
\\
\>[0]\<%
\\
\>[0]\AgdaComment{executeStackVerify\ (compareNaturals\ pbkh\ (param\ .hash\ pbk)\ ∷\ pbk\ ∷\ stack₁)}\<%
\\
\>[0]\AgdaComment{\ensuremath{\mathbin{{>}\mkern-8.5mu{>}\mkern-2mu{=}}}\ executeStackCheckSig\ msg₁}\<%
\\
\>[0]\<%
\\
\>[0]\AgdaComment{--\ older\ version}\<%
\\
\>[0]\AgdaComment{lift2Maybe\ (executeStackCheckSig\ msg₁)}\<%
\\
\>[0]\AgdaComment{(executeStackVerify}\<%
\\
\>[0]\AgdaComment{\ (compareNaturals\ pbkh\ (hashFun\ pbk)\ ∷\ pbk\ ∷\ stack₁))}\<%
\\
\>[0]\<%
\\
\>[0]\AgdaComment{\ -\}}\<%
\\
\\[\AgdaEmptyExtraSkip]%
\\[\AgdaEmptyExtraSkip]%
\>[0]\<%
\end{code}
} 

\newcommand{\stackVerificationPtoPKHsymbolicExecutionstackempty}{
\begin{code}%
\>[0]\AgdaFunction{stackfunP2PKHNonEmptyStack}\AgdaSpace{}%
\AgdaSymbol{:}\AgdaSpace{}%
\AgdaSymbol{(}\AgdaBound{pubKeyHash}\AgdaSpace{}%
\AgdaSymbol{:}\AgdaSpace{}%
\AgdaDatatype{ℕ}\AgdaSymbol{)(}\AgdaBound{msg₁}\AgdaSpace{}%
\AgdaSymbol{:}\AgdaSpace{}%
\AgdaDatatype{Msg}\AgdaSymbol{)(}\AgdaBound{pbk}\AgdaSpace{}%
\AgdaSymbol{:}\AgdaSpace{}%
\AgdaDatatype{ℕ}\AgdaSymbol{)(}\AgdaBound{stack₂}\AgdaSpace{}%
\AgdaSymbol{:}\AgdaSpace{}%
\AgdaFunction{Stack}\AgdaSymbol{)}\AgdaSpace{}%
\AgdaSymbol{→}\AgdaSpace{}%
\AgdaDatatype{Maybe}\AgdaSpace{}%
\AgdaFunction{Stack}\<%
\\
\>[0]\AgdaFunction{stackfunP2PKHNonEmptyStack}\AgdaSpace{}%
\AgdaBound{pubKeyHash}\AgdaSpace{}%
\AgdaBound{msg₁}\AgdaSpace{}%
\AgdaBound{pbk}\AgdaSpace{}%
\AgdaBound{stack₂}\<%
\\
\>[0][@{}l@{\AgdaIndent{0}}]%
\>[14]\AgdaSymbol{=}%
\>[317I]\AgdaFunction{executeStackVerify}\AgdaSpace{}%
\AgdaSymbol{(}\AgdaFunction{compareNaturals}\AgdaSpace{}%
\AgdaBound{pubKeyHash}\AgdaSpace{}%
\AgdaSymbol{(}\AgdaFunction{hashFun}\AgdaSpace{}%
\AgdaBound{pbk}\AgdaSymbol{)}\AgdaSpace{}%
\AgdaOperator{\AgdaInductiveConstructor{∷}}\AgdaSpace{}%
\AgdaBound{pbk}\AgdaSpace{}%
\AgdaOperator{\AgdaInductiveConstructor{∷}}\AgdaSpace{}%
\AgdaBound{stack₂}\AgdaSymbol{)}\<%
\\
\>[317I][@{}l@{\AgdaIndent{0}}]%
\>[17]\AgdaOperator{\AgdaFunction{\ensuremath{\mathbin{{>}\mkern-8.5mu{>}\mkern-2mu{=}}}}}\AgdaSpace{}%
\AgdaFunction{executeStackCheckSig}\AgdaSpace{}%
\AgdaBound{msg₁}\<%
\\
\\[\AgdaEmptyExtraSkip]%
\>[0]\<%
\end{code}
} 

\AgdaHide{
\begin{code}%
\>[0]\<%
\\
\>[0]\AgdaComment{\{-}\<%
\\
\>[0]\AgdaComment{and\ check\ that\ this\ is\ correct}\<%
\\
\>[0]\AgdaComment{-\}}\<%
\\
\\[\AgdaEmptyExtraSkip]%
\>[0]\AgdaFunction{stackfunP2PKHemptyNonEmptyStackCorrect}\AgdaSpace{}%
\AgdaSymbol{:}\AgdaSpace{}%
\AgdaSymbol{(}\AgdaBound{pubKeyHash}\AgdaSpace{}%
\AgdaSymbol{:}\AgdaSpace{}%
\AgdaDatatype{ℕ}\AgdaSymbol{)(}\AgdaBound{time₁}\AgdaSpace{}%
\AgdaSymbol{:}\AgdaSpace{}%
\AgdaFunction{Time}\AgdaSymbol{)(}\AgdaBound{msg₁}\AgdaSpace{}%
\AgdaSymbol{:}\AgdaSpace{}%
\AgdaDatatype{Msg}\AgdaSymbol{)(}\AgdaBound{pbk}\AgdaSpace{}%
\AgdaSymbol{:}\AgdaSpace{}%
\AgdaDatatype{ℕ}\AgdaSymbol{)(}\AgdaBound{stack₂}\AgdaSpace{}%
\AgdaSymbol{:}\AgdaSpace{}%
\AgdaFunction{Stack}\AgdaSymbol{)}\<%
\\
\>[0][@{}l@{\AgdaIndent{0}}]%
\>[8]\AgdaSymbol{→}\AgdaSpace{}%
\AgdaOperator{\AgdaFunction{⟦}}\AgdaSpace{}%
\AgdaFunction{scriptP2PKHᵇ}\AgdaSpace{}%
\AgdaBound{pubKeyHash}%
\>[37]\AgdaOperator{\AgdaFunction{⟧ˢ}}\AgdaSpace{}%
\AgdaBound{time₁}\AgdaSpace{}%
\AgdaBound{msg₁}\AgdaSpace{}%
\AgdaSymbol{(}\AgdaBound{pbk}\AgdaSpace{}%
\AgdaOperator{\AgdaInductiveConstructor{∷}}\AgdaSpace{}%
\AgdaBound{stack₂}\AgdaSymbol{)}\AgdaSpace{}%
\AgdaOperator{\AgdaDatatype{≡}}\AgdaSpace{}%
\AgdaFunction{stackfunP2PKHNonEmptyStack}\AgdaSpace{}%
\AgdaBound{pubKeyHash}\AgdaSpace{}%
\AgdaBound{msg₁}%
\>[112]\AgdaBound{pbk}\AgdaSpace{}%
\AgdaBound{stack₂}\<%
\\
\>[0]\AgdaFunction{stackfunP2PKHemptyNonEmptyStackCorrect}\AgdaSpace{}%
\AgdaBound{pubKeyHash}\AgdaSpace{}%
\AgdaBound{time₁}\AgdaSpace{}%
\AgdaBound{msg₁}\AgdaSpace{}%
\AgdaBound{pbk}\AgdaSpace{}%
\AgdaBound{stack₂}\AgdaSpace{}%
\AgdaSymbol{=}\AgdaSpace{}%
\AgdaInductiveConstructor{refl}\<%
\\
\\[\AgdaEmptyExtraSkip]%
\\[\AgdaEmptyExtraSkip]%
\\[\AgdaEmptyExtraSkip]%
\>[0]\AgdaComment{\{-\ We\ see\ now\ the\ case\ distinction\ depends\ on}\<%
\\
\>[0]\AgdaComment{\ \ \ compres\ :=\ compareNaturals\ pbkh\ (hashFun\ pbk)}\<%
\\
\>[0]\<%
\\
\>[0]\AgdaComment{since}\<%
\\
\>[0]\<%
\\
\>[0]\AgdaComment{executeStackVerify\ \ \ \ (compres\ ∷\ \ pbk\ ∷\ stack₁)}\<%
\\
\>[0]\<%
\\
\>[0]\AgdaComment{will\ depend\ on\ whether\ compres\ is\ 0\ or\ \ suc\ x'}\<%
\\
\>[0]\<%
\\
\>[0]\AgdaComment{so\ we\ abstract\ from}\<%
\\
\>[0]\<%
\\
\>[0]\AgdaComment{compres\ =\ compareNaturals\ pubKeyHash\ (hashFun\ pbk)}\<%
\\
\>[0]\<%
\\
\>[0]\AgdaComment{-\}}\<%
\\
\\[\AgdaEmptyExtraSkip]%
\>[0]\AgdaComment{--\ This\ function\ will\ be\ repeated\ in\ stackVerificationP2PKHextractedProgram.agda}\<%
\\
\>[0]\AgdaComment{--\ and\ therefore\ is\ kept\ private\ in\ this\ section\ in\ order\ to\ avoid\ a\ conflict}\<%
\\
\>[0]\AgdaComment{--\ \textbackslash{}stackVerificationPtoPKHsymbolicExecutionabstract}\<%
\end{code}
} 

\newcommand{\stackVerificationPtoPKHsymbolicExecutionabstract}{
\begin{code}%
\>[0]\AgdaFunction{p2PKHNonEmptyStackAbstr'}\AgdaSpace{}%
\AgdaSymbol{:}\AgdaSpace{}%
\AgdaSymbol{(}\AgdaBound{msg₁}\AgdaSpace{}%
\AgdaSymbol{:}\AgdaSpace{}%
\AgdaDatatype{Msg}\AgdaSymbol{)(}\AgdaBound{pbk}\AgdaSpace{}%
\AgdaSymbol{:}\AgdaSpace{}%
\AgdaDatatype{ℕ}\AgdaSymbol{)(}\AgdaBound{stack₁}\AgdaSpace{}%
\AgdaSymbol{:}\AgdaSpace{}%
\AgdaFunction{Stack}\AgdaSymbol{)(}\AgdaBound{cmp}\AgdaSpace{}%
\AgdaSymbol{:}\AgdaSpace{}%
\AgdaDatatype{ℕ}\AgdaSymbol{)}\AgdaSpace{}%
\AgdaSymbol{→}\AgdaSpace{}%
\AgdaDatatype{Maybe}\AgdaSpace{}%
\AgdaFunction{Stack}\<%
\\
\>[0]\AgdaFunction{p2PKHNonEmptyStackAbstr'}\AgdaSpace{}%
\AgdaBound{msg₁}\AgdaSpace{}%
\AgdaBound{pbk}\AgdaSpace{}%
\AgdaBound{stack₁}\AgdaSpace{}%
\AgdaBound{cmp}\AgdaSpace{}%
\AgdaSymbol{=}%
\>[48]\AgdaFunction{executeStackVerify}\AgdaSpace{}%
\AgdaSymbol{(}\AgdaBound{cmp}\AgdaSpace{}%
\AgdaOperator{\AgdaInductiveConstructor{∷}}%
\>[75]\AgdaBound{pbk}\AgdaSpace{}%
\AgdaOperator{\AgdaInductiveConstructor{∷}}\AgdaSpace{}%
\AgdaBound{stack₁}\AgdaSymbol{)}\AgdaSpace{}%
\AgdaOperator{\AgdaFunction{\ensuremath{\mathbin{{>}\mkern-8.5mu{>}\mkern-2mu{=}}}}}\<%
\\
\>[48]\AgdaFunction{executeStackCheckSig}\AgdaSpace{}%
\AgdaBound{msg₁}\<%
\end{code}
} 

\AgdaHide{
\begin{code}%
\>[0]\<%
\\
\>[0]\AgdaComment{--\ \textbackslash{}stackVerificationPtoPKHsymbolicExecutionabstract}\<%
\end{code}
} 

\newcommand{\stackVerificationPtoPKHsymbolicExecutionabstractUsingVar}{
\begin{code}%
\>[0]\AgdaFunction{abstrFun}\AgdaSpace{}%
\AgdaSymbol{:}\AgdaSpace{}%
\AgdaSymbol{(}\AgdaBound{stack₁}\AgdaSpace{}%
\AgdaSymbol{:}\AgdaSpace{}%
\AgdaFunction{Stack}\AgdaSymbol{)(}\AgdaBound{cmp}\AgdaSpace{}%
\AgdaSymbol{:}\AgdaSpace{}%
\AgdaDatatype{ℕ}\AgdaSymbol{)}\AgdaSpace{}%
\AgdaSymbol{→}\AgdaSpace{}%
\AgdaDatatype{Maybe}\AgdaSpace{}%
\AgdaFunction{Stack}\<%
\\
\>[0]\AgdaFunction{abstrFun}\AgdaSpace{}%
\AgdaBound{stack₁}\AgdaSpace{}%
\AgdaBound{cmp}\AgdaSpace{}%
\AgdaSymbol{=}%
\>[23]\AgdaKeyword{do}%
\>[27]\AgdaBound{stack₅}\AgdaSpace{}%
\AgdaOperator{\AgdaFunction{←}}\AgdaSpace{}%
\AgdaFunction{executeStackVerify}\AgdaSpace{}%
\AgdaSymbol{(}\AgdaBound{cmp}\AgdaSpace{}%
\AgdaOperator{\AgdaInductiveConstructor{∷}}%
\>[63]\AgdaPostulate{pbk}\AgdaSpace{}%
\AgdaOperator{\AgdaInductiveConstructor{∷}}\AgdaSpace{}%
\AgdaBound{stack₁}\AgdaSymbol{)}\<%
\\
\>[27]\AgdaFunction{executeStackCheckSig}\AgdaSpace{}%
\AgdaPostulate{msg₁}\AgdaSpace{}%
\AgdaBound{stack₅}\<%
\end{code}
} 

\AgdaHide{
\begin{code}%
\>[0]\<%
\\
\\[\AgdaEmptyExtraSkip]%
\\[\AgdaEmptyExtraSkip]%
\>[0]\AgdaFunction{stackfunP2PKHNonEmptyStackNormalFormUsingAbstractedFun}\AgdaSpace{}%
\AgdaSymbol{:}\AgdaSpace{}%
\AgdaDatatype{Maybe}\AgdaSpace{}%
\AgdaFunction{Stack}\<%
\\
\>[0]\AgdaFunction{stackfunP2PKHNonEmptyStackNormalFormUsingAbstractedFun}\AgdaSpace{}%
\AgdaSymbol{=}\<%
\\
\>[0]\AgdaComment{--\ \textbackslash{}stackVerificationPtoPKHsymbolicExecutionnonEmptyStackNormalFormWithAbstractedFun}\<%
\end{code}
} 

\newcommand{\stackVerificationPtoPKHsymbolicExecutionnonEmptyStackNormalFormWithAbstractedFun}{
\begin{code}[inline]%
\>[0][@{}l@{\AgdaIndent{1}}]%
\>[2]\AgdaFunction{abstrFun}\AgdaSpace{}%
\AgdaPostulate{stack₁}\AgdaSpace{}%
\AgdaSymbol{(}\AgdaFunction{compareNaturals}\AgdaSpace{}%
\AgdaPostulate{pbkh}\AgdaSpace{}%
\AgdaSymbol{(}\AgdaFunction{hashFun}\AgdaSpace{}%
\AgdaPostulate{pbk}\AgdaSymbol{))}\<%
\end{code}
} 

\AgdaHide{
\begin{code}%
\>[0]\<%
\\
\>[0]\AgdaFunction{stackfunP2PKHNonEmptyStackNormalFormUsingAbstractedFunTest}\AgdaSpace{}%
\AgdaSymbol{:}\<%
\\
\>[0][@{}l@{\AgdaIndent{0}}]%
\>[4]\AgdaFunction{stackfunP2PKHNonEmptyStackNormalForm}\AgdaSpace{}%
\AgdaOperator{\AgdaDatatype{≡}}\AgdaSpace{}%
\AgdaFunction{stackfunP2PKHNonEmptyStackNormalFormUsingAbstractedFun}\<%
\\
\>[0]\AgdaFunction{stackfunP2PKHNonEmptyStackNormalFormUsingAbstractedFunTest}%
\>[60]\AgdaSymbol{=}\AgdaSpace{}%
\AgdaInductiveConstructor{refl}\<%
\\
\\[\AgdaEmptyExtraSkip]%
\>[0]\AgdaComment{\{-\ and\ we\ show\ that\ this\ is\ the\ right\ function}\<%
\\
\>[0]\AgdaComment{-\}}\<%
\\
\\[\AgdaEmptyExtraSkip]%
\>[0]\AgdaComment{--\ This\ function\ will\ be\ repeated\ in\ stackVerificationP2PKHextractedProgram.agda}\<%
\\
\>[0]\AgdaComment{--\ and\ therefore\ is\ kept\ private\ in\ this\ section\ in\ order\ to\ avoid\ a\ conflict}\<%
\\
\>[0]\AgdaKeyword{private}\<%
\\
\>[0][@{}l@{\AgdaIndent{0}}]%
\>[2]\AgdaFunction{stackfunP2PKHNonEmptyStackAbstractedCor}\AgdaSpace{}%
\AgdaSymbol{:}%
\>[45]\AgdaSymbol{(}\AgdaBound{pubKeyHash}\AgdaSpace{}%
\AgdaSymbol{:}\AgdaSpace{}%
\AgdaDatatype{ℕ}\AgdaSymbol{)(}\AgdaBound{time₁}\AgdaSpace{}%
\AgdaSymbol{:}\AgdaSpace{}%
\AgdaFunction{Time}\AgdaSymbol{)(}\AgdaBound{msg₁}\AgdaSpace{}%
\AgdaSymbol{:}\AgdaSpace{}%
\AgdaDatatype{Msg}\AgdaSymbol{)(}\AgdaBound{pbk}\AgdaSpace{}%
\AgdaSymbol{:}\AgdaSpace{}%
\AgdaDatatype{ℕ}\AgdaSymbol{)(}\AgdaBound{stack₂}\AgdaSpace{}%
\AgdaSymbol{:}\AgdaSpace{}%
\AgdaFunction{Stack}\AgdaSymbol{)}\<%
\\
\>[2][@{}l@{\AgdaIndent{0}}]%
\>[18]\AgdaSymbol{→}%
\>[428I]\AgdaOperator{\AgdaFunction{⟦}}\AgdaSpace{}%
\AgdaFunction{scriptP2PKHᵇ}\AgdaSpace{}%
\AgdaBound{pubKeyHash}\AgdaSpace{}%
\AgdaOperator{\AgdaFunction{⟧ˢ}}\AgdaSpace{}%
\AgdaBound{time₁}\AgdaSpace{}%
\AgdaBound{msg₁}\AgdaSpace{}%
\AgdaSymbol{(}\AgdaBound{pbk}\AgdaSpace{}%
\AgdaOperator{\AgdaInductiveConstructor{∷}}\AgdaSpace{}%
\AgdaBound{stack₂}\AgdaSymbol{)}\<%
\\
\>[428I][@{}l@{\AgdaIndent{0}}]%
\>[21]\AgdaOperator{\AgdaDatatype{≡}}%
\>[437I]\AgdaFunction{p2PKHNonEmptyStackAbstr'}%
\>[49]\AgdaBound{msg₁}\AgdaSpace{}%
\AgdaBound{pbk}\AgdaSpace{}%
\AgdaBound{stack₂}\<%
\\
\>[437I][@{}l@{\AgdaIndent{0}}]%
\>[29]\AgdaSymbol{(}\AgdaFunction{compareNaturals}\AgdaSpace{}%
\AgdaBound{pubKeyHash}\AgdaSpace{}%
\AgdaSymbol{(}\AgdaFunction{hashFun}\AgdaSpace{}%
\AgdaBound{pbk}\AgdaSymbol{))}\<%
\\
\>[2]\AgdaFunction{stackfunP2PKHNonEmptyStackAbstractedCor}\AgdaSpace{}%
\AgdaBound{pubKeyHash}\AgdaSpace{}%
\AgdaBound{time₁}\AgdaSpace{}%
\AgdaBound{msg₁}\AgdaSpace{}%
\AgdaBound{pbk}\AgdaSpace{}%
\AgdaBound{stack₂}\AgdaSpace{}%
\AgdaSymbol{=}\AgdaSpace{}%
\AgdaInductiveConstructor{refl}\<%
\\
\\[\AgdaEmptyExtraSkip]%
\\[\AgdaEmptyExtraSkip]%
\>[0]\AgdaComment{\{-\ Now\ we\ investigate\ what\ \ p2PKHNonEmptyStackAbstr'}\<%
\\
\>[0]\AgdaComment{When\ looking\ at\ it\ and\ see\ that}\<%
\\
\>[0]\<%
\\
\>[0]\AgdaComment{\ p2PKHNonEmptyStackAbstr'\ msg₁\ pbk\ stack₂\ cmp}\<%
\\
\>[0]\<%
\\
\>[0]\AgdaComment{will\ execute}\<%
\\
\>[0]\AgdaComment{executeStackVerify\ (cmp\ ∷\ \ pbk\ ∷\ stack₂)}\<%
\\
\>[0]\AgdaComment{which\ will\ in\ turn\ make\ a\ case\ disctintion\ on\ whether\ cmp\ is\ 0\ or\ \ not\ zero}\<%
\\
\>[0]\<%
\\
\>[0]\AgdaComment{(that\ corresponds\ to\ what\ the\ original\ function\ does\ because\ it\ makes\ this\ comparison}\<%
\\
\>[0]\AgdaComment{\ \ \ compareNaturals\ pubKeyHash\ (hashFun\ pbk)}\<%
\\
\>[0]\AgdaComment{\ \ which\ checks\ \ whether\ the\ pbk\ provided\ \ by\ the\ user\ hashes\ to\ the\ pubKeyHash\ of\ the\ locking\ script}\<%
\\
\>[0]\AgdaComment{\ \ \ \ if\ it\ \ is\ 0\ it\ should\ fail,\ and\ if\ it\ is\ 1\ it\ should\ succeed.}\<%
\\
\>[0]\<%
\\
\>[0]\AgdaComment{So\ lets\ make\ the\ test}\<%
\\
\>[0]\AgdaComment{-\}}\<%
\\
\\[\AgdaEmptyExtraSkip]%
\>[0]\AgdaFunction{testStackfunP2PKHNonEmptyStackAbstractedCompre0}\AgdaSpace{}%
\AgdaSymbol{:}\AgdaSpace{}%
\AgdaDatatype{Maybe}\AgdaSpace{}%
\AgdaFunction{Stack}\<%
\\
\>[0]\AgdaFunction{testStackfunP2PKHNonEmptyStackAbstractedCompre0}\AgdaSpace{}%
\AgdaSymbol{=}\AgdaSpace{}%
\AgdaFunction{p2PKHNonEmptyStackAbstr'}\AgdaSpace{}%
\AgdaPostulate{msg₁}\AgdaSpace{}%
\AgdaPostulate{pbk}\AgdaSpace{}%
\AgdaPostulate{stack₁}\AgdaSpace{}%
\AgdaNumber{0}\<%
\\
\\[\AgdaEmptyExtraSkip]%
\>[0]\AgdaComment{\{-\ if\ we\ evaluate\ it\ we\ get}\<%
\\
\>[0]\<%
\\
\>[0]\AgdaComment{nothing}\<%
\\
\>[0]\AgdaComment{-\}}\<%
\\
\\[\AgdaEmptyExtraSkip]%
\>[0]\AgdaComment{--\ we\ evaluate\ now}\<%
\\
\>[0]\AgdaFunction{abstrFunZeroCase}\AgdaSpace{}%
\AgdaSymbol{:}\AgdaSpace{}%
\AgdaDatatype{Maybe}\AgdaSpace{}%
\AgdaFunction{Stack}\<%
\\
\>[0]\AgdaFunction{abstrFunZeroCase}\AgdaSpace{}%
\AgdaSymbol{=}\<%
\\
\>[0]\AgdaComment{--\ \textbackslash{}stackVerificationPtoPKHsymbolicExecutionabstrFunZeroCase}\<%
\end{code}
} 

\newcommand{\stackVerificationPtoPKHsymbolicExecutionabstrFunZeroCase}{
\begin{code}[inline]%
\>[0][@{}l@{\AgdaIndent{1}}]%
\>[1]\AgdaFunction{abstrFun}\AgdaSpace{}%
\AgdaPostulate{stack₁}\AgdaSpace{}%
\AgdaNumber{0}\<%
\end{code}
} 

\AgdaHide{
\begin{code}%
\>[0]\<%
\\
\\[\AgdaEmptyExtraSkip]%
\>[0]\AgdaComment{\{-}\<%
\\
\>[0]\<%
\\
\>[0]\AgdaComment{We\ \ show\ now\ this\ is\ always\ the\ case}\<%
\\
\>[0]\<%
\\
\>[0]\AgdaComment{-\}}\<%
\\
\\[\AgdaEmptyExtraSkip]%
\>[0]\AgdaComment{--\ This\ function\ will\ be\ repeated\ in\ stackVerificationP2PKHextractedProgram.agda}\<%
\\
\>[0]\AgdaComment{--\ and\ therefore\ is\ kept\ private\ in\ this\ section\ in\ order\ to\ avoid\ a\ conflict}\<%
\\
\>[0]\AgdaKeyword{private}\<%
\\
\>[0][@{}l@{\AgdaIndent{0}}]%
\>[2]\AgdaFunction{stackfunP2PKHNonEmptyStackAbstractedCorCompr0IsNothing}\AgdaSpace{}%
\AgdaSymbol{:}\AgdaSpace{}%
\AgdaSymbol{(}\AgdaBound{msg₁}\AgdaSpace{}%
\AgdaSymbol{:}\AgdaSpace{}%
\AgdaDatatype{Msg}\AgdaSymbol{)(}\AgdaBound{pbk}\AgdaSpace{}%
\AgdaSymbol{:}\AgdaSpace{}%
\AgdaDatatype{ℕ}\AgdaSymbol{)(}\AgdaBound{stack₂}\AgdaSpace{}%
\AgdaSymbol{:}\AgdaSpace{}%
\AgdaFunction{Stack}\AgdaSymbol{)}\<%
\\
\>[2][@{}l@{\AgdaIndent{0}}]%
\>[8]\AgdaSymbol{→}%
\>[11]\AgdaFunction{p2PKHNonEmptyStackAbstr'}\AgdaSpace{}%
\AgdaBound{msg₁}\AgdaSpace{}%
\AgdaBound{pbk}\AgdaSpace{}%
\AgdaBound{stack₂}\AgdaSpace{}%
\AgdaNumber{0}\AgdaSpace{}%
\AgdaOperator{\AgdaDatatype{≡}}\AgdaSpace{}%
\AgdaInductiveConstructor{nothing}\<%
\\
\>[2]\AgdaFunction{stackfunP2PKHNonEmptyStackAbstractedCorCompr0IsNothing}\AgdaSpace{}%
\AgdaBound{msg₁}\AgdaSpace{}%
\AgdaBound{pbk}\AgdaSpace{}%
\AgdaBound{stack₂}\AgdaSpace{}%
\AgdaSymbol{=}\AgdaSpace{}%
\AgdaInductiveConstructor{refl}\<%
\\
\\[\AgdaEmptyExtraSkip]%
\\[\AgdaEmptyExtraSkip]%
\>[0]\AgdaComment{\{-\ Now\ we\ look\ at\ what\ happens\ \ if\ the\ value\ is\ non\ zero\ -\}}\<%
\\
\\[\AgdaEmptyExtraSkip]%
\\[\AgdaEmptyExtraSkip]%
\>[0]\AgdaFunction{testStackfunP2PKHNonEmptyStackAbstractedCompreSucCase}\AgdaSpace{}%
\AgdaSymbol{:}\AgdaSpace{}%
\AgdaDatatype{Maybe}\AgdaSpace{}%
\AgdaFunction{Stack}\<%
\\
\>[0]\AgdaFunction{testStackfunP2PKHNonEmptyStackAbstractedCompreSucCase}\AgdaSpace{}%
\AgdaSymbol{=}\AgdaSpace{}%
\AgdaFunction{p2PKHNonEmptyStackAbstr'}\AgdaSpace{}%
\AgdaPostulate{msg₁}\AgdaSpace{}%
\AgdaPostulate{pbk}\AgdaSpace{}%
\AgdaPostulate{stack₁}\AgdaSpace{}%
\AgdaSymbol{(}\AgdaInductiveConstructor{suc}\AgdaSpace{}%
\AgdaPostulate{x₁}\AgdaSymbol{)}\<%
\\
\\[\AgdaEmptyExtraSkip]%
\\[\AgdaEmptyExtraSkip]%
\>[0]\AgdaComment{--\ We\ evaluate}\<%
\\
\\[\AgdaEmptyExtraSkip]%
\>[0]\AgdaComment{--\ we\ evaluate\ now}\<%
\\
\>[0]\AgdaFunction{abstrFunSucCase}\AgdaSpace{}%
\AgdaSymbol{:}\AgdaSpace{}%
\AgdaDatatype{Maybe}\AgdaSpace{}%
\AgdaFunction{Stack}\<%
\\
\>[0]\AgdaFunction{abstrFunSucCase}\AgdaSpace{}%
\AgdaSymbol{=}\<%
\\
\>[0]\AgdaComment{--\ \textbackslash{}stackVerificationPtoPKHsymbolicExecutionabstrFunSucCase}\<%
\end{code}
} 

\newcommand{\stackVerificationPtoPKHsymbolicExecutionabstrFunSucCase}{
\begin{code}[inline]%
\>[0][@{}l@{\AgdaIndent{1}}]%
\>[1]\AgdaFunction{abstrFun}\AgdaSpace{}%
\AgdaPostulate{stack₁}\AgdaSpace{}%
\AgdaSymbol{(}\AgdaInductiveConstructor{suc}\AgdaSpace{}%
\AgdaPostulate{x₁}\AgdaSymbol{)}\<%
\end{code}
} 

\AgdaHide{
\begin{code}%
\>[0]\<%
\\
\>[0]\AgdaComment{--\ we\ obtain}\<%
\\
\\[\AgdaEmptyExtraSkip]%
\>[0]\AgdaFunction{abstrFunSucCaseNormal}\AgdaSpace{}%
\AgdaSymbol{:}\AgdaSpace{}%
\AgdaDatatype{Maybe}\AgdaSpace{}%
\AgdaFunction{Stack}\<%
\\
\>[0]\AgdaFunction{abstrFunSucCaseNormal}\AgdaSpace{}%
\AgdaSymbol{=}\<%
\\
\>[0]\AgdaComment{--\ \textbackslash{}stackVerificationPtoPKHsymbolicExecutionabstrFunSucCaseNormal}\<%
\end{code}
} 

\newcommand{\stackVerificationPtoPKHsymbolicExecutionabstrFunSucCaseNormal}{
\begin{code}[inline]%
\>[0][@{}l@{\AgdaIndent{1}}]%
\>[1]\AgdaFunction{executeStackCheckSig}\AgdaSpace{}%
\AgdaPostulate{msg₁}\AgdaSpace{}%
\AgdaSymbol{(}\AgdaPostulate{pbk}\AgdaSpace{}%
\AgdaOperator{\AgdaInductiveConstructor{∷}}\AgdaSpace{}%
\AgdaPostulate{stack₁}\AgdaSymbol{)}\<%
\end{code}
} 

\AgdaHide{
\begin{code}%
\>[0]\<%
\\
\>[0]\AgdaComment{\{-}\<%
\\
\>[0]\AgdaComment{executeStackCheckSig\ msg₁\ (pbk\ ∷\ stack₁)}\<%
\\
\>[0]\AgdaComment{-\}}\<%
\\
\\[\AgdaEmptyExtraSkip]%
\\[\AgdaEmptyExtraSkip]%
\>[0]\AgdaComment{\{-\ if\ we\ evalute}\<%
\\
\>[0]\AgdaComment{\ \ \ \ \ testStackfunP2PKHNonEmptyStackAbstractedCompreSucCase}\<%
\\
\>[0]\AgdaComment{\ \ \ \ we\ get}\<%
\\
\>[0]\<%
\\
\>[0]\AgdaComment{executeStackCheckSig\ msg₁\ (pbk\ ∷\ stack₁)}\<%
\\
\>[0]\<%
\\
\>[0]\<%
\\
\>[0]\AgdaComment{This\ corresponds\ to\ the\ situation\ where}\<%
\\
\>[0]\AgdaComment{the\ original\ stack₁\ was\ non\ empty,\ and\ the\ comparision\ of\ the\ pbk\ with\ the\ pbkhash}\<%
\\
\>[0]\AgdaComment{got\ a\ result\ >\ 0}\<%
\\
\>[0]\<%
\\
\>[0]\<%
\\
\>[0]\<%
\\
\>[0]\AgdaComment{If\ we\ look\ at}\<%
\\
\>[0]\AgdaComment{executeStackCheckSig}\<%
\\
\>[0]\<%
\\
\>[0]\AgdaComment{we\ see\ that\ it\ gives\ nothing\ when\ the\ stack\ has\ height\ <\ 2}\<%
\\
\>[0]\AgdaComment{and\ otherwise\ does\ something,}\<%
\\
\>[0]\<%
\\
\>[0]\AgdaComment{so\ we\ can\ \ make\ a\ case\ distinction\ on\ whether\ in}\<%
\\
\>[0]\<%
\\
\>[0]\AgdaComment{p2PKHNonEmptyStackAbstr'\ msg₁\ pbk\ stack₁\ x}\<%
\\
\>[0]\<%
\\
\>[0]\AgdaComment{stack₁\ is\ []\ or\ nonempty}\<%
\\
\>[0]\<%
\\
\>[0]\AgdaComment{So\ lets\ look\ at\ the\ easy\ case\ []}\<%
\\
\>[0]\<%
\\
\>[0]\AgdaComment{\ \ \ \ -\}}\<%
\\
\\[\AgdaEmptyExtraSkip]%
\\[\AgdaEmptyExtraSkip]%
\>[0]\AgdaComment{--\ we\ evaluate\ now}\<%
\\
\>[0]\AgdaFunction{abstrFunSucCaseEmpty}\AgdaSpace{}%
\AgdaSymbol{:}\AgdaSpace{}%
\AgdaDatatype{Maybe}\AgdaSpace{}%
\AgdaFunction{Stack}\<%
\\
\>[0]\AgdaFunction{abstrFunSucCaseEmpty}\AgdaSpace{}%
\AgdaSymbol{=}\<%
\\
\>[0]\AgdaComment{--\ \textbackslash{}stackVerificationPtoPKHsymbolicExecutionabstrFunSucCaseEmpty}\<%
\end{code}
} 

\newcommand{\stackVerificationPtoPKHsymbolicExecutionabstrFunSucCaseEmpty}{
\begin{code}[inline]%
\>[0][@{}l@{\AgdaIndent{1}}]%
\>[1]\AgdaFunction{abstrFun}\AgdaSpace{}%
\AgdaInductiveConstructor{[]}\AgdaSpace{}%
\AgdaSymbol{(}\AgdaInductiveConstructor{suc}\AgdaSpace{}%
\AgdaPostulate{x₁}\AgdaSymbol{)}\<%
\end{code}
} 

\AgdaHide{
\begin{code}%
\>[0]\<%
\\
\>[0]\AgdaComment{--\ and\ obtain\ nothing}\<%
\\
\>[0]\AgdaFunction{abstrFunSucCaseEmptyCheck}\AgdaSpace{}%
\AgdaSymbol{:}%
\>[29]\AgdaFunction{abstrFunSucCaseEmpty}\AgdaSpace{}%
\AgdaOperator{\AgdaDatatype{≡}}\AgdaSpace{}%
\AgdaInductiveConstructor{nothing}\<%
\\
\>[0]\AgdaFunction{abstrFunSucCaseEmptyCheck}\AgdaSpace{}%
\AgdaSymbol{=}\AgdaSpace{}%
\AgdaInductiveConstructor{refl}\<%
\\
\\[\AgdaEmptyExtraSkip]%
\>[0]\AgdaComment{--\ we\ evaluate\ now}\<%
\\
\>[0]\AgdaFunction{abstrFunSucCaseNonEmpty}\AgdaSpace{}%
\AgdaSymbol{:}\AgdaSpace{}%
\AgdaDatatype{Maybe}\AgdaSpace{}%
\AgdaFunction{Stack}\<%
\\
\>[0]\AgdaFunction{abstrFunSucCaseNonEmpty}\AgdaSpace{}%
\AgdaSymbol{=}\<%
\\
\>[0]\AgdaComment{--\ \textbackslash{}stackVerificationPtoPKHsymbolicExecutionabstrFunSucCaseNonEmpty}\<%
\end{code}
} 

\newcommand{\stackVerificationPtoPKHsymbolicExecutionabstrFunSucCaseNonEmpty}{
\begin{code}[inline]%
\>[0][@{}l@{\AgdaIndent{1}}]%
\>[1]\AgdaFunction{abstrFun}\AgdaSpace{}%
\AgdaSymbol{(}\AgdaPostulate{sig₁}\AgdaSpace{}%
\AgdaOperator{\AgdaInductiveConstructor{∷}}\AgdaSpace{}%
\AgdaPostulate{stack₁}\AgdaSymbol{)}\AgdaSpace{}%
\AgdaSymbol{(}\AgdaInductiveConstructor{suc}\AgdaSpace{}%
\AgdaPostulate{x₁}\AgdaSymbol{)}\<%
\end{code}
} 

\AgdaHide{
\begin{code}%
\>[0]\<%
\\
\\[\AgdaEmptyExtraSkip]%
\>[0]\AgdaFunction{abstrFunSucCaseNonEmptyNormal}\AgdaSpace{}%
\AgdaSymbol{:}\AgdaSpace{}%
\AgdaDatatype{Maybe}\AgdaSpace{}%
\AgdaFunction{Stack}\<%
\\
\>[0]\AgdaFunction{abstrFunSucCaseNonEmptyNormal}\AgdaSpace{}%
\AgdaSymbol{=}\<%
\\
\>[0]\AgdaComment{--\ \textbackslash{}stackVerificationPtoPKHsymbolicExecutionabstrFunSucCaseNonEmptyNormal}\<%
\end{code}
} 

\newcommand{\stackVerificationPtoPKHsymbolicExecutionabstrFunSucCaseNonEmptyNormal}{
\begin{code}[inline]%
\>[0][@{}l@{\AgdaIndent{1}}]%
\>[1]\AgdaInductiveConstructor{just}\AgdaSpace{}%
\AgdaSymbol{(}\AgdaFunction{boolToNat}\AgdaSpace{}%
\AgdaSymbol{(}\AgdaFunction{isSigned}\AgdaSpace{}%
\AgdaPostulate{msg₁}\AgdaSpace{}%
\AgdaPostulate{sig₁}\AgdaSpace{}%
\AgdaPostulate{pbk}\AgdaSymbol{)}\AgdaSpace{}%
\AgdaOperator{\AgdaInductiveConstructor{∷}}\AgdaSpace{}%
\AgdaPostulate{stack₁}\AgdaSymbol{)}\<%
\end{code}
} 

\AgdaHide{
\begin{code}%
\>[0]\<%
\\
\>[0]\AgdaFunction{abstrFunSucCaseNonEmptyCheck}\AgdaSpace{}%
\AgdaSymbol{:}\AgdaSpace{}%
\AgdaFunction{abstrFunSucCaseNonEmpty}\AgdaSpace{}%
\AgdaOperator{\AgdaDatatype{≡}}\AgdaSpace{}%
\AgdaFunction{abstrFunSucCaseNonEmptyNormal}\<%
\\
\>[0]\AgdaFunction{abstrFunSucCaseNonEmptyCheck}\AgdaSpace{}%
\AgdaSymbol{=}\AgdaSpace{}%
\AgdaInductiveConstructor{refl}\<%
\\
\>[0]\<%
\end{code}
} 

\newcommand{\stackVerificationPtoPKHsymbolicExecutioneasycase}{
\begin{code}%
\>[0]\AgdaFunction{testStackfunP2PKHNonEmptyStackAbstractedCompreSucEmpty}\AgdaSpace{}%
\AgdaSymbol{:}\AgdaSpace{}%
\AgdaDatatype{Maybe}\AgdaSpace{}%
\AgdaFunction{Stack}\<%
\\
\>[0]\AgdaFunction{testStackfunP2PKHNonEmptyStackAbstractedCompreSucEmpty}\AgdaSpace{}%
\AgdaSymbol{=}\<%
\\
\>[0][@{}l@{\AgdaIndent{0}}]%
\>[27]\AgdaFunction{p2PKHNonEmptyStackAbstr'}\AgdaSpace{}%
\AgdaPostulate{msg₁}\AgdaSpace{}%
\AgdaPostulate{pbk}\AgdaSpace{}%
\AgdaInductiveConstructor{[]}\AgdaSpace{}%
\AgdaSymbol{(}\AgdaInductiveConstructor{suc}\AgdaSpace{}%
\AgdaPostulate{x₁}\AgdaSymbol{)}\<%
\end{code}
} 

\AgdaHide{
\begin{code}%
\>[0]\<%
\\
\>[0]\AgdaComment{\{-\ if\ we\ evaluate\ it\ we\ get\ \ result}\<%
\\
\>[0]\<%
\\
\>[0]\AgdaComment{nothing}\<%
\\
\>[0]\<%
\\
\>[0]\AgdaComment{we\ check\ that\ this\ always\ holds}\<%
\\
\>[0]\AgdaComment{-\}}\<%
\\
\\[\AgdaEmptyExtraSkip]%
\\[\AgdaEmptyExtraSkip]%
\>[0]\AgdaFunction{stackfunP2PKHNonEmptyStackAbstractedCorComprSucStackEmpty}\AgdaSpace{}%
\AgdaSymbol{:}\AgdaSpace{}%
\AgdaSymbol{(}\AgdaBound{msg₁}\AgdaSpace{}%
\AgdaSymbol{:}\AgdaSpace{}%
\AgdaDatatype{Msg}\AgdaSymbol{)(}\AgdaBound{pbk}\AgdaSpace{}%
\AgdaSymbol{:}\AgdaSpace{}%
\AgdaDatatype{ℕ}\AgdaSymbol{)(}\AgdaBound{x}\AgdaSpace{}%
\AgdaSymbol{:}\AgdaSpace{}%
\AgdaDatatype{ℕ}\AgdaSymbol{)}\<%
\\
\>[0][@{}l@{\AgdaIndent{0}}]%
\>[13]\AgdaSymbol{→}\AgdaSpace{}%
\AgdaFunction{p2PKHNonEmptyStackAbstr'}\AgdaSpace{}%
\AgdaBound{msg₁}\AgdaSpace{}%
\AgdaBound{pbk}\AgdaSpace{}%
\AgdaInductiveConstructor{[]}\AgdaSpace{}%
\AgdaSymbol{(}\AgdaInductiveConstructor{suc}\AgdaSpace{}%
\AgdaBound{x}\AgdaSymbol{)}\AgdaSpace{}%
\AgdaOperator{\AgdaDatatype{≡}}\AgdaSpace{}%
\AgdaInductiveConstructor{nothing}\<%
\\
\>[0]\AgdaFunction{stackfunP2PKHNonEmptyStackAbstractedCorComprSucStackEmpty}\AgdaSpace{}%
\AgdaBound{msg₁}\AgdaSpace{}%
\AgdaBound{pbk}\AgdaSpace{}%
\AgdaBound{x}%
\>[70]\AgdaSymbol{=}\AgdaSpace{}%
\AgdaInductiveConstructor{refl}\<%
\\
\\[\AgdaEmptyExtraSkip]%
\>[0]\AgdaComment{\{-\ \ Intermezzo:\ we\ can\ see\ that\ the\ fanction\ gives\ always\ nothing\ if\ the\ stack\ is\ empty}\<%
\\
\>[0]\AgdaComment{\ \ \ independent\ of\ the\ result}\<%
\\
\>[0]\<%
\\
\>[0]\AgdaComment{But\ this\ result\ is\ not\ really\ needed\ (so\ can\ be\ probably\ be\ ommitted\ in\ the\ paper)}\<%
\\
\>[0]\AgdaComment{\ \ \ -\}}\<%
\\
\\[\AgdaEmptyExtraSkip]%
\>[0]\AgdaComment{--\ This\ function\ will\ be\ repeated\ in\ stackVerificationP2PKHextractedProgram.agda}\<%
\\
\>[0]\AgdaComment{--\ and\ therefore\ is\ kept\ private\ in\ this\ section\ in\ order\ to\ avoid\ a\ conflict}\<%
\\
\>[0]\AgdaKeyword{private}\<%
\\
\>[0][@{}l@{\AgdaIndent{0}}]%
\>[2]\AgdaFunction{stackfunP2PKHNonEmptyStackAbstractedCorEmptysNothing}\AgdaSpace{}%
\AgdaSymbol{:}\AgdaSpace{}%
\AgdaSymbol{(}\AgdaBound{msg₁}\AgdaSpace{}%
\AgdaSymbol{:}\AgdaSpace{}%
\AgdaDatatype{Msg}\AgdaSymbol{)(}\AgdaBound{pbk}\AgdaSpace{}%
\AgdaSymbol{:}\AgdaSpace{}%
\AgdaDatatype{ℕ}\AgdaSymbol{)(}\AgdaBound{x}\AgdaSpace{}%
\AgdaSymbol{:}\AgdaSpace{}%
\AgdaDatatype{ℕ}\AgdaSymbol{)}\<%
\\
\>[2][@{}l@{\AgdaIndent{0}}]%
\>[8]\AgdaSymbol{→}%
\>[11]\AgdaFunction{p2PKHNonEmptyStackAbstr'}\AgdaSpace{}%
\AgdaBound{msg₁}\AgdaSpace{}%
\AgdaBound{pbk}\AgdaSpace{}%
\AgdaInductiveConstructor{[]}\AgdaSpace{}%
\AgdaBound{x}\AgdaSpace{}%
\AgdaOperator{\AgdaDatatype{≡}}\AgdaSpace{}%
\AgdaInductiveConstructor{nothing}\<%
\\
\\[\AgdaEmptyExtraSkip]%
\>[2]\AgdaFunction{stackfunP2PKHNonEmptyStackAbstractedCorEmptysNothing}\AgdaSpace{}%
\AgdaBound{msg₁}\AgdaSpace{}%
\AgdaBound{pbk₁}\AgdaSpace{}%
\AgdaInductiveConstructor{zero}\AgdaSpace{}%
\AgdaSymbol{=}\AgdaSpace{}%
\AgdaInductiveConstructor{refl}\<%
\\
\>[2]\AgdaFunction{stackfunP2PKHNonEmptyStackAbstractedCorEmptysNothing}\AgdaSpace{}%
\AgdaBound{msg₁}\AgdaSpace{}%
\AgdaBound{pbk₁}\AgdaSpace{}%
\AgdaSymbol{(}\AgdaInductiveConstructor{suc}\AgdaSpace{}%
\AgdaBound{x}\AgdaSymbol{)}\AgdaSpace{}%
\AgdaSymbol{=}\AgdaSpace{}%
\AgdaInductiveConstructor{refl}\<%
\\
\\[\AgdaEmptyExtraSkip]%
\>[0]\AgdaComment{\{-\ Now\ we\ look\ at\ what\ happens\ if\ we\ have\ non\ empty\ stack₁\ and\ comparision\ >\ 0}\<%
\\
\>[0]\AgdaComment{-\}}\<%
\\
\\[\AgdaEmptyExtraSkip]%
\>[0]\AgdaComment{\{-\ if\ we\ evaluate\ it\ we\ get}\<%
\\
\>[0]\<%
\\
\>[0]\AgdaComment{just\ (boolToNat\ (isSigned\ \ msg₁\ sig₁\ pbk)\ ∷\ stack₁)}\<%
\\
\>[0]\<%
\\
\>[0]\<%
\\
\>[0]\<%
\\
\>[0]\AgdaComment{and\ we\ show\ that\ this\ is\ the\ case}\<%
\\
\>[0]\<%
\\
\>[0]\AgdaComment{-\}}\<%
\\
\>[0]\<%
\end{code}
} 

\newcommand{\stackVerificationPtoPKHsymbolicExecutionstacknonempty}{
\begin{code}%
\>[0]\AgdaFunction{testStackfunP2PKHNonEmptyStackAbstractedCompreSucNonEmpty}\AgdaSpace{}%
\AgdaSymbol{:}\AgdaSpace{}%
\AgdaDatatype{Maybe}\AgdaSpace{}%
\AgdaFunction{Stack}\<%
\\
\>[0]\AgdaFunction{testStackfunP2PKHNonEmptyStackAbstractedCompreSucNonEmpty}\AgdaSpace{}%
\AgdaSymbol{=}\AgdaSpace{}%
\AgdaFunction{p2PKHNonEmptyStackAbstr'}\AgdaSpace{}%
\AgdaPostulate{msg₁}\AgdaSpace{}%
\AgdaPostulate{pbk}\AgdaSpace{}%
\AgdaSymbol{(}\AgdaPostulate{sig₁}%
\>[101]\AgdaOperator{\AgdaInductiveConstructor{∷}}\AgdaSpace{}%
\AgdaPostulate{stack₁}\AgdaSymbol{)}\AgdaSpace{}%
\AgdaSymbol{(}\AgdaInductiveConstructor{suc}\AgdaSpace{}%
\AgdaPostulate{x₁}\AgdaSymbol{)}\<%
\end{code}
} 

\AgdaHide{
\begin{code}%
\>[0]\<%
\\
\\[\AgdaEmptyExtraSkip]%
\>[0]\AgdaFunction{stackfunP2PKHNonEmptyStackAbstractedCorComprSucStackNonEmptyCor}\AgdaSpace{}%
\AgdaSymbol{:}\<%
\\
\>[0][@{}l@{\AgdaIndent{0}}]%
\>[7]\AgdaSymbol{(}\AgdaBound{msg₁}\AgdaSpace{}%
\AgdaSymbol{:}\AgdaSpace{}%
\AgdaDatatype{Msg}\AgdaSymbol{)(}\AgdaBound{pbk}\AgdaSpace{}%
\AgdaSymbol{:}\AgdaSpace{}%
\AgdaDatatype{ℕ}\AgdaSymbol{)(}\AgdaBound{x}\AgdaSpace{}%
\AgdaSymbol{:}\AgdaSpace{}%
\AgdaDatatype{ℕ}\AgdaSymbol{)(}\AgdaBound{sig₁}\AgdaSpace{}%
\AgdaSymbol{:}\AgdaSpace{}%
\AgdaDatatype{ℕ}\AgdaSymbol{)(}\AgdaBound{stack₂}\AgdaSpace{}%
\AgdaSymbol{:}\AgdaSpace{}%
\AgdaFunction{Stack}\AgdaSymbol{)}\<%
\\
\>[7]\AgdaSymbol{→}%
\>[623I]\AgdaFunction{p2PKHNonEmptyStackAbstr'}\AgdaSpace{}%
\AgdaBound{msg₁}\AgdaSpace{}%
\AgdaBound{pbk}\AgdaSpace{}%
\AgdaSymbol{(}\AgdaBound{sig₁}%
\>[50]\AgdaOperator{\AgdaInductiveConstructor{∷}}\AgdaSpace{}%
\AgdaBound{stack₂}\AgdaSymbol{)}\AgdaSpace{}%
\AgdaSymbol{(}\AgdaInductiveConstructor{suc}\AgdaSpace{}%
\AgdaBound{x}\AgdaSymbol{)}\<%
\\
\>[623I][@{}l@{\AgdaIndent{0}}]%
\>[10]\AgdaOperator{\AgdaDatatype{≡}}%
\>[13]\AgdaInductiveConstructor{just}\AgdaSpace{}%
\AgdaSymbol{(}\AgdaFunction{boolToNat}\AgdaSpace{}%
\AgdaSymbol{(}\AgdaFunction{isSigned}%
\>[40]\AgdaBound{msg₁}\AgdaSpace{}%
\AgdaBound{sig₁}\AgdaSpace{}%
\AgdaBound{pbk}\AgdaSymbol{)}\AgdaSpace{}%
\AgdaOperator{\AgdaInductiveConstructor{∷}}\AgdaSpace{}%
\AgdaBound{stack₂}\AgdaSymbol{)}\<%
\\
\>[0]\AgdaFunction{stackfunP2PKHNonEmptyStackAbstractedCorComprSucStackNonEmptyCor}\AgdaSpace{}%
\AgdaBound{msg₂}\AgdaSpace{}%
\AgdaBound{pbk₁}\AgdaSpace{}%
\AgdaBound{sig₁}\AgdaSpace{}%
\AgdaBound{x}\AgdaSpace{}%
\AgdaBound{stack₃}\AgdaSpace{}%
\AgdaSymbol{=}\AgdaSpace{}%
\AgdaInductiveConstructor{refl}\<%
\\
\\[\AgdaEmptyExtraSkip]%
\>[0]\AgdaComment{\{-\ the\ following\ can\ be\ ommitted\ probably}\<%
\\
\>[0]\AgdaComment{\ \ \ since\ we\ have\ digged\ out\ the\ function\ completely}\<%
\\
\>[0]\AgdaComment{\ \ \ but\ here\ is\ anyway\ a\ theorem\ that\ the\ original\ function\ gives\ you\ nothing}\<%
\\
\>[0]\AgdaComment{\ \ \ \ \ \ \ \ \ if\ the\ stack\ has\ hight\ 1\ only}\<%
\\
\>[0]\<%
\\
\>[0]\AgdaComment{-\}}\<%
\\
\\[\AgdaEmptyExtraSkip]%
\\[\AgdaEmptyExtraSkip]%
\>[0]\AgdaComment{\{-\ this\ function\ is\ obolete\ but\ an\ interesting\ observation\ -\}}\<%
\\
\\[\AgdaEmptyExtraSkip]%
\>[0]\AgdaFunction{stackfunP2PKHemptySingleStackIsNothing}\AgdaSpace{}%
\AgdaSymbol{:}\AgdaSpace{}%
\AgdaSymbol{(}\AgdaBound{pubKeyHash}\AgdaSpace{}%
\AgdaSymbol{:}\AgdaSpace{}%
\AgdaDatatype{ℕ}\AgdaSymbol{)(}\AgdaBound{time₁}\AgdaSpace{}%
\AgdaSymbol{:}\AgdaSpace{}%
\AgdaFunction{Time}\AgdaSymbol{)(}\AgdaBound{msg₁}\AgdaSpace{}%
\AgdaSymbol{:}\AgdaSpace{}%
\AgdaDatatype{Msg}\AgdaSymbol{)(}\AgdaBound{pbk}\AgdaSpace{}%
\AgdaSymbol{:}\AgdaSpace{}%
\AgdaDatatype{ℕ}\AgdaSymbol{)}\<%
\\
\>[0][@{}l@{\AgdaIndent{0}}]%
\>[8]\AgdaSymbol{→}\AgdaSpace{}%
\AgdaOperator{\AgdaFunction{⟦}}\AgdaSpace{}%
\AgdaFunction{scriptP2PKHᵇ}\AgdaSpace{}%
\AgdaBound{pubKeyHash}\AgdaSpace{}%
\AgdaOperator{\AgdaFunction{⟧ˢ}}\AgdaSpace{}%
\AgdaBound{time₁}\AgdaSpace{}%
\AgdaBound{msg₁}\AgdaSpace{}%
\AgdaSymbol{(}\AgdaBound{pbk}\AgdaSpace{}%
\AgdaOperator{\AgdaInductiveConstructor{∷}}\AgdaSpace{}%
\AgdaInductiveConstructor{[]}\AgdaSymbol{)}\AgdaSpace{}%
\AgdaOperator{\AgdaDatatype{≡}}\AgdaSpace{}%
\AgdaInductiveConstructor{nothing}\<%
\\
\>[0]\AgdaFunction{stackfunP2PKHemptySingleStackIsNothing}%
\>[40]\AgdaBound{pubKeyHash}\AgdaSpace{}%
\AgdaBound{time₁}\AgdaSpace{}%
\AgdaBound{msg₁}\AgdaSpace{}%
\AgdaBound{pbk}\<%
\\
\>[0][@{}l@{\AgdaIndent{0}}]%
\>[2]\AgdaSymbol{=}%
\>[5]\AgdaOperator{\AgdaFunction{⟦}}%
\>[667I]\AgdaFunction{scriptP2PKHᵇ}\AgdaSpace{}%
\AgdaBound{pubKeyHash}\AgdaSpace{}%
\AgdaOperator{\AgdaFunction{⟧ˢ}}\AgdaSpace{}%
\AgdaBound{time₁}\AgdaSpace{}%
\AgdaBound{msg₁}\AgdaSpace{}%
\AgdaSymbol{(}\AgdaBound{pbk}\AgdaSpace{}%
\AgdaOperator{\AgdaInductiveConstructor{∷}}\AgdaSpace{}%
\AgdaInductiveConstructor{[]}\AgdaSymbol{)}\<%
\\
\>[667I][@{}l@{\AgdaIndent{0}}]%
\>[15]\AgdaFunction{≡⟨}\AgdaSpace{}%
\AgdaFunction{stackfunP2PKHNonEmptyStackAbstractedCor}\AgdaSpace{}%
\AgdaBound{pubKeyHash}\AgdaSpace{}%
\AgdaBound{time₁}\AgdaSpace{}%
\AgdaBound{msg₁}\AgdaSpace{}%
\AgdaBound{pbk}\AgdaSpace{}%
\AgdaInductiveConstructor{[]}%
\>[89]\AgdaFunction{⟩}\<%
\\
\>[5]\AgdaFunction{p2PKHNonEmptyStackAbstr'}\AgdaSpace{}%
\AgdaBound{msg₁}\AgdaSpace{}%
\AgdaBound{pbk}\AgdaSpace{}%
\AgdaInductiveConstructor{[]}\AgdaSpace{}%
\AgdaSymbol{(}\AgdaFunction{compareNaturals}\AgdaSpace{}%
\AgdaBound{pubKeyHash}\AgdaSpace{}%
\AgdaSymbol{(}\AgdaFunction{hashFun}\AgdaSpace{}%
\AgdaBound{pbk}\AgdaSymbol{))}\<%
\\
\>[5][@{}l@{\AgdaIndent{0}}]%
\>[15]\AgdaFunction{≡⟨}\AgdaSpace{}%
\AgdaFunction{stackfunP2PKHNonEmptyStackAbstractedCorEmptysNothing}\AgdaSpace{}%
\AgdaBound{msg₁}\AgdaSpace{}%
\AgdaBound{pbk}\AgdaSpace{}%
\AgdaSymbol{(}\AgdaFunction{compareNaturals}\AgdaSpace{}%
\AgdaBound{pubKeyHash}\AgdaSpace{}%
\AgdaSymbol{(}\AgdaFunction{hashFun}\AgdaSpace{}%
\AgdaBound{pbk}\AgdaSymbol{))}%
\>[125]\AgdaFunction{⟩}\<%
\\
\>[5]\AgdaInductiveConstructor{nothing}\<%
\\
\>[5]\AgdaOperator{\AgdaFunction{∎}}\<%
\\
\\[\AgdaEmptyExtraSkip]%
\>[0]\AgdaFunction{abstrFunSucCaseNonEmptyNormalSubTerm1}\AgdaSpace{}%
\AgdaSymbol{:}\AgdaSpace{}%
\AgdaDatatype{ℕ}\<%
\\
\>[0]\AgdaFunction{abstrFunSucCaseNonEmptyNormalSubTerm1}\AgdaSpace{}%
\AgdaSymbol{=}\<%
\\
\>[0]\AgdaComment{--\ \textbackslash{}stackVerificationPtoPKHsymbolicExecutionabstrFunSucCaseNonEmptySubTermOne}\<%
\end{code}
} 

\newcommand{\stackVerificationPtoPKHsymbolicExecutionabstrFunSucCaseNonEmptySubTermOne}{
\begin{code}[inline]%
\>[0][@{}l@{\AgdaIndent{1}}]%
\>[1]\AgdaFunction{boolToNat}\AgdaSpace{}%
\AgdaSymbol{(}\AgdaFunction{isSigned}\AgdaSpace{}%
\AgdaPostulate{msg₁}\AgdaSpace{}%
\AgdaPostulate{sig₁}\AgdaSpace{}%
\AgdaPostulate{pbk}\AgdaSymbol{)}\<%
\end{code}
} 

\AgdaHide{
\begin{code}%
\>[0]\<%
\\
\>[0]\AgdaFunction{abstrFunSucCaseNonEmptyNormalSubTerm2}\AgdaSpace{}%
\AgdaSymbol{:}\AgdaSpace{}%
\AgdaDatatype{Bool}\<%
\\
\>[0]\AgdaFunction{abstrFunSucCaseNonEmptyNormalSubTerm2}\AgdaSpace{}%
\AgdaSymbol{=}\<%
\\
\>[0]\AgdaComment{--\ \textbackslash{}stackVerificationPtoPKHsymbolicExecutionabstrFunSucCaseNonEmptySubTermTwo}\<%
\end{code}
} 

\newcommand{\stackVerificationPtoPKHsymbolicExecutionabstrFunSucCaseNonEmptySubTermTwo}{
\begin{code}[inline]%
\>[0][@{}l@{\AgdaIndent{1}}]%
\>[1]\AgdaFunction{isSigned}\AgdaSpace{}%
\AgdaPostulate{msg₁}\AgdaSpace{}%
\AgdaPostulate{sig₁}\AgdaSpace{}%
\AgdaPostulate{pbk}\<%
\end{code}
} 

\AgdaHide{
\begin{code}%
\>[0]\<%
\\
\\[\AgdaEmptyExtraSkip]%
\>[0]\AgdaComment{--\ uncomment\ to\ be\ able\ to\ evaluate}\<%
\\
\>[0]\AgdaComment{--\ test\ :\ Set}\<%
\\
\>[0]\AgdaComment{--\ test\ =\ \{!!\}}\<%
\end{code}
} 


\AgdaHide{
\begin{code}%
\>[0]\AgdaComment{--\ \textbackslash{}stackVerificationPtoPKHUsingEqualityOfPrograms}\<%
\\
\>[0]\AgdaKeyword{open}\AgdaSpace{}%
\AgdaKeyword{import}\AgdaSpace{}%
\AgdaModule{basicBitcoinDataType}\<%
\\
\\[\AgdaEmptyExtraSkip]%
\>[0]\AgdaKeyword{module}\AgdaSpace{}%
\AgdaModule{verificationStackScripts.stackVerificationP2PKHUsingEqualityOfPrograms}\AgdaSymbol{(}\AgdaBound{param}\AgdaSpace{}%
\AgdaSymbol{:}\AgdaSpace{}%
\AgdaRecord{GlobalParameters}\AgdaSymbol{)}%
\>[105]\AgdaKeyword{where}\<%
\\
\\[\AgdaEmptyExtraSkip]%
\\[\AgdaEmptyExtraSkip]%
\\[\AgdaEmptyExtraSkip]%
\>[0]\AgdaKeyword{open}\AgdaSpace{}%
\AgdaKeyword{import}\AgdaSpace{}%
\AgdaModule{Data.Nat}%
\>[22]\AgdaKeyword{hiding}\AgdaSpace{}%
\AgdaSymbol{(}\AgdaOperator{\AgdaDatatype{\AgdaUnderscore{}≤\AgdaUnderscore{}}}\AgdaSymbol{)}\<%
\\
\>[0]\AgdaKeyword{open}\AgdaSpace{}%
\AgdaKeyword{import}\AgdaSpace{}%
\AgdaModule{Data.List}\AgdaSpace{}%
\AgdaKeyword{hiding}\AgdaSpace{}%
\AgdaSymbol{(}\AgdaOperator{\AgdaFunction{\AgdaUnderscore{}\ensuremath{+\!\!+}\AgdaUnderscore{}}}\AgdaSymbol{)}\<%
\\
\>[0]\AgdaKeyword{open}\AgdaSpace{}%
\AgdaKeyword{import}\AgdaSpace{}%
\AgdaModule{Data.Unit}%
\>[23]\AgdaKeyword{hiding}\AgdaSpace{}%
\AgdaSymbol{(}\AgdaOperator{\AgdaRecord{\AgdaUnderscore{}≤\AgdaUnderscore{}}}\AgdaSymbol{)}\<%
\\
\>[0]\AgdaKeyword{open}\AgdaSpace{}%
\AgdaKeyword{import}\AgdaSpace{}%
\AgdaModule{Data.Empty}\<%
\\
\>[0]\AgdaKeyword{open}\AgdaSpace{}%
\AgdaKeyword{import}\AgdaSpace{}%
\AgdaModule{Data.Sum}\<%
\\
\>[0]\AgdaKeyword{open}\AgdaSpace{}%
\AgdaKeyword{import}\AgdaSpace{}%
\AgdaModule{Data.Bool}%
\>[23]\AgdaKeyword{hiding}\AgdaSpace{}%
\AgdaSymbol{(}\AgdaOperator{\AgdaDatatype{\AgdaUnderscore{}≤\AgdaUnderscore{}}}\AgdaSpace{}%
\AgdaSymbol{;}\AgdaSpace{}%
\AgdaOperator{\AgdaFunction{if\AgdaUnderscore{}then\AgdaUnderscore{}else\AgdaUnderscore{}}}\AgdaSpace{}%
\AgdaSymbol{)}\AgdaSpace{}%
\AgdaKeyword{renaming}\AgdaSpace{}%
\AgdaSymbol{(}\AgdaOperator{\AgdaFunction{\AgdaUnderscore{}∧\AgdaUnderscore{}}}\AgdaSpace{}%
\AgdaSymbol{to}\AgdaSpace{}%
\AgdaOperator{\AgdaFunction{\AgdaUnderscore{}∧b\AgdaUnderscore{}}}\AgdaSpace{}%
\AgdaSymbol{;}\AgdaSpace{}%
\AgdaOperator{\AgdaFunction{\AgdaUnderscore{}∨\AgdaUnderscore{}}}\AgdaSpace{}%
\AgdaSymbol{to}\AgdaSpace{}%
\AgdaOperator{\AgdaFunction{\AgdaUnderscore{}∨b\AgdaUnderscore{}}}\AgdaSpace{}%
\AgdaSymbol{;}\AgdaSpace{}%
\AgdaFunction{T}\AgdaSpace{}%
\AgdaSymbol{to}\AgdaSpace{}%
\AgdaFunction{True}\AgdaSymbol{)}\<%
\\
\>[0]\AgdaKeyword{open}\AgdaSpace{}%
\AgdaKeyword{import}\AgdaSpace{}%
\AgdaModule{Data.Product}\AgdaSpace{}%
\AgdaKeyword{renaming}\AgdaSpace{}%
\AgdaSymbol{(}\AgdaOperator{\AgdaInductiveConstructor{\AgdaUnderscore{},\AgdaUnderscore{}}}\AgdaSpace{}%
\AgdaSymbol{to}\AgdaSpace{}%
\AgdaOperator{\AgdaInductiveConstructor{\AgdaUnderscore{},,\AgdaUnderscore{}}}\AgdaSpace{}%
\AgdaSymbol{)}\<%
\\
\>[0]\AgdaKeyword{open}\AgdaSpace{}%
\AgdaKeyword{import}\AgdaSpace{}%
\AgdaModule{Data.Nat.Base}\AgdaSpace{}%
\AgdaKeyword{hiding}\AgdaSpace{}%
\AgdaSymbol{(}\AgdaOperator{\AgdaDatatype{\AgdaUnderscore{}≤\AgdaUnderscore{}}}\AgdaSymbol{)}\<%
\\
\>[0]\AgdaKeyword{open}\AgdaSpace{}%
\AgdaKeyword{import}\AgdaSpace{}%
\AgdaModule{Data.List.NonEmpty}\AgdaSpace{}%
\AgdaKeyword{hiding}\AgdaSpace{}%
\AgdaSymbol{(}\AgdaField{head}\AgdaSpace{}%
\AgdaSymbol{;}\AgdaSpace{}%
\AgdaOperator{\AgdaFunction{[\AgdaUnderscore{}]}}\AgdaSpace{}%
\AgdaSymbol{)}\<%
\\
\>[0]\AgdaKeyword{open}\AgdaSpace{}%
\AgdaKeyword{import}\AgdaSpace{}%
\AgdaModule{Data.Maybe}\<%
\\
\>[0]\AgdaKeyword{open}\AgdaSpace{}%
\AgdaKeyword{import}\AgdaSpace{}%
\AgdaModule{Relation.Nullary}\<%
\\
\\[\AgdaEmptyExtraSkip]%
\>[0]\AgdaKeyword{import}\AgdaSpace{}%
\AgdaModule{Relation.Binary.PropositionalEquality}\AgdaSpace{}%
\AgdaSymbol{as}\AgdaSpace{}%
\AgdaModule{Eq}\<%
\\
\>[0]\AgdaKeyword{open}\AgdaSpace{}%
\AgdaModule{Eq}\AgdaSpace{}%
\AgdaKeyword{using}\AgdaSpace{}%
\AgdaSymbol{(}\AgdaOperator{\AgdaDatatype{\AgdaUnderscore{}≡\AgdaUnderscore{}}}\AgdaSymbol{;}\AgdaSpace{}%
\AgdaInductiveConstructor{refl}\AgdaSymbol{;}\AgdaSpace{}%
\AgdaFunction{cong}\AgdaSymbol{;}\AgdaSpace{}%
\AgdaKeyword{module}\AgdaSpace{}%
\AgdaModule{≡-Reasoning}\AgdaSymbol{;}\AgdaSpace{}%
\AgdaFunction{sym}\AgdaSymbol{)}\<%
\\
\>[0]\AgdaKeyword{open}\AgdaSpace{}%
\AgdaModule{≡-Reasoning}\<%
\\
\>[0]\AgdaKeyword{open}\AgdaSpace{}%
\AgdaKeyword{import}\AgdaSpace{}%
\AgdaModule{Agda.Builtin.Equality}\<%
\\
\\[\AgdaEmptyExtraSkip]%
\\[\AgdaEmptyExtraSkip]%
\>[0]\AgdaComment{--our\ libraries}\<%
\\
\>[0]\AgdaKeyword{open}\AgdaSpace{}%
\AgdaKeyword{import}\AgdaSpace{}%
\AgdaModule{libraries.listLib}\<%
\\
\>[0]\AgdaKeyword{open}\AgdaSpace{}%
\AgdaKeyword{import}\AgdaSpace{}%
\AgdaModule{libraries.equalityLib}\<%
\\
\>[0]\AgdaKeyword{open}\AgdaSpace{}%
\AgdaKeyword{import}\AgdaSpace{}%
\AgdaModule{libraries.natLib}\<%
\\
\>[0]\AgdaKeyword{open}\AgdaSpace{}%
\AgdaKeyword{import}\AgdaSpace{}%
\AgdaModule{libraries.boolLib}\<%
\\
\>[0]\AgdaKeyword{open}\AgdaSpace{}%
\AgdaKeyword{import}\AgdaSpace{}%
\AgdaModule{libraries.emptyLib}\<%
\\
\>[0]\AgdaKeyword{open}\AgdaSpace{}%
\AgdaKeyword{import}\AgdaSpace{}%
\AgdaModule{libraries.andLib}\<%
\\
\>[0]\AgdaKeyword{open}\AgdaSpace{}%
\AgdaKeyword{import}\AgdaSpace{}%
\AgdaModule{libraries.miscLib}\<%
\\
\>[0]\AgdaKeyword{open}\AgdaSpace{}%
\AgdaKeyword{import}\AgdaSpace{}%
\AgdaModule{libraries.maybeLib}\<%
\\
\\[\AgdaEmptyExtraSkip]%
\>[0]\AgdaKeyword{open}\AgdaSpace{}%
\AgdaKeyword{import}\AgdaSpace{}%
\AgdaModule{stack}\<%
\\
\>[0]\AgdaKeyword{open}\AgdaSpace{}%
\AgdaKeyword{import}\AgdaSpace{}%
\AgdaModule{stackPredicate}\<%
\\
\>[0]\AgdaKeyword{open}\AgdaSpace{}%
\AgdaKeyword{import}\AgdaSpace{}%
\AgdaModule{semanticBasicOperations}\AgdaSpace{}%
\AgdaBound{param}\<%
\\
\>[0]\AgdaKeyword{open}\AgdaSpace{}%
\AgdaKeyword{import}\AgdaSpace{}%
\AgdaModule{stackSemanticsInstructions}\AgdaSpace{}%
\AgdaBound{param}\<%
\\
\>[0]\AgdaKeyword{open}\AgdaSpace{}%
\AgdaKeyword{import}\AgdaSpace{}%
\AgdaModule{hoareTripleStack}\AgdaSpace{}%
\AgdaBound{param}\<%
\\
\>[0]\AgdaKeyword{open}\AgdaSpace{}%
\AgdaKeyword{import}\AgdaSpace{}%
\AgdaModule{instruction}\<%
\\
\>[0]\AgdaKeyword{open}\AgdaSpace{}%
\AgdaKeyword{import}\AgdaSpace{}%
\AgdaModule{verificationP2PKHbasic}\AgdaSpace{}%
\AgdaBound{param}\<%
\\
\>[0]\AgdaKeyword{open}\AgdaSpace{}%
\AgdaKeyword{import}\AgdaSpace{}%
\AgdaModule{verificationStackScripts.stackState}\<%
\\
\>[0]\AgdaKeyword{open}\AgdaSpace{}%
\AgdaKeyword{import}\AgdaSpace{}%
\AgdaModule{verificationStackScripts.sPredicate}\<%
\\
\>[0]\AgdaKeyword{open}\AgdaSpace{}%
\AgdaKeyword{import}\AgdaSpace{}%
\AgdaModule{verificationStackScripts.stackHoareTriple}\AgdaSpace{}%
\AgdaBound{param}\<%
\\
\>[0]\AgdaKeyword{open}\AgdaSpace{}%
\AgdaKeyword{import}\AgdaSpace{}%
\AgdaModule{verificationStackScripts.stackVerificationLemmas}\AgdaSpace{}%
\AgdaBound{param}\<%
\\
\>[0]\AgdaKeyword{open}\AgdaSpace{}%
\AgdaKeyword{import}\AgdaSpace{}%
\AgdaModule{verificationStackScripts.stackSemanticsInstructionsBasic}\AgdaSpace{}%
\AgdaBound{param}\<%
\\
\>[0]\AgdaKeyword{open}\AgdaSpace{}%
\AgdaKeyword{import}\AgdaSpace{}%
\AgdaModule{verificationStackScripts.semanticsStackInstructions}\AgdaSpace{}%
\AgdaBound{param}\<%
\\
\>[0]\AgdaKeyword{open}\AgdaSpace{}%
\AgdaKeyword{import}\AgdaSpace{}%
\AgdaModule{verificationStackScripts.stackVerificationP2PKH}\AgdaSpace{}%
\AgdaBound{param}\<%
\\
\>[0]\AgdaKeyword{open}\AgdaSpace{}%
\AgdaKeyword{import}\AgdaSpace{}%
\AgdaModule{verificationStackScripts.stackVerificationP2PKHindexed}\AgdaSpace{}%
\AgdaBound{param}\<%
\\
\>[0]\AgdaKeyword{open}\AgdaSpace{}%
\AgdaKeyword{import}\AgdaSpace{}%
\AgdaModule{verificationStackScripts.stackVerificationP2PKHextractedProgram}\AgdaSpace{}%
\AgdaBound{param}\<%
\\
\>[0]\AgdaKeyword{open}\AgdaSpace{}%
\AgdaKeyword{import}\AgdaSpace{}%
\AgdaModule{verificationStackScripts.hoareTripleStackBasic}\AgdaSpace{}%
\AgdaBound{param}\<%
\\
\>[0]\AgdaKeyword{open}\AgdaSpace{}%
\AgdaKeyword{import}\AgdaSpace{}%
\AgdaModule{verificationStackScripts.stackVerificationLemmasPart2}\AgdaSpace{}%
\AgdaBound{param}\<%
\\
\>[0]\AgdaComment{-------------------------------------------------------}\<%
\\
\>[0]\AgdaComment{--\ The\ symbolic\ execution\ has\ been\ moved\ to}\<%
\\
\>[0]\AgdaComment{--}\<%
\\
\>[0]\AgdaComment{--\ \ \ \ \ stackVerificationP2PKHsymbolicExecution.agda}\<%
\\
\>[0]\AgdaComment{--}\<%
\\
\>[0]\AgdaComment{--\ The\ extracted\ program\ obtained\ by\ the\ symbolic\ execution\ has\ been\ moved\ to}\<%
\\
\>[0]\AgdaComment{--}\<%
\\
\>[0]\AgdaComment{--\ \ stackVerificationP2PKHextractedProgram.agda}\<%
\\
\>[0]\AgdaComment{--}\<%
\\
\>[0]\AgdaComment{------------------------------------------------------------------------------}\<%
\\
\\[\AgdaEmptyExtraSkip]%
\\[\AgdaEmptyExtraSkip]%
\\[\AgdaEmptyExtraSkip]%
\>[0]\AgdaComment{\{-\ \ \ Discussion\ with\ Arnold:}\<%
\\
\>[0]\AgdaComment{\ \ \ \ \ \ \ maybe\ better\ not\ to\ create\ the\ optimized\ program\ because\ we\ obtained\ this\ program\ in\ a\ systematic\ way}\<%
\\
\>[0]\AgdaComment{\ \ \ \ \ \ \ and\ can\ from\ it\ read\ off\ the\ weakest\ pre\ condition\ already}\<%
\\
\>[0]\AgdaComment{\ \ \ \ \ \ \ Discussion\ of\ further\ improvement\ (permutative\ conversions\ in\ a\ systematic\ way)\ for\ future\ work}\<%
\\
\>[0]\AgdaComment{-\}}\<%
\\
\\[\AgdaEmptyExtraSkip]%
\\[\AgdaEmptyExtraSkip]%
\>[0]\AgdaComment{--\ \textbackslash{}stackVerificationPtoPKHUsingEqualityOfProgramsptopkhNonEmptyStackAbstr}\<%
\end{code}
} 

\newcommand{\stackVerificationPtoPKHUsingEqualityOfProgramsptopkhNonEmptyStackAbstr}{
\begin{code}%
\>[0]\AgdaFunction{p2PKHNonEmptyStackAbstr}\AgdaSpace{}%
\AgdaSymbol{:}\AgdaSpace{}%
\AgdaSymbol{(}\AgdaBound{msg₁}\AgdaSpace{}%
\AgdaSymbol{:}\AgdaSpace{}%
\AgdaDatatype{Msg}\AgdaSymbol{)(}\AgdaBound{pbk}\AgdaSpace{}%
\AgdaSymbol{:}\AgdaSpace{}%
\AgdaDatatype{ℕ}\AgdaSymbol{)(}\AgdaBound{stack₁}\AgdaSpace{}%
\AgdaSymbol{:}\AgdaSpace{}%
\AgdaFunction{Stack}\AgdaSymbol{)(}\AgdaBound{cmp}\AgdaSpace{}%
\AgdaSymbol{:}\AgdaSpace{}%
\AgdaDatatype{ℕ}\AgdaSymbol{)}\AgdaSpace{}%
\AgdaSymbol{→}\AgdaSpace{}%
\AgdaDatatype{Maybe}\AgdaSpace{}%
\AgdaFunction{Stack}\<%
\\
\>[0]\AgdaFunction{p2PKHNonEmptyStackAbstr}\AgdaSpace{}%
\AgdaBound{msg₁}\AgdaSpace{}%
\AgdaBound{pbk}\AgdaSpace{}%
\AgdaBound{stack₁}\AgdaSpace{}%
\AgdaBound{cmp}\AgdaSpace{}%
\AgdaSymbol{=}%
\>[47]\AgdaFunction{executeStackVerify}\AgdaSpace{}%
\AgdaSymbol{(}\AgdaBound{cmp}\AgdaSpace{}%
\AgdaOperator{\AgdaInductiveConstructor{∷}}%
\>[74]\AgdaBound{pbk}\AgdaSpace{}%
\AgdaOperator{\AgdaInductiveConstructor{∷}}\AgdaSpace{}%
\AgdaBound{stack₁}\AgdaSymbol{)}\AgdaSpace{}%
\AgdaOperator{\AgdaFunction{\ensuremath{\mathbin{{>}\mkern-8.5mu{>}\mkern-2mu{=}}}}}\<%
\\
\>[47]\AgdaFunction{executeStackCheckSig}\AgdaSpace{}%
\AgdaBound{msg₁}\<%
\end{code}
} 

\AgdaHide{
\begin{code}%
\>[0]\<%
\\
\>[0]\AgdaFunction{stackfunP2PKHNonEmptyStackAbstractedCorCompr0IsNothing}\AgdaSpace{}%
\AgdaSymbol{:}\AgdaSpace{}%
\AgdaSymbol{(}\AgdaBound{msg₁}\AgdaSpace{}%
\AgdaSymbol{:}\AgdaSpace{}%
\AgdaDatatype{Msg}\AgdaSymbol{)(}\AgdaBound{pbk}\AgdaSpace{}%
\AgdaSymbol{:}\AgdaSpace{}%
\AgdaDatatype{ℕ}\AgdaSymbol{)(}\AgdaBound{stack₁}\AgdaSpace{}%
\AgdaSymbol{:}\AgdaSpace{}%
\AgdaFunction{Stack}\AgdaSymbol{)}\<%
\\
\>[0][@{}l@{\AgdaIndent{0}}]%
\>[6]\AgdaSymbol{→}%
\>[9]\AgdaFunction{p2PKHNonEmptyStackAbstr}\AgdaSpace{}%
\AgdaBound{msg₁}\AgdaSpace{}%
\AgdaBound{pbk}\AgdaSpace{}%
\AgdaBound{stack₁}\AgdaSpace{}%
\AgdaNumber{0}\AgdaSpace{}%
\AgdaOperator{\AgdaDatatype{≡}}\AgdaSpace{}%
\AgdaInductiveConstructor{nothing}\<%
\\
\>[0]\AgdaFunction{stackfunP2PKHNonEmptyStackAbstractedCorCompr0IsNothing}\AgdaSpace{}%
\AgdaBound{msg₁}\AgdaSpace{}%
\AgdaBound{pbk}\AgdaSpace{}%
\AgdaBound{stack₁}\AgdaSpace{}%
\AgdaSymbol{=}\AgdaSpace{}%
\AgdaInductiveConstructor{refl}\<%
\\
\\[\AgdaEmptyExtraSkip]%
\\[\AgdaEmptyExtraSkip]%
\>[0]\AgdaFunction{stackfunP2PKHNonEmptyStackAbstractedCorEmptysNothing}\AgdaSpace{}%
\AgdaSymbol{:}\AgdaSpace{}%
\AgdaSymbol{(}\AgdaBound{msg₁}\AgdaSpace{}%
\AgdaSymbol{:}\AgdaSpace{}%
\AgdaDatatype{Msg}\AgdaSymbol{)(}\AgdaBound{pbk}\AgdaSpace{}%
\AgdaSymbol{:}\AgdaSpace{}%
\AgdaDatatype{ℕ}\AgdaSymbol{)(}\AgdaBound{x}\AgdaSpace{}%
\AgdaSymbol{:}\AgdaSpace{}%
\AgdaDatatype{ℕ}\AgdaSymbol{)}\<%
\\
\>[0][@{}l@{\AgdaIndent{0}}]%
\>[6]\AgdaSymbol{→}%
\>[9]\AgdaFunction{p2PKHNonEmptyStackAbstr}\AgdaSpace{}%
\AgdaBound{msg₁}\AgdaSpace{}%
\AgdaBound{pbk}\AgdaSpace{}%
\AgdaInductiveConstructor{[]}\AgdaSpace{}%
\AgdaBound{x}\AgdaSpace{}%
\AgdaOperator{\AgdaDatatype{≡}}\AgdaSpace{}%
\AgdaInductiveConstructor{nothing}\<%
\\
\\[\AgdaEmptyExtraSkip]%
\>[0]\AgdaFunction{stackfunP2PKHNonEmptyStackAbstractedCorEmptysNothing}\AgdaSpace{}%
\AgdaBound{msg₁}\AgdaSpace{}%
\AgdaBound{pbk₁}\AgdaSpace{}%
\AgdaInductiveConstructor{zero}\AgdaSpace{}%
\AgdaSymbol{=}\AgdaSpace{}%
\AgdaInductiveConstructor{refl}\<%
\\
\>[0]\AgdaFunction{stackfunP2PKHNonEmptyStackAbstractedCorEmptysNothing}\AgdaSpace{}%
\AgdaBound{msg₁}\AgdaSpace{}%
\AgdaBound{pbk₁}\AgdaSpace{}%
\AgdaSymbol{(}\AgdaInductiveConstructor{suc}\AgdaSpace{}%
\AgdaBound{x}\AgdaSymbol{)}\AgdaSpace{}%
\AgdaSymbol{=}\AgdaSpace{}%
\AgdaInductiveConstructor{refl}\<%
\\
\\[\AgdaEmptyExtraSkip]%
\>[0]\AgdaFunction{stackfunP2PKHNonEmptyStackAbstractedCor}\AgdaSpace{}%
\AgdaSymbol{:}%
\>[43]\AgdaSymbol{(}\AgdaBound{pubKeyHash}\AgdaSpace{}%
\AgdaSymbol{:}\AgdaSpace{}%
\AgdaDatatype{ℕ}\AgdaSymbol{)(}\AgdaBound{time₁}\AgdaSpace{}%
\AgdaSymbol{:}\AgdaSpace{}%
\AgdaFunction{Time}\AgdaSymbol{)(}\AgdaBound{msg₁}\AgdaSpace{}%
\AgdaSymbol{:}\AgdaSpace{}%
\AgdaDatatype{Msg}\AgdaSymbol{)(}\AgdaBound{pbk}\AgdaSpace{}%
\AgdaSymbol{:}\AgdaSpace{}%
\AgdaDatatype{ℕ}\AgdaSymbol{)(}\AgdaBound{stack₁}\AgdaSpace{}%
\AgdaSymbol{:}\AgdaSpace{}%
\AgdaFunction{Stack}\AgdaSymbol{)}\<%
\\
\>[0][@{}l@{\AgdaIndent{0}}]%
\>[16]\AgdaSymbol{→}%
\>[217I]\AgdaOperator{\AgdaFunction{⟦}}\AgdaSpace{}%
\AgdaFunction{scriptP2PKHᵇ}\AgdaSpace{}%
\AgdaBound{pubKeyHash}\AgdaSpace{}%
\AgdaOperator{\AgdaFunction{⟧ˢ}}\AgdaSpace{}%
\AgdaBound{time₁}\AgdaSpace{}%
\AgdaBound{msg₁}\AgdaSpace{}%
\AgdaSymbol{(}\AgdaBound{pbk}\AgdaSpace{}%
\AgdaOperator{\AgdaInductiveConstructor{∷}}\AgdaSpace{}%
\AgdaBound{stack₁}\AgdaSymbol{)}\<%
\\
\>[217I][@{}l@{\AgdaIndent{0}}]%
\>[19]\AgdaOperator{\AgdaDatatype{≡}}%
\>[226I]\AgdaFunction{p2PKHNonEmptyStackAbstr}%
\>[46]\AgdaBound{msg₁}\AgdaSpace{}%
\AgdaBound{pbk}\AgdaSpace{}%
\AgdaBound{stack₁}\<%
\\
\>[226I][@{}l@{\AgdaIndent{0}}]%
\>[27]\AgdaSymbol{(}\AgdaFunction{compareNaturals}\AgdaSpace{}%
\AgdaBound{pubKeyHash}\AgdaSpace{}%
\AgdaSymbol{(}\AgdaFunction{hashFun}\AgdaSpace{}%
\AgdaBound{pbk}\AgdaSymbol{))}\<%
\\
\>[0]\AgdaFunction{stackfunP2PKHNonEmptyStackAbstractedCor}\AgdaSpace{}%
\AgdaBound{pubKeyHash}\AgdaSpace{}%
\AgdaBound{time₁}\AgdaSpace{}%
\AgdaBound{msg₁}\AgdaSpace{}%
\AgdaBound{pbk}\AgdaSpace{}%
\AgdaBound{stack₁}\AgdaSpace{}%
\AgdaSymbol{=}\AgdaSpace{}%
\AgdaInductiveConstructor{refl}\<%
\\
\\[\AgdaEmptyExtraSkip]%
\\[\AgdaEmptyExtraSkip]%
\>[0]\AgdaFunction{p2pkhFunctionDecodedAux1Cor}\AgdaSpace{}%
\AgdaSymbol{:}\AgdaSpace{}%
\AgdaSymbol{(}\AgdaBound{pbk}\AgdaSpace{}%
\AgdaSymbol{:}\AgdaSpace{}%
\AgdaDatatype{ℕ}\AgdaSymbol{)(}\AgdaBound{msg₁}\AgdaSpace{}%
\AgdaSymbol{:}\AgdaSpace{}%
\AgdaDatatype{Msg}\AgdaSymbol{)(}\AgdaBound{stack₁}\AgdaSpace{}%
\AgdaSymbol{:}\AgdaSpace{}%
\AgdaFunction{Stack}\AgdaSymbol{)(}\AgdaBound{cpRes}\AgdaSpace{}%
\AgdaSymbol{:}\AgdaSpace{}%
\AgdaDatatype{ℕ}\AgdaSymbol{)}\<%
\\
\>[0][@{}l@{\AgdaIndent{0}}]%
\>[1]\AgdaSymbol{→}\AgdaSpace{}%
\AgdaFunction{p2PKHNonEmptyStackAbstr}\AgdaSpace{}%
\AgdaBound{msg₁}\AgdaSpace{}%
\AgdaBound{pbk}\AgdaSpace{}%
\AgdaBound{stack₁}\AgdaSpace{}%
\AgdaBound{cpRes}\AgdaSpace{}%
\AgdaOperator{\AgdaDatatype{≡}}\AgdaSpace{}%
\AgdaFunction{p2pkhFunctionDecodedAux1}\AgdaSpace{}%
\AgdaBound{pbk}\AgdaSpace{}%
\AgdaBound{msg₁}\AgdaSpace{}%
\AgdaBound{stack₁}\AgdaSpace{}%
\AgdaBound{cpRes}\<%
\\
\\[\AgdaEmptyExtraSkip]%
\>[0]\AgdaFunction{p2pkhFunctionDecodedAux1Cor}\AgdaSpace{}%
\AgdaBound{pbk₁}\AgdaSpace{}%
\AgdaBound{msg₁}\AgdaSpace{}%
\AgdaInductiveConstructor{[]}\AgdaSpace{}%
\AgdaBound{cpRes}\AgdaSpace{}%
\AgdaSymbol{=}\AgdaSpace{}%
\AgdaFunction{stackfunP2PKHNonEmptyStackAbstractedCorEmptysNothing}\AgdaSpace{}%
\AgdaBound{msg₁}\AgdaSpace{}%
\AgdaBound{pbk₁}\AgdaSpace{}%
\AgdaBound{cpRes}\<%
\\
\>[0]\AgdaFunction{p2pkhFunctionDecodedAux1Cor}\AgdaSpace{}%
\AgdaBound{pbk₁}\AgdaSpace{}%
\AgdaBound{msg₁}\AgdaSpace{}%
\AgdaSymbol{(}\AgdaBound{x}\AgdaSpace{}%
\AgdaOperator{\AgdaInductiveConstructor{∷}}\AgdaSpace{}%
\AgdaBound{stack₁}\AgdaSymbol{)}\AgdaSpace{}%
\AgdaInductiveConstructor{zero}\AgdaSpace{}%
\AgdaSymbol{=}\AgdaSpace{}%
\AgdaInductiveConstructor{refl}\<%
\\
\>[0]\AgdaFunction{p2pkhFunctionDecodedAux1Cor}\AgdaSpace{}%
\AgdaBound{pbk₁}\AgdaSpace{}%
\AgdaBound{msg₁}\AgdaSpace{}%
\AgdaSymbol{(}\AgdaBound{x}\AgdaSpace{}%
\AgdaOperator{\AgdaInductiveConstructor{∷}}\AgdaSpace{}%
\AgdaBound{stack₁}\AgdaSymbol{)}\AgdaSpace{}%
\AgdaSymbol{(}\AgdaInductiveConstructor{suc}\AgdaSpace{}%
\AgdaBound{cpRes}\AgdaSymbol{)}\AgdaSpace{}%
\AgdaSymbol{=}\AgdaSpace{}%
\AgdaInductiveConstructor{refl}\<%
\\
\\[\AgdaEmptyExtraSkip]%
\>[0]\AgdaComment{--\ \textbackslash{}stackVerificationPtoPKHUsingEqualityOfProgramsptopkhFunctionDecodedcor}\<%
\end{code}
} 

\newcommand{\stackVerificationPtoPKHUsingEqualityOfProgramsptopkhFunctionDecodedcor}{
\begin{code}%
\>[0]\AgdaFunction{p2pkhFunctionDecodedcor}\AgdaSpace{}%
\AgdaSymbol{:}\AgdaSpace{}%
\AgdaSymbol{(}\AgdaBound{time₁}\AgdaSpace{}%
\AgdaSymbol{:}\AgdaSpace{}%
\AgdaDatatype{ℕ}\AgdaSymbol{)}\AgdaSpace{}%
\AgdaSymbol{(}\AgdaBound{pbkh}\AgdaSpace{}%
\AgdaSymbol{:}\AgdaSpace{}%
\AgdaDatatype{ℕ}\AgdaSymbol{)(}\AgdaBound{msg₁}\AgdaSpace{}%
\AgdaSymbol{:}\AgdaSpace{}%
\AgdaDatatype{Msg}\AgdaSymbol{)(}\AgdaBound{stack₁}\AgdaSpace{}%
\AgdaSymbol{:}\AgdaSpace{}%
\AgdaFunction{Stack}\AgdaSymbol{)}\<%
\\
\>[0][@{}l@{\AgdaIndent{0}}]%
\>[1]\AgdaSymbol{→}\AgdaSpace{}%
\AgdaOperator{\AgdaFunction{⟦}}\AgdaSpace{}%
\AgdaFunction{scriptP2PKHᵇ}\AgdaSpace{}%
\AgdaBound{pbkh}\AgdaSpace{}%
\AgdaOperator{\AgdaFunction{⟧ˢ}}\AgdaSpace{}%
\AgdaBound{time₁}\AgdaSpace{}%
\AgdaBound{msg₁}\AgdaSpace{}%
\AgdaBound{stack₁}%
\>[45]\AgdaOperator{\AgdaDatatype{≡}}\AgdaSpace{}%
\AgdaFunction{p2pkhFunctionDecoded}\AgdaSpace{}%
\AgdaBound{pbkh}%
\>[74]\AgdaBound{msg₁}\AgdaSpace{}%
\AgdaBound{stack₁}\<%
\end{code}
} 

\AgdaHide{
\begin{code}%
\>[0]\AgdaFunction{p2pkhFunctionDecodedcor}\AgdaSpace{}%
\AgdaBound{time₁}\AgdaSpace{}%
\AgdaBound{pbkh}\AgdaSpace{}%
\AgdaBound{msg₁}\AgdaSpace{}%
\AgdaInductiveConstructor{[]}\AgdaSpace{}%
\AgdaSymbol{=}\AgdaSpace{}%
\AgdaInductiveConstructor{refl}\<%
\\
\>[0]\AgdaFunction{p2pkhFunctionDecodedcor}\AgdaSpace{}%
\AgdaBound{time₁}\AgdaSpace{}%
\AgdaBound{pbkh}\AgdaSpace{}%
\AgdaBound{msg₁}\AgdaSpace{}%
\AgdaSymbol{(}\AgdaBound{pbk}\AgdaSpace{}%
\AgdaOperator{\AgdaInductiveConstructor{∷}}\AgdaSpace{}%
\AgdaBound{stack₁}\AgdaSymbol{)}\AgdaSpace{}%
\AgdaSymbol{=}\<%
\\
\>[0][@{}l@{\AgdaIndent{0}}]%
\>[7]\AgdaOperator{\AgdaFunction{⟦}}%
\>[320I]\AgdaFunction{scriptP2PKHᵇ}\AgdaSpace{}%
\AgdaBound{pbkh}\AgdaSpace{}%
\AgdaOperator{\AgdaFunction{⟧ˢ}}\AgdaSpace{}%
\AgdaBound{time₁}\AgdaSpace{}%
\AgdaBound{msg₁}\AgdaSpace{}%
\AgdaSymbol{(}\AgdaBound{pbk}\AgdaSpace{}%
\AgdaOperator{\AgdaInductiveConstructor{∷}}\AgdaSpace{}%
\AgdaBound{stack₁}\AgdaSymbol{)}\<%
\\
\>[320I][@{}l@{\AgdaIndent{0}}]%
\>[10]\AgdaFunction{≡⟨}\AgdaSpace{}%
\AgdaFunction{stackfunP2PKHNonEmptyStackAbstractedCor}\AgdaSpace{}%
\AgdaBound{pbkh}\AgdaSpace{}%
\AgdaBound{time₁}\AgdaSpace{}%
\AgdaBound{msg₁}\AgdaSpace{}%
\AgdaBound{pbk}\AgdaSpace{}%
\AgdaBound{stack₁}\AgdaSpace{}%
\AgdaFunction{⟩}\<%
\\
\>[7]\AgdaFunction{p2PKHNonEmptyStackAbstr}%
\>[32]\AgdaBound{msg₁}\AgdaSpace{}%
\AgdaBound{pbk}\AgdaSpace{}%
\AgdaBound{stack₁}\AgdaSpace{}%
\AgdaSymbol{(}\AgdaFunction{compareNaturals}\AgdaSpace{}%
\AgdaBound{pbkh}\AgdaSpace{}%
\AgdaSymbol{(}\AgdaFunction{hashFun}\AgdaSpace{}%
\AgdaBound{pbk}\AgdaSymbol{))}\<%
\\
\>[7][@{}l@{\AgdaIndent{0}}]%
\>[10]\AgdaFunction{≡⟨}\AgdaSpace{}%
\AgdaFunction{p2pkhFunctionDecodedAux1Cor}\AgdaSpace{}%
\AgdaBound{pbk}\AgdaSpace{}%
\AgdaBound{msg₁}\AgdaSpace{}%
\AgdaBound{stack₁}\AgdaSpace{}%
\AgdaSymbol{(}\AgdaFunction{compareNaturals}\AgdaSpace{}%
\AgdaBound{pbkh}\AgdaSpace{}%
\AgdaSymbol{(}\AgdaFunction{hashFun}\AgdaSpace{}%
\AgdaBound{pbk}\AgdaSymbol{))}\AgdaSpace{}%
\AgdaFunction{⟩}\<%
\\
\>[7]\AgdaFunction{p2pkhFunctionDecodedAux1}\AgdaSpace{}%
\AgdaBound{pbk}\AgdaSpace{}%
\AgdaBound{msg₁}\AgdaSpace{}%
\AgdaBound{stack₁}%
\>[51]\AgdaSymbol{(}\AgdaFunction{compareNaturals}\AgdaSpace{}%
\AgdaBound{pbkh}\AgdaSpace{}%
\AgdaSymbol{(}\AgdaFunction{hashFun}\AgdaSpace{}%
\AgdaBound{pbk}\AgdaSymbol{))}\<%
\\
\>[7]\AgdaOperator{\AgdaFunction{∎}}\<%
\\
\\[\AgdaEmptyExtraSkip]%
\>[0]\AgdaComment{\{-\ Now\ we\ just\ verify\ the\ \ \ hoare\ triple\ for\ the\ function\ we\ have\ found\ we\ have\ decoded}\<%
\\
\>[0]\<%
\\
\>[0]\AgdaComment{-\}}\<%
\\
\\[\AgdaEmptyExtraSkip]%
\>[0]\AgdaFunction{lemmaPHKcoraux3}%
\>[356I]\AgdaSymbol{:}\AgdaSpace{}%
\AgdaSymbol{(}\AgdaBound{x₁}\AgdaSpace{}%
\AgdaSymbol{:}\AgdaSpace{}%
\AgdaDatatype{ℕ}\AgdaSymbol{)(}\AgdaBound{time}\AgdaSpace{}%
\AgdaSymbol{:}\AgdaSpace{}%
\AgdaFunction{Time}\AgdaSymbol{)}\AgdaSpace{}%
\AgdaSymbol{(}\AgdaBound{msg₁}\AgdaSpace{}%
\AgdaSymbol{:}\AgdaSpace{}%
\AgdaDatatype{Msg}\AgdaSymbol{)}\AgdaSpace{}%
\AgdaSymbol{(}\AgdaBound{x₂}\AgdaSpace{}%
\AgdaSymbol{:}\AgdaSpace{}%
\AgdaDatatype{ℕ}\AgdaSymbol{)(}\AgdaBound{s}\AgdaSpace{}%
\AgdaSymbol{:}\AgdaSpace{}%
\AgdaFunction{Stack}\AgdaSymbol{)}\AgdaSpace{}%
\AgdaSymbol{(}\AgdaBound{x}\AgdaSpace{}%
\AgdaSymbol{:}\AgdaSpace{}%
\AgdaDatatype{ℕ}\AgdaSymbol{)}\AgdaSpace{}%
\AgdaSymbol{→}\<%
\\
\>[356I][@{}l@{\AgdaIndent{0}}]%
\>[17]\AgdaFunction{liftPred2Maybe}\AgdaSpace{}%
\AgdaSymbol{(λ}\AgdaSpace{}%
\AgdaBound{s₁}\AgdaSpace{}%
\AgdaSymbol{→}\AgdaSpace{}%
\AgdaFunction{acceptStateˢ}\AgdaSpace{}%
\AgdaBound{time}\AgdaSpace{}%
\AgdaBound{msg₁}\AgdaSpace{}%
\AgdaBound{s₁}\AgdaSymbol{)}\<%
\\
\>[17][@{}l@{\AgdaIndent{0}}]%
\>[18]\AgdaSymbol{(}\AgdaFunction{p2pkhFunctionDecodedAux1}\AgdaSpace{}%
\AgdaBound{x₁}\AgdaSpace{}%
\AgdaBound{msg₁}\AgdaSpace{}%
\AgdaSymbol{(}\AgdaBound{x₂}\AgdaSpace{}%
\AgdaOperator{\AgdaInductiveConstructor{∷}}\AgdaSpace{}%
\AgdaBound{s}\AgdaSymbol{)}\AgdaSpace{}%
\AgdaBound{x}\AgdaSymbol{)}\<%
\\
\>[18]\AgdaSymbol{→}\AgdaSpace{}%
\AgdaOperator{\AgdaFunction{¬}}\AgdaSpace{}%
\AgdaSymbol{(}\AgdaBound{x}\AgdaSpace{}%
\AgdaOperator{\AgdaDatatype{≡}}\AgdaSpace{}%
\AgdaNumber{0}\AgdaSpace{}%
\AgdaSymbol{)}\<%
\\
\>[0]\AgdaFunction{lemmaPHKcoraux3}\AgdaSpace{}%
\AgdaBound{x₁}\AgdaSpace{}%
\AgdaBound{time}\AgdaSpace{}%
\AgdaBound{msg₁}\AgdaSpace{}%
\AgdaBound{x₂}\AgdaSpace{}%
\AgdaBound{s}\AgdaSpace{}%
\AgdaInductiveConstructor{zero}\AgdaSpace{}%
\AgdaSymbol{()}\AgdaSpace{}%
\AgdaBound{x₄}\<%
\\
\>[0]\AgdaFunction{lemmaPHKcoraux3}\AgdaSpace{}%
\AgdaBound{x₁}\AgdaSpace{}%
\AgdaBound{time}\AgdaSpace{}%
\AgdaBound{msg₁}\AgdaSpace{}%
\AgdaBound{x₂}\AgdaSpace{}%
\AgdaBound{s}\AgdaSpace{}%
\AgdaSymbol{(}\AgdaInductiveConstructor{suc}\AgdaSpace{}%
\AgdaBound{x}\AgdaSymbol{)}\AgdaSpace{}%
\AgdaBound{x₃}\AgdaSpace{}%
\AgdaSymbol{()}\<%
\\
\\[\AgdaEmptyExtraSkip]%
\>[0]\AgdaFunction{lemmaCompareNat2}\AgdaSpace{}%
\AgdaSymbol{:}\AgdaSpace{}%
\AgdaSymbol{(}\AgdaSpace{}%
\AgdaBound{x}\AgdaSpace{}%
\AgdaBound{y}\AgdaSpace{}%
\AgdaSymbol{:}\AgdaSpace{}%
\AgdaDatatype{ℕ}\AgdaSpace{}%
\AgdaSymbol{)}\AgdaSpace{}%
\AgdaSymbol{→}\AgdaSpace{}%
\AgdaOperator{\AgdaFunction{¬}}\AgdaSpace{}%
\AgdaSymbol{(}\AgdaFunction{compareNaturals}\AgdaSpace{}%
\AgdaBound{x}\AgdaSpace{}%
\AgdaBound{y}\AgdaSpace{}%
\AgdaOperator{\AgdaDatatype{≡}}\AgdaSpace{}%
\AgdaNumber{0}\AgdaSpace{}%
\AgdaSymbol{)}\AgdaSpace{}%
\AgdaSymbol{→}\AgdaSpace{}%
\AgdaBound{x}\AgdaSpace{}%
\AgdaOperator{\AgdaDatatype{≡}}\AgdaSpace{}%
\AgdaBound{y}\<%
\\
\>[0]\AgdaFunction{lemmaCompareNat2}\AgdaSpace{}%
\AgdaInductiveConstructor{zero}\AgdaSpace{}%
\AgdaInductiveConstructor{zero}\AgdaSpace{}%
\AgdaBound{p}\AgdaSpace{}%
\AgdaSymbol{=}\AgdaSpace{}%
\AgdaInductiveConstructor{refl}\<%
\\
\>[0]\AgdaFunction{lemmaCompareNat2}\AgdaSpace{}%
\AgdaInductiveConstructor{zero}\AgdaSpace{}%
\AgdaSymbol{(}\AgdaInductiveConstructor{suc}\AgdaSpace{}%
\AgdaBound{y}\AgdaSymbol{)}\AgdaSpace{}%
\AgdaBound{p}\AgdaSpace{}%
\AgdaSymbol{=}\AgdaSpace{}%
\AgdaFunction{efq}\AgdaSpace{}%
\AgdaSymbol{(}\AgdaBound{p}\AgdaSpace{}%
\AgdaInductiveConstructor{refl}\AgdaSymbol{)}\<%
\\
\>[0]\AgdaFunction{lemmaCompareNat2}\AgdaSpace{}%
\AgdaSymbol{(}\AgdaInductiveConstructor{suc}\AgdaSpace{}%
\AgdaBound{x}\AgdaSymbol{)}\AgdaSpace{}%
\AgdaInductiveConstructor{zero}\AgdaSpace{}%
\AgdaBound{p}\AgdaSpace{}%
\AgdaSymbol{=}\AgdaSpace{}%
\AgdaFunction{efq}\AgdaSpace{}%
\AgdaSymbol{(}\AgdaBound{p}\AgdaSpace{}%
\AgdaInductiveConstructor{refl}\AgdaSymbol{)}\<%
\\
\>[0]\AgdaFunction{lemmaCompareNat2}\AgdaSpace{}%
\AgdaSymbol{(}\AgdaInductiveConstructor{suc}\AgdaSpace{}%
\AgdaBound{x}\AgdaSymbol{)}\AgdaSpace{}%
\AgdaSymbol{(}\AgdaInductiveConstructor{suc}\AgdaSpace{}%
\AgdaBound{y}\AgdaSymbol{)}\AgdaSpace{}%
\AgdaBound{p}\AgdaSpace{}%
\AgdaSymbol{=}\AgdaSpace{}%
\AgdaFunction{cong}\AgdaSpace{}%
\AgdaInductiveConstructor{suc}\AgdaSpace{}%
\AgdaSymbol{(}\AgdaFunction{lemmaCompareNat2}\AgdaSpace{}%
\AgdaBound{x}\AgdaSpace{}%
\AgdaBound{y}\AgdaSpace{}%
\AgdaBound{p}\AgdaSymbol{)}\<%
\\
\\[\AgdaEmptyExtraSkip]%
\\[\AgdaEmptyExtraSkip]%
\>[0]\AgdaFunction{lemmaPHKcoraux2}%
\>[461I]\AgdaSymbol{:}\AgdaSpace{}%
\AgdaSymbol{(}\AgdaBound{pbk}\AgdaSpace{}%
\AgdaSymbol{:}\AgdaSpace{}%
\AgdaDatatype{ℕ}\AgdaSymbol{)(}\AgdaBound{time}\AgdaSpace{}%
\AgdaSymbol{:}\AgdaSpace{}%
\AgdaFunction{Time}\AgdaSymbol{)}\AgdaSpace{}%
\AgdaSymbol{(}\AgdaBound{msg₁}\AgdaSpace{}%
\AgdaSymbol{:}\AgdaSpace{}%
\AgdaDatatype{Msg}\AgdaSymbol{)}\AgdaSpace{}%
\AgdaSymbol{(}\AgdaBound{sig}\AgdaSpace{}%
\AgdaSymbol{:}\AgdaSpace{}%
\AgdaDatatype{ℕ}\AgdaSymbol{)(}\AgdaBound{s}\AgdaSpace{}%
\AgdaSymbol{:}\AgdaSpace{}%
\AgdaFunction{Stack}\AgdaSymbol{)}\AgdaSpace{}%
\AgdaSymbol{(}\AgdaBound{cpRes}\AgdaSpace{}%
\AgdaSymbol{:}\AgdaSpace{}%
\AgdaDatatype{ℕ}\AgdaSymbol{)}\AgdaSpace{}%
\AgdaSymbol{→}\<%
\\
\>[461I][@{}l@{\AgdaIndent{0}}]%
\>[17]\AgdaFunction{liftPred2Maybe}\AgdaSpace{}%
\AgdaSymbol{(λ}\AgdaSpace{}%
\AgdaBound{s₁}\AgdaSpace{}%
\AgdaSymbol{→}\AgdaSpace{}%
\AgdaFunction{acceptStateˢ}\AgdaSpace{}%
\AgdaBound{time}\AgdaSpace{}%
\AgdaBound{msg₁}\AgdaSpace{}%
\AgdaBound{s₁}\AgdaSymbol{)}\<%
\\
\>[17][@{}l@{\AgdaIndent{0}}]%
\>[18]\AgdaSymbol{(}\AgdaFunction{p2pkhFunctionDecodedAux1}\AgdaSpace{}%
\AgdaBound{pbk}\AgdaSpace{}%
\AgdaBound{msg₁}\AgdaSpace{}%
\AgdaSymbol{(}\AgdaBound{sig}\AgdaSpace{}%
\AgdaOperator{\AgdaInductiveConstructor{∷}}\AgdaSpace{}%
\AgdaBound{s}\AgdaSymbol{)}\AgdaSpace{}%
\AgdaBound{cpRes}\AgdaSymbol{)}\<%
\\
\>[18]\AgdaSymbol{→}\AgdaSpace{}%
\AgdaFunction{NotFalse}\AgdaSpace{}%
\AgdaSymbol{(}\AgdaFunction{boolToNat}\AgdaSpace{}%
\AgdaSymbol{(}\AgdaFunction{isSigned}%
\>[51]\AgdaBound{msg₁}\AgdaSpace{}%
\AgdaBound{sig}\AgdaSpace{}%
\AgdaBound{pbk}\AgdaSymbol{))}\<%
\\
\>[0]\AgdaFunction{lemmaPHKcoraux2}\AgdaSpace{}%
\AgdaBound{pbk}\AgdaSpace{}%
\AgdaBound{time}\AgdaSpace{}%
\AgdaBound{msg₁}\AgdaSpace{}%
\AgdaBound{sig}\AgdaSpace{}%
\AgdaBound{s}\AgdaSpace{}%
\AgdaSymbol{(}\AgdaInductiveConstructor{suc}\AgdaSpace{}%
\AgdaBound{cpRes}\AgdaSymbol{)}\AgdaSpace{}%
\AgdaBound{p}\AgdaSpace{}%
\AgdaSymbol{=}\AgdaSpace{}%
\AgdaBound{p}\<%
\\
\\[\AgdaEmptyExtraSkip]%
\\[\AgdaEmptyExtraSkip]%
\>[0]\AgdaComment{\{-\ This\ lemma\ \ is\ actually\ for\ the\ paper\ 1}\<%
\\
\>[0]\AgdaComment{\ \ \ and\ follows\ because}\<%
\\
\>[0]\AgdaComment{\ \ \ \ ⟦\ scriptP2PKHᵇ\ pbkh\ ⟧stack\ \ =\ \ (λ\ time\ msg₁\ s\ →\ p2pkhFunctionDecoded\ pbkh\ msg₁\ s)}\<%
\\
\>[0]\AgdaComment{-\}}\<%
\\
\\[\AgdaEmptyExtraSkip]%
\>[0]\AgdaComment{--\ \textbackslash{}stackVerificationPtoPKHUsingEqualityOfProgramslemmaPTKHcoraux}\<%
\end{code}
} 

\newcommand{\stackVerificationPtoPKHUsingEqualityOfProgramslemmaPTKHcoraux}{
\begin{code}%
\>[0]\AgdaFunction{lemmaPTKHcoraux}\AgdaSpace{}%
\AgdaSymbol{:}\AgdaSpace{}%
\AgdaSymbol{(}\AgdaBound{pbkh}\AgdaSpace{}%
\AgdaSymbol{:}\AgdaSpace{}%
\AgdaDatatype{ℕ}\AgdaSymbol{)}\AgdaSpace{}%
\AgdaSymbol{→}%
\>[32]\AgdaOperator{\AgdaRecord{<}}\AgdaSpace{}%
\AgdaFunction{weakestPreConditionP2PKHˢ}\AgdaSpace{}%
\AgdaBound{pbkh}\AgdaSpace{}%
\AgdaOperator{\AgdaRecord{>gˢ}}\<%
\\
\>[32]\AgdaSymbol{(λ}\AgdaSpace{}%
\AgdaBound{time}\AgdaSpace{}%
\AgdaBound{msg₁}\AgdaSpace{}%
\AgdaBound{s}\AgdaSpace{}%
\AgdaSymbol{→}\AgdaSpace{}%
\AgdaFunction{p2pkhFunctionDecoded}\AgdaSpace{}%
\AgdaBound{pbkh}\AgdaSpace{}%
\AgdaBound{msg₁}\AgdaSpace{}%
\AgdaBound{s}\AgdaSymbol{)}\<%
\\
\>[32]\AgdaOperator{\AgdaRecord{<}}\AgdaSpace{}%
\AgdaFunction{acceptStateˢ}\AgdaSpace{}%
\AgdaOperator{\AgdaRecord{>}}\<%
\end{code}
} 

\AgdaHide{
\begin{code}%
\>[0]\AgdaFunction{lemmaPTKHcoraux}\AgdaSpace{}%
\AgdaDottedPattern{\AgdaSymbol{.(}}\AgdaDottedPattern{\AgdaFunction{hashFun}}\AgdaSpace{}%
\AgdaDottedPattern{\AgdaBound{pbk}}\AgdaDottedPattern{\AgdaSymbol{)}}\AgdaSpace{}%
\AgdaSymbol{.}\AgdaField{==>stg}\AgdaSpace{}%
\AgdaBound{time}\AgdaSpace{}%
\AgdaBound{msg₁}\AgdaSpace{}%
\AgdaSymbol{(}\AgdaBound{pbk}\AgdaSpace{}%
\AgdaOperator{\AgdaInductiveConstructor{∷}}\AgdaSpace{}%
\AgdaBound{sig}\AgdaSpace{}%
\AgdaOperator{\AgdaInductiveConstructor{∷}}\AgdaSpace{}%
\AgdaBound{s}\AgdaSymbol{)}\AgdaSpace{}%
\AgdaSymbol{(}\AgdaInductiveConstructor{conj}\AgdaSpace{}%
\AgdaInductiveConstructor{refl}\AgdaSpace{}%
\AgdaBound{and4}\AgdaSymbol{)}\<%
\\
\>[0][@{}l@{\AgdaIndent{0}}]%
\>[6]\AgdaKeyword{rewrite}\AgdaSpace{}%
\AgdaSymbol{(}\AgdaFunction{lemmaCompareNat}\AgdaSpace{}%
\AgdaSymbol{(}\AgdaFunction{hashFun}\AgdaSpace{}%
\AgdaBound{pbk}\AgdaSymbol{))}\<%
\\
\>[6]\AgdaSymbol{=}\AgdaSpace{}%
\AgdaFunction{boolToNatNotFalseLemma}\AgdaSpace{}%
\AgdaSymbol{(}\AgdaFunction{isSigned}%
\>[42]\AgdaBound{msg₁}\AgdaSpace{}%
\AgdaBound{sig}\AgdaSpace{}%
\AgdaBound{pbk}\AgdaSymbol{)}\AgdaSpace{}%
\AgdaBound{and4}\<%
\\
\>[0]\AgdaFunction{lemmaPTKHcoraux}\AgdaSpace{}%
\AgdaBound{pbkh}\AgdaSpace{}%
\AgdaSymbol{.}\AgdaField{<==stg}\AgdaSpace{}%
\AgdaBound{time}\AgdaSpace{}%
\AgdaBound{msg₁}\AgdaSpace{}%
\AgdaSymbol{(}\AgdaBound{pbk}\AgdaSpace{}%
\AgdaOperator{\AgdaInductiveConstructor{∷}}\AgdaSpace{}%
\AgdaBound{sig}\AgdaSpace{}%
\AgdaOperator{\AgdaInductiveConstructor{∷}}\AgdaSpace{}%
\AgdaBound{s}\AgdaSymbol{)}\AgdaSpace{}%
\AgdaBound{x}\<%
\\
\>[0][@{}l@{\AgdaIndent{0}}]%
\>[4]\AgdaSymbol{=}\AgdaSpace{}%
\AgdaInductiveConstructor{conj}%
\>[557I]\AgdaSymbol{(}\AgdaFunction{sym}%
\>[558I]\AgdaSymbol{(}\AgdaFunction{lemmaCompareNat2}\AgdaSpace{}%
\AgdaBound{pbkh}\AgdaSpace{}%
\AgdaSymbol{(}\AgdaFunction{hashFun}\AgdaSpace{}%
\AgdaBound{pbk}\AgdaSymbol{)}\<%
\\
\>[558I][@{}l@{\AgdaIndent{0}}]%
\>[17]\AgdaSymbol{(}\AgdaFunction{lemmaPHKcoraux3}\AgdaSpace{}%
\AgdaBound{pbk}\AgdaSpace{}%
\AgdaBound{time}\AgdaSpace{}%
\AgdaBound{msg₁}\AgdaSpace{}%
\AgdaBound{sig}\AgdaSpace{}%
\AgdaBound{s}\AgdaSpace{}%
\AgdaSymbol{(}\AgdaFunction{compareNaturals}\AgdaSpace{}%
\AgdaBound{pbkh}\AgdaSpace{}%
\AgdaSymbol{(}\AgdaFunction{hashFun}\AgdaSpace{}%
\AgdaBound{pbk}\AgdaSymbol{))}\AgdaSpace{}%
\AgdaBound{x}\AgdaSymbol{)))}\<%
\\
\>[.][@{}l@{}]\<[557I]%
\>[11]\AgdaSymbol{(}\AgdaFunction{boolToNatNotFalseLemma2}\AgdaSpace{}%
\AgdaSymbol{(}\AgdaFunction{isSigned}%
\>[47]\AgdaBound{msg₁}\AgdaSpace{}%
\AgdaBound{sig}\AgdaSpace{}%
\AgdaBound{pbk}\AgdaSymbol{)}\<%
\\
\>[11][@{}l@{\AgdaIndent{0}}]%
\>[13]\AgdaSymbol{(}\AgdaFunction{lemmaPHKcoraux2}\AgdaSpace{}%
\AgdaBound{pbk}\AgdaSpace{}%
\AgdaBound{time}\AgdaSpace{}%
\AgdaBound{msg₁}\AgdaSpace{}%
\AgdaBound{sig}\AgdaSpace{}%
\AgdaBound{s}\AgdaSpace{}%
\AgdaSymbol{((}\AgdaFunction{compareNaturals}\AgdaSpace{}%
\AgdaBound{pbkh}\AgdaSpace{}%
\AgdaSymbol{(}\AgdaFunction{hashFun}\AgdaSpace{}%
\AgdaBound{pbk}\AgdaSymbol{)))}\AgdaSpace{}%
\AgdaBound{x}\AgdaSymbol{))}\<%
\\
\\[\AgdaEmptyExtraSkip]%
\\[\AgdaEmptyExtraSkip]%
\>[0]\AgdaFunction{LemmaPTPKHcor}\AgdaSpace{}%
\AgdaSymbol{:}\AgdaSpace{}%
\AgdaSymbol{(}\AgdaBound{pubKeyHash}\AgdaSpace{}%
\AgdaSymbol{:}\AgdaSpace{}%
\AgdaDatatype{ℕ}\AgdaSymbol{)}\<%
\\
\>[0][@{}l@{\AgdaIndent{0}}]%
\>[2]\AgdaSymbol{→}%
\>[6]\AgdaOperator{\AgdaFunction{<}}%
\>[9]\AgdaFunction{weakestPreConditionP2PKHˢ}\AgdaSpace{}%
\AgdaBound{pubKeyHash}\AgdaSpace{}%
\AgdaOperator{\AgdaFunction{>stackb}}\<%
\\
\>[6][@{}l@{\AgdaIndent{0}}]%
\>[8]\AgdaFunction{scriptP2PKHᵇ}\AgdaSpace{}%
\AgdaBound{pubKeyHash}\<%
\\
\>[6]\AgdaOperator{\AgdaFunction{<}}\AgdaSpace{}%
\AgdaFunction{acceptStateˢ}%
\>[23]\AgdaOperator{\AgdaFunction{>}}\<%
\\
\>[0]\AgdaFunction{LemmaPTPKHcor}\AgdaSpace{}%
\AgdaBound{pbkh}\<%
\\
\>[0][@{}l@{\AgdaIndent{0}}]%
\>[4]\AgdaSymbol{=}%
\>[594I]\AgdaFunction{lemmaTransferHoareTripleStack}\AgdaSpace{}%
\AgdaSymbol{(}\AgdaFunction{weakestPreConditionP2PKHˢ}\AgdaSpace{}%
\AgdaBound{pbkh}\AgdaSymbol{)}\AgdaSpace{}%
\AgdaFunction{acceptStateˢ}\<%
\\
\>[594I][@{}l@{\AgdaIndent{0}}]%
\>[8]\AgdaSymbol{(λ}\AgdaSpace{}%
\AgdaBound{time}\AgdaSpace{}%
\AgdaBound{msg}\AgdaSpace{}%
\AgdaBound{s}\AgdaSpace{}%
\AgdaSymbol{→}\AgdaSpace{}%
\AgdaFunction{p2pkhFunctionDecoded}\AgdaSpace{}%
\AgdaBound{pbkh}\AgdaSpace{}%
\AgdaBound{msg}\AgdaSpace{}%
\AgdaBound{s}\AgdaSpace{}%
\AgdaSymbol{)}\<%
\\
\>[8]\AgdaOperator{\AgdaFunction{⟦}}\AgdaSpace{}%
\AgdaFunction{scriptP2PKH}\AgdaSpace{}%
\AgdaBound{pbkh}\AgdaSpace{}%
\AgdaOperator{\AgdaFunction{⟧stack}}\<%
\\
\>[8]\AgdaSymbol{(λ}\AgdaSpace{}%
\AgdaBound{t}\AgdaSpace{}%
\AgdaBound{m}\AgdaSpace{}%
\AgdaBound{s}\AgdaSpace{}%
\AgdaSymbol{→}\AgdaSpace{}%
\AgdaFunction{sym}\AgdaSpace{}%
\AgdaSymbol{(}\AgdaFunction{p2pkhFunctionDecodedcor}\AgdaSpace{}%
\AgdaBound{t}\AgdaSpace{}%
\AgdaBound{pbkh}\AgdaSpace{}%
\AgdaBound{m}\AgdaSpace{}%
\AgdaBound{s}\AgdaSymbol{))}\<%
\\
\>[8]\AgdaSymbol{(}\AgdaFunction{lemmaPTKHcoraux}\AgdaSpace{}%
\AgdaBound{pbkh}\AgdaSymbol{)}\<%
\\
\\[\AgdaEmptyExtraSkip]%
\\[\AgdaEmptyExtraSkip]%
\>[0]\AgdaComment{--\ \textbackslash{}stackVerificationPtoPKHUsingEqualityOfProgramstheoremPTPKHcor}\<%
\end{code}
} 

\newcommand{\stackVerificationPtoPKHUsingEqualityOfProgramstheoremPTPKHcor}{
\begin{code}%
\>[0]\AgdaFunction{theoPTPKHcor}\AgdaSpace{}%
\AgdaSymbol{:}%
\>[16]\AgdaSymbol{(}\AgdaBound{pbkh}\AgdaSpace{}%
\AgdaSymbol{:}\AgdaSpace{}%
\AgdaDatatype{ℕ}\AgdaSymbol{)}\AgdaSpace{}%
\AgdaSymbol{→}\AgdaSpace{}%
\AgdaOperator{\AgdaRecord{<}}\AgdaSpace{}%
\AgdaFunction{wPreCondP2PKH}\AgdaSpace{}%
\AgdaBound{pbkh}\AgdaSpace{}%
\AgdaOperator{\AgdaRecord{>ⁱᶠᶠ}}\AgdaSpace{}%
\AgdaFunction{scriptP2PKHᵇ}\AgdaSpace{}%
\AgdaBound{pbkh}\AgdaSpace{}%
\AgdaOperator{\AgdaRecord{<}}\AgdaSpace{}%
\AgdaFunction{acceptState}\AgdaSpace{}%
\AgdaOperator{\AgdaRecord{>}}\<%
\end{code}
} 

\AgdaHide{
\begin{code}%
\>[0]\AgdaFunction{theoPTPKHcor}\AgdaSpace{}%
\AgdaBound{pbkh}\AgdaSpace{}%
\AgdaSymbol{=}\<%
\\
\>[0][@{}l@{\AgdaIndent{0}}]%
\>[3]\AgdaFunction{hoareTripleStack2HoareTriple}\AgdaSpace{}%
\AgdaSymbol{(}\AgdaFunction{scriptP2PKHᵇ}\AgdaSpace{}%
\AgdaBound{pbkh}\AgdaSymbol{)}\<%
\\
\>[3][@{}l@{\AgdaIndent{0}}]%
\>[6]\AgdaSymbol{(}\AgdaFunction{wPreCondP2PKHˢ}\AgdaSpace{}%
\AgdaBound{pbkh}\AgdaSymbol{)}\AgdaSpace{}%
\AgdaFunction{acceptStateˢ}\AgdaSpace{}%
\AgdaSymbol{(}\AgdaFunction{LemmaPTPKHcor}\AgdaSpace{}%
\AgdaBound{pbkh}\AgdaSymbol{)}\<%
\\
\\[\AgdaEmptyExtraSkip]%
\\[\AgdaEmptyExtraSkip]%
\\[\AgdaEmptyExtraSkip]%
\\[\AgdaEmptyExtraSkip]%
\\[\AgdaEmptyExtraSkip]%
\\[\AgdaEmptyExtraSkip]%
\\[\AgdaEmptyExtraSkip]%
\>[0]\AgdaComment{--\ Some\ test\ cases\ used\ in\ the\ development}\<%
\\
\>[0]\AgdaComment{--\ shows\ how\ to\ check\ for\ each\ step\ \ what\ the\ functions\ are\ and\ how\ they\ compute}\<%
\\
\\[\AgdaEmptyExtraSkip]%
\>[0]\AgdaComment{-------------------------------------------\ Tests\ ------------------------------------------------------------}\<%
\\
\\[\AgdaEmptyExtraSkip]%
\\[\AgdaEmptyExtraSkip]%
\\[\AgdaEmptyExtraSkip]%
\\[\AgdaEmptyExtraSkip]%
\\[\AgdaEmptyExtraSkip]%
\\[\AgdaEmptyExtraSkip]%
\>[0]\AgdaComment{\{-}\<%
\\
\>[0]\AgdaComment{\ \ stackfunP2PKHemptyNotCorrectPbkIsNothing\ :\ (pubKeyHash\ :\ ℕ)(time₁\ :\ Time)(msg₁\ :\ Msg)(stack₁\ :\ State1)}\<%
\\
\>[0]\AgdaComment{\ \ \ \ \ \ \ \ →\ (\ neg\ (pubKeyHash\ ≡\ hashFun\ pbk)\ \ )}\<%
\\
\>[0]\AgdaComment{\ \ \ \ \ \ \ \ →\ ⟦\ scriptP2PKHᵇ\ pubKeyHash\ ⟧ˢ\ time₁\ msg₁\ (pbk\ ∷\ [])\ ≡\ nothing}\<%
\\
\>[0]\AgdaComment{\ \ stackfunP2PKHemptyNotCorrectPbkIsNothing\ pubKeyHash\ time₁\ msg₁\ stack₁\ =\ \{!!\}}\<%
\\
\>[0]\<%
\\
\>[0]\AgdaComment{-\}}\<%
\end{code}
} 


\AgdaHide{
\begin{code}%
\>[0]\<%
\\
\>[0]\AgdaKeyword{open}\AgdaSpace{}%
\AgdaKeyword{import}\AgdaSpace{}%
\AgdaModule{basicBitcoinDataType}\<%
\\
\>[0]\AgdaKeyword{module}\AgdaSpace{}%
\AgdaModule{paperTypes2021PostProceed.demoEqualityReasoning}\AgdaSpace{}%
\AgdaSymbol{(}\AgdaBound{param}\AgdaSpace{}%
\AgdaSymbol{:}\AgdaSpace{}%
\AgdaRecord{GlobalParameters}\AgdaSymbol{)}\AgdaSpace{}%
\AgdaKeyword{where}\<%
\\
\\[\AgdaEmptyExtraSkip]%
\>[0]\AgdaKeyword{open}\AgdaSpace{}%
\AgdaKeyword{import}\AgdaSpace{}%
\AgdaModule{Data.List.Base}\AgdaSpace{}%
\AgdaKeyword{hiding}\AgdaSpace{}%
\AgdaSymbol{(}\AgdaOperator{\AgdaFunction{\AgdaUnderscore{}\ensuremath{+\!\!+}\AgdaUnderscore{}}}\AgdaSpace{}%
\AgdaSymbol{)}\<%
\\
\>[0]\AgdaKeyword{open}\AgdaSpace{}%
\AgdaKeyword{import}\AgdaSpace{}%
\AgdaModule{Data.Nat}%
\>[22]\AgdaKeyword{renaming}\AgdaSpace{}%
\AgdaSymbol{(}\AgdaOperator{\AgdaDatatype{\AgdaUnderscore{}≤\AgdaUnderscore{}}}\AgdaSpace{}%
\AgdaSymbol{to}\AgdaSpace{}%
\AgdaOperator{\AgdaDatatype{\AgdaUnderscore{}≤'\AgdaUnderscore{}}}\AgdaSpace{}%
\AgdaSymbol{;}\AgdaSpace{}%
\AgdaOperator{\AgdaFunction{\AgdaUnderscore{}<\AgdaUnderscore{}}}\AgdaSpace{}%
\AgdaSymbol{to}\AgdaSpace{}%
\AgdaOperator{\AgdaFunction{\AgdaUnderscore{}<'\AgdaUnderscore{}}}\AgdaSymbol{)}\<%
\\
\>[0]\AgdaKeyword{open}\AgdaSpace{}%
\AgdaKeyword{import}\AgdaSpace{}%
\AgdaModule{Data.List}\AgdaSpace{}%
\AgdaKeyword{hiding}\AgdaSpace{}%
\AgdaSymbol{(}\AgdaOperator{\AgdaFunction{\AgdaUnderscore{}\ensuremath{+\!\!+}\AgdaUnderscore{}}}%
\>[36]\AgdaSymbol{)}\<%
\\
\>[0]\AgdaKeyword{open}\AgdaSpace{}%
\AgdaKeyword{import}\AgdaSpace{}%
\AgdaModule{Data.Sum}\<%
\\
\>[0]\AgdaKeyword{open}\AgdaSpace{}%
\AgdaKeyword{import}\AgdaSpace{}%
\AgdaModule{Data.Unit}%
\>[23]\AgdaKeyword{hiding}\AgdaSpace{}%
\AgdaSymbol{(}\AgdaOperator{\AgdaRecord{\AgdaUnderscore{}≤\AgdaUnderscore{}}}\AgdaSymbol{)}\<%
\\
\>[0]\AgdaKeyword{open}\AgdaSpace{}%
\AgdaKeyword{import}\AgdaSpace{}%
\AgdaModule{Data.Empty}\<%
\\
\>[0]\AgdaKeyword{open}\AgdaSpace{}%
\AgdaKeyword{import}\AgdaSpace{}%
\AgdaModule{Data.Maybe}\<%
\\
\>[0]\AgdaKeyword{open}\AgdaSpace{}%
\AgdaKeyword{import}\AgdaSpace{}%
\AgdaModule{Data.Bool}%
\>[23]\AgdaKeyword{hiding}\AgdaSpace{}%
\AgdaSymbol{(}\AgdaOperator{\AgdaDatatype{\AgdaUnderscore{}≤\AgdaUnderscore{}}}\AgdaSpace{}%
\AgdaSymbol{;}\AgdaSpace{}%
\AgdaOperator{\AgdaDatatype{\AgdaUnderscore{}<\AgdaUnderscore{}}}\AgdaSpace{}%
\AgdaSymbol{;}\AgdaSpace{}%
\AgdaOperator{\AgdaFunction{if\AgdaUnderscore{}then\AgdaUnderscore{}else\AgdaUnderscore{}}}%
\>[58]\AgdaSymbol{)}%
\>[61]\AgdaKeyword{renaming}\AgdaSpace{}%
\AgdaSymbol{(}\AgdaOperator{\AgdaFunction{\AgdaUnderscore{}∧\AgdaUnderscore{}}}\AgdaSpace{}%
\AgdaSymbol{to}\AgdaSpace{}%
\AgdaOperator{\AgdaFunction{\AgdaUnderscore{}∧b\AgdaUnderscore{}}}\AgdaSpace{}%
\AgdaSymbol{;}\AgdaSpace{}%
\AgdaOperator{\AgdaFunction{\AgdaUnderscore{}∨\AgdaUnderscore{}}}\AgdaSpace{}%
\AgdaSymbol{to}\AgdaSpace{}%
\AgdaOperator{\AgdaFunction{\AgdaUnderscore{}∨b\AgdaUnderscore{}}}\AgdaSpace{}%
\AgdaSymbol{;}\AgdaSpace{}%
\AgdaFunction{T}\AgdaSpace{}%
\AgdaSymbol{to}\AgdaSpace{}%
\AgdaFunction{True}\AgdaSymbol{)}\<%
\\
\>[0]\AgdaKeyword{open}\AgdaSpace{}%
\AgdaKeyword{import}\AgdaSpace{}%
\AgdaModule{Data.Product}\AgdaSpace{}%
\AgdaKeyword{renaming}\AgdaSpace{}%
\AgdaSymbol{(}\AgdaOperator{\AgdaInductiveConstructor{\AgdaUnderscore{},\AgdaUnderscore{}}}\AgdaSpace{}%
\AgdaSymbol{to}\AgdaSpace{}%
\AgdaOperator{\AgdaInductiveConstructor{\AgdaUnderscore{},,\AgdaUnderscore{}}}\AgdaSpace{}%
\AgdaSymbol{)}\<%
\\
\>[0]\AgdaKeyword{open}\AgdaSpace{}%
\AgdaKeyword{import}\AgdaSpace{}%
\AgdaModule{Data.Nat.Base}\AgdaSpace{}%
\AgdaKeyword{hiding}\AgdaSpace{}%
\AgdaSymbol{(}\AgdaOperator{\AgdaDatatype{\AgdaUnderscore{}≤\AgdaUnderscore{}}}\AgdaSpace{}%
\AgdaSymbol{;}\AgdaSpace{}%
\AgdaOperator{\AgdaFunction{\AgdaUnderscore{}<\AgdaUnderscore{}}}\AgdaSymbol{)}\<%
\\
\>[0]\AgdaKeyword{open}\AgdaSpace{}%
\AgdaKeyword{import}\AgdaSpace{}%
\AgdaModule{Data.List.NonEmpty}\AgdaSpace{}%
\AgdaKeyword{hiding}\AgdaSpace{}%
\AgdaSymbol{(}\AgdaField{head}\AgdaSymbol{;}\AgdaSpace{}%
\AgdaOperator{\AgdaFunction{[\AgdaUnderscore{}]}}\AgdaSymbol{)}\<%
\\
\>[0]\AgdaKeyword{open}\AgdaSpace{}%
\AgdaKeyword{import}\AgdaSpace{}%
\AgdaModule{Data.Nat}\AgdaSpace{}%
\AgdaKeyword{using}\AgdaSpace{}%
\AgdaSymbol{(}\AgdaDatatype{ℕ}\AgdaSymbol{;}\AgdaSpace{}%
\AgdaOperator{\AgdaPrimitive{\AgdaUnderscore{}+\AgdaUnderscore{}}}\AgdaSymbol{;}\AgdaSpace{}%
\AgdaOperator{\AgdaFunction{\AgdaUnderscore{}≥\AgdaUnderscore{}}}\AgdaSymbol{;}\AgdaSpace{}%
\AgdaOperator{\AgdaFunction{\AgdaUnderscore{}>\AgdaUnderscore{}}}\AgdaSymbol{;}\AgdaSpace{}%
\AgdaInductiveConstructor{zero}\AgdaSymbol{;}\AgdaSpace{}%
\AgdaInductiveConstructor{suc}\AgdaSymbol{;}\AgdaSpace{}%
\AgdaInductiveConstructor{s≤s}\AgdaSymbol{;}\AgdaSpace{}%
\AgdaInductiveConstructor{z≤n}\AgdaSymbol{)}\<%
\\
\>[0]\AgdaKeyword{import}\AgdaSpace{}%
\AgdaModule{Relation.Binary.PropositionalEquality}\AgdaSpace{}%
\AgdaSymbol{as}\AgdaSpace{}%
\AgdaModule{Eq}\<%
\\
\>[0]\AgdaKeyword{open}\AgdaSpace{}%
\AgdaModule{Eq}\AgdaSpace{}%
\AgdaKeyword{using}\AgdaSpace{}%
\AgdaSymbol{(}\AgdaOperator{\AgdaDatatype{\AgdaUnderscore{}≡\AgdaUnderscore{}}}\AgdaSymbol{;}\AgdaSpace{}%
\AgdaInductiveConstructor{refl}\AgdaSymbol{;}\AgdaSpace{}%
\AgdaFunction{cong}\AgdaSymbol{;}\AgdaSpace{}%
\AgdaKeyword{module}\AgdaSpace{}%
\AgdaModule{≡-Reasoning}\AgdaSymbol{;}\AgdaSpace{}%
\AgdaFunction{sym}\AgdaSymbol{)}\<%
\\
\>[0]\AgdaKeyword{open}\AgdaSpace{}%
\AgdaModule{≡-Reasoning}\<%
\\
\>[0]\AgdaKeyword{open}\AgdaSpace{}%
\AgdaKeyword{import}\AgdaSpace{}%
\AgdaModule{Agda.Builtin.Equality}\<%
\\
\\[\AgdaEmptyExtraSkip]%
\\[\AgdaEmptyExtraSkip]%
\>[0]\AgdaComment{--our\ libraries}\<%
\\
\>[0]\AgdaKeyword{open}\AgdaSpace{}%
\AgdaKeyword{import}\AgdaSpace{}%
\AgdaModule{libraries.listLib}\<%
\\
\>[0]\AgdaKeyword{open}\AgdaSpace{}%
\AgdaKeyword{import}\AgdaSpace{}%
\AgdaModule{libraries.emptyLib}\<%
\\
\>[0]\AgdaKeyword{open}\AgdaSpace{}%
\AgdaKeyword{import}\AgdaSpace{}%
\AgdaModule{libraries.natLib}\<%
\\
\>[0]\AgdaKeyword{open}\AgdaSpace{}%
\AgdaKeyword{import}\AgdaSpace{}%
\AgdaModule{libraries.boolLib}\<%
\\
\>[0]\AgdaKeyword{open}\AgdaSpace{}%
\AgdaKeyword{import}\AgdaSpace{}%
\AgdaModule{libraries.equalityLib}\<%
\\
\>[0]\AgdaKeyword{open}\AgdaSpace{}%
\AgdaKeyword{import}\AgdaSpace{}%
\AgdaModule{libraries.andLib}\<%
\\
\>[0]\AgdaKeyword{open}\AgdaSpace{}%
\AgdaKeyword{import}\AgdaSpace{}%
\AgdaModule{libraries.miscLib}\<%
\\
\>[0]\AgdaKeyword{open}\AgdaSpace{}%
\AgdaKeyword{import}\AgdaSpace{}%
\AgdaModule{libraries.maybeLib}\<%
\\
\>[0]\AgdaKeyword{open}\AgdaSpace{}%
\AgdaKeyword{import}\AgdaSpace{}%
\AgdaModule{stack}\<%
\\
\>[0]\AgdaKeyword{open}\AgdaSpace{}%
\AgdaKeyword{import}\AgdaSpace{}%
\AgdaModule{stackPredicate}\<%
\\
\>[0]\AgdaKeyword{open}\AgdaSpace{}%
\AgdaKeyword{import}\AgdaSpace{}%
\AgdaModule{semanticBasicOperations}\AgdaSpace{}%
\AgdaBound{param}\<%
\\
\>[0]\AgdaKeyword{open}\AgdaSpace{}%
\AgdaKeyword{import}\AgdaSpace{}%
\AgdaModule{instructionBasic}\<%
\\
\>[0]\AgdaKeyword{open}\AgdaSpace{}%
\AgdaKeyword{import}\AgdaSpace{}%
\AgdaModule{verificationMultiSig}\AgdaSpace{}%
\AgdaBound{param}\<%
\\
\>[0]\AgdaKeyword{open}\AgdaSpace{}%
\AgdaKeyword{import}\AgdaSpace{}%
\AgdaModule{verificationStackScripts.semanticsStackInstructions}\AgdaSpace{}%
\AgdaBound{param}\<%
\\
\>[0]\AgdaKeyword{open}\AgdaSpace{}%
\AgdaKeyword{import}\AgdaSpace{}%
\AgdaModule{verificationStackScripts.stackVerificationLemmas}\AgdaSpace{}%
\AgdaBound{param}\<%
\\
\>[0]\AgdaKeyword{open}\AgdaSpace{}%
\AgdaKeyword{import}\AgdaSpace{}%
\AgdaModule{verificationStackScripts.stackHoareTriple}\AgdaSpace{}%
\AgdaBound{param}\<%
\\
\>[0]\AgdaKeyword{open}\AgdaSpace{}%
\AgdaKeyword{import}\AgdaSpace{}%
\AgdaModule{verificationStackScripts.sPredicate}\<%
\\
\>[0]\AgdaKeyword{open}\AgdaSpace{}%
\AgdaKeyword{import}\AgdaSpace{}%
\AgdaModule{verificationStackScripts.hoareTripleStackBasic}\AgdaSpace{}%
\AgdaBound{param}\<%
\\
\>[0]\AgdaKeyword{open}\AgdaSpace{}%
\AgdaKeyword{import}\AgdaSpace{}%
\AgdaModule{verificationStackScripts.stackState}\<%
\\
\>[0]\AgdaKeyword{open}\AgdaSpace{}%
\AgdaKeyword{import}\AgdaSpace{}%
\AgdaModule{verificationStackScripts.stackSemanticsInstructionsBasic}\AgdaSpace{}%
\AgdaBound{param}\<%
\\
\>[0]\AgdaKeyword{open}\AgdaSpace{}%
\AgdaKeyword{import}\AgdaSpace{}%
\AgdaModule{verificationStackScripts.verificationMultiSigBasic}\AgdaSpace{}%
\AgdaBound{param}\<%
\\
\\[\AgdaEmptyExtraSkip]%
\\[\AgdaEmptyExtraSkip]%
\>[0]\AgdaComment{--\ postulate\ time₁\ :\ Time}\<%
\\
\>[0]\AgdaComment{--\ postulate\ pbk1\ pbk2\ pbk3\ pbk4\ :\ ℕ}\<%
\\
\\[\AgdaEmptyExtraSkip]%
\>[0]\AgdaKeyword{postulate}\<%
\end{code}
} 

\newcommand{\demoEqualityReasoningprecondition}{
\begin{code}%
\>[0][@{}l@{\AgdaIndent{1}}]%
\>[2]\AgdaPostulate{precondition}\AgdaSpace{}%
\AgdaSymbol{:}\AgdaSpace{}%
\AgdaFunction{StackStatePred}\<%
\end{code}
} 

\AgdaHide{
\begin{code}%
\>[0]\<%
\end{code}
} 

\newcommand{\demoEqualityReasoningpostcondition}{
\begin{code}%
\>[0][@{}l@{\AgdaIndent{1}}]%
\>[2]\AgdaPostulate{postcondition}\AgdaSpace{}%
\AgdaSymbol{:}\AgdaSpace{}%
\AgdaFunction{StackStatePred}\<%
\end{code}
} 

\AgdaHide{
\begin{code}%
\>[0]\<%
\end{code}
} 

\newcommand{\demoEqualityReasoningfsintermediateCond}{
\begin{code}%
\>[0][@{}l@{\AgdaIndent{1}}]%
\>[2]\AgdaPostulate{intermediateCond1}\AgdaSpace{}%
\AgdaSymbol{:}\AgdaSpace{}%
\AgdaFunction{StackStatePred}\<%
\end{code}
} 

\AgdaHide{
\begin{code}%
\>[0]\<%
\end{code}
} 

\newcommand{\demoEqualityReasoningseintermediateCond}{
\begin{code}%
\>[0][@{}l@{\AgdaIndent{1}}]%
\>[2]\AgdaPostulate{intermediateCond2}\AgdaSpace{}%
\AgdaSymbol{:}\AgdaSpace{}%
\AgdaFunction{StackStatePred}\<%
\end{code}
} 

\AgdaHide{
\begin{code}%
\>[0]\<%
\end{code}
} 

\newcommand{\demoEqualityReasoningtrintermediateCond}{
\begin{code}%
\>[0][@{}l@{\AgdaIndent{1}}]%
\>[2]\AgdaPostulate{intermediateCond3}\AgdaSpace{}%
\AgdaSymbol{:}\AgdaSpace{}%
\AgdaFunction{StackStatePred}\<%
\end{code}
} 

\AgdaHide{
\begin{code}%
\>[0]\<%
\end{code}
} 

\newcommand{\demoEqualityReasoningprogone}{
\begin{code}%
\>[0][@{}l@{\AgdaIndent{1}}]%
\>[2]\AgdaPostulate{prog1}\AgdaSpace{}%
\AgdaSymbol{:}\AgdaSpace{}%
\AgdaFunction{BitcoinScriptBasic}\<%
\end{code}
} 

\newcommand{\demoEqualityReasoningprogtwo}{
\begin{code}%
\>[2]\AgdaPostulate{prog2}\AgdaSpace{}%
\AgdaPostulate{prog3}\AgdaSpace{}%
\AgdaSymbol{:}\AgdaSpace{}%
\AgdaFunction{BitcoinScriptBasic}\<%
\end{code}
} 

\AgdaHide{
\begin{code}%
\>[0]\<%
\\
\>[0]\<%
\end{code}
} 

\newcommand{\demoEqualityReasoningproofs}{
\begin{code}%
\>[0][@{}l@{\AgdaIndent{1}}]%
\>[2]\AgdaPostulate{proof1}\AgdaSpace{}%
\AgdaSymbol{:}\AgdaSpace{}%
\AgdaOperator{\AgdaRecord{<}}\AgdaSpace{}%
\AgdaPostulate{precondition}\AgdaSpace{}%
\AgdaOperator{\AgdaRecord{>ⁱᶠᶠ}}\AgdaSpace{}%
\AgdaPostulate{prog1}%
\>[38]\AgdaOperator{\AgdaRecord{<}}\AgdaSpace{}%
\AgdaPostulate{intermediateCond1}\AgdaSpace{}%
\AgdaOperator{\AgdaRecord{>}}\<%
\\
\>[2]\AgdaPostulate{proof2}\AgdaSpace{}%
\AgdaSymbol{:}\AgdaSpace{}%
\AgdaOperator{\AgdaRecord{<}}\AgdaSpace{}%
\AgdaPostulate{intermediateCond1}\AgdaSpace{}%
\AgdaOperator{\AgdaRecord{>ⁱᶠᶠ}}\AgdaSpace{}%
\AgdaPostulate{prog2}\AgdaSpace{}%
\AgdaOperator{\AgdaRecord{<}}\AgdaSpace{}%
\AgdaPostulate{intermediateCond2}\AgdaSpace{}%
\AgdaOperator{\AgdaRecord{>}}\<%
\\
\>[2]\AgdaPostulate{proof3}\AgdaSpace{}%
\AgdaSymbol{:}\AgdaSpace{}%
\AgdaPostulate{intermediateCond2}\AgdaSpace{}%
\AgdaOperator{\AgdaRecord{<=>ᵖ}}\AgdaSpace{}%
\AgdaPostulate{intermediateCond3}\<%
\\
\>[2]\AgdaPostulate{proof4}\AgdaSpace{}%
\AgdaSymbol{:}\AgdaSpace{}%
\AgdaOperator{\AgdaRecord{<}}\AgdaSpace{}%
\AgdaPostulate{intermediateCond3}\AgdaSpace{}%
\AgdaOperator{\AgdaRecord{>ⁱᶠᶠ}}\AgdaSpace{}%
\AgdaPostulate{prog3}\AgdaSpace{}%
\AgdaOperator{\AgdaRecord{<}}\AgdaSpace{}%
\AgdaPostulate{postcondition}\AgdaSpace{}%
\AgdaOperator{\AgdaRecord{>}}\<%
\end{code}
} 

\AgdaHide{
\begin{code}%
\>[0]\<%
\end{code}
} 

\newcommand{\demoEqualityReasoninglemma}{
\begin{code}%
\>[0]\AgdaFunction{theorem}\AgdaSpace{}%
\AgdaSymbol{:}\AgdaSpace{}%
\AgdaOperator{\AgdaRecord{<}}\AgdaSpace{}%
\AgdaPostulate{precondition}\AgdaSpace{}%
\AgdaOperator{\AgdaRecord{>ⁱᶠᶠ}}\AgdaSpace{}%
\AgdaPostulate{prog1}\AgdaSpace{}%
\AgdaOperator{\AgdaFunction{\ensuremath{+\!\!+}}}\AgdaSpace{}%
\AgdaSymbol{(}\AgdaPostulate{prog2}\AgdaSpace{}%
\AgdaOperator{\AgdaFunction{\ensuremath{+\!\!+}}}\AgdaSpace{}%
\AgdaPostulate{prog3}\AgdaSymbol{)}%
\>[57]\AgdaOperator{\AgdaRecord{<}}%
\>[60]\AgdaPostulate{postcondition}\AgdaSpace{}%
\AgdaOperator{\AgdaRecord{>}}\<%
\\
\>[0]\AgdaFunction{theorem}\AgdaSpace{}%
\AgdaSymbol{=}%
\>[11]\AgdaPostulate{precondition}%
\>[30]\AgdaFunction{<><>⟨}%
\>[37]\AgdaPostulate{prog1}%
\>[45]\AgdaFunction{⟩⟨}%
\>[49]\AgdaPostulate{proof1}%
\>[57]\AgdaFunction{⟩}\<%
\\
\>[11]\AgdaPostulate{intermediateCond1}%
\>[30]\AgdaFunction{<><>⟨}%
\>[37]\AgdaPostulate{prog2}%
\>[45]\AgdaFunction{⟩⟨}%
\>[49]\AgdaPostulate{proof2}%
\>[57]\AgdaFunction{⟩}\<%
\\
\>[11]\AgdaPostulate{intermediateCond2}%
\>[30]\AgdaFunction{<=>⟨}%
\>[37]\AgdaPostulate{proof3}%
\>[45]\AgdaFunction{⟩}\<%
\\
\>[11]\AgdaPostulate{intermediateCond3}%
\>[30]\AgdaFunction{<><>⟨}%
\>[37]\AgdaPostulate{prog3}%
\>[45]\AgdaFunction{⟩⟨}%
\>[49]\AgdaPostulate{proof4}%
\>[57]\AgdaFunction{⟩ᵉ}%
\>[62]\AgdaPostulate{postcondition}%
\>[77]\AgdaOperator{\AgdaFunction{∎p}}\<%
\end{code}
} 

\AgdaHide{
\begin{code}%
\>[0]\<%
\end{code}
} 


\AgdaHide{
\begin{code}%
\>[0]\AgdaKeyword{module}\AgdaSpace{}%
\AgdaModule{verificationStackScripts.stackState}\AgdaSpace{}%
\AgdaKeyword{where}\<%
\\
\\[\AgdaEmptyExtraSkip]%
\>[0]\AgdaKeyword{open}\AgdaSpace{}%
\AgdaKeyword{import}\AgdaSpace{}%
\AgdaModule{Data.Nat}%
\>[22]\AgdaKeyword{hiding}\AgdaSpace{}%
\AgdaSymbol{(}\AgdaOperator{\AgdaDatatype{\AgdaUnderscore{}≤\AgdaUnderscore{}}}\AgdaSymbol{)}\<%
\\
\>[0]\AgdaKeyword{open}\AgdaSpace{}%
\AgdaKeyword{import}\AgdaSpace{}%
\AgdaModule{Data.List}\AgdaSpace{}%
\AgdaKeyword{hiding}\AgdaSpace{}%
\AgdaSymbol{(}\AgdaOperator{\AgdaFunction{\AgdaUnderscore{}\ensuremath{+\!\!+}\AgdaUnderscore{}}}\AgdaSymbol{)}\<%
\\
\>[0]\AgdaKeyword{open}\AgdaSpace{}%
\AgdaKeyword{import}\AgdaSpace{}%
\AgdaModule{Data.Unit}%
\>[23]\AgdaKeyword{hiding}\AgdaSpace{}%
\AgdaSymbol{(}\AgdaOperator{\AgdaRecord{\AgdaUnderscore{}≤\AgdaUnderscore{}}}\AgdaSymbol{)}\<%
\\
\>[0]\AgdaKeyword{open}\AgdaSpace{}%
\AgdaKeyword{import}\AgdaSpace{}%
\AgdaModule{Data.Empty}\<%
\\
\>[0]\AgdaKeyword{open}\AgdaSpace{}%
\AgdaKeyword{import}\AgdaSpace{}%
\AgdaModule{Data.Maybe}\<%
\\
\>[0]\AgdaKeyword{open}\AgdaSpace{}%
\AgdaKeyword{import}\AgdaSpace{}%
\AgdaModule{Data.Bool}%
\>[23]\AgdaKeyword{hiding}\AgdaSpace{}%
\AgdaSymbol{(}\AgdaOperator{\AgdaDatatype{\AgdaUnderscore{}≤\AgdaUnderscore{}}}\AgdaSpace{}%
\AgdaSymbol{;}\AgdaSpace{}%
\AgdaOperator{\AgdaFunction{if\AgdaUnderscore{}then\AgdaUnderscore{}else\AgdaUnderscore{}}}\AgdaSpace{}%
\AgdaSymbol{)}\AgdaSpace{}%
\AgdaKeyword{renaming}\AgdaSpace{}%
\AgdaSymbol{(}\AgdaOperator{\AgdaFunction{\AgdaUnderscore{}∧\AgdaUnderscore{}}}\AgdaSpace{}%
\AgdaSymbol{to}\AgdaSpace{}%
\AgdaOperator{\AgdaFunction{\AgdaUnderscore{}∧b\AgdaUnderscore{}}}\AgdaSpace{}%
\AgdaSymbol{;}\AgdaSpace{}%
\AgdaOperator{\AgdaFunction{\AgdaUnderscore{}∨\AgdaUnderscore{}}}\AgdaSpace{}%
\AgdaSymbol{to}\AgdaSpace{}%
\AgdaOperator{\AgdaFunction{\AgdaUnderscore{}∨b\AgdaUnderscore{}}}\AgdaSpace{}%
\AgdaSymbol{;}\AgdaSpace{}%
\AgdaFunction{T}\AgdaSpace{}%
\AgdaSymbol{to}\AgdaSpace{}%
\AgdaFunction{True}\AgdaSymbol{)}\<%
\\
\>[0]\AgdaKeyword{open}\AgdaSpace{}%
\AgdaKeyword{import}\AgdaSpace{}%
\AgdaModule{Data.Bool.Base}\AgdaSpace{}%
\AgdaKeyword{hiding}\AgdaSpace{}%
\AgdaSymbol{(}\AgdaOperator{\AgdaDatatype{\AgdaUnderscore{}≤\AgdaUnderscore{}}}\AgdaSpace{}%
\AgdaSymbol{;}\AgdaSpace{}%
\AgdaOperator{\AgdaFunction{if\AgdaUnderscore{}then\AgdaUnderscore{}else\AgdaUnderscore{}}}\AgdaSpace{}%
\AgdaSymbol{)}\AgdaSpace{}%
\AgdaKeyword{renaming}\AgdaSpace{}%
\AgdaSymbol{(}\AgdaOperator{\AgdaFunction{\AgdaUnderscore{}∧\AgdaUnderscore{}}}\AgdaSpace{}%
\AgdaSymbol{to}\AgdaSpace{}%
\AgdaOperator{\AgdaFunction{\AgdaUnderscore{}∧b\AgdaUnderscore{}}}\AgdaSpace{}%
\AgdaSymbol{;}\AgdaSpace{}%
\AgdaOperator{\AgdaFunction{\AgdaUnderscore{}∨\AgdaUnderscore{}}}\AgdaSpace{}%
\AgdaSymbol{to}\AgdaSpace{}%
\AgdaOperator{\AgdaFunction{\AgdaUnderscore{}∨b\AgdaUnderscore{}}}\AgdaSpace{}%
\AgdaSymbol{;}\AgdaSpace{}%
\AgdaFunction{T}\AgdaSpace{}%
\AgdaSymbol{to}\AgdaSpace{}%
\AgdaFunction{True}\AgdaSymbol{)}\<%
\\
\>[0]\AgdaKeyword{open}\AgdaSpace{}%
\AgdaKeyword{import}\AgdaSpace{}%
\AgdaModule{Data.Product}\AgdaSpace{}%
\AgdaKeyword{renaming}\AgdaSpace{}%
\AgdaSymbol{(}\AgdaOperator{\AgdaInductiveConstructor{\AgdaUnderscore{},\AgdaUnderscore{}}}\AgdaSpace{}%
\AgdaSymbol{to}\AgdaSpace{}%
\AgdaOperator{\AgdaInductiveConstructor{\AgdaUnderscore{},,\AgdaUnderscore{}}}\AgdaSpace{}%
\AgdaSymbol{)}\<%
\\
\>[0]\AgdaKeyword{open}\AgdaSpace{}%
\AgdaKeyword{import}\AgdaSpace{}%
\AgdaModule{Data.Nat.Base}\AgdaSpace{}%
\AgdaKeyword{hiding}\AgdaSpace{}%
\AgdaSymbol{(}\AgdaOperator{\AgdaDatatype{\AgdaUnderscore{}≤\AgdaUnderscore{}}}\AgdaSymbol{)}\<%
\\
\>[0]\AgdaKeyword{open}\AgdaSpace{}%
\AgdaKeyword{import}\AgdaSpace{}%
\AgdaModule{Data.List.NonEmpty}\AgdaSpace{}%
\AgdaKeyword{hiding}\AgdaSpace{}%
\AgdaSymbol{(}\AgdaField{head}\AgdaSymbol{)}\<%
\\
\>[0]\AgdaKeyword{import}\AgdaSpace{}%
\AgdaModule{Relation.Binary.PropositionalEquality}\AgdaSpace{}%
\AgdaSymbol{as}\AgdaSpace{}%
\AgdaModule{Eq}\<%
\\
\>[0]\AgdaKeyword{open}\AgdaSpace{}%
\AgdaModule{Eq}\AgdaSpace{}%
\AgdaKeyword{using}\AgdaSpace{}%
\AgdaSymbol{(}\AgdaOperator{\AgdaDatatype{\AgdaUnderscore{}≡\AgdaUnderscore{}}}\AgdaSymbol{;}\AgdaSpace{}%
\AgdaInductiveConstructor{refl}\AgdaSymbol{;}\AgdaSpace{}%
\AgdaFunction{cong}\AgdaSymbol{;}\AgdaSpace{}%
\AgdaKeyword{module}\AgdaSpace{}%
\AgdaModule{≡-Reasoning}\AgdaSymbol{;}\AgdaSpace{}%
\AgdaFunction{sym}\AgdaSymbol{)}\<%
\\
\>[0]\AgdaKeyword{open}\AgdaSpace{}%
\AgdaModule{≡-Reasoning}\<%
\\
\>[0]\AgdaKeyword{open}\AgdaSpace{}%
\AgdaKeyword{import}\AgdaSpace{}%
\AgdaModule{Agda.Builtin.Equality}\<%
\\
\\[\AgdaEmptyExtraSkip]%
\>[0]\AgdaComment{--our\ libraries}\<%
\\
\>[0]\AgdaKeyword{open}\AgdaSpace{}%
\AgdaKeyword{import}\AgdaSpace{}%
\AgdaModule{libraries.listLib}\<%
\\
\>[0]\AgdaKeyword{open}\AgdaSpace{}%
\AgdaKeyword{import}\AgdaSpace{}%
\AgdaModule{libraries.natLib}\<%
\\
\>[0]\AgdaKeyword{open}\AgdaSpace{}%
\AgdaKeyword{import}\AgdaSpace{}%
\AgdaModule{libraries.boolLib}\<%
\\
\>[0]\AgdaKeyword{open}\AgdaSpace{}%
\AgdaKeyword{import}\AgdaSpace{}%
\AgdaModule{libraries.andLib}\<%
\\
\>[0]\AgdaKeyword{open}\AgdaSpace{}%
\AgdaKeyword{import}\AgdaSpace{}%
\AgdaModule{libraries.miscLib}\<%
\\
\>[0]\AgdaKeyword{open}\AgdaSpace{}%
\AgdaKeyword{import}\AgdaSpace{}%
\AgdaModule{libraries.maybeLib}\<%
\\
\\[\AgdaEmptyExtraSkip]%
\>[0]\AgdaKeyword{open}\AgdaSpace{}%
\AgdaKeyword{import}\AgdaSpace{}%
\AgdaModule{basicBitcoinDataType}\<%
\\
\>[0]\AgdaKeyword{open}\AgdaSpace{}%
\AgdaKeyword{import}\AgdaSpace{}%
\AgdaModule{stack}\<%
\\
\>[0]\<%
\end{code}
} 

\newcommand{\stackStateStackState}{
\begin{code}%
\>[0]\AgdaKeyword{record}%
\>[98I]\AgdaRecord{StackState}%
\>[19]\AgdaSymbol{:}\AgdaSpace{}%
\AgdaPrimitive{Set}%
\>[26]\AgdaKeyword{where}\<%
\\
\>[.][@{}l@{}]\<[98I]%
\>[7]\AgdaKeyword{constructor}\AgdaSpace{}%
\AgdaOperator{\AgdaInductiveConstructor{⟨\AgdaUnderscore{},\AgdaUnderscore{},\AgdaUnderscore{}⟩}}\<%
\\
\>[7]\agdaField%
\>[14]\AgdaField{currentTime}\AgdaSpace{}%
\AgdaSymbol{:}%
\>[29]\AgdaFunction{Time}%
\>[35]\AgdaComment{--\ current\ time,\ i.e.\ time\ when\ the\ the\ smart\ contract\ is\ executed}\<%
\\
\>[14]\AgdaField{msg}%
\>[26]\AgdaSymbol{:}%
\>[29]\AgdaDatatype{Msg}\<%
\\
\>[14]\AgdaField{stack}%
\>[26]\AgdaSymbol{:}%
\>[29]\AgdaFunction{Stack}\<%
\end{code}
} 

\AgdaHide{
\begin{code}%
\>[0]\<%
\\
\>[0]\AgdaKeyword{open}\AgdaSpace{}%
\AgdaModule{StackState}\AgdaSpace{}%
\AgdaKeyword{public}\<%
\\
\\[\AgdaEmptyExtraSkip]%
\\[\AgdaEmptyExtraSkip]%
\>[0]\AgdaKeyword{record}%
\>[104I]\AgdaRecord{StackStateWithMaybe}%
\>[28]\AgdaSymbol{:}\AgdaSpace{}%
\AgdaPrimitive{Set}%
\>[35]\AgdaKeyword{where}\<%
\\
\>[.][@{}l@{}]\<[104I]%
\>[7]\AgdaKeyword{constructor}\AgdaSpace{}%
\AgdaOperator{\AgdaInductiveConstructor{⟨\AgdaUnderscore{},\AgdaUnderscore{},\AgdaUnderscore{}⟩}}\<%
\\
\>[7]\agdaField\<%
\\
\>[7][@{}l@{\AgdaIndent{0}}]%
\>[9]\AgdaField{currentTime}\AgdaSpace{}%
\AgdaSymbol{:}\AgdaSpace{}%
\AgdaFunction{Time}%
\>[29]\AgdaComment{--\ current\ time,\ i.e.\ time\ when\ the}\<%
\\
\>[29]\AgdaComment{--\ the\ smart\ contract\ is\ executed}\<%
\\
\>[9]\AgdaField{msg}\AgdaSpace{}%
\AgdaSymbol{:}\AgdaSpace{}%
\AgdaDatatype{Msg}\<%
\\
\>[9]\AgdaField{maybeStack}\AgdaSpace{}%
\AgdaSymbol{:}\AgdaSpace{}%
\AgdaDatatype{Maybe}\AgdaSpace{}%
\AgdaFunction{Stack}\<%
\\
\\[\AgdaEmptyExtraSkip]%
\>[0]\AgdaKeyword{open}\AgdaSpace{}%
\AgdaModule{StackStateWithMaybe}\AgdaSpace{}%
\AgdaKeyword{public}\<%
\\
\\[\AgdaEmptyExtraSkip]%
\\[\AgdaEmptyExtraSkip]%
\>[0]\AgdaFunction{stackState2WithMaybe}\AgdaSpace{}%
\AgdaSymbol{:}\AgdaSpace{}%
\AgdaRecord{StackStateWithMaybe}\AgdaSpace{}%
\AgdaSymbol{→}\AgdaSpace{}%
\AgdaDatatype{Maybe}\AgdaSpace{}%
\AgdaRecord{StackState}\<%
\\
\>[0]\AgdaFunction{stackState2WithMaybe}\AgdaSpace{}%
\AgdaOperator{\AgdaInductiveConstructor{⟨}}\AgdaSpace{}%
\AgdaBound{currentTime₁}\AgdaSpace{}%
\AgdaOperator{\AgdaInductiveConstructor{,}}\AgdaSpace{}%
\AgdaBound{msg₁}\AgdaSpace{}%
\AgdaOperator{\AgdaInductiveConstructor{,}}\AgdaSpace{}%
\AgdaInductiveConstructor{just}\AgdaSpace{}%
\AgdaBound{x}%
\>[53]\AgdaOperator{\AgdaInductiveConstructor{⟩}}\AgdaSpace{}%
\AgdaSymbol{=}\AgdaSpace{}%
\AgdaInductiveConstructor{just}\AgdaSpace{}%
\AgdaOperator{\AgdaInductiveConstructor{⟨}}\AgdaSpace{}%
\AgdaBound{currentTime₁}\AgdaSpace{}%
\AgdaOperator{\AgdaInductiveConstructor{,}}\AgdaSpace{}%
\AgdaBound{msg₁}\AgdaSpace{}%
\AgdaOperator{\AgdaInductiveConstructor{,}}\AgdaSpace{}%
\AgdaBound{x}%
\>[89]\AgdaOperator{\AgdaInductiveConstructor{⟩}}\<%
\\
\>[0]\AgdaFunction{stackState2WithMaybe}\AgdaSpace{}%
\AgdaOperator{\AgdaInductiveConstructor{⟨}}\AgdaSpace{}%
\AgdaBound{currentTime₁}\AgdaSpace{}%
\AgdaOperator{\AgdaInductiveConstructor{,}}\AgdaSpace{}%
\AgdaBound{msg₁}\AgdaSpace{}%
\AgdaOperator{\AgdaInductiveConstructor{,}}\AgdaSpace{}%
\AgdaInductiveConstructor{nothing}\AgdaSpace{}%
\AgdaOperator{\AgdaInductiveConstructor{⟩}}\AgdaSpace{}%
\AgdaSymbol{=}\AgdaSpace{}%
\AgdaInductiveConstructor{nothing}\<%
\\
\\[\AgdaEmptyExtraSkip]%
\\[\AgdaEmptyExtraSkip]%
\\[\AgdaEmptyExtraSkip]%
\\[\AgdaEmptyExtraSkip]%
\>[0]\AgdaKeyword{mutual}\<%
\\
\\[\AgdaEmptyExtraSkip]%
\>[0][@{}l@{\AgdaIndent{0}}]%
\>[2]\AgdaFunction{liftStackToStateTransformerAux'}\AgdaSpace{}%
\AgdaSymbol{:}\AgdaSpace{}%
\AgdaDatatype{Maybe}\AgdaSpace{}%
\AgdaFunction{Stack}\AgdaSpace{}%
\AgdaSymbol{→}\AgdaSpace{}%
\AgdaRecord{StackState}\AgdaSpace{}%
\AgdaSymbol{→}\AgdaSpace{}%
\AgdaRecord{StackStateWithMaybe}\<%
\\
\>[2]\AgdaFunction{liftStackToStateTransformerAux'}\AgdaSpace{}%
\AgdaBound{maybest}\AgdaSpace{}%
\AgdaOperator{\AgdaInductiveConstructor{⟨}}\AgdaSpace{}%
\AgdaBound{currentTime₁}\AgdaSpace{}%
\AgdaOperator{\AgdaInductiveConstructor{,}}\AgdaSpace{}%
\AgdaBound{msg₁}\AgdaSpace{}%
\AgdaOperator{\AgdaInductiveConstructor{,}}\AgdaSpace{}%
\AgdaBound{stack₁}%
\>[74]\AgdaOperator{\AgdaInductiveConstructor{⟩}}\AgdaSpace{}%
\AgdaSymbol{=}\AgdaSpace{}%
\AgdaOperator{\AgdaInductiveConstructor{⟨}}\AgdaSpace{}%
\AgdaBound{currentTime₁}\AgdaSpace{}%
\AgdaOperator{\AgdaInductiveConstructor{,}}\AgdaSpace{}%
\AgdaBound{msg₁}\AgdaSpace{}%
\AgdaOperator{\AgdaInductiveConstructor{,}}\AgdaSpace{}%
\AgdaBound{maybest}%
\>[111]\AgdaOperator{\AgdaInductiveConstructor{⟩}}\<%
\\
\\[\AgdaEmptyExtraSkip]%
\\[\AgdaEmptyExtraSkip]%
\\[\AgdaEmptyExtraSkip]%
\\[\AgdaEmptyExtraSkip]%
\>[0]\AgdaFunction{exeTransformerDepIfStack'}\AgdaSpace{}%
\AgdaSymbol{:}\AgdaSpace{}%
\AgdaSymbol{(}\AgdaSpace{}%
\AgdaRecord{StackState}\AgdaSpace{}%
\AgdaSymbol{→}\AgdaSpace{}%
\AgdaRecord{StackStateWithMaybe}\AgdaSpace{}%
\AgdaSymbol{)}\AgdaSpace{}%
\AgdaSymbol{→}%
\>[68]\AgdaRecord{StackState}\AgdaSpace{}%
\AgdaSymbol{→}\AgdaSpace{}%
\AgdaDatatype{Maybe}\AgdaSpace{}%
\AgdaRecord{StackState}\<%
\\
\>[0]\AgdaFunction{exeTransformerDepIfStack'}\AgdaSpace{}%
\AgdaBound{\,f\,}\AgdaSpace{}%
\AgdaBound{s}\AgdaSpace{}%
\AgdaSymbol{=}\AgdaSpace{}%
\AgdaFunction{stackState2WithMaybe}\AgdaSpace{}%
\AgdaSymbol{(}\AgdaBound{\,f\,}\AgdaSpace{}%
\AgdaBound{s}\AgdaSymbol{)}\<%
\\
\\[\AgdaEmptyExtraSkip]%
\\[\AgdaEmptyExtraSkip]%
\>[0]\AgdaFunction{stackTransform2StackStateTransform}\AgdaSpace{}%
\AgdaSymbol{:}\AgdaSpace{}%
\AgdaFunction{StackTransformer}%
\>[55]\AgdaSymbol{→}\AgdaSpace{}%
\AgdaRecord{StackState}\AgdaSpace{}%
\AgdaSymbol{→}\AgdaSpace{}%
\AgdaDatatype{Maybe}\AgdaSpace{}%
\AgdaRecord{StackState}\<%
\\
\>[0]\AgdaFunction{stackTransform2StackStateTransform}\AgdaSpace{}%
\AgdaBound{\,f\,}\AgdaSpace{}%
\AgdaBound{s}\<%
\\
\>[0][@{}l@{\AgdaIndent{0}}]%
\>[5]\AgdaSymbol{=}\AgdaSpace{}%
\AgdaFunction{stackState2WithMaybe}\AgdaSpace{}%
\AgdaSymbol{((}\AgdaFunction{liftStackToStateTransformerAux'}\AgdaSpace{}%
\AgdaSymbol{(}\AgdaBound{\,f\,}\AgdaSpace{}%
\AgdaSymbol{(}\AgdaBound{s}\AgdaSpace{}%
\AgdaSymbol{.}\AgdaField{currentTime}\AgdaSymbol{)}\AgdaSpace{}%
\AgdaSymbol{(}\AgdaBound{s}\AgdaSpace{}%
\AgdaSymbol{.}\AgdaField{msg}\AgdaSymbol{)}\AgdaSpace{}%
\AgdaSymbol{(}\AgdaBound{s}\AgdaSpace{}%
\AgdaSymbol{.}\AgdaField{stack}\AgdaSymbol{)))}\AgdaSpace{}%
\AgdaBound{s}\AgdaSpace{}%
\AgdaSymbol{)}\<%
\\
\\[\AgdaEmptyExtraSkip]%
\\[\AgdaEmptyExtraSkip]%
\>[0]\AgdaFunction{liftStackToStackStateTransformer'}\AgdaSpace{}%
\AgdaSymbol{:}\AgdaSpace{}%
\AgdaSymbol{(}\AgdaFunction{Stack}\AgdaSpace{}%
\AgdaSymbol{→}\AgdaSpace{}%
\AgdaDatatype{Maybe}\AgdaSpace{}%
\AgdaFunction{Stack}\AgdaSymbol{)}%
\>[59]\AgdaSymbol{→}\AgdaSpace{}%
\AgdaRecord{StackState}\AgdaSpace{}%
\AgdaSymbol{→}\AgdaSpace{}%
\AgdaDatatype{Maybe}\AgdaSpace{}%
\AgdaRecord{StackState}\<%
\\
\>[0]\AgdaFunction{liftStackToStackStateTransformer'}\AgdaSpace{}%
\AgdaBound{\,f\,}\AgdaSpace{}%
\AgdaSymbol{=}\AgdaSpace{}%
\AgdaFunction{stackTransform2StackStateTransform}\AgdaSpace{}%
\AgdaSymbol{(λ}\AgdaSpace{}%
\AgdaBound{time}\AgdaSpace{}%
\AgdaBound{msg}\AgdaSpace{}%
\AgdaSymbol{→}\AgdaSpace{}%
\AgdaBound{\,f\,}\AgdaSymbol{)}\<%
\\
\\[\AgdaEmptyExtraSkip]%
\\[\AgdaEmptyExtraSkip]%
\>[0]\AgdaFunction{liftTimeStackToStateTransformer'}\AgdaSpace{}%
\AgdaSymbol{:}\AgdaSpace{}%
\AgdaSymbol{(}\AgdaFunction{Time}\AgdaSpace{}%
\AgdaSymbol{→}\AgdaSpace{}%
\AgdaFunction{Stack}\AgdaSpace{}%
\AgdaSymbol{→}\AgdaSpace{}%
\AgdaDatatype{Maybe}\AgdaSpace{}%
\AgdaFunction{Stack}\AgdaSymbol{)}%
\>[65]\AgdaSymbol{→}\AgdaSpace{}%
\AgdaRecord{StackState}\AgdaSpace{}%
\AgdaSymbol{→}\AgdaSpace{}%
\AgdaDatatype{Maybe}\AgdaSpace{}%
\AgdaRecord{StackState}\<%
\\
\>[0]\AgdaFunction{liftTimeStackToStateTransformer'}\AgdaSpace{}%
\AgdaBound{\,f\,}\AgdaSpace{}%
\AgdaSymbol{=}\AgdaSpace{}%
\AgdaFunction{stackTransform2StackStateTransform}\AgdaSpace{}%
\AgdaSymbol{(λ}\AgdaSpace{}%
\AgdaBound{time}\AgdaSpace{}%
\AgdaBound{msg}\AgdaSpace{}%
\AgdaSymbol{→}\AgdaSpace{}%
\AgdaBound{\,f\,}\AgdaSpace{}%
\AgdaBound{time}\AgdaSymbol{)}\<%
\\
\\[\AgdaEmptyExtraSkip]%
\>[0]\AgdaFunction{liftMsgStackToStateTransformer'}\AgdaSpace{}%
\AgdaSymbol{:}\AgdaSpace{}%
\AgdaSymbol{(}\AgdaDatatype{Msg}\AgdaSpace{}%
\AgdaSymbol{→}\AgdaSpace{}%
\AgdaFunction{Stack}\AgdaSpace{}%
\AgdaSymbol{→}\AgdaSpace{}%
\AgdaDatatype{Maybe}\AgdaSpace{}%
\AgdaFunction{Stack}\AgdaSymbol{)}%
\>[63]\AgdaSymbol{→}\AgdaSpace{}%
\AgdaRecord{StackState}\AgdaSpace{}%
\AgdaSymbol{→}\AgdaSpace{}%
\AgdaDatatype{Maybe}\AgdaSpace{}%
\AgdaRecord{StackState}\<%
\\
\>[0]\AgdaFunction{liftMsgStackToStateTransformer'}\AgdaSpace{}%
\AgdaBound{\,f\,}\AgdaSpace{}%
\AgdaSymbol{=}\AgdaSpace{}%
\AgdaFunction{stackTransform2StackStateTransform}\AgdaSpace{}%
\AgdaSymbol{(λ}\AgdaSpace{}%
\AgdaBound{time}\AgdaSpace{}%
\AgdaBound{msg}\AgdaSpace{}%
\AgdaSymbol{→}\AgdaSpace{}%
\AgdaBound{\,f\,}\AgdaSpace{}%
\AgdaBound{msg}\AgdaSymbol{)}\<%
\\
\\[\AgdaEmptyExtraSkip]%
\\[\AgdaEmptyExtraSkip]%
\\[\AgdaEmptyExtraSkip]%
\\[\AgdaEmptyExtraSkip]%
\>[0]\AgdaFunction{msgToMStackToIfStackToMState}\AgdaSpace{}%
\AgdaSymbol{:}\AgdaSpace{}%
\AgdaFunction{Time}\AgdaSpace{}%
\AgdaSymbol{→}\AgdaSpace{}%
\AgdaDatatype{Msg}%
\>[43]\AgdaSymbol{→}\AgdaSpace{}%
\AgdaDatatype{Maybe}\AgdaSpace{}%
\AgdaFunction{Stack}\AgdaSpace{}%
\AgdaSymbol{→}\AgdaSpace{}%
\AgdaDatatype{Maybe}\AgdaSpace{}%
\AgdaRecord{StackState}\<%
\\
\>[0]\AgdaFunction{msgToMStackToIfStackToMState}\AgdaSpace{}%
\AgdaBound{time}\AgdaSpace{}%
\AgdaBound{msg}\AgdaSpace{}%
\AgdaInductiveConstructor{nothing}\AgdaSpace{}%
\AgdaSymbol{=}\AgdaSpace{}%
\AgdaInductiveConstructor{nothing}\<%
\\
\>[0]\AgdaFunction{msgToMStackToIfStackToMState}\AgdaSpace{}%
\AgdaBound{time}\AgdaSpace{}%
\AgdaBound{msg}\AgdaSpace{}%
\AgdaSymbol{(}\AgdaInductiveConstructor{just}\AgdaSpace{}%
\AgdaBound{x}\AgdaSymbol{)}\AgdaSpace{}%
\AgdaSymbol{=}\AgdaSpace{}%
\AgdaInductiveConstructor{just}\AgdaSpace{}%
\AgdaOperator{\AgdaInductiveConstructor{⟨}}\AgdaSpace{}%
\AgdaBound{time}\AgdaSpace{}%
\AgdaOperator{\AgdaInductiveConstructor{,}}\AgdaSpace{}%
\AgdaBound{msg}\AgdaSpace{}%
\AgdaOperator{\AgdaInductiveConstructor{,}}\AgdaSpace{}%
\AgdaBound{x}\AgdaSpace{}%
\AgdaOperator{\AgdaInductiveConstructor{⟩}}\<%
\\
\\[\AgdaEmptyExtraSkip]%
\\[\AgdaEmptyExtraSkip]%
\>[0]\AgdaFunction{liftStackFun2StackState}\AgdaSpace{}%
\AgdaSymbol{:}\AgdaSpace{}%
\AgdaSymbol{(}\AgdaFunction{Time}\AgdaSpace{}%
\AgdaSymbol{→}\AgdaSpace{}%
\AgdaDatatype{Msg}\AgdaSpace{}%
\AgdaSymbol{→}\AgdaSpace{}%
\AgdaFunction{Stack}\AgdaSpace{}%
\AgdaSymbol{→}\AgdaSpace{}%
\AgdaDatatype{Maybe}\AgdaSpace{}%
\AgdaFunction{Stack}\AgdaSymbol{)}\AgdaSpace{}%
\AgdaSymbol{→}\AgdaSpace{}%
\AgdaRecord{StackState}\AgdaSpace{}%
\AgdaSymbol{→}\AgdaSpace{}%
\AgdaDatatype{Maybe}\AgdaSpace{}%
\AgdaRecord{StackState}\<%
\\
\>[0]\AgdaFunction{liftStackFun2StackState}\AgdaSpace{}%
\AgdaBound{\,f\,}\AgdaSpace{}%
\AgdaOperator{\AgdaInductiveConstructor{⟨}}\AgdaSpace{}%
\AgdaBound{currentTime₁}\AgdaSpace{}%
\AgdaOperator{\AgdaInductiveConstructor{,}}\AgdaSpace{}%
\AgdaBound{msg₁}\AgdaSpace{}%
\AgdaOperator{\AgdaInductiveConstructor{,}}\AgdaSpace{}%
\AgdaBound{stack₁}\AgdaSpace{}%
\AgdaOperator{\AgdaInductiveConstructor{⟩}}\AgdaSpace{}%
\AgdaSymbol{=}\<%
\\
\>[0][@{}l@{\AgdaIndent{0}}]%
\>[4]\AgdaFunction{stackState2WithMaybe}\AgdaSpace{}%
\AgdaOperator{\AgdaInductiveConstructor{⟨}}\AgdaSpace{}%
\AgdaBound{currentTime₁}\AgdaSpace{}%
\AgdaOperator{\AgdaInductiveConstructor{,}}\AgdaSpace{}%
\AgdaBound{msg₁}\AgdaSpace{}%
\AgdaOperator{\AgdaInductiveConstructor{,}}\AgdaSymbol{(}\AgdaBound{\,f\,}\AgdaSpace{}%
\AgdaBound{currentTime₁}%
\>[65]\AgdaBound{msg₁}%
\>[71]\AgdaBound{stack₁}\AgdaSymbol{)}\AgdaSpace{}%
\AgdaOperator{\AgdaInductiveConstructor{⟩}}\<%
\\
\\[\AgdaEmptyExtraSkip]%
\>[0]\AgdaComment{\{-
semanticBasicOperations.executeStackEquality\ param\ (x₁\ .stack)\ ⟩

msgToMStackToIfStackToMState\ currentTime₁\ msg₁
\ \ \ \ \ \ \ \ \ \ \ \ \ \ \ \ \ \ \ \ \ \ \ \ \ \ \ \ \ \ \ \ \ \ \ \ \ \ \ \ \ \ \ \ \ \ \ \ \ \ \ \ \ \ \ \ \ \ \ \ \ \ \ \ \ \ \ \ \ \ (f\ currentTime₁\ \ msg₁\ \ stack₁)
-\}}\<%
\end{code}
} 


\AgdaHide{
\begin{code}%
\>[0]\AgdaKeyword{module}\AgdaSpace{}%
\AgdaModule{paperTypes2021PostProceed.maybeDef}\AgdaSpace{}%
\AgdaKeyword{where}\<%
\\
\\[\AgdaEmptyExtraSkip]%
\>[0]\<%
\end{code}
} 

\newcommand{\maybeDefMaybe}{
\begin{code}%
\>[0]\AgdaKeyword{data}\AgdaSpace{}%
\AgdaDatatype{Maybe}\AgdaSpace{}%
\AgdaSymbol{(}\AgdaBound{X}\AgdaSpace{}%
\AgdaSymbol{:}\AgdaSpace{}%
\AgdaPrimitive{Set}\AgdaSymbol{)}\AgdaSpace{}%
\AgdaSymbol{:}\AgdaSpace{}%
\AgdaPrimitive{Set}\AgdaSpace{}%
\AgdaKeyword{where}%
\>[34]\AgdaInductiveConstructor{nothing}\AgdaSpace{}%
\AgdaSymbol{:}%
\>[45]\AgdaDatatype{Maybe}\AgdaSpace{}%
\AgdaBound{X}\AgdaSymbol{;}\AgdaSpace{}%
\AgdaInductiveConstructor{just}%
\>[62]\AgdaSymbol{:}%
\>[65]\AgdaBound{X}\AgdaSpace{}%
\AgdaSymbol{→}\AgdaSpace{}%
\AgdaDatatype{Maybe}\AgdaSpace{}%
\AgdaBound{X}\<%
\end{code}
} 

\AgdaHide{
\begin{code}%
\>[0]\<%
\\
\\[\AgdaEmptyExtraSkip]%
\>[0]\AgdaFunction{return}\AgdaSpace{}%
\AgdaSymbol{:}\AgdaSpace{}%
\AgdaSymbol{\{}\AgdaBound{A}\AgdaSpace{}%
\AgdaSymbol{:}\AgdaSpace{}%
\AgdaPrimitive{Set}\AgdaSymbol{\}}\AgdaSpace{}%
\AgdaSymbol{→}\AgdaSpace{}%
\AgdaBound{A}\AgdaSpace{}%
\AgdaSymbol{→}\AgdaSpace{}%
\AgdaDatatype{Maybe}\AgdaSpace{}%
\AgdaBound{A}\<%
\\
\>[0]\AgdaFunction{return}\AgdaSpace{}%
\AgdaSymbol{=}\AgdaSpace{}%
\AgdaInductiveConstructor{just}\<%
\\
\\[\AgdaEmptyExtraSkip]%
\\[\AgdaEmptyExtraSkip]%
\>[0]\AgdaOperator{\AgdaFunction{\AgdaUnderscore{}\ensuremath{\mathbin{{>}\mkern-8.5mu{>}\mkern-2mu{=}}}\AgdaUnderscore{}}}\AgdaSpace{}%
\AgdaSymbol{:}\AgdaSpace{}%
\AgdaSymbol{\{}\AgdaBound{A}\AgdaSpace{}%
\AgdaBound{B}\AgdaSpace{}%
\AgdaSymbol{:}\AgdaSpace{}%
\AgdaPrimitive{Set}\AgdaSymbol{\}}\AgdaSpace{}%
\AgdaSymbol{→}\AgdaSpace{}%
\AgdaDatatype{Maybe}\AgdaSpace{}%
\AgdaBound{A}\AgdaSpace{}%
\AgdaSymbol{→}\AgdaSpace{}%
\AgdaSymbol{(}\AgdaBound{A}\AgdaSpace{}%
\AgdaSymbol{→}\AgdaSpace{}%
\AgdaDatatype{Maybe}\AgdaSpace{}%
\AgdaBound{B}\AgdaSymbol{)}\AgdaSpace{}%
\AgdaSymbol{→}\AgdaSpace{}%
\AgdaDatatype{Maybe}\AgdaSpace{}%
\AgdaBound{B}\<%
\\
\>[0]\AgdaInductiveConstructor{nothing}\AgdaSpace{}%
\AgdaOperator{\AgdaFunction{\ensuremath{\mathbin{{>}\mkern-8.5mu{>}\mkern-2mu{=}}}}}\AgdaSpace{}%
\AgdaBound{q}\AgdaSpace{}%
\AgdaSymbol{=}\AgdaSpace{}%
\AgdaInductiveConstructor{nothing}\<%
\\
\>[0]\AgdaInductiveConstructor{just}\AgdaSpace{}%
\AgdaBound{x}\AgdaSpace{}%
\AgdaOperator{\AgdaFunction{\ensuremath{\mathbin{{>}\mkern-8.5mu{>}\mkern-2mu{=}}}}}\AgdaSpace{}%
\AgdaBound{q}\AgdaSpace{}%
\AgdaSymbol{=}\AgdaSpace{}%
\AgdaBound{q}\AgdaSpace{}%
\AgdaBound{x}\<%
\end{code}
} 


\AgdaHide{
\begin{code}%
\>[0]\AgdaComment{--\textbackslash{}multisig}\<%
\\
\\[\AgdaEmptyExtraSkip]%
\>[0]\AgdaComment{\{-\ OPTIONS\ --allow-unsolved-metas\ \#-\}}\<%
\\
\\[\AgdaEmptyExtraSkip]%
\>[0]\AgdaKeyword{open}\AgdaSpace{}%
\AgdaKeyword{import}\AgdaSpace{}%
\AgdaModule{basicBitcoinDataType}\<%
\\
\\[\AgdaEmptyExtraSkip]%
\>[0]\AgdaKeyword{module}\AgdaSpace{}%
\AgdaModule{verificationMultiSig}\AgdaSpace{}%
\AgdaSymbol{(}\AgdaBound{param}\AgdaSpace{}%
\AgdaSymbol{:}\AgdaSpace{}%
\AgdaRecord{GlobalParameters}\AgdaSymbol{)}\AgdaSpace{}%
\AgdaKeyword{where}\<%
\\
\\[\AgdaEmptyExtraSkip]%
\\[\AgdaEmptyExtraSkip]%
\>[0]\AgdaKeyword{open}\AgdaSpace{}%
\AgdaKeyword{import}\AgdaSpace{}%
\AgdaModule{Data.List.Base}\AgdaSpace{}%
\AgdaKeyword{hiding}\AgdaSpace{}%
\AgdaSymbol{(}\AgdaOperator{\AgdaFunction{\AgdaUnderscore{}\ensuremath{+\!\!+}\AgdaUnderscore{}}}\AgdaSpace{}%
\AgdaSymbol{)}\<%
\\
\>[0]\AgdaKeyword{open}\AgdaSpace{}%
\AgdaKeyword{import}\AgdaSpace{}%
\AgdaModule{Data.Nat}%
\>[22]\AgdaKeyword{renaming}\AgdaSpace{}%
\AgdaSymbol{(}\AgdaOperator{\AgdaDatatype{\AgdaUnderscore{}≤\AgdaUnderscore{}}}\AgdaSpace{}%
\AgdaSymbol{to}\AgdaSpace{}%
\AgdaOperator{\AgdaDatatype{\AgdaUnderscore{}≤'\AgdaUnderscore{}}}\AgdaSymbol{)}\AgdaSpace{}%
\AgdaComment{--\ \ \AgdaUnderscore{}<\AgdaUnderscore{}\ to\ \AgdaUnderscore{}<'\AgdaUnderscore{})}\<%
\\
\>[0]\AgdaKeyword{open}\AgdaSpace{}%
\AgdaKeyword{import}\AgdaSpace{}%
\AgdaModule{Data.List}\AgdaSpace{}%
\AgdaKeyword{hiding}\AgdaSpace{}%
\AgdaSymbol{(}\AgdaOperator{\AgdaFunction{\AgdaUnderscore{}\ensuremath{+\!\!+}\AgdaUnderscore{}}}%
\>[36]\AgdaSymbol{)}\<%
\\
\>[0]\AgdaKeyword{open}\AgdaSpace{}%
\AgdaKeyword{import}\AgdaSpace{}%
\AgdaModule{Data.Sum}\<%
\\
\>[0]\AgdaKeyword{open}\AgdaSpace{}%
\AgdaKeyword{import}\AgdaSpace{}%
\AgdaModule{Data.Unit}%
\>[23]\AgdaKeyword{hiding}\AgdaSpace{}%
\AgdaSymbol{(}\AgdaOperator{\AgdaRecord{\AgdaUnderscore{}≤\AgdaUnderscore{}}}\AgdaSymbol{)}\<%
\\
\>[0]\AgdaKeyword{open}\AgdaSpace{}%
\AgdaKeyword{import}\AgdaSpace{}%
\AgdaModule{Data.Empty}\<%
\\
\>[0]\AgdaKeyword{open}\AgdaSpace{}%
\AgdaKeyword{import}\AgdaSpace{}%
\AgdaModule{Data.Maybe}\<%
\\
\>[0]\AgdaComment{--open\ import\ Data.List.Base}\<%
\\
\>[0]\AgdaKeyword{open}\AgdaSpace{}%
\AgdaKeyword{import}\AgdaSpace{}%
\AgdaModule{Data.Bool}%
\>[23]\AgdaKeyword{hiding}\AgdaSpace{}%
\AgdaSymbol{(}\AgdaOperator{\AgdaDatatype{\AgdaUnderscore{}≤\AgdaUnderscore{}}}\AgdaSpace{}%
\AgdaSymbol{;}\AgdaSpace{}%
\AgdaOperator{\AgdaDatatype{\AgdaUnderscore{}<\AgdaUnderscore{}}}\AgdaSpace{}%
\AgdaSymbol{;}\AgdaSpace{}%
\AgdaOperator{\AgdaFunction{if\AgdaUnderscore{}then\AgdaUnderscore{}else\AgdaUnderscore{}}}%
\>[58]\AgdaSymbol{)}%
\>[61]\AgdaKeyword{renaming}\AgdaSpace{}%
\AgdaSymbol{(}\AgdaOperator{\AgdaFunction{\AgdaUnderscore{}∧\AgdaUnderscore{}}}\AgdaSpace{}%
\AgdaSymbol{to}\AgdaSpace{}%
\AgdaOperator{\AgdaFunction{\AgdaUnderscore{}∧b\AgdaUnderscore{}}}\AgdaSpace{}%
\AgdaSymbol{;}\AgdaSpace{}%
\AgdaOperator{\AgdaFunction{\AgdaUnderscore{}∨\AgdaUnderscore{}}}\AgdaSpace{}%
\AgdaSymbol{to}\AgdaSpace{}%
\AgdaOperator{\AgdaFunction{\AgdaUnderscore{}∨b\AgdaUnderscore{}}}\AgdaSpace{}%
\AgdaSymbol{;}\AgdaSpace{}%
\AgdaFunction{T}\AgdaSpace{}%
\AgdaSymbol{to}\AgdaSpace{}%
\AgdaFunction{True}\AgdaSymbol{)}\<%
\\
\>[0]\AgdaKeyword{open}\AgdaSpace{}%
\AgdaKeyword{import}\AgdaSpace{}%
\AgdaModule{Data.Product}\AgdaSpace{}%
\AgdaKeyword{renaming}\AgdaSpace{}%
\AgdaSymbol{(}\AgdaOperator{\AgdaInductiveConstructor{\AgdaUnderscore{},\AgdaUnderscore{}}}\AgdaSpace{}%
\AgdaSymbol{to}\AgdaSpace{}%
\AgdaOperator{\AgdaInductiveConstructor{\AgdaUnderscore{},,\AgdaUnderscore{}}}\AgdaSpace{}%
\AgdaSymbol{)}\<%
\\
\>[0]\AgdaKeyword{open}\AgdaSpace{}%
\AgdaKeyword{import}\AgdaSpace{}%
\AgdaModule{Data.Nat.Base}\AgdaSpace{}%
\AgdaKeyword{hiding}\AgdaSpace{}%
\AgdaSymbol{(}\AgdaOperator{\AgdaDatatype{\AgdaUnderscore{}≤\AgdaUnderscore{}}}\AgdaSpace{}%
\AgdaSymbol{;}\AgdaSpace{}%
\AgdaOperator{\AgdaFunction{\AgdaUnderscore{}<\AgdaUnderscore{}}}\AgdaSymbol{)}\<%
\\
\>[0]\AgdaKeyword{open}\AgdaSpace{}%
\AgdaKeyword{import}\AgdaSpace{}%
\AgdaModule{Data.List.NonEmpty}\AgdaSpace{}%
\AgdaKeyword{hiding}\AgdaSpace{}%
\AgdaSymbol{(}\AgdaField{head}\AgdaSymbol{;}\AgdaSpace{}%
\AgdaOperator{\AgdaFunction{[\AgdaUnderscore{}]}}\AgdaSymbol{;}\AgdaSpace{}%
\AgdaFunction{length}\AgdaSymbol{)}\<%
\\
\>[0]\AgdaKeyword{open}\AgdaSpace{}%
\AgdaKeyword{import}\AgdaSpace{}%
\AgdaModule{Data.Nat}\AgdaSpace{}%
\AgdaKeyword{using}\AgdaSpace{}%
\AgdaSymbol{(}\AgdaDatatype{ℕ}\AgdaSymbol{;}\AgdaSpace{}%
\AgdaOperator{\AgdaPrimitive{\AgdaUnderscore{}+\AgdaUnderscore{}}}\AgdaSymbol{;}\AgdaSpace{}%
\AgdaOperator{\AgdaFunction{\AgdaUnderscore{}≥\AgdaUnderscore{}}}\AgdaSymbol{;}\AgdaSpace{}%
\AgdaOperator{\AgdaFunction{\AgdaUnderscore{}>\AgdaUnderscore{}}}\AgdaSymbol{;}\AgdaSpace{}%
\AgdaInductiveConstructor{zero}\AgdaSymbol{;}\AgdaSpace{}%
\AgdaInductiveConstructor{suc}\AgdaSymbol{;}\AgdaSpace{}%
\AgdaInductiveConstructor{s≤s}\AgdaSymbol{;}\AgdaSpace{}%
\AgdaInductiveConstructor{z≤n}\AgdaSymbol{)}\<%
\\
\>[0]\AgdaKeyword{import}\AgdaSpace{}%
\AgdaModule{Relation.Binary.PropositionalEquality}\AgdaSpace{}%
\AgdaSymbol{as}\AgdaSpace{}%
\AgdaModule{Eq}\<%
\\
\>[0]\AgdaKeyword{open}\AgdaSpace{}%
\AgdaModule{Eq}\AgdaSpace{}%
\AgdaKeyword{using}\AgdaSpace{}%
\AgdaSymbol{(}\AgdaOperator{\AgdaDatatype{\AgdaUnderscore{}≡\AgdaUnderscore{}}}\AgdaSymbol{;}\AgdaSpace{}%
\AgdaInductiveConstructor{refl}\AgdaSymbol{;}\AgdaSpace{}%
\AgdaFunction{cong}\AgdaSymbol{;}\AgdaSpace{}%
\AgdaKeyword{module}\AgdaSpace{}%
\AgdaModule{≡-Reasoning}\AgdaSymbol{;}\AgdaSpace{}%
\AgdaFunction{sym}\AgdaSymbol{)}\<%
\\
\>[0]\AgdaKeyword{open}\AgdaSpace{}%
\AgdaModule{≡-Reasoning}\<%
\\
\>[0]\AgdaKeyword{open}\AgdaSpace{}%
\AgdaKeyword{import}\AgdaSpace{}%
\AgdaModule{Agda.Builtin.Equality}\<%
\\
\>[0]\AgdaComment{--open\ import\ Agda.Builtin.Equality.Rewrite}\<%
\\
\\[\AgdaEmptyExtraSkip]%
\>[0]\AgdaKeyword{open}\AgdaSpace{}%
\AgdaKeyword{import}\AgdaSpace{}%
\AgdaModule{libraries.listLib}\<%
\\
\>[0]\AgdaKeyword{open}\AgdaSpace{}%
\AgdaKeyword{import}\AgdaSpace{}%
\AgdaModule{libraries.emptyLib}\<%
\\
\>[0]\AgdaKeyword{open}\AgdaSpace{}%
\AgdaKeyword{import}\AgdaSpace{}%
\AgdaModule{libraries.natLib}\<%
\\
\>[0]\AgdaKeyword{open}\AgdaSpace{}%
\AgdaKeyword{import}\AgdaSpace{}%
\AgdaModule{libraries.boolLib}\<%
\\
\>[0]\AgdaKeyword{open}\AgdaSpace{}%
\AgdaKeyword{import}\AgdaSpace{}%
\AgdaModule{libraries.equalityLib}\<%
\\
\\[\AgdaEmptyExtraSkip]%
\\[\AgdaEmptyExtraSkip]%
\\[\AgdaEmptyExtraSkip]%
\>[0]\AgdaKeyword{open}\AgdaSpace{}%
\AgdaKeyword{import}\AgdaSpace{}%
\AgdaModule{libraries.andLib}\<%
\\
\>[0]\AgdaKeyword{open}\AgdaSpace{}%
\AgdaKeyword{import}\AgdaSpace{}%
\AgdaModule{libraries.miscLib}\<%
\\
\>[0]\AgdaKeyword{open}\AgdaSpace{}%
\AgdaKeyword{import}\AgdaSpace{}%
\AgdaModule{libraries.maybeLib}\<%
\\
\\[\AgdaEmptyExtraSkip]%
\>[0]\AgdaKeyword{open}\AgdaSpace{}%
\AgdaKeyword{import}\AgdaSpace{}%
\AgdaModule{stack}\<%
\\
\>[0]\AgdaKeyword{open}\AgdaSpace{}%
\AgdaKeyword{import}\AgdaSpace{}%
\AgdaModule{stackPredicate}\<%
\\
\>[0]\AgdaKeyword{open}\AgdaSpace{}%
\AgdaKeyword{import}\AgdaSpace{}%
\AgdaModule{instructionBasic}\<%
\\
\>[0]\AgdaKeyword{open}\AgdaSpace{}%
\AgdaKeyword{import}\AgdaSpace{}%
\AgdaModule{semanticBasicOperations}\AgdaSpace{}%
\AgdaBound{param}\<%
\\
\>[0]\AgdaKeyword{open}\AgdaSpace{}%
\AgdaKeyword{import}\AgdaSpace{}%
\AgdaModule{stackSemanticsInstructions}\AgdaSpace{}%
\AgdaBound{param}\<%
\\
\>[0]\AgdaKeyword{open}\AgdaSpace{}%
\AgdaKeyword{import}\AgdaSpace{}%
\AgdaModule{hoareTripleStack}\AgdaSpace{}%
\AgdaBound{param}\<%
\\
\\[\AgdaEmptyExtraSkip]%
\>[0]\AgdaComment{-------------------------------------------------------}\<%
\\
\>[0]\AgdaComment{--\ Verification\ of\ the\ Multisig\ script\ both\ for\ active\ and\ nonactive\ ifStack}\<%
\\
\>[0]\AgdaComment{--\ \ theoremCorrectnessMultiSig-2-3\ \ \ proves\ correctness\ for\ active\ ifStack}\<%
\\
\>[0]\AgdaComment{--\ \ theoremCorrectnessMultiSig-2-3-nonactive\ proves\ correctness\ for\ nonactive\ ifStack}\<%
\\
\>[0]\AgdaComment{-----------------------------------------------------------------}\<%
\\
\\[\AgdaEmptyExtraSkip]%
\\[\AgdaEmptyExtraSkip]%
\\[\AgdaEmptyExtraSkip]%
\\[\AgdaEmptyExtraSkip]%
\\[\AgdaEmptyExtraSkip]%
\\[\AgdaEmptyExtraSkip]%
\>[0]\AgdaFunction{weakestPreCondMultiSig-2-3-bas}\AgdaSpace{}%
\AgdaSymbol{:}%
\>[125I]\AgdaSymbol{(}\AgdaBound{pbk1}\AgdaSpace{}%
\AgdaBound{pbk2}\AgdaSpace{}%
\AgdaBound{pbk3}\AgdaSpace{}%
\AgdaSymbol{:}%
\>[52]\AgdaDatatype{ℕ}\AgdaSymbol{)}\<%
\\
\>[125I][@{}l@{\AgdaIndent{0}}]%
\>[38]\AgdaSymbol{→}\AgdaSpace{}%
\AgdaFunction{StackPredicate}\<%
\\
\>[0]\AgdaFunction{weakestPreCondMultiSig-2-3-bas}\AgdaSpace{}%
\AgdaBound{pbk1}\AgdaSpace{}%
\AgdaBound{pbk2}\AgdaSpace{}%
\AgdaBound{pbk3}\AgdaSpace{}%
\AgdaBound{time}\AgdaSpace{}%
\AgdaBound{msg₁}\AgdaSpace{}%
\AgdaInductiveConstructor{[]}\AgdaSpace{}%
\AgdaSymbol{=}\AgdaSpace{}%
\AgdaDatatype{⊥}\<%
\\
\>[0]\AgdaFunction{weakestPreCondMultiSig-2-3-bas}\AgdaSpace{}%
\AgdaBound{pbk1}\AgdaSpace{}%
\AgdaBound{pbk2}\AgdaSpace{}%
\AgdaBound{pbk3}\AgdaSpace{}%
\AgdaBound{time}\AgdaSpace{}%
\AgdaBound{msg₁}\AgdaSpace{}%
\AgdaSymbol{(}\AgdaBound{x}\AgdaSpace{}%
\AgdaOperator{\AgdaInductiveConstructor{∷}}\AgdaSpace{}%
\AgdaInductiveConstructor{[]}\AgdaSymbol{)}\AgdaSpace{}%
\AgdaSymbol{=}\AgdaSpace{}%
\AgdaDatatype{⊥}\<%
\\
\>[0]\AgdaFunction{weakestPreCondMultiSig-2-3-bas}\AgdaSpace{}%
\AgdaBound{pbk1}\AgdaSpace{}%
\AgdaBound{pbk2}\AgdaSpace{}%
\AgdaBound{pbk3}\AgdaSpace{}%
\AgdaBound{time}\AgdaSpace{}%
\AgdaBound{msg₁}\AgdaSpace{}%
\AgdaSymbol{(}\AgdaBound{x}\AgdaSpace{}%
\AgdaOperator{\AgdaInductiveConstructor{∷}}\AgdaSpace{}%
\AgdaBound{y}\AgdaSpace{}%
\AgdaOperator{\AgdaInductiveConstructor{∷}}\AgdaSpace{}%
\AgdaInductiveConstructor{[]}\AgdaSymbol{)}\AgdaSpace{}%
\AgdaSymbol{=}\AgdaSpace{}%
\AgdaDatatype{⊥}\<%
\\
\>[0]\AgdaFunction{weakestPreCondMultiSig-2-3-bas}\AgdaSpace{}%
\AgdaBound{pbk1}\AgdaSpace{}%
\AgdaBound{pbk2}\AgdaSpace{}%
\AgdaBound{pbk3}\AgdaSpace{}%
\AgdaBound{time}\AgdaSpace{}%
\AgdaBound{msg₁}\AgdaSpace{}%
\AgdaSymbol{(}\AgdaBound{sig2}\AgdaSpace{}%
\AgdaOperator{\AgdaInductiveConstructor{∷}}\AgdaSpace{}%
\AgdaBound{sig1}\AgdaSpace{}%
\AgdaOperator{\AgdaInductiveConstructor{∷}}\AgdaSpace{}%
\AgdaBound{dummy}\AgdaSpace{}%
\AgdaOperator{\AgdaInductiveConstructor{∷}}\AgdaSpace{}%
\AgdaBound{stack₁}\AgdaSymbol{)}\AgdaSpace{}%
\AgdaSymbol{=}\<%
\\
\>[0][@{}l@{\AgdaIndent{0}}]%
\>[9]\AgdaSymbol{(}\AgdaSpace{}%
\AgdaSymbol{(}\AgdaFunction{IsSigned}\AgdaSpace{}%
\AgdaBound{msg₁}\AgdaSpace{}%
\AgdaBound{sig1}\AgdaSpace{}%
\AgdaBound{pbk1}\AgdaSpace{}%
\AgdaOperator{\AgdaRecord{∧}}%
\>[39]\AgdaFunction{IsSigned}\AgdaSpace{}%
\AgdaBound{msg₁}\AgdaSpace{}%
\AgdaBound{sig2}\AgdaSpace{}%
\AgdaBound{pbk2}\AgdaSymbol{)}\AgdaSpace{}%
\AgdaOperator{\AgdaDatatype{⊎}}\<%
\\
\>[9][@{}l@{\AgdaIndent{0}}]%
\>[10]\AgdaSymbol{(}\AgdaFunction{IsSigned}\AgdaSpace{}%
\AgdaBound{msg₁}\AgdaSpace{}%
\AgdaBound{sig1}\AgdaSpace{}%
\AgdaBound{pbk1}\AgdaSpace{}%
\AgdaOperator{\AgdaRecord{∧}}%
\>[38]\AgdaFunction{IsSigned}\AgdaSpace{}%
\AgdaBound{msg₁}\AgdaSpace{}%
\AgdaBound{sig2}\AgdaSpace{}%
\AgdaBound{pbk3}\AgdaSymbol{)}\AgdaSpace{}%
\AgdaOperator{\AgdaDatatype{⊎}}\<%
\\
\>[10]\AgdaSymbol{(}\AgdaFunction{IsSigned}\AgdaSpace{}%
\AgdaBound{msg₁}\AgdaSpace{}%
\AgdaBound{sig1}\AgdaSpace{}%
\AgdaBound{pbk2}\AgdaSpace{}%
\AgdaOperator{\AgdaRecord{∧}}%
\>[38]\AgdaFunction{IsSigned}\AgdaSpace{}%
\AgdaBound{msg₁}\AgdaSpace{}%
\AgdaBound{sig2}\AgdaSpace{}%
\AgdaBound{pbk3}\AgdaSpace{}%
\AgdaSymbol{))}\<%
\\
\\[\AgdaEmptyExtraSkip]%
\\[\AgdaEmptyExtraSkip]%
\>[0]\AgdaFunction{multiSigScript-2-3-b}\AgdaSpace{}%
\AgdaSymbol{:}\AgdaSpace{}%
\AgdaSymbol{(}\AgdaBound{pbk1}\AgdaSpace{}%
\AgdaBound{pbk2}\AgdaSpace{}%
\AgdaBound{pbk3}\AgdaSpace{}%
\AgdaSymbol{:}%
\>[42]\AgdaDatatype{ℕ}\AgdaSymbol{)}\AgdaSpace{}%
\AgdaSymbol{→}\AgdaSpace{}%
\AgdaFunction{BitcoinScriptBasic}\<%
\\
\>[0]\AgdaFunction{multiSigScript-2-3-b}\AgdaSpace{}%
\AgdaBound{pbk1}\AgdaSpace{}%
\AgdaBound{pbk2}\AgdaSpace{}%
\AgdaBound{pbk3}\AgdaSpace{}%
\AgdaSymbol{=}\AgdaSpace{}%
\AgdaSymbol{(}\AgdaInductiveConstructor{opPush}\AgdaSpace{}%
\AgdaNumber{2}\AgdaSymbol{)}\AgdaSpace{}%
\AgdaOperator{\AgdaInductiveConstructor{∷}}\AgdaSpace{}%
\AgdaSymbol{(}\AgdaInductiveConstructor{opPush}\AgdaSpace{}%
\AgdaBound{pbk1}\AgdaSymbol{)}\AgdaSpace{}%
\AgdaOperator{\AgdaInductiveConstructor{∷}}%
\>[68]\AgdaSymbol{(}\AgdaInductiveConstructor{opPush}\AgdaSpace{}%
\AgdaBound{pbk2}\AgdaSymbol{)}\AgdaSpace{}%
\AgdaOperator{\AgdaInductiveConstructor{∷}}%
\>[85]\AgdaSymbol{(}\AgdaInductiveConstructor{opPush}\AgdaSpace{}%
\AgdaBound{pbk3}\AgdaSymbol{)}\AgdaSpace{}%
\AgdaOperator{\AgdaInductiveConstructor{∷}}%
\>[102]\AgdaSymbol{(}\AgdaInductiveConstructor{opPush}\AgdaSpace{}%
\AgdaNumber{3}\AgdaSymbol{)}\AgdaSpace{}%
\AgdaOperator{\AgdaInductiveConstructor{∷}}%
\>[117]\AgdaInductiveConstructor{opMultiSig}\AgdaSpace{}%
\AgdaOperator{\AgdaInductiveConstructor{∷}}\AgdaSpace{}%
\AgdaInductiveConstructor{[]}\<%
\\
\\[\AgdaEmptyExtraSkip]%
\>[0]\AgdaFunction{lemmaHoareTripleStackGeAux'1}%
\>[223I]\AgdaSymbol{:}%
\>[224I]\AgdaSymbol{(}\AgdaBound{msg₂}\AgdaSpace{}%
\AgdaSymbol{:}\AgdaSpace{}%
\AgdaDatatype{Msg}\AgdaSymbol{)}\<%
\\
\>[.][@{}l@{}]\<[224I]%
\>[31]\AgdaSymbol{(}\AgdaBound{pbk1}\AgdaSpace{}%
\AgdaBound{pbk2}\AgdaSpace{}%
\AgdaBound{pbk3}\AgdaSpace{}%
\AgdaBound{sig1}\AgdaSpace{}%
\AgdaBound{sig2}\AgdaSpace{}%
\AgdaSymbol{:}\AgdaSpace{}%
\AgdaDatatype{ℕ}\AgdaSymbol{)}\<%
\\
\>[31]\AgdaSymbol{→}\AgdaSpace{}%
\AgdaFunction{True}\AgdaSpace{}%
\AgdaSymbol{(}\AgdaFunction{isSigned}%
\>[49]\AgdaBound{msg₂}\AgdaSpace{}%
\AgdaBound{sig1}\AgdaSpace{}%
\AgdaBound{pbk1}\AgdaSymbol{)}\<%
\\
\>[223I][@{}l@{\AgdaIndent{0}}]%
\>[30]\AgdaSymbol{→}%
\>[33]\AgdaFunction{True}\AgdaSpace{}%
\AgdaSymbol{(}\AgdaFunction{isSigned}%
\>[49]\AgdaBound{msg₂}\AgdaSpace{}%
\AgdaBound{sig2}\AgdaSpace{}%
\AgdaBound{pbk2}\AgdaSymbol{)}\<%
\\
\>[30][@{}l@{\AgdaIndent{0}}]%
\>[31]\AgdaSymbol{→}\AgdaSpace{}%
\AgdaFunction{True}\AgdaSpace{}%
\AgdaSymbol{(}\AgdaFunction{compareSigsMultiSig}\AgdaSpace{}%
\AgdaBound{msg₂}\AgdaSpace{}%
\AgdaSymbol{(}\AgdaSpace{}%
\AgdaBound{sig1}\AgdaSpace{}%
\AgdaOperator{\AgdaInductiveConstructor{∷}}\AgdaSpace{}%
\AgdaBound{sig2}\AgdaSpace{}%
\AgdaOperator{\AgdaInductiveConstructor{∷}}\AgdaSpace{}%
\AgdaInductiveConstructor{[]}\AgdaSymbol{)}\AgdaSpace{}%
\AgdaSymbol{(}\AgdaBound{pbk1}\AgdaSpace{}%
\AgdaOperator{\AgdaInductiveConstructor{∷}}\AgdaSpace{}%
\AgdaBound{pbk2}\AgdaSpace{}%
\AgdaOperator{\AgdaInductiveConstructor{∷}}\AgdaSpace{}%
\AgdaBound{pbk3}\AgdaSpace{}%
\AgdaOperator{\AgdaInductiveConstructor{∷}}\AgdaSpace{}%
\AgdaInductiveConstructor{[]}\AgdaSymbol{))}\<%
\\
\\[\AgdaEmptyExtraSkip]%
\>[0]\AgdaFunction{lemmaHoareTripleStackGeAux'1}\AgdaSpace{}%
\AgdaBound{msg₂}\AgdaSpace{}%
\AgdaBound{pbk1}\AgdaSpace{}%
\AgdaBound{pbk2}\AgdaSpace{}%
\AgdaBound{pbk3}\AgdaSpace{}%
\AgdaBound{sig1}\AgdaSpace{}%
\AgdaBound{sig2}\AgdaSpace{}%
\AgdaBound{x}\AgdaSpace{}%
\AgdaBound{x₁}\AgdaSpace{}%
\AgdaKeyword{with}%
\>[70]\AgdaSymbol{(}\AgdaFunction{isSigned}%
\>[81]\AgdaBound{msg₂}\AgdaSpace{}%
\AgdaBound{sig1}\AgdaSpace{}%
\AgdaBound{pbk1}\AgdaSymbol{)}\<%
\\
\>[0]\AgdaFunction{lemmaHoareTripleStackGeAux'1}\AgdaSpace{}%
\AgdaBound{msg₂}\AgdaSpace{}%
\AgdaBound{pbk1}\AgdaSpace{}%
\AgdaBound{pbk2}\AgdaSpace{}%
\AgdaBound{pbk3}\AgdaSpace{}%
\AgdaBound{sig1}\AgdaSpace{}%
\AgdaBound{sig2}\AgdaSpace{}%
\AgdaBound{x}\AgdaSpace{}%
\AgdaBound{x₁}\AgdaSpace{}%
\AgdaSymbol{|}\AgdaSpace{}%
\AgdaInductiveConstructor{true}%
\>[72]\AgdaKeyword{with}\AgdaSpace{}%
\AgdaSymbol{(}\AgdaFunction{isSigned}%
\>[88]\AgdaBound{msg₂}\AgdaSpace{}%
\AgdaBound{sig2}\AgdaSpace{}%
\AgdaBound{pbk2}\AgdaSymbol{)}\<%
\\
\>[0]\AgdaFunction{lemmaHoareTripleStackGeAux'1}\AgdaSpace{}%
\AgdaBound{msg₂}\AgdaSpace{}%
\AgdaBound{pbk1}\AgdaSpace{}%
\AgdaBound{pbk2}\AgdaSpace{}%
\AgdaBound{pbk3}\AgdaSpace{}%
\AgdaBound{sig1}\AgdaSpace{}%
\AgdaBound{sig2}\AgdaSpace{}%
\AgdaBound{x}\AgdaSpace{}%
\AgdaBound{x₁}\AgdaSpace{}%
\AgdaSymbol{|}\AgdaSpace{}%
\AgdaInductiveConstructor{true}\AgdaSpace{}%
\AgdaSymbol{|}\AgdaSpace{}%
\AgdaInductiveConstructor{true}\AgdaSpace{}%
\AgdaSymbol{=}\AgdaSpace{}%
\AgdaInductiveConstructor{tt}\<%
\\
\\[\AgdaEmptyExtraSkip]%
\\[\AgdaEmptyExtraSkip]%
\\[\AgdaEmptyExtraSkip]%
\\[\AgdaEmptyExtraSkip]%
\>[0]\AgdaFunction{lemmaHoareTripleStackGeAux'2}%
\>[294I]\AgdaSymbol{:}%
\>[295I]\AgdaSymbol{(}\AgdaBound{msg₂}\AgdaSpace{}%
\AgdaSymbol{:}\AgdaSpace{}%
\AgdaDatatype{Msg}\AgdaSymbol{)}\<%
\\
\>[.][@{}l@{}]\<[295I]%
\>[31]\AgdaSymbol{(}\AgdaBound{pbk1}\AgdaSpace{}%
\AgdaBound{pbk2}\AgdaSpace{}%
\AgdaBound{pbk3}\AgdaSpace{}%
\AgdaBound{sig1}\AgdaSpace{}%
\AgdaBound{sig2}\AgdaSpace{}%
\AgdaSymbol{:}\AgdaSpace{}%
\AgdaDatatype{ℕ}\AgdaSymbol{)}\<%
\\
\>[31]\AgdaSymbol{→}\AgdaSpace{}%
\AgdaFunction{True}%
\>[39]\AgdaSymbol{(}\AgdaFunction{isSigned}%
\>[50]\AgdaBound{msg₂}\AgdaSpace{}%
\AgdaBound{sig1}\AgdaSpace{}%
\AgdaBound{pbk1}\AgdaSymbol{)}\<%
\\
\>[294I][@{}l@{\AgdaIndent{0}}]%
\>[30]\AgdaSymbol{→}%
\>[33]\AgdaFunction{True}\AgdaSpace{}%
\AgdaSymbol{(}\AgdaFunction{isSigned}%
\>[49]\AgdaBound{msg₂}\AgdaSpace{}%
\AgdaBound{sig2}\AgdaSpace{}%
\AgdaBound{pbk3}\AgdaSymbol{)}\<%
\\
\>[30][@{}l@{\AgdaIndent{0}}]%
\>[31]\AgdaSymbol{→}\AgdaSpace{}%
\AgdaFunction{True}\AgdaSpace{}%
\AgdaSymbol{(}\AgdaFunction{compareSigsMultiSig}\AgdaSpace{}%
\AgdaBound{msg₂}\AgdaSpace{}%
\AgdaSymbol{(}\AgdaSpace{}%
\AgdaBound{sig1}\AgdaSpace{}%
\AgdaOperator{\AgdaInductiveConstructor{∷}}\AgdaSpace{}%
\AgdaBound{sig2}\AgdaSpace{}%
\AgdaOperator{\AgdaInductiveConstructor{∷}}\AgdaSpace{}%
\AgdaInductiveConstructor{[]}\AgdaSymbol{)}\AgdaSpace{}%
\AgdaSymbol{(}\AgdaBound{pbk1}\AgdaSpace{}%
\AgdaOperator{\AgdaInductiveConstructor{∷}}\AgdaSpace{}%
\AgdaBound{pbk2}\AgdaSpace{}%
\AgdaOperator{\AgdaInductiveConstructor{∷}}\AgdaSpace{}%
\AgdaBound{pbk3}\AgdaSpace{}%
\AgdaOperator{\AgdaInductiveConstructor{∷}}\AgdaSpace{}%
\AgdaInductiveConstructor{[]}\AgdaSymbol{))}\<%
\\
\>[0]\AgdaFunction{lemmaHoareTripleStackGeAux'2}\AgdaSpace{}%
\AgdaBound{msg₂}\AgdaSpace{}%
\AgdaBound{pbk1}\AgdaSpace{}%
\AgdaBound{pbk2}\AgdaSpace{}%
\AgdaBound{pbk3}\AgdaSpace{}%
\AgdaBound{sig1}\AgdaSpace{}%
\AgdaBound{sig2}\AgdaSpace{}%
\AgdaBound{x}\AgdaSpace{}%
\AgdaBound{x₁}\AgdaSpace{}%
\AgdaKeyword{with}\AgdaSpace{}%
\AgdaSymbol{(}\AgdaFunction{isSigned}%
\>[80]\AgdaBound{msg₂}\AgdaSpace{}%
\AgdaBound{sig1}\AgdaSpace{}%
\AgdaBound{pbk1}\AgdaSymbol{)}\<%
\\
\>[0]\AgdaFunction{lemmaHoareTripleStackGeAux'2}\AgdaSpace{}%
\AgdaBound{msg₂}\AgdaSpace{}%
\AgdaBound{pbk1}\AgdaSpace{}%
\AgdaBound{pbk2}\AgdaSpace{}%
\AgdaBound{pbk3}\AgdaSpace{}%
\AgdaBound{sig1}\AgdaSpace{}%
\AgdaBound{sig2}\AgdaSpace{}%
\AgdaBound{x}\AgdaSpace{}%
\AgdaBound{x₁}\AgdaSpace{}%
\AgdaSymbol{|}\AgdaSpace{}%
\AgdaInductiveConstructor{true}\AgdaSpace{}%
\AgdaKeyword{with}\AgdaSpace{}%
\AgdaSymbol{(}\AgdaFunction{isSigned}%
\>[87]\AgdaBound{msg₂}\AgdaSpace{}%
\AgdaBound{sig2}\AgdaSpace{}%
\AgdaBound{pbk2}\AgdaSymbol{)}\<%
\\
\>[0]\AgdaFunction{lemmaHoareTripleStackGeAux'2}\AgdaSpace{}%
\AgdaBound{msg₂}\AgdaSpace{}%
\AgdaBound{pbk1}\AgdaSpace{}%
\AgdaBound{pbk2}\AgdaSpace{}%
\AgdaBound{pbk3}\AgdaSpace{}%
\AgdaBound{sig1}\AgdaSpace{}%
\AgdaBound{sig2}\AgdaSpace{}%
\AgdaBound{x}\AgdaSpace{}%
\AgdaBound{x₁}\AgdaSpace{}%
\AgdaSymbol{|}\AgdaSpace{}%
\AgdaInductiveConstructor{true}\AgdaSpace{}%
\AgdaSymbol{|}\AgdaSpace{}%
\AgdaInductiveConstructor{false}\AgdaSpace{}%
\AgdaKeyword{with}\AgdaSpace{}%
\AgdaSymbol{(}\AgdaFunction{isSigned}%
\>[95]\AgdaBound{msg₂}\AgdaSpace{}%
\AgdaBound{sig2}\AgdaSpace{}%
\AgdaBound{pbk3}\AgdaSymbol{)}\<%
\\
\>[0]\AgdaFunction{lemmaHoareTripleStackGeAux'2}\AgdaSpace{}%
\AgdaBound{msg₂}\AgdaSpace{}%
\AgdaBound{pbk1}\AgdaSpace{}%
\AgdaBound{pbk2}\AgdaSpace{}%
\AgdaBound{pbk3}\AgdaSpace{}%
\AgdaBound{sig1}\AgdaSpace{}%
\AgdaBound{sig2}\AgdaSpace{}%
\AgdaBound{x}\AgdaSpace{}%
\AgdaBound{x₁}\AgdaSpace{}%
\AgdaSymbol{|}\AgdaSpace{}%
\AgdaInductiveConstructor{true}\AgdaSpace{}%
\AgdaSymbol{|}\AgdaSpace{}%
\AgdaInductiveConstructor{false}\AgdaSpace{}%
\AgdaSymbol{|}\AgdaSpace{}%
\AgdaInductiveConstructor{true}\AgdaSpace{}%
\AgdaSymbol{=}\AgdaSpace{}%
\AgdaInductiveConstructor{tt}\<%
\\
\>[0]\AgdaFunction{lemmaHoareTripleStackGeAux'2}\AgdaSpace{}%
\AgdaBound{msg₂}\AgdaSpace{}%
\AgdaBound{pbk1}\AgdaSpace{}%
\AgdaBound{pbk2}\AgdaSpace{}%
\AgdaBound{pbk3}\AgdaSpace{}%
\AgdaBound{sig1}\AgdaSpace{}%
\AgdaBound{sig2}\AgdaSpace{}%
\AgdaBound{x}\AgdaSpace{}%
\AgdaBound{x₁}\AgdaSpace{}%
\AgdaSymbol{|}\AgdaSpace{}%
\AgdaInductiveConstructor{true}\AgdaSpace{}%
\AgdaSymbol{|}\AgdaSpace{}%
\AgdaInductiveConstructor{true}\AgdaSpace{}%
\AgdaSymbol{=}\AgdaSpace{}%
\AgdaInductiveConstructor{tt}\<%
\\
\\[\AgdaEmptyExtraSkip]%
\\[\AgdaEmptyExtraSkip]%
\\[\AgdaEmptyExtraSkip]%
\>[0]\AgdaFunction{lemmaHoareTripleStackGeAux'3}%
\>[398I]\AgdaSymbol{:}%
\>[399I]\AgdaSymbol{(}\AgdaBound{msg₂}\AgdaSpace{}%
\AgdaSymbol{:}\AgdaSpace{}%
\AgdaDatatype{Msg}\AgdaSymbol{)}\<%
\\
\>[.][@{}l@{}]\<[399I]%
\>[31]\AgdaSymbol{(}\AgdaBound{pbk1}\AgdaSpace{}%
\AgdaBound{pbk2}\AgdaSpace{}%
\AgdaBound{pbk3}\AgdaSpace{}%
\AgdaBound{sig1}\AgdaSpace{}%
\AgdaBound{sig2}\AgdaSpace{}%
\AgdaSymbol{:}\AgdaSpace{}%
\AgdaDatatype{ℕ}\AgdaSymbol{)}\<%
\\
\>[31]\AgdaSymbol{→}\AgdaSpace{}%
\AgdaFunction{True}\AgdaSpace{}%
\AgdaSymbol{(}\AgdaFunction{isSigned}%
\>[49]\AgdaBound{msg₂}\AgdaSpace{}%
\AgdaBound{sig1}\AgdaSpace{}%
\AgdaBound{pbk2}\AgdaSymbol{)}\<%
\\
\>[398I][@{}l@{\AgdaIndent{0}}]%
\>[30]\AgdaSymbol{→}%
\>[33]\AgdaFunction{True}\AgdaSpace{}%
\AgdaSymbol{(}\AgdaFunction{isSigned}%
\>[49]\AgdaBound{msg₂}\AgdaSpace{}%
\AgdaBound{sig2}\AgdaSpace{}%
\AgdaBound{pbk3}\AgdaSymbol{)}\<%
\\
\>[30][@{}l@{\AgdaIndent{0}}]%
\>[31]\AgdaSymbol{→}\AgdaSpace{}%
\AgdaFunction{True}\AgdaSpace{}%
\AgdaSymbol{(}\AgdaFunction{compareSigsMultiSig}\AgdaSpace{}%
\AgdaBound{msg₂}\AgdaSpace{}%
\AgdaSymbol{(}\AgdaSpace{}%
\AgdaBound{sig1}\AgdaSpace{}%
\AgdaOperator{\AgdaInductiveConstructor{∷}}\AgdaSpace{}%
\AgdaBound{sig2}\AgdaSpace{}%
\AgdaOperator{\AgdaInductiveConstructor{∷}}\AgdaSpace{}%
\AgdaInductiveConstructor{[]}\AgdaSymbol{)}\AgdaSpace{}%
\AgdaSymbol{(}\AgdaBound{pbk1}\AgdaSpace{}%
\AgdaOperator{\AgdaInductiveConstructor{∷}}\AgdaSpace{}%
\AgdaBound{pbk2}\AgdaSpace{}%
\AgdaOperator{\AgdaInductiveConstructor{∷}}\AgdaSpace{}%
\AgdaBound{pbk3}\AgdaSpace{}%
\AgdaOperator{\AgdaInductiveConstructor{∷}}\AgdaSpace{}%
\AgdaInductiveConstructor{[]}\AgdaSymbol{))}\<%
\\
\>[0]\AgdaFunction{lemmaHoareTripleStackGeAux'3}\AgdaSpace{}%
\AgdaBound{msg₂}\AgdaSpace{}%
\AgdaBound{pbk1}\AgdaSpace{}%
\AgdaBound{pbk2}\AgdaSpace{}%
\AgdaBound{pbk3}\AgdaSpace{}%
\AgdaBound{sig1}\AgdaSpace{}%
\AgdaBound{sig2}\AgdaSpace{}%
\AgdaBound{x}\AgdaSpace{}%
\AgdaBound{x₁}%
\>[65]\AgdaKeyword{with}\AgdaSpace{}%
\AgdaSymbol{(}\AgdaFunction{isSigned}%
\>[81]\AgdaBound{msg₂}\AgdaSpace{}%
\AgdaBound{sig1}\AgdaSpace{}%
\AgdaBound{pbk1}\AgdaSpace{}%
\AgdaSymbol{)}\<%
\\
\>[0]\AgdaFunction{lemmaHoareTripleStackGeAux'3}\AgdaSpace{}%
\AgdaBound{msg₂}\AgdaSpace{}%
\AgdaBound{pbk1}\AgdaSpace{}%
\AgdaBound{pbk2}\AgdaSpace{}%
\AgdaBound{pbk3}\AgdaSpace{}%
\AgdaBound{sig1}\AgdaSpace{}%
\AgdaBound{sig2}\AgdaSpace{}%
\AgdaBound{x}\AgdaSpace{}%
\AgdaBound{x₁}\AgdaSpace{}%
\AgdaSymbol{|}\AgdaSpace{}%
\AgdaInductiveConstructor{false}\AgdaSpace{}%
\AgdaKeyword{with}\AgdaSpace{}%
\AgdaSymbol{(}\AgdaFunction{isSigned}%
\>[88]\AgdaBound{msg₂}\AgdaSpace{}%
\AgdaBound{sig1}\AgdaSpace{}%
\AgdaBound{pbk2}\AgdaSymbol{)}\<%
\\
\>[0]\AgdaFunction{lemmaHoareTripleStackGeAux'3}\AgdaSpace{}%
\AgdaBound{msg₂}\AgdaSpace{}%
\AgdaBound{pbk1}\AgdaSpace{}%
\AgdaBound{pbk2}\AgdaSpace{}%
\AgdaBound{pbk3}\AgdaSpace{}%
\AgdaBound{sig1}\AgdaSpace{}%
\AgdaBound{sig2}\AgdaSpace{}%
\AgdaBound{x}\AgdaSpace{}%
\AgdaBound{x₁}\AgdaSpace{}%
\AgdaSymbol{|}\AgdaSpace{}%
\AgdaInductiveConstructor{false}\AgdaSpace{}%
\AgdaSymbol{|}\AgdaSpace{}%
\AgdaInductiveConstructor{true}\AgdaSpace{}%
\AgdaKeyword{with}\AgdaSpace{}%
\AgdaFunction{isSigned}%
\>[94]\AgdaBound{msg₂}\AgdaSpace{}%
\AgdaBound{sig2}\AgdaSpace{}%
\AgdaBound{pbk3}\<%
\\
\>[0]\AgdaFunction{lemmaHoareTripleStackGeAux'3}\AgdaSpace{}%
\AgdaBound{msg₂}\AgdaSpace{}%
\AgdaBound{pbk1}\AgdaSpace{}%
\AgdaBound{pbk2}\AgdaSpace{}%
\AgdaBound{pbk3}\AgdaSpace{}%
\AgdaBound{sig1}\AgdaSpace{}%
\AgdaBound{sig2}\AgdaSpace{}%
\AgdaBound{x}\AgdaSpace{}%
\AgdaBound{x₁}\AgdaSpace{}%
\AgdaSymbol{|}\AgdaSpace{}%
\AgdaInductiveConstructor{false}\AgdaSpace{}%
\AgdaSymbol{|}\AgdaSpace{}%
\AgdaInductiveConstructor{true}\AgdaSpace{}%
\AgdaSymbol{|}\AgdaSpace{}%
\AgdaInductiveConstructor{true}\AgdaSpace{}%
\AgdaSymbol{=}\AgdaSpace{}%
\AgdaInductiveConstructor{tt}\<%
\\
\>[0]\AgdaFunction{lemmaHoareTripleStackGeAux'3}\AgdaSpace{}%
\AgdaBound{msg₂}\AgdaSpace{}%
\AgdaBound{pbk1}\AgdaSpace{}%
\AgdaBound{pbk2}\AgdaSpace{}%
\AgdaBound{pbk3}\AgdaSpace{}%
\AgdaBound{sig1}\AgdaSpace{}%
\AgdaBound{sig2}\AgdaSpace{}%
\AgdaBound{x}\AgdaSpace{}%
\AgdaBound{x₁}\AgdaSpace{}%
\AgdaSymbol{|}\AgdaSpace{}%
\AgdaInductiveConstructor{true}\AgdaSpace{}%
\AgdaKeyword{with}\AgdaSpace{}%
\AgdaSymbol{(}\AgdaFunction{isSigned}%
\>[87]\AgdaBound{msg₂}\AgdaSpace{}%
\AgdaBound{sig2}\AgdaSpace{}%
\AgdaBound{pbk2}\AgdaSymbol{)}\<%
\\
\>[0]\AgdaFunction{lemmaHoareTripleStackGeAux'3}\AgdaSpace{}%
\AgdaBound{msg₂}\AgdaSpace{}%
\AgdaBound{pbk1}\AgdaSpace{}%
\AgdaBound{pbk2}\AgdaSpace{}%
\AgdaBound{pbk3}\AgdaSpace{}%
\AgdaBound{sig1}\AgdaSpace{}%
\AgdaBound{sig2}\AgdaSpace{}%
\AgdaBound{x}\AgdaSpace{}%
\AgdaBound{x₁}\AgdaSpace{}%
\AgdaSymbol{|}\AgdaSpace{}%
\AgdaInductiveConstructor{true}\AgdaSpace{}%
\AgdaSymbol{|}\AgdaSpace{}%
\AgdaInductiveConstructor{false}\AgdaSpace{}%
\AgdaKeyword{with}\AgdaSpace{}%
\AgdaSymbol{(}\AgdaFunction{isSigned}%
\>[95]\AgdaBound{msg₂}\AgdaSpace{}%
\AgdaBound{sig2}\AgdaSpace{}%
\AgdaBound{pbk3}\AgdaSymbol{)}\<%
\\
\>[0]\AgdaFunction{lemmaHoareTripleStackGeAux'3}\AgdaSpace{}%
\AgdaBound{msg₂}\AgdaSpace{}%
\AgdaBound{pbk1}\AgdaSpace{}%
\AgdaBound{pbk2}\AgdaSpace{}%
\AgdaBound{pbk3}\AgdaSpace{}%
\AgdaBound{sig1}\AgdaSpace{}%
\AgdaBound{sig2}\AgdaSpace{}%
\AgdaBound{x}\AgdaSpace{}%
\AgdaBound{x₁}\AgdaSpace{}%
\AgdaSymbol{|}\AgdaSpace{}%
\AgdaInductiveConstructor{true}\AgdaSpace{}%
\AgdaSymbol{|}\AgdaSpace{}%
\AgdaInductiveConstructor{false}\AgdaSpace{}%
\AgdaSymbol{|}\AgdaSpace{}%
\AgdaInductiveConstructor{true}\AgdaSpace{}%
\AgdaSymbol{=}\AgdaSpace{}%
\AgdaInductiveConstructor{tt}\<%
\\
\>[0]\AgdaFunction{lemmaHoareTripleStackGeAux'3}\AgdaSpace{}%
\AgdaBound{msg₂}\AgdaSpace{}%
\AgdaBound{pbk1}\AgdaSpace{}%
\AgdaBound{pbk2}\AgdaSpace{}%
\AgdaBound{pbk3}\AgdaSpace{}%
\AgdaBound{sig1}\AgdaSpace{}%
\AgdaBound{sig2}\AgdaSpace{}%
\AgdaBound{x}\AgdaSpace{}%
\AgdaBound{x₁}\AgdaSpace{}%
\AgdaSymbol{|}\AgdaSpace{}%
\AgdaInductiveConstructor{true}\AgdaSpace{}%
\AgdaSymbol{|}\AgdaSpace{}%
\AgdaInductiveConstructor{true}\AgdaSpace{}%
\AgdaSymbol{=}\AgdaSpace{}%
\AgdaInductiveConstructor{tt}\<%
\\
\\[\AgdaEmptyExtraSkip]%
\>[0]\AgdaComment{\{-}\<%
\\
\>[0]\AgdaComment{For\ the\ other\ direction\ we\ get\ the\ goal}\<%
\\
\>[0]\<%
\\
\>[0]\<%
\\
\>[0]\AgdaComment{Goal:\ (True\ (isSigned\ \ msg₂\ sig1\ pbk1\ )\ ∧}\<%
\\
\>[0]\AgdaComment{\ \ \ \ \ \ \ True\ (isSigned\ \ msg₂\ sig2\ pbk2))}\<%
\\
\>[0]\AgdaComment{\ \ \ \ \ \ ⊎}\<%
\\
\>[0]\AgdaComment{\ \ \ \ \ \ (True\ (isSigned\ \ msg₂\ sig1\ pbk1\ )\ ∧}\<%
\\
\>[0]\AgdaComment{\ \ \ \ \ \ \ True\ (isSigned\ \ msg₂\ sig2\ pbk3))}\<%
\\
\>[0]\AgdaComment{\ \ \ \ \ \ ⊎}\<%
\\
\>[0]\AgdaComment{\ \ \ \ \ \ (True\ (isSigned\ \ msg₂\ sig1\ pbk2)\ ∧}\<%
\\
\>[0]\AgdaComment{\ \ \ \ \ \ \ True\ (isSigned\ \ msg₂\ sig2\ pbk3))}\<%
\\
\>[0]\AgdaComment{————————————————————————————————————————————————————————————}\<%
\\
\>[0]\AgdaComment{x\ \ \ \ \ :\ True}\<%
\\
\>[0]\AgdaComment{\ \ \ \ \ \ \ \ (notFalse}\<%
\\
\>[0]\AgdaComment{\ \ \ \ \ \ \ \ \ (boolToNat}\<%
\\
\>[0]\AgdaComment{\ \ \ \ \ \ \ \ \ \ (compareSigsMultiSigAux\ msg₂\ (sig2\ ∷\ [])\ (pbk2\ ∷\ pbk3\ ∷\ [])\ sig1}\<%
\\
\>[0]\AgdaComment{\ \ \ \ \ \ \ \ \ \ \ (isSigned\ \ msg₂\ sig1\ pbk1\ ))))}\<%
\\
\>[0]\<%
\\
\>[0]\<%
\\
\>[0]\<%
\\
\>[0]\AgdaComment{Again\ we\ use\ the\ boolToNatNotFalseLemma2\ lemma\ to\ get\ rid\ of\ the}\<%
\\
\>[0]\AgdaComment{True(notFalse\ (boolToNat}\<%
\\
\>[0]\<%
\\
\>[0]\AgdaComment{and\ show}\<%
\\
\>[0]\AgdaComment{lemmaHoareTripleStackGeAux'4}\<%
\\
\>[0]\<%
\\
\>[0]\AgdaComment{we\ see\ that\ now\ the\ x\ ends\ with}\<%
\\
\>[0]\AgdaComment{(isSigned\ \ msg₂\ sig1\ pbk1\ )}\<%
\\
\>[0]\AgdaComment{so\ we\ start\ with\ the\ with\ pattern\ with\ a\ case\ distinction\ on\ this}\<%
\\
\>[0]\AgdaComment{and\ continue\ making\ casedistintion\ on\ the\ last\ formula\ in\ x}\<%
\\
\>[0]\<%
\\
\>[0]\<%
\\
\>[0]\<%
\\
\>[0]\AgdaComment{-\}}\<%
\\
\\[\AgdaEmptyExtraSkip]%
\>[0]\AgdaFunction{lemmaHoareTripleStackGeAux'4}\AgdaSpace{}%
\AgdaSymbol{:}\AgdaSpace{}%
\AgdaSymbol{(}\AgdaBound{msg₂}\AgdaSpace{}%
\AgdaSymbol{:}\AgdaSpace{}%
\AgdaDatatype{Msg}\AgdaSymbol{)}\<%
\\
\>[0][@{}l@{\AgdaIndent{0}}]%
\>[9]\AgdaSymbol{(}\AgdaBound{pbk1}\AgdaSpace{}%
\AgdaBound{pbk2}\AgdaSpace{}%
\AgdaBound{pbk3}\AgdaSpace{}%
\AgdaBound{sig1}\AgdaSpace{}%
\AgdaBound{sig2}\AgdaSpace{}%
\AgdaSymbol{:}\AgdaSpace{}%
\AgdaDatatype{ℕ}\AgdaSymbol{)}\<%
\\
\>[9][@{}l@{\AgdaIndent{0}}]%
\>[10]\AgdaSymbol{→}\AgdaSpace{}%
\AgdaFunction{True}%
\>[560I]\AgdaSymbol{(}\AgdaFunction{compareSigsMultiSigAux}\AgdaSpace{}%
\AgdaBound{msg₂}\AgdaSpace{}%
\AgdaSymbol{(}\AgdaBound{sig2}\AgdaSpace{}%
\AgdaOperator{\AgdaInductiveConstructor{∷}}\AgdaSpace{}%
\AgdaInductiveConstructor{[]}\AgdaSymbol{)}\AgdaSpace{}%
\AgdaSymbol{(}\AgdaBound{pbk2}\AgdaSpace{}%
\AgdaOperator{\AgdaInductiveConstructor{∷}}\AgdaSpace{}%
\AgdaBound{pbk3}\AgdaSpace{}%
\AgdaOperator{\AgdaInductiveConstructor{∷}}\AgdaSpace{}%
\AgdaInductiveConstructor{[]}\AgdaSymbol{)}%
\>[78]\AgdaBound{sig1}\<%
\\
\>[.][@{}l@{}]\<[560I]%
\>[17]\AgdaSymbol{(}\AgdaFunction{isSigned}%
\>[28]\AgdaBound{msg₂}\AgdaSpace{}%
\AgdaBound{sig1}\AgdaSpace{}%
\AgdaBound{pbk1}\AgdaSpace{}%
\AgdaSymbol{))}\<%
\\
\>[10]\AgdaSymbol{→}%
\>[16]\AgdaSymbol{(}\AgdaFunction{True}\AgdaSpace{}%
\AgdaSymbol{(}\AgdaFunction{isSigned}%
\>[33]\AgdaBound{msg₂}\AgdaSpace{}%
\AgdaBound{sig1}\AgdaSpace{}%
\AgdaBound{pbk1}\AgdaSpace{}%
\AgdaSymbol{)}\AgdaSpace{}%
\AgdaOperator{\AgdaRecord{∧}}\AgdaSpace{}%
\AgdaFunction{True}\AgdaSpace{}%
\AgdaSymbol{(}\AgdaFunction{isSigned}%
\>[68]\AgdaBound{msg₂}\AgdaSpace{}%
\AgdaBound{sig2}\AgdaSpace{}%
\AgdaBound{pbk2}\AgdaSymbol{))}\<%
\\
\>[10][@{}l@{\AgdaIndent{0}}]%
\>[13]\AgdaOperator{\AgdaDatatype{⊎}}%
\>[16]\AgdaSymbol{(}\AgdaFunction{True}\AgdaSpace{}%
\AgdaSymbol{(}\AgdaFunction{isSigned}%
\>[33]\AgdaBound{msg₂}\AgdaSpace{}%
\AgdaBound{sig1}\AgdaSpace{}%
\AgdaBound{pbk1}\AgdaSpace{}%
\AgdaSymbol{)}\AgdaSpace{}%
\AgdaOperator{\AgdaRecord{∧}}%
\>[53]\AgdaFunction{True}\AgdaSpace{}%
\AgdaSymbol{(}\AgdaFunction{isSigned}%
\>[69]\AgdaBound{msg₂}\AgdaSpace{}%
\AgdaBound{sig2}\AgdaSpace{}%
\AgdaBound{pbk3}\AgdaSymbol{))}\<%
\\
\>[13]\AgdaOperator{\AgdaDatatype{⊎}}%
\>[16]\AgdaSymbol{(}\AgdaFunction{True}\AgdaSpace{}%
\AgdaSymbol{(}\AgdaFunction{isSigned}%
\>[33]\AgdaBound{msg₂}\AgdaSpace{}%
\AgdaBound{sig1}\AgdaSpace{}%
\AgdaBound{pbk2}\AgdaSymbol{)}\AgdaSpace{}%
\AgdaOperator{\AgdaRecord{∧}}\AgdaSpace{}%
\AgdaFunction{True}\AgdaSpace{}%
\AgdaSymbol{(}\AgdaFunction{isSigned}%
\>[67]\AgdaBound{msg₂}\AgdaSpace{}%
\AgdaBound{sig2}\AgdaSpace{}%
\AgdaBound{pbk3}\AgdaSymbol{))}\<%
\\
\>[0]\AgdaFunction{lemmaHoareTripleStackGeAux'4}\AgdaSpace{}%
\AgdaBound{msg₂}\AgdaSpace{}%
\AgdaBound{pbk1}\AgdaSpace{}%
\AgdaBound{pbk2}\AgdaSpace{}%
\AgdaBound{pbk3}\AgdaSpace{}%
\AgdaBound{sig1}\AgdaSpace{}%
\AgdaBound{sig2}\AgdaSpace{}%
\AgdaSymbol{\AgdaUnderscore{}}\AgdaSpace{}%
\AgdaKeyword{with}\AgdaSpace{}%
\AgdaSymbol{(}\AgdaFunction{isSigned}%
\>[77]\AgdaBound{msg₂}\AgdaSpace{}%
\AgdaBound{sig1}\AgdaSpace{}%
\AgdaBound{pbk1}\AgdaSpace{}%
\AgdaSymbol{)}\<%
\\
\>[0]\AgdaFunction{lemmaHoareTripleStackGeAux'4}\AgdaSpace{}%
\AgdaBound{msg₂}\AgdaSpace{}%
\AgdaBound{pbk1}\AgdaSpace{}%
\AgdaBound{pbk2}\AgdaSpace{}%
\AgdaBound{pbk3}\AgdaSpace{}%
\AgdaBound{sig1}\AgdaSpace{}%
\AgdaBound{sig2}\AgdaSpace{}%
\AgdaSymbol{\AgdaUnderscore{}}\AgdaSpace{}%
\AgdaSymbol{|}\AgdaSpace{}%
\AgdaInductiveConstructor{false}\AgdaSpace{}%
\AgdaKeyword{with}\AgdaSpace{}%
\AgdaSymbol{(}\AgdaFunction{isSigned}%
\>[85]\AgdaBound{msg₂}\AgdaSpace{}%
\AgdaBound{sig1}\AgdaSpace{}%
\AgdaBound{pbk2}\AgdaSymbol{)}\<%
\\
\>[0]\AgdaFunction{lemmaHoareTripleStackGeAux'4}\AgdaSpace{}%
\AgdaBound{msg₂}\AgdaSpace{}%
\AgdaBound{pbk1}\AgdaSpace{}%
\AgdaBound{pbk2}\AgdaSpace{}%
\AgdaBound{pbk3}\AgdaSpace{}%
\AgdaBound{sig1}\AgdaSpace{}%
\AgdaBound{sig2}\AgdaSpace{}%
\AgdaSymbol{\AgdaUnderscore{}}\AgdaSpace{}%
\AgdaSymbol{|}\AgdaSpace{}%
\AgdaInductiveConstructor{false}\AgdaSpace{}%
\AgdaSymbol{|}\AgdaSpace{}%
\AgdaInductiveConstructor{false}\AgdaSpace{}%
\AgdaKeyword{with}\AgdaSpace{}%
\AgdaSymbol{(}\AgdaFunction{isSigned}%
\>[93]\AgdaBound{msg₂}\AgdaSpace{}%
\AgdaBound{sig1}\AgdaSpace{}%
\AgdaBound{pbk3}\AgdaSymbol{)}\<%
\\
\>[0]\AgdaFunction{lemmaHoareTripleStackGeAux'4}\AgdaSpace{}%
\AgdaBound{msg₂}\AgdaSpace{}%
\AgdaBound{pbk1}\AgdaSpace{}%
\AgdaBound{pbk2}\AgdaSpace{}%
\AgdaBound{pbk3}\AgdaSpace{}%
\AgdaBound{sig1}\AgdaSpace{}%
\AgdaBound{sig2}\AgdaSpace{}%
\AgdaSymbol{()}\AgdaSpace{}%
\AgdaSymbol{|}\AgdaSpace{}%
\AgdaInductiveConstructor{false}\AgdaSpace{}%
\AgdaSymbol{|}\AgdaSpace{}%
\AgdaInductiveConstructor{false}\AgdaSpace{}%
\AgdaSymbol{|}\AgdaSpace{}%
\AgdaInductiveConstructor{false}\<%
\\
\>[0]\AgdaFunction{lemmaHoareTripleStackGeAux'4}\AgdaSpace{}%
\AgdaBound{msg₂}\AgdaSpace{}%
\AgdaBound{pbk1}\AgdaSpace{}%
\AgdaBound{pbk2}\AgdaSpace{}%
\AgdaBound{pbk3}\AgdaSpace{}%
\AgdaBound{sig1}\AgdaSpace{}%
\AgdaBound{sig2}\AgdaSpace{}%
\AgdaSymbol{()}\AgdaSpace{}%
\AgdaSymbol{|}\AgdaSpace{}%
\AgdaInductiveConstructor{false}\AgdaSpace{}%
\AgdaSymbol{|}\AgdaSpace{}%
\AgdaInductiveConstructor{false}\AgdaSpace{}%
\AgdaSymbol{|}\AgdaSpace{}%
\AgdaInductiveConstructor{true}\<%
\\
\>[0]\AgdaFunction{lemmaHoareTripleStackGeAux'4}\AgdaSpace{}%
\AgdaBound{msg₂}\AgdaSpace{}%
\AgdaBound{pbk1}\AgdaSpace{}%
\AgdaBound{pbk2}\AgdaSpace{}%
\AgdaBound{pbk3}\AgdaSpace{}%
\AgdaBound{sig1}\AgdaSpace{}%
\AgdaBound{sig2}\AgdaSpace{}%
\AgdaSymbol{\AgdaUnderscore{}}\AgdaSpace{}%
\AgdaSymbol{|}\AgdaSpace{}%
\AgdaInductiveConstructor{false}\AgdaSpace{}%
\AgdaSymbol{|}\AgdaSpace{}%
\AgdaInductiveConstructor{true}\AgdaSpace{}%
\AgdaKeyword{with}\AgdaSpace{}%
\AgdaSymbol{(}\AgdaFunction{isSigned}%
\>[92]\AgdaBound{msg₂}\AgdaSpace{}%
\AgdaBound{sig2}\AgdaSpace{}%
\AgdaBound{pbk3}\AgdaSymbol{)}\<%
\\
\>[0]\AgdaFunction{lemmaHoareTripleStackGeAux'4}\AgdaSpace{}%
\AgdaBound{msg₂}\AgdaSpace{}%
\AgdaBound{pbk1}\AgdaSpace{}%
\AgdaBound{pbk2}\AgdaSpace{}%
\AgdaBound{pbk3}\AgdaSpace{}%
\AgdaBound{sig1}\AgdaSpace{}%
\AgdaBound{sig2}\AgdaSpace{}%
\AgdaSymbol{\AgdaUnderscore{}}\AgdaSpace{}%
\AgdaSymbol{|}\AgdaSpace{}%
\AgdaInductiveConstructor{false}\AgdaSpace{}%
\AgdaSymbol{|}\AgdaSpace{}%
\AgdaInductiveConstructor{true}\AgdaSpace{}%
\AgdaSymbol{|}\AgdaSpace{}%
\AgdaInductiveConstructor{true}\AgdaSpace{}%
\AgdaSymbol{=}\AgdaSpace{}%
\AgdaInductiveConstructor{inj₂}\AgdaSpace{}%
\AgdaSymbol{(}\AgdaInductiveConstructor{inj₂}\AgdaSpace{}%
\AgdaSymbol{(}\AgdaInductiveConstructor{conj}\AgdaSpace{}%
\AgdaInductiveConstructor{tt}\AgdaSpace{}%
\AgdaInductiveConstructor{tt}\AgdaSymbol{))}\<%
\\
\>[0]\AgdaFunction{lemmaHoareTripleStackGeAux'4}\AgdaSpace{}%
\AgdaBound{msg₂}\AgdaSpace{}%
\AgdaBound{pbk1}\AgdaSpace{}%
\AgdaBound{pbk2}\AgdaSpace{}%
\AgdaBound{pbk3}\AgdaSpace{}%
\AgdaBound{sig1}\AgdaSpace{}%
\AgdaBound{sig2}\AgdaSpace{}%
\AgdaSymbol{\AgdaUnderscore{}}\AgdaSpace{}%
\AgdaSymbol{|}\AgdaSpace{}%
\AgdaInductiveConstructor{true}%
\>[69]\AgdaKeyword{with}\AgdaSpace{}%
\AgdaSymbol{(}\AgdaFunction{isSigned}%
\>[85]\AgdaBound{msg₂}\AgdaSpace{}%
\AgdaBound{sig2}\AgdaSpace{}%
\AgdaBound{pbk2}\AgdaSymbol{)}\<%
\\
\>[0]\AgdaFunction{lemmaHoareTripleStackGeAux'4}\AgdaSpace{}%
\AgdaBound{msg₂}\AgdaSpace{}%
\AgdaBound{pbk1}\AgdaSpace{}%
\AgdaBound{pbk2}\AgdaSpace{}%
\AgdaBound{pbk3}\AgdaSpace{}%
\AgdaBound{sig1}\AgdaSpace{}%
\AgdaBound{sig2}\AgdaSpace{}%
\AgdaSymbol{\AgdaUnderscore{}}\AgdaSpace{}%
\AgdaSymbol{|}\AgdaSpace{}%
\AgdaInductiveConstructor{true}\AgdaSpace{}%
\AgdaSymbol{|}\AgdaSpace{}%
\AgdaInductiveConstructor{false}\AgdaSpace{}%
\AgdaKeyword{with}\AgdaSpace{}%
\AgdaSymbol{(}\AgdaFunction{isSigned}%
\>[92]\AgdaBound{msg₂}\AgdaSpace{}%
\AgdaBound{sig2}\AgdaSpace{}%
\AgdaBound{pbk3}\AgdaSymbol{)}\<%
\\
\>[0]\AgdaFunction{lemmaHoareTripleStackGeAux'4}\AgdaSpace{}%
\AgdaBound{msg₂}\AgdaSpace{}%
\AgdaBound{pbk1}\AgdaSpace{}%
\AgdaBound{pbk2}\AgdaSpace{}%
\AgdaBound{pbk3}\AgdaSpace{}%
\AgdaBound{sig1}\AgdaSpace{}%
\AgdaBound{sig2}\AgdaSpace{}%
\AgdaSymbol{\AgdaUnderscore{}}\AgdaSpace{}%
\AgdaSymbol{|}\AgdaSpace{}%
\AgdaInductiveConstructor{true}\AgdaSpace{}%
\AgdaSymbol{|}\AgdaSpace{}%
\AgdaInductiveConstructor{false}\AgdaSpace{}%
\AgdaSymbol{|}\AgdaSpace{}%
\AgdaInductiveConstructor{true}\AgdaSpace{}%
\AgdaSymbol{=}\AgdaSpace{}%
\AgdaInductiveConstructor{inj₂}\AgdaSpace{}%
\AgdaSymbol{(}\AgdaInductiveConstructor{inj₁}\AgdaSpace{}%
\AgdaSymbol{(}\AgdaInductiveConstructor{conj}\AgdaSpace{}%
\AgdaInductiveConstructor{tt}\AgdaSpace{}%
\AgdaInductiveConstructor{tt}\AgdaSymbol{))}\<%
\\
\>[0]\AgdaFunction{lemmaHoareTripleStackGeAux'4}\AgdaSpace{}%
\AgdaBound{msg₂}\AgdaSpace{}%
\AgdaBound{pbk1}\AgdaSpace{}%
\AgdaBound{pbk2}\AgdaSpace{}%
\AgdaBound{pbk3}\AgdaSpace{}%
\AgdaBound{sig1}\AgdaSpace{}%
\AgdaBound{sig2}\AgdaSpace{}%
\AgdaSymbol{\AgdaUnderscore{}}\AgdaSpace{}%
\AgdaSymbol{|}\AgdaSpace{}%
\AgdaInductiveConstructor{true}\AgdaSpace{}%
\AgdaSymbol{|}\AgdaSpace{}%
\AgdaInductiveConstructor{true}\AgdaSpace{}%
\AgdaSymbol{=}\AgdaSpace{}%
\AgdaInductiveConstructor{inj₁}\AgdaSpace{}%
\AgdaSymbol{(}\AgdaInductiveConstructor{conj}\AgdaSpace{}%
\AgdaInductiveConstructor{tt}\AgdaSpace{}%
\AgdaInductiveConstructor{tt}\AgdaSymbol{)}\<%
\\
\\[\AgdaEmptyExtraSkip]%
\\[\AgdaEmptyExtraSkip]%
\\[\AgdaEmptyExtraSkip]%
\\[\AgdaEmptyExtraSkip]%
\>[0]\AgdaComment{--\textbackslash{}multisigweakestPreCondTwoFour}\<%
\end{code}
} 

\newcommand{\multisigweakestPreCondTwoFour}{
\begin{code}%
\>[0]\AgdaFunction{weakestPreCondMultiSig-2-4ˢ}\AgdaSpace{}%
\AgdaSymbol{:}\AgdaSpace{}%
\AgdaSymbol{(}\AgdaBound{pbk1}\AgdaSpace{}%
\AgdaBound{pbk2}\AgdaSpace{}%
\AgdaBound{pbk3}\AgdaSpace{}%
\AgdaBound{pbk4}\AgdaSpace{}%
\AgdaSymbol{:}%
\>[54]\AgdaDatatype{ℕ}\AgdaSymbol{)}\AgdaSpace{}%
\AgdaSymbol{→}\AgdaSpace{}%
\AgdaFunction{StackPredicate}\<%
\\
\>[0]\AgdaFunction{weakestPreCondMultiSig-2-4ˢ}\AgdaSpace{}%
\AgdaBound{pbk1}\AgdaSpace{}%
\AgdaBound{pbk2}\AgdaSpace{}%
\AgdaBound{pbk3}\AgdaSpace{}%
\AgdaBound{pbk4}\AgdaSpace{}%
\AgdaBound{time}%
\>[54]\AgdaBound{msg₁}\AgdaSpace{}%
\AgdaInductiveConstructor{[]}\AgdaSpace{}%
\AgdaSymbol{=}\AgdaSpace{}%
\AgdaDatatype{⊥}\<%
\\
\>[0]\AgdaFunction{weakestPreCondMultiSig-2-4ˢ}\AgdaSpace{}%
\AgdaBound{pbk1}\AgdaSpace{}%
\AgdaBound{pbk2}\AgdaSpace{}%
\AgdaBound{pbk3}\AgdaSpace{}%
\AgdaBound{pbk4}\AgdaSpace{}%
\AgdaBound{time}%
\>[54]\AgdaBound{msg₁}\AgdaSpace{}%
\AgdaSymbol{(}\AgdaBound{x}\AgdaSpace{}%
\AgdaOperator{\AgdaInductiveConstructor{∷}}\AgdaSpace{}%
\AgdaInductiveConstructor{[]}\AgdaSymbol{)}\AgdaSpace{}%
\AgdaSymbol{=}\AgdaSpace{}%
\AgdaDatatype{⊥}\<%
\\
\>[0]\AgdaFunction{weakestPreCondMultiSig-2-4ˢ}\AgdaSpace{}%
\AgdaBound{pbk1}\AgdaSpace{}%
\AgdaBound{pbk2}\AgdaSpace{}%
\AgdaBound{pbk3}\AgdaSpace{}%
\AgdaBound{pbk4}\AgdaSpace{}%
\AgdaBound{time}%
\>[54]\AgdaBound{msg₁}\AgdaSpace{}%
\AgdaSymbol{(}\AgdaBound{x}\AgdaSpace{}%
\AgdaOperator{\AgdaInductiveConstructor{∷}}\AgdaSpace{}%
\AgdaBound{y}\AgdaSpace{}%
\AgdaOperator{\AgdaInductiveConstructor{∷}}\AgdaSpace{}%
\AgdaInductiveConstructor{[]}\AgdaSymbol{)}\AgdaSpace{}%
\AgdaSymbol{=}\AgdaSpace{}%
\AgdaDatatype{⊥}\<%
\\
\>[0]\AgdaFunction{weakestPreCondMultiSig-2-4ˢ}\AgdaSpace{}%
\AgdaBound{pbk1}\AgdaSpace{}%
\AgdaBound{pbk2}\AgdaSpace{}%
\AgdaBound{pbk3}\AgdaSpace{}%
\AgdaBound{pbk4}\AgdaSpace{}%
\AgdaBound{time}%
\>[54]\AgdaBound{msg₁}\AgdaSpace{}%
\AgdaSymbol{(}\AgdaSpace{}%
\AgdaBound{sig2}\AgdaSpace{}%
\AgdaOperator{\AgdaInductiveConstructor{∷}}\AgdaSpace{}%
\AgdaBound{sig1}\AgdaSpace{}%
\AgdaOperator{\AgdaInductiveConstructor{∷}}\AgdaSpace{}%
\AgdaBound{dummy}\AgdaSpace{}%
\AgdaOperator{\AgdaInductiveConstructor{∷}}\AgdaSpace{}%
\AgdaBound{stack₁}\AgdaSymbol{)}\AgdaSpace{}%
\AgdaSymbol{=}\<%
\\
\>[0][@{}l@{\AgdaIndent{0}}]%
\>[7]\AgdaSymbol{(}%
\>[10]\AgdaSymbol{(}\AgdaFunction{IsSigned}\AgdaSpace{}%
\AgdaBound{msg₁}\AgdaSpace{}%
\AgdaBound{sig1}\AgdaSpace{}%
\AgdaBound{pbk1}\AgdaSpace{}%
\AgdaOperator{\AgdaRecord{∧}}%
\>[38]\AgdaFunction{IsSigned}\AgdaSpace{}%
\AgdaBound{msg₁}\AgdaSpace{}%
\AgdaBound{sig2}\AgdaSpace{}%
\AgdaBound{pbk2}\AgdaSymbol{)}\AgdaSpace{}%
\AgdaOperator{\AgdaDatatype{⊎}}\<%
\\
\>[10]\AgdaSymbol{(}\AgdaFunction{IsSigned}\AgdaSpace{}%
\AgdaBound{msg₁}\AgdaSpace{}%
\AgdaBound{sig1}\AgdaSpace{}%
\AgdaBound{pbk1}\AgdaSpace{}%
\AgdaOperator{\AgdaRecord{∧}}%
\>[38]\AgdaFunction{IsSigned}\AgdaSpace{}%
\AgdaBound{msg₁}\AgdaSpace{}%
\AgdaBound{sig2}\AgdaSpace{}%
\AgdaBound{pbk3}\AgdaSymbol{)}\AgdaSpace{}%
\AgdaOperator{\AgdaDatatype{⊎}}\<%
\\
\>[10]\AgdaSymbol{(}\AgdaFunction{IsSigned}\AgdaSpace{}%
\AgdaBound{msg₁}\AgdaSpace{}%
\AgdaBound{sig1}\AgdaSpace{}%
\AgdaBound{pbk1}\AgdaSpace{}%
\AgdaOperator{\AgdaRecord{∧}}%
\>[38]\AgdaFunction{IsSigned}\AgdaSpace{}%
\AgdaBound{msg₁}\AgdaSpace{}%
\AgdaBound{sig2}\AgdaSpace{}%
\AgdaBound{pbk4}\AgdaSymbol{)}\AgdaSpace{}%
\AgdaOperator{\AgdaDatatype{⊎}}\<%
\\
\>[10]\AgdaSymbol{(}\AgdaFunction{IsSigned}\AgdaSpace{}%
\AgdaBound{msg₁}\AgdaSpace{}%
\AgdaBound{sig1}\AgdaSpace{}%
\AgdaBound{pbk2}\AgdaSpace{}%
\AgdaOperator{\AgdaRecord{∧}}%
\>[38]\AgdaFunction{IsSigned}\AgdaSpace{}%
\AgdaBound{msg₁}\AgdaSpace{}%
\AgdaBound{sig2}\AgdaSpace{}%
\AgdaBound{pbk3}\AgdaSymbol{)}\AgdaSpace{}%
\AgdaOperator{\AgdaDatatype{⊎}}\<%
\\
\>[10]\AgdaSymbol{(}\AgdaFunction{IsSigned}\AgdaSpace{}%
\AgdaBound{msg₁}\AgdaSpace{}%
\AgdaBound{sig1}\AgdaSpace{}%
\AgdaBound{pbk2}\AgdaSpace{}%
\AgdaOperator{\AgdaRecord{∧}}%
\>[38]\AgdaFunction{IsSigned}\AgdaSpace{}%
\AgdaBound{msg₁}\AgdaSpace{}%
\AgdaBound{sig2}\AgdaSpace{}%
\AgdaBound{pbk4}\AgdaSymbol{)}\AgdaSpace{}%
\AgdaOperator{\AgdaDatatype{⊎}}\<%
\\
\>[10]\AgdaSymbol{(}\AgdaFunction{IsSigned}\AgdaSpace{}%
\AgdaBound{msg₁}\AgdaSpace{}%
\AgdaBound{sig1}\AgdaSpace{}%
\AgdaBound{pbk3}\AgdaSpace{}%
\AgdaOperator{\AgdaRecord{∧}}%
\>[38]\AgdaFunction{IsSigned}\AgdaSpace{}%
\AgdaBound{msg₁}\AgdaSpace{}%
\AgdaBound{sig2}\AgdaSpace{}%
\AgdaBound{pbk4}\AgdaSymbol{))}\<%
\end{code}
} 

\AgdaHide{
\begin{code}%
\>[0]\<%
\\
\\[\AgdaEmptyExtraSkip]%
\\[\AgdaEmptyExtraSkip]%
\\[\AgdaEmptyExtraSkip]%
\>[0]\AgdaFunction{HoareTripleStackGeAux'}\AgdaSpace{}%
\AgdaSymbol{:}\AgdaSpace{}%
\AgdaSymbol{(}\AgdaBound{msg₁}\AgdaSpace{}%
\AgdaSymbol{:}\AgdaSpace{}%
\AgdaDatatype{Msg}\AgdaSymbol{)(}\AgdaBound{pbk1}\AgdaSpace{}%
\AgdaBound{pbk2}\AgdaSpace{}%
\AgdaBound{pbk3}\AgdaSpace{}%
\AgdaSymbol{:}\AgdaSpace{}%
\AgdaDatatype{ℕ}\AgdaSymbol{)}\AgdaSpace{}%
\AgdaSymbol{→}\<%
\\
\>[0][@{}l@{\AgdaIndent{0}}]%
\>[17]\AgdaOperator{\AgdaRecord{<}}\AgdaSpace{}%
\AgdaSymbol{(}\AgdaFunction{weakestPreCondMultiSig-2-3-bas}\AgdaSpace{}%
\AgdaBound{pbk1}\AgdaSpace{}%
\AgdaBound{pbk2}\AgdaSpace{}%
\AgdaBound{pbk3}\AgdaSymbol{)}\AgdaSpace{}%
\AgdaOperator{\AgdaRecord{>gˢ}}\<%
\\
\>[17]\AgdaSymbol{(λ}%
\>[873I]\AgdaBound{time₁}\AgdaSpace{}%
\AgdaBound{msg₁}\AgdaSpace{}%
\AgdaBound{stack}\AgdaSpace{}%
\AgdaSymbol{→}\<%
\\
\>[.][@{}l@{}]\<[873I]%
\>[20]\AgdaFunction{executeMultiSig3}\AgdaSpace{}%
\AgdaBound{msg₁}\AgdaSpace{}%
\AgdaSymbol{(}\AgdaBound{pbk1}\AgdaSpace{}%
\AgdaOperator{\AgdaInductiveConstructor{∷}}\AgdaSpace{}%
\AgdaBound{pbk2}\AgdaSpace{}%
\AgdaOperator{\AgdaInductiveConstructor{∷}}\AgdaSpace{}%
\AgdaBound{pbk3}\AgdaSpace{}%
\AgdaOperator{\AgdaInductiveConstructor{∷}}\AgdaSpace{}%
\AgdaInductiveConstructor{[]}\AgdaSymbol{)}\AgdaSpace{}%
\AgdaNumber{2}\AgdaSpace{}%
\AgdaBound{stack}\AgdaSpace{}%
\AgdaInductiveConstructor{[]}\AgdaSymbol{)}\<%
\\
\>[17]\AgdaOperator{\AgdaRecord{<}}\AgdaSpace{}%
\AgdaSymbol{(λ}\AgdaSpace{}%
\AgdaBound{time₁}\AgdaSpace{}%
\AgdaBound{msg₁}\AgdaSpace{}%
\AgdaBound{stack}\AgdaSpace{}%
\AgdaSymbol{→}\AgdaSpace{}%
\AgdaFunction{acceptStateˢ}\AgdaSpace{}%
\AgdaBound{time₁}\AgdaSpace{}%
\AgdaBound{msg₁}\AgdaSpace{}%
\AgdaBound{stack}\AgdaSymbol{)}\AgdaSpace{}%
\AgdaOperator{\AgdaRecord{>}}\<%
\\
\>[0]\AgdaFunction{HoareTripleStackGeAux'}\AgdaSpace{}%
\AgdaBound{msg₁}\AgdaSpace{}%
\AgdaBound{pbk1}\AgdaSpace{}%
\AgdaBound{pbk2}\AgdaSpace{}%
\AgdaBound{pbk3}\AgdaSpace{}%
\AgdaSymbol{.}\AgdaField{==>stg}\AgdaSpace{}%
\AgdaBound{time}\AgdaSpace{}%
\AgdaBound{msg₂}\AgdaSpace{}%
\AgdaSymbol{(}\AgdaBound{sig2}\AgdaSpace{}%
\AgdaOperator{\AgdaInductiveConstructor{∷}}\AgdaSpace{}%
\AgdaBound{sig1}\AgdaSpace{}%
\AgdaOperator{\AgdaInductiveConstructor{∷}}\AgdaSpace{}%
\AgdaBound{dummy}\AgdaSpace{}%
\AgdaOperator{\AgdaInductiveConstructor{∷}}\AgdaSpace{}%
\AgdaBound{s}\AgdaSymbol{)}\AgdaSpace{}%
\AgdaSymbol{(}\AgdaInductiveConstructor{inj₁}\AgdaSpace{}%
\AgdaSymbol{(}\AgdaInductiveConstructor{conj}\AgdaSpace{}%
\AgdaBound{and3}\AgdaSpace{}%
\AgdaBound{and4}\AgdaSymbol{))}\<%
\\
\>[0][@{}l@{\AgdaIndent{0}}]%
\>[10]\AgdaSymbol{=}%
\>[916I]\AgdaFunction{boolToNatNotFalseLemma}\AgdaSpace{}%
\AgdaSymbol{(}\AgdaFunction{compareSigsMultiSigAux}\AgdaSpace{}%
\AgdaBound{msg₂}\AgdaSpace{}%
\AgdaSymbol{(}\AgdaBound{sig2}\AgdaSpace{}%
\AgdaOperator{\AgdaInductiveConstructor{∷}}\AgdaSpace{}%
\AgdaInductiveConstructor{[]}\AgdaSymbol{)}\AgdaSpace{}%
\AgdaSymbol{(}\AgdaBound{pbk2}\AgdaSpace{}%
\AgdaOperator{\AgdaInductiveConstructor{∷}}\AgdaSpace{}%
\AgdaBound{pbk3}\AgdaSpace{}%
\AgdaOperator{\AgdaInductiveConstructor{∷}}\AgdaSpace{}%
\AgdaInductiveConstructor{[]}\AgdaSymbol{)}\AgdaSpace{}%
\AgdaBound{sig1}\<%
\\
\>[.][@{}l@{}]\<[916I]%
\>[12]\AgdaSymbol{(}\AgdaFunction{isSigned}%
\>[23]\AgdaBound{msg₂}\AgdaSpace{}%
\AgdaBound{sig1}\AgdaSpace{}%
\AgdaBound{pbk1}\AgdaSymbol{))}\AgdaSpace{}%
\AgdaSymbol{(}\AgdaFunction{lemmaHoareTripleStackGeAux'1}\AgdaSpace{}%
\AgdaBound{msg₂}\AgdaSpace{}%
\AgdaBound{pbk1}\AgdaSpace{}%
\AgdaBound{pbk2}\AgdaSpace{}%
\AgdaBound{pbk3}\AgdaSpace{}%
\AgdaBound{sig1}\AgdaSpace{}%
\AgdaBound{sig2}\AgdaSpace{}%
\AgdaBound{and3}\AgdaSpace{}%
\AgdaBound{and4}\AgdaSymbol{)}\<%
\\
\\[\AgdaEmptyExtraSkip]%
\>[0]\AgdaFunction{HoareTripleStackGeAux'}\AgdaSpace{}%
\AgdaBound{msg₁}\AgdaSpace{}%
\AgdaBound{pbk1}\AgdaSpace{}%
\AgdaBound{pbk2}\AgdaSpace{}%
\AgdaBound{pbk3}\AgdaSpace{}%
\AgdaSymbol{.}\AgdaField{==>stg}\AgdaSpace{}%
\AgdaBound{time}\AgdaSpace{}%
\AgdaBound{msg₂}\AgdaSpace{}%
\AgdaSymbol{(}\AgdaBound{sig2}\AgdaSpace{}%
\AgdaOperator{\AgdaInductiveConstructor{∷}}\AgdaSpace{}%
\AgdaBound{sig1}\AgdaSpace{}%
\AgdaOperator{\AgdaInductiveConstructor{∷}}\AgdaSpace{}%
\AgdaBound{dummy}\AgdaSpace{}%
\AgdaOperator{\AgdaInductiveConstructor{∷}}\AgdaSpace{}%
\AgdaBound{s}\AgdaSymbol{)}\AgdaSpace{}%
\AgdaSymbol{(}\AgdaInductiveConstructor{inj₂}\AgdaSpace{}%
\AgdaSymbol{(}\AgdaInductiveConstructor{inj₁}\AgdaSpace{}%
\AgdaSymbol{(}\AgdaInductiveConstructor{conj}\AgdaSpace{}%
\AgdaBound{and3}\AgdaSpace{}%
\AgdaBound{and4}\AgdaSymbol{)))}\<%
\\
\>[0][@{}l@{\AgdaIndent{0}}]%
\>[10]\AgdaSymbol{=}%
\>[958I]\AgdaFunction{boolToNatNotFalseLemma}\AgdaSpace{}%
\AgdaSymbol{(}\AgdaFunction{compareSigsMultiSigAux}\AgdaSpace{}%
\AgdaBound{msg₂}\AgdaSpace{}%
\AgdaSymbol{(}\AgdaBound{sig2}\AgdaSpace{}%
\AgdaOperator{\AgdaInductiveConstructor{∷}}\AgdaSpace{}%
\AgdaInductiveConstructor{[]}\AgdaSymbol{)}\AgdaSpace{}%
\AgdaSymbol{(}\AgdaBound{pbk2}\AgdaSpace{}%
\AgdaOperator{\AgdaInductiveConstructor{∷}}\AgdaSpace{}%
\AgdaBound{pbk3}\AgdaSpace{}%
\AgdaOperator{\AgdaInductiveConstructor{∷}}\AgdaSpace{}%
\AgdaInductiveConstructor{[]}\AgdaSymbol{)}\AgdaSpace{}%
\AgdaBound{sig1}\<%
\\
\>[958I][@{}l@{\AgdaIndent{0}}]%
\>[13]\AgdaSymbol{(}\AgdaFunction{isSigned}%
\>[24]\AgdaBound{msg₂}\AgdaSpace{}%
\AgdaBound{sig1}\AgdaSpace{}%
\AgdaBound{pbk1}\AgdaSymbol{))}\AgdaSpace{}%
\AgdaSymbol{(}\AgdaFunction{lemmaHoareTripleStackGeAux'2}%
\>[72]\AgdaBound{msg₂}\AgdaSpace{}%
\AgdaBound{pbk1}\AgdaSpace{}%
\AgdaBound{pbk2}\AgdaSpace{}%
\AgdaBound{pbk3}\AgdaSpace{}%
\AgdaBound{sig1}\AgdaSpace{}%
\AgdaBound{sig2}\AgdaSpace{}%
\AgdaBound{and3}\AgdaSpace{}%
\AgdaBound{and4}\AgdaSymbol{)}\<%
\\
\\[\AgdaEmptyExtraSkip]%
\>[0]\AgdaFunction{HoareTripleStackGeAux'}\AgdaSpace{}%
\AgdaBound{msg₁}\AgdaSpace{}%
\AgdaBound{pbk1}\AgdaSpace{}%
\AgdaBound{pbk2}\AgdaSpace{}%
\AgdaBound{pbk3}\AgdaSpace{}%
\AgdaSymbol{.}\AgdaField{==>stg}\AgdaSpace{}%
\AgdaBound{time}\AgdaSpace{}%
\AgdaBound{msg₂}\AgdaSpace{}%
\AgdaSymbol{(}\AgdaBound{sig2}\AgdaSpace{}%
\AgdaOperator{\AgdaInductiveConstructor{∷}}\AgdaSpace{}%
\AgdaBound{sig1}\AgdaSpace{}%
\AgdaOperator{\AgdaInductiveConstructor{∷}}\AgdaSpace{}%
\AgdaBound{dummy}\AgdaSpace{}%
\AgdaOperator{\AgdaInductiveConstructor{∷}}\AgdaSpace{}%
\AgdaBound{s}\AgdaSymbol{)}\AgdaSpace{}%
\AgdaSymbol{(}\AgdaInductiveConstructor{inj₂}\AgdaSpace{}%
\AgdaSymbol{(}\AgdaInductiveConstructor{inj₂}\AgdaSpace{}%
\AgdaSymbol{(}\AgdaInductiveConstructor{conj}\AgdaSpace{}%
\AgdaBound{and1}\AgdaSpace{}%
\AgdaBound{and2}\AgdaSymbol{)))}\<%
\\
\>[0][@{}l@{\AgdaIndent{0}}]%
\>[10]\AgdaSymbol{=}\AgdaSpace{}%
\AgdaFunction{boolToNatNotFalseLemma}\AgdaSpace{}%
\AgdaSymbol{(}\AgdaFunction{compareSigsMultiSigAux}\AgdaSpace{}%
\AgdaBound{msg₂}\AgdaSpace{}%
\AgdaSymbol{(}\AgdaBound{sig2}\AgdaSpace{}%
\AgdaOperator{\AgdaInductiveConstructor{∷}}\AgdaSpace{}%
\AgdaInductiveConstructor{[]}\AgdaSymbol{)}\AgdaSpace{}%
\AgdaSymbol{(}\AgdaBound{pbk2}\AgdaSpace{}%
\AgdaOperator{\AgdaInductiveConstructor{∷}}\AgdaSpace{}%
\AgdaBound{pbk3}\AgdaSpace{}%
\AgdaOperator{\AgdaInductiveConstructor{∷}}\AgdaSpace{}%
\AgdaInductiveConstructor{[]}\AgdaSymbol{)}\AgdaSpace{}%
\AgdaBound{sig1}\<%
\\
\>[10][@{}l@{\AgdaIndent{0}}]%
\>[11]\AgdaSymbol{(}\AgdaFunction{isSigned}%
\>[22]\AgdaBound{msg₂}\AgdaSpace{}%
\AgdaBound{sig1}\AgdaSpace{}%
\AgdaBound{pbk1}\AgdaSpace{}%
\AgdaSymbol{))}\AgdaSpace{}%
\AgdaSymbol{(}\AgdaFunction{lemmaHoareTripleStackGeAux'3}\AgdaSpace{}%
\AgdaBound{msg₂}\AgdaSpace{}%
\AgdaBound{pbk1}\AgdaSpace{}%
\AgdaBound{pbk2}\AgdaSpace{}%
\AgdaBound{pbk3}\AgdaSpace{}%
\AgdaBound{sig1}\AgdaSpace{}%
\AgdaBound{sig2}\AgdaSpace{}%
\AgdaBound{and1}\AgdaSpace{}%
\AgdaBound{and2}\AgdaSymbol{)}\<%
\\
\\[\AgdaEmptyExtraSkip]%
\>[0]\AgdaFunction{HoareTripleStackGeAux'}\AgdaSpace{}%
\AgdaBound{msg₁}\AgdaSpace{}%
\AgdaBound{pbk1}\AgdaSpace{}%
\AgdaBound{pbk2}\AgdaSpace{}%
\AgdaBound{pbk3}\AgdaSpace{}%
\AgdaComment{\{-numsig\ stack\ time₁-\}}\AgdaSpace{}%
\AgdaSymbol{.}\AgdaField{<==stg}\AgdaSpace{}%
\AgdaBound{time}\AgdaSpace{}%
\AgdaBound{msg₂}\AgdaSpace{}%
\AgdaSymbol{(}\AgdaBound{sig2}\AgdaSpace{}%
\AgdaOperator{\AgdaInductiveConstructor{∷}}\AgdaSpace{}%
\AgdaBound{sig1}\AgdaSpace{}%
\AgdaOperator{\AgdaInductiveConstructor{∷}}\AgdaSpace{}%
\AgdaBound{dummy}\AgdaSpace{}%
\AgdaOperator{\AgdaInductiveConstructor{∷}}\AgdaSpace{}%
\AgdaBound{s}\AgdaSymbol{)}\AgdaSpace{}%
\AgdaBound{x}\<%
\\
\>[0][@{}l@{\AgdaIndent{0}}]%
\>[4]\AgdaSymbol{=}%
\>[1039I]\AgdaFunction{lemmaHoareTripleStackGeAux'4}\AgdaSpace{}%
\AgdaBound{msg₂}\AgdaSpace{}%
\AgdaBound{pbk1}\AgdaSpace{}%
\AgdaBound{pbk2}\AgdaSpace{}%
\AgdaBound{pbk3}\AgdaSpace{}%
\AgdaBound{sig1}\AgdaSpace{}%
\AgdaBound{sig2}\<%
\\
\>[1039I][@{}l@{\AgdaIndent{0}}]%
\>[8]\AgdaSymbol{(}\AgdaFunction{boolToNatNotFalseLemma2}\AgdaSpace{}%
\AgdaSymbol{(}\AgdaFunction{compareSigsMultiSigAux}\AgdaSpace{}%
\AgdaBound{msg₂}\AgdaSpace{}%
\AgdaSymbol{(}\AgdaBound{sig2}\AgdaSpace{}%
\AgdaOperator{\AgdaInductiveConstructor{∷}}\AgdaSpace{}%
\AgdaInductiveConstructor{[]}\AgdaSymbol{)}\AgdaSpace{}%
\AgdaSymbol{(}\AgdaBound{pbk2}\AgdaSpace{}%
\AgdaOperator{\AgdaInductiveConstructor{∷}}\AgdaSpace{}%
\AgdaBound{pbk3}\AgdaSpace{}%
\AgdaOperator{\AgdaInductiveConstructor{∷}}\AgdaSpace{}%
\AgdaInductiveConstructor{[]}\AgdaSymbol{)}\AgdaSpace{}%
\AgdaBound{sig1}\<%
\\
\>[8][@{}l@{\AgdaIndent{0}}]%
\>[11]\AgdaSymbol{(}\AgdaFunction{isSigned}%
\>[22]\AgdaBound{msg₂}\AgdaSpace{}%
\AgdaBound{sig1}\AgdaSpace{}%
\AgdaBound{pbk1}\AgdaSpace{}%
\AgdaSymbol{))}\AgdaSpace{}%
\AgdaBound{x}\AgdaSymbol{)}\<%
\\
\\[\AgdaEmptyExtraSkip]%
\\[\AgdaEmptyExtraSkip]%
\>[0]\AgdaFunction{lemmaHoareTripleStackGeAux'7}%
\>[1061I]\AgdaSymbol{:}%
\>[1062I]\AgdaSymbol{(}\AgdaBound{msg₂}\AgdaSpace{}%
\AgdaSymbol{:}\AgdaSpace{}%
\AgdaDatatype{Msg}\AgdaSymbol{)}\<%
\\
\>[.][@{}l@{}]\<[1062I]%
\>[31]\AgdaSymbol{(}\AgdaBound{pbk1}\AgdaSpace{}%
\AgdaBound{pbk2}\AgdaSpace{}%
\AgdaBound{pbk3}\AgdaSpace{}%
\AgdaBound{pbk4}\AgdaSpace{}%
\AgdaBound{sig1}\AgdaSpace{}%
\AgdaBound{sig2}\AgdaSpace{}%
\AgdaSymbol{:}\AgdaSpace{}%
\AgdaDatatype{ℕ}\AgdaSymbol{)}\<%
\\
\>[31]\AgdaSymbol{→}\AgdaSpace{}%
\AgdaFunction{True}\AgdaSpace{}%
\AgdaSymbol{(}\AgdaFunction{isSigned}%
\>[49]\AgdaBound{msg₂}\AgdaSpace{}%
\AgdaBound{sig1}\AgdaSpace{}%
\AgdaBound{pbk1}\AgdaSymbol{)}\<%
\\
\>[1061I][@{}l@{\AgdaIndent{0}}]%
\>[30]\AgdaSymbol{→}%
\>[33]\AgdaFunction{True}\AgdaSpace{}%
\AgdaSymbol{(}\AgdaFunction{isSigned}%
\>[49]\AgdaBound{msg₂}\AgdaSpace{}%
\AgdaBound{sig2}\AgdaSpace{}%
\AgdaBound{pbk2}\AgdaSymbol{)}\<%
\\
\>[30][@{}l@{\AgdaIndent{0}}]%
\>[31]\AgdaSymbol{→}\AgdaSpace{}%
\AgdaFunction{True}\AgdaSpace{}%
\AgdaSymbol{(}\AgdaFunction{compareSigsMultiSig}\AgdaSpace{}%
\AgdaBound{msg₂}\AgdaSpace{}%
\AgdaSymbol{(}\AgdaSpace{}%
\AgdaBound{sig1}\AgdaSpace{}%
\AgdaOperator{\AgdaInductiveConstructor{∷}}\AgdaSpace{}%
\AgdaBound{sig2}\AgdaSpace{}%
\AgdaOperator{\AgdaInductiveConstructor{∷}}\AgdaSpace{}%
\AgdaInductiveConstructor{[]}\AgdaSymbol{)}\AgdaSpace{}%
\AgdaSymbol{(}\AgdaBound{pbk1}\AgdaSpace{}%
\AgdaOperator{\AgdaInductiveConstructor{∷}}\AgdaSpace{}%
\AgdaBound{pbk2}\AgdaSpace{}%
\AgdaOperator{\AgdaInductiveConstructor{∷}}\AgdaSpace{}%
\AgdaBound{pbk3}\AgdaSpace{}%
\AgdaOperator{\AgdaInductiveConstructor{∷}}\AgdaSpace{}%
\AgdaBound{pbk4}\AgdaSpace{}%
\AgdaOperator{\AgdaInductiveConstructor{∷}}%
\>[114]\AgdaInductiveConstructor{[]}\AgdaSymbol{))}\<%
\\
\\[\AgdaEmptyExtraSkip]%
\>[0]\AgdaFunction{lemmaHoareTripleStackGeAux'7}\AgdaSpace{}%
\AgdaBound{msg₂}\AgdaSpace{}%
\AgdaBound{pbk1}\AgdaSpace{}%
\AgdaBound{pbk2}\AgdaSpace{}%
\AgdaBound{pbk3}\AgdaSpace{}%
\AgdaBound{pbk4}\AgdaSpace{}%
\AgdaBound{sig1}\AgdaSpace{}%
\AgdaBound{sig2}\AgdaSpace{}%
\AgdaBound{x}\AgdaSpace{}%
\AgdaBound{x₁}\AgdaSpace{}%
\AgdaKeyword{with}%
\>[75]\AgdaSymbol{(}\AgdaFunction{isSigned}%
\>[86]\AgdaBound{msg₂}\AgdaSpace{}%
\AgdaBound{sig1}\AgdaSpace{}%
\AgdaBound{pbk1}\AgdaSymbol{)}\<%
\\
\>[0]\AgdaFunction{lemmaHoareTripleStackGeAux'7}\AgdaSpace{}%
\AgdaBound{msg₂}\AgdaSpace{}%
\AgdaBound{pbk1}\AgdaSpace{}%
\AgdaBound{pbk2}\AgdaSpace{}%
\AgdaBound{pbk3}\AgdaSpace{}%
\AgdaBound{pbk4}\AgdaSpace{}%
\AgdaBound{sig1}\AgdaSpace{}%
\AgdaBound{sig2}\AgdaSpace{}%
\AgdaBound{x}\AgdaSpace{}%
\AgdaBound{x₁}\AgdaSpace{}%
\AgdaSymbol{|}\AgdaSpace{}%
\AgdaInductiveConstructor{true}\AgdaSpace{}%
\AgdaKeyword{with}\AgdaSpace{}%
\AgdaSymbol{(}\AgdaFunction{isSigned}%
\>[92]\AgdaBound{msg₂}\AgdaSpace{}%
\AgdaBound{sig2}\AgdaSpace{}%
\AgdaBound{pbk2}\AgdaSymbol{)}\<%
\\
\>[0]\AgdaFunction{lemmaHoareTripleStackGeAux'7}\AgdaSpace{}%
\AgdaBound{msg₂}\AgdaSpace{}%
\AgdaBound{pbk1}\AgdaSpace{}%
\AgdaBound{pbk2}\AgdaSpace{}%
\AgdaBound{pbk3}\AgdaSpace{}%
\AgdaBound{pbk4}\AgdaSpace{}%
\AgdaBound{sig1}\AgdaSpace{}%
\AgdaBound{sig2}\AgdaSpace{}%
\AgdaBound{x}\AgdaSpace{}%
\AgdaBound{x₁}\AgdaSpace{}%
\AgdaSymbol{|}\AgdaSpace{}%
\AgdaInductiveConstructor{true}\AgdaSpace{}%
\AgdaSymbol{|}\AgdaSpace{}%
\AgdaInductiveConstructor{true}\AgdaSpace{}%
\AgdaSymbol{=}\AgdaSpace{}%
\AgdaInductiveConstructor{tt}\<%
\\
\\[\AgdaEmptyExtraSkip]%
\\[\AgdaEmptyExtraSkip]%
\>[0]\AgdaFunction{lemmaHoareTripleStackGeAux'8}%
\>[1138I]\AgdaSymbol{:}%
\>[1139I]\AgdaSymbol{(}\AgdaBound{msg₂}\AgdaSpace{}%
\AgdaSymbol{:}\AgdaSpace{}%
\AgdaDatatype{Msg}\AgdaSymbol{)}\<%
\\
\>[.][@{}l@{}]\<[1139I]%
\>[31]\AgdaSymbol{(}\AgdaBound{pbk1}\AgdaSpace{}%
\AgdaBound{pbk2}\AgdaSpace{}%
\AgdaBound{pbk3}\AgdaSpace{}%
\AgdaBound{pbk4}\AgdaSpace{}%
\AgdaBound{sig1}\AgdaSpace{}%
\AgdaBound{sig2}\AgdaSpace{}%
\AgdaSymbol{:}\AgdaSpace{}%
\AgdaDatatype{ℕ}\AgdaSymbol{)}\<%
\\
\>[31]\AgdaSymbol{→}\AgdaSpace{}%
\AgdaFunction{True}\AgdaSpace{}%
\AgdaSymbol{(}\AgdaFunction{isSigned}%
\>[49]\AgdaBound{msg₂}\AgdaSpace{}%
\AgdaBound{sig1}\AgdaSpace{}%
\AgdaBound{pbk1}\AgdaSymbol{)}\<%
\\
\>[1138I][@{}l@{\AgdaIndent{0}}]%
\>[30]\AgdaSymbol{→}%
\>[33]\AgdaFunction{True}\AgdaSpace{}%
\AgdaSymbol{(}\AgdaFunction{isSigned}%
\>[49]\AgdaBound{msg₂}\AgdaSpace{}%
\AgdaBound{sig2}\AgdaSpace{}%
\AgdaBound{pbk3}\AgdaSymbol{)}\<%
\\
\>[30][@{}l@{\AgdaIndent{0}}]%
\>[31]\AgdaSymbol{→}\AgdaSpace{}%
\AgdaFunction{True}\AgdaSpace{}%
\AgdaSymbol{(}\AgdaFunction{compareSigsMultiSig}\AgdaSpace{}%
\AgdaBound{msg₂}\AgdaSpace{}%
\AgdaSymbol{(}\AgdaSpace{}%
\AgdaBound{sig1}\AgdaSpace{}%
\AgdaOperator{\AgdaInductiveConstructor{∷}}\AgdaSpace{}%
\AgdaBound{sig2}\AgdaSpace{}%
\AgdaOperator{\AgdaInductiveConstructor{∷}}\AgdaSpace{}%
\AgdaInductiveConstructor{[]}\AgdaSymbol{)}\AgdaSpace{}%
\AgdaSymbol{(}\AgdaBound{pbk1}\AgdaSpace{}%
\AgdaOperator{\AgdaInductiveConstructor{∷}}\AgdaSpace{}%
\AgdaBound{pbk2}\AgdaSpace{}%
\AgdaOperator{\AgdaInductiveConstructor{∷}}\AgdaSpace{}%
\AgdaBound{pbk3}\AgdaSpace{}%
\AgdaOperator{\AgdaInductiveConstructor{∷}}\AgdaSpace{}%
\AgdaBound{pbk4}\AgdaSpace{}%
\AgdaOperator{\AgdaInductiveConstructor{∷}}%
\>[114]\AgdaInductiveConstructor{[]}\AgdaSymbol{))}\<%
\\
\\[\AgdaEmptyExtraSkip]%
\>[0]\AgdaFunction{lemmaHoareTripleStackGeAux'8}\AgdaSpace{}%
\AgdaBound{msg₂}\AgdaSpace{}%
\AgdaBound{pbk1}\AgdaSpace{}%
\AgdaBound{pbk2}\AgdaSpace{}%
\AgdaBound{pbk3}\AgdaSpace{}%
\AgdaBound{pbk4}\AgdaSpace{}%
\AgdaBound{sig1}\AgdaSpace{}%
\AgdaBound{sig2}\AgdaSpace{}%
\AgdaBound{x}\AgdaSpace{}%
\AgdaBound{x₁}\AgdaSpace{}%
\AgdaKeyword{with}%
\>[75]\AgdaSymbol{(}\AgdaFunction{isSigned}%
\>[86]\AgdaBound{msg₂}\AgdaSpace{}%
\AgdaBound{sig2}\AgdaSpace{}%
\AgdaBound{pbk1}\AgdaSymbol{)}\<%
\\
\>[0]\AgdaFunction{lemmaHoareTripleStackGeAux'8}\AgdaSpace{}%
\AgdaBound{msg₂}\AgdaSpace{}%
\AgdaBound{pbk1}\AgdaSpace{}%
\AgdaBound{pbk2}\AgdaSpace{}%
\AgdaBound{pbk3}\AgdaSpace{}%
\AgdaBound{pbk4}\AgdaSpace{}%
\AgdaBound{sig1}\AgdaSpace{}%
\AgdaBound{sig2}\AgdaSpace{}%
\AgdaBound{x}\AgdaSpace{}%
\AgdaBound{x₁}\AgdaSpace{}%
\AgdaSymbol{|}\AgdaSpace{}%
\AgdaInductiveConstructor{false}\AgdaSpace{}%
\AgdaKeyword{with}\AgdaSpace{}%
\AgdaSymbol{(}\AgdaFunction{isSigned}%
\>[93]\AgdaBound{msg₂}\AgdaSpace{}%
\AgdaBound{sig1}\AgdaSpace{}%
\AgdaBound{pbk1}\AgdaSymbol{)}\<%
\\
\>[0]\AgdaFunction{lemmaHoareTripleStackGeAux'8}\AgdaSpace{}%
\AgdaBound{msg₂}\AgdaSpace{}%
\AgdaBound{pbk1}\AgdaSpace{}%
\AgdaBound{pbk2}\AgdaSpace{}%
\AgdaBound{pbk3}\AgdaSpace{}%
\AgdaBound{pbk4}\AgdaSpace{}%
\AgdaBound{sig1}\AgdaSpace{}%
\AgdaBound{sig2}\AgdaSpace{}%
\AgdaBound{x}\AgdaSpace{}%
\AgdaBound{x₁}\AgdaSpace{}%
\AgdaSymbol{|}\AgdaSpace{}%
\AgdaInductiveConstructor{false}\AgdaSpace{}%
\AgdaSymbol{|}\AgdaSpace{}%
\AgdaInductiveConstructor{true}\AgdaSpace{}%
\AgdaKeyword{with}%
\>[90]\AgdaSymbol{(}\AgdaFunction{isSigned}%
\>[101]\AgdaBound{msg₂}\AgdaSpace{}%
\AgdaBound{sig2}\AgdaSpace{}%
\AgdaBound{pbk2}\AgdaSymbol{)}\<%
\\
\>[0]\AgdaFunction{lemmaHoareTripleStackGeAux'8}\AgdaSpace{}%
\AgdaBound{msg₂}\AgdaSpace{}%
\AgdaBound{pbk1}\AgdaSpace{}%
\AgdaBound{pbk2}\AgdaSpace{}%
\AgdaBound{pbk3}\AgdaSpace{}%
\AgdaBound{pbk4}\AgdaSpace{}%
\AgdaBound{sig1}\AgdaSpace{}%
\AgdaBound{sig2}\AgdaSpace{}%
\AgdaBound{x}\AgdaSpace{}%
\AgdaBound{x₁}\AgdaSpace{}%
\AgdaSymbol{|}\AgdaSpace{}%
\AgdaInductiveConstructor{false}\AgdaSpace{}%
\AgdaSymbol{|}\AgdaSpace{}%
\AgdaInductiveConstructor{true}\AgdaSpace{}%
\AgdaSymbol{|}\AgdaSpace{}%
\AgdaInductiveConstructor{false}\AgdaSpace{}%
\AgdaKeyword{with}\AgdaSpace{}%
\AgdaSymbol{(}\AgdaFunction{isSigned}%
\>[108]\AgdaBound{msg₂}\AgdaSpace{}%
\AgdaBound{sig2}\AgdaSpace{}%
\AgdaBound{pbk3}\AgdaSymbol{)}\<%
\\
\>[0]\AgdaFunction{lemmaHoareTripleStackGeAux'8}\AgdaSpace{}%
\AgdaBound{msg₂}\AgdaSpace{}%
\AgdaBound{pbk1}\AgdaSpace{}%
\AgdaBound{pbk2}\AgdaSpace{}%
\AgdaBound{pbk3}\AgdaSpace{}%
\AgdaBound{pbk4}\AgdaSpace{}%
\AgdaBound{sig1}\AgdaSpace{}%
\AgdaBound{sig2}\AgdaSpace{}%
\AgdaBound{x}\AgdaSpace{}%
\AgdaBound{x₁}\AgdaSpace{}%
\AgdaSymbol{|}\AgdaSpace{}%
\AgdaInductiveConstructor{false}\AgdaSpace{}%
\AgdaSymbol{|}\AgdaSpace{}%
\AgdaInductiveConstructor{true}\AgdaSpace{}%
\AgdaSymbol{|}\AgdaSpace{}%
\AgdaInductiveConstructor{false}\AgdaSpace{}%
\AgdaSymbol{|}\AgdaSpace{}%
\AgdaInductiveConstructor{true}\AgdaSpace{}%
\AgdaSymbol{=}\AgdaSpace{}%
\AgdaInductiveConstructor{tt}\<%
\\
\>[0]\AgdaFunction{lemmaHoareTripleStackGeAux'8}\AgdaSpace{}%
\AgdaBound{msg₂}\AgdaSpace{}%
\AgdaBound{pbk1}\AgdaSpace{}%
\AgdaBound{pbk2}\AgdaSpace{}%
\AgdaBound{pbk3}\AgdaSpace{}%
\AgdaBound{pbk4}\AgdaSpace{}%
\AgdaBound{sig1}\AgdaSpace{}%
\AgdaBound{sig2}\AgdaSpace{}%
\AgdaBound{x}\AgdaSpace{}%
\AgdaBound{x₁}\AgdaSpace{}%
\AgdaSymbol{|}\AgdaSpace{}%
\AgdaInductiveConstructor{false}\AgdaSpace{}%
\AgdaSymbol{|}\AgdaSpace{}%
\AgdaInductiveConstructor{true}\AgdaSpace{}%
\AgdaSymbol{|}\AgdaSpace{}%
\AgdaInductiveConstructor{true}\AgdaSpace{}%
\AgdaSymbol{=}\AgdaSpace{}%
\AgdaInductiveConstructor{tt}\<%
\\
\>[0]\AgdaFunction{lemmaHoareTripleStackGeAux'8}\AgdaSpace{}%
\AgdaBound{msg₂}\AgdaSpace{}%
\AgdaBound{pbk1}\AgdaSpace{}%
\AgdaBound{pbk2}\AgdaSpace{}%
\AgdaBound{pbk3}\AgdaSpace{}%
\AgdaBound{pbk4}\AgdaSpace{}%
\AgdaBound{sig1}\AgdaSpace{}%
\AgdaBound{sig2}\AgdaSpace{}%
\AgdaBound{x}\AgdaSpace{}%
\AgdaBound{x₁}\AgdaSpace{}%
\AgdaSymbol{|}\AgdaSpace{}%
\AgdaInductiveConstructor{true}%
\>[77]\AgdaKeyword{with}\AgdaSpace{}%
\AgdaSymbol{(}\AgdaFunction{isSigned}%
\>[93]\AgdaBound{msg₂}\AgdaSpace{}%
\AgdaBound{sig1}\AgdaSpace{}%
\AgdaBound{pbk1}\AgdaSymbol{)}\<%
\\
\>[0]\AgdaFunction{lemmaHoareTripleStackGeAux'8}\AgdaSpace{}%
\AgdaBound{msg₂}\AgdaSpace{}%
\AgdaBound{pbk1}\AgdaSpace{}%
\AgdaBound{pbk2}\AgdaSpace{}%
\AgdaBound{pbk3}\AgdaSpace{}%
\AgdaBound{pbk4}\AgdaSpace{}%
\AgdaBound{sig1}\AgdaSpace{}%
\AgdaBound{sig2}\AgdaSpace{}%
\AgdaBound{x}\AgdaSpace{}%
\AgdaBound{x₁}\AgdaSpace{}%
\AgdaSymbol{|}\AgdaSpace{}%
\AgdaInductiveConstructor{true}\AgdaSpace{}%
\AgdaSymbol{|}\AgdaSpace{}%
\AgdaInductiveConstructor{true}\AgdaSpace{}%
\AgdaKeyword{with}%
\>[89]\AgdaSymbol{(}\AgdaFunction{isSigned}%
\>[100]\AgdaBound{msg₂}\AgdaSpace{}%
\AgdaBound{sig2}\AgdaSpace{}%
\AgdaBound{pbk2}\AgdaSymbol{)}\<%
\\
\>[0]\AgdaFunction{lemmaHoareTripleStackGeAux'8}\AgdaSpace{}%
\AgdaBound{msg₂}\AgdaSpace{}%
\AgdaBound{pbk1}\AgdaSpace{}%
\AgdaBound{pbk2}\AgdaSpace{}%
\AgdaBound{pbk3}\AgdaSpace{}%
\AgdaBound{pbk4}\AgdaSpace{}%
\AgdaBound{sig1}\AgdaSpace{}%
\AgdaBound{sig2}\AgdaSpace{}%
\AgdaBound{x}\AgdaSpace{}%
\AgdaBound{x₁}\AgdaSpace{}%
\AgdaSymbol{|}\AgdaSpace{}%
\AgdaInductiveConstructor{true}\AgdaSpace{}%
\AgdaSymbol{|}\AgdaSpace{}%
\AgdaInductiveConstructor{true}\AgdaSpace{}%
\AgdaSymbol{|}\AgdaSpace{}%
\AgdaInductiveConstructor{false}\AgdaSpace{}%
\AgdaKeyword{with}%
\>[97]\AgdaSymbol{(}\AgdaFunction{isSigned}%
\>[108]\AgdaBound{msg₂}\AgdaSpace{}%
\AgdaBound{sig2}\AgdaSpace{}%
\AgdaBound{pbk3}\AgdaSymbol{)}\<%
\\
\>[0]\AgdaFunction{lemmaHoareTripleStackGeAux'8}\AgdaSpace{}%
\AgdaBound{msg₂}\AgdaSpace{}%
\AgdaBound{pbk1}\AgdaSpace{}%
\AgdaBound{pbk2}\AgdaSpace{}%
\AgdaBound{pbk3}\AgdaSpace{}%
\AgdaBound{pbk4}\AgdaSpace{}%
\AgdaBound{sig1}\AgdaSpace{}%
\AgdaBound{sig2}\AgdaSpace{}%
\AgdaBound{x}\AgdaSpace{}%
\AgdaBound{x₁}\AgdaSpace{}%
\AgdaSymbol{|}\AgdaSpace{}%
\AgdaInductiveConstructor{true}\AgdaSpace{}%
\AgdaSymbol{|}\AgdaSpace{}%
\AgdaInductiveConstructor{true}\AgdaSpace{}%
\AgdaSymbol{|}\AgdaSpace{}%
\AgdaInductiveConstructor{false}\AgdaSpace{}%
\AgdaSymbol{|}\AgdaSpace{}%
\AgdaInductiveConstructor{true}\AgdaSpace{}%
\AgdaSymbol{=}\AgdaSpace{}%
\AgdaInductiveConstructor{tt}\<%
\\
\>[0]\AgdaFunction{lemmaHoareTripleStackGeAux'8}\AgdaSpace{}%
\AgdaBound{msg₂}\AgdaSpace{}%
\AgdaBound{pbk1}\AgdaSpace{}%
\AgdaBound{pbk2}\AgdaSpace{}%
\AgdaBound{pbk3}\AgdaSpace{}%
\AgdaBound{pbk4}\AgdaSpace{}%
\AgdaBound{sig1}\AgdaSpace{}%
\AgdaBound{sig2}\AgdaSpace{}%
\AgdaBound{x}\AgdaSpace{}%
\AgdaBound{x₁}\AgdaSpace{}%
\AgdaSymbol{|}\AgdaSpace{}%
\AgdaInductiveConstructor{true}\AgdaSpace{}%
\AgdaSymbol{|}\AgdaSpace{}%
\AgdaInductiveConstructor{true}\AgdaSpace{}%
\AgdaSymbol{|}\AgdaSpace{}%
\AgdaInductiveConstructor{true}\AgdaSpace{}%
\AgdaSymbol{=}\AgdaSpace{}%
\AgdaInductiveConstructor{tt}\<%
\\
\\[\AgdaEmptyExtraSkip]%
\\[\AgdaEmptyExtraSkip]%
\>[0]\AgdaFunction{lemmaHoareTripleStackGeAux'9}%
\>[1355I]\AgdaSymbol{:}%
\>[1356I]\AgdaSymbol{(}\AgdaBound{msg₂}\AgdaSpace{}%
\AgdaSymbol{:}\AgdaSpace{}%
\AgdaDatatype{Msg}\AgdaSymbol{)}\<%
\\
\>[.][@{}l@{}]\<[1356I]%
\>[31]\AgdaSymbol{(}\AgdaBound{pbk1}\AgdaSpace{}%
\AgdaBound{pbk2}\AgdaSpace{}%
\AgdaBound{pbk3}\AgdaSpace{}%
\AgdaBound{pbk4}\AgdaSpace{}%
\AgdaBound{sig1}\AgdaSpace{}%
\AgdaBound{sig2}\AgdaSpace{}%
\AgdaSymbol{:}\AgdaSpace{}%
\AgdaDatatype{ℕ}\AgdaSymbol{)}\<%
\\
\>[31]\AgdaSymbol{→}\AgdaSpace{}%
\AgdaFunction{True}\AgdaSpace{}%
\AgdaSymbol{(}\AgdaFunction{isSigned}%
\>[49]\AgdaBound{msg₂}\AgdaSpace{}%
\AgdaBound{sig1}\AgdaSpace{}%
\AgdaBound{pbk1}\AgdaSymbol{)}\<%
\\
\>[1355I][@{}l@{\AgdaIndent{0}}]%
\>[30]\AgdaSymbol{→}%
\>[33]\AgdaFunction{True}\AgdaSpace{}%
\AgdaSymbol{(}\AgdaFunction{isSigned}%
\>[49]\AgdaBound{msg₂}\AgdaSpace{}%
\AgdaBound{sig2}\AgdaSpace{}%
\AgdaBound{pbk4}\AgdaSymbol{)}\<%
\\
\>[30][@{}l@{\AgdaIndent{0}}]%
\>[31]\AgdaSymbol{→}\AgdaSpace{}%
\AgdaFunction{True}\AgdaSpace{}%
\AgdaSymbol{(}\AgdaFunction{compareSigsMultiSig}\AgdaSpace{}%
\AgdaBound{msg₂}\AgdaSpace{}%
\AgdaSymbol{(}\AgdaSpace{}%
\AgdaBound{sig1}\AgdaSpace{}%
\AgdaOperator{\AgdaInductiveConstructor{∷}}\AgdaSpace{}%
\AgdaBound{sig2}\AgdaSpace{}%
\AgdaOperator{\AgdaInductiveConstructor{∷}}\AgdaSpace{}%
\AgdaInductiveConstructor{[]}\AgdaSymbol{)}\AgdaSpace{}%
\AgdaSymbol{(}\AgdaBound{pbk1}\AgdaSpace{}%
\AgdaOperator{\AgdaInductiveConstructor{∷}}\AgdaSpace{}%
\AgdaBound{pbk2}\AgdaSpace{}%
\AgdaOperator{\AgdaInductiveConstructor{∷}}\AgdaSpace{}%
\AgdaBound{pbk3}\AgdaSpace{}%
\AgdaOperator{\AgdaInductiveConstructor{∷}}\AgdaSpace{}%
\AgdaBound{pbk4}\AgdaSpace{}%
\AgdaOperator{\AgdaInductiveConstructor{∷}}%
\>[114]\AgdaInductiveConstructor{[]}\AgdaSymbol{))}\<%
\\
\\[\AgdaEmptyExtraSkip]%
\>[0]\AgdaFunction{lemmaHoareTripleStackGeAux'9}\AgdaSpace{}%
\AgdaBound{msg₂}\AgdaSpace{}%
\AgdaBound{pbk1}\AgdaSpace{}%
\AgdaBound{pbk2}\AgdaSpace{}%
\AgdaBound{pbk3}\AgdaSpace{}%
\AgdaBound{pbk4}\AgdaSpace{}%
\AgdaBound{sig1}\AgdaSpace{}%
\AgdaBound{sig2}\AgdaSpace{}%
\AgdaBound{x}\AgdaSpace{}%
\AgdaBound{x₁}\AgdaSpace{}%
\AgdaKeyword{with}%
\>[75]\AgdaSymbol{(}\AgdaFunction{isSigned}%
\>[86]\AgdaBound{msg₂}\AgdaSpace{}%
\AgdaBound{sig1}\AgdaSpace{}%
\AgdaBound{pbk1}\AgdaSymbol{)}\<%
\\
\>[0]\AgdaFunction{lemmaHoareTripleStackGeAux'9}\AgdaSpace{}%
\AgdaBound{msg₂}\AgdaSpace{}%
\AgdaBound{pbk1}\AgdaSpace{}%
\AgdaBound{pbk2}\AgdaSpace{}%
\AgdaBound{pbk3}\AgdaSpace{}%
\AgdaBound{pbk4}\AgdaSpace{}%
\AgdaBound{sig1}\AgdaSpace{}%
\AgdaBound{sig2}\AgdaSpace{}%
\AgdaBound{x}\AgdaSpace{}%
\AgdaBound{x₁}\AgdaSpace{}%
\AgdaSymbol{|}\AgdaSpace{}%
\AgdaInductiveConstructor{true}\AgdaSpace{}%
\AgdaKeyword{with}%
\>[82]\AgdaSymbol{(}\AgdaFunction{isSigned}%
\>[93]\AgdaBound{msg₂}\AgdaSpace{}%
\AgdaBound{sig2}\AgdaSpace{}%
\AgdaBound{pbk2}\AgdaSymbol{)}\<%
\\
\>[0]\AgdaFunction{lemmaHoareTripleStackGeAux'9}\AgdaSpace{}%
\AgdaBound{msg₂}\AgdaSpace{}%
\AgdaBound{pbk1}\AgdaSpace{}%
\AgdaBound{pbk2}\AgdaSpace{}%
\AgdaBound{pbk3}\AgdaSpace{}%
\AgdaBound{pbk4}\AgdaSpace{}%
\AgdaBound{sig1}\AgdaSpace{}%
\AgdaBound{sig2}\AgdaSpace{}%
\AgdaBound{x}\AgdaSpace{}%
\AgdaBound{x₁}\AgdaSpace{}%
\AgdaSymbol{|}\AgdaSpace{}%
\AgdaInductiveConstructor{true}\AgdaSpace{}%
\AgdaSymbol{|}\AgdaSpace{}%
\AgdaInductiveConstructor{false}\AgdaSpace{}%
\AgdaKeyword{with}%
\>[90]\AgdaSymbol{(}\AgdaFunction{isSigned}%
\>[101]\AgdaBound{msg₂}\AgdaSpace{}%
\AgdaBound{sig2}\AgdaSpace{}%
\AgdaBound{pbk3}\AgdaSymbol{)}\<%
\\
\>[0]\AgdaFunction{lemmaHoareTripleStackGeAux'9}\AgdaSpace{}%
\AgdaBound{msg₂}\AgdaSpace{}%
\AgdaBound{pbk1}\AgdaSpace{}%
\AgdaBound{pbk2}\AgdaSpace{}%
\AgdaBound{pbk3}\AgdaSpace{}%
\AgdaBound{pbk4}\AgdaSpace{}%
\AgdaBound{sig1}\AgdaSpace{}%
\AgdaBound{sig2}\AgdaSpace{}%
\AgdaBound{x}\AgdaSpace{}%
\AgdaBound{x₁}\AgdaSpace{}%
\AgdaSymbol{|}\AgdaSpace{}%
\AgdaInductiveConstructor{true}\AgdaSpace{}%
\AgdaSymbol{|}\AgdaSpace{}%
\AgdaInductiveConstructor{false}\AgdaSpace{}%
\AgdaSymbol{|}\AgdaSpace{}%
\AgdaInductiveConstructor{false}\AgdaSpace{}%
\AgdaKeyword{with}%
\>[98]\AgdaSymbol{(}\AgdaFunction{isSigned}%
\>[109]\AgdaBound{msg₂}\AgdaSpace{}%
\AgdaBound{sig2}\AgdaSpace{}%
\AgdaBound{pbk4}\AgdaSymbol{)}\<%
\\
\>[0]\AgdaFunction{lemmaHoareTripleStackGeAux'9}\AgdaSpace{}%
\AgdaBound{msg₂}\AgdaSpace{}%
\AgdaBound{pbk1}\AgdaSpace{}%
\AgdaBound{pbk2}\AgdaSpace{}%
\AgdaBound{pbk3}\AgdaSpace{}%
\AgdaBound{pbk4}\AgdaSpace{}%
\AgdaBound{sig1}\AgdaSpace{}%
\AgdaBound{sig2}\AgdaSpace{}%
\AgdaBound{x}\AgdaSpace{}%
\AgdaBound{x₁}\AgdaSpace{}%
\AgdaSymbol{|}\AgdaSpace{}%
\AgdaInductiveConstructor{true}\AgdaSpace{}%
\AgdaSymbol{|}\AgdaSpace{}%
\AgdaInductiveConstructor{false}\AgdaSpace{}%
\AgdaSymbol{|}\AgdaSpace{}%
\AgdaInductiveConstructor{false}\AgdaSpace{}%
\AgdaSymbol{|}\AgdaSpace{}%
\AgdaInductiveConstructor{true}\AgdaSpace{}%
\AgdaSymbol{=}\AgdaSpace{}%
\AgdaInductiveConstructor{tt}\<%
\\
\>[0]\AgdaFunction{lemmaHoareTripleStackGeAux'9}\AgdaSpace{}%
\AgdaBound{msg₂}\AgdaSpace{}%
\AgdaBound{pbk1}\AgdaSpace{}%
\AgdaBound{pbk2}\AgdaSpace{}%
\AgdaBound{pbk3}\AgdaSpace{}%
\AgdaBound{pbk4}\AgdaSpace{}%
\AgdaBound{sig1}\AgdaSpace{}%
\AgdaBound{sig2}\AgdaSpace{}%
\AgdaBound{x}\AgdaSpace{}%
\AgdaBound{x₁}\AgdaSpace{}%
\AgdaSymbol{|}\AgdaSpace{}%
\AgdaInductiveConstructor{true}\AgdaSpace{}%
\AgdaSymbol{|}\AgdaSpace{}%
\AgdaInductiveConstructor{false}\AgdaSpace{}%
\AgdaSymbol{|}\AgdaSpace{}%
\AgdaInductiveConstructor{true}\AgdaSpace{}%
\AgdaSymbol{=}\AgdaSpace{}%
\AgdaInductiveConstructor{tt}\<%
\\
\>[0]\AgdaFunction{lemmaHoareTripleStackGeAux'9}\AgdaSpace{}%
\AgdaBound{msg₂}\AgdaSpace{}%
\AgdaBound{pbk1}\AgdaSpace{}%
\AgdaBound{pbk2}\AgdaSpace{}%
\AgdaBound{pbk3}\AgdaSpace{}%
\AgdaBound{pbk4}\AgdaSpace{}%
\AgdaBound{sig1}\AgdaSpace{}%
\AgdaBound{sig2}\AgdaSpace{}%
\AgdaBound{x}\AgdaSpace{}%
\AgdaBound{x₁}\AgdaSpace{}%
\AgdaSymbol{|}\AgdaSpace{}%
\AgdaInductiveConstructor{true}\AgdaSpace{}%
\AgdaSymbol{|}\AgdaSpace{}%
\AgdaInductiveConstructor{true}\AgdaSpace{}%
\AgdaSymbol{=}\AgdaSpace{}%
\AgdaInductiveConstructor{tt}\<%
\\
\\[\AgdaEmptyExtraSkip]%
\\[\AgdaEmptyExtraSkip]%
\\[\AgdaEmptyExtraSkip]%
\\[\AgdaEmptyExtraSkip]%
\>[0]\AgdaFunction{lemmaHoareTripleStackGeAux'10}%
\>[1501I]\AgdaSymbol{:}\AgdaSpace{}%
\AgdaSymbol{(}\AgdaBound{msg₂}\AgdaSpace{}%
\AgdaSymbol{:}\AgdaSpace{}%
\AgdaDatatype{Msg}\AgdaSymbol{)}\<%
\\
\>[1501I][@{}l@{\AgdaIndent{0}}]%
\>[31]\AgdaSymbol{(}\AgdaBound{pbk1}\AgdaSpace{}%
\AgdaBound{pbk2}\AgdaSpace{}%
\AgdaBound{pbk3}\AgdaSpace{}%
\AgdaBound{pbk4}\AgdaSpace{}%
\AgdaBound{sig1}\AgdaSpace{}%
\AgdaBound{sig2}\AgdaSpace{}%
\AgdaSymbol{:}\AgdaSpace{}%
\AgdaDatatype{ℕ}\AgdaSymbol{)}\<%
\\
\>[31]\AgdaSymbol{→}\AgdaSpace{}%
\AgdaFunction{True}\AgdaSpace{}%
\AgdaSymbol{(}\AgdaFunction{isSigned}%
\>[49]\AgdaBound{msg₂}\AgdaSpace{}%
\AgdaBound{sig1}\AgdaSpace{}%
\AgdaBound{pbk2}\AgdaSymbol{)}\<%
\\
\>[.][@{}l@{}]\<[1501I]%
\>[30]\AgdaSymbol{→}%
\>[33]\AgdaFunction{True}\AgdaSpace{}%
\AgdaSymbol{(}\AgdaFunction{isSigned}%
\>[49]\AgdaBound{msg₂}\AgdaSpace{}%
\AgdaBound{sig2}\AgdaSpace{}%
\AgdaBound{pbk3}\AgdaSymbol{)}\<%
\\
\>[30][@{}l@{\AgdaIndent{0}}]%
\>[31]\AgdaSymbol{→}\AgdaSpace{}%
\AgdaFunction{True}\AgdaSpace{}%
\AgdaSymbol{(}\AgdaFunction{compareSigsMultiSig}\AgdaSpace{}%
\AgdaBound{msg₂}\AgdaSpace{}%
\AgdaSymbol{(}\AgdaSpace{}%
\AgdaBound{sig1}\AgdaSpace{}%
\AgdaOperator{\AgdaInductiveConstructor{∷}}\AgdaSpace{}%
\AgdaBound{sig2}\AgdaSpace{}%
\AgdaOperator{\AgdaInductiveConstructor{∷}}\AgdaSpace{}%
\AgdaInductiveConstructor{[]}\AgdaSymbol{)}\AgdaSpace{}%
\AgdaSymbol{(}\AgdaBound{pbk1}\AgdaSpace{}%
\AgdaOperator{\AgdaInductiveConstructor{∷}}\AgdaSpace{}%
\AgdaBound{pbk2}\AgdaSpace{}%
\AgdaOperator{\AgdaInductiveConstructor{∷}}\AgdaSpace{}%
\AgdaBound{pbk3}\AgdaSpace{}%
\AgdaOperator{\AgdaInductiveConstructor{∷}}\AgdaSpace{}%
\AgdaBound{pbk4}\AgdaSpace{}%
\AgdaOperator{\AgdaInductiveConstructor{∷}}%
\>[114]\AgdaInductiveConstructor{[]}\AgdaSymbol{))}\<%
\\
\\[\AgdaEmptyExtraSkip]%
\>[0]\AgdaFunction{lemmaHoareTripleStackGeAux'10}\AgdaSpace{}%
\AgdaBound{msg₂}\AgdaSpace{}%
\AgdaBound{pbk1}\AgdaSpace{}%
\AgdaBound{pbk2}\AgdaSpace{}%
\AgdaBound{pbk3}\AgdaSpace{}%
\AgdaBound{pbk4}\AgdaSpace{}%
\AgdaBound{sig1}\AgdaSpace{}%
\AgdaBound{sig2}\AgdaSpace{}%
\AgdaBound{x}\AgdaSpace{}%
\AgdaBound{x₁}\AgdaSpace{}%
\AgdaKeyword{with}%
\>[76]\AgdaSymbol{(}\AgdaFunction{isSigned}%
\>[87]\AgdaBound{msg₂}\AgdaSpace{}%
\AgdaBound{sig1}\AgdaSpace{}%
\AgdaBound{pbk1}\AgdaSymbol{)}\<%
\\
\>[0]\AgdaFunction{lemmaHoareTripleStackGeAux'10}\AgdaSpace{}%
\AgdaBound{msg₂}\AgdaSpace{}%
\AgdaBound{pbk1}\AgdaSpace{}%
\AgdaBound{pbk2}\AgdaSpace{}%
\AgdaBound{pbk3}\AgdaSpace{}%
\AgdaBound{pbk4}\AgdaSpace{}%
\AgdaBound{sig1}\AgdaSpace{}%
\AgdaBound{sig2}\AgdaSpace{}%
\AgdaBound{x}\AgdaSpace{}%
\AgdaBound{x₁}\AgdaSpace{}%
\AgdaSymbol{|}\AgdaSpace{}%
\AgdaInductiveConstructor{false}\AgdaSpace{}%
\AgdaKeyword{with}\AgdaSpace{}%
\AgdaSymbol{(}\AgdaFunction{isSigned}%
\>[94]\AgdaBound{msg₂}\AgdaSpace{}%
\AgdaBound{sig1}\AgdaSpace{}%
\AgdaBound{pbk2}\AgdaSymbol{)}\<%
\\
\>[0]\AgdaFunction{lemmaHoareTripleStackGeAux'10}\AgdaSpace{}%
\AgdaBound{msg₂}\AgdaSpace{}%
\AgdaBound{pbk1}\AgdaSpace{}%
\AgdaBound{pbk2}\AgdaSpace{}%
\AgdaBound{pbk3}\AgdaSpace{}%
\AgdaBound{pbk4}\AgdaSpace{}%
\AgdaBound{sig1}\AgdaSpace{}%
\AgdaBound{sig2}\AgdaSpace{}%
\AgdaBound{x}\AgdaSpace{}%
\AgdaBound{x₁}\AgdaSpace{}%
\AgdaSymbol{|}\AgdaSpace{}%
\AgdaInductiveConstructor{false}\AgdaSpace{}%
\AgdaSymbol{|}\AgdaSpace{}%
\AgdaInductiveConstructor{true}\AgdaSpace{}%
\AgdaKeyword{with}\AgdaSpace{}%
\AgdaSymbol{(}\AgdaFunction{isSigned}%
\>[101]\AgdaBound{msg₂}\AgdaSpace{}%
\AgdaBound{sig2}\AgdaSpace{}%
\AgdaBound{pbk3}\AgdaSymbol{)}\<%
\\
\>[0]\AgdaFunction{lemmaHoareTripleStackGeAux'10}\AgdaSpace{}%
\AgdaBound{msg₂}\AgdaSpace{}%
\AgdaBound{pbk1}\AgdaSpace{}%
\AgdaBound{pbk2}\AgdaSpace{}%
\AgdaBound{pbk3}\AgdaSpace{}%
\AgdaBound{pbk4}\AgdaSpace{}%
\AgdaBound{sig1}\AgdaSpace{}%
\AgdaBound{sig2}\AgdaSpace{}%
\AgdaBound{x}\AgdaSpace{}%
\AgdaBound{x₁}\AgdaSpace{}%
\AgdaSymbol{|}\AgdaSpace{}%
\AgdaInductiveConstructor{false}\AgdaSpace{}%
\AgdaSymbol{|}\AgdaSpace{}%
\AgdaInductiveConstructor{true}\AgdaSpace{}%
\AgdaSymbol{|}\AgdaSpace{}%
\AgdaInductiveConstructor{true}\AgdaSpace{}%
\AgdaSymbol{=}\AgdaSpace{}%
\AgdaInductiveConstructor{tt}\<%
\\
\>[0]\AgdaFunction{lemmaHoareTripleStackGeAux'10}\AgdaSpace{}%
\AgdaBound{msg₂}\AgdaSpace{}%
\AgdaBound{pbk1}\AgdaSpace{}%
\AgdaBound{pbk2}\AgdaSpace{}%
\AgdaBound{pbk3}\AgdaSpace{}%
\AgdaBound{pbk4}\AgdaSpace{}%
\AgdaBound{sig1}\AgdaSpace{}%
\AgdaBound{sig2}\AgdaSpace{}%
\AgdaBound{x}\AgdaSpace{}%
\AgdaBound{x₁}\AgdaSpace{}%
\AgdaSymbol{|}\AgdaSpace{}%
\AgdaInductiveConstructor{true}\AgdaSpace{}%
\AgdaKeyword{with}%
\>[83]\AgdaSymbol{(}\AgdaFunction{isSigned}%
\>[94]\AgdaBound{msg₂}\AgdaSpace{}%
\AgdaBound{sig2}\AgdaSpace{}%
\AgdaBound{pbk2}\AgdaSymbol{)}\<%
\\
\>[0]\AgdaFunction{lemmaHoareTripleStackGeAux'10}\AgdaSpace{}%
\AgdaBound{msg₂}\AgdaSpace{}%
\AgdaBound{pbk1}\AgdaSpace{}%
\AgdaBound{pbk2}\AgdaSpace{}%
\AgdaBound{pbk3}\AgdaSpace{}%
\AgdaBound{pbk4}\AgdaSpace{}%
\AgdaBound{sig1}\AgdaSpace{}%
\AgdaBound{sig2}\AgdaSpace{}%
\AgdaBound{x}\AgdaSpace{}%
\AgdaBound{x₁}\AgdaSpace{}%
\AgdaSymbol{|}\AgdaSpace{}%
\AgdaInductiveConstructor{true}\AgdaSpace{}%
\AgdaSymbol{|}\AgdaSpace{}%
\AgdaInductiveConstructor{false}\AgdaSpace{}%
\AgdaKeyword{with}\AgdaSpace{}%
\AgdaSymbol{(}\AgdaFunction{isSigned}%
\>[101]\AgdaBound{msg₂}\AgdaSpace{}%
\AgdaBound{sig2}\AgdaSpace{}%
\AgdaBound{pbk3}\AgdaSymbol{)}\<%
\\
\>[0]\AgdaFunction{lemmaHoareTripleStackGeAux'10}\AgdaSpace{}%
\AgdaBound{msg₂}\AgdaSpace{}%
\AgdaBound{pbk1}\AgdaSpace{}%
\AgdaBound{pbk2}\AgdaSpace{}%
\AgdaBound{pbk3}\AgdaSpace{}%
\AgdaBound{pbk4}\AgdaSpace{}%
\AgdaBound{sig1}\AgdaSpace{}%
\AgdaBound{sig2}\AgdaSpace{}%
\AgdaBound{x}\AgdaSpace{}%
\AgdaBound{x₁}\AgdaSpace{}%
\AgdaSymbol{|}\AgdaSpace{}%
\AgdaInductiveConstructor{true}\AgdaSpace{}%
\AgdaSymbol{|}\AgdaSpace{}%
\AgdaInductiveConstructor{false}\AgdaSpace{}%
\AgdaSymbol{|}\AgdaSpace{}%
\AgdaInductiveConstructor{true}\AgdaSpace{}%
\AgdaSymbol{=}\AgdaSpace{}%
\AgdaInductiveConstructor{tt}\<%
\\
\>[0]\AgdaFunction{lemmaHoareTripleStackGeAux'10}\AgdaSpace{}%
\AgdaBound{msg₂}\AgdaSpace{}%
\AgdaBound{pbk1}\AgdaSpace{}%
\AgdaBound{pbk2}\AgdaSpace{}%
\AgdaBound{pbk3}\AgdaSpace{}%
\AgdaBound{pbk4}\AgdaSpace{}%
\AgdaBound{sig1}\AgdaSpace{}%
\AgdaBound{sig2}\AgdaSpace{}%
\AgdaBound{x}\AgdaSpace{}%
\AgdaBound{x₁}\AgdaSpace{}%
\AgdaSymbol{|}\AgdaSpace{}%
\AgdaInductiveConstructor{true}\AgdaSpace{}%
\AgdaSymbol{|}\AgdaSpace{}%
\AgdaInductiveConstructor{true}\AgdaSpace{}%
\AgdaSymbol{=}\AgdaSpace{}%
\AgdaInductiveConstructor{tt}\<%
\\
\\[\AgdaEmptyExtraSkip]%
\\[\AgdaEmptyExtraSkip]%
\>[0]\AgdaFunction{lemmaHoareTripleStackGeAux'11}%
\>[1660I]\AgdaSymbol{:}\AgdaSpace{}%
\AgdaSymbol{(}\AgdaBound{msg₂}\AgdaSpace{}%
\AgdaSymbol{:}\AgdaSpace{}%
\AgdaDatatype{Msg}\AgdaSymbol{)}\<%
\\
\>[1660I][@{}l@{\AgdaIndent{0}}]%
\>[31]\AgdaSymbol{(}\AgdaBound{pbk1}\AgdaSpace{}%
\AgdaBound{pbk2}\AgdaSpace{}%
\AgdaBound{pbk3}\AgdaSpace{}%
\AgdaBound{pbk4}\AgdaSpace{}%
\AgdaBound{sig1}\AgdaSpace{}%
\AgdaBound{sig2}\AgdaSpace{}%
\AgdaSymbol{:}\AgdaSpace{}%
\AgdaDatatype{ℕ}\AgdaSymbol{)}\<%
\\
\>[31]\AgdaSymbol{→}\AgdaSpace{}%
\AgdaFunction{True}\AgdaSpace{}%
\AgdaSymbol{(}\AgdaFunction{isSigned}%
\>[49]\AgdaBound{msg₂}\AgdaSpace{}%
\AgdaBound{sig1}\AgdaSpace{}%
\AgdaBound{pbk2}\AgdaSymbol{)}\<%
\\
\>[.][@{}l@{}]\<[1660I]%
\>[30]\AgdaSymbol{→}%
\>[33]\AgdaFunction{True}\AgdaSpace{}%
\AgdaSymbol{(}\AgdaFunction{isSigned}%
\>[49]\AgdaBound{msg₂}\AgdaSpace{}%
\AgdaBound{sig2}\AgdaSpace{}%
\AgdaBound{pbk4}\AgdaSymbol{)}\<%
\\
\>[30][@{}l@{\AgdaIndent{0}}]%
\>[31]\AgdaSymbol{→}\AgdaSpace{}%
\AgdaFunction{True}\AgdaSpace{}%
\AgdaSymbol{(}\AgdaFunction{compareSigsMultiSig}\AgdaSpace{}%
\AgdaBound{msg₂}\AgdaSpace{}%
\AgdaSymbol{(}\AgdaSpace{}%
\AgdaBound{sig1}\AgdaSpace{}%
\AgdaOperator{\AgdaInductiveConstructor{∷}}\AgdaSpace{}%
\AgdaBound{sig2}\AgdaSpace{}%
\AgdaOperator{\AgdaInductiveConstructor{∷}}\AgdaSpace{}%
\AgdaInductiveConstructor{[]}\AgdaSymbol{)}\AgdaSpace{}%
\AgdaSymbol{(}\AgdaBound{pbk1}\AgdaSpace{}%
\AgdaOperator{\AgdaInductiveConstructor{∷}}\AgdaSpace{}%
\AgdaBound{pbk2}\AgdaSpace{}%
\AgdaOperator{\AgdaInductiveConstructor{∷}}\AgdaSpace{}%
\AgdaBound{pbk3}\AgdaSpace{}%
\AgdaOperator{\AgdaInductiveConstructor{∷}}\AgdaSpace{}%
\AgdaBound{pbk4}\AgdaSpace{}%
\AgdaOperator{\AgdaInductiveConstructor{∷}}%
\>[114]\AgdaInductiveConstructor{[]}\AgdaSymbol{))}\<%
\\
\\[\AgdaEmptyExtraSkip]%
\>[0]\AgdaFunction{lemmaHoareTripleStackGeAux'11}\AgdaSpace{}%
\AgdaBound{msg₂}\AgdaSpace{}%
\AgdaBound{pbk1}\AgdaSpace{}%
\AgdaBound{pbk2}\AgdaSpace{}%
\AgdaBound{pbk3}\AgdaSpace{}%
\AgdaBound{pbk4}\AgdaSpace{}%
\AgdaBound{sig1}\AgdaSpace{}%
\AgdaBound{sig2}\AgdaSpace{}%
\AgdaBound{x}\AgdaSpace{}%
\AgdaBound{x₁}\AgdaSpace{}%
\AgdaKeyword{with}%
\>[76]\AgdaSymbol{(}\AgdaFunction{isSigned}%
\>[87]\AgdaBound{msg₂}\AgdaSpace{}%
\AgdaBound{sig1}\AgdaSpace{}%
\AgdaBound{pbk1}\AgdaSymbol{)}\<%
\\
\>[0]\AgdaFunction{lemmaHoareTripleStackGeAux'11}\AgdaSpace{}%
\AgdaBound{msg₂}\AgdaSpace{}%
\AgdaBound{pbk1}\AgdaSpace{}%
\AgdaBound{pbk2}\AgdaSpace{}%
\AgdaBound{pbk3}\AgdaSpace{}%
\AgdaBound{pbk4}\AgdaSpace{}%
\AgdaBound{sig1}\AgdaSpace{}%
\AgdaBound{sig2}\AgdaSpace{}%
\AgdaBound{x}\AgdaSpace{}%
\AgdaBound{x₁}\AgdaSpace{}%
\AgdaSymbol{|}\AgdaSpace{}%
\AgdaInductiveConstructor{false}\AgdaSpace{}%
\AgdaKeyword{with}\AgdaSpace{}%
\AgdaSymbol{(}\AgdaFunction{isSigned}%
\>[94]\AgdaBound{msg₂}\AgdaSpace{}%
\AgdaBound{sig1}\AgdaSpace{}%
\AgdaBound{pbk2}\AgdaSymbol{)}\<%
\\
\>[0]\AgdaFunction{lemmaHoareTripleStackGeAux'11}\AgdaSpace{}%
\AgdaBound{msg₂}\AgdaSpace{}%
\AgdaBound{pbk1}\AgdaSpace{}%
\AgdaBound{pbk2}\AgdaSpace{}%
\AgdaBound{pbk3}\AgdaSpace{}%
\AgdaBound{pbk4}\AgdaSpace{}%
\AgdaBound{sig1}\AgdaSpace{}%
\AgdaBound{sig2}\AgdaSpace{}%
\AgdaBound{x}\AgdaSpace{}%
\AgdaBound{x₁}\AgdaSpace{}%
\AgdaSymbol{|}\AgdaSpace{}%
\AgdaInductiveConstructor{false}\AgdaSpace{}%
\AgdaSymbol{|}\AgdaSpace{}%
\AgdaInductiveConstructor{true}\AgdaSpace{}%
\AgdaKeyword{with}%
\>[92]\AgdaSymbol{(}\AgdaFunction{isSigned}%
\>[103]\AgdaBound{msg₂}\AgdaSpace{}%
\AgdaBound{sig2}\AgdaSpace{}%
\AgdaBound{pbk3}\AgdaSymbol{)}\<%
\\
\>[0]\AgdaFunction{lemmaHoareTripleStackGeAux'11}\AgdaSpace{}%
\AgdaBound{msg₂}\AgdaSpace{}%
\AgdaBound{pbk1}\AgdaSpace{}%
\AgdaBound{pbk2}\AgdaSpace{}%
\AgdaBound{pbk3}\AgdaSpace{}%
\AgdaBound{pbk4}\AgdaSpace{}%
\AgdaBound{sig1}\AgdaSpace{}%
\AgdaBound{sig2}\AgdaSpace{}%
\AgdaBound{x}\AgdaSpace{}%
\AgdaBound{x₁}\AgdaSpace{}%
\AgdaSymbol{|}\AgdaSpace{}%
\AgdaInductiveConstructor{false}\AgdaSpace{}%
\AgdaSymbol{|}\AgdaSpace{}%
\AgdaInductiveConstructor{true}\AgdaSpace{}%
\AgdaSymbol{|}\AgdaSpace{}%
\AgdaInductiveConstructor{false}\AgdaSpace{}%
\AgdaKeyword{with}%
\>[99]\AgdaSymbol{(}\AgdaFunction{isSigned}%
\>[110]\AgdaBound{msg₂}\AgdaSpace{}%
\AgdaBound{sig2}\AgdaSpace{}%
\AgdaBound{pbk4}\AgdaSymbol{)}\<%
\\
\>[0]\AgdaFunction{lemmaHoareTripleStackGeAux'11}\AgdaSpace{}%
\AgdaBound{msg₂}\AgdaSpace{}%
\AgdaBound{pbk1}\AgdaSpace{}%
\AgdaBound{pbk2}\AgdaSpace{}%
\AgdaBound{pbk3}\AgdaSpace{}%
\AgdaBound{pbk4}\AgdaSpace{}%
\AgdaBound{sig1}\AgdaSpace{}%
\AgdaBound{sig2}\AgdaSpace{}%
\AgdaBound{x}\AgdaSpace{}%
\AgdaBound{x₁}\AgdaSpace{}%
\AgdaSymbol{|}\AgdaSpace{}%
\AgdaInductiveConstructor{false}\AgdaSpace{}%
\AgdaSymbol{|}\AgdaSpace{}%
\AgdaInductiveConstructor{true}\AgdaSpace{}%
\AgdaSymbol{|}\AgdaSpace{}%
\AgdaInductiveConstructor{false}\AgdaSpace{}%
\AgdaSymbol{|}\AgdaSpace{}%
\AgdaInductiveConstructor{true}\AgdaSpace{}%
\AgdaSymbol{=}\AgdaSpace{}%
\AgdaInductiveConstructor{tt}\<%
\\
\>[0]\AgdaFunction{lemmaHoareTripleStackGeAux'11}\AgdaSpace{}%
\AgdaBound{msg₂}\AgdaSpace{}%
\AgdaBound{pbk1}\AgdaSpace{}%
\AgdaBound{pbk2}\AgdaSpace{}%
\AgdaBound{pbk3}\AgdaSpace{}%
\AgdaBound{pbk4}\AgdaSpace{}%
\AgdaBound{sig1}\AgdaSpace{}%
\AgdaBound{sig2}\AgdaSpace{}%
\AgdaBound{x}\AgdaSpace{}%
\AgdaBound{x₁}\AgdaSpace{}%
\AgdaSymbol{|}\AgdaSpace{}%
\AgdaInductiveConstructor{false}\AgdaSpace{}%
\AgdaSymbol{|}\AgdaSpace{}%
\AgdaInductiveConstructor{true}\AgdaSpace{}%
\AgdaSymbol{|}\AgdaSpace{}%
\AgdaInductiveConstructor{true}\AgdaSpace{}%
\AgdaSymbol{=}\AgdaSpace{}%
\AgdaInductiveConstructor{tt}\<%
\\
\>[0]\AgdaFunction{lemmaHoareTripleStackGeAux'11}\AgdaSpace{}%
\AgdaBound{msg₂}\AgdaSpace{}%
\AgdaBound{pbk1}\AgdaSpace{}%
\AgdaBound{pbk2}\AgdaSpace{}%
\AgdaBound{pbk3}\AgdaSpace{}%
\AgdaBound{pbk4}\AgdaSpace{}%
\AgdaBound{sig1}\AgdaSpace{}%
\AgdaBound{sig2}\AgdaSpace{}%
\AgdaBound{x}\AgdaSpace{}%
\AgdaBound{x₁}\AgdaSpace{}%
\AgdaSymbol{|}\AgdaSpace{}%
\AgdaInductiveConstructor{true}\AgdaSpace{}%
\AgdaKeyword{with}\AgdaSpace{}%
\AgdaSymbol{(}\AgdaFunction{isSigned}%
\>[93]\AgdaBound{msg₂}\AgdaSpace{}%
\AgdaBound{sig2}\AgdaSpace{}%
\AgdaBound{pbk2}\AgdaSymbol{)}\<%
\\
\>[0]\AgdaFunction{lemmaHoareTripleStackGeAux'11}\AgdaSpace{}%
\AgdaBound{msg₂}\AgdaSpace{}%
\AgdaBound{pbk1}\AgdaSpace{}%
\AgdaBound{pbk2}\AgdaSpace{}%
\AgdaBound{pbk3}\AgdaSpace{}%
\AgdaBound{pbk4}\AgdaSpace{}%
\AgdaBound{sig1}\AgdaSpace{}%
\AgdaBound{sig2}\AgdaSpace{}%
\AgdaBound{x}\AgdaSpace{}%
\AgdaBound{x₁}\AgdaSpace{}%
\AgdaSymbol{|}\AgdaSpace{}%
\AgdaInductiveConstructor{true}\AgdaSpace{}%
\AgdaSymbol{|}\AgdaSpace{}%
\AgdaInductiveConstructor{false}\AgdaSpace{}%
\AgdaKeyword{with}\AgdaSpace{}%
\AgdaSymbol{(}\AgdaFunction{isSigned}%
\>[101]\AgdaBound{msg₂}\AgdaSpace{}%
\AgdaBound{sig2}\AgdaSpace{}%
\AgdaBound{pbk3}\AgdaSymbol{)}\<%
\\
\>[0]\AgdaFunction{lemmaHoareTripleStackGeAux'11}\AgdaSpace{}%
\AgdaBound{msg₂}\AgdaSpace{}%
\AgdaBound{pbk1}\AgdaSpace{}%
\AgdaBound{pbk2}\AgdaSpace{}%
\AgdaBound{pbk3}\AgdaSpace{}%
\AgdaBound{pbk4}\AgdaSpace{}%
\AgdaBound{sig1}\AgdaSpace{}%
\AgdaBound{sig2}\AgdaSpace{}%
\AgdaBound{x}\AgdaSpace{}%
\AgdaBound{x₁}\AgdaSpace{}%
\AgdaSymbol{|}\AgdaSpace{}%
\AgdaInductiveConstructor{true}\AgdaSpace{}%
\AgdaSymbol{|}\AgdaSpace{}%
\AgdaInductiveConstructor{false}\AgdaSpace{}%
\AgdaSymbol{|}\AgdaSpace{}%
\AgdaInductiveConstructor{false}\AgdaSpace{}%
\AgdaKeyword{with}\AgdaSpace{}%
\AgdaSymbol{(}\AgdaFunction{isSigned}%
\>[109]\AgdaBound{msg₂}\AgdaSpace{}%
\AgdaBound{sig2}\AgdaSpace{}%
\AgdaBound{pbk4}\AgdaSymbol{)}\<%
\\
\>[0]\AgdaFunction{lemmaHoareTripleStackGeAux'11}\AgdaSpace{}%
\AgdaBound{msg₂}\AgdaSpace{}%
\AgdaBound{pbk1}\AgdaSpace{}%
\AgdaBound{pbk2}\AgdaSpace{}%
\AgdaBound{pbk3}\AgdaSpace{}%
\AgdaBound{pbk4}\AgdaSpace{}%
\AgdaBound{sig1}\AgdaSpace{}%
\AgdaBound{sig2}\AgdaSpace{}%
\AgdaBound{x}\AgdaSpace{}%
\AgdaBound{x₁}\AgdaSpace{}%
\AgdaSymbol{|}\AgdaSpace{}%
\AgdaInductiveConstructor{true}\AgdaSpace{}%
\AgdaSymbol{|}\AgdaSpace{}%
\AgdaInductiveConstructor{false}\AgdaSpace{}%
\AgdaSymbol{|}\AgdaSpace{}%
\AgdaInductiveConstructor{false}\AgdaSpace{}%
\AgdaSymbol{|}\AgdaSpace{}%
\AgdaInductiveConstructor{true}\AgdaSpace{}%
\AgdaSymbol{=}\AgdaSpace{}%
\AgdaInductiveConstructor{tt}\<%
\\
\>[0]\AgdaFunction{lemmaHoareTripleStackGeAux'11}\AgdaSpace{}%
\AgdaBound{msg₂}\AgdaSpace{}%
\AgdaBound{pbk1}\AgdaSpace{}%
\AgdaBound{pbk2}\AgdaSpace{}%
\AgdaBound{pbk3}\AgdaSpace{}%
\AgdaBound{pbk4}\AgdaSpace{}%
\AgdaBound{sig1}\AgdaSpace{}%
\AgdaBound{sig2}\AgdaSpace{}%
\AgdaBound{x}\AgdaSpace{}%
\AgdaBound{x₁}\AgdaSpace{}%
\AgdaSymbol{|}\AgdaSpace{}%
\AgdaInductiveConstructor{true}\AgdaSpace{}%
\AgdaSymbol{|}\AgdaSpace{}%
\AgdaInductiveConstructor{false}\AgdaSpace{}%
\AgdaSymbol{|}\AgdaSpace{}%
\AgdaInductiveConstructor{true}\AgdaSpace{}%
\AgdaSymbol{=}\AgdaSpace{}%
\AgdaInductiveConstructor{tt}\<%
\\
\>[0]\AgdaFunction{lemmaHoareTripleStackGeAux'11}\AgdaSpace{}%
\AgdaBound{msg₂}\AgdaSpace{}%
\AgdaBound{pbk1}\AgdaSpace{}%
\AgdaBound{pbk2}\AgdaSpace{}%
\AgdaBound{pbk3}\AgdaSpace{}%
\AgdaBound{pbk4}\AgdaSpace{}%
\AgdaBound{sig1}\AgdaSpace{}%
\AgdaBound{sig2}\AgdaSpace{}%
\AgdaBound{x}\AgdaSpace{}%
\AgdaBound{x₁}\AgdaSpace{}%
\AgdaSymbol{|}\AgdaSpace{}%
\AgdaInductiveConstructor{true}\AgdaSpace{}%
\AgdaSymbol{|}\AgdaSpace{}%
\AgdaInductiveConstructor{true}\AgdaSpace{}%
\AgdaSymbol{=}\AgdaSpace{}%
\AgdaInductiveConstructor{tt}\<%
\\
\\[\AgdaEmptyExtraSkip]%
\>[0]\AgdaFunction{lemmaHoareTripleStackGeAux'12}%
\>[1894I]\AgdaSymbol{:}\AgdaSpace{}%
\AgdaSymbol{(}\AgdaBound{msg₂}\AgdaSpace{}%
\AgdaSymbol{:}\AgdaSpace{}%
\AgdaDatatype{Msg}\AgdaSymbol{)}\<%
\\
\>[1894I][@{}l@{\AgdaIndent{0}}]%
\>[31]\AgdaSymbol{(}\AgdaBound{pbk1}\AgdaSpace{}%
\AgdaBound{pbk2}\AgdaSpace{}%
\AgdaBound{pbk3}\AgdaSpace{}%
\AgdaBound{pbk4}\AgdaSpace{}%
\AgdaBound{sig1}\AgdaSpace{}%
\AgdaBound{sig2}\AgdaSpace{}%
\AgdaSymbol{:}\AgdaSpace{}%
\AgdaDatatype{ℕ}\AgdaSymbol{)}\<%
\\
\>[31]\AgdaSymbol{→}\AgdaSpace{}%
\AgdaFunction{True}\AgdaSpace{}%
\AgdaSymbol{(}\AgdaFunction{isSigned}%
\>[49]\AgdaBound{msg₂}\AgdaSpace{}%
\AgdaBound{sig1}\AgdaSpace{}%
\AgdaBound{pbk3}\AgdaSymbol{)}\<%
\\
\>[.][@{}l@{}]\<[1894I]%
\>[30]\AgdaSymbol{→}%
\>[33]\AgdaFunction{True}\AgdaSpace{}%
\AgdaSymbol{(}\AgdaFunction{isSigned}%
\>[49]\AgdaBound{msg₂}\AgdaSpace{}%
\AgdaBound{sig2}\AgdaSpace{}%
\AgdaBound{pbk4}\AgdaSymbol{)}\<%
\\
\>[30][@{}l@{\AgdaIndent{0}}]%
\>[31]\AgdaSymbol{→}\AgdaSpace{}%
\AgdaFunction{True}\AgdaSpace{}%
\AgdaSymbol{(}\AgdaFunction{compareSigsMultiSig}\AgdaSpace{}%
\AgdaBound{msg₂}\AgdaSpace{}%
\AgdaSymbol{(}\AgdaSpace{}%
\AgdaBound{sig1}\AgdaSpace{}%
\AgdaOperator{\AgdaInductiveConstructor{∷}}\AgdaSpace{}%
\AgdaBound{sig2}\AgdaSpace{}%
\AgdaOperator{\AgdaInductiveConstructor{∷}}\AgdaSpace{}%
\AgdaInductiveConstructor{[]}\AgdaSymbol{)}\AgdaSpace{}%
\AgdaSymbol{(}\AgdaBound{pbk1}\AgdaSpace{}%
\AgdaOperator{\AgdaInductiveConstructor{∷}}\AgdaSpace{}%
\AgdaBound{pbk2}\AgdaSpace{}%
\AgdaOperator{\AgdaInductiveConstructor{∷}}\AgdaSpace{}%
\AgdaBound{pbk3}\AgdaSpace{}%
\AgdaOperator{\AgdaInductiveConstructor{∷}}\AgdaSpace{}%
\AgdaBound{pbk4}\AgdaSpace{}%
\AgdaOperator{\AgdaInductiveConstructor{∷}}%
\>[114]\AgdaInductiveConstructor{[]}\AgdaSymbol{))}\<%
\\
\\[\AgdaEmptyExtraSkip]%
\>[0]\AgdaFunction{lemmaHoareTripleStackGeAux'12}\AgdaSpace{}%
\AgdaBound{msg₂}\AgdaSpace{}%
\AgdaBound{pbk1}\AgdaSpace{}%
\AgdaBound{pbk2}\AgdaSpace{}%
\AgdaBound{pbk3}\AgdaSpace{}%
\AgdaBound{pbk4}\AgdaSpace{}%
\AgdaBound{sig1}\AgdaSpace{}%
\AgdaBound{sig2}\AgdaSpace{}%
\AgdaBound{x}\AgdaSpace{}%
\AgdaBound{x₁}\AgdaSpace{}%
\AgdaKeyword{with}%
\>[76]\AgdaSymbol{(}\AgdaFunction{isSigned}%
\>[87]\AgdaBound{msg₂}\AgdaSpace{}%
\AgdaBound{sig1}\AgdaSpace{}%
\AgdaBound{pbk1}\AgdaSymbol{)}\<%
\\
\>[0]\AgdaFunction{lemmaHoareTripleStackGeAux'12}\AgdaSpace{}%
\AgdaBound{msg₂}\AgdaSpace{}%
\AgdaBound{pbk1}\AgdaSpace{}%
\AgdaBound{pbk2}\AgdaSpace{}%
\AgdaBound{pbk3}\AgdaSpace{}%
\AgdaBound{pbk4}\AgdaSpace{}%
\AgdaBound{sig1}\AgdaSpace{}%
\AgdaBound{sig2}\AgdaSpace{}%
\AgdaBound{x}\AgdaSpace{}%
\AgdaBound{x₁}\AgdaSpace{}%
\AgdaSymbol{|}\AgdaSpace{}%
\AgdaInductiveConstructor{false}\AgdaSpace{}%
\AgdaKeyword{with}%
\>[84]\AgdaSymbol{(}\AgdaFunction{isSigned}%
\>[95]\AgdaBound{msg₂}\AgdaSpace{}%
\AgdaBound{sig1}\AgdaSpace{}%
\AgdaBound{pbk2}\AgdaSymbol{)}\<%
\\
\>[0]\AgdaFunction{lemmaHoareTripleStackGeAux'12}\AgdaSpace{}%
\AgdaBound{msg₂}\AgdaSpace{}%
\AgdaBound{pbk1}\AgdaSpace{}%
\AgdaBound{pbk2}\AgdaSpace{}%
\AgdaBound{pbk3}\AgdaSpace{}%
\AgdaBound{pbk4}\AgdaSpace{}%
\AgdaBound{sig1}\AgdaSpace{}%
\AgdaBound{sig2}\AgdaSpace{}%
\AgdaBound{x}\AgdaSpace{}%
\AgdaBound{x₁}\AgdaSpace{}%
\AgdaSymbol{|}\AgdaSpace{}%
\AgdaInductiveConstructor{false}\AgdaSpace{}%
\AgdaSymbol{|}\AgdaSpace{}%
\AgdaInductiveConstructor{false}\AgdaSpace{}%
\AgdaKeyword{with}%
\>[92]\AgdaSymbol{(}\AgdaFunction{isSigned}%
\>[103]\AgdaBound{msg₂}\AgdaSpace{}%
\AgdaBound{sig1}\AgdaSpace{}%
\AgdaBound{pbk3}\AgdaSymbol{)}\<%
\\
\>[0]\AgdaFunction{lemmaHoareTripleStackGeAux'12}\AgdaSpace{}%
\AgdaBound{msg₂}\AgdaSpace{}%
\AgdaBound{pbk1}\AgdaSpace{}%
\AgdaBound{pbk2}\AgdaSpace{}%
\AgdaBound{pbk3}\AgdaSpace{}%
\AgdaBound{pbk4}\AgdaSpace{}%
\AgdaBound{sig1}\AgdaSpace{}%
\AgdaBound{sig2}\AgdaSpace{}%
\AgdaBound{x}\AgdaSpace{}%
\AgdaBound{x₁}\AgdaSpace{}%
\AgdaSymbol{|}\AgdaSpace{}%
\AgdaInductiveConstructor{false}\AgdaSpace{}%
\AgdaSymbol{|}\AgdaSpace{}%
\AgdaInductiveConstructor{false}\AgdaSpace{}%
\AgdaSymbol{|}\AgdaSpace{}%
\AgdaInductiveConstructor{true}\AgdaSpace{}%
\AgdaKeyword{with}%
\>[99]\AgdaSymbol{(}\AgdaFunction{isSigned}%
\>[110]\AgdaBound{msg₂}\AgdaSpace{}%
\AgdaBound{sig2}\AgdaSpace{}%
\AgdaBound{pbk4}\AgdaSymbol{)}\<%
\\
\>[0]\AgdaFunction{lemmaHoareTripleStackGeAux'12}\AgdaSpace{}%
\AgdaBound{msg₂}\AgdaSpace{}%
\AgdaBound{pbk1}\AgdaSpace{}%
\AgdaBound{pbk2}\AgdaSpace{}%
\AgdaBound{pbk3}\AgdaSpace{}%
\AgdaBound{pbk4}\AgdaSpace{}%
\AgdaBound{sig1}\AgdaSpace{}%
\AgdaBound{sig2}\AgdaSpace{}%
\AgdaBound{x}\AgdaSpace{}%
\AgdaBound{x₁}\AgdaSpace{}%
\AgdaSymbol{|}\AgdaSpace{}%
\AgdaInductiveConstructor{false}\AgdaSpace{}%
\AgdaSymbol{|}\AgdaSpace{}%
\AgdaInductiveConstructor{false}\AgdaSpace{}%
\AgdaSymbol{|}\AgdaSpace{}%
\AgdaInductiveConstructor{true}\AgdaSpace{}%
\AgdaSymbol{|}\AgdaSpace{}%
\AgdaInductiveConstructor{true}\AgdaSpace{}%
\AgdaSymbol{=}\AgdaSpace{}%
\AgdaInductiveConstructor{tt}\<%
\\
\>[0]\AgdaFunction{lemmaHoareTripleStackGeAux'12}\AgdaSpace{}%
\AgdaBound{msg₂}\AgdaSpace{}%
\AgdaBound{pbk1}\AgdaSpace{}%
\AgdaBound{pbk2}\AgdaSpace{}%
\AgdaBound{pbk3}\AgdaSpace{}%
\AgdaBound{pbk4}\AgdaSpace{}%
\AgdaBound{sig1}\AgdaSpace{}%
\AgdaBound{sig2}\AgdaSpace{}%
\AgdaBound{x}\AgdaSpace{}%
\AgdaBound{x₁}\AgdaSpace{}%
\AgdaSymbol{|}\AgdaSpace{}%
\AgdaInductiveConstructor{false}\AgdaSpace{}%
\AgdaSymbol{|}\AgdaSpace{}%
\AgdaInductiveConstructor{true}\AgdaSpace{}%
\AgdaKeyword{with}\AgdaSpace{}%
\AgdaSymbol{(}\AgdaFunction{isSigned}%
\>[101]\AgdaBound{msg₂}\AgdaSpace{}%
\AgdaBound{sig2}\AgdaSpace{}%
\AgdaBound{pbk3}\AgdaSymbol{)}\<%
\\
\>[0]\AgdaFunction{lemmaHoareTripleStackGeAux'12}\AgdaSpace{}%
\AgdaBound{msg₂}\AgdaSpace{}%
\AgdaBound{pbk1}\AgdaSpace{}%
\AgdaBound{pbk2}\AgdaSpace{}%
\AgdaBound{pbk3}\AgdaSpace{}%
\AgdaBound{pbk4}\AgdaSpace{}%
\AgdaBound{sig1}\AgdaSpace{}%
\AgdaBound{sig2}\AgdaSpace{}%
\AgdaBound{x}\AgdaSpace{}%
\AgdaBound{x₁}\AgdaSpace{}%
\AgdaSymbol{|}\AgdaSpace{}%
\AgdaInductiveConstructor{false}\AgdaSpace{}%
\AgdaSymbol{|}\AgdaSpace{}%
\AgdaInductiveConstructor{true}\AgdaSpace{}%
\AgdaSymbol{|}\AgdaSpace{}%
\AgdaInductiveConstructor{false}\AgdaSpace{}%
\AgdaKeyword{with}%
\>[99]\AgdaSymbol{(}\AgdaFunction{isSigned}%
\>[110]\AgdaBound{msg₂}\AgdaSpace{}%
\AgdaBound{sig2}\AgdaSpace{}%
\AgdaBound{pbk4}\AgdaSymbol{)}\<%
\\
\>[0]\AgdaFunction{lemmaHoareTripleStackGeAux'12}\AgdaSpace{}%
\AgdaBound{msg₂}\AgdaSpace{}%
\AgdaBound{pbk1}\AgdaSpace{}%
\AgdaBound{pbk2}\AgdaSpace{}%
\AgdaBound{pbk3}\AgdaSpace{}%
\AgdaBound{pbk4}\AgdaSpace{}%
\AgdaBound{sig1}\AgdaSpace{}%
\AgdaBound{sig2}\AgdaSpace{}%
\AgdaBound{x}\AgdaSpace{}%
\AgdaBound{x₁}\AgdaSpace{}%
\AgdaSymbol{|}\AgdaSpace{}%
\AgdaInductiveConstructor{false}\AgdaSpace{}%
\AgdaSymbol{|}\AgdaSpace{}%
\AgdaInductiveConstructor{true}\AgdaSpace{}%
\AgdaSymbol{|}\AgdaSpace{}%
\AgdaInductiveConstructor{false}\AgdaSpace{}%
\AgdaSymbol{|}\AgdaSpace{}%
\AgdaInductiveConstructor{true}\AgdaSpace{}%
\AgdaSymbol{=}\AgdaSpace{}%
\AgdaInductiveConstructor{tt}\<%
\\
\>[0]\AgdaFunction{lemmaHoareTripleStackGeAux'12}\AgdaSpace{}%
\AgdaBound{msg₂}\AgdaSpace{}%
\AgdaBound{pbk1}\AgdaSpace{}%
\AgdaBound{pbk2}\AgdaSpace{}%
\AgdaBound{pbk3}\AgdaSpace{}%
\AgdaBound{pbk4}\AgdaSpace{}%
\AgdaBound{sig1}\AgdaSpace{}%
\AgdaBound{sig2}\AgdaSpace{}%
\AgdaBound{x}\AgdaSpace{}%
\AgdaBound{x₁}\AgdaSpace{}%
\AgdaSymbol{|}\AgdaSpace{}%
\AgdaInductiveConstructor{false}\AgdaSpace{}%
\AgdaSymbol{|}\AgdaSpace{}%
\AgdaInductiveConstructor{true}\AgdaSpace{}%
\AgdaSymbol{|}\AgdaSpace{}%
\AgdaInductiveConstructor{true}\AgdaSpace{}%
\AgdaSymbol{=}\AgdaSpace{}%
\AgdaInductiveConstructor{tt}\<%
\\
\>[0]\AgdaFunction{lemmaHoareTripleStackGeAux'12}\AgdaSpace{}%
\AgdaBound{msg₂}\AgdaSpace{}%
\AgdaBound{pbk1}\AgdaSpace{}%
\AgdaBound{pbk2}\AgdaSpace{}%
\AgdaBound{pbk3}\AgdaSpace{}%
\AgdaBound{pbk4}\AgdaSpace{}%
\AgdaBound{sig1}\AgdaSpace{}%
\AgdaBound{sig2}\AgdaSpace{}%
\AgdaBound{x}\AgdaSpace{}%
\AgdaBound{x₁}\AgdaSpace{}%
\AgdaSymbol{|}\AgdaSpace{}%
\AgdaInductiveConstructor{true}\AgdaSpace{}%
\AgdaKeyword{with}%
\>[83]\AgdaSymbol{(}\AgdaFunction{isSigned}%
\>[94]\AgdaBound{msg₂}\AgdaSpace{}%
\AgdaBound{sig2}\AgdaSpace{}%
\AgdaBound{pbk2}\AgdaSymbol{)}\<%
\\
\>[0]\AgdaFunction{lemmaHoareTripleStackGeAux'12}\AgdaSpace{}%
\AgdaBound{msg₂}\AgdaSpace{}%
\AgdaBound{pbk1}\AgdaSpace{}%
\AgdaBound{pbk2}\AgdaSpace{}%
\AgdaBound{pbk3}\AgdaSpace{}%
\AgdaBound{pbk4}\AgdaSpace{}%
\AgdaBound{sig1}\AgdaSpace{}%
\AgdaBound{sig2}\AgdaSpace{}%
\AgdaBound{x}\AgdaSpace{}%
\AgdaBound{x₁}\AgdaSpace{}%
\AgdaSymbol{|}\AgdaSpace{}%
\AgdaInductiveConstructor{true}\AgdaSpace{}%
\AgdaSymbol{|}\AgdaSpace{}%
\AgdaInductiveConstructor{false}\AgdaSpace{}%
\AgdaKeyword{with}\AgdaSpace{}%
\AgdaSymbol{(}\AgdaFunction{isSigned}%
\>[101]\AgdaBound{msg₂}\AgdaSpace{}%
\AgdaBound{sig2}\AgdaSpace{}%
\AgdaBound{pbk3}\AgdaSymbol{)}\<%
\\
\>[0]\AgdaFunction{lemmaHoareTripleStackGeAux'12}\AgdaSpace{}%
\AgdaBound{msg₂}\AgdaSpace{}%
\AgdaBound{pbk1}\AgdaSpace{}%
\AgdaBound{pbk2}\AgdaSpace{}%
\AgdaBound{pbk3}\AgdaSpace{}%
\AgdaBound{pbk4}\AgdaSpace{}%
\AgdaBound{sig1}\AgdaSpace{}%
\AgdaBound{sig2}\AgdaSpace{}%
\AgdaBound{x}\AgdaSpace{}%
\AgdaBound{x₁}\AgdaSpace{}%
\AgdaSymbol{|}\AgdaSpace{}%
\AgdaInductiveConstructor{true}\AgdaSpace{}%
\AgdaSymbol{|}\AgdaSpace{}%
\AgdaInductiveConstructor{false}\AgdaSpace{}%
\AgdaSymbol{|}\AgdaSpace{}%
\AgdaInductiveConstructor{false}\AgdaSpace{}%
\AgdaKeyword{with}%
\>[99]\AgdaSymbol{(}\AgdaFunction{isSigned}%
\>[110]\AgdaBound{msg₂}\AgdaSpace{}%
\AgdaBound{sig2}\AgdaSpace{}%
\AgdaBound{pbk4}\AgdaSymbol{)}\<%
\\
\>[0]\AgdaFunction{lemmaHoareTripleStackGeAux'12}\AgdaSpace{}%
\AgdaBound{msg₂}\AgdaSpace{}%
\AgdaBound{pbk1}\AgdaSpace{}%
\AgdaBound{pbk2}\AgdaSpace{}%
\AgdaBound{pbk3}\AgdaSpace{}%
\AgdaBound{pbk4}\AgdaSpace{}%
\AgdaBound{sig1}\AgdaSpace{}%
\AgdaBound{sig2}\AgdaSpace{}%
\AgdaBound{x}\AgdaSpace{}%
\AgdaBound{x₁}\AgdaSpace{}%
\AgdaSymbol{|}\AgdaSpace{}%
\AgdaInductiveConstructor{true}\AgdaSpace{}%
\AgdaSymbol{|}\AgdaSpace{}%
\AgdaInductiveConstructor{false}\AgdaSpace{}%
\AgdaSymbol{|}\AgdaSpace{}%
\AgdaInductiveConstructor{false}\AgdaSpace{}%
\AgdaSymbol{|}\AgdaSpace{}%
\AgdaInductiveConstructor{true}\AgdaSpace{}%
\AgdaSymbol{=}\AgdaSpace{}%
\AgdaInductiveConstructor{tt}\<%
\\
\>[0]\AgdaFunction{lemmaHoareTripleStackGeAux'12}\AgdaSpace{}%
\AgdaBound{msg₂}\AgdaSpace{}%
\AgdaBound{pbk1}\AgdaSpace{}%
\AgdaBound{pbk2}\AgdaSpace{}%
\AgdaBound{pbk3}\AgdaSpace{}%
\AgdaBound{pbk4}\AgdaSpace{}%
\AgdaBound{sig1}\AgdaSpace{}%
\AgdaBound{sig2}\AgdaSpace{}%
\AgdaBound{x}\AgdaSpace{}%
\AgdaBound{x₁}\AgdaSpace{}%
\AgdaSymbol{|}\AgdaSpace{}%
\AgdaInductiveConstructor{true}\AgdaSpace{}%
\AgdaSymbol{|}\AgdaSpace{}%
\AgdaInductiveConstructor{false}\AgdaSpace{}%
\AgdaSymbol{|}\AgdaSpace{}%
\AgdaInductiveConstructor{true}\AgdaSpace{}%
\AgdaSymbol{=}\AgdaSpace{}%
\AgdaInductiveConstructor{tt}\<%
\\
\>[0]\AgdaFunction{lemmaHoareTripleStackGeAux'12}\AgdaSpace{}%
\AgdaBound{msg₂}\AgdaSpace{}%
\AgdaBound{pbk1}\AgdaSpace{}%
\AgdaBound{pbk2}\AgdaSpace{}%
\AgdaBound{pbk3}\AgdaSpace{}%
\AgdaBound{pbk4}\AgdaSpace{}%
\AgdaBound{sig1}\AgdaSpace{}%
\AgdaBound{sig2}\AgdaSpace{}%
\AgdaBound{x}\AgdaSpace{}%
\AgdaBound{x₁}\AgdaSpace{}%
\AgdaSymbol{|}\AgdaSpace{}%
\AgdaInductiveConstructor{true}\AgdaSpace{}%
\AgdaSymbol{|}\AgdaSpace{}%
\AgdaInductiveConstructor{true}\AgdaSpace{}%
\AgdaSymbol{=}\AgdaSpace{}%
\AgdaInductiveConstructor{tt}\<%
\\
\\[\AgdaEmptyExtraSkip]%
\>[0]\AgdaFunction{lemmaHoareTripleStackGeAux'Comb2-4}\AgdaSpace{}%
\AgdaSymbol{:}\AgdaSpace{}%
\AgdaSymbol{(}\AgdaBound{msg₂}\AgdaSpace{}%
\AgdaSymbol{:}\AgdaSpace{}%
\AgdaDatatype{Msg}\AgdaSymbol{)}\<%
\\
\>[0][@{}l@{\AgdaIndent{0}}]%
\>[9]\AgdaSymbol{(}\AgdaBound{pbk1}\AgdaSpace{}%
\AgdaBound{pbk2}\AgdaSpace{}%
\AgdaBound{pbk3}\AgdaSpace{}%
\AgdaBound{pbk4}\AgdaSpace{}%
\AgdaBound{sig1}\AgdaSpace{}%
\AgdaBound{sig2}\AgdaSpace{}%
\AgdaSymbol{:}\AgdaSpace{}%
\AgdaDatatype{ℕ}\AgdaSymbol{)}\<%
\\
\>[9][@{}l@{\AgdaIndent{0}}]%
\>[10]\AgdaSymbol{→}\AgdaSpace{}%
\AgdaFunction{True}%
\>[2191I]\AgdaSymbol{(}\AgdaFunction{compareSigsMultiSigAux}\AgdaSpace{}%
\AgdaBound{msg₂}\AgdaSpace{}%
\AgdaSymbol{(}\AgdaBound{sig2}\AgdaSpace{}%
\AgdaOperator{\AgdaInductiveConstructor{∷}}\AgdaSpace{}%
\AgdaInductiveConstructor{[]}\AgdaSymbol{)}\AgdaSpace{}%
\AgdaSymbol{(}\AgdaBound{pbk2}\AgdaSpace{}%
\AgdaOperator{\AgdaInductiveConstructor{∷}}\AgdaSpace{}%
\AgdaBound{pbk3}\AgdaSpace{}%
\AgdaOperator{\AgdaInductiveConstructor{∷}}%
\>[74]\AgdaBound{pbk4}\AgdaSpace{}%
\AgdaOperator{\AgdaInductiveConstructor{∷}}\AgdaSpace{}%
\AgdaInductiveConstructor{[]}\AgdaSymbol{)}%
\>[86]\AgdaBound{sig1}\<%
\\
\>[.][@{}l@{}]\<[2191I]%
\>[17]\AgdaSymbol{(}\AgdaFunction{isSigned}%
\>[28]\AgdaBound{msg₂}\AgdaSpace{}%
\AgdaBound{sig1}\AgdaSpace{}%
\AgdaBound{pbk1}\AgdaSpace{}%
\AgdaSymbol{))}\<%
\\
\>[10]\AgdaSymbol{→}%
\>[13]\AgdaSymbol{(}\AgdaFunction{True}\AgdaSpace{}%
\AgdaSymbol{(}\AgdaFunction{isSigned}%
\>[30]\AgdaBound{msg₂}\AgdaSpace{}%
\AgdaBound{sig1}\AgdaSpace{}%
\AgdaBound{pbk1}\AgdaSymbol{)}\AgdaSpace{}%
\AgdaOperator{\AgdaRecord{∧}}\AgdaSpace{}%
\AgdaFunction{True}\AgdaSpace{}%
\AgdaSymbol{(}\AgdaFunction{isSigned}%
\>[64]\AgdaBound{msg₂}\AgdaSpace{}%
\AgdaBound{sig2}\AgdaSpace{}%
\AgdaBound{pbk2}\AgdaSymbol{))}\AgdaSpace{}%
\AgdaOperator{\AgdaDatatype{⊎}}\<%
\\
\>[13]\AgdaSymbol{(}\AgdaFunction{True}\AgdaSpace{}%
\AgdaSymbol{(}\AgdaFunction{isSigned}%
\>[30]\AgdaBound{msg₂}\AgdaSpace{}%
\AgdaBound{sig1}\AgdaSpace{}%
\AgdaBound{pbk1}\AgdaSymbol{)}\AgdaSpace{}%
\AgdaOperator{\AgdaRecord{∧}}\AgdaSpace{}%
\AgdaFunction{True}\AgdaSpace{}%
\AgdaSymbol{(}\AgdaFunction{isSigned}%
\>[64]\AgdaBound{msg₂}\AgdaSpace{}%
\AgdaBound{sig2}\AgdaSpace{}%
\AgdaBound{pbk3}\AgdaSymbol{))}\AgdaSpace{}%
\AgdaOperator{\AgdaDatatype{⊎}}\<%
\\
\>[13]\AgdaSymbol{(}\AgdaFunction{True}\AgdaSpace{}%
\AgdaSymbol{(}\AgdaFunction{isSigned}%
\>[30]\AgdaBound{msg₂}\AgdaSpace{}%
\AgdaBound{sig1}\AgdaSpace{}%
\AgdaBound{pbk1}\AgdaSymbol{)}\AgdaSpace{}%
\AgdaOperator{\AgdaRecord{∧}}\AgdaSpace{}%
\AgdaFunction{True}\AgdaSpace{}%
\AgdaSymbol{(}\AgdaFunction{isSigned}%
\>[64]\AgdaBound{msg₂}\AgdaSpace{}%
\AgdaBound{sig2}\AgdaSpace{}%
\AgdaBound{pbk4}\AgdaSymbol{))}\AgdaSpace{}%
\AgdaOperator{\AgdaDatatype{⊎}}\<%
\\
\>[13]\AgdaSymbol{(}\AgdaFunction{True}\AgdaSpace{}%
\AgdaSymbol{(}\AgdaFunction{isSigned}%
\>[30]\AgdaBound{msg₂}\AgdaSpace{}%
\AgdaBound{sig1}\AgdaSpace{}%
\AgdaBound{pbk2}\AgdaSymbol{)}\AgdaSpace{}%
\AgdaOperator{\AgdaRecord{∧}}\AgdaSpace{}%
\AgdaFunction{True}\AgdaSpace{}%
\AgdaSymbol{(}\AgdaFunction{isSigned}%
\>[64]\AgdaBound{msg₂}\AgdaSpace{}%
\AgdaBound{sig2}\AgdaSpace{}%
\AgdaBound{pbk3}\AgdaSymbol{))}\AgdaSpace{}%
\AgdaOperator{\AgdaDatatype{⊎}}\<%
\\
\>[13]\AgdaSymbol{(}\AgdaFunction{True}\AgdaSpace{}%
\AgdaSymbol{(}\AgdaFunction{isSigned}%
\>[30]\AgdaBound{msg₂}\AgdaSpace{}%
\AgdaBound{sig1}\AgdaSpace{}%
\AgdaBound{pbk2}\AgdaSymbol{)}\AgdaSpace{}%
\AgdaOperator{\AgdaRecord{∧}}\AgdaSpace{}%
\AgdaFunction{True}\AgdaSpace{}%
\AgdaSymbol{(}\AgdaFunction{isSigned}%
\>[64]\AgdaBound{msg₂}\AgdaSpace{}%
\AgdaBound{sig2}\AgdaSpace{}%
\AgdaBound{pbk4}\AgdaSymbol{))}\AgdaSpace{}%
\AgdaOperator{\AgdaDatatype{⊎}}\<%
\\
\>[13]\AgdaSymbol{(}\AgdaFunction{True}\AgdaSpace{}%
\AgdaSymbol{(}\AgdaFunction{isSigned}%
\>[30]\AgdaBound{msg₂}\AgdaSpace{}%
\AgdaBound{sig1}\AgdaSpace{}%
\AgdaBound{pbk3}\AgdaSymbol{)}\AgdaSpace{}%
\AgdaOperator{\AgdaRecord{∧}}\AgdaSpace{}%
\AgdaFunction{True}\AgdaSpace{}%
\AgdaSymbol{(}\AgdaFunction{isSigned}%
\>[64]\AgdaBound{msg₂}\AgdaSpace{}%
\AgdaBound{sig2}\AgdaSpace{}%
\AgdaBound{pbk4}\AgdaSymbol{))}\<%
\\
\>[0]\AgdaFunction{lemmaHoareTripleStackGeAux'Comb2-4}\AgdaSpace{}%
\AgdaBound{msg₂}\AgdaSpace{}%
\AgdaBound{pbk1}\AgdaSpace{}%
\AgdaBound{pbk2}\AgdaSpace{}%
\AgdaBound{pbk3}\AgdaSpace{}%
\AgdaBound{pbk4}\AgdaSpace{}%
\AgdaBound{sig1}\AgdaSpace{}%
\AgdaBound{sig2}\AgdaSpace{}%
\AgdaBound{x}\AgdaSpace{}%
\AgdaKeyword{with}\AgdaSpace{}%
\AgdaSymbol{(}\AgdaFunction{isSigned}%
\>[88]\AgdaBound{msg₂}\AgdaSpace{}%
\AgdaBound{sig1}\AgdaSpace{}%
\AgdaBound{pbk1}\AgdaSymbol{)}\<%
\\
\>[0]\AgdaFunction{lemmaHoareTripleStackGeAux'Comb2-4}\AgdaSpace{}%
\AgdaBound{msg₂}\AgdaSpace{}%
\AgdaBound{pbk1}\AgdaSpace{}%
\AgdaBound{pbk2}\AgdaSpace{}%
\AgdaBound{pbk3}\AgdaSpace{}%
\AgdaBound{pbk4}\AgdaSpace{}%
\AgdaBound{sig1}\AgdaSpace{}%
\AgdaBound{sig2}\AgdaSpace{}%
\AgdaBound{x}\AgdaSpace{}%
\AgdaSymbol{|}\AgdaSpace{}%
\AgdaInductiveConstructor{false}\AgdaSpace{}%
\AgdaKeyword{with}\AgdaSpace{}%
\AgdaSymbol{(}\AgdaFunction{isSigned}%
\>[96]\AgdaBound{msg₂}\AgdaSpace{}%
\AgdaBound{sig1}\AgdaSpace{}%
\AgdaBound{pbk2}\AgdaSymbol{)}\<%
\\
\>[0]\AgdaFunction{lemmaHoareTripleStackGeAux'Comb2-4}\AgdaSpace{}%
\AgdaBound{msg₂}\AgdaSpace{}%
\AgdaBound{pbk1}\AgdaSpace{}%
\AgdaBound{pbk2}\AgdaSpace{}%
\AgdaBound{pbk3}\AgdaSpace{}%
\AgdaBound{pbk4}\AgdaSpace{}%
\AgdaBound{sig1}\AgdaSpace{}%
\AgdaBound{sig2}\AgdaSpace{}%
\AgdaBound{x}\AgdaSpace{}%
\AgdaSymbol{|}\AgdaSpace{}%
\AgdaInductiveConstructor{false}\AgdaSpace{}%
\AgdaSymbol{|}\AgdaSpace{}%
\AgdaInductiveConstructor{false}\AgdaSpace{}%
\AgdaKeyword{with}\AgdaSpace{}%
\AgdaSymbol{(}\AgdaFunction{isSigned}%
\>[104]\AgdaBound{msg₂}\AgdaSpace{}%
\AgdaBound{sig1}\AgdaSpace{}%
\AgdaBound{pbk3}\AgdaSymbol{)}\<%
\\
\>[0]\AgdaFunction{lemmaHoareTripleStackGeAux'Comb2-4}\AgdaSpace{}%
\AgdaBound{msg₂}\AgdaSpace{}%
\AgdaBound{pbk1}\AgdaSpace{}%
\AgdaBound{pbk2}\AgdaSpace{}%
\AgdaBound{pbk3}\AgdaSpace{}%
\AgdaBound{pbk4}\AgdaSpace{}%
\AgdaBound{sig1}\AgdaSpace{}%
\AgdaBound{sig2}\AgdaSpace{}%
\AgdaBound{x}\AgdaSpace{}%
\AgdaSymbol{|}\AgdaSpace{}%
\AgdaInductiveConstructor{false}\AgdaSpace{}%
\AgdaSymbol{|}\AgdaSpace{}%
\AgdaInductiveConstructor{false}\AgdaSpace{}%
\AgdaSymbol{|}\AgdaSpace{}%
\AgdaInductiveConstructor{false}\AgdaSpace{}%
\AgdaKeyword{with}\AgdaSpace{}%
\AgdaSymbol{(}\AgdaFunction{isSigned}%
\>[112]\AgdaBound{msg₂}\AgdaSpace{}%
\AgdaBound{sig1}\AgdaSpace{}%
\AgdaBound{pbk4}\AgdaSymbol{)}\<%
\\
\>[0]\AgdaFunction{lemmaHoareTripleStackGeAux'Comb2-4}\AgdaSpace{}%
\AgdaBound{msg₂}\AgdaSpace{}%
\AgdaBound{pbk1}\AgdaSpace{}%
\AgdaBound{pbk2}\AgdaSpace{}%
\AgdaBound{pbk3}\AgdaSpace{}%
\AgdaBound{pbk4}\AgdaSpace{}%
\AgdaBound{sig1}\AgdaSpace{}%
\AgdaBound{sig2}\AgdaSpace{}%
\AgdaSymbol{()}\AgdaSpace{}%
\AgdaSymbol{|}\AgdaSpace{}%
\AgdaInductiveConstructor{false}\AgdaSpace{}%
\AgdaSymbol{|}\AgdaSpace{}%
\AgdaInductiveConstructor{false}\AgdaSpace{}%
\AgdaSymbol{|}\AgdaSpace{}%
\AgdaInductiveConstructor{false}\AgdaSpace{}%
\AgdaSymbol{|}\AgdaSpace{}%
\AgdaInductiveConstructor{false}\<%
\\
\>[0]\AgdaFunction{lemmaHoareTripleStackGeAux'Comb2-4}\AgdaSpace{}%
\AgdaBound{msg₂}\AgdaSpace{}%
\AgdaBound{pbk1}\AgdaSpace{}%
\AgdaBound{pbk2}\AgdaSpace{}%
\AgdaBound{pbk3}\AgdaSpace{}%
\AgdaBound{pbk4}\AgdaSpace{}%
\AgdaBound{sig1}\AgdaSpace{}%
\AgdaBound{sig2}\AgdaSpace{}%
\AgdaSymbol{()}\AgdaSpace{}%
\AgdaSymbol{|}\AgdaSpace{}%
\AgdaInductiveConstructor{false}\AgdaSpace{}%
\AgdaSymbol{|}\AgdaSpace{}%
\AgdaInductiveConstructor{false}\AgdaSpace{}%
\AgdaSymbol{|}\AgdaSpace{}%
\AgdaInductiveConstructor{false}\AgdaSpace{}%
\AgdaSymbol{|}\AgdaSpace{}%
\AgdaInductiveConstructor{true}\<%
\\
\>[0]\AgdaFunction{lemmaHoareTripleStackGeAux'Comb2-4}\AgdaSpace{}%
\AgdaBound{msg₂}\AgdaSpace{}%
\AgdaBound{pbk1}\AgdaSpace{}%
\AgdaBound{pbk2}\AgdaSpace{}%
\AgdaBound{pbk3}\AgdaSpace{}%
\AgdaBound{pbk4}\AgdaSpace{}%
\AgdaBound{sig1}\AgdaSpace{}%
\AgdaBound{sig2}\AgdaSpace{}%
\AgdaBound{x}\AgdaSpace{}%
\AgdaSymbol{|}\AgdaSpace{}%
\AgdaInductiveConstructor{false}\AgdaSpace{}%
\AgdaSymbol{|}\AgdaSpace{}%
\AgdaInductiveConstructor{false}\AgdaSpace{}%
\AgdaSymbol{|}\AgdaSpace{}%
\AgdaInductiveConstructor{true}\AgdaSpace{}%
\AgdaKeyword{with}\AgdaSpace{}%
\AgdaSymbol{(}\AgdaFunction{isSigned}%
\>[111]\AgdaBound{msg₂}\AgdaSpace{}%
\AgdaBound{sig2}\AgdaSpace{}%
\AgdaBound{pbk4}\AgdaSymbol{)}\<%
\\
\>[0]\AgdaFunction{lemmaHoareTripleStackGeAux'Comb2-4}\AgdaSpace{}%
\AgdaBound{msg₂}\AgdaSpace{}%
\AgdaBound{pbk1}\AgdaSpace{}%
\AgdaBound{pbk2}\AgdaSpace{}%
\AgdaBound{pbk3}\AgdaSpace{}%
\AgdaBound{pbk4}\AgdaSpace{}%
\AgdaBound{sig1}\AgdaSpace{}%
\AgdaBound{sig2}\AgdaSpace{}%
\AgdaInductiveConstructor{tt}\AgdaSpace{}%
\AgdaSymbol{|}\AgdaSpace{}%
\AgdaInductiveConstructor{false}\AgdaSpace{}%
\AgdaSymbol{|}\AgdaSpace{}%
\AgdaInductiveConstructor{false}\AgdaSpace{}%
\AgdaSymbol{|}\AgdaSpace{}%
\AgdaInductiveConstructor{true}\AgdaSpace{}%
\AgdaSymbol{|}\AgdaSpace{}%
\AgdaInductiveConstructor{true}\AgdaSpace{}%
\AgdaKeyword{with}%
\>[109]\AgdaSymbol{(}\AgdaFunction{isSigned}%
\>[120]\AgdaBound{msg₂}\AgdaSpace{}%
\AgdaBound{sig2}\AgdaSpace{}%
\AgdaBound{pbk3}\AgdaSymbol{)}\<%
\\
\>[0]\AgdaFunction{lemmaHoareTripleStackGeAux'Comb2-4}\AgdaSpace{}%
\AgdaBound{msg₂}\AgdaSpace{}%
\AgdaBound{pbk1}\AgdaSpace{}%
\AgdaBound{pbk2}\AgdaSpace{}%
\AgdaBound{pbk3}\AgdaSpace{}%
\AgdaBound{pbk4}\AgdaSpace{}%
\AgdaBound{sig1}\AgdaSpace{}%
\AgdaBound{sig2}\AgdaSpace{}%
\AgdaInductiveConstructor{tt}\AgdaSpace{}%
\AgdaSymbol{|}\AgdaSpace{}%
\AgdaInductiveConstructor{false}\AgdaSpace{}%
\AgdaSymbol{|}\AgdaSpace{}%
\AgdaInductiveConstructor{false}\AgdaSpace{}%
\AgdaSymbol{|}\AgdaSpace{}%
\AgdaInductiveConstructor{true}\AgdaSpace{}%
\AgdaSymbol{|}\AgdaSpace{}%
\AgdaInductiveConstructor{true}\AgdaSpace{}%
\AgdaSymbol{|}\AgdaSpace{}%
\AgdaInductiveConstructor{false}\AgdaSpace{}%
\AgdaKeyword{with}\AgdaSpace{}%
\AgdaSymbol{(}\AgdaFunction{isSigned}%
\>[127]\AgdaBound{msg₂}\AgdaSpace{}%
\AgdaBound{sig2}\AgdaSpace{}%
\AgdaBound{pbk2}\AgdaSymbol{)}\<%
\\
\>[0]\AgdaFunction{lemmaHoareTripleStackGeAux'Comb2-4}\AgdaSpace{}%
\AgdaBound{msg₂}\AgdaSpace{}%
\AgdaBound{pbk1}\AgdaSpace{}%
\AgdaBound{pbk2}\AgdaSpace{}%
\AgdaBound{pbk3}\AgdaSpace{}%
\AgdaBound{pbk4}\AgdaSpace{}%
\AgdaBound{sig1}\AgdaSpace{}%
\AgdaBound{sig2}\AgdaSpace{}%
\AgdaInductiveConstructor{tt}\AgdaSpace{}%
\AgdaSymbol{|}\AgdaSpace{}%
\AgdaInductiveConstructor{false}\AgdaSpace{}%
\AgdaSymbol{|}\AgdaSpace{}%
\AgdaInductiveConstructor{false}\AgdaSpace{}%
\AgdaSymbol{|}\AgdaSpace{}%
\AgdaInductiveConstructor{true}\AgdaSpace{}%
\AgdaSymbol{|}\AgdaSpace{}%
\AgdaInductiveConstructor{true}\AgdaSpace{}%
\AgdaSymbol{|}\AgdaSpace{}%
\AgdaInductiveConstructor{false}\AgdaSpace{}%
\AgdaSymbol{|}\AgdaSpace{}%
\AgdaInductiveConstructor{false}\AgdaSpace{}%
\AgdaSymbol{=}\AgdaSpace{}%
\AgdaInductiveConstructor{inj₂}\AgdaSpace{}%
\AgdaSymbol{(}\AgdaInductiveConstructor{inj₂}\AgdaSpace{}%
\AgdaSymbol{(}\AgdaInductiveConstructor{inj₂}\AgdaSpace{}%
\AgdaSymbol{(}\AgdaInductiveConstructor{inj₂}\AgdaSpace{}%
\AgdaSymbol{(}\AgdaInductiveConstructor{inj₂}\AgdaSpace{}%
\AgdaSymbol{(}\AgdaInductiveConstructor{conj}\AgdaSpace{}%
\AgdaInductiveConstructor{tt}\AgdaSpace{}%
\AgdaInductiveConstructor{tt}\AgdaSymbol{)))))}\<%
\\
\>[0]\AgdaFunction{lemmaHoareTripleStackGeAux'Comb2-4}\AgdaSpace{}%
\AgdaBound{msg₂}\AgdaSpace{}%
\AgdaBound{pbk1}\AgdaSpace{}%
\AgdaBound{pbk2}\AgdaSpace{}%
\AgdaBound{pbk3}\AgdaSpace{}%
\AgdaBound{pbk4}\AgdaSpace{}%
\AgdaBound{sig1}\AgdaSpace{}%
\AgdaBound{sig2}\AgdaSpace{}%
\AgdaInductiveConstructor{tt}\AgdaSpace{}%
\AgdaSymbol{|}\AgdaSpace{}%
\AgdaInductiveConstructor{false}\AgdaSpace{}%
\AgdaSymbol{|}\AgdaSpace{}%
\AgdaInductiveConstructor{false}\AgdaSpace{}%
\AgdaSymbol{|}\AgdaSpace{}%
\AgdaInductiveConstructor{true}\AgdaSpace{}%
\AgdaSymbol{|}\AgdaSpace{}%
\AgdaInductiveConstructor{true}\AgdaSpace{}%
\AgdaSymbol{|}\AgdaSpace{}%
\AgdaInductiveConstructor{false}\AgdaSpace{}%
\AgdaSymbol{|}\AgdaSpace{}%
\AgdaInductiveConstructor{true}\AgdaSpace{}%
\AgdaSymbol{=}\AgdaSpace{}%
\AgdaInductiveConstructor{inj₂}\AgdaSpace{}%
\AgdaSymbol{(}\AgdaInductiveConstructor{inj₂}\AgdaSpace{}%
\AgdaSymbol{(}\AgdaInductiveConstructor{inj₂}\AgdaSpace{}%
\AgdaSymbol{(}\AgdaInductiveConstructor{inj₂}\AgdaSpace{}%
\AgdaSymbol{(}\AgdaInductiveConstructor{inj₂}\AgdaSpace{}%
\AgdaSymbol{(}\AgdaInductiveConstructor{conj}\AgdaSpace{}%
\AgdaInductiveConstructor{tt}\AgdaSpace{}%
\AgdaInductiveConstructor{tt}\AgdaSymbol{)))))}\<%
\\
\>[0]\AgdaFunction{lemmaHoareTripleStackGeAux'Comb2-4}\AgdaSpace{}%
\AgdaBound{msg₂}\AgdaSpace{}%
\AgdaBound{pbk1}\AgdaSpace{}%
\AgdaBound{pbk2}\AgdaSpace{}%
\AgdaBound{pbk3}\AgdaSpace{}%
\AgdaBound{pbk4}\AgdaSpace{}%
\AgdaBound{sig1}\AgdaSpace{}%
\AgdaBound{sig2}\AgdaSpace{}%
\AgdaInductiveConstructor{tt}\AgdaSpace{}%
\AgdaSymbol{|}\AgdaSpace{}%
\AgdaInductiveConstructor{false}\AgdaSpace{}%
\AgdaSymbol{|}\AgdaSpace{}%
\AgdaInductiveConstructor{false}\AgdaSpace{}%
\AgdaSymbol{|}\AgdaSpace{}%
\AgdaInductiveConstructor{true}\AgdaSpace{}%
\AgdaSymbol{|}\AgdaSpace{}%
\AgdaInductiveConstructor{true}\AgdaSpace{}%
\AgdaSymbol{|}\AgdaSpace{}%
\AgdaInductiveConstructor{true}\AgdaSpace{}%
\AgdaKeyword{with}\AgdaSpace{}%
\AgdaSymbol{(}\AgdaFunction{isSigned}%
\>[126]\AgdaBound{msg₂}\AgdaSpace{}%
\AgdaBound{sig2}\AgdaSpace{}%
\AgdaBound{pbk2}\AgdaSymbol{)}\<%
\\
\>[0]\AgdaFunction{lemmaHoareTripleStackGeAux'Comb2-4}\AgdaSpace{}%
\AgdaBound{msg₂}\AgdaSpace{}%
\AgdaBound{pbk1}\AgdaSpace{}%
\AgdaBound{pbk2}\AgdaSpace{}%
\AgdaBound{pbk3}\AgdaSpace{}%
\AgdaBound{pbk4}\AgdaSpace{}%
\AgdaBound{sig1}\AgdaSpace{}%
\AgdaBound{sig2}\AgdaSpace{}%
\AgdaInductiveConstructor{tt}\AgdaSpace{}%
\AgdaSymbol{|}\AgdaSpace{}%
\AgdaInductiveConstructor{false}\AgdaSpace{}%
\AgdaSymbol{|}\AgdaSpace{}%
\AgdaInductiveConstructor{false}\AgdaSpace{}%
\AgdaSymbol{|}\AgdaSpace{}%
\AgdaInductiveConstructor{true}\AgdaSpace{}%
\AgdaSymbol{|}\AgdaSpace{}%
\AgdaInductiveConstructor{true}\AgdaSpace{}%
\AgdaSymbol{|}\AgdaSpace{}%
\AgdaInductiveConstructor{true}\AgdaSpace{}%
\AgdaSymbol{|}\AgdaSpace{}%
\AgdaInductiveConstructor{false}\AgdaSpace{}%
\AgdaSymbol{=}\AgdaSpace{}%
\AgdaInductiveConstructor{inj₂}\AgdaSpace{}%
\AgdaSymbol{(}\AgdaInductiveConstructor{inj₂}\AgdaSpace{}%
\AgdaSymbol{(}\AgdaInductiveConstructor{inj₂}\AgdaSpace{}%
\AgdaSymbol{(}\AgdaInductiveConstructor{inj₂}\AgdaSpace{}%
\AgdaSymbol{(}\AgdaInductiveConstructor{inj₂}\AgdaSpace{}%
\AgdaSymbol{(}\AgdaInductiveConstructor{conj}\AgdaSpace{}%
\AgdaInductiveConstructor{tt}\AgdaSpace{}%
\AgdaInductiveConstructor{tt}\AgdaSymbol{)))))}\<%
\\
\>[0]\AgdaFunction{lemmaHoareTripleStackGeAux'Comb2-4}\AgdaSpace{}%
\AgdaBound{msg₂}\AgdaSpace{}%
\AgdaBound{pbk1}\AgdaSpace{}%
\AgdaBound{pbk2}\AgdaSpace{}%
\AgdaBound{pbk3}\AgdaSpace{}%
\AgdaBound{pbk4}\AgdaSpace{}%
\AgdaBound{sig1}\AgdaSpace{}%
\AgdaBound{sig2}\AgdaSpace{}%
\AgdaInductiveConstructor{tt}\AgdaSpace{}%
\AgdaSymbol{|}\AgdaSpace{}%
\AgdaInductiveConstructor{false}\AgdaSpace{}%
\AgdaSymbol{|}\AgdaSpace{}%
\AgdaInductiveConstructor{false}\AgdaSpace{}%
\AgdaSymbol{|}\AgdaSpace{}%
\AgdaInductiveConstructor{true}\AgdaSpace{}%
\AgdaSymbol{|}\AgdaSpace{}%
\AgdaInductiveConstructor{true}\AgdaSpace{}%
\AgdaSymbol{|}\AgdaSpace{}%
\AgdaInductiveConstructor{true}\AgdaSpace{}%
\AgdaSymbol{|}\AgdaSpace{}%
\AgdaInductiveConstructor{true}\AgdaSpace{}%
\AgdaSymbol{=}\AgdaSpace{}%
\AgdaInductiveConstructor{inj₂}\AgdaSpace{}%
\AgdaSymbol{(}\AgdaInductiveConstructor{inj₂}\AgdaSpace{}%
\AgdaSymbol{(}\AgdaInductiveConstructor{inj₂}\AgdaSpace{}%
\AgdaSymbol{(}\AgdaInductiveConstructor{inj₂}\AgdaSpace{}%
\AgdaSymbol{(}\AgdaInductiveConstructor{inj₂}\AgdaSpace{}%
\AgdaSymbol{(}\AgdaInductiveConstructor{conj}\AgdaSpace{}%
\AgdaInductiveConstructor{tt}\AgdaSpace{}%
\AgdaInductiveConstructor{tt}\AgdaSymbol{)))))}\<%
\\
\>[0]\AgdaFunction{lemmaHoareTripleStackGeAux'Comb2-4}\AgdaSpace{}%
\AgdaBound{msg₂}\AgdaSpace{}%
\AgdaBound{pbk1}\AgdaSpace{}%
\AgdaBound{pbk2}\AgdaSpace{}%
\AgdaBound{pbk3}\AgdaSpace{}%
\AgdaBound{pbk4}\AgdaSpace{}%
\AgdaBound{sig1}\AgdaSpace{}%
\AgdaBound{sig2}\AgdaSpace{}%
\AgdaBound{x}\AgdaSpace{}%
\AgdaSymbol{|}\AgdaSpace{}%
\AgdaInductiveConstructor{false}\AgdaSpace{}%
\AgdaSymbol{|}\AgdaSpace{}%
\AgdaInductiveConstructor{true}\AgdaSpace{}%
\AgdaKeyword{with}\AgdaSpace{}%
\AgdaSymbol{(}\AgdaFunction{isSigned}%
\>[103]\AgdaBound{msg₂}\AgdaSpace{}%
\AgdaBound{sig2}\AgdaSpace{}%
\AgdaBound{pbk3}\AgdaSymbol{)}\<%
\\
\>[0]\AgdaFunction{lemmaHoareTripleStackGeAux'Comb2-4}\AgdaSpace{}%
\AgdaBound{msg₂}\AgdaSpace{}%
\AgdaBound{pbk1}\AgdaSpace{}%
\AgdaBound{pbk2}\AgdaSpace{}%
\AgdaBound{pbk3}\AgdaSpace{}%
\AgdaBound{pbk4}\AgdaSpace{}%
\AgdaBound{sig1}\AgdaSpace{}%
\AgdaBound{sig2}\AgdaSpace{}%
\AgdaBound{x}\AgdaSpace{}%
\AgdaSymbol{|}\AgdaSpace{}%
\AgdaInductiveConstructor{false}\AgdaSpace{}%
\AgdaSymbol{|}\AgdaSpace{}%
\AgdaInductiveConstructor{true}\AgdaSpace{}%
\AgdaSymbol{|}\AgdaSpace{}%
\AgdaInductiveConstructor{false}\AgdaSpace{}%
\AgdaKeyword{with}%
\>[101]\AgdaSymbol{(}\AgdaFunction{isSigned}%
\>[112]\AgdaBound{msg₂}\AgdaSpace{}%
\AgdaBound{sig2}\AgdaSpace{}%
\AgdaBound{pbk4}\AgdaSymbol{)}\<%
\\
\>[0]\AgdaFunction{lemmaHoareTripleStackGeAux'Comb2-4}\AgdaSpace{}%
\AgdaBound{msg₂}\AgdaSpace{}%
\AgdaBound{pbk1}\AgdaSpace{}%
\AgdaBound{pbk2}\AgdaSpace{}%
\AgdaBound{pbk3}\AgdaSpace{}%
\AgdaBound{pbk4}\AgdaSpace{}%
\AgdaBound{sig1}\AgdaSpace{}%
\AgdaBound{sig2}\AgdaSpace{}%
\AgdaBound{x}\AgdaSpace{}%
\AgdaSymbol{|}\AgdaSpace{}%
\AgdaInductiveConstructor{false}\AgdaSpace{}%
\AgdaSymbol{|}\AgdaSpace{}%
\AgdaInductiveConstructor{true}\AgdaSpace{}%
\AgdaSymbol{|}\AgdaSpace{}%
\AgdaInductiveConstructor{false}\AgdaSpace{}%
\AgdaSymbol{|}\AgdaSpace{}%
\AgdaInductiveConstructor{true}\AgdaSpace{}%
\AgdaKeyword{with}\AgdaSpace{}%
\AgdaFunction{isSigned}%
\>[117]\AgdaBound{msg₂}\AgdaSpace{}%
\AgdaBound{sig1}\AgdaSpace{}%
\AgdaBound{pbk3}\<%
\\
\>[0]\AgdaFunction{lemmaHoareTripleStackGeAux'Comb2-4}\AgdaSpace{}%
\AgdaBound{msg₂}\AgdaSpace{}%
\AgdaBound{pbk1}\AgdaSpace{}%
\AgdaBound{pbk2}\AgdaSpace{}%
\AgdaBound{pbk3}\AgdaSpace{}%
\AgdaBound{pbk4}\AgdaSpace{}%
\AgdaBound{sig1}\AgdaSpace{}%
\AgdaBound{sig2}\AgdaSpace{}%
\AgdaBound{x}\AgdaSpace{}%
\AgdaSymbol{|}\AgdaSpace{}%
\AgdaInductiveConstructor{false}\AgdaSpace{}%
\AgdaSymbol{|}\AgdaSpace{}%
\AgdaInductiveConstructor{true}\AgdaSpace{}%
\AgdaSymbol{|}\AgdaSpace{}%
\AgdaInductiveConstructor{false}\AgdaSpace{}%
\AgdaSymbol{|}\AgdaSpace{}%
\AgdaInductiveConstructor{true}\AgdaSpace{}%
\AgdaSymbol{|}\AgdaSpace{}%
\AgdaInductiveConstructor{false}\AgdaSpace{}%
\AgdaKeyword{with}\AgdaSpace{}%
\AgdaSymbol{(}\AgdaFunction{isSigned}%
\>[126]\AgdaBound{msg₂}\AgdaSpace{}%
\AgdaBound{sig2}\AgdaSpace{}%
\AgdaBound{pbk2}\AgdaSymbol{)}\<%
\\
\>[0]\AgdaFunction{lemmaHoareTripleStackGeAux'Comb2-4}\AgdaSpace{}%
\AgdaBound{msg₂}\AgdaSpace{}%
\AgdaBound{pbk1}\AgdaSpace{}%
\AgdaBound{pbk2}\AgdaSpace{}%
\AgdaBound{pbk3}\AgdaSpace{}%
\AgdaBound{pbk4}\AgdaSpace{}%
\AgdaBound{sig1}\AgdaSpace{}%
\AgdaBound{sig2}\AgdaSpace{}%
\AgdaBound{x}\AgdaSpace{}%
\AgdaSymbol{|}\AgdaSpace{}%
\AgdaInductiveConstructor{false}\AgdaSpace{}%
\AgdaSymbol{|}\AgdaSpace{}%
\AgdaInductiveConstructor{true}\AgdaSpace{}%
\AgdaSymbol{|}\AgdaSpace{}%
\AgdaInductiveConstructor{false}\AgdaSpace{}%
\AgdaSymbol{|}\AgdaSpace{}%
\AgdaInductiveConstructor{true}\AgdaSpace{}%
\AgdaSymbol{|}\AgdaSpace{}%
\AgdaInductiveConstructor{false}\AgdaSpace{}%
\AgdaSymbol{|}\AgdaSpace{}%
\AgdaInductiveConstructor{false}\AgdaSpace{}%
\AgdaSymbol{=}\AgdaSpace{}%
\AgdaInductiveConstructor{inj₂}\AgdaSpace{}%
\AgdaSymbol{(}\AgdaInductiveConstructor{inj₂}\AgdaSpace{}%
\AgdaSymbol{(}\AgdaInductiveConstructor{inj₂}\AgdaSpace{}%
\AgdaSymbol{(}\AgdaInductiveConstructor{inj₂}\AgdaSpace{}%
\AgdaSymbol{(}\AgdaInductiveConstructor{inj₁}\AgdaSpace{}%
\AgdaSymbol{(}\AgdaInductiveConstructor{conj}\AgdaSpace{}%
\AgdaInductiveConstructor{tt}\AgdaSpace{}%
\AgdaInductiveConstructor{tt}\AgdaSymbol{)))))}\<%
\\
\>[0]\AgdaFunction{lemmaHoareTripleStackGeAux'Comb2-4}\AgdaSpace{}%
\AgdaBound{msg₂}\AgdaSpace{}%
\AgdaBound{pbk1}\AgdaSpace{}%
\AgdaBound{pbk2}\AgdaSpace{}%
\AgdaBound{pbk3}\AgdaSpace{}%
\AgdaBound{pbk4}\AgdaSpace{}%
\AgdaBound{sig1}\AgdaSpace{}%
\AgdaBound{sig2}\AgdaSpace{}%
\AgdaBound{x}\AgdaSpace{}%
\AgdaSymbol{|}\AgdaSpace{}%
\AgdaInductiveConstructor{false}\AgdaSpace{}%
\AgdaSymbol{|}\AgdaSpace{}%
\AgdaInductiveConstructor{true}\AgdaSpace{}%
\AgdaSymbol{|}\AgdaSpace{}%
\AgdaInductiveConstructor{false}\AgdaSpace{}%
\AgdaSymbol{|}\AgdaSpace{}%
\AgdaInductiveConstructor{true}\AgdaSpace{}%
\AgdaSymbol{|}\AgdaSpace{}%
\AgdaInductiveConstructor{false}\AgdaSpace{}%
\AgdaSymbol{|}\AgdaSpace{}%
\AgdaInductiveConstructor{true}\AgdaSpace{}%
\AgdaSymbol{=}\AgdaSpace{}%
\AgdaInductiveConstructor{inj₂}\AgdaSpace{}%
\AgdaSymbol{(}\AgdaInductiveConstructor{inj₂}\AgdaSpace{}%
\AgdaSymbol{(}\AgdaInductiveConstructor{inj₂}\AgdaSpace{}%
\AgdaSymbol{(}\AgdaInductiveConstructor{inj₂}\AgdaSpace{}%
\AgdaSymbol{(}\AgdaInductiveConstructor{inj₁}\AgdaSpace{}%
\AgdaSymbol{(}\AgdaInductiveConstructor{conj}\AgdaSpace{}%
\AgdaInductiveConstructor{tt}\AgdaSpace{}%
\AgdaInductiveConstructor{tt}\AgdaSymbol{)))))}\<%
\\
\>[0]\AgdaFunction{lemmaHoareTripleStackGeAux'Comb2-4}\AgdaSpace{}%
\AgdaBound{msg₂}\AgdaSpace{}%
\AgdaBound{pbk1}\AgdaSpace{}%
\AgdaBound{pbk2}\AgdaSpace{}%
\AgdaBound{pbk3}\AgdaSpace{}%
\AgdaBound{pbk4}\AgdaSpace{}%
\AgdaBound{sig1}\AgdaSpace{}%
\AgdaBound{sig2}\AgdaSpace{}%
\AgdaBound{x}\AgdaSpace{}%
\AgdaSymbol{|}\AgdaSpace{}%
\AgdaInductiveConstructor{false}\AgdaSpace{}%
\AgdaSymbol{|}\AgdaSpace{}%
\AgdaInductiveConstructor{true}\AgdaSpace{}%
\AgdaSymbol{|}\AgdaSpace{}%
\AgdaInductiveConstructor{false}\AgdaSpace{}%
\AgdaSymbol{|}\AgdaSpace{}%
\AgdaInductiveConstructor{true}\AgdaSpace{}%
\AgdaSymbol{|}\AgdaSpace{}%
\AgdaInductiveConstructor{true}\AgdaSpace{}%
\AgdaKeyword{with}\AgdaSpace{}%
\AgdaSymbol{(}\AgdaFunction{isSigned}%
\>[125]\AgdaBound{msg₂}\AgdaSpace{}%
\AgdaBound{sig2}\AgdaSpace{}%
\AgdaBound{pbk2}\AgdaSymbol{)}\<%
\\
\>[0]\AgdaFunction{lemmaHoareTripleStackGeAux'Comb2-4}\AgdaSpace{}%
\AgdaBound{msg₂}\AgdaSpace{}%
\AgdaBound{pbk1}\AgdaSpace{}%
\AgdaBound{pbk2}\AgdaSpace{}%
\AgdaBound{pbk3}\AgdaSpace{}%
\AgdaBound{pbk4}\AgdaSpace{}%
\AgdaBound{sig1}\AgdaSpace{}%
\AgdaBound{sig2}\AgdaSpace{}%
\AgdaBound{x}\AgdaSpace{}%
\AgdaSymbol{|}\AgdaSpace{}%
\AgdaInductiveConstructor{false}\AgdaSpace{}%
\AgdaSymbol{|}\AgdaSpace{}%
\AgdaInductiveConstructor{true}\AgdaSpace{}%
\AgdaSymbol{|}\AgdaSpace{}%
\AgdaInductiveConstructor{false}\AgdaSpace{}%
\AgdaSymbol{|}\AgdaSpace{}%
\AgdaInductiveConstructor{true}\AgdaSpace{}%
\AgdaSymbol{|}\AgdaSpace{}%
\AgdaInductiveConstructor{true}\AgdaSpace{}%
\AgdaSymbol{|}\AgdaSpace{}%
\AgdaInductiveConstructor{false}\AgdaSpace{}%
\AgdaSymbol{=}\AgdaSpace{}%
\AgdaInductiveConstructor{inj₂}\AgdaSpace{}%
\AgdaSymbol{(}\AgdaInductiveConstructor{inj₂}\AgdaSpace{}%
\AgdaSymbol{(}\AgdaInductiveConstructor{inj₂}\AgdaSpace{}%
\AgdaSymbol{(}\AgdaInductiveConstructor{inj₂}\AgdaSpace{}%
\AgdaSymbol{(}\AgdaInductiveConstructor{inj₁}\AgdaSpace{}%
\AgdaSymbol{(}\AgdaInductiveConstructor{conj}\AgdaSpace{}%
\AgdaInductiveConstructor{tt}\AgdaSpace{}%
\AgdaInductiveConstructor{tt}\AgdaSymbol{)))))}\<%
\\
\>[0]\AgdaFunction{lemmaHoareTripleStackGeAux'Comb2-4}\AgdaSpace{}%
\AgdaBound{msg₂}\AgdaSpace{}%
\AgdaBound{pbk1}\AgdaSpace{}%
\AgdaBound{pbk2}\AgdaSpace{}%
\AgdaBound{pbk3}\AgdaSpace{}%
\AgdaBound{pbk4}\AgdaSpace{}%
\AgdaBound{sig1}\AgdaSpace{}%
\AgdaBound{sig2}\AgdaSpace{}%
\AgdaBound{x}\AgdaSpace{}%
\AgdaSymbol{|}\AgdaSpace{}%
\AgdaInductiveConstructor{false}\AgdaSpace{}%
\AgdaSymbol{|}\AgdaSpace{}%
\AgdaInductiveConstructor{true}\AgdaSpace{}%
\AgdaSymbol{|}\AgdaSpace{}%
\AgdaInductiveConstructor{false}\AgdaSpace{}%
\AgdaSymbol{|}\AgdaSpace{}%
\AgdaInductiveConstructor{true}\AgdaSpace{}%
\AgdaSymbol{|}\AgdaSpace{}%
\AgdaInductiveConstructor{true}\AgdaSpace{}%
\AgdaSymbol{|}\AgdaSpace{}%
\AgdaInductiveConstructor{true}\AgdaSpace{}%
\AgdaSymbol{=}\AgdaSpace{}%
\AgdaInductiveConstructor{inj₂}\AgdaSpace{}%
\AgdaSymbol{(}\AgdaInductiveConstructor{inj₂}\AgdaSpace{}%
\AgdaSymbol{(}\AgdaInductiveConstructor{inj₂}\AgdaSpace{}%
\AgdaSymbol{(}\AgdaInductiveConstructor{inj₂}\AgdaSpace{}%
\AgdaSymbol{(}\AgdaInductiveConstructor{inj₁}\AgdaSpace{}%
\AgdaSymbol{(}\AgdaInductiveConstructor{conj}\AgdaSpace{}%
\AgdaInductiveConstructor{tt}\AgdaSpace{}%
\AgdaInductiveConstructor{tt}\AgdaSymbol{)))))}\<%
\\
\>[0]\AgdaFunction{lemmaHoareTripleStackGeAux'Comb2-4}\AgdaSpace{}%
\AgdaBound{msg₂}\AgdaSpace{}%
\AgdaBound{pbk1}\AgdaSpace{}%
\AgdaBound{pbk2}\AgdaSpace{}%
\AgdaBound{pbk3}\AgdaSpace{}%
\AgdaBound{pbk4}\AgdaSpace{}%
\AgdaBound{sig1}\AgdaSpace{}%
\AgdaBound{sig2}\AgdaSpace{}%
\AgdaBound{x}\AgdaSpace{}%
\AgdaSymbol{|}\AgdaSpace{}%
\AgdaInductiveConstructor{false}\AgdaSpace{}%
\AgdaSymbol{|}\AgdaSpace{}%
\AgdaInductiveConstructor{true}\AgdaSpace{}%
\AgdaSymbol{|}\AgdaSpace{}%
\AgdaInductiveConstructor{true}\AgdaSpace{}%
\AgdaKeyword{with}\AgdaSpace{}%
\AgdaSymbol{(}\AgdaFunction{isSigned}%
\>[110]\AgdaBound{msg₂}\AgdaSpace{}%
\AgdaBound{sig2}\AgdaSpace{}%
\AgdaBound{pbk4}\AgdaSymbol{)}\<%
\\
\>[0]\AgdaFunction{lemmaHoareTripleStackGeAux'Comb2-4}\AgdaSpace{}%
\AgdaBound{msg₂}\AgdaSpace{}%
\AgdaBound{pbk1}\AgdaSpace{}%
\AgdaBound{pbk2}\AgdaSpace{}%
\AgdaBound{pbk3}\AgdaSpace{}%
\AgdaBound{pbk4}\AgdaSpace{}%
\AgdaBound{sig1}\AgdaSpace{}%
\AgdaBound{sig2}\AgdaSpace{}%
\AgdaBound{x}\AgdaSpace{}%
\AgdaSymbol{|}\AgdaSpace{}%
\AgdaInductiveConstructor{false}\AgdaSpace{}%
\AgdaSymbol{|}\AgdaSpace{}%
\AgdaInductiveConstructor{true}\AgdaSpace{}%
\AgdaSymbol{|}\AgdaSpace{}%
\AgdaInductiveConstructor{true}\AgdaSpace{}%
\AgdaSymbol{|}\AgdaSpace{}%
\AgdaInductiveConstructor{false}\AgdaSpace{}%
\AgdaKeyword{with}\AgdaSpace{}%
\AgdaSymbol{(}\AgdaFunction{isSigned}%
\>[118]\AgdaBound{msg₂}\AgdaSpace{}%
\AgdaBound{sig1}\AgdaSpace{}%
\AgdaBound{pbk3}\AgdaSymbol{)}\<%
\\
\>[0]\AgdaFunction{lemmaHoareTripleStackGeAux'Comb2-4}\AgdaSpace{}%
\AgdaBound{msg₂}\AgdaSpace{}%
\AgdaBound{pbk1}\AgdaSpace{}%
\AgdaBound{pbk2}\AgdaSpace{}%
\AgdaBound{pbk3}\AgdaSpace{}%
\AgdaBound{pbk4}\AgdaSpace{}%
\AgdaBound{sig1}\AgdaSpace{}%
\AgdaBound{sig2}\AgdaSpace{}%
\AgdaBound{x}\AgdaSpace{}%
\AgdaSymbol{|}\AgdaSpace{}%
\AgdaInductiveConstructor{false}\AgdaSpace{}%
\AgdaSymbol{|}\AgdaSpace{}%
\AgdaInductiveConstructor{true}\AgdaSpace{}%
\AgdaSymbol{|}\AgdaSpace{}%
\AgdaInductiveConstructor{true}\AgdaSpace{}%
\AgdaSymbol{|}\AgdaSpace{}%
\AgdaInductiveConstructor{false}\AgdaSpace{}%
\AgdaSymbol{|}\AgdaSpace{}%
\AgdaInductiveConstructor{false}\AgdaSpace{}%
\AgdaKeyword{with}\AgdaSpace{}%
\AgdaSymbol{(}\AgdaFunction{isSigned}%
\>[126]\AgdaBound{msg₂}\AgdaSpace{}%
\AgdaBound{sig2}\AgdaSpace{}%
\AgdaBound{pbk2}\AgdaSymbol{)}\<%
\\
\>[0]\AgdaFunction{lemmaHoareTripleStackGeAux'Comb2-4}\AgdaSpace{}%
\AgdaBound{msg₂}\AgdaSpace{}%
\AgdaBound{pbk1}\AgdaSpace{}%
\AgdaBound{pbk2}\AgdaSpace{}%
\AgdaBound{pbk3}\AgdaSpace{}%
\AgdaBound{pbk4}\AgdaSpace{}%
\AgdaBound{sig1}\AgdaSpace{}%
\AgdaBound{sig2}\AgdaSpace{}%
\AgdaBound{x}\AgdaSpace{}%
\AgdaSymbol{|}\AgdaSpace{}%
\AgdaInductiveConstructor{false}\AgdaSpace{}%
\AgdaSymbol{|}\AgdaSpace{}%
\AgdaInductiveConstructor{true}\AgdaSpace{}%
\AgdaSymbol{|}\AgdaSpace{}%
\AgdaInductiveConstructor{true}\AgdaSpace{}%
\AgdaSymbol{|}\AgdaSpace{}%
\AgdaInductiveConstructor{false}\AgdaSpace{}%
\AgdaSymbol{|}\AgdaSpace{}%
\AgdaInductiveConstructor{false}\AgdaSpace{}%
\AgdaSymbol{|}\AgdaSpace{}%
\AgdaInductiveConstructor{false}\AgdaSpace{}%
\AgdaSymbol{=}\AgdaSpace{}%
\AgdaInductiveConstructor{inj₂}\AgdaSpace{}%
\AgdaSymbol{(}\AgdaInductiveConstructor{inj₂}\AgdaSpace{}%
\AgdaSymbol{(}\AgdaInductiveConstructor{inj₂}\AgdaSpace{}%
\AgdaSymbol{(}\AgdaInductiveConstructor{inj₁}\AgdaSpace{}%
\AgdaSymbol{(}\AgdaInductiveConstructor{conj}\AgdaSpace{}%
\AgdaInductiveConstructor{tt}\AgdaSpace{}%
\AgdaInductiveConstructor{tt}\AgdaSymbol{))))}\<%
\\
\>[0]\AgdaFunction{lemmaHoareTripleStackGeAux'Comb2-4}\AgdaSpace{}%
\AgdaBound{msg₂}\AgdaSpace{}%
\AgdaBound{pbk1}\AgdaSpace{}%
\AgdaBound{pbk2}\AgdaSpace{}%
\AgdaBound{pbk3}\AgdaSpace{}%
\AgdaBound{pbk4}\AgdaSpace{}%
\AgdaBound{sig1}\AgdaSpace{}%
\AgdaBound{sig2}\AgdaSpace{}%
\AgdaBound{x}\AgdaSpace{}%
\AgdaSymbol{|}\AgdaSpace{}%
\AgdaInductiveConstructor{false}\AgdaSpace{}%
\AgdaSymbol{|}\AgdaSpace{}%
\AgdaInductiveConstructor{true}\AgdaSpace{}%
\AgdaSymbol{|}\AgdaSpace{}%
\AgdaInductiveConstructor{true}\AgdaSpace{}%
\AgdaSymbol{|}\AgdaSpace{}%
\AgdaInductiveConstructor{false}\AgdaSpace{}%
\AgdaSymbol{|}\AgdaSpace{}%
\AgdaInductiveConstructor{false}\AgdaSpace{}%
\AgdaSymbol{|}\AgdaSpace{}%
\AgdaInductiveConstructor{true}\AgdaSpace{}%
\AgdaSymbol{=}\AgdaSpace{}%
\AgdaInductiveConstructor{inj₂}\AgdaSpace{}%
\AgdaSymbol{(}\AgdaInductiveConstructor{inj₂}\AgdaSpace{}%
\AgdaSymbol{(}\AgdaInductiveConstructor{inj₂}\AgdaSpace{}%
\AgdaSymbol{(}\AgdaInductiveConstructor{inj₁}\AgdaSpace{}%
\AgdaSymbol{(}\AgdaInductiveConstructor{conj}\AgdaSpace{}%
\AgdaInductiveConstructor{tt}\AgdaSpace{}%
\AgdaInductiveConstructor{tt}\AgdaSymbol{))))}\<%
\\
\>[0]\AgdaFunction{lemmaHoareTripleStackGeAux'Comb2-4}\AgdaSpace{}%
\AgdaBound{msg₂}\AgdaSpace{}%
\AgdaBound{pbk1}\AgdaSpace{}%
\AgdaBound{pbk2}\AgdaSpace{}%
\AgdaBound{pbk3}\AgdaSpace{}%
\AgdaBound{pbk4}\AgdaSpace{}%
\AgdaBound{sig1}\AgdaSpace{}%
\AgdaBound{sig2}\AgdaSpace{}%
\AgdaBound{x}\AgdaSpace{}%
\AgdaSymbol{|}\AgdaSpace{}%
\AgdaInductiveConstructor{false}\AgdaSpace{}%
\AgdaSymbol{|}\AgdaSpace{}%
\AgdaInductiveConstructor{true}\AgdaSpace{}%
\AgdaSymbol{|}\AgdaSpace{}%
\AgdaInductiveConstructor{true}\AgdaSpace{}%
\AgdaSymbol{|}\AgdaSpace{}%
\AgdaInductiveConstructor{false}\AgdaSpace{}%
\AgdaSymbol{|}\AgdaSpace{}%
\AgdaInductiveConstructor{true}\AgdaSpace{}%
\AgdaSymbol{=}\AgdaSpace{}%
\AgdaInductiveConstructor{inj₂}\AgdaSpace{}%
\AgdaSymbol{(}\AgdaInductiveConstructor{inj₂}\AgdaSpace{}%
\AgdaSymbol{(}\AgdaInductiveConstructor{inj₂}\AgdaSpace{}%
\AgdaSymbol{(}\AgdaInductiveConstructor{inj₁}\AgdaSpace{}%
\AgdaSymbol{(}\AgdaInductiveConstructor{conj}\AgdaSpace{}%
\AgdaInductiveConstructor{tt}\AgdaSpace{}%
\AgdaInductiveConstructor{tt}\AgdaSymbol{))))}\<%
\\
\>[0]\AgdaFunction{lemmaHoareTripleStackGeAux'Comb2-4}\AgdaSpace{}%
\AgdaBound{msg₂}\AgdaSpace{}%
\AgdaBound{pbk1}\AgdaSpace{}%
\AgdaBound{pbk2}\AgdaSpace{}%
\AgdaBound{pbk3}\AgdaSpace{}%
\AgdaBound{pbk4}\AgdaSpace{}%
\AgdaBound{sig1}\AgdaSpace{}%
\AgdaBound{sig2}\AgdaSpace{}%
\AgdaBound{x}\AgdaSpace{}%
\AgdaSymbol{|}\AgdaSpace{}%
\AgdaInductiveConstructor{false}\AgdaSpace{}%
\AgdaSymbol{|}\AgdaSpace{}%
\AgdaInductiveConstructor{true}\AgdaSpace{}%
\AgdaSymbol{|}\AgdaSpace{}%
\AgdaInductiveConstructor{true}\AgdaSpace{}%
\AgdaSymbol{|}\AgdaSpace{}%
\AgdaInductiveConstructor{true}\AgdaSpace{}%
\AgdaSymbol{=}\AgdaSpace{}%
\AgdaInductiveConstructor{inj₂}\AgdaSpace{}%
\AgdaSymbol{(}\AgdaInductiveConstructor{inj₂}\AgdaSpace{}%
\AgdaSymbol{(}\AgdaInductiveConstructor{inj₂}\AgdaSpace{}%
\AgdaSymbol{(}\AgdaInductiveConstructor{inj₁}\AgdaSpace{}%
\AgdaSymbol{(}\AgdaInductiveConstructor{conj}\AgdaSpace{}%
\AgdaInductiveConstructor{tt}\AgdaSpace{}%
\AgdaInductiveConstructor{tt}\AgdaSymbol{))))}\<%
\\
\>[0]\AgdaFunction{lemmaHoareTripleStackGeAux'Comb2-4}\AgdaSpace{}%
\AgdaBound{msg₂}\AgdaSpace{}%
\AgdaBound{pbk1}\AgdaSpace{}%
\AgdaBound{pbk2}\AgdaSpace{}%
\AgdaBound{pbk3}\AgdaSpace{}%
\AgdaBound{pbk4}\AgdaSpace{}%
\AgdaBound{sig1}\AgdaSpace{}%
\AgdaBound{sig2}\AgdaSpace{}%
\AgdaBound{x}\AgdaSpace{}%
\AgdaSymbol{|}\AgdaSpace{}%
\AgdaInductiveConstructor{true}\AgdaSpace{}%
\AgdaKeyword{with}\AgdaSpace{}%
\AgdaSymbol{(}\AgdaFunction{isSigned}%
\>[95]\AgdaBound{msg₂}\AgdaSpace{}%
\AgdaBound{sig2}\AgdaSpace{}%
\AgdaBound{pbk2}\AgdaSymbol{)}\<%
\\
\>[0]\AgdaFunction{lemmaHoareTripleStackGeAux'Comb2-4}\AgdaSpace{}%
\AgdaBound{msg₂}\AgdaSpace{}%
\AgdaBound{pbk1}\AgdaSpace{}%
\AgdaBound{pbk2}\AgdaSpace{}%
\AgdaBound{pbk3}\AgdaSpace{}%
\AgdaBound{pbk4}\AgdaSpace{}%
\AgdaBound{sig1}\AgdaSpace{}%
\AgdaBound{sig2}\AgdaSpace{}%
\AgdaBound{x}\AgdaSpace{}%
\AgdaSymbol{|}\AgdaSpace{}%
\AgdaInductiveConstructor{true}\AgdaSpace{}%
\AgdaSymbol{|}\AgdaSpace{}%
\AgdaInductiveConstructor{false}\AgdaSpace{}%
\AgdaKeyword{with}\AgdaSpace{}%
\AgdaSymbol{(}\AgdaFunction{isSigned}%
\>[103]\AgdaBound{msg₂}\AgdaSpace{}%
\AgdaBound{sig2}\AgdaSpace{}%
\AgdaBound{pbk3}\AgdaSymbol{)}\<%
\\
\>[0]\AgdaFunction{lemmaHoareTripleStackGeAux'Comb2-4}\AgdaSpace{}%
\AgdaBound{msg₂}\AgdaSpace{}%
\AgdaBound{pbk1}\AgdaSpace{}%
\AgdaBound{pbk2}\AgdaSpace{}%
\AgdaBound{pbk3}\AgdaSpace{}%
\AgdaBound{pbk4}\AgdaSpace{}%
\AgdaBound{sig1}\AgdaSpace{}%
\AgdaBound{sig2}\AgdaSpace{}%
\AgdaBound{x}\AgdaSpace{}%
\AgdaSymbol{|}\AgdaSpace{}%
\AgdaInductiveConstructor{true}\AgdaSpace{}%
\AgdaSymbol{|}\AgdaSpace{}%
\AgdaInductiveConstructor{false}\AgdaSpace{}%
\AgdaSymbol{|}\AgdaSpace{}%
\AgdaInductiveConstructor{false}\AgdaSpace{}%
\AgdaKeyword{with}\AgdaSpace{}%
\AgdaSymbol{(}\AgdaFunction{isSigned}%
\>[111]\AgdaBound{msg₂}\AgdaSpace{}%
\AgdaBound{sig2}\AgdaSpace{}%
\AgdaBound{pbk4}\AgdaSymbol{)}\<%
\\
\>[0]\AgdaFunction{lemmaHoareTripleStackGeAux'Comb2-4}\AgdaSpace{}%
\AgdaBound{msg₂}\AgdaSpace{}%
\AgdaBound{pbk1}\AgdaSpace{}%
\AgdaBound{pbk2}\AgdaSpace{}%
\AgdaBound{pbk3}\AgdaSpace{}%
\AgdaBound{pbk4}\AgdaSpace{}%
\AgdaBound{sig1}\AgdaSpace{}%
\AgdaBound{sig2}\AgdaSpace{}%
\AgdaBound{x}\AgdaSpace{}%
\AgdaSymbol{|}\AgdaSpace{}%
\AgdaInductiveConstructor{true}\AgdaSpace{}%
\AgdaSymbol{|}\AgdaSpace{}%
\AgdaInductiveConstructor{false}\AgdaSpace{}%
\AgdaSymbol{|}\AgdaSpace{}%
\AgdaInductiveConstructor{false}\AgdaSpace{}%
\AgdaSymbol{|}\AgdaSpace{}%
\AgdaInductiveConstructor{true}\AgdaSpace{}%
\AgdaSymbol{=}\AgdaSpace{}%
\AgdaInductiveConstructor{inj₂}\AgdaSpace{}%
\AgdaSymbol{(}\AgdaInductiveConstructor{inj₂}\AgdaSpace{}%
\AgdaSymbol{(}\AgdaInductiveConstructor{inj₁}\AgdaSpace{}%
\AgdaSymbol{(}\AgdaInductiveConstructor{conj}\AgdaSpace{}%
\AgdaInductiveConstructor{tt}\AgdaSpace{}%
\AgdaInductiveConstructor{tt}\AgdaSymbol{)))}\<%
\\
\>[0]\AgdaFunction{lemmaHoareTripleStackGeAux'Comb2-4}\AgdaSpace{}%
\AgdaBound{msg₂}\AgdaSpace{}%
\AgdaBound{pbk1}\AgdaSpace{}%
\AgdaBound{pbk2}\AgdaSpace{}%
\AgdaBound{pbk3}\AgdaSpace{}%
\AgdaBound{pbk4}\AgdaSpace{}%
\AgdaBound{sig1}\AgdaSpace{}%
\AgdaBound{sig2}\AgdaSpace{}%
\AgdaBound{x}\AgdaSpace{}%
\AgdaSymbol{|}\AgdaSpace{}%
\AgdaInductiveConstructor{true}\AgdaSpace{}%
\AgdaSymbol{|}\AgdaSpace{}%
\AgdaInductiveConstructor{false}\AgdaSpace{}%
\AgdaSymbol{|}\AgdaSpace{}%
\AgdaInductiveConstructor{true}\AgdaSpace{}%
\AgdaSymbol{=}\AgdaSpace{}%
\AgdaInductiveConstructor{inj₂}\AgdaSpace{}%
\AgdaSymbol{(}\AgdaInductiveConstructor{inj₁}\AgdaSpace{}%
\AgdaSymbol{(}\AgdaInductiveConstructor{conj}\AgdaSpace{}%
\AgdaInductiveConstructor{tt}\AgdaSpace{}%
\AgdaInductiveConstructor{tt}\AgdaSymbol{))}\<%
\\
\>[0]\AgdaFunction{lemmaHoareTripleStackGeAux'Comb2-4}\AgdaSpace{}%
\AgdaBound{msg₂}\AgdaSpace{}%
\AgdaBound{pbk1}\AgdaSpace{}%
\AgdaBound{pbk2}\AgdaSpace{}%
\AgdaBound{pbk3}\AgdaSpace{}%
\AgdaBound{pbk4}\AgdaSpace{}%
\AgdaBound{sig1}\AgdaSpace{}%
\AgdaBound{sig2}\AgdaSpace{}%
\AgdaBound{x}\AgdaSpace{}%
\AgdaSymbol{|}\AgdaSpace{}%
\AgdaInductiveConstructor{true}\AgdaSpace{}%
\AgdaSymbol{|}\AgdaSpace{}%
\AgdaInductiveConstructor{true}\AgdaSpace{}%
\AgdaSymbol{=}\AgdaSpace{}%
\AgdaInductiveConstructor{inj₁}\AgdaSpace{}%
\AgdaSymbol{(}\AgdaInductiveConstructor{conj}\AgdaSpace{}%
\AgdaInductiveConstructor{tt}\AgdaSpace{}%
\AgdaInductiveConstructor{tt}\AgdaSymbol{)}\<%
\\
\\[\AgdaEmptyExtraSkip]%
\\[\AgdaEmptyExtraSkip]%
\\[\AgdaEmptyExtraSkip]%
\\[\AgdaEmptyExtraSkip]%
\\[\AgdaEmptyExtraSkip]%
\>[0]\AgdaFunction{lemmaHoareTripleStackGeAux'14}%
\>[3034I]\AgdaSymbol{:}\AgdaSpace{}%
\AgdaSymbol{(}\AgdaBound{msg}\AgdaSpace{}%
\AgdaSymbol{:}\AgdaSpace{}%
\AgdaDatatype{Msg}\AgdaSymbol{)}\<%
\\
\>[3034I][@{}l@{\AgdaIndent{0}}]%
\>[31]\AgdaSymbol{(}\AgdaBound{pbk1}\AgdaSpace{}%
\AgdaBound{pbk2}\AgdaSpace{}%
\AgdaBound{pbk3}\AgdaSpace{}%
\AgdaBound{pbk4}\AgdaSpace{}%
\AgdaBound{pbk5}\AgdaSpace{}%
\AgdaBound{sig1}\AgdaSpace{}%
\AgdaBound{sig2}\AgdaSpace{}%
\AgdaBound{sig3}\AgdaSpace{}%
\AgdaSymbol{:}\AgdaSpace{}%
\AgdaDatatype{ℕ}\AgdaSymbol{)}\<%
\\
\>[.][@{}l@{}]\<[3034I]%
\>[30]\AgdaSymbol{→}\AgdaSpace{}%
\AgdaFunction{True}\AgdaSpace{}%
\AgdaSymbol{(}\AgdaFunction{isSigned}%
\>[48]\AgdaBound{msg}\AgdaSpace{}%
\AgdaBound{sig1}\AgdaSpace{}%
\AgdaBound{pbk1}\AgdaSymbol{)}\<%
\\
\>[30]\AgdaSymbol{→}\AgdaSpace{}%
\AgdaFunction{True}\AgdaSpace{}%
\AgdaSymbol{(}\AgdaFunction{isSigned}%
\>[48]\AgdaBound{msg}\AgdaSpace{}%
\AgdaBound{sig3}\AgdaSpace{}%
\AgdaBound{pbk3}\AgdaSymbol{)}\<%
\\
\>[30]\AgdaSymbol{→}\AgdaSpace{}%
\AgdaFunction{True}\AgdaSpace{}%
\AgdaSymbol{(}\AgdaFunction{isSigned}%
\>[48]\AgdaBound{msg}\AgdaSpace{}%
\AgdaBound{sig2}\AgdaSpace{}%
\AgdaBound{pbk2}\AgdaSymbol{)}\<%
\\
\>[30]\AgdaSymbol{→}\AgdaSpace{}%
\AgdaFunction{True}\AgdaSpace{}%
\AgdaSymbol{(}\AgdaFunction{compareSigsMultiSig}\AgdaSpace{}%
\AgdaBound{msg}\AgdaSpace{}%
\AgdaSymbol{(}\AgdaSpace{}%
\AgdaBound{sig1}\AgdaSpace{}%
\AgdaOperator{\AgdaInductiveConstructor{∷}}\AgdaSpace{}%
\AgdaBound{sig2}\AgdaSpace{}%
\AgdaOperator{\AgdaInductiveConstructor{∷}}\AgdaSpace{}%
\AgdaBound{sig3}\AgdaSpace{}%
\AgdaOperator{\AgdaInductiveConstructor{∷}}\AgdaSpace{}%
\AgdaInductiveConstructor{[]}\AgdaSymbol{)}\AgdaSpace{}%
\AgdaSymbol{(}\AgdaBound{pbk1}\AgdaSpace{}%
\AgdaOperator{\AgdaInductiveConstructor{∷}}\AgdaSpace{}%
\AgdaBound{pbk2}\AgdaSpace{}%
\AgdaOperator{\AgdaInductiveConstructor{∷}}\AgdaSpace{}%
\AgdaBound{pbk3}\AgdaSpace{}%
\AgdaOperator{\AgdaInductiveConstructor{∷}}\AgdaSpace{}%
\AgdaBound{pbk4}\AgdaSpace{}%
\AgdaOperator{\AgdaInductiveConstructor{∷}}\AgdaSpace{}%
\AgdaBound{pbk5}\AgdaSpace{}%
\AgdaOperator{\AgdaInductiveConstructor{∷}}\AgdaSpace{}%
\AgdaInductiveConstructor{[]}\AgdaSymbol{))}\<%
\\
\>[0]\AgdaFunction{lemmaHoareTripleStackGeAux'14}\AgdaSpace{}%
\AgdaBound{msg}\AgdaSpace{}%
\AgdaBound{pbk1}\AgdaSpace{}%
\AgdaBound{pbk2}\AgdaSpace{}%
\AgdaBound{pbk3}\AgdaSpace{}%
\AgdaBound{pbk4}\AgdaSpace{}%
\AgdaBound{pbk5}\AgdaSpace{}%
\AgdaBound{sig1}\AgdaSpace{}%
\AgdaBound{sig2}\AgdaSpace{}%
\AgdaBound{sig3}\AgdaSpace{}%
\AgdaBound{x}\AgdaSpace{}%
\AgdaBound{x₁}\AgdaSpace{}%
\AgdaBound{x₂}\AgdaSpace{}%
\AgdaKeyword{with}\AgdaSpace{}%
\AgdaSymbol{(}\AgdaFunction{isSigned}%
\>[98]\AgdaBound{msg}\AgdaSpace{}%
\AgdaBound{sig1}\AgdaSpace{}%
\AgdaBound{pbk1}\AgdaSymbol{)}\<%
\\
\>[0]\AgdaFunction{lemmaHoareTripleStackGeAux'14}\AgdaSpace{}%
\AgdaBound{msg}\AgdaSpace{}%
\AgdaBound{pbk1}\AgdaSpace{}%
\AgdaBound{pbk2}\AgdaSpace{}%
\AgdaBound{pbk3}\AgdaSpace{}%
\AgdaBound{pbk4}\AgdaSpace{}%
\AgdaBound{pbk5}\AgdaSpace{}%
\AgdaBound{sig1}\AgdaSpace{}%
\AgdaBound{sig2}\AgdaSpace{}%
\AgdaBound{sig3}\AgdaSpace{}%
\AgdaBound{x}\AgdaSpace{}%
\AgdaBound{x₁}\AgdaSpace{}%
\AgdaBound{x₂}\AgdaSpace{}%
\AgdaSymbol{|}\AgdaSpace{}%
\AgdaInductiveConstructor{true}%
\>[90]\AgdaKeyword{with}\AgdaSpace{}%
\AgdaSymbol{(}\AgdaFunction{isSigned}%
\>[106]\AgdaBound{msg}\AgdaSpace{}%
\AgdaBound{sig2}\AgdaSpace{}%
\AgdaBound{pbk2}\AgdaSymbol{)}\<%
\\
\>[0]\AgdaFunction{lemmaHoareTripleStackGeAux'14}\AgdaSpace{}%
\AgdaBound{msg}\AgdaSpace{}%
\AgdaBound{pbk1}\AgdaSpace{}%
\AgdaBound{pbk2}\AgdaSpace{}%
\AgdaBound{pbk3}\AgdaSpace{}%
\AgdaBound{pbk4}\AgdaSpace{}%
\AgdaBound{pbk5}\AgdaSpace{}%
\AgdaBound{sig1}\AgdaSpace{}%
\AgdaBound{sig2}\AgdaSpace{}%
\AgdaBound{sig3}\AgdaSpace{}%
\AgdaBound{x}\AgdaSpace{}%
\AgdaBound{x₁}\AgdaSpace{}%
\AgdaBound{x₂}\AgdaSpace{}%
\AgdaSymbol{|}\AgdaSpace{}%
\AgdaInductiveConstructor{true}\AgdaSpace{}%
\AgdaSymbol{|}\AgdaSpace{}%
\AgdaInductiveConstructor{true}\AgdaSpace{}%
\AgdaKeyword{with}\AgdaSpace{}%
\AgdaSymbol{(}\AgdaFunction{isSigned}%
\>[112]\AgdaBound{msg}\AgdaSpace{}%
\AgdaBound{sig3}\AgdaSpace{}%
\AgdaBound{pbk3}\AgdaSymbol{)}\<%
\\
\>[0]\AgdaFunction{lemmaHoareTripleStackGeAux'14}\AgdaSpace{}%
\AgdaBound{msg}\AgdaSpace{}%
\AgdaBound{pbk1}\AgdaSpace{}%
\AgdaBound{pbk2}\AgdaSpace{}%
\AgdaBound{pbk3}\AgdaSpace{}%
\AgdaBound{pbk4}\AgdaSpace{}%
\AgdaBound{pbk5}\AgdaSpace{}%
\AgdaBound{sig1}\AgdaSpace{}%
\AgdaBound{sig2}\AgdaSpace{}%
\AgdaBound{sig3}\AgdaSpace{}%
\AgdaBound{x}\AgdaSpace{}%
\AgdaBound{x₁}\AgdaSpace{}%
\AgdaBound{x₂}\AgdaSpace{}%
\AgdaSymbol{|}\AgdaSpace{}%
\AgdaInductiveConstructor{true}\AgdaSpace{}%
\AgdaSymbol{|}\AgdaSpace{}%
\AgdaInductiveConstructor{true}\AgdaSpace{}%
\AgdaSymbol{|}\AgdaSpace{}%
\AgdaInductiveConstructor{true}\AgdaSpace{}%
\AgdaSymbol{=}\AgdaSpace{}%
\AgdaInductiveConstructor{tt}\<%
\\
\\[\AgdaEmptyExtraSkip]%
\\[\AgdaEmptyExtraSkip]%
\>[0]\AgdaFunction{lemmaHoareTripleStackGeAux'15}%
\>[3154I]\AgdaSymbol{:}\AgdaSpace{}%
\AgdaSymbol{(}\AgdaBound{msg}\AgdaSpace{}%
\AgdaSymbol{:}\AgdaSpace{}%
\AgdaDatatype{Msg}\AgdaSymbol{)}\<%
\\
\>[3154I][@{}l@{\AgdaIndent{0}}]%
\>[31]\AgdaSymbol{(}\AgdaBound{pbk1}\AgdaSpace{}%
\AgdaBound{pbk2}\AgdaSpace{}%
\AgdaBound{pbk3}\AgdaSpace{}%
\AgdaBound{pbk4}\AgdaSpace{}%
\AgdaBound{pbk5}\AgdaSpace{}%
\AgdaBound{sig1}\AgdaSpace{}%
\AgdaBound{sig2}\AgdaSpace{}%
\AgdaBound{sig3}\AgdaSpace{}%
\AgdaSymbol{:}\AgdaSpace{}%
\AgdaDatatype{ℕ}\AgdaSymbol{)}\<%
\\
\>[31]\AgdaSymbol{→}\AgdaSpace{}%
\AgdaFunction{True}\AgdaSpace{}%
\AgdaSymbol{(}\AgdaFunction{isSigned}%
\>[49]\AgdaBound{msg}\AgdaSpace{}%
\AgdaBound{sig3}\AgdaSpace{}%
\AgdaBound{pbk4}\AgdaSymbol{)}\<%
\\
\>[.][@{}l@{}]\<[3154I]%
\>[30]\AgdaSymbol{→}%
\>[33]\AgdaFunction{True}\AgdaSpace{}%
\AgdaSymbol{(}\AgdaFunction{isSigned}%
\>[49]\AgdaBound{msg}\AgdaSpace{}%
\AgdaBound{sig2}\AgdaSpace{}%
\AgdaBound{pbk2}\AgdaSymbol{)}\<%
\\
\>[30]\AgdaSymbol{→}%
\>[33]\AgdaFunction{True}\AgdaSpace{}%
\AgdaSymbol{(}\AgdaFunction{isSigned}%
\>[49]\AgdaBound{msg}\AgdaSpace{}%
\AgdaBound{sig1}\AgdaSpace{}%
\AgdaBound{pbk1}\AgdaSymbol{)}\<%
\\
\>[30][@{}l@{\AgdaIndent{0}}]%
\>[31]\AgdaSymbol{→}\AgdaSpace{}%
\AgdaFunction{True}\AgdaSpace{}%
\AgdaSymbol{(}\AgdaFunction{compareSigsMultiSig}\AgdaSpace{}%
\AgdaBound{msg}\AgdaSpace{}%
\AgdaSymbol{(}\AgdaSpace{}%
\AgdaBound{sig1}\AgdaSpace{}%
\AgdaOperator{\AgdaInductiveConstructor{∷}}\AgdaSpace{}%
\AgdaBound{sig2}\AgdaSpace{}%
\AgdaOperator{\AgdaInductiveConstructor{∷}}\AgdaSpace{}%
\AgdaBound{sig3}\AgdaSpace{}%
\AgdaOperator{\AgdaInductiveConstructor{∷}}\AgdaSpace{}%
\AgdaInductiveConstructor{[]}\AgdaSymbol{)}\AgdaSpace{}%
\AgdaSymbol{(}\AgdaBound{pbk1}\AgdaSpace{}%
\AgdaOperator{\AgdaInductiveConstructor{∷}}\AgdaSpace{}%
\AgdaBound{pbk2}\AgdaSpace{}%
\AgdaOperator{\AgdaInductiveConstructor{∷}}\AgdaSpace{}%
\AgdaBound{pbk3}\AgdaSpace{}%
\AgdaOperator{\AgdaInductiveConstructor{∷}}\AgdaSpace{}%
\AgdaBound{pbk4}\AgdaSpace{}%
\AgdaOperator{\AgdaInductiveConstructor{∷}}\AgdaSpace{}%
\AgdaBound{pbk5}\AgdaSpace{}%
\AgdaOperator{\AgdaInductiveConstructor{∷}}\AgdaSpace{}%
\AgdaInductiveConstructor{[]}\AgdaSymbol{))}\<%
\\
\>[0]\AgdaFunction{lemmaHoareTripleStackGeAux'15}\AgdaSpace{}%
\AgdaBound{msg₁}\AgdaSpace{}%
\AgdaBound{pbk1}\AgdaSpace{}%
\AgdaBound{pbk2}\AgdaSpace{}%
\AgdaBound{pbk3}\AgdaSpace{}%
\AgdaBound{pbk4}\AgdaSpace{}%
\AgdaBound{pbk5}\AgdaSpace{}%
\AgdaBound{sig1}\AgdaSpace{}%
\AgdaBound{sig2}\AgdaSpace{}%
\AgdaBound{sig3}\AgdaSpace{}%
\AgdaBound{x}\AgdaSpace{}%
\AgdaBound{x₁}\AgdaSpace{}%
\AgdaBound{x₂}\AgdaSpace{}%
\AgdaKeyword{with}\AgdaSpace{}%
\AgdaSymbol{(}\AgdaFunction{isSigned}%
\>[99]\AgdaBound{msg₁}\AgdaSpace{}%
\AgdaBound{sig1}\AgdaSpace{}%
\AgdaBound{pbk1}\AgdaSymbol{)}\<%
\\
\>[0]\AgdaFunction{lemmaHoareTripleStackGeAux'15}\AgdaSpace{}%
\AgdaBound{msg₁}\AgdaSpace{}%
\AgdaBound{pbk1}\AgdaSpace{}%
\AgdaBound{pbk2}\AgdaSpace{}%
\AgdaBound{pbk3}\AgdaSpace{}%
\AgdaBound{pbk4}\AgdaSpace{}%
\AgdaBound{pbk5}\AgdaSpace{}%
\AgdaBound{sig1}\AgdaSpace{}%
\AgdaBound{sig2}\AgdaSpace{}%
\AgdaBound{sig3}\AgdaSpace{}%
\AgdaBound{x}\AgdaSpace{}%
\AgdaBound{x₁}\AgdaSpace{}%
\AgdaBound{x₂}\AgdaSpace{}%
\AgdaSymbol{|}\AgdaSpace{}%
\AgdaInductiveConstructor{true}\AgdaSpace{}%
\AgdaKeyword{with}\AgdaSpace{}%
\AgdaSymbol{(}\AgdaFunction{isSigned}%
\>[106]\AgdaBound{msg₁}\AgdaSpace{}%
\AgdaBound{sig2}\AgdaSpace{}%
\AgdaBound{pbk2}\AgdaSymbol{)}\<%
\\
\>[0]\AgdaFunction{lemmaHoareTripleStackGeAux'15}\AgdaSpace{}%
\AgdaBound{msg₁}\AgdaSpace{}%
\AgdaBound{pbk1}\AgdaSpace{}%
\AgdaBound{pbk2}\AgdaSpace{}%
\AgdaBound{pbk3}\AgdaSpace{}%
\AgdaBound{pbk4}\AgdaSpace{}%
\AgdaBound{pbk5}\AgdaSpace{}%
\AgdaBound{sig1}\AgdaSpace{}%
\AgdaBound{sig2}\AgdaSpace{}%
\AgdaBound{sig3}\AgdaSpace{}%
\AgdaBound{x}\AgdaSpace{}%
\AgdaBound{x₁}\AgdaSpace{}%
\AgdaBound{x₂}\AgdaSpace{}%
\AgdaSymbol{|}\AgdaSpace{}%
\AgdaInductiveConstructor{true}\AgdaSpace{}%
\AgdaSymbol{|}\AgdaSpace{}%
\AgdaInductiveConstructor{true}\AgdaSpace{}%
\AgdaKeyword{with}\AgdaSpace{}%
\AgdaSymbol{(}\AgdaFunction{isSigned}%
\>[113]\AgdaBound{msg₁}\AgdaSpace{}%
\AgdaBound{sig3}\AgdaSpace{}%
\AgdaBound{pbk3}\AgdaSymbol{)}\<%
\\
\>[0]\AgdaFunction{lemmaHoareTripleStackGeAux'15}\AgdaSpace{}%
\AgdaBound{msg₁}\AgdaSpace{}%
\AgdaBound{pbk1}\AgdaSpace{}%
\AgdaBound{pbk2}\AgdaSpace{}%
\AgdaBound{pbk3}\AgdaSpace{}%
\AgdaBound{pbk4}\AgdaSpace{}%
\AgdaBound{pbk5}\AgdaSpace{}%
\AgdaBound{sig1}\AgdaSpace{}%
\AgdaBound{sig2}\AgdaSpace{}%
\AgdaBound{sig3}\AgdaSpace{}%
\AgdaBound{x}\AgdaSpace{}%
\AgdaBound{x₁}\AgdaSpace{}%
\AgdaBound{x₂}\AgdaSpace{}%
\AgdaSymbol{|}\AgdaSpace{}%
\AgdaInductiveConstructor{true}\AgdaSpace{}%
\AgdaSymbol{|}\AgdaSpace{}%
\AgdaInductiveConstructor{true}\AgdaSpace{}%
\AgdaSymbol{|}\AgdaSpace{}%
\AgdaInductiveConstructor{false}\AgdaSpace{}%
\AgdaKeyword{with}%
\>[111]\AgdaSymbol{(}\AgdaFunction{isSigned}%
\>[122]\AgdaBound{msg₁}\AgdaSpace{}%
\AgdaBound{sig3}\AgdaSpace{}%
\AgdaBound{pbk4}\AgdaSymbol{)}\<%
\\
\>[0]\AgdaFunction{lemmaHoareTripleStackGeAux'15}\AgdaSpace{}%
\AgdaBound{msg₁}\AgdaSpace{}%
\AgdaBound{pbk1}\AgdaSpace{}%
\AgdaBound{pbk2}\AgdaSpace{}%
\AgdaBound{pbk3}\AgdaSpace{}%
\AgdaBound{pbk4}\AgdaSpace{}%
\AgdaBound{pbk5}\AgdaSpace{}%
\AgdaBound{sig1}\AgdaSpace{}%
\AgdaBound{sig2}\AgdaSpace{}%
\AgdaBound{sig3}\AgdaSpace{}%
\AgdaBound{x}\AgdaSpace{}%
\AgdaBound{x₁}\AgdaSpace{}%
\AgdaBound{x₂}\AgdaSpace{}%
\AgdaSymbol{|}\AgdaSpace{}%
\AgdaInductiveConstructor{true}\AgdaSpace{}%
\AgdaSymbol{|}\AgdaSpace{}%
\AgdaInductiveConstructor{true}\AgdaSpace{}%
\AgdaSymbol{|}\AgdaSpace{}%
\AgdaInductiveConstructor{false}\AgdaSpace{}%
\AgdaSymbol{|}\AgdaSpace{}%
\AgdaInductiveConstructor{true}\AgdaSpace{}%
\AgdaSymbol{=}\AgdaSpace{}%
\AgdaInductiveConstructor{tt}\<%
\\
\>[0]\AgdaFunction{lemmaHoareTripleStackGeAux'15}\AgdaSpace{}%
\AgdaBound{msg₁}\AgdaSpace{}%
\AgdaBound{pbk1}\AgdaSpace{}%
\AgdaBound{pbk2}\AgdaSpace{}%
\AgdaBound{pbk3}\AgdaSpace{}%
\AgdaBound{pbk4}\AgdaSpace{}%
\AgdaBound{pbk5}\AgdaSpace{}%
\AgdaBound{sig1}\AgdaSpace{}%
\AgdaBound{sig2}\AgdaSpace{}%
\AgdaBound{sig3}\AgdaSpace{}%
\AgdaBound{x}\AgdaSpace{}%
\AgdaBound{x₁}\AgdaSpace{}%
\AgdaBound{x₂}\AgdaSpace{}%
\AgdaSymbol{|}\AgdaSpace{}%
\AgdaInductiveConstructor{true}\AgdaSpace{}%
\AgdaSymbol{|}\AgdaSpace{}%
\AgdaInductiveConstructor{true}\AgdaSpace{}%
\AgdaSymbol{|}\AgdaSpace{}%
\AgdaInductiveConstructor{true}\AgdaSpace{}%
\AgdaSymbol{=}\AgdaSpace{}%
\AgdaInductiveConstructor{tt}\<%
\\
\\[\AgdaEmptyExtraSkip]%
\\[\AgdaEmptyExtraSkip]%
\>[0]\AgdaFunction{lemmaHoareTripleStackGeAux'16}%
\>[3316I]\AgdaSymbol{:}\AgdaSpace{}%
\AgdaSymbol{(}\AgdaBound{msg}\AgdaSpace{}%
\AgdaSymbol{:}\AgdaSpace{}%
\AgdaDatatype{Msg}\AgdaSymbol{)}\<%
\\
\>[3316I][@{}l@{\AgdaIndent{0}}]%
\>[31]\AgdaSymbol{(}\AgdaBound{pbk1}\AgdaSpace{}%
\AgdaBound{pbk2}\AgdaSpace{}%
\AgdaBound{pbk3}\AgdaSpace{}%
\AgdaBound{pbk4}\AgdaSpace{}%
\AgdaBound{pbk5}\AgdaSpace{}%
\AgdaBound{sig1}\AgdaSpace{}%
\AgdaBound{sig2}\AgdaSpace{}%
\AgdaBound{sig3}\AgdaSpace{}%
\AgdaSymbol{:}\AgdaSpace{}%
\AgdaDatatype{ℕ}\AgdaSymbol{)}\<%
\\
\>[31]\AgdaSymbol{→}\AgdaSpace{}%
\AgdaFunction{True}\AgdaSpace{}%
\AgdaSymbol{(}\AgdaFunction{isSigned}%
\>[49]\AgdaBound{msg}\AgdaSpace{}%
\AgdaBound{sig3}\AgdaSpace{}%
\AgdaBound{pbk5}\AgdaSymbol{)}\<%
\\
\>[.][@{}l@{}]\<[3316I]%
\>[30]\AgdaSymbol{→}%
\>[33]\AgdaFunction{True}\AgdaSpace{}%
\AgdaSymbol{(}\AgdaFunction{isSigned}%
\>[49]\AgdaBound{msg}\AgdaSpace{}%
\AgdaBound{sig2}\AgdaSpace{}%
\AgdaBound{pbk2}\AgdaSymbol{)}\<%
\\
\>[30][@{}l@{\AgdaIndent{0}}]%
\>[31]\AgdaSymbol{→}%
\>[34]\AgdaFunction{True}\AgdaSpace{}%
\AgdaSymbol{(}\AgdaFunction{isSigned}%
\>[50]\AgdaBound{msg}\AgdaSpace{}%
\AgdaBound{sig1}\AgdaSpace{}%
\AgdaBound{pbk1}\AgdaSymbol{)}\<%
\\
\>[31]\AgdaSymbol{→}\AgdaSpace{}%
\AgdaFunction{True}\AgdaSpace{}%
\AgdaSymbol{(}\AgdaFunction{compareSigsMultiSig}\AgdaSpace{}%
\AgdaBound{msg}\AgdaSpace{}%
\AgdaSymbol{(}\AgdaSpace{}%
\AgdaBound{sig1}\AgdaSpace{}%
\AgdaOperator{\AgdaInductiveConstructor{∷}}\AgdaSpace{}%
\AgdaBound{sig2}\AgdaSpace{}%
\AgdaOperator{\AgdaInductiveConstructor{∷}}\AgdaSpace{}%
\AgdaBound{sig3}\AgdaSpace{}%
\AgdaOperator{\AgdaInductiveConstructor{∷}}\AgdaSpace{}%
\AgdaInductiveConstructor{[]}\AgdaSymbol{)}\AgdaSpace{}%
\AgdaSymbol{(}\AgdaBound{pbk1}\AgdaSpace{}%
\AgdaOperator{\AgdaInductiveConstructor{∷}}\AgdaSpace{}%
\AgdaBound{pbk2}\AgdaSpace{}%
\AgdaOperator{\AgdaInductiveConstructor{∷}}\AgdaSpace{}%
\AgdaBound{pbk3}\AgdaSpace{}%
\AgdaOperator{\AgdaInductiveConstructor{∷}}\AgdaSpace{}%
\AgdaBound{pbk4}\AgdaSpace{}%
\AgdaOperator{\AgdaInductiveConstructor{∷}}\AgdaSpace{}%
\AgdaBound{pbk5}\AgdaSpace{}%
\AgdaOperator{\AgdaInductiveConstructor{∷}}\AgdaSpace{}%
\AgdaInductiveConstructor{[]}\AgdaSymbol{))}\<%
\\
\>[0]\AgdaFunction{lemmaHoareTripleStackGeAux'16}\AgdaSpace{}%
\AgdaBound{msg₁}\AgdaSpace{}%
\AgdaBound{pbk1}\AgdaSpace{}%
\AgdaBound{pbk2}\AgdaSpace{}%
\AgdaBound{pbk3}\AgdaSpace{}%
\AgdaBound{pbk4}\AgdaSpace{}%
\AgdaBound{pbk5}\AgdaSpace{}%
\AgdaBound{sig1}\AgdaSpace{}%
\AgdaBound{sig2}\AgdaSpace{}%
\AgdaBound{sig3}\AgdaSpace{}%
\AgdaBound{x}\AgdaSpace{}%
\AgdaBound{x₁}\AgdaSpace{}%
\AgdaBound{x₂}\AgdaSpace{}%
\AgdaKeyword{with}\AgdaSpace{}%
\AgdaSymbol{(}\AgdaFunction{isSigned}%
\>[99]\AgdaBound{msg₁}\AgdaSpace{}%
\AgdaBound{sig1}\AgdaSpace{}%
\AgdaBound{pbk1}\AgdaSymbol{)}\<%
\\
\>[0]\AgdaFunction{lemmaHoareTripleStackGeAux'16}\AgdaSpace{}%
\AgdaBound{msg₁}\AgdaSpace{}%
\AgdaBound{pbk1}\AgdaSpace{}%
\AgdaBound{pbk2}\AgdaSpace{}%
\AgdaBound{pbk3}\AgdaSpace{}%
\AgdaBound{pbk4}\AgdaSpace{}%
\AgdaBound{pbk5}\AgdaSpace{}%
\AgdaBound{sig1}\AgdaSpace{}%
\AgdaBound{sig2}\AgdaSpace{}%
\AgdaBound{sig3}\AgdaSpace{}%
\AgdaBound{x}\AgdaSpace{}%
\AgdaBound{x₁}\AgdaSpace{}%
\AgdaBound{x₂}\AgdaSpace{}%
\AgdaSymbol{|}\AgdaSpace{}%
\AgdaInductiveConstructor{true}\AgdaSpace{}%
\AgdaKeyword{with}\AgdaSpace{}%
\AgdaSymbol{(}\AgdaFunction{isSigned}%
\>[106]\AgdaBound{msg₁}\AgdaSpace{}%
\AgdaBound{sig2}\AgdaSpace{}%
\AgdaBound{pbk2}\AgdaSymbol{)}\<%
\\
\>[0]\AgdaFunction{lemmaHoareTripleStackGeAux'16}\AgdaSpace{}%
\AgdaBound{msg₁}\AgdaSpace{}%
\AgdaBound{pbk1}\AgdaSpace{}%
\AgdaBound{pbk2}\AgdaSpace{}%
\AgdaBound{pbk3}\AgdaSpace{}%
\AgdaBound{pbk4}\AgdaSpace{}%
\AgdaBound{pbk5}\AgdaSpace{}%
\AgdaBound{sig1}\AgdaSpace{}%
\AgdaBound{sig2}\AgdaSpace{}%
\AgdaBound{sig3}\AgdaSpace{}%
\AgdaBound{x}\AgdaSpace{}%
\AgdaBound{x₁}\AgdaSpace{}%
\AgdaBound{x₂}\AgdaSpace{}%
\AgdaSymbol{|}\AgdaSpace{}%
\AgdaInductiveConstructor{true}\AgdaSpace{}%
\AgdaSymbol{|}\AgdaSpace{}%
\AgdaInductiveConstructor{true}\AgdaSpace{}%
\AgdaKeyword{with}%
\>[103]\AgdaSymbol{(}\AgdaFunction{isSigned}%
\>[114]\AgdaBound{msg₁}\AgdaSpace{}%
\AgdaBound{sig3}\AgdaSpace{}%
\AgdaBound{pbk3}\AgdaSymbol{)}\<%
\\
\>[0]\AgdaFunction{lemmaHoareTripleStackGeAux'16}\AgdaSpace{}%
\AgdaBound{msg₁}\AgdaSpace{}%
\AgdaBound{pbk1}\AgdaSpace{}%
\AgdaBound{pbk2}\AgdaSpace{}%
\AgdaBound{pbk3}\AgdaSpace{}%
\AgdaBound{pbk4}\AgdaSpace{}%
\AgdaBound{pbk5}\AgdaSpace{}%
\AgdaBound{sig1}\AgdaSpace{}%
\AgdaBound{sig2}\AgdaSpace{}%
\AgdaBound{sig3}\AgdaSpace{}%
\AgdaBound{x}\AgdaSpace{}%
\AgdaBound{x₁}\AgdaSpace{}%
\AgdaBound{x₂}\AgdaSpace{}%
\AgdaSymbol{|}\AgdaSpace{}%
\AgdaInductiveConstructor{true}\AgdaSpace{}%
\AgdaSymbol{|}\AgdaSpace{}%
\AgdaInductiveConstructor{true}\AgdaSpace{}%
\AgdaSymbol{|}\AgdaSpace{}%
\AgdaInductiveConstructor{false}\AgdaSpace{}%
\AgdaKeyword{with}%
\>[111]\AgdaSymbol{(}\AgdaFunction{isSigned}%
\>[122]\AgdaBound{msg₁}\AgdaSpace{}%
\AgdaBound{sig3}\AgdaSpace{}%
\AgdaBound{pbk4}\AgdaSymbol{)}\<%
\\
\>[0]\AgdaFunction{lemmaHoareTripleStackGeAux'16}\AgdaSpace{}%
\AgdaBound{msg₁}\AgdaSpace{}%
\AgdaBound{pbk1}\AgdaSpace{}%
\AgdaBound{pbk2}\AgdaSpace{}%
\AgdaBound{pbk3}\AgdaSpace{}%
\AgdaBound{pbk4}\AgdaSpace{}%
\AgdaBound{pbk5}\AgdaSpace{}%
\AgdaBound{sig1}\AgdaSpace{}%
\AgdaBound{sig2}\AgdaSpace{}%
\AgdaBound{sig3}\AgdaSpace{}%
\AgdaBound{x}\AgdaSpace{}%
\AgdaBound{x₁}\AgdaSpace{}%
\AgdaBound{x₂}\AgdaSpace{}%
\AgdaSymbol{|}\AgdaSpace{}%
\AgdaInductiveConstructor{true}\AgdaSpace{}%
\AgdaSymbol{|}\AgdaSpace{}%
\AgdaInductiveConstructor{true}\AgdaSpace{}%
\AgdaSymbol{|}\AgdaSpace{}%
\AgdaInductiveConstructor{false}\AgdaSpace{}%
\AgdaSymbol{|}\AgdaSpace{}%
\AgdaInductiveConstructor{false}\AgdaSpace{}%
\AgdaKeyword{with}\AgdaSpace{}%
\AgdaSymbol{(}\AgdaFunction{isSigned}%
\>[129]\AgdaBound{msg₁}\AgdaSpace{}%
\AgdaBound{sig3}\AgdaSpace{}%
\AgdaBound{pbk5}\AgdaSymbol{)}\<%
\\
\>[0]\AgdaFunction{lemmaHoareTripleStackGeAux'16}\AgdaSpace{}%
\AgdaBound{msg₁}\AgdaSpace{}%
\AgdaBound{pbk1}\AgdaSpace{}%
\AgdaBound{pbk2}\AgdaSpace{}%
\AgdaBound{pbk3}\AgdaSpace{}%
\AgdaBound{pbk4}\AgdaSpace{}%
\AgdaBound{pbk5}\AgdaSpace{}%
\AgdaBound{sig1}\AgdaSpace{}%
\AgdaBound{sig2}\AgdaSpace{}%
\AgdaBound{sig3}\AgdaSpace{}%
\AgdaBound{x}\AgdaSpace{}%
\AgdaBound{x₁}\AgdaSpace{}%
\AgdaBound{x₂}\AgdaSpace{}%
\AgdaSymbol{|}\AgdaSpace{}%
\AgdaInductiveConstructor{true}\AgdaSpace{}%
\AgdaSymbol{|}\AgdaSpace{}%
\AgdaInductiveConstructor{true}\AgdaSpace{}%
\AgdaSymbol{|}\AgdaSpace{}%
\AgdaInductiveConstructor{false}\AgdaSpace{}%
\AgdaSymbol{|}\AgdaSpace{}%
\AgdaInductiveConstructor{false}\AgdaSpace{}%
\AgdaSymbol{|}\AgdaSpace{}%
\AgdaInductiveConstructor{true}\AgdaSpace{}%
\AgdaSymbol{=}\AgdaSpace{}%
\AgdaInductiveConstructor{tt}\<%
\\
\>[0]\AgdaFunction{lemmaHoareTripleStackGeAux'16}\AgdaSpace{}%
\AgdaBound{msg₁}\AgdaSpace{}%
\AgdaBound{pbk1}\AgdaSpace{}%
\AgdaBound{pbk2}\AgdaSpace{}%
\AgdaBound{pbk3}\AgdaSpace{}%
\AgdaBound{pbk4}\AgdaSpace{}%
\AgdaBound{pbk5}\AgdaSpace{}%
\AgdaBound{sig1}\AgdaSpace{}%
\AgdaBound{sig2}\AgdaSpace{}%
\AgdaBound{sig3}\AgdaSpace{}%
\AgdaBound{x}\AgdaSpace{}%
\AgdaBound{x₁}\AgdaSpace{}%
\AgdaBound{x₂}\AgdaSpace{}%
\AgdaSymbol{|}\AgdaSpace{}%
\AgdaInductiveConstructor{true}\AgdaSpace{}%
\AgdaSymbol{|}\AgdaSpace{}%
\AgdaInductiveConstructor{true}\AgdaSpace{}%
\AgdaSymbol{|}\AgdaSpace{}%
\AgdaInductiveConstructor{false}\AgdaSpace{}%
\AgdaSymbol{|}\AgdaSpace{}%
\AgdaInductiveConstructor{true}\AgdaSpace{}%
\AgdaSymbol{=}\AgdaSpace{}%
\AgdaInductiveConstructor{tt}\<%
\\
\>[0]\AgdaFunction{lemmaHoareTripleStackGeAux'16}\AgdaSpace{}%
\AgdaBound{msg₁}\AgdaSpace{}%
\AgdaBound{pbk1}\AgdaSpace{}%
\AgdaBound{pbk2}\AgdaSpace{}%
\AgdaBound{pbk3}\AgdaSpace{}%
\AgdaBound{pbk4}\AgdaSpace{}%
\AgdaBound{pbk5}\AgdaSpace{}%
\AgdaBound{sig1}\AgdaSpace{}%
\AgdaBound{sig2}\AgdaSpace{}%
\AgdaBound{sig3}\AgdaSpace{}%
\AgdaBound{x}\AgdaSpace{}%
\AgdaBound{x₁}\AgdaSpace{}%
\AgdaBound{x₂}\AgdaSpace{}%
\AgdaSymbol{|}\AgdaSpace{}%
\AgdaInductiveConstructor{true}\AgdaSpace{}%
\AgdaSymbol{|}\AgdaSpace{}%
\AgdaInductiveConstructor{true}\AgdaSpace{}%
\AgdaSymbol{|}\AgdaSpace{}%
\AgdaInductiveConstructor{true}\AgdaSpace{}%
\AgdaSymbol{=}\AgdaSpace{}%
\AgdaInductiveConstructor{tt}\<%
\\
\\[\AgdaEmptyExtraSkip]%
\\[\AgdaEmptyExtraSkip]%
\>[0]\AgdaFunction{lemmaHoareTripleStackGeAux'17}%
\>[3525I]\AgdaSymbol{:}\AgdaSpace{}%
\AgdaSymbol{(}\AgdaBound{msg}\AgdaSpace{}%
\AgdaSymbol{:}\AgdaSpace{}%
\AgdaDatatype{Msg}\AgdaSymbol{)}\<%
\\
\>[3525I][@{}l@{\AgdaIndent{0}}]%
\>[31]\AgdaSymbol{(}\AgdaBound{pbk1}\AgdaSpace{}%
\AgdaBound{pbk2}\AgdaSpace{}%
\AgdaBound{pbk3}\AgdaSpace{}%
\AgdaBound{pbk4}\AgdaSpace{}%
\AgdaBound{pbk5}\AgdaSpace{}%
\AgdaBound{sig1}\AgdaSpace{}%
\AgdaBound{sig2}\AgdaSpace{}%
\AgdaBound{sig3}\AgdaSpace{}%
\AgdaSymbol{:}\AgdaSpace{}%
\AgdaDatatype{ℕ}\AgdaSymbol{)}\<%
\\
\>[31]\AgdaSymbol{→}\AgdaSpace{}%
\AgdaFunction{True}\AgdaSpace{}%
\AgdaSymbol{(}\AgdaFunction{isSigned}%
\>[49]\AgdaBound{msg}\AgdaSpace{}%
\AgdaBound{sig3}\AgdaSpace{}%
\AgdaBound{pbk4}\AgdaSymbol{)}\<%
\\
\>[.][@{}l@{}]\<[3525I]%
\>[30]\AgdaSymbol{→}%
\>[33]\AgdaFunction{True}\AgdaSpace{}%
\AgdaSymbol{(}\AgdaFunction{isSigned}%
\>[49]\AgdaBound{msg}\AgdaSpace{}%
\AgdaBound{sig2}\AgdaSpace{}%
\AgdaBound{pbk3}\AgdaSymbol{)}\<%
\\
\>[30][@{}l@{\AgdaIndent{0}}]%
\>[31]\AgdaSymbol{→}%
\>[34]\AgdaFunction{True}\AgdaSpace{}%
\AgdaSymbol{(}\AgdaFunction{isSigned}%
\>[50]\AgdaBound{msg}\AgdaSpace{}%
\AgdaBound{sig1}\AgdaSpace{}%
\AgdaBound{pbk1}\AgdaSymbol{)}\<%
\\
\>[31]\AgdaSymbol{→}\AgdaSpace{}%
\AgdaFunction{True}\AgdaSpace{}%
\AgdaSymbol{(}\AgdaFunction{compareSigsMultiSig}\AgdaSpace{}%
\AgdaBound{msg}\AgdaSpace{}%
\AgdaSymbol{(}\AgdaSpace{}%
\AgdaBound{sig1}\AgdaSpace{}%
\AgdaOperator{\AgdaInductiveConstructor{∷}}\AgdaSpace{}%
\AgdaBound{sig2}\AgdaSpace{}%
\AgdaOperator{\AgdaInductiveConstructor{∷}}\AgdaSpace{}%
\AgdaBound{sig3}\AgdaSpace{}%
\AgdaOperator{\AgdaInductiveConstructor{∷}}\AgdaSpace{}%
\AgdaInductiveConstructor{[]}\AgdaSymbol{)}\AgdaSpace{}%
\AgdaSymbol{(}\AgdaBound{pbk1}\AgdaSpace{}%
\AgdaOperator{\AgdaInductiveConstructor{∷}}\AgdaSpace{}%
\AgdaBound{pbk2}\AgdaSpace{}%
\AgdaOperator{\AgdaInductiveConstructor{∷}}\AgdaSpace{}%
\AgdaBound{pbk3}\AgdaSpace{}%
\AgdaOperator{\AgdaInductiveConstructor{∷}}\AgdaSpace{}%
\AgdaBound{pbk4}\AgdaSpace{}%
\AgdaOperator{\AgdaInductiveConstructor{∷}}\AgdaSpace{}%
\AgdaBound{pbk5}\AgdaSpace{}%
\AgdaOperator{\AgdaInductiveConstructor{∷}}\AgdaSpace{}%
\AgdaInductiveConstructor{[]}\AgdaSymbol{))}\<%
\\
\>[0]\AgdaFunction{lemmaHoareTripleStackGeAux'17}\AgdaSpace{}%
\AgdaBound{msg₁}\AgdaSpace{}%
\AgdaBound{pbk1}\AgdaSpace{}%
\AgdaBound{pbk2}\AgdaSpace{}%
\AgdaBound{pbk3}\AgdaSpace{}%
\AgdaBound{pbk4}\AgdaSpace{}%
\AgdaBound{pbk5}\AgdaSpace{}%
\AgdaBound{sig1}\AgdaSpace{}%
\AgdaBound{sig2}\AgdaSpace{}%
\AgdaBound{sig3}\AgdaSpace{}%
\AgdaBound{x}\AgdaSpace{}%
\AgdaBound{x₁}\AgdaSpace{}%
\AgdaBound{x₂}\AgdaSpace{}%
\AgdaKeyword{with}\AgdaSpace{}%
\AgdaSymbol{(}\AgdaFunction{isSigned}%
\>[99]\AgdaBound{msg₁}\AgdaSpace{}%
\AgdaBound{sig1}\AgdaSpace{}%
\AgdaBound{pbk1}\AgdaSymbol{)}\<%
\\
\>[0]\AgdaFunction{lemmaHoareTripleStackGeAux'17}\AgdaSpace{}%
\AgdaBound{msg₁}\AgdaSpace{}%
\AgdaBound{pbk1}\AgdaSpace{}%
\AgdaBound{pbk2}\AgdaSpace{}%
\AgdaBound{pbk3}\AgdaSpace{}%
\AgdaBound{pbk4}\AgdaSpace{}%
\AgdaBound{pbk5}\AgdaSpace{}%
\AgdaBound{sig1}\AgdaSpace{}%
\AgdaBound{sig2}\AgdaSpace{}%
\AgdaBound{sig3}\AgdaSpace{}%
\AgdaBound{x}\AgdaSpace{}%
\AgdaBound{x₁}\AgdaSpace{}%
\AgdaBound{x₂}\AgdaSpace{}%
\AgdaSymbol{|}\AgdaSpace{}%
\AgdaInductiveConstructor{true}\AgdaSpace{}%
\AgdaKeyword{with}%
\>[96]\AgdaSymbol{(}\AgdaFunction{isSigned}%
\>[107]\AgdaBound{msg₁}\AgdaSpace{}%
\AgdaBound{sig2}\AgdaSpace{}%
\AgdaBound{pbk2}\AgdaSymbol{)}\<%
\\
\>[0]\AgdaFunction{lemmaHoareTripleStackGeAux'17}\AgdaSpace{}%
\AgdaBound{msg₁}\AgdaSpace{}%
\AgdaBound{pbk1}\AgdaSpace{}%
\AgdaBound{pbk2}\AgdaSpace{}%
\AgdaBound{pbk3}\AgdaSpace{}%
\AgdaBound{pbk4}\AgdaSpace{}%
\AgdaBound{pbk5}\AgdaSpace{}%
\AgdaBound{sig1}\AgdaSpace{}%
\AgdaBound{sig2}\AgdaSpace{}%
\AgdaBound{sig3}\AgdaSpace{}%
\AgdaBound{x}\AgdaSpace{}%
\AgdaBound{x₁}\AgdaSpace{}%
\AgdaBound{x₂}\AgdaSpace{}%
\AgdaSymbol{|}\AgdaSpace{}%
\AgdaInductiveConstructor{true}\AgdaSpace{}%
\AgdaSymbol{|}\AgdaSpace{}%
\AgdaInductiveConstructor{false}\AgdaSpace{}%
\AgdaKeyword{with}\AgdaSpace{}%
\AgdaSymbol{(}\AgdaFunction{isSigned}%
\>[114]\AgdaBound{msg₁}\AgdaSpace{}%
\AgdaBound{sig2}\AgdaSpace{}%
\AgdaBound{pbk3}\AgdaSymbol{)}\<%
\\
\>[0]\AgdaFunction{lemmaHoareTripleStackGeAux'17}\AgdaSpace{}%
\AgdaBound{msg₁}\AgdaSpace{}%
\AgdaBound{pbk1}\AgdaSpace{}%
\AgdaBound{pbk2}\AgdaSpace{}%
\AgdaBound{pbk3}\AgdaSpace{}%
\AgdaBound{pbk4}\AgdaSpace{}%
\AgdaBound{pbk5}\AgdaSpace{}%
\AgdaBound{sig1}\AgdaSpace{}%
\AgdaBound{sig2}\AgdaSpace{}%
\AgdaBound{sig3}\AgdaSpace{}%
\AgdaBound{x}\AgdaSpace{}%
\AgdaBound{x₁}\AgdaSpace{}%
\AgdaBound{x₂}\AgdaSpace{}%
\AgdaSymbol{|}\AgdaSpace{}%
\AgdaInductiveConstructor{true}\AgdaSpace{}%
\AgdaSymbol{|}\AgdaSpace{}%
\AgdaInductiveConstructor{false}\AgdaSpace{}%
\AgdaSymbol{|}\AgdaSpace{}%
\AgdaInductiveConstructor{true}\AgdaSpace{}%
\AgdaKeyword{with}%
\>[111]\AgdaSymbol{(}\AgdaFunction{isSigned}%
\>[122]\AgdaBound{msg₁}\AgdaSpace{}%
\AgdaBound{sig3}\AgdaSpace{}%
\AgdaBound{pbk4}\AgdaSymbol{)}\<%
\\
\>[0]\AgdaFunction{lemmaHoareTripleStackGeAux'17}\AgdaSpace{}%
\AgdaBound{msg₁}\AgdaSpace{}%
\AgdaBound{pbk1}\AgdaSpace{}%
\AgdaBound{pbk2}\AgdaSpace{}%
\AgdaBound{pbk3}\AgdaSpace{}%
\AgdaBound{pbk4}\AgdaSpace{}%
\AgdaBound{pbk5}\AgdaSpace{}%
\AgdaBound{sig1}\AgdaSpace{}%
\AgdaBound{sig2}\AgdaSpace{}%
\AgdaBound{sig3}\AgdaSpace{}%
\AgdaBound{x}\AgdaSpace{}%
\AgdaBound{x₁}\AgdaSpace{}%
\AgdaBound{x₂}\AgdaSpace{}%
\AgdaSymbol{|}\AgdaSpace{}%
\AgdaInductiveConstructor{true}\AgdaSpace{}%
\AgdaSymbol{|}\AgdaSpace{}%
\AgdaInductiveConstructor{false}\AgdaSpace{}%
\AgdaSymbol{|}\AgdaSpace{}%
\AgdaInductiveConstructor{true}\AgdaSpace{}%
\AgdaSymbol{|}\AgdaSpace{}%
\AgdaInductiveConstructor{true}\AgdaSpace{}%
\AgdaSymbol{=}\AgdaSpace{}%
\AgdaInductiveConstructor{tt}\<%
\\
\>[0]\AgdaFunction{lemmaHoareTripleStackGeAux'17}\AgdaSpace{}%
\AgdaBound{msg₁}\AgdaSpace{}%
\AgdaBound{pbk1}\AgdaSpace{}%
\AgdaBound{pbk2}\AgdaSpace{}%
\AgdaBound{pbk3}\AgdaSpace{}%
\AgdaBound{pbk4}\AgdaSpace{}%
\AgdaBound{pbk5}\AgdaSpace{}%
\AgdaBound{sig1}\AgdaSpace{}%
\AgdaBound{sig2}\AgdaSpace{}%
\AgdaBound{sig3}\AgdaSpace{}%
\AgdaBound{x}\AgdaSpace{}%
\AgdaBound{x₁}\AgdaSpace{}%
\AgdaBound{x₂}\AgdaSpace{}%
\AgdaSymbol{|}\AgdaSpace{}%
\AgdaInductiveConstructor{true}\AgdaSpace{}%
\AgdaSymbol{|}\AgdaSpace{}%
\AgdaInductiveConstructor{true}\AgdaSpace{}%
\AgdaKeyword{with}\AgdaSpace{}%
\AgdaSymbol{(}\AgdaFunction{isSigned}%
\>[113]\AgdaBound{msg₁}\AgdaSpace{}%
\AgdaBound{sig3}\AgdaSpace{}%
\AgdaBound{pbk3}\AgdaSymbol{)}\<%
\\
\>[0]\AgdaFunction{lemmaHoareTripleStackGeAux'17}\AgdaSpace{}%
\AgdaBound{msg₁}\AgdaSpace{}%
\AgdaBound{pbk1}\AgdaSpace{}%
\AgdaBound{pbk2}\AgdaSpace{}%
\AgdaBound{pbk3}\AgdaSpace{}%
\AgdaBound{pbk4}\AgdaSpace{}%
\AgdaBound{pbk5}\AgdaSpace{}%
\AgdaBound{sig1}\AgdaSpace{}%
\AgdaBound{sig2}\AgdaSpace{}%
\AgdaBound{sig3}\AgdaSpace{}%
\AgdaBound{x}\AgdaSpace{}%
\AgdaBound{x₁}\AgdaSpace{}%
\AgdaBound{x₂}\AgdaSpace{}%
\AgdaSymbol{|}\AgdaSpace{}%
\AgdaInductiveConstructor{true}\AgdaSpace{}%
\AgdaSymbol{|}\AgdaSpace{}%
\AgdaInductiveConstructor{true}\AgdaSpace{}%
\AgdaSymbol{|}\AgdaSpace{}%
\AgdaInductiveConstructor{false}\AgdaSpace{}%
\AgdaKeyword{with}%
\>[111]\AgdaSymbol{(}\AgdaFunction{isSigned}%
\>[122]\AgdaBound{msg₁}\AgdaSpace{}%
\AgdaBound{sig3}\AgdaSpace{}%
\AgdaBound{pbk4}\AgdaSymbol{)}\<%
\\
\>[0]\AgdaFunction{lemmaHoareTripleStackGeAux'17}\AgdaSpace{}%
\AgdaBound{msg₁}\AgdaSpace{}%
\AgdaBound{pbk1}\AgdaSpace{}%
\AgdaBound{pbk2}\AgdaSpace{}%
\AgdaBound{pbk3}\AgdaSpace{}%
\AgdaBound{pbk4}\AgdaSpace{}%
\AgdaBound{pbk5}\AgdaSpace{}%
\AgdaBound{sig1}\AgdaSpace{}%
\AgdaBound{sig2}\AgdaSpace{}%
\AgdaBound{sig3}\AgdaSpace{}%
\AgdaBound{x}\AgdaSpace{}%
\AgdaBound{x₁}\AgdaSpace{}%
\AgdaBound{x₂}\AgdaSpace{}%
\AgdaSymbol{|}\AgdaSpace{}%
\AgdaInductiveConstructor{true}\AgdaSpace{}%
\AgdaSymbol{|}\AgdaSpace{}%
\AgdaInductiveConstructor{true}\AgdaSpace{}%
\AgdaSymbol{|}\AgdaSpace{}%
\AgdaInductiveConstructor{false}\AgdaSpace{}%
\AgdaSymbol{|}\AgdaSpace{}%
\AgdaInductiveConstructor{true}\AgdaSpace{}%
\AgdaSymbol{=}\AgdaSpace{}%
\AgdaInductiveConstructor{tt}\<%
\\
\>[0]\AgdaFunction{lemmaHoareTripleStackGeAux'17}\AgdaSpace{}%
\AgdaBound{msg₁}\AgdaSpace{}%
\AgdaBound{pbk1}\AgdaSpace{}%
\AgdaBound{pbk2}\AgdaSpace{}%
\AgdaBound{pbk3}\AgdaSpace{}%
\AgdaBound{pbk4}\AgdaSpace{}%
\AgdaBound{pbk5}\AgdaSpace{}%
\AgdaBound{sig1}\AgdaSpace{}%
\AgdaBound{sig2}\AgdaSpace{}%
\AgdaBound{sig3}\AgdaSpace{}%
\AgdaBound{x}\AgdaSpace{}%
\AgdaBound{x₁}\AgdaSpace{}%
\AgdaBound{x₂}\AgdaSpace{}%
\AgdaSymbol{|}\AgdaSpace{}%
\AgdaInductiveConstructor{true}\AgdaSpace{}%
\AgdaSymbol{|}\AgdaSpace{}%
\AgdaInductiveConstructor{true}\AgdaSpace{}%
\AgdaSymbol{|}\AgdaSpace{}%
\AgdaInductiveConstructor{true}\AgdaSpace{}%
\AgdaSymbol{=}\AgdaSpace{}%
\AgdaInductiveConstructor{tt}\<%
\\
\\[\AgdaEmptyExtraSkip]%
\\[\AgdaEmptyExtraSkip]%
\>[0]\AgdaFunction{lemmaHoareTripleStackGeAux'18}%
\>[3749I]\AgdaSymbol{:}\AgdaSpace{}%
\AgdaSymbol{(}\AgdaBound{msg}\AgdaSpace{}%
\AgdaSymbol{:}\AgdaSpace{}%
\AgdaDatatype{Msg}\AgdaSymbol{)}\<%
\\
\>[3749I][@{}l@{\AgdaIndent{0}}]%
\>[31]\AgdaSymbol{(}\AgdaBound{pbk1}\AgdaSpace{}%
\AgdaBound{pbk2}\AgdaSpace{}%
\AgdaBound{pbk3}\AgdaSpace{}%
\AgdaBound{pbk4}\AgdaSpace{}%
\AgdaBound{pbk5}\AgdaSpace{}%
\AgdaBound{sig1}\AgdaSpace{}%
\AgdaBound{sig2}\AgdaSpace{}%
\AgdaBound{sig3}\AgdaSpace{}%
\AgdaSymbol{:}\AgdaSpace{}%
\AgdaDatatype{ℕ}\AgdaSymbol{)}\<%
\\
\>[31]\AgdaSymbol{→}\AgdaSpace{}%
\AgdaFunction{True}\AgdaSpace{}%
\AgdaSymbol{(}\AgdaFunction{isSigned}%
\>[49]\AgdaBound{msg}\AgdaSpace{}%
\AgdaBound{sig3}\AgdaSpace{}%
\AgdaBound{pbk5}\AgdaSymbol{)}\<%
\\
\>[.][@{}l@{}]\<[3749I]%
\>[30]\AgdaSymbol{→}%
\>[33]\AgdaFunction{True}\AgdaSpace{}%
\AgdaSymbol{(}\AgdaFunction{isSigned}%
\>[49]\AgdaBound{msg}\AgdaSpace{}%
\AgdaBound{sig2}\AgdaSpace{}%
\AgdaBound{pbk3}\AgdaSymbol{)}\<%
\\
\>[30]\AgdaSymbol{→}%
\>[33]\AgdaFunction{True}\AgdaSpace{}%
\AgdaSymbol{(}\AgdaFunction{isSigned}%
\>[49]\AgdaBound{msg}\AgdaSpace{}%
\AgdaBound{sig1}\AgdaSpace{}%
\AgdaBound{pbk1}\AgdaSymbol{)}\<%
\\
\>[30][@{}l@{\AgdaIndent{0}}]%
\>[31]\AgdaSymbol{→}\AgdaSpace{}%
\AgdaFunction{True}\AgdaSpace{}%
\AgdaSymbol{(}\AgdaFunction{compareSigsMultiSig}\AgdaSpace{}%
\AgdaBound{msg}\AgdaSpace{}%
\AgdaSymbol{(}\AgdaSpace{}%
\AgdaBound{sig1}\AgdaSpace{}%
\AgdaOperator{\AgdaInductiveConstructor{∷}}\AgdaSpace{}%
\AgdaBound{sig2}\AgdaSpace{}%
\AgdaOperator{\AgdaInductiveConstructor{∷}}\AgdaSpace{}%
\AgdaBound{sig3}\AgdaSpace{}%
\AgdaOperator{\AgdaInductiveConstructor{∷}}\AgdaSpace{}%
\AgdaInductiveConstructor{[]}\AgdaSymbol{)}\AgdaSpace{}%
\AgdaSymbol{(}\AgdaBound{pbk1}\AgdaSpace{}%
\AgdaOperator{\AgdaInductiveConstructor{∷}}\AgdaSpace{}%
\AgdaBound{pbk2}\AgdaSpace{}%
\AgdaOperator{\AgdaInductiveConstructor{∷}}\AgdaSpace{}%
\AgdaBound{pbk3}\AgdaSpace{}%
\AgdaOperator{\AgdaInductiveConstructor{∷}}\AgdaSpace{}%
\AgdaBound{pbk4}\AgdaSpace{}%
\AgdaOperator{\AgdaInductiveConstructor{∷}}\AgdaSpace{}%
\AgdaBound{pbk5}\AgdaSpace{}%
\AgdaOperator{\AgdaInductiveConstructor{∷}}\AgdaSpace{}%
\AgdaInductiveConstructor{[]}\AgdaSymbol{))}\<%
\\
\>[0]\AgdaFunction{lemmaHoareTripleStackGeAux'18}\AgdaSpace{}%
\AgdaBound{msg₁}\AgdaSpace{}%
\AgdaBound{pbk1}\AgdaSpace{}%
\AgdaBound{pbk2}\AgdaSpace{}%
\AgdaBound{pbk3}\AgdaSpace{}%
\AgdaBound{pbk4}\AgdaSpace{}%
\AgdaBound{pbk5}\AgdaSpace{}%
\AgdaBound{sig1}\AgdaSpace{}%
\AgdaBound{sig2}\AgdaSpace{}%
\AgdaBound{sig3}\AgdaSpace{}%
\AgdaBound{x}\AgdaSpace{}%
\AgdaBound{x₁}\AgdaSpace{}%
\AgdaBound{x₂}\AgdaSpace{}%
\AgdaKeyword{with}\AgdaSpace{}%
\AgdaSymbol{(}\AgdaFunction{isSigned}%
\>[99]\AgdaBound{msg₁}\AgdaSpace{}%
\AgdaBound{sig1}\AgdaSpace{}%
\AgdaBound{pbk1}\AgdaSymbol{)}\<%
\\
\>[0]\AgdaFunction{lemmaHoareTripleStackGeAux'18}\AgdaSpace{}%
\AgdaBound{msg₁}\AgdaSpace{}%
\AgdaBound{pbk1}\AgdaSpace{}%
\AgdaBound{pbk2}\AgdaSpace{}%
\AgdaBound{pbk3}\AgdaSpace{}%
\AgdaBound{pbk4}\AgdaSpace{}%
\AgdaBound{pbk5}\AgdaSpace{}%
\AgdaBound{sig1}\AgdaSpace{}%
\AgdaBound{sig2}\AgdaSpace{}%
\AgdaBound{sig3}\AgdaSpace{}%
\AgdaBound{x}\AgdaSpace{}%
\AgdaBound{x₁}\AgdaSpace{}%
\AgdaBound{x₂}\AgdaSpace{}%
\AgdaSymbol{|}\AgdaSpace{}%
\AgdaInductiveConstructor{true}\AgdaSpace{}%
\AgdaKeyword{with}\AgdaSpace{}%
\AgdaSymbol{(}\AgdaFunction{isSigned}%
\>[106]\AgdaBound{msg₁}\AgdaSpace{}%
\AgdaBound{sig2}\AgdaSpace{}%
\AgdaBound{pbk2}\AgdaSymbol{)}\<%
\\
\>[0]\AgdaFunction{lemmaHoareTripleStackGeAux'18}\AgdaSpace{}%
\AgdaBound{msg₁}\AgdaSpace{}%
\AgdaBound{pbk1}\AgdaSpace{}%
\AgdaBound{pbk2}\AgdaSpace{}%
\AgdaBound{pbk3}\AgdaSpace{}%
\AgdaBound{pbk4}\AgdaSpace{}%
\AgdaBound{pbk5}\AgdaSpace{}%
\AgdaBound{sig1}\AgdaSpace{}%
\AgdaBound{sig2}\AgdaSpace{}%
\AgdaBound{sig3}\AgdaSpace{}%
\AgdaBound{x}\AgdaSpace{}%
\AgdaBound{x₁}\AgdaSpace{}%
\AgdaBound{x₂}\AgdaSpace{}%
\AgdaSymbol{|}\AgdaSpace{}%
\AgdaInductiveConstructor{true}\AgdaSpace{}%
\AgdaSymbol{|}\AgdaSpace{}%
\AgdaInductiveConstructor{false}\AgdaSpace{}%
\AgdaKeyword{with}%
\>[104]\AgdaSymbol{(}\AgdaFunction{isSigned}%
\>[115]\AgdaBound{msg₁}\AgdaSpace{}%
\AgdaBound{sig2}\AgdaSpace{}%
\AgdaBound{pbk3}\AgdaSymbol{)}\<%
\\
\>[0]\AgdaFunction{lemmaHoareTripleStackGeAux'18}\AgdaSpace{}%
\AgdaBound{msg₁}\AgdaSpace{}%
\AgdaBound{pbk1}\AgdaSpace{}%
\AgdaBound{pbk2}\AgdaSpace{}%
\AgdaBound{pbk3}\AgdaSpace{}%
\AgdaBound{pbk4}\AgdaSpace{}%
\AgdaBound{pbk5}\AgdaSpace{}%
\AgdaBound{sig1}\AgdaSpace{}%
\AgdaBound{sig2}\AgdaSpace{}%
\AgdaBound{sig3}\AgdaSpace{}%
\AgdaBound{x}\AgdaSpace{}%
\AgdaBound{x₁}\AgdaSpace{}%
\AgdaBound{x₂}\AgdaSpace{}%
\AgdaSymbol{|}\AgdaSpace{}%
\AgdaInductiveConstructor{true}\AgdaSpace{}%
\AgdaSymbol{|}\AgdaSpace{}%
\AgdaInductiveConstructor{false}\AgdaSpace{}%
\AgdaSymbol{|}\AgdaSpace{}%
\AgdaInductiveConstructor{true}\AgdaSpace{}%
\AgdaKeyword{with}\AgdaSpace{}%
\AgdaSymbol{(}\AgdaFunction{isSigned}%
\>[121]\AgdaBound{msg₁}\AgdaSpace{}%
\AgdaBound{sig3}\AgdaSpace{}%
\AgdaBound{pbk4}\AgdaSymbol{)}\<%
\\
\>[0]\AgdaFunction{lemmaHoareTripleStackGeAux'18}\AgdaSpace{}%
\AgdaBound{msg₁}\AgdaSpace{}%
\AgdaBound{pbk1}\AgdaSpace{}%
\AgdaBound{pbk2}\AgdaSpace{}%
\AgdaBound{pbk3}\AgdaSpace{}%
\AgdaBound{pbk4}\AgdaSpace{}%
\AgdaBound{pbk5}\AgdaSpace{}%
\AgdaBound{sig1}\AgdaSpace{}%
\AgdaBound{sig2}\AgdaSpace{}%
\AgdaBound{sig3}\AgdaSpace{}%
\AgdaBound{x}\AgdaSpace{}%
\AgdaBound{x₁}\AgdaSpace{}%
\AgdaBound{x₂}\AgdaSpace{}%
\AgdaSymbol{|}\AgdaSpace{}%
\AgdaInductiveConstructor{true}\AgdaSpace{}%
\AgdaSymbol{|}\AgdaSpace{}%
\AgdaInductiveConstructor{false}\AgdaSpace{}%
\AgdaSymbol{|}\AgdaSpace{}%
\AgdaInductiveConstructor{true}\AgdaSpace{}%
\AgdaSymbol{|}\AgdaSpace{}%
\AgdaInductiveConstructor{false}\AgdaSpace{}%
\AgdaKeyword{with}\AgdaSpace{}%
\AgdaSymbol{(}\AgdaFunction{isSigned}%
\>[129]\AgdaBound{msg₁}\AgdaSpace{}%
\AgdaBound{sig3}\AgdaSpace{}%
\AgdaBound{pbk5}\AgdaSymbol{)}\<%
\\
\>[0]\AgdaFunction{lemmaHoareTripleStackGeAux'18}\AgdaSpace{}%
\AgdaBound{msg₁}\AgdaSpace{}%
\AgdaBound{pbk1}\AgdaSpace{}%
\AgdaBound{pbk2}\AgdaSpace{}%
\AgdaBound{pbk3}\AgdaSpace{}%
\AgdaBound{pbk4}\AgdaSpace{}%
\AgdaBound{pbk5}\AgdaSpace{}%
\AgdaBound{sig1}\AgdaSpace{}%
\AgdaBound{sig2}\AgdaSpace{}%
\AgdaBound{sig3}\AgdaSpace{}%
\AgdaBound{x}\AgdaSpace{}%
\AgdaBound{x₁}\AgdaSpace{}%
\AgdaBound{x₂}\AgdaSpace{}%
\AgdaSymbol{|}\AgdaSpace{}%
\AgdaInductiveConstructor{true}\AgdaSpace{}%
\AgdaSymbol{|}\AgdaSpace{}%
\AgdaInductiveConstructor{false}\AgdaSpace{}%
\AgdaSymbol{|}\AgdaSpace{}%
\AgdaInductiveConstructor{true}\AgdaSpace{}%
\AgdaSymbol{|}\AgdaSpace{}%
\AgdaInductiveConstructor{false}\AgdaSpace{}%
\AgdaSymbol{|}\AgdaSpace{}%
\AgdaInductiveConstructor{true}\AgdaSpace{}%
\AgdaSymbol{=}\AgdaSpace{}%
\AgdaInductiveConstructor{tt}\<%
\\
\>[0]\AgdaFunction{lemmaHoareTripleStackGeAux'18}\AgdaSpace{}%
\AgdaBound{msg₁}\AgdaSpace{}%
\AgdaBound{pbk1}\AgdaSpace{}%
\AgdaBound{pbk2}\AgdaSpace{}%
\AgdaBound{pbk3}\AgdaSpace{}%
\AgdaBound{pbk4}\AgdaSpace{}%
\AgdaBound{pbk5}\AgdaSpace{}%
\AgdaBound{sig1}\AgdaSpace{}%
\AgdaBound{sig2}\AgdaSpace{}%
\AgdaBound{sig3}\AgdaSpace{}%
\AgdaBound{x}\AgdaSpace{}%
\AgdaBound{x₁}\AgdaSpace{}%
\AgdaBound{x₂}\AgdaSpace{}%
\AgdaSymbol{|}\AgdaSpace{}%
\AgdaInductiveConstructor{true}\AgdaSpace{}%
\AgdaSymbol{|}\AgdaSpace{}%
\AgdaInductiveConstructor{false}\AgdaSpace{}%
\AgdaSymbol{|}\AgdaSpace{}%
\AgdaInductiveConstructor{true}\AgdaSpace{}%
\AgdaSymbol{|}\AgdaSpace{}%
\AgdaInductiveConstructor{true}\AgdaSpace{}%
\AgdaSymbol{=}\AgdaSpace{}%
\AgdaInductiveConstructor{tt}\<%
\\
\>[0]\AgdaFunction{lemmaHoareTripleStackGeAux'18}\AgdaSpace{}%
\AgdaBound{msg₁}\AgdaSpace{}%
\AgdaBound{pbk1}\AgdaSpace{}%
\AgdaBound{pbk2}\AgdaSpace{}%
\AgdaBound{pbk3}\AgdaSpace{}%
\AgdaBound{pbk4}\AgdaSpace{}%
\AgdaBound{pbk5}\AgdaSpace{}%
\AgdaBound{sig1}\AgdaSpace{}%
\AgdaBound{sig2}\AgdaSpace{}%
\AgdaBound{sig3}\AgdaSpace{}%
\AgdaBound{x}\AgdaSpace{}%
\AgdaBound{x₁}\AgdaSpace{}%
\AgdaBound{x₂}\AgdaSpace{}%
\AgdaSymbol{|}\AgdaSpace{}%
\AgdaInductiveConstructor{true}\AgdaSpace{}%
\AgdaSymbol{|}\AgdaSpace{}%
\AgdaInductiveConstructor{true}\AgdaSpace{}%
\AgdaKeyword{with}\AgdaSpace{}%
\AgdaSymbol{(}\AgdaFunction{isSigned}%
\>[113]\AgdaBound{msg₁}\AgdaSpace{}%
\AgdaBound{sig3}\AgdaSpace{}%
\AgdaBound{pbk3}\AgdaSymbol{)}\<%
\\
\>[0]\AgdaFunction{lemmaHoareTripleStackGeAux'18}\AgdaSpace{}%
\AgdaBound{msg₁}\AgdaSpace{}%
\AgdaBound{pbk1}\AgdaSpace{}%
\AgdaBound{pbk2}\AgdaSpace{}%
\AgdaBound{pbk3}\AgdaSpace{}%
\AgdaBound{pbk4}\AgdaSpace{}%
\AgdaBound{pbk5}\AgdaSpace{}%
\AgdaBound{sig1}\AgdaSpace{}%
\AgdaBound{sig2}\AgdaSpace{}%
\AgdaBound{sig3}\AgdaSpace{}%
\AgdaBound{x}\AgdaSpace{}%
\AgdaBound{x₁}\AgdaSpace{}%
\AgdaBound{x₂}\AgdaSpace{}%
\AgdaSymbol{|}\AgdaSpace{}%
\AgdaInductiveConstructor{true}\AgdaSpace{}%
\AgdaSymbol{|}\AgdaSpace{}%
\AgdaInductiveConstructor{true}\AgdaSpace{}%
\AgdaSymbol{|}\AgdaSpace{}%
\AgdaInductiveConstructor{false}\AgdaSpace{}%
\AgdaKeyword{with}\AgdaSpace{}%
\AgdaSymbol{(}\AgdaFunction{isSigned}%
\>[121]\AgdaBound{msg₁}\AgdaSpace{}%
\AgdaBound{sig3}\AgdaSpace{}%
\AgdaBound{pbk4}\AgdaSymbol{)}\<%
\\
\>[0]\AgdaFunction{lemmaHoareTripleStackGeAux'18}\AgdaSpace{}%
\AgdaBound{msg₁}\AgdaSpace{}%
\AgdaBound{pbk1}\AgdaSpace{}%
\AgdaBound{pbk2}\AgdaSpace{}%
\AgdaBound{pbk3}\AgdaSpace{}%
\AgdaBound{pbk4}\AgdaSpace{}%
\AgdaBound{pbk5}\AgdaSpace{}%
\AgdaBound{sig1}\AgdaSpace{}%
\AgdaBound{sig2}\AgdaSpace{}%
\AgdaBound{sig3}\AgdaSpace{}%
\AgdaBound{x}\AgdaSpace{}%
\AgdaBound{x₁}\AgdaSpace{}%
\AgdaBound{x₂}\AgdaSpace{}%
\AgdaSymbol{|}\AgdaSpace{}%
\AgdaInductiveConstructor{true}\AgdaSpace{}%
\AgdaSymbol{|}\AgdaSpace{}%
\AgdaInductiveConstructor{true}\AgdaSpace{}%
\AgdaSymbol{|}\AgdaSpace{}%
\AgdaInductiveConstructor{false}\AgdaSpace{}%
\AgdaSymbol{|}\AgdaSpace{}%
\AgdaInductiveConstructor{false}\AgdaSpace{}%
\AgdaKeyword{with}\AgdaSpace{}%
\AgdaSymbol{(}\AgdaFunction{isSigned}%
\>[129]\AgdaBound{msg₁}\AgdaSpace{}%
\AgdaBound{sig3}\AgdaSpace{}%
\AgdaBound{pbk5}\AgdaSymbol{)}\<%
\\
\>[0]\AgdaFunction{lemmaHoareTripleStackGeAux'18}\AgdaSpace{}%
\AgdaBound{msg₁}\AgdaSpace{}%
\AgdaBound{pbk1}\AgdaSpace{}%
\AgdaBound{pbk2}\AgdaSpace{}%
\AgdaBound{pbk3}\AgdaSpace{}%
\AgdaBound{pbk4}\AgdaSpace{}%
\AgdaBound{pbk5}\AgdaSpace{}%
\AgdaBound{sig1}\AgdaSpace{}%
\AgdaBound{sig2}\AgdaSpace{}%
\AgdaBound{sig3}\AgdaSpace{}%
\AgdaBound{x}\AgdaSpace{}%
\AgdaBound{x₁}\AgdaSpace{}%
\AgdaBound{x₂}\AgdaSpace{}%
\AgdaSymbol{|}\AgdaSpace{}%
\AgdaInductiveConstructor{true}\AgdaSpace{}%
\AgdaSymbol{|}\AgdaSpace{}%
\AgdaInductiveConstructor{true}\AgdaSpace{}%
\AgdaSymbol{|}\AgdaSpace{}%
\AgdaInductiveConstructor{false}\AgdaSpace{}%
\AgdaSymbol{|}\AgdaSpace{}%
\AgdaInductiveConstructor{false}\AgdaSpace{}%
\AgdaSymbol{|}\AgdaSpace{}%
\AgdaInductiveConstructor{true}\AgdaSpace{}%
\AgdaSymbol{=}\AgdaSpace{}%
\AgdaInductiveConstructor{tt}\<%
\\
\>[0]\AgdaFunction{lemmaHoareTripleStackGeAux'18}\AgdaSpace{}%
\AgdaBound{msg₁}\AgdaSpace{}%
\AgdaBound{pbk1}\AgdaSpace{}%
\AgdaBound{pbk2}\AgdaSpace{}%
\AgdaBound{pbk3}\AgdaSpace{}%
\AgdaBound{pbk4}\AgdaSpace{}%
\AgdaBound{pbk5}\AgdaSpace{}%
\AgdaBound{sig1}\AgdaSpace{}%
\AgdaBound{sig2}\AgdaSpace{}%
\AgdaBound{sig3}\AgdaSpace{}%
\AgdaBound{x}\AgdaSpace{}%
\AgdaBound{x₁}\AgdaSpace{}%
\AgdaBound{x₂}\AgdaSpace{}%
\AgdaSymbol{|}\AgdaSpace{}%
\AgdaInductiveConstructor{true}\AgdaSpace{}%
\AgdaSymbol{|}\AgdaSpace{}%
\AgdaInductiveConstructor{true}\AgdaSpace{}%
\AgdaSymbol{|}\AgdaSpace{}%
\AgdaInductiveConstructor{false}\AgdaSpace{}%
\AgdaSymbol{|}\AgdaSpace{}%
\AgdaInductiveConstructor{true}\AgdaSpace{}%
\AgdaSymbol{=}\AgdaSpace{}%
\AgdaInductiveConstructor{tt}\<%
\\
\>[0]\AgdaFunction{lemmaHoareTripleStackGeAux'18}\AgdaSpace{}%
\AgdaBound{msg₁}\AgdaSpace{}%
\AgdaBound{pbk1}\AgdaSpace{}%
\AgdaBound{pbk2}\AgdaSpace{}%
\AgdaBound{pbk3}\AgdaSpace{}%
\AgdaBound{pbk4}\AgdaSpace{}%
\AgdaBound{pbk5}\AgdaSpace{}%
\AgdaBound{sig1}\AgdaSpace{}%
\AgdaBound{sig2}\AgdaSpace{}%
\AgdaBound{sig3}\AgdaSpace{}%
\AgdaBound{x}\AgdaSpace{}%
\AgdaBound{x₁}\AgdaSpace{}%
\AgdaBound{x₂}\AgdaSpace{}%
\AgdaSymbol{|}\AgdaSpace{}%
\AgdaInductiveConstructor{true}\AgdaSpace{}%
\AgdaSymbol{|}\AgdaSpace{}%
\AgdaInductiveConstructor{true}\AgdaSpace{}%
\AgdaSymbol{|}\AgdaSpace{}%
\AgdaInductiveConstructor{true}\AgdaSpace{}%
\AgdaSymbol{=}\AgdaSpace{}%
\AgdaInductiveConstructor{tt}\<%
\\
\\[\AgdaEmptyExtraSkip]%
\\[\AgdaEmptyExtraSkip]%
\>[0]\AgdaFunction{lemmaHoareTripleStackGeAux'19}%
\>[4071I]\AgdaSymbol{:}\AgdaSpace{}%
\AgdaSymbol{(}\AgdaBound{msg}\AgdaSpace{}%
\AgdaSymbol{:}\AgdaSpace{}%
\AgdaDatatype{Msg}\AgdaSymbol{)}\<%
\\
\>[4071I][@{}l@{\AgdaIndent{0}}]%
\>[31]\AgdaSymbol{(}\AgdaBound{pbk1}\AgdaSpace{}%
\AgdaBound{pbk2}\AgdaSpace{}%
\AgdaBound{pbk3}\AgdaSpace{}%
\AgdaBound{pbk4}\AgdaSpace{}%
\AgdaBound{pbk5}\AgdaSpace{}%
\AgdaBound{sig1}\AgdaSpace{}%
\AgdaBound{sig2}\AgdaSpace{}%
\AgdaBound{sig3}\AgdaSpace{}%
\AgdaSymbol{:}\AgdaSpace{}%
\AgdaDatatype{ℕ}\AgdaSymbol{)}\<%
\\
\>[31]\AgdaSymbol{→}\AgdaSpace{}%
\AgdaFunction{True}\AgdaSpace{}%
\AgdaSymbol{(}\AgdaFunction{isSigned}%
\>[49]\AgdaBound{msg}\AgdaSpace{}%
\AgdaBound{sig3}\AgdaSpace{}%
\AgdaBound{pbk5}\AgdaSymbol{)}\<%
\\
\>[.][@{}l@{}]\<[4071I]%
\>[30]\AgdaSymbol{→}%
\>[33]\AgdaFunction{True}\AgdaSpace{}%
\AgdaSymbol{(}\AgdaFunction{isSigned}%
\>[49]\AgdaBound{msg}\AgdaSpace{}%
\AgdaBound{sig2}\AgdaSpace{}%
\AgdaBound{pbk4}\AgdaSymbol{)}\<%
\\
\>[30][@{}l@{\AgdaIndent{0}}]%
\>[31]\AgdaSymbol{→}%
\>[34]\AgdaFunction{True}\AgdaSpace{}%
\AgdaSymbol{(}\AgdaFunction{isSigned}%
\>[50]\AgdaBound{msg}\AgdaSpace{}%
\AgdaBound{sig1}\AgdaSpace{}%
\AgdaBound{pbk1}\AgdaSymbol{)}\<%
\\
\>[31]\AgdaSymbol{→}\AgdaSpace{}%
\AgdaFunction{True}\AgdaSpace{}%
\AgdaSymbol{(}\AgdaFunction{compareSigsMultiSig}\AgdaSpace{}%
\AgdaBound{msg}\AgdaSpace{}%
\AgdaSymbol{(}\AgdaSpace{}%
\AgdaBound{sig1}\AgdaSpace{}%
\AgdaOperator{\AgdaInductiveConstructor{∷}}\AgdaSpace{}%
\AgdaBound{sig2}\AgdaSpace{}%
\AgdaOperator{\AgdaInductiveConstructor{∷}}\AgdaSpace{}%
\AgdaBound{sig3}\AgdaSpace{}%
\AgdaOperator{\AgdaInductiveConstructor{∷}}\AgdaSpace{}%
\AgdaInductiveConstructor{[]}\AgdaSymbol{)}\AgdaSpace{}%
\AgdaSymbol{(}\AgdaBound{pbk1}\AgdaSpace{}%
\AgdaOperator{\AgdaInductiveConstructor{∷}}\AgdaSpace{}%
\AgdaBound{pbk2}\AgdaSpace{}%
\AgdaOperator{\AgdaInductiveConstructor{∷}}\AgdaSpace{}%
\AgdaBound{pbk3}\AgdaSpace{}%
\AgdaOperator{\AgdaInductiveConstructor{∷}}\AgdaSpace{}%
\AgdaBound{pbk4}\AgdaSpace{}%
\AgdaOperator{\AgdaInductiveConstructor{∷}}\AgdaSpace{}%
\AgdaBound{pbk5}\AgdaSpace{}%
\AgdaOperator{\AgdaInductiveConstructor{∷}}\AgdaSpace{}%
\AgdaInductiveConstructor{[]}\AgdaSymbol{))}\<%
\\
\>[0]\AgdaFunction{lemmaHoareTripleStackGeAux'19}\AgdaSpace{}%
\AgdaBound{msg₁}\AgdaSpace{}%
\AgdaBound{pbk1}\AgdaSpace{}%
\AgdaBound{pbk2}\AgdaSpace{}%
\AgdaBound{pbk3}\AgdaSpace{}%
\AgdaBound{pbk4}\AgdaSpace{}%
\AgdaBound{pbk5}\AgdaSpace{}%
\AgdaBound{sig1}\AgdaSpace{}%
\AgdaBound{sig2}\AgdaSpace{}%
\AgdaBound{sig3}\AgdaSpace{}%
\AgdaBound{x}\AgdaSpace{}%
\AgdaBound{x₁}\AgdaSpace{}%
\AgdaBound{x₂}\AgdaSpace{}%
\AgdaKeyword{with}%
\>[89]\AgdaSymbol{(}\AgdaFunction{isSigned}%
\>[100]\AgdaBound{msg₁}\AgdaSpace{}%
\AgdaBound{sig1}\AgdaSpace{}%
\AgdaBound{pbk1}\AgdaSymbol{)}\<%
\\
\>[0]\AgdaFunction{lemmaHoareTripleStackGeAux'19}\AgdaSpace{}%
\AgdaBound{msg₁}\AgdaSpace{}%
\AgdaBound{pbk1}\AgdaSpace{}%
\AgdaBound{pbk2}\AgdaSpace{}%
\AgdaBound{pbk3}\AgdaSpace{}%
\AgdaBound{pbk4}\AgdaSpace{}%
\AgdaBound{pbk5}\AgdaSpace{}%
\AgdaBound{sig1}\AgdaSpace{}%
\AgdaBound{sig2}\AgdaSpace{}%
\AgdaBound{sig3}\AgdaSpace{}%
\AgdaBound{x}\AgdaSpace{}%
\AgdaBound{x₁}\AgdaSpace{}%
\AgdaBound{x₂}\AgdaSpace{}%
\AgdaSymbol{|}\AgdaSpace{}%
\AgdaInductiveConstructor{true}\AgdaSpace{}%
\AgdaKeyword{with}\AgdaSpace{}%
\AgdaSymbol{(}\AgdaFunction{isSigned}%
\>[106]\AgdaBound{msg₁}\AgdaSpace{}%
\AgdaBound{sig2}\AgdaSpace{}%
\AgdaBound{pbk2}\AgdaSymbol{)}\<%
\\
\>[0]\AgdaFunction{lemmaHoareTripleStackGeAux'19}\AgdaSpace{}%
\AgdaBound{msg₁}\AgdaSpace{}%
\AgdaBound{pbk1}\AgdaSpace{}%
\AgdaBound{pbk2}\AgdaSpace{}%
\AgdaBound{pbk3}\AgdaSpace{}%
\AgdaBound{pbk4}\AgdaSpace{}%
\AgdaBound{pbk5}\AgdaSpace{}%
\AgdaBound{sig1}\AgdaSpace{}%
\AgdaBound{sig2}\AgdaSpace{}%
\AgdaBound{sig3}\AgdaSpace{}%
\AgdaBound{x}\AgdaSpace{}%
\AgdaBound{x₁}\AgdaSpace{}%
\AgdaBound{x₂}\AgdaSpace{}%
\AgdaSymbol{|}\AgdaSpace{}%
\AgdaInductiveConstructor{true}\AgdaSpace{}%
\AgdaSymbol{|}\AgdaSpace{}%
\AgdaInductiveConstructor{false}\AgdaSpace{}%
\AgdaKeyword{with}%
\>[104]\AgdaSymbol{(}\AgdaFunction{isSigned}%
\>[115]\AgdaBound{msg₁}\AgdaSpace{}%
\AgdaBound{sig2}\AgdaSpace{}%
\AgdaBound{pbk3}\AgdaSymbol{)}\<%
\\
\>[0]\AgdaFunction{lemmaHoareTripleStackGeAux'19}\AgdaSpace{}%
\AgdaBound{msg₁}\AgdaSpace{}%
\AgdaBound{pbk1}\AgdaSpace{}%
\AgdaBound{pbk2}\AgdaSpace{}%
\AgdaBound{pbk3}\AgdaSpace{}%
\AgdaBound{pbk4}\AgdaSpace{}%
\AgdaBound{pbk5}\AgdaSpace{}%
\AgdaBound{sig1}\AgdaSpace{}%
\AgdaBound{sig2}\AgdaSpace{}%
\AgdaBound{sig3}\AgdaSpace{}%
\AgdaBound{x}\AgdaSpace{}%
\AgdaBound{x₁}\AgdaSpace{}%
\AgdaBound{x₂}\AgdaSpace{}%
\AgdaSymbol{|}\AgdaSpace{}%
\AgdaInductiveConstructor{true}\AgdaSpace{}%
\AgdaSymbol{|}\AgdaSpace{}%
\AgdaInductiveConstructor{false}\AgdaSpace{}%
\AgdaSymbol{|}\AgdaSpace{}%
\AgdaInductiveConstructor{false}\AgdaSpace{}%
\AgdaKeyword{with}\AgdaSpace{}%
\AgdaSymbol{(}\AgdaFunction{isSigned}%
\>[122]\AgdaBound{msg₁}\AgdaSpace{}%
\AgdaBound{sig2}\AgdaSpace{}%
\AgdaBound{pbk4}\AgdaSymbol{)}\<%
\\
\>[0]\AgdaFunction{lemmaHoareTripleStackGeAux'19}\AgdaSpace{}%
\AgdaBound{msg₁}\AgdaSpace{}%
\AgdaBound{pbk1}\AgdaSpace{}%
\AgdaBound{pbk2}\AgdaSpace{}%
\AgdaBound{pbk3}\AgdaSpace{}%
\AgdaBound{pbk4}\AgdaSpace{}%
\AgdaBound{pbk5}\AgdaSpace{}%
\AgdaBound{sig1}\AgdaSpace{}%
\AgdaBound{sig2}\AgdaSpace{}%
\AgdaBound{sig3}\AgdaSpace{}%
\AgdaBound{x}\AgdaSpace{}%
\AgdaBound{x₁}\AgdaSpace{}%
\AgdaBound{x₂}\AgdaSpace{}%
\AgdaSymbol{|}\AgdaSpace{}%
\AgdaInductiveConstructor{true}\AgdaSpace{}%
\AgdaSymbol{|}\AgdaSpace{}%
\AgdaInductiveConstructor{false}\AgdaSpace{}%
\AgdaSymbol{|}\AgdaSpace{}%
\AgdaInductiveConstructor{false}\AgdaSpace{}%
\AgdaSymbol{|}\AgdaSpace{}%
\AgdaInductiveConstructor{true}\AgdaSpace{}%
\AgdaKeyword{with}%
\>[119]\AgdaSymbol{(}\AgdaFunction{isSigned}%
\>[130]\AgdaBound{msg₁}\AgdaSpace{}%
\AgdaBound{sig3}\AgdaSpace{}%
\AgdaBound{pbk5}\AgdaSymbol{)}\<%
\\
\>[0]\AgdaFunction{lemmaHoareTripleStackGeAux'19}\AgdaSpace{}%
\AgdaBound{msg₁}\AgdaSpace{}%
\AgdaBound{pbk1}\AgdaSpace{}%
\AgdaBound{pbk2}\AgdaSpace{}%
\AgdaBound{pbk3}\AgdaSpace{}%
\AgdaBound{pbk4}\AgdaSpace{}%
\AgdaBound{pbk5}\AgdaSpace{}%
\AgdaBound{sig1}\AgdaSpace{}%
\AgdaBound{sig2}\AgdaSpace{}%
\AgdaBound{sig3}\AgdaSpace{}%
\AgdaBound{x}\AgdaSpace{}%
\AgdaBound{x₁}\AgdaSpace{}%
\AgdaBound{x₂}\AgdaSpace{}%
\AgdaSymbol{|}\AgdaSpace{}%
\AgdaInductiveConstructor{true}\AgdaSpace{}%
\AgdaSymbol{|}\AgdaSpace{}%
\AgdaInductiveConstructor{false}\AgdaSpace{}%
\AgdaSymbol{|}\AgdaSpace{}%
\AgdaInductiveConstructor{false}\AgdaSpace{}%
\AgdaSymbol{|}\AgdaSpace{}%
\AgdaInductiveConstructor{true}\AgdaSpace{}%
\AgdaSymbol{|}\AgdaSpace{}%
\AgdaInductiveConstructor{true}\AgdaSpace{}%
\AgdaSymbol{=}\AgdaSpace{}%
\AgdaInductiveConstructor{tt}\<%
\\
\>[0]\AgdaFunction{lemmaHoareTripleStackGeAux'19}\AgdaSpace{}%
\AgdaBound{msg₁}\AgdaSpace{}%
\AgdaBound{pbk1}\AgdaSpace{}%
\AgdaBound{pbk2}\AgdaSpace{}%
\AgdaBound{pbk3}\AgdaSpace{}%
\AgdaBound{pbk4}\AgdaSpace{}%
\AgdaBound{pbk5}\AgdaSpace{}%
\AgdaBound{sig1}\AgdaSpace{}%
\AgdaBound{sig2}\AgdaSpace{}%
\AgdaBound{sig3}\AgdaSpace{}%
\AgdaBound{x}\AgdaSpace{}%
\AgdaBound{x₁}\AgdaSpace{}%
\AgdaBound{x₂}\AgdaSpace{}%
\AgdaSymbol{|}\AgdaSpace{}%
\AgdaInductiveConstructor{true}\AgdaSpace{}%
\AgdaSymbol{|}\AgdaSpace{}%
\AgdaInductiveConstructor{false}\AgdaSpace{}%
\AgdaSymbol{|}\AgdaSpace{}%
\AgdaInductiveConstructor{true}\AgdaSpace{}%
\AgdaKeyword{with}\AgdaSpace{}%
\AgdaSymbol{(}\AgdaFunction{isSigned}%
\>[121]\AgdaBound{msg₁}\AgdaSpace{}%
\AgdaBound{sig3}\AgdaSpace{}%
\AgdaBound{pbk4}\AgdaSymbol{)}\<%
\\
\>[0]\AgdaFunction{lemmaHoareTripleStackGeAux'19}\AgdaSpace{}%
\AgdaBound{msg₁}\AgdaSpace{}%
\AgdaBound{pbk1}\AgdaSpace{}%
\AgdaBound{pbk2}\AgdaSpace{}%
\AgdaBound{pbk3}\AgdaSpace{}%
\AgdaBound{pbk4}\AgdaSpace{}%
\AgdaBound{pbk5}\AgdaSpace{}%
\AgdaBound{sig1}\AgdaSpace{}%
\AgdaBound{sig2}\AgdaSpace{}%
\AgdaBound{sig3}\AgdaSpace{}%
\AgdaBound{x}\AgdaSpace{}%
\AgdaBound{x₁}\AgdaSpace{}%
\AgdaBound{x₂}\AgdaSpace{}%
\AgdaSymbol{|}\AgdaSpace{}%
\AgdaInductiveConstructor{true}\AgdaSpace{}%
\AgdaSymbol{|}\AgdaSpace{}%
\AgdaInductiveConstructor{false}\AgdaSpace{}%
\AgdaSymbol{|}\AgdaSpace{}%
\AgdaInductiveConstructor{true}\AgdaSpace{}%
\AgdaSymbol{|}\AgdaSpace{}%
\AgdaInductiveConstructor{false}\AgdaSpace{}%
\AgdaKeyword{with}%
\>[120]\AgdaSymbol{(}\AgdaFunction{isSigned}%
\>[131]\AgdaBound{msg₁}\AgdaSpace{}%
\AgdaBound{sig3}\AgdaSpace{}%
\AgdaBound{pbk5}\AgdaSymbol{)}\<%
\\
\>[0]\AgdaFunction{lemmaHoareTripleStackGeAux'19}\AgdaSpace{}%
\AgdaBound{msg₁}\AgdaSpace{}%
\AgdaBound{pbk1}\AgdaSpace{}%
\AgdaBound{pbk2}\AgdaSpace{}%
\AgdaBound{pbk3}\AgdaSpace{}%
\AgdaBound{pbk4}\AgdaSpace{}%
\AgdaBound{pbk5}\AgdaSpace{}%
\AgdaBound{sig1}\AgdaSpace{}%
\AgdaBound{sig2}\AgdaSpace{}%
\AgdaBound{sig3}\AgdaSpace{}%
\AgdaBound{x}\AgdaSpace{}%
\AgdaBound{x₁}\AgdaSpace{}%
\AgdaBound{x₂}\AgdaSpace{}%
\AgdaSymbol{|}\AgdaSpace{}%
\AgdaInductiveConstructor{true}\AgdaSpace{}%
\AgdaSymbol{|}\AgdaSpace{}%
\AgdaInductiveConstructor{false}\AgdaSpace{}%
\AgdaSymbol{|}\AgdaSpace{}%
\AgdaInductiveConstructor{true}\AgdaSpace{}%
\AgdaSymbol{|}\AgdaSpace{}%
\AgdaInductiveConstructor{false}\AgdaSpace{}%
\AgdaSymbol{|}\AgdaSpace{}%
\AgdaInductiveConstructor{true}\AgdaSpace{}%
\AgdaSymbol{=}\AgdaSpace{}%
\AgdaInductiveConstructor{tt}\<%
\\
\>[0]\AgdaFunction{lemmaHoareTripleStackGeAux'19}\AgdaSpace{}%
\AgdaBound{msg₁}\AgdaSpace{}%
\AgdaBound{pbk1}\AgdaSpace{}%
\AgdaBound{pbk2}\AgdaSpace{}%
\AgdaBound{pbk3}\AgdaSpace{}%
\AgdaBound{pbk4}\AgdaSpace{}%
\AgdaBound{pbk5}\AgdaSpace{}%
\AgdaBound{sig1}\AgdaSpace{}%
\AgdaBound{sig2}\AgdaSpace{}%
\AgdaBound{sig3}\AgdaSpace{}%
\AgdaBound{x}\AgdaSpace{}%
\AgdaBound{x₁}\AgdaSpace{}%
\AgdaBound{x₂}\AgdaSpace{}%
\AgdaSymbol{|}\AgdaSpace{}%
\AgdaInductiveConstructor{true}\AgdaSpace{}%
\AgdaSymbol{|}\AgdaSpace{}%
\AgdaInductiveConstructor{false}\AgdaSpace{}%
\AgdaSymbol{|}\AgdaSpace{}%
\AgdaInductiveConstructor{true}\AgdaSpace{}%
\AgdaSymbol{|}\AgdaSpace{}%
\AgdaInductiveConstructor{true}\AgdaSpace{}%
\AgdaSymbol{=}\AgdaSpace{}%
\AgdaInductiveConstructor{tt}\<%
\\
\>[0]\AgdaFunction{lemmaHoareTripleStackGeAux'19}\AgdaSpace{}%
\AgdaBound{msg₁}\AgdaSpace{}%
\AgdaBound{pbk1}\AgdaSpace{}%
\AgdaBound{pbk2}\AgdaSpace{}%
\AgdaBound{pbk3}\AgdaSpace{}%
\AgdaBound{pbk4}\AgdaSpace{}%
\AgdaBound{pbk5}\AgdaSpace{}%
\AgdaBound{sig1}\AgdaSpace{}%
\AgdaBound{sig2}\AgdaSpace{}%
\AgdaBound{sig3}\AgdaSpace{}%
\AgdaBound{x}\AgdaSpace{}%
\AgdaBound{x₁}\AgdaSpace{}%
\AgdaBound{x₂}\AgdaSpace{}%
\AgdaSymbol{|}\AgdaSpace{}%
\AgdaInductiveConstructor{true}\AgdaSpace{}%
\AgdaSymbol{|}\AgdaSpace{}%
\AgdaInductiveConstructor{true}\AgdaSpace{}%
\AgdaKeyword{with}\AgdaSpace{}%
\AgdaSymbol{(}\AgdaFunction{isSigned}%
\>[113]\AgdaBound{msg₁}\AgdaSpace{}%
\AgdaBound{sig3}\AgdaSpace{}%
\AgdaBound{pbk3}\AgdaSymbol{)}\<%
\\
\>[0]\AgdaFunction{lemmaHoareTripleStackGeAux'19}\AgdaSpace{}%
\AgdaBound{msg₁}\AgdaSpace{}%
\AgdaBound{pbk1}\AgdaSpace{}%
\AgdaBound{pbk2}\AgdaSpace{}%
\AgdaBound{pbk3}\AgdaSpace{}%
\AgdaBound{pbk4}\AgdaSpace{}%
\AgdaBound{pbk5}\AgdaSpace{}%
\AgdaBound{sig1}\AgdaSpace{}%
\AgdaBound{sig2}\AgdaSpace{}%
\AgdaBound{sig3}\AgdaSpace{}%
\AgdaBound{x}\AgdaSpace{}%
\AgdaBound{x₁}\AgdaSpace{}%
\AgdaBound{x₂}\AgdaSpace{}%
\AgdaSymbol{|}\AgdaSpace{}%
\AgdaInductiveConstructor{true}\AgdaSpace{}%
\AgdaSymbol{|}\AgdaSpace{}%
\AgdaInductiveConstructor{true}\AgdaSpace{}%
\AgdaSymbol{|}\AgdaSpace{}%
\AgdaInductiveConstructor{false}\AgdaSpace{}%
\AgdaKeyword{with}\AgdaSpace{}%
\AgdaSymbol{(}\AgdaFunction{isSigned}%
\>[121]\AgdaBound{msg₁}\AgdaSpace{}%
\AgdaBound{sig3}\AgdaSpace{}%
\AgdaBound{pbk4}\AgdaSymbol{)}\<%
\\
\>[0]\AgdaFunction{lemmaHoareTripleStackGeAux'19}\AgdaSpace{}%
\AgdaBound{msg₁}\AgdaSpace{}%
\AgdaBound{pbk1}\AgdaSpace{}%
\AgdaBound{pbk2}\AgdaSpace{}%
\AgdaBound{pbk3}\AgdaSpace{}%
\AgdaBound{pbk4}\AgdaSpace{}%
\AgdaBound{pbk5}\AgdaSpace{}%
\AgdaBound{sig1}\AgdaSpace{}%
\AgdaBound{sig2}\AgdaSpace{}%
\AgdaBound{sig3}\AgdaSpace{}%
\AgdaBound{x}\AgdaSpace{}%
\AgdaBound{x₁}\AgdaSpace{}%
\AgdaBound{x₂}\AgdaSpace{}%
\AgdaSymbol{|}\AgdaSpace{}%
\AgdaInductiveConstructor{true}\AgdaSpace{}%
\AgdaSymbol{|}\AgdaSpace{}%
\AgdaInductiveConstructor{true}\AgdaSpace{}%
\AgdaSymbol{|}\AgdaSpace{}%
\AgdaInductiveConstructor{false}\AgdaSpace{}%
\AgdaSymbol{|}\AgdaSpace{}%
\AgdaInductiveConstructor{false}\AgdaSpace{}%
\AgdaKeyword{with}%
\>[119]\AgdaSymbol{(}\AgdaFunction{isSigned}%
\>[130]\AgdaBound{msg₁}\AgdaSpace{}%
\AgdaBound{sig3}\AgdaSpace{}%
\AgdaBound{pbk5}\AgdaSymbol{)}\<%
\\
\>[0]\AgdaFunction{lemmaHoareTripleStackGeAux'19}\AgdaSpace{}%
\AgdaBound{msg₁}\AgdaSpace{}%
\AgdaBound{pbk1}\AgdaSpace{}%
\AgdaBound{pbk2}\AgdaSpace{}%
\AgdaBound{pbk3}\AgdaSpace{}%
\AgdaBound{pbk4}\AgdaSpace{}%
\AgdaBound{pbk5}\AgdaSpace{}%
\AgdaBound{sig1}\AgdaSpace{}%
\AgdaBound{sig2}\AgdaSpace{}%
\AgdaBound{sig3}\AgdaSpace{}%
\AgdaBound{x}\AgdaSpace{}%
\AgdaBound{x₁}\AgdaSpace{}%
\AgdaBound{x₂}\AgdaSpace{}%
\AgdaSymbol{|}\AgdaSpace{}%
\AgdaInductiveConstructor{true}\AgdaSpace{}%
\AgdaSymbol{|}\AgdaSpace{}%
\AgdaInductiveConstructor{true}\AgdaSpace{}%
\AgdaSymbol{|}\AgdaSpace{}%
\AgdaInductiveConstructor{false}\AgdaSpace{}%
\AgdaSymbol{|}\AgdaSpace{}%
\AgdaInductiveConstructor{false}\AgdaSpace{}%
\AgdaSymbol{|}\AgdaSpace{}%
\AgdaInductiveConstructor{true}\AgdaSpace{}%
\AgdaSymbol{=}\AgdaSpace{}%
\AgdaInductiveConstructor{tt}\<%
\\
\>[0]\AgdaFunction{lemmaHoareTripleStackGeAux'19}\AgdaSpace{}%
\AgdaBound{msg₁}\AgdaSpace{}%
\AgdaBound{pbk1}\AgdaSpace{}%
\AgdaBound{pbk2}\AgdaSpace{}%
\AgdaBound{pbk3}\AgdaSpace{}%
\AgdaBound{pbk4}\AgdaSpace{}%
\AgdaBound{pbk5}\AgdaSpace{}%
\AgdaBound{sig1}\AgdaSpace{}%
\AgdaBound{sig2}\AgdaSpace{}%
\AgdaBound{sig3}\AgdaSpace{}%
\AgdaBound{x}\AgdaSpace{}%
\AgdaBound{x₁}\AgdaSpace{}%
\AgdaBound{x₂}\AgdaSpace{}%
\AgdaSymbol{|}\AgdaSpace{}%
\AgdaInductiveConstructor{true}\AgdaSpace{}%
\AgdaSymbol{|}\AgdaSpace{}%
\AgdaInductiveConstructor{true}\AgdaSpace{}%
\AgdaSymbol{|}\AgdaSpace{}%
\AgdaInductiveConstructor{false}\AgdaSpace{}%
\AgdaSymbol{|}\AgdaSpace{}%
\AgdaInductiveConstructor{true}\AgdaSpace{}%
\AgdaSymbol{=}\AgdaSpace{}%
\AgdaInductiveConstructor{tt}\<%
\\
\>[0]\AgdaFunction{lemmaHoareTripleStackGeAux'19}\AgdaSpace{}%
\AgdaBound{msg₁}\AgdaSpace{}%
\AgdaBound{pbk1}\AgdaSpace{}%
\AgdaBound{pbk2}\AgdaSpace{}%
\AgdaBound{pbk3}\AgdaSpace{}%
\AgdaBound{pbk4}\AgdaSpace{}%
\AgdaBound{pbk5}\AgdaSpace{}%
\AgdaBound{sig1}\AgdaSpace{}%
\AgdaBound{sig2}\AgdaSpace{}%
\AgdaBound{sig3}\AgdaSpace{}%
\AgdaBound{x}\AgdaSpace{}%
\AgdaBound{x₁}\AgdaSpace{}%
\AgdaBound{x₂}\AgdaSpace{}%
\AgdaSymbol{|}\AgdaSpace{}%
\AgdaInductiveConstructor{true}\AgdaSpace{}%
\AgdaSymbol{|}\AgdaSpace{}%
\AgdaInductiveConstructor{true}\AgdaSpace{}%
\AgdaSymbol{|}\AgdaSpace{}%
\AgdaInductiveConstructor{true}\AgdaSpace{}%
\AgdaSymbol{=}\AgdaSpace{}%
\AgdaInductiveConstructor{tt}\<%
\\
\\[\AgdaEmptyExtraSkip]%
\>[0]\AgdaFunction{lemmaHoareTripleStackGeAux'20}%
\>[4459I]\AgdaSymbol{:}\AgdaSpace{}%
\AgdaSymbol{(}\AgdaBound{msg}\AgdaSpace{}%
\AgdaSymbol{:}\AgdaSpace{}%
\AgdaDatatype{Msg}\AgdaSymbol{)}\<%
\\
\>[4459I][@{}l@{\AgdaIndent{0}}]%
\>[31]\AgdaSymbol{(}\AgdaBound{pbk1}\AgdaSpace{}%
\AgdaBound{pbk2}\AgdaSpace{}%
\AgdaBound{pbk3}\AgdaSpace{}%
\AgdaBound{pbk4}\AgdaSpace{}%
\AgdaBound{pbk5}\AgdaSpace{}%
\AgdaBound{sig1}\AgdaSpace{}%
\AgdaBound{sig2}\AgdaSpace{}%
\AgdaBound{sig3}\AgdaSpace{}%
\AgdaSymbol{:}\AgdaSpace{}%
\AgdaDatatype{ℕ}\AgdaSymbol{)}\<%
\\
\>[31]\AgdaSymbol{→}\AgdaSpace{}%
\AgdaFunction{True}\AgdaSpace{}%
\AgdaSymbol{(}\AgdaFunction{isSigned}%
\>[49]\AgdaBound{msg}\AgdaSpace{}%
\AgdaBound{sig3}\AgdaSpace{}%
\AgdaBound{pbk4}\AgdaSymbol{)}\<%
\\
\>[.][@{}l@{}]\<[4459I]%
\>[30]\AgdaSymbol{→}%
\>[33]\AgdaFunction{True}\AgdaSpace{}%
\AgdaSymbol{(}\AgdaFunction{isSigned}%
\>[49]\AgdaBound{msg}\AgdaSpace{}%
\AgdaBound{sig2}\AgdaSpace{}%
\AgdaBound{pbk3}\AgdaSymbol{)}\<%
\\
\>[30]\AgdaSymbol{→}%
\>[33]\AgdaFunction{True}\AgdaSpace{}%
\AgdaSymbol{(}\AgdaFunction{isSigned}%
\>[49]\AgdaBound{msg}\AgdaSpace{}%
\AgdaBound{sig1}\AgdaSpace{}%
\AgdaBound{pbk2}\AgdaSymbol{)}\<%
\\
\>[30][@{}l@{\AgdaIndent{0}}]%
\>[31]\AgdaSymbol{→}\AgdaSpace{}%
\AgdaFunction{True}\AgdaSpace{}%
\AgdaSymbol{(}\AgdaFunction{compareSigsMultiSig}\AgdaSpace{}%
\AgdaBound{msg}\AgdaSpace{}%
\AgdaSymbol{(}\AgdaSpace{}%
\AgdaBound{sig1}\AgdaSpace{}%
\AgdaOperator{\AgdaInductiveConstructor{∷}}\AgdaSpace{}%
\AgdaBound{sig2}\AgdaSpace{}%
\AgdaOperator{\AgdaInductiveConstructor{∷}}\AgdaSpace{}%
\AgdaBound{sig3}\AgdaSpace{}%
\AgdaOperator{\AgdaInductiveConstructor{∷}}\AgdaSpace{}%
\AgdaInductiveConstructor{[]}\AgdaSymbol{)}\AgdaSpace{}%
\AgdaSymbol{(}\AgdaBound{pbk1}\AgdaSpace{}%
\AgdaOperator{\AgdaInductiveConstructor{∷}}\AgdaSpace{}%
\AgdaBound{pbk2}\AgdaSpace{}%
\AgdaOperator{\AgdaInductiveConstructor{∷}}\AgdaSpace{}%
\AgdaBound{pbk3}\AgdaSpace{}%
\AgdaOperator{\AgdaInductiveConstructor{∷}}\AgdaSpace{}%
\AgdaBound{pbk4}\AgdaSpace{}%
\AgdaOperator{\AgdaInductiveConstructor{∷}}\AgdaSpace{}%
\AgdaBound{pbk5}\AgdaSpace{}%
\AgdaOperator{\AgdaInductiveConstructor{∷}}\AgdaSpace{}%
\AgdaInductiveConstructor{[]}\AgdaSymbol{))}\<%
\\
\>[0]\AgdaFunction{lemmaHoareTripleStackGeAux'20}\AgdaSpace{}%
\AgdaBound{msg₁}\AgdaSpace{}%
\AgdaBound{pbk1}\AgdaSpace{}%
\AgdaBound{pbk2}\AgdaSpace{}%
\AgdaBound{pbk3}\AgdaSpace{}%
\AgdaBound{pbk4}\AgdaSpace{}%
\AgdaBound{pbk5}\AgdaSpace{}%
\AgdaBound{sig1}\AgdaSpace{}%
\AgdaBound{sig2}\AgdaSpace{}%
\AgdaBound{sig3}\AgdaSpace{}%
\AgdaBound{x}\AgdaSpace{}%
\AgdaBound{x₁}\AgdaSpace{}%
\AgdaBound{x₂}\AgdaSpace{}%
\AgdaKeyword{with}\AgdaSpace{}%
\AgdaSymbol{(}\AgdaFunction{isSigned}%
\>[99]\AgdaBound{msg₁}\AgdaSpace{}%
\AgdaBound{sig1}\AgdaSpace{}%
\AgdaBound{pbk1}\AgdaSymbol{)}\<%
\\
\>[0]\AgdaFunction{lemmaHoareTripleStackGeAux'20}\AgdaSpace{}%
\AgdaBound{msg₁}\AgdaSpace{}%
\AgdaBound{pbk1}\AgdaSpace{}%
\AgdaBound{pbk2}\AgdaSpace{}%
\AgdaBound{pbk3}\AgdaSpace{}%
\AgdaBound{pbk4}\AgdaSpace{}%
\AgdaBound{pbk5}\AgdaSpace{}%
\AgdaBound{sig1}\AgdaSpace{}%
\AgdaBound{sig2}\AgdaSpace{}%
\AgdaBound{sig3}\AgdaSpace{}%
\AgdaBound{x}\AgdaSpace{}%
\AgdaBound{x₁}\AgdaSpace{}%
\AgdaBound{x₂}\AgdaSpace{}%
\AgdaSymbol{|}\AgdaSpace{}%
\AgdaInductiveConstructor{false}\AgdaSpace{}%
\AgdaKeyword{with}\AgdaSpace{}%
\AgdaSymbol{(}\AgdaFunction{isSigned}%
\>[107]\AgdaBound{msg₁}\AgdaSpace{}%
\AgdaBound{sig1}\AgdaSpace{}%
\AgdaBound{pbk2}\AgdaSymbol{)}\<%
\\
\>[0]\AgdaFunction{lemmaHoareTripleStackGeAux'20}\AgdaSpace{}%
\AgdaBound{msg₁}\AgdaSpace{}%
\AgdaBound{pbk1}\AgdaSpace{}%
\AgdaBound{pbk2}\AgdaSpace{}%
\AgdaBound{pbk3}\AgdaSpace{}%
\AgdaBound{pbk4}\AgdaSpace{}%
\AgdaBound{pbk5}\AgdaSpace{}%
\AgdaBound{sig1}\AgdaSpace{}%
\AgdaBound{sig2}\AgdaSpace{}%
\AgdaBound{sig3}\AgdaSpace{}%
\AgdaBound{x}\AgdaSpace{}%
\AgdaBound{x₁}\AgdaSpace{}%
\AgdaBound{x₂}\AgdaSpace{}%
\AgdaSymbol{|}\AgdaSpace{}%
\AgdaInductiveConstructor{false}\AgdaSpace{}%
\AgdaSymbol{|}\AgdaSpace{}%
\AgdaInductiveConstructor{true}\AgdaSpace{}%
\AgdaKeyword{with}%
\>[104]\AgdaSymbol{(}\AgdaFunction{isSigned}%
\>[115]\AgdaBound{msg₁}\AgdaSpace{}%
\AgdaBound{sig2}\AgdaSpace{}%
\AgdaBound{pbk3}\AgdaSymbol{)}\<%
\\
\>[0]\AgdaFunction{lemmaHoareTripleStackGeAux'20}\AgdaSpace{}%
\AgdaBound{msg₁}\AgdaSpace{}%
\AgdaBound{pbk1}\AgdaSpace{}%
\AgdaBound{pbk2}\AgdaSpace{}%
\AgdaBound{pbk3}\AgdaSpace{}%
\AgdaBound{pbk4}\AgdaSpace{}%
\AgdaBound{pbk5}\AgdaSpace{}%
\AgdaBound{sig1}\AgdaSpace{}%
\AgdaBound{sig2}\AgdaSpace{}%
\AgdaBound{sig3}\AgdaSpace{}%
\AgdaBound{x}\AgdaSpace{}%
\AgdaBound{x₁}\AgdaSpace{}%
\AgdaBound{x₂}\AgdaSpace{}%
\AgdaSymbol{|}\AgdaSpace{}%
\AgdaInductiveConstructor{false}\AgdaSpace{}%
\AgdaSymbol{|}\AgdaSpace{}%
\AgdaInductiveConstructor{true}\AgdaSpace{}%
\AgdaSymbol{|}\AgdaSpace{}%
\AgdaInductiveConstructor{true}\AgdaSpace{}%
\AgdaKeyword{with}\AgdaSpace{}%
\AgdaSymbol{(}\AgdaFunction{isSigned}%
\>[121]\AgdaBound{msg₁}\AgdaSpace{}%
\AgdaBound{sig3}\AgdaSpace{}%
\AgdaBound{pbk4}\AgdaSymbol{)}\<%
\\
\>[0]\AgdaFunction{lemmaHoareTripleStackGeAux'20}\AgdaSpace{}%
\AgdaBound{msg₁}\AgdaSpace{}%
\AgdaBound{pbk1}\AgdaSpace{}%
\AgdaBound{pbk2}\AgdaSpace{}%
\AgdaBound{pbk3}\AgdaSpace{}%
\AgdaBound{pbk4}\AgdaSpace{}%
\AgdaBound{pbk5}\AgdaSpace{}%
\AgdaBound{sig1}\AgdaSpace{}%
\AgdaBound{sig2}\AgdaSpace{}%
\AgdaBound{sig3}\AgdaSpace{}%
\AgdaBound{x}\AgdaSpace{}%
\AgdaBound{x₁}\AgdaSpace{}%
\AgdaBound{x₂}\AgdaSpace{}%
\AgdaSymbol{|}\AgdaSpace{}%
\AgdaInductiveConstructor{false}\AgdaSpace{}%
\AgdaSymbol{|}\AgdaSpace{}%
\AgdaInductiveConstructor{true}\AgdaSpace{}%
\AgdaSymbol{|}\AgdaSpace{}%
\AgdaInductiveConstructor{true}\AgdaSpace{}%
\AgdaSymbol{|}\AgdaSpace{}%
\AgdaInductiveConstructor{true}\AgdaSpace{}%
\AgdaSymbol{=}\AgdaSpace{}%
\AgdaInductiveConstructor{tt}\<%
\\
\>[0]\AgdaFunction{lemmaHoareTripleStackGeAux'20}\AgdaSpace{}%
\AgdaBound{msg₁}\AgdaSpace{}%
\AgdaBound{pbk1}\AgdaSpace{}%
\AgdaBound{pbk2}\AgdaSpace{}%
\AgdaBound{pbk3}\AgdaSpace{}%
\AgdaBound{pbk4}\AgdaSpace{}%
\AgdaBound{pbk5}\AgdaSpace{}%
\AgdaBound{sig1}\AgdaSpace{}%
\AgdaBound{sig2}\AgdaSpace{}%
\AgdaBound{sig3}\AgdaSpace{}%
\AgdaBound{x}\AgdaSpace{}%
\AgdaBound{x₁}\AgdaSpace{}%
\AgdaBound{x₂}\AgdaSpace{}%
\AgdaSymbol{|}\AgdaSpace{}%
\AgdaInductiveConstructor{true}\AgdaSpace{}%
\AgdaKeyword{with}\AgdaSpace{}%
\AgdaSymbol{(}\AgdaFunction{isSigned}%
\>[106]\AgdaBound{msg₁}\AgdaSpace{}%
\AgdaBound{sig2}\AgdaSpace{}%
\AgdaBound{pbk2}\AgdaSymbol{)}\<%
\\
\>[0]\AgdaFunction{lemmaHoareTripleStackGeAux'20}\AgdaSpace{}%
\AgdaBound{msg₁}\AgdaSpace{}%
\AgdaBound{pbk1}\AgdaSpace{}%
\AgdaBound{pbk2}\AgdaSpace{}%
\AgdaBound{pbk3}\AgdaSpace{}%
\AgdaBound{pbk4}\AgdaSpace{}%
\AgdaBound{pbk5}\AgdaSpace{}%
\AgdaBound{sig1}\AgdaSpace{}%
\AgdaBound{sig2}\AgdaSpace{}%
\AgdaBound{sig3}\AgdaSpace{}%
\AgdaBound{x}\AgdaSpace{}%
\AgdaBound{x₁}\AgdaSpace{}%
\AgdaBound{x₂}\AgdaSpace{}%
\AgdaSymbol{|}\AgdaSpace{}%
\AgdaInductiveConstructor{true}\AgdaSpace{}%
\AgdaSymbol{|}\AgdaSpace{}%
\AgdaInductiveConstructor{false}\AgdaSpace{}%
\AgdaKeyword{with}\AgdaSpace{}%
\AgdaSymbol{(}\AgdaFunction{isSigned}%
\>[114]\AgdaBound{msg₁}\AgdaSpace{}%
\AgdaBound{sig2}\AgdaSpace{}%
\AgdaBound{pbk3}\AgdaSymbol{)}\<%
\\
\>[0]\AgdaFunction{lemmaHoareTripleStackGeAux'20}\AgdaSpace{}%
\AgdaBound{msg₁}\AgdaSpace{}%
\AgdaBound{pbk1}\AgdaSpace{}%
\AgdaBound{pbk2}\AgdaSpace{}%
\AgdaBound{pbk3}\AgdaSpace{}%
\AgdaBound{pbk4}\AgdaSpace{}%
\AgdaBound{pbk5}\AgdaSpace{}%
\AgdaBound{sig1}\AgdaSpace{}%
\AgdaBound{sig2}\AgdaSpace{}%
\AgdaBound{sig3}\AgdaSpace{}%
\AgdaBound{x}\AgdaSpace{}%
\AgdaBound{x₁}\AgdaSpace{}%
\AgdaBound{x₂}\AgdaSpace{}%
\AgdaSymbol{|}\AgdaSpace{}%
\AgdaInductiveConstructor{true}\AgdaSpace{}%
\AgdaSymbol{|}\AgdaSpace{}%
\AgdaInductiveConstructor{false}\AgdaSpace{}%
\AgdaSymbol{|}\AgdaSpace{}%
\AgdaInductiveConstructor{true}\AgdaSpace{}%
\AgdaKeyword{with}\AgdaSpace{}%
\AgdaSymbol{(}\AgdaFunction{isSigned}%
\>[121]\AgdaBound{msg₁}\AgdaSpace{}%
\AgdaBound{sig3}\AgdaSpace{}%
\AgdaBound{pbk4}\AgdaSymbol{)}\<%
\\
\>[0]\AgdaFunction{lemmaHoareTripleStackGeAux'20}\AgdaSpace{}%
\AgdaBound{msg₁}\AgdaSpace{}%
\AgdaBound{pbk1}\AgdaSpace{}%
\AgdaBound{pbk2}\AgdaSpace{}%
\AgdaBound{pbk3}\AgdaSpace{}%
\AgdaBound{pbk4}\AgdaSpace{}%
\AgdaBound{pbk5}\AgdaSpace{}%
\AgdaBound{sig1}\AgdaSpace{}%
\AgdaBound{sig2}\AgdaSpace{}%
\AgdaBound{sig3}\AgdaSpace{}%
\AgdaBound{x}\AgdaSpace{}%
\AgdaBound{x₁}\AgdaSpace{}%
\AgdaBound{x₂}\AgdaSpace{}%
\AgdaSymbol{|}\AgdaSpace{}%
\AgdaInductiveConstructor{true}\AgdaSpace{}%
\AgdaSymbol{|}\AgdaSpace{}%
\AgdaInductiveConstructor{false}\AgdaSpace{}%
\AgdaSymbol{|}\AgdaSpace{}%
\AgdaInductiveConstructor{true}\AgdaSpace{}%
\AgdaSymbol{|}\AgdaSpace{}%
\AgdaInductiveConstructor{true}\AgdaSpace{}%
\AgdaSymbol{=}\AgdaSpace{}%
\AgdaInductiveConstructor{tt}\<%
\\
\>[0]\AgdaFunction{lemmaHoareTripleStackGeAux'20}\AgdaSpace{}%
\AgdaBound{msg₁}\AgdaSpace{}%
\AgdaBound{pbk1}\AgdaSpace{}%
\AgdaBound{pbk2}\AgdaSpace{}%
\AgdaBound{pbk3}\AgdaSpace{}%
\AgdaBound{pbk4}\AgdaSpace{}%
\AgdaBound{pbk5}\AgdaSpace{}%
\AgdaBound{sig1}\AgdaSpace{}%
\AgdaBound{sig2}\AgdaSpace{}%
\AgdaBound{sig3}\AgdaSpace{}%
\AgdaBound{x}\AgdaSpace{}%
\AgdaBound{x₁}\AgdaSpace{}%
\AgdaBound{x₂}\AgdaSpace{}%
\AgdaSymbol{|}\AgdaSpace{}%
\AgdaInductiveConstructor{true}\AgdaSpace{}%
\AgdaSymbol{|}\AgdaSpace{}%
\AgdaInductiveConstructor{true}\AgdaSpace{}%
\AgdaKeyword{with}\AgdaSpace{}%
\AgdaSymbol{(}\AgdaFunction{isSigned}%
\>[113]\AgdaBound{msg₁}\AgdaSpace{}%
\AgdaBound{sig3}\AgdaSpace{}%
\AgdaBound{pbk3}\AgdaSymbol{)}\<%
\\
\>[0]\AgdaFunction{lemmaHoareTripleStackGeAux'20}\AgdaSpace{}%
\AgdaBound{msg₁}\AgdaSpace{}%
\AgdaBound{pbk1}\AgdaSpace{}%
\AgdaBound{pbk2}\AgdaSpace{}%
\AgdaBound{pbk3}\AgdaSpace{}%
\AgdaBound{pbk4}\AgdaSpace{}%
\AgdaBound{pbk5}\AgdaSpace{}%
\AgdaBound{sig1}\AgdaSpace{}%
\AgdaBound{sig2}\AgdaSpace{}%
\AgdaBound{sig3}\AgdaSpace{}%
\AgdaBound{x}\AgdaSpace{}%
\AgdaBound{x₁}\AgdaSpace{}%
\AgdaBound{x₂}\AgdaSpace{}%
\AgdaSymbol{|}\AgdaSpace{}%
\AgdaInductiveConstructor{true}\AgdaSpace{}%
\AgdaSymbol{|}\AgdaSpace{}%
\AgdaInductiveConstructor{true}\AgdaSpace{}%
\AgdaSymbol{|}\AgdaSpace{}%
\AgdaInductiveConstructor{false}\AgdaSpace{}%
\AgdaKeyword{with}\AgdaSpace{}%
\AgdaSymbol{(}\AgdaFunction{isSigned}%
\>[121]\AgdaBound{msg₁}\AgdaSpace{}%
\AgdaBound{sig3}\AgdaSpace{}%
\AgdaBound{pbk4}\AgdaSymbol{)}\<%
\\
\>[0]\AgdaFunction{lemmaHoareTripleStackGeAux'20}\AgdaSpace{}%
\AgdaBound{msg₁}\AgdaSpace{}%
\AgdaBound{pbk1}\AgdaSpace{}%
\AgdaBound{pbk2}\AgdaSpace{}%
\AgdaBound{pbk3}\AgdaSpace{}%
\AgdaBound{pbk4}\AgdaSpace{}%
\AgdaBound{pbk5}\AgdaSpace{}%
\AgdaBound{sig1}\AgdaSpace{}%
\AgdaBound{sig2}\AgdaSpace{}%
\AgdaBound{sig3}\AgdaSpace{}%
\AgdaBound{x}\AgdaSpace{}%
\AgdaBound{x₁}\AgdaSpace{}%
\AgdaBound{x₂}\AgdaSpace{}%
\AgdaSymbol{|}\AgdaSpace{}%
\AgdaInductiveConstructor{true}\AgdaSpace{}%
\AgdaSymbol{|}\AgdaSpace{}%
\AgdaInductiveConstructor{true}\AgdaSpace{}%
\AgdaSymbol{|}\AgdaSpace{}%
\AgdaInductiveConstructor{false}\AgdaSpace{}%
\AgdaSymbol{|}\AgdaSpace{}%
\AgdaInductiveConstructor{true}\AgdaSpace{}%
\AgdaSymbol{=}\AgdaSpace{}%
\AgdaInductiveConstructor{tt}\<%
\\
\>[0]\AgdaFunction{lemmaHoareTripleStackGeAux'20}\AgdaSpace{}%
\AgdaBound{msg₁}\AgdaSpace{}%
\AgdaBound{pbk1}\AgdaSpace{}%
\AgdaBound{pbk2}\AgdaSpace{}%
\AgdaBound{pbk3}\AgdaSpace{}%
\AgdaBound{pbk4}\AgdaSpace{}%
\AgdaBound{pbk5}\AgdaSpace{}%
\AgdaBound{sig1}\AgdaSpace{}%
\AgdaBound{sig2}\AgdaSpace{}%
\AgdaBound{sig3}\AgdaSpace{}%
\AgdaBound{x}\AgdaSpace{}%
\AgdaBound{x₁}\AgdaSpace{}%
\AgdaBound{x₂}\AgdaSpace{}%
\AgdaSymbol{|}\AgdaSpace{}%
\AgdaInductiveConstructor{true}\AgdaSpace{}%
\AgdaSymbol{|}\AgdaSpace{}%
\AgdaInductiveConstructor{true}\AgdaSpace{}%
\AgdaSymbol{|}\AgdaSpace{}%
\AgdaInductiveConstructor{true}\AgdaSpace{}%
\AgdaSymbol{=}\AgdaSpace{}%
\AgdaInductiveConstructor{tt}\<%
\\
\\[\AgdaEmptyExtraSkip]%
\\[\AgdaEmptyExtraSkip]%
\\[\AgdaEmptyExtraSkip]%
\>[0]\AgdaFunction{lemmaHoareTripleStackGeAux'21}%
\>[4767I]\AgdaSymbol{:}\AgdaSpace{}%
\AgdaSymbol{(}\AgdaBound{msg}\AgdaSpace{}%
\AgdaSymbol{:}\AgdaSpace{}%
\AgdaDatatype{Msg}\AgdaSymbol{)}\<%
\\
\>[4767I][@{}l@{\AgdaIndent{0}}]%
\>[31]\AgdaSymbol{(}\AgdaBound{pbk1}\AgdaSpace{}%
\AgdaBound{pbk2}\AgdaSpace{}%
\AgdaBound{pbk3}\AgdaSpace{}%
\AgdaBound{pbk4}\AgdaSpace{}%
\AgdaBound{pbk5}\AgdaSpace{}%
\AgdaBound{sig1}\AgdaSpace{}%
\AgdaBound{sig2}\AgdaSpace{}%
\AgdaBound{sig3}\AgdaSpace{}%
\AgdaSymbol{:}\AgdaSpace{}%
\AgdaDatatype{ℕ}\AgdaSymbol{)}\<%
\\
\>[31]\AgdaSymbol{→}\AgdaSpace{}%
\AgdaFunction{True}\AgdaSpace{}%
\AgdaSymbol{(}\AgdaFunction{isSigned}%
\>[49]\AgdaBound{msg}\AgdaSpace{}%
\AgdaBound{sig3}\AgdaSpace{}%
\AgdaBound{pbk5}\AgdaSymbol{)}\<%
\\
\>[.][@{}l@{}]\<[4767I]%
\>[30]\AgdaSymbol{→}%
\>[33]\AgdaFunction{True}\AgdaSpace{}%
\AgdaSymbol{(}\AgdaFunction{isSigned}%
\>[49]\AgdaBound{msg}\AgdaSpace{}%
\AgdaBound{sig2}\AgdaSpace{}%
\AgdaBound{pbk3}\AgdaSymbol{)}\<%
\\
\>[30]\AgdaSymbol{→}%
\>[33]\AgdaFunction{True}\AgdaSpace{}%
\AgdaSymbol{(}\AgdaFunction{isSigned}%
\>[49]\AgdaBound{msg}\AgdaSpace{}%
\AgdaBound{sig1}\AgdaSpace{}%
\AgdaBound{pbk2}\AgdaSymbol{)}\<%
\\
\>[30][@{}l@{\AgdaIndent{0}}]%
\>[31]\AgdaSymbol{→}\AgdaSpace{}%
\AgdaFunction{True}\AgdaSpace{}%
\AgdaSymbol{(}\AgdaFunction{compareSigsMultiSig}\AgdaSpace{}%
\AgdaBound{msg}\AgdaSpace{}%
\AgdaSymbol{(}\AgdaSpace{}%
\AgdaBound{sig1}\AgdaSpace{}%
\AgdaOperator{\AgdaInductiveConstructor{∷}}\AgdaSpace{}%
\AgdaBound{sig2}\AgdaSpace{}%
\AgdaOperator{\AgdaInductiveConstructor{∷}}\AgdaSpace{}%
\AgdaBound{sig3}\AgdaSpace{}%
\AgdaOperator{\AgdaInductiveConstructor{∷}}\AgdaSpace{}%
\AgdaInductiveConstructor{[]}\AgdaSymbol{)}\AgdaSpace{}%
\AgdaSymbol{(}\AgdaBound{pbk1}\AgdaSpace{}%
\AgdaOperator{\AgdaInductiveConstructor{∷}}\AgdaSpace{}%
\AgdaBound{pbk2}\AgdaSpace{}%
\AgdaOperator{\AgdaInductiveConstructor{∷}}\AgdaSpace{}%
\AgdaBound{pbk3}\AgdaSpace{}%
\AgdaOperator{\AgdaInductiveConstructor{∷}}\AgdaSpace{}%
\AgdaBound{pbk4}\AgdaSpace{}%
\AgdaOperator{\AgdaInductiveConstructor{∷}}\AgdaSpace{}%
\AgdaBound{pbk5}\AgdaSpace{}%
\AgdaOperator{\AgdaInductiveConstructor{∷}}\AgdaSpace{}%
\AgdaInductiveConstructor{[]}\AgdaSymbol{))}\<%
\\
\>[0]\AgdaFunction{lemmaHoareTripleStackGeAux'21}\AgdaSpace{}%
\AgdaBound{msg₁}\AgdaSpace{}%
\AgdaBound{pbk1}\AgdaSpace{}%
\AgdaBound{pbk2}\AgdaSpace{}%
\AgdaBound{pbk3}\AgdaSpace{}%
\AgdaBound{pbk4}\AgdaSpace{}%
\AgdaBound{pbk5}\AgdaSpace{}%
\AgdaBound{sig1}\AgdaSpace{}%
\AgdaBound{sig2}\AgdaSpace{}%
\AgdaBound{sig3}\AgdaSpace{}%
\AgdaBound{x}\AgdaSpace{}%
\AgdaBound{x₁}\AgdaSpace{}%
\AgdaBound{x₂}\AgdaSpace{}%
\AgdaKeyword{with}\AgdaSpace{}%
\AgdaSymbol{(}\AgdaFunction{isSigned}%
\>[99]\AgdaBound{msg₁}\AgdaSpace{}%
\AgdaBound{sig1}\AgdaSpace{}%
\AgdaBound{pbk1}\AgdaSymbol{)}\<%
\\
\>[0]\AgdaFunction{lemmaHoareTripleStackGeAux'21}\AgdaSpace{}%
\AgdaBound{msg₁}\AgdaSpace{}%
\AgdaBound{pbk1}\AgdaSpace{}%
\AgdaBound{pbk2}\AgdaSpace{}%
\AgdaBound{pbk3}\AgdaSpace{}%
\AgdaBound{pbk4}\AgdaSpace{}%
\AgdaBound{pbk5}\AgdaSpace{}%
\AgdaBound{sig1}\AgdaSpace{}%
\AgdaBound{sig2}\AgdaSpace{}%
\AgdaBound{sig3}\AgdaSpace{}%
\AgdaBound{x}\AgdaSpace{}%
\AgdaBound{x₁}\AgdaSpace{}%
\AgdaBound{x₂}\AgdaSpace{}%
\AgdaSymbol{|}\AgdaSpace{}%
\AgdaInductiveConstructor{false}\AgdaSpace{}%
\AgdaKeyword{with}\AgdaSpace{}%
\AgdaSymbol{(}\AgdaFunction{isSigned}%
\>[107]\AgdaBound{msg₁}\AgdaSpace{}%
\AgdaBound{sig1}\AgdaSpace{}%
\AgdaBound{pbk2}\AgdaSymbol{)}\<%
\\
\>[0]\AgdaFunction{lemmaHoareTripleStackGeAux'21}\AgdaSpace{}%
\AgdaBound{msg₁}\AgdaSpace{}%
\AgdaBound{pbk1}\AgdaSpace{}%
\AgdaBound{pbk2}\AgdaSpace{}%
\AgdaBound{pbk3}\AgdaSpace{}%
\AgdaBound{pbk4}\AgdaSpace{}%
\AgdaBound{pbk5}\AgdaSpace{}%
\AgdaBound{sig1}\AgdaSpace{}%
\AgdaBound{sig2}\AgdaSpace{}%
\AgdaBound{sig3}\AgdaSpace{}%
\AgdaBound{x}\AgdaSpace{}%
\AgdaBound{x₁}\AgdaSpace{}%
\AgdaBound{x₂}\AgdaSpace{}%
\AgdaSymbol{|}\AgdaSpace{}%
\AgdaInductiveConstructor{false}\AgdaSpace{}%
\AgdaSymbol{|}\AgdaSpace{}%
\AgdaInductiveConstructor{true}\AgdaSpace{}%
\AgdaKeyword{with}\AgdaSpace{}%
\AgdaSymbol{(}\AgdaFunction{isSigned}%
\>[114]\AgdaBound{msg₁}\AgdaSpace{}%
\AgdaBound{sig2}\AgdaSpace{}%
\AgdaBound{pbk3}\AgdaSymbol{)}\<%
\\
\>[0]\AgdaFunction{lemmaHoareTripleStackGeAux'21}\AgdaSpace{}%
\AgdaBound{msg₁}\AgdaSpace{}%
\AgdaBound{pbk1}\AgdaSpace{}%
\AgdaBound{pbk2}\AgdaSpace{}%
\AgdaBound{pbk3}\AgdaSpace{}%
\AgdaBound{pbk4}\AgdaSpace{}%
\AgdaBound{pbk5}\AgdaSpace{}%
\AgdaBound{sig1}\AgdaSpace{}%
\AgdaBound{sig2}\AgdaSpace{}%
\AgdaBound{sig3}\AgdaSpace{}%
\AgdaBound{x}\AgdaSpace{}%
\AgdaBound{x₁}\AgdaSpace{}%
\AgdaBound{x₂}\AgdaSpace{}%
\AgdaSymbol{|}\AgdaSpace{}%
\AgdaInductiveConstructor{false}\AgdaSpace{}%
\AgdaSymbol{|}\AgdaSpace{}%
\AgdaInductiveConstructor{true}\AgdaSpace{}%
\AgdaSymbol{|}\AgdaSpace{}%
\AgdaInductiveConstructor{true}\AgdaSpace{}%
\AgdaKeyword{with}%
\>[111]\AgdaSymbol{(}\AgdaFunction{isSigned}%
\>[122]\AgdaBound{msg₁}\AgdaSpace{}%
\AgdaBound{sig3}\AgdaSpace{}%
\AgdaBound{pbk4}\AgdaSymbol{)}\<%
\\
\>[0]\AgdaFunction{lemmaHoareTripleStackGeAux'21}\AgdaSpace{}%
\AgdaBound{msg₁}\AgdaSpace{}%
\AgdaBound{pbk1}\AgdaSpace{}%
\AgdaBound{pbk2}\AgdaSpace{}%
\AgdaBound{pbk3}\AgdaSpace{}%
\AgdaBound{pbk4}\AgdaSpace{}%
\AgdaBound{pbk5}\AgdaSpace{}%
\AgdaBound{sig1}\AgdaSpace{}%
\AgdaBound{sig2}\AgdaSpace{}%
\AgdaBound{sig3}\AgdaSpace{}%
\AgdaBound{x}\AgdaSpace{}%
\AgdaBound{x₁}\AgdaSpace{}%
\AgdaBound{x₂}\AgdaSpace{}%
\AgdaSymbol{|}\AgdaSpace{}%
\AgdaInductiveConstructor{false}\AgdaSpace{}%
\AgdaSymbol{|}\AgdaSpace{}%
\AgdaInductiveConstructor{true}\AgdaSpace{}%
\AgdaSymbol{|}\AgdaSpace{}%
\AgdaInductiveConstructor{true}\AgdaSpace{}%
\AgdaSymbol{|}\AgdaSpace{}%
\AgdaInductiveConstructor{false}\AgdaSpace{}%
\AgdaKeyword{with}%
\>[119]\AgdaSymbol{(}\AgdaFunction{isSigned}%
\>[130]\AgdaBound{msg₁}\AgdaSpace{}%
\AgdaBound{sig3}\AgdaSpace{}%
\AgdaBound{pbk5}\AgdaSymbol{)}\<%
\\
\>[0]\AgdaFunction{lemmaHoareTripleStackGeAux'21}\AgdaSpace{}%
\AgdaBound{msg₁}\AgdaSpace{}%
\AgdaBound{pbk1}\AgdaSpace{}%
\AgdaBound{pbk2}\AgdaSpace{}%
\AgdaBound{pbk3}\AgdaSpace{}%
\AgdaBound{pbk4}\AgdaSpace{}%
\AgdaBound{pbk5}\AgdaSpace{}%
\AgdaBound{sig1}\AgdaSpace{}%
\AgdaBound{sig2}\AgdaSpace{}%
\AgdaBound{sig3}\AgdaSpace{}%
\AgdaBound{x}\AgdaSpace{}%
\AgdaBound{x₁}\AgdaSpace{}%
\AgdaBound{x₂}\AgdaSpace{}%
\AgdaSymbol{|}\AgdaSpace{}%
\AgdaInductiveConstructor{false}\AgdaSpace{}%
\AgdaSymbol{|}\AgdaSpace{}%
\AgdaInductiveConstructor{true}\AgdaSpace{}%
\AgdaSymbol{|}\AgdaSpace{}%
\AgdaInductiveConstructor{true}\AgdaSpace{}%
\AgdaSymbol{|}\AgdaSpace{}%
\AgdaInductiveConstructor{false}\AgdaSpace{}%
\AgdaSymbol{|}\AgdaSpace{}%
\AgdaInductiveConstructor{true}\AgdaSpace{}%
\AgdaSymbol{=}\AgdaSpace{}%
\AgdaInductiveConstructor{tt}\<%
\\
\>[0]\AgdaFunction{lemmaHoareTripleStackGeAux'21}\AgdaSpace{}%
\AgdaBound{msg₁}\AgdaSpace{}%
\AgdaBound{pbk1}\AgdaSpace{}%
\AgdaBound{pbk2}\AgdaSpace{}%
\AgdaBound{pbk3}\AgdaSpace{}%
\AgdaBound{pbk4}\AgdaSpace{}%
\AgdaBound{pbk5}\AgdaSpace{}%
\AgdaBound{sig1}\AgdaSpace{}%
\AgdaBound{sig2}\AgdaSpace{}%
\AgdaBound{sig3}\AgdaSpace{}%
\AgdaBound{x}\AgdaSpace{}%
\AgdaBound{x₁}\AgdaSpace{}%
\AgdaBound{x₂}\AgdaSpace{}%
\AgdaSymbol{|}\AgdaSpace{}%
\AgdaInductiveConstructor{false}\AgdaSpace{}%
\AgdaSymbol{|}\AgdaSpace{}%
\AgdaInductiveConstructor{true}\AgdaSpace{}%
\AgdaSymbol{|}\AgdaSpace{}%
\AgdaInductiveConstructor{true}\AgdaSpace{}%
\AgdaSymbol{|}\AgdaSpace{}%
\AgdaInductiveConstructor{true}\AgdaSpace{}%
\AgdaSymbol{=}\AgdaSpace{}%
\AgdaInductiveConstructor{tt}\<%
\\
\>[0]\AgdaFunction{lemmaHoareTripleStackGeAux'21}\AgdaSpace{}%
\AgdaBound{msg₁}\AgdaSpace{}%
\AgdaBound{pbk1}\AgdaSpace{}%
\AgdaBound{pbk2}\AgdaSpace{}%
\AgdaBound{pbk3}\AgdaSpace{}%
\AgdaBound{pbk4}\AgdaSpace{}%
\AgdaBound{pbk5}\AgdaSpace{}%
\AgdaBound{sig1}\AgdaSpace{}%
\AgdaBound{sig2}\AgdaSpace{}%
\AgdaBound{sig3}\AgdaSpace{}%
\AgdaBound{x}\AgdaSpace{}%
\AgdaBound{x₁}\AgdaSpace{}%
\AgdaBound{x₂}\AgdaSpace{}%
\AgdaSymbol{|}\AgdaSpace{}%
\AgdaInductiveConstructor{true}\AgdaSpace{}%
\AgdaKeyword{with}%
\>[96]\AgdaSymbol{(}\AgdaFunction{isSigned}%
\>[107]\AgdaBound{msg₁}\AgdaSpace{}%
\AgdaBound{sig2}\AgdaSpace{}%
\AgdaBound{pbk2}\AgdaSymbol{)}\<%
\\
\>[0]\AgdaFunction{lemmaHoareTripleStackGeAux'21}\AgdaSpace{}%
\AgdaBound{msg₁}\AgdaSpace{}%
\AgdaBound{pbk1}\AgdaSpace{}%
\AgdaBound{pbk2}\AgdaSpace{}%
\AgdaBound{pbk3}\AgdaSpace{}%
\AgdaBound{pbk4}\AgdaSpace{}%
\AgdaBound{pbk5}\AgdaSpace{}%
\AgdaBound{sig1}\AgdaSpace{}%
\AgdaBound{sig2}\AgdaSpace{}%
\AgdaBound{sig3}\AgdaSpace{}%
\AgdaBound{x}\AgdaSpace{}%
\AgdaBound{x₁}\AgdaSpace{}%
\AgdaBound{x₂}\AgdaSpace{}%
\AgdaSymbol{|}\AgdaSpace{}%
\AgdaInductiveConstructor{true}\AgdaSpace{}%
\AgdaSymbol{|}\AgdaSpace{}%
\AgdaInductiveConstructor{false}\AgdaSpace{}%
\AgdaKeyword{with}\AgdaSpace{}%
\AgdaSymbol{(}\AgdaFunction{isSigned}%
\>[114]\AgdaBound{msg₁}\AgdaSpace{}%
\AgdaBound{sig2}\AgdaSpace{}%
\AgdaBound{pbk3}\AgdaSymbol{)}\<%
\\
\>[0]\AgdaFunction{lemmaHoareTripleStackGeAux'21}\AgdaSpace{}%
\AgdaBound{msg₁}\AgdaSpace{}%
\AgdaBound{pbk1}\AgdaSpace{}%
\AgdaBound{pbk2}\AgdaSpace{}%
\AgdaBound{pbk3}\AgdaSpace{}%
\AgdaBound{pbk4}\AgdaSpace{}%
\AgdaBound{pbk5}\AgdaSpace{}%
\AgdaBound{sig1}\AgdaSpace{}%
\AgdaBound{sig2}\AgdaSpace{}%
\AgdaBound{sig3}\AgdaSpace{}%
\AgdaBound{x}\AgdaSpace{}%
\AgdaBound{x₁}\AgdaSpace{}%
\AgdaBound{x₂}\AgdaSpace{}%
\AgdaSymbol{|}\AgdaSpace{}%
\AgdaInductiveConstructor{true}\AgdaSpace{}%
\AgdaSymbol{|}\AgdaSpace{}%
\AgdaInductiveConstructor{false}\AgdaSpace{}%
\AgdaSymbol{|}\AgdaSpace{}%
\AgdaInductiveConstructor{true}\AgdaSpace{}%
\AgdaKeyword{with}\AgdaSpace{}%
\AgdaSymbol{(}\AgdaFunction{isSigned}%
\>[121]\AgdaBound{msg₁}\AgdaSpace{}%
\AgdaBound{sig3}\AgdaSpace{}%
\AgdaBound{pbk4}\AgdaSymbol{)}\<%
\\
\>[0]\AgdaFunction{lemmaHoareTripleStackGeAux'21}\AgdaSpace{}%
\AgdaBound{msg₁}\AgdaSpace{}%
\AgdaBound{pbk1}\AgdaSpace{}%
\AgdaBound{pbk2}\AgdaSpace{}%
\AgdaBound{pbk3}\AgdaSpace{}%
\AgdaBound{pbk4}\AgdaSpace{}%
\AgdaBound{pbk5}\AgdaSpace{}%
\AgdaBound{sig1}\AgdaSpace{}%
\AgdaBound{sig2}\AgdaSpace{}%
\AgdaBound{sig3}\AgdaSpace{}%
\AgdaBound{x}\AgdaSpace{}%
\AgdaBound{x₁}\AgdaSpace{}%
\AgdaBound{x₂}\AgdaSpace{}%
\AgdaSymbol{|}\AgdaSpace{}%
\AgdaInductiveConstructor{true}\AgdaSpace{}%
\AgdaSymbol{|}\AgdaSpace{}%
\AgdaInductiveConstructor{false}\AgdaSpace{}%
\AgdaSymbol{|}\AgdaSpace{}%
\AgdaInductiveConstructor{true}\AgdaSpace{}%
\AgdaSymbol{|}\AgdaSpace{}%
\AgdaInductiveConstructor{false}\AgdaSpace{}%
\AgdaKeyword{with}\AgdaSpace{}%
\AgdaSymbol{(}\AgdaFunction{isSigned}%
\>[129]\AgdaBound{msg₁}\AgdaSpace{}%
\AgdaBound{sig3}\AgdaSpace{}%
\AgdaBound{pbk5}\AgdaSymbol{)}\<%
\\
\>[0]\AgdaFunction{lemmaHoareTripleStackGeAux'21}\AgdaSpace{}%
\AgdaBound{msg₁}\AgdaSpace{}%
\AgdaBound{pbk1}\AgdaSpace{}%
\AgdaBound{pbk2}\AgdaSpace{}%
\AgdaBound{pbk3}\AgdaSpace{}%
\AgdaBound{pbk4}\AgdaSpace{}%
\AgdaBound{pbk5}\AgdaSpace{}%
\AgdaBound{sig1}\AgdaSpace{}%
\AgdaBound{sig2}\AgdaSpace{}%
\AgdaBound{sig3}\AgdaSpace{}%
\AgdaBound{x}\AgdaSpace{}%
\AgdaBound{x₁}\AgdaSpace{}%
\AgdaBound{x₂}\AgdaSpace{}%
\AgdaSymbol{|}\AgdaSpace{}%
\AgdaInductiveConstructor{true}\AgdaSpace{}%
\AgdaSymbol{|}\AgdaSpace{}%
\AgdaInductiveConstructor{false}\AgdaSpace{}%
\AgdaSymbol{|}\AgdaSpace{}%
\AgdaInductiveConstructor{true}\AgdaSpace{}%
\AgdaSymbol{|}\AgdaSpace{}%
\AgdaInductiveConstructor{false}\AgdaSpace{}%
\AgdaSymbol{|}\AgdaSpace{}%
\AgdaInductiveConstructor{true}\AgdaSpace{}%
\AgdaSymbol{=}\AgdaSpace{}%
\AgdaInductiveConstructor{tt}\<%
\\
\>[0]\AgdaFunction{lemmaHoareTripleStackGeAux'21}\AgdaSpace{}%
\AgdaBound{msg₁}\AgdaSpace{}%
\AgdaBound{pbk1}\AgdaSpace{}%
\AgdaBound{pbk2}\AgdaSpace{}%
\AgdaBound{pbk3}\AgdaSpace{}%
\AgdaBound{pbk4}\AgdaSpace{}%
\AgdaBound{pbk5}\AgdaSpace{}%
\AgdaBound{sig1}\AgdaSpace{}%
\AgdaBound{sig2}\AgdaSpace{}%
\AgdaBound{sig3}\AgdaSpace{}%
\AgdaBound{x}\AgdaSpace{}%
\AgdaBound{x₁}\AgdaSpace{}%
\AgdaBound{x₂}\AgdaSpace{}%
\AgdaSymbol{|}\AgdaSpace{}%
\AgdaInductiveConstructor{true}\AgdaSpace{}%
\AgdaSymbol{|}\AgdaSpace{}%
\AgdaInductiveConstructor{false}\AgdaSpace{}%
\AgdaSymbol{|}\AgdaSpace{}%
\AgdaInductiveConstructor{true}\AgdaSpace{}%
\AgdaSymbol{|}\AgdaSpace{}%
\AgdaInductiveConstructor{true}\AgdaSpace{}%
\AgdaSymbol{=}\AgdaSpace{}%
\AgdaInductiveConstructor{tt}\<%
\\
\>[0]\AgdaFunction{lemmaHoareTripleStackGeAux'21}\AgdaSpace{}%
\AgdaBound{msg₁}\AgdaSpace{}%
\AgdaBound{pbk1}\AgdaSpace{}%
\AgdaBound{pbk2}\AgdaSpace{}%
\AgdaBound{pbk3}\AgdaSpace{}%
\AgdaBound{pbk4}\AgdaSpace{}%
\AgdaBound{pbk5}\AgdaSpace{}%
\AgdaBound{sig1}\AgdaSpace{}%
\AgdaBound{sig2}\AgdaSpace{}%
\AgdaBound{sig3}\AgdaSpace{}%
\AgdaBound{x}\AgdaSpace{}%
\AgdaBound{x₁}\AgdaSpace{}%
\AgdaBound{x₂}\AgdaSpace{}%
\AgdaSymbol{|}\AgdaSpace{}%
\AgdaInductiveConstructor{true}\AgdaSpace{}%
\AgdaSymbol{|}\AgdaSpace{}%
\AgdaInductiveConstructor{true}\AgdaSpace{}%
\AgdaKeyword{with}\AgdaSpace{}%
\AgdaSymbol{(}\AgdaFunction{isSigned}%
\>[113]\AgdaBound{msg₁}\AgdaSpace{}%
\AgdaBound{sig3}\AgdaSpace{}%
\AgdaBound{pbk3}\AgdaSymbol{)}\<%
\\
\>[0]\AgdaFunction{lemmaHoareTripleStackGeAux'21}\AgdaSpace{}%
\AgdaBound{msg₁}\AgdaSpace{}%
\AgdaBound{pbk1}\AgdaSpace{}%
\AgdaBound{pbk2}\AgdaSpace{}%
\AgdaBound{pbk3}\AgdaSpace{}%
\AgdaBound{pbk4}\AgdaSpace{}%
\AgdaBound{pbk5}\AgdaSpace{}%
\AgdaBound{sig1}\AgdaSpace{}%
\AgdaBound{sig2}\AgdaSpace{}%
\AgdaBound{sig3}\AgdaSpace{}%
\AgdaBound{x}\AgdaSpace{}%
\AgdaBound{x₁}\AgdaSpace{}%
\AgdaBound{x₂}\AgdaSpace{}%
\AgdaSymbol{|}\AgdaSpace{}%
\AgdaInductiveConstructor{true}\AgdaSpace{}%
\AgdaSymbol{|}\AgdaSpace{}%
\AgdaInductiveConstructor{true}\AgdaSpace{}%
\AgdaSymbol{|}\AgdaSpace{}%
\AgdaInductiveConstructor{false}\AgdaSpace{}%
\AgdaKeyword{with}\AgdaSpace{}%
\AgdaSymbol{(}\AgdaFunction{isSigned}%
\>[121]\AgdaBound{msg₁}\AgdaSpace{}%
\AgdaBound{sig3}\AgdaSpace{}%
\AgdaBound{pbk4}\AgdaSymbol{)}\<%
\\
\>[0]\AgdaFunction{lemmaHoareTripleStackGeAux'21}\AgdaSpace{}%
\AgdaBound{msg₁}\AgdaSpace{}%
\AgdaBound{pbk1}\AgdaSpace{}%
\AgdaBound{pbk2}\AgdaSpace{}%
\AgdaBound{pbk3}\AgdaSpace{}%
\AgdaBound{pbk4}\AgdaSpace{}%
\AgdaBound{pbk5}\AgdaSpace{}%
\AgdaBound{sig1}\AgdaSpace{}%
\AgdaBound{sig2}\AgdaSpace{}%
\AgdaBound{sig3}\AgdaSpace{}%
\AgdaBound{x}\AgdaSpace{}%
\AgdaBound{x₁}\AgdaSpace{}%
\AgdaBound{x₂}\AgdaSpace{}%
\AgdaSymbol{|}\AgdaSpace{}%
\AgdaInductiveConstructor{true}\AgdaSpace{}%
\AgdaSymbol{|}\AgdaSpace{}%
\AgdaInductiveConstructor{true}\AgdaSpace{}%
\AgdaSymbol{|}\AgdaSpace{}%
\AgdaInductiveConstructor{false}\AgdaSpace{}%
\AgdaSymbol{|}\AgdaSpace{}%
\AgdaInductiveConstructor{false}\AgdaSpace{}%
\AgdaKeyword{with}\AgdaSpace{}%
\AgdaSymbol{(}\AgdaFunction{isSigned}%
\>[129]\AgdaBound{msg₁}\AgdaSpace{}%
\AgdaBound{sig3}\AgdaSpace{}%
\AgdaBound{pbk5}\AgdaSymbol{)}\<%
\\
\>[0]\AgdaFunction{lemmaHoareTripleStackGeAux'21}\AgdaSpace{}%
\AgdaBound{msg₁}\AgdaSpace{}%
\AgdaBound{pbk1}\AgdaSpace{}%
\AgdaBound{pbk2}\AgdaSpace{}%
\AgdaBound{pbk3}\AgdaSpace{}%
\AgdaBound{pbk4}\AgdaSpace{}%
\AgdaBound{pbk5}\AgdaSpace{}%
\AgdaBound{sig1}\AgdaSpace{}%
\AgdaBound{sig2}\AgdaSpace{}%
\AgdaBound{sig3}\AgdaSpace{}%
\AgdaBound{x}\AgdaSpace{}%
\AgdaBound{x₁}\AgdaSpace{}%
\AgdaBound{x₂}\AgdaSpace{}%
\AgdaSymbol{|}\AgdaSpace{}%
\AgdaInductiveConstructor{true}\AgdaSpace{}%
\AgdaSymbol{|}\AgdaSpace{}%
\AgdaInductiveConstructor{true}\AgdaSpace{}%
\AgdaSymbol{|}\AgdaSpace{}%
\AgdaInductiveConstructor{false}\AgdaSpace{}%
\AgdaSymbol{|}\AgdaSpace{}%
\AgdaInductiveConstructor{false}\AgdaSpace{}%
\AgdaSymbol{|}\AgdaSpace{}%
\AgdaInductiveConstructor{true}\AgdaSpace{}%
\AgdaSymbol{=}\AgdaSpace{}%
\AgdaInductiveConstructor{tt}\<%
\\
\>[0]\AgdaFunction{lemmaHoareTripleStackGeAux'21}\AgdaSpace{}%
\AgdaBound{msg₁}\AgdaSpace{}%
\AgdaBound{pbk1}\AgdaSpace{}%
\AgdaBound{pbk2}\AgdaSpace{}%
\AgdaBound{pbk3}\AgdaSpace{}%
\AgdaBound{pbk4}\AgdaSpace{}%
\AgdaBound{pbk5}\AgdaSpace{}%
\AgdaBound{sig1}\AgdaSpace{}%
\AgdaBound{sig2}\AgdaSpace{}%
\AgdaBound{sig3}\AgdaSpace{}%
\AgdaBound{x}\AgdaSpace{}%
\AgdaBound{x₁}\AgdaSpace{}%
\AgdaBound{x₂}\AgdaSpace{}%
\AgdaSymbol{|}\AgdaSpace{}%
\AgdaInductiveConstructor{true}\AgdaSpace{}%
\AgdaSymbol{|}\AgdaSpace{}%
\AgdaInductiveConstructor{true}\AgdaSpace{}%
\AgdaSymbol{|}\AgdaSpace{}%
\AgdaInductiveConstructor{false}\AgdaSpace{}%
\AgdaSymbol{|}\AgdaSpace{}%
\AgdaInductiveConstructor{true}\AgdaSpace{}%
\AgdaSymbol{=}\AgdaSpace{}%
\AgdaInductiveConstructor{tt}\<%
\\
\>[0]\AgdaFunction{lemmaHoareTripleStackGeAux'21}\AgdaSpace{}%
\AgdaBound{msg₁}\AgdaSpace{}%
\AgdaBound{pbk1}\AgdaSpace{}%
\AgdaBound{pbk2}\AgdaSpace{}%
\AgdaBound{pbk3}\AgdaSpace{}%
\AgdaBound{pbk4}\AgdaSpace{}%
\AgdaBound{pbk5}\AgdaSpace{}%
\AgdaBound{sig1}\AgdaSpace{}%
\AgdaBound{sig2}\AgdaSpace{}%
\AgdaBound{sig3}\AgdaSpace{}%
\AgdaBound{x}\AgdaSpace{}%
\AgdaBound{x₁}\AgdaSpace{}%
\AgdaBound{x₂}\AgdaSpace{}%
\AgdaSymbol{|}\AgdaSpace{}%
\AgdaInductiveConstructor{true}\AgdaSpace{}%
\AgdaSymbol{|}\AgdaSpace{}%
\AgdaInductiveConstructor{true}\AgdaSpace{}%
\AgdaSymbol{|}\AgdaSpace{}%
\AgdaInductiveConstructor{true}\AgdaSpace{}%
\AgdaSymbol{=}\AgdaSpace{}%
\AgdaInductiveConstructor{tt}\<%
\\
\\[\AgdaEmptyExtraSkip]%
\\[\AgdaEmptyExtraSkip]%
\\[\AgdaEmptyExtraSkip]%
\\[\AgdaEmptyExtraSkip]%
\\[\AgdaEmptyExtraSkip]%
\>[0]\AgdaFunction{lemmaHoareTripleStackGeAux'22}%
\>[5217I]\AgdaSymbol{:}\AgdaSpace{}%
\AgdaSymbol{(}\AgdaBound{msg}\AgdaSpace{}%
\AgdaSymbol{:}\AgdaSpace{}%
\AgdaDatatype{Msg}\AgdaSymbol{)}\<%
\\
\>[5217I][@{}l@{\AgdaIndent{0}}]%
\>[31]\AgdaSymbol{(}\AgdaBound{pbk1}\AgdaSpace{}%
\AgdaBound{pbk2}\AgdaSpace{}%
\AgdaBound{pbk3}\AgdaSpace{}%
\AgdaBound{pbk4}\AgdaSpace{}%
\AgdaBound{pbk5}\AgdaSpace{}%
\AgdaBound{sig1}\AgdaSpace{}%
\AgdaBound{sig2}\AgdaSpace{}%
\AgdaBound{sig3}\AgdaSpace{}%
\AgdaSymbol{:}\AgdaSpace{}%
\AgdaDatatype{ℕ}\AgdaSymbol{)}\<%
\\
\>[31]\AgdaSymbol{→}\AgdaSpace{}%
\AgdaFunction{True}\AgdaSpace{}%
\AgdaSymbol{(}\AgdaFunction{isSigned}%
\>[49]\AgdaBound{msg}\AgdaSpace{}%
\AgdaBound{sig3}\AgdaSpace{}%
\AgdaBound{pbk5}\AgdaSymbol{)}\<%
\\
\>[.][@{}l@{}]\<[5217I]%
\>[30]\AgdaSymbol{→}%
\>[33]\AgdaFunction{True}\AgdaSpace{}%
\AgdaSymbol{(}\AgdaFunction{isSigned}%
\>[49]\AgdaBound{msg}\AgdaSpace{}%
\AgdaBound{sig2}\AgdaSpace{}%
\AgdaBound{pbk4}\AgdaSymbol{)}\<%
\\
\>[30]\AgdaSymbol{→}%
\>[33]\AgdaFunction{True}\AgdaSpace{}%
\AgdaSymbol{(}\AgdaFunction{isSigned}%
\>[49]\AgdaBound{msg}\AgdaSpace{}%
\AgdaBound{sig1}\AgdaSpace{}%
\AgdaBound{pbk2}\AgdaSymbol{)}\<%
\\
\>[30][@{}l@{\AgdaIndent{0}}]%
\>[31]\AgdaSymbol{→}\AgdaSpace{}%
\AgdaFunction{True}\AgdaSpace{}%
\AgdaSymbol{(}\AgdaFunction{compareSigsMultiSig}\AgdaSpace{}%
\AgdaBound{msg}\AgdaSpace{}%
\AgdaSymbol{(}\AgdaSpace{}%
\AgdaBound{sig1}\AgdaSpace{}%
\AgdaOperator{\AgdaInductiveConstructor{∷}}\AgdaSpace{}%
\AgdaBound{sig2}\AgdaSpace{}%
\AgdaOperator{\AgdaInductiveConstructor{∷}}\AgdaSpace{}%
\AgdaBound{sig3}\AgdaSpace{}%
\AgdaOperator{\AgdaInductiveConstructor{∷}}\AgdaSpace{}%
\AgdaInductiveConstructor{[]}\AgdaSymbol{)}\AgdaSpace{}%
\AgdaSymbol{(}\AgdaBound{pbk1}\AgdaSpace{}%
\AgdaOperator{\AgdaInductiveConstructor{∷}}\AgdaSpace{}%
\AgdaBound{pbk2}\AgdaSpace{}%
\AgdaOperator{\AgdaInductiveConstructor{∷}}\AgdaSpace{}%
\AgdaBound{pbk3}\AgdaSpace{}%
\AgdaOperator{\AgdaInductiveConstructor{∷}}\AgdaSpace{}%
\AgdaBound{pbk4}\AgdaSpace{}%
\AgdaOperator{\AgdaInductiveConstructor{∷}}\AgdaSpace{}%
\AgdaBound{pbk5}\AgdaSpace{}%
\AgdaOperator{\AgdaInductiveConstructor{∷}}\AgdaSpace{}%
\AgdaInductiveConstructor{[]}\AgdaSymbol{))}\<%
\\
\>[0]\AgdaFunction{lemmaHoareTripleStackGeAux'22}\AgdaSpace{}%
\AgdaBound{msg₁}\AgdaSpace{}%
\AgdaBound{pbk1}\AgdaSpace{}%
\AgdaBound{pbk2}\AgdaSpace{}%
\AgdaBound{pbk3}\AgdaSpace{}%
\AgdaBound{pbk4}\AgdaSpace{}%
\AgdaBound{pbk5}\AgdaSpace{}%
\AgdaBound{sig1}\AgdaSpace{}%
\AgdaBound{sig2}\AgdaSpace{}%
\AgdaBound{sig3}\AgdaSpace{}%
\AgdaBound{x}\AgdaSpace{}%
\AgdaBound{x₁}\AgdaSpace{}%
\AgdaBound{x₂}\AgdaSpace{}%
\AgdaKeyword{with}\AgdaSpace{}%
\AgdaSymbol{(}\AgdaFunction{isSigned}%
\>[99]\AgdaBound{msg₁}\AgdaSpace{}%
\AgdaBound{sig1}\AgdaSpace{}%
\AgdaBound{pbk1}\AgdaSymbol{)}\<%
\\
\>[0]\AgdaFunction{lemmaHoareTripleStackGeAux'22}\AgdaSpace{}%
\AgdaBound{msg₁}\AgdaSpace{}%
\AgdaBound{pbk1}\AgdaSpace{}%
\AgdaBound{pbk2}\AgdaSpace{}%
\AgdaBound{pbk3}\AgdaSpace{}%
\AgdaBound{pbk4}\AgdaSpace{}%
\AgdaBound{pbk5}\AgdaSpace{}%
\AgdaBound{sig1}\AgdaSpace{}%
\AgdaBound{sig2}\AgdaSpace{}%
\AgdaBound{sig3}\AgdaSpace{}%
\AgdaBound{x}\AgdaSpace{}%
\AgdaBound{x₁}\AgdaSpace{}%
\AgdaBound{x₂}\AgdaSpace{}%
\AgdaSymbol{|}\AgdaSpace{}%
\AgdaInductiveConstructor{false}\AgdaSpace{}%
\AgdaKeyword{with}%
\>[97]\AgdaSymbol{(}\AgdaFunction{isSigned}%
\>[108]\AgdaBound{msg₁}\AgdaSpace{}%
\AgdaBound{sig1}\AgdaSpace{}%
\AgdaBound{pbk2}\AgdaSymbol{)}\<%
\\
\>[0]\AgdaFunction{lemmaHoareTripleStackGeAux'22}\AgdaSpace{}%
\AgdaBound{msg₁}\AgdaSpace{}%
\AgdaBound{pbk1}\AgdaSpace{}%
\AgdaBound{pbk2}\AgdaSpace{}%
\AgdaBound{pbk3}\AgdaSpace{}%
\AgdaBound{pbk4}\AgdaSpace{}%
\AgdaBound{pbk5}\AgdaSpace{}%
\AgdaBound{sig1}\AgdaSpace{}%
\AgdaBound{sig2}\AgdaSpace{}%
\AgdaBound{sig3}\AgdaSpace{}%
\AgdaBound{x}\AgdaSpace{}%
\AgdaBound{x₁}\AgdaSpace{}%
\AgdaBound{x₂}\AgdaSpace{}%
\AgdaSymbol{|}\AgdaSpace{}%
\AgdaInductiveConstructor{false}\AgdaSpace{}%
\AgdaSymbol{|}\AgdaSpace{}%
\AgdaInductiveConstructor{true}\AgdaSpace{}%
\AgdaKeyword{with}\AgdaSpace{}%
\AgdaSymbol{(}\AgdaFunction{isSigned}%
\>[114]\AgdaBound{msg₁}\AgdaSpace{}%
\AgdaBound{sig2}\AgdaSpace{}%
\AgdaBound{pbk3}\AgdaSymbol{)}\<%
\\
\>[0]\AgdaFunction{lemmaHoareTripleStackGeAux'22}\AgdaSpace{}%
\AgdaBound{msg₁}\AgdaSpace{}%
\AgdaBound{pbk1}\AgdaSpace{}%
\AgdaBound{pbk2}\AgdaSpace{}%
\AgdaBound{pbk3}\AgdaSpace{}%
\AgdaBound{pbk4}\AgdaSpace{}%
\AgdaBound{pbk5}\AgdaSpace{}%
\AgdaBound{sig1}\AgdaSpace{}%
\AgdaBound{sig2}\AgdaSpace{}%
\AgdaBound{sig3}\AgdaSpace{}%
\AgdaBound{x}\AgdaSpace{}%
\AgdaBound{x₁}\AgdaSpace{}%
\AgdaBound{x₂}\AgdaSpace{}%
\AgdaSymbol{|}\AgdaSpace{}%
\AgdaInductiveConstructor{false}\AgdaSpace{}%
\AgdaSymbol{|}\AgdaSpace{}%
\AgdaInductiveConstructor{true}\AgdaSpace{}%
\AgdaSymbol{|}\AgdaSpace{}%
\AgdaInductiveConstructor{false}\AgdaSpace{}%
\AgdaKeyword{with}%
\>[112]\AgdaSymbol{(}\AgdaFunction{isSigned}%
\>[123]\AgdaBound{msg₁}\AgdaSpace{}%
\AgdaBound{sig2}\AgdaSpace{}%
\AgdaBound{pbk4}\AgdaSymbol{)}\<%
\\
\>[0]\AgdaFunction{lemmaHoareTripleStackGeAux'22}\AgdaSpace{}%
\AgdaBound{msg₁}\AgdaSpace{}%
\AgdaBound{pbk1}\AgdaSpace{}%
\AgdaBound{pbk2}\AgdaSpace{}%
\AgdaBound{pbk3}\AgdaSpace{}%
\AgdaBound{pbk4}\AgdaSpace{}%
\AgdaBound{pbk5}\AgdaSpace{}%
\AgdaBound{sig1}\AgdaSpace{}%
\AgdaBound{sig2}\AgdaSpace{}%
\AgdaBound{sig3}\AgdaSpace{}%
\AgdaBound{x}\AgdaSpace{}%
\AgdaBound{x₁}\AgdaSpace{}%
\AgdaBound{x₂}\AgdaSpace{}%
\AgdaSymbol{|}\AgdaSpace{}%
\AgdaInductiveConstructor{false}\AgdaSpace{}%
\AgdaSymbol{|}\AgdaSpace{}%
\AgdaInductiveConstructor{true}\AgdaSpace{}%
\AgdaSymbol{|}\AgdaSpace{}%
\AgdaInductiveConstructor{false}\AgdaSpace{}%
\AgdaSymbol{|}\AgdaSpace{}%
\AgdaInductiveConstructor{true}\AgdaSpace{}%
\AgdaKeyword{with}\AgdaSpace{}%
\AgdaSymbol{(}\AgdaFunction{isSigned}%
\>[129]\AgdaBound{msg₁}\AgdaSpace{}%
\AgdaBound{sig3}\AgdaSpace{}%
\AgdaBound{pbk5}\AgdaSymbol{)}\<%
\\
\>[0]\AgdaFunction{lemmaHoareTripleStackGeAux'22}\AgdaSpace{}%
\AgdaBound{msg₁}\AgdaSpace{}%
\AgdaBound{pbk1}\AgdaSpace{}%
\AgdaBound{pbk2}\AgdaSpace{}%
\AgdaBound{pbk3}\AgdaSpace{}%
\AgdaBound{pbk4}\AgdaSpace{}%
\AgdaBound{pbk5}\AgdaSpace{}%
\AgdaBound{sig1}\AgdaSpace{}%
\AgdaBound{sig2}\AgdaSpace{}%
\AgdaBound{sig3}\AgdaSpace{}%
\AgdaBound{x}\AgdaSpace{}%
\AgdaBound{x₁}\AgdaSpace{}%
\AgdaBound{x₂}\AgdaSpace{}%
\AgdaSymbol{|}\AgdaSpace{}%
\AgdaInductiveConstructor{false}\AgdaSpace{}%
\AgdaSymbol{|}\AgdaSpace{}%
\AgdaInductiveConstructor{true}\AgdaSpace{}%
\AgdaSymbol{|}\AgdaSpace{}%
\AgdaInductiveConstructor{false}\AgdaSpace{}%
\AgdaSymbol{|}\AgdaSpace{}%
\AgdaInductiveConstructor{true}\AgdaSpace{}%
\AgdaSymbol{|}\AgdaSpace{}%
\AgdaInductiveConstructor{true}\AgdaSpace{}%
\AgdaSymbol{=}\AgdaSpace{}%
\AgdaInductiveConstructor{tt}\<%
\\
\>[0]\AgdaFunction{lemmaHoareTripleStackGeAux'22}\AgdaSpace{}%
\AgdaBound{msg₁}\AgdaSpace{}%
\AgdaBound{pbk1}\AgdaSpace{}%
\AgdaBound{pbk2}\AgdaSpace{}%
\AgdaBound{pbk3}\AgdaSpace{}%
\AgdaBound{pbk4}\AgdaSpace{}%
\AgdaBound{pbk5}\AgdaSpace{}%
\AgdaBound{sig1}\AgdaSpace{}%
\AgdaBound{sig2}\AgdaSpace{}%
\AgdaBound{sig3}\AgdaSpace{}%
\AgdaBound{x}\AgdaSpace{}%
\AgdaBound{x₁}\AgdaSpace{}%
\AgdaBound{x₂}\AgdaSpace{}%
\AgdaSymbol{|}\AgdaSpace{}%
\AgdaInductiveConstructor{false}\AgdaSpace{}%
\AgdaSymbol{|}\AgdaSpace{}%
\AgdaInductiveConstructor{true}\AgdaSpace{}%
\AgdaSymbol{|}\AgdaSpace{}%
\AgdaInductiveConstructor{true}\AgdaSpace{}%
\AgdaKeyword{with}\AgdaSpace{}%
\AgdaSymbol{(}\AgdaFunction{isSigned}%
\>[121]\AgdaBound{msg₁}\AgdaSpace{}%
\AgdaBound{sig3}\AgdaSpace{}%
\AgdaBound{pbk4}\AgdaSymbol{)}\<%
\\
\>[0]\AgdaFunction{lemmaHoareTripleStackGeAux'22}\AgdaSpace{}%
\AgdaBound{msg₁}\AgdaSpace{}%
\AgdaBound{pbk1}\AgdaSpace{}%
\AgdaBound{pbk2}\AgdaSpace{}%
\AgdaBound{pbk3}\AgdaSpace{}%
\AgdaBound{pbk4}\AgdaSpace{}%
\AgdaBound{pbk5}\AgdaSpace{}%
\AgdaBound{sig1}\AgdaSpace{}%
\AgdaBound{sig2}\AgdaSpace{}%
\AgdaBound{sig3}\AgdaSpace{}%
\AgdaBound{x}\AgdaSpace{}%
\AgdaBound{x₁}\AgdaSpace{}%
\AgdaBound{x₂}\AgdaSpace{}%
\AgdaSymbol{|}\AgdaSpace{}%
\AgdaInductiveConstructor{false}\AgdaSpace{}%
\AgdaSymbol{|}\AgdaSpace{}%
\AgdaInductiveConstructor{true}\AgdaSpace{}%
\AgdaSymbol{|}\AgdaSpace{}%
\AgdaInductiveConstructor{true}\AgdaSpace{}%
\AgdaSymbol{|}\AgdaSpace{}%
\AgdaInductiveConstructor{false}\AgdaSpace{}%
\AgdaKeyword{with}\AgdaSpace{}%
\AgdaSymbol{(}\AgdaFunction{isSigned}%
\>[129]\AgdaBound{msg₁}\AgdaSpace{}%
\AgdaBound{sig3}\AgdaSpace{}%
\AgdaBound{pbk5}\AgdaSymbol{)}\<%
\\
\>[0]\AgdaFunction{lemmaHoareTripleStackGeAux'22}\AgdaSpace{}%
\AgdaBound{msg₁}\AgdaSpace{}%
\AgdaBound{pbk1}\AgdaSpace{}%
\AgdaBound{pbk2}\AgdaSpace{}%
\AgdaBound{pbk3}\AgdaSpace{}%
\AgdaBound{pbk4}\AgdaSpace{}%
\AgdaBound{pbk5}\AgdaSpace{}%
\AgdaBound{sig1}\AgdaSpace{}%
\AgdaBound{sig2}\AgdaSpace{}%
\AgdaBound{sig3}\AgdaSpace{}%
\AgdaBound{x}\AgdaSpace{}%
\AgdaBound{x₁}\AgdaSpace{}%
\AgdaBound{x₂}\AgdaSpace{}%
\AgdaSymbol{|}\AgdaSpace{}%
\AgdaInductiveConstructor{false}\AgdaSpace{}%
\AgdaSymbol{|}\AgdaSpace{}%
\AgdaInductiveConstructor{true}\AgdaSpace{}%
\AgdaSymbol{|}\AgdaSpace{}%
\AgdaInductiveConstructor{true}\AgdaSpace{}%
\AgdaSymbol{|}\AgdaSpace{}%
\AgdaInductiveConstructor{false}\AgdaSpace{}%
\AgdaSymbol{|}\AgdaSpace{}%
\AgdaInductiveConstructor{true}\AgdaSpace{}%
\AgdaSymbol{=}\AgdaSpace{}%
\AgdaInductiveConstructor{tt}\<%
\\
\>[0]\AgdaFunction{lemmaHoareTripleStackGeAux'22}\AgdaSpace{}%
\AgdaBound{msg₁}\AgdaSpace{}%
\AgdaBound{pbk1}\AgdaSpace{}%
\AgdaBound{pbk2}\AgdaSpace{}%
\AgdaBound{pbk3}\AgdaSpace{}%
\AgdaBound{pbk4}\AgdaSpace{}%
\AgdaBound{pbk5}\AgdaSpace{}%
\AgdaBound{sig1}\AgdaSpace{}%
\AgdaBound{sig2}\AgdaSpace{}%
\AgdaBound{sig3}\AgdaSpace{}%
\AgdaBound{x}\AgdaSpace{}%
\AgdaBound{x₁}\AgdaSpace{}%
\AgdaBound{x₂}\AgdaSpace{}%
\AgdaSymbol{|}\AgdaSpace{}%
\AgdaInductiveConstructor{false}\AgdaSpace{}%
\AgdaSymbol{|}\AgdaSpace{}%
\AgdaInductiveConstructor{true}\AgdaSpace{}%
\AgdaSymbol{|}\AgdaSpace{}%
\AgdaInductiveConstructor{true}\AgdaSpace{}%
\AgdaSymbol{|}\AgdaSpace{}%
\AgdaInductiveConstructor{true}\AgdaSpace{}%
\AgdaSymbol{=}\AgdaSpace{}%
\AgdaInductiveConstructor{tt}\<%
\\
\>[0]\AgdaFunction{lemmaHoareTripleStackGeAux'22}\AgdaSpace{}%
\AgdaBound{msg₁}\AgdaSpace{}%
\AgdaBound{pbk1}\AgdaSpace{}%
\AgdaBound{pbk2}\AgdaSpace{}%
\AgdaBound{pbk3}\AgdaSpace{}%
\AgdaBound{pbk4}\AgdaSpace{}%
\AgdaBound{pbk5}\AgdaSpace{}%
\AgdaBound{sig1}\AgdaSpace{}%
\AgdaBound{sig2}\AgdaSpace{}%
\AgdaBound{sig3}\AgdaSpace{}%
\AgdaBound{x}\AgdaSpace{}%
\AgdaBound{x₁}\AgdaSpace{}%
\AgdaBound{x₂}\AgdaSpace{}%
\AgdaSymbol{|}\AgdaSpace{}%
\AgdaInductiveConstructor{true}\AgdaSpace{}%
\AgdaKeyword{with}\AgdaSpace{}%
\AgdaSymbol{(}\AgdaFunction{isSigned}%
\>[106]\AgdaBound{msg₁}\AgdaSpace{}%
\AgdaBound{sig2}\AgdaSpace{}%
\AgdaBound{pbk2}\AgdaSymbol{)}\<%
\\
\>[0]\AgdaFunction{lemmaHoareTripleStackGeAux'22}\AgdaSpace{}%
\AgdaBound{msg₁}\AgdaSpace{}%
\AgdaBound{pbk1}\AgdaSpace{}%
\AgdaBound{pbk2}\AgdaSpace{}%
\AgdaBound{pbk3}\AgdaSpace{}%
\AgdaBound{pbk4}\AgdaSpace{}%
\AgdaBound{pbk5}\AgdaSpace{}%
\AgdaBound{sig1}\AgdaSpace{}%
\AgdaBound{sig2}\AgdaSpace{}%
\AgdaBound{sig3}\AgdaSpace{}%
\AgdaBound{x}\AgdaSpace{}%
\AgdaBound{x₁}\AgdaSpace{}%
\AgdaBound{x₂}\AgdaSpace{}%
\AgdaSymbol{|}\AgdaSpace{}%
\AgdaInductiveConstructor{true}\AgdaSpace{}%
\AgdaSymbol{|}\AgdaSpace{}%
\AgdaInductiveConstructor{false}\AgdaSpace{}%
\AgdaKeyword{with}%
\>[104]\AgdaSymbol{(}\AgdaFunction{isSigned}%
\>[115]\AgdaBound{msg₁}\AgdaSpace{}%
\AgdaBound{sig2}\AgdaSpace{}%
\AgdaBound{pbk3}\AgdaSymbol{)}\<%
\\
\>[0]\AgdaFunction{lemmaHoareTripleStackGeAux'22}\AgdaSpace{}%
\AgdaBound{msg₁}\AgdaSpace{}%
\AgdaBound{pbk1}\AgdaSpace{}%
\AgdaBound{pbk2}\AgdaSpace{}%
\AgdaBound{pbk3}\AgdaSpace{}%
\AgdaBound{pbk4}\AgdaSpace{}%
\AgdaBound{pbk5}\AgdaSpace{}%
\AgdaBound{sig1}\AgdaSpace{}%
\AgdaBound{sig2}\AgdaSpace{}%
\AgdaBound{sig3}\AgdaSpace{}%
\AgdaBound{x}\AgdaSpace{}%
\AgdaBound{x₁}\AgdaSpace{}%
\AgdaBound{x₂}\AgdaSpace{}%
\AgdaSymbol{|}\AgdaSpace{}%
\AgdaInductiveConstructor{true}\AgdaSpace{}%
\AgdaSymbol{|}\AgdaSpace{}%
\AgdaInductiveConstructor{false}\AgdaSpace{}%
\AgdaSymbol{|}\AgdaSpace{}%
\AgdaInductiveConstructor{false}\AgdaSpace{}%
\AgdaKeyword{with}\AgdaSpace{}%
\AgdaSymbol{(}\AgdaFunction{isSigned}%
\>[122]\AgdaBound{msg₁}\AgdaSpace{}%
\AgdaBound{sig2}\AgdaSpace{}%
\AgdaBound{pbk4}\AgdaSymbol{)}\<%
\\
\>[0]\AgdaFunction{lemmaHoareTripleStackGeAux'22}\AgdaSpace{}%
\AgdaBound{msg₁}\AgdaSpace{}%
\AgdaBound{pbk1}\AgdaSpace{}%
\AgdaBound{pbk2}\AgdaSpace{}%
\AgdaBound{pbk3}\AgdaSpace{}%
\AgdaBound{pbk4}\AgdaSpace{}%
\AgdaBound{pbk5}\AgdaSpace{}%
\AgdaBound{sig1}\AgdaSpace{}%
\AgdaBound{sig2}\AgdaSpace{}%
\AgdaBound{sig3}\AgdaSpace{}%
\AgdaBound{x}\AgdaSpace{}%
\AgdaBound{x₁}\AgdaSpace{}%
\AgdaBound{x₂}\AgdaSpace{}%
\AgdaSymbol{|}\AgdaSpace{}%
\AgdaInductiveConstructor{true}\AgdaSpace{}%
\AgdaSymbol{|}\AgdaSpace{}%
\AgdaInductiveConstructor{false}\AgdaSpace{}%
\AgdaSymbol{|}\AgdaSpace{}%
\AgdaInductiveConstructor{false}\AgdaSpace{}%
\AgdaSymbol{|}\AgdaSpace{}%
\AgdaInductiveConstructor{true}\AgdaSpace{}%
\AgdaKeyword{with}%
\>[119]\AgdaSymbol{(}\AgdaFunction{isSigned}%
\>[130]\AgdaBound{msg₁}\AgdaSpace{}%
\AgdaBound{sig3}\AgdaSpace{}%
\AgdaBound{pbk5}\AgdaSymbol{)}\<%
\\
\>[0]\AgdaFunction{lemmaHoareTripleStackGeAux'22}\AgdaSpace{}%
\AgdaBound{msg₁}\AgdaSpace{}%
\AgdaBound{pbk1}\AgdaSpace{}%
\AgdaBound{pbk2}\AgdaSpace{}%
\AgdaBound{pbk3}\AgdaSpace{}%
\AgdaBound{pbk4}\AgdaSpace{}%
\AgdaBound{pbk5}\AgdaSpace{}%
\AgdaBound{sig1}\AgdaSpace{}%
\AgdaBound{sig2}\AgdaSpace{}%
\AgdaBound{sig3}\AgdaSpace{}%
\AgdaBound{x}\AgdaSpace{}%
\AgdaBound{x₁}\AgdaSpace{}%
\AgdaBound{x₂}\AgdaSpace{}%
\AgdaSymbol{|}\AgdaSpace{}%
\AgdaInductiveConstructor{true}\AgdaSpace{}%
\AgdaSymbol{|}\AgdaSpace{}%
\AgdaInductiveConstructor{false}\AgdaSpace{}%
\AgdaSymbol{|}\AgdaSpace{}%
\AgdaInductiveConstructor{false}\AgdaSpace{}%
\AgdaSymbol{|}\AgdaSpace{}%
\AgdaInductiveConstructor{true}\AgdaSpace{}%
\AgdaSymbol{|}\AgdaSpace{}%
\AgdaInductiveConstructor{true}\AgdaSpace{}%
\AgdaSymbol{=}\AgdaSpace{}%
\AgdaInductiveConstructor{tt}\<%
\\
\>[0]\AgdaFunction{lemmaHoareTripleStackGeAux'22}\AgdaSpace{}%
\AgdaBound{msg₁}\AgdaSpace{}%
\AgdaBound{pbk1}\AgdaSpace{}%
\AgdaBound{pbk2}\AgdaSpace{}%
\AgdaBound{pbk3}\AgdaSpace{}%
\AgdaBound{pbk4}\AgdaSpace{}%
\AgdaBound{pbk5}\AgdaSpace{}%
\AgdaBound{sig1}\AgdaSpace{}%
\AgdaBound{sig2}\AgdaSpace{}%
\AgdaBound{sig3}\AgdaSpace{}%
\AgdaBound{x}\AgdaSpace{}%
\AgdaBound{x₁}\AgdaSpace{}%
\AgdaBound{x₂}\AgdaSpace{}%
\AgdaSymbol{|}\AgdaSpace{}%
\AgdaInductiveConstructor{true}\AgdaSpace{}%
\AgdaSymbol{|}\AgdaSpace{}%
\AgdaInductiveConstructor{false}\AgdaSpace{}%
\AgdaSymbol{|}\AgdaSpace{}%
\AgdaInductiveConstructor{true}\AgdaSpace{}%
\AgdaKeyword{with}\AgdaSpace{}%
\AgdaSymbol{(}\AgdaFunction{isSigned}%
\>[121]\AgdaBound{msg₁}\AgdaSpace{}%
\AgdaBound{sig3}\AgdaSpace{}%
\AgdaBound{pbk4}\AgdaSymbol{)}\<%
\\
\>[0]\AgdaFunction{lemmaHoareTripleStackGeAux'22}\AgdaSpace{}%
\AgdaBound{msg₁}\AgdaSpace{}%
\AgdaBound{pbk1}\AgdaSpace{}%
\AgdaBound{pbk2}\AgdaSpace{}%
\AgdaBound{pbk3}\AgdaSpace{}%
\AgdaBound{pbk4}\AgdaSpace{}%
\AgdaBound{pbk5}\AgdaSpace{}%
\AgdaBound{sig1}\AgdaSpace{}%
\AgdaBound{sig2}\AgdaSpace{}%
\AgdaBound{sig3}\AgdaSpace{}%
\AgdaBound{x}\AgdaSpace{}%
\AgdaBound{x₁}\AgdaSpace{}%
\AgdaBound{x₂}\AgdaSpace{}%
\AgdaSymbol{|}\AgdaSpace{}%
\AgdaInductiveConstructor{true}\AgdaSpace{}%
\AgdaSymbol{|}\AgdaSpace{}%
\AgdaInductiveConstructor{false}\AgdaSpace{}%
\AgdaSymbol{|}\AgdaSpace{}%
\AgdaInductiveConstructor{true}\AgdaSpace{}%
\AgdaSymbol{|}\AgdaSpace{}%
\AgdaInductiveConstructor{false}\AgdaSpace{}%
\AgdaKeyword{with}%
\>[119]\AgdaSymbol{(}\AgdaFunction{isSigned}%
\>[130]\AgdaBound{msg₁}\AgdaSpace{}%
\AgdaBound{sig3}\AgdaSpace{}%
\AgdaBound{pbk5}\AgdaSymbol{)}\<%
\\
\>[0]\AgdaFunction{lemmaHoareTripleStackGeAux'22}\AgdaSpace{}%
\AgdaBound{msg₁}\AgdaSpace{}%
\AgdaBound{pbk1}\AgdaSpace{}%
\AgdaBound{pbk2}\AgdaSpace{}%
\AgdaBound{pbk3}\AgdaSpace{}%
\AgdaBound{pbk4}\AgdaSpace{}%
\AgdaBound{pbk5}\AgdaSpace{}%
\AgdaBound{sig1}\AgdaSpace{}%
\AgdaBound{sig2}\AgdaSpace{}%
\AgdaBound{sig3}\AgdaSpace{}%
\AgdaBound{x}\AgdaSpace{}%
\AgdaBound{x₁}\AgdaSpace{}%
\AgdaBound{x₂}\AgdaSpace{}%
\AgdaSymbol{|}\AgdaSpace{}%
\AgdaInductiveConstructor{true}\AgdaSpace{}%
\AgdaSymbol{|}\AgdaSpace{}%
\AgdaInductiveConstructor{false}\AgdaSpace{}%
\AgdaSymbol{|}\AgdaSpace{}%
\AgdaInductiveConstructor{true}\AgdaSpace{}%
\AgdaSymbol{|}\AgdaSpace{}%
\AgdaInductiveConstructor{false}\AgdaSpace{}%
\AgdaSymbol{|}\AgdaSpace{}%
\AgdaInductiveConstructor{true}\AgdaSpace{}%
\AgdaSymbol{=}\AgdaSpace{}%
\AgdaInductiveConstructor{tt}\<%
\\
\>[0]\AgdaFunction{lemmaHoareTripleStackGeAux'22}\AgdaSpace{}%
\AgdaBound{msg₁}\AgdaSpace{}%
\AgdaBound{pbk1}\AgdaSpace{}%
\AgdaBound{pbk2}\AgdaSpace{}%
\AgdaBound{pbk3}\AgdaSpace{}%
\AgdaBound{pbk4}\AgdaSpace{}%
\AgdaBound{pbk5}\AgdaSpace{}%
\AgdaBound{sig1}\AgdaSpace{}%
\AgdaBound{sig2}\AgdaSpace{}%
\AgdaBound{sig3}\AgdaSpace{}%
\AgdaBound{x}\AgdaSpace{}%
\AgdaBound{x₁}\AgdaSpace{}%
\AgdaBound{x₂}\AgdaSpace{}%
\AgdaSymbol{|}\AgdaSpace{}%
\AgdaInductiveConstructor{true}\AgdaSpace{}%
\AgdaSymbol{|}\AgdaSpace{}%
\AgdaInductiveConstructor{false}\AgdaSpace{}%
\AgdaSymbol{|}\AgdaSpace{}%
\AgdaInductiveConstructor{true}\AgdaSpace{}%
\AgdaSymbol{|}\AgdaSpace{}%
\AgdaInductiveConstructor{true}\AgdaSpace{}%
\AgdaSymbol{=}\AgdaSpace{}%
\AgdaInductiveConstructor{tt}\<%
\\
\>[0]\AgdaFunction{lemmaHoareTripleStackGeAux'22}\AgdaSpace{}%
\AgdaBound{msg₁}\AgdaSpace{}%
\AgdaBound{pbk1}\AgdaSpace{}%
\AgdaBound{pbk2}\AgdaSpace{}%
\AgdaBound{pbk3}\AgdaSpace{}%
\AgdaBound{pbk4}\AgdaSpace{}%
\AgdaBound{pbk5}\AgdaSpace{}%
\AgdaBound{sig1}\AgdaSpace{}%
\AgdaBound{sig2}\AgdaSpace{}%
\AgdaBound{sig3}\AgdaSpace{}%
\AgdaBound{x}\AgdaSpace{}%
\AgdaBound{x₁}\AgdaSpace{}%
\AgdaBound{x₂}\AgdaSpace{}%
\AgdaSymbol{|}\AgdaSpace{}%
\AgdaInductiveConstructor{true}\AgdaSpace{}%
\AgdaSymbol{|}\AgdaSpace{}%
\AgdaInductiveConstructor{true}\AgdaSpace{}%
\AgdaKeyword{with}\AgdaSpace{}%
\AgdaSymbol{(}\AgdaFunction{isSigned}%
\>[113]\AgdaBound{msg₁}\AgdaSpace{}%
\AgdaBound{sig3}\AgdaSpace{}%
\AgdaBound{pbk3}\AgdaSymbol{)}\<%
\\
\>[0]\AgdaFunction{lemmaHoareTripleStackGeAux'22}\AgdaSpace{}%
\AgdaBound{msg₁}\AgdaSpace{}%
\AgdaBound{pbk1}\AgdaSpace{}%
\AgdaBound{pbk2}\AgdaSpace{}%
\AgdaBound{pbk3}\AgdaSpace{}%
\AgdaBound{pbk4}\AgdaSpace{}%
\AgdaBound{pbk5}\AgdaSpace{}%
\AgdaBound{sig1}\AgdaSpace{}%
\AgdaBound{sig2}\AgdaSpace{}%
\AgdaBound{sig3}\AgdaSpace{}%
\AgdaBound{x}\AgdaSpace{}%
\AgdaBound{x₁}\AgdaSpace{}%
\AgdaBound{x₂}\AgdaSpace{}%
\AgdaSymbol{|}\AgdaSpace{}%
\AgdaInductiveConstructor{true}\AgdaSpace{}%
\AgdaSymbol{|}\AgdaSpace{}%
\AgdaInductiveConstructor{true}\AgdaSpace{}%
\AgdaSymbol{|}\AgdaSpace{}%
\AgdaInductiveConstructor{false}\AgdaSpace{}%
\AgdaKeyword{with}\AgdaSpace{}%
\AgdaSymbol{(}\AgdaFunction{isSigned}%
\>[121]\AgdaBound{msg₁}\AgdaSpace{}%
\AgdaBound{sig3}\AgdaSpace{}%
\AgdaBound{pbk4}\AgdaSymbol{)}\<%
\\
\>[0]\AgdaFunction{lemmaHoareTripleStackGeAux'22}\AgdaSpace{}%
\AgdaBound{msg₁}\AgdaSpace{}%
\AgdaBound{pbk1}\AgdaSpace{}%
\AgdaBound{pbk2}\AgdaSpace{}%
\AgdaBound{pbk3}\AgdaSpace{}%
\AgdaBound{pbk4}\AgdaSpace{}%
\AgdaBound{pbk5}\AgdaSpace{}%
\AgdaBound{sig1}\AgdaSpace{}%
\AgdaBound{sig2}\AgdaSpace{}%
\AgdaBound{sig3}\AgdaSpace{}%
\AgdaBound{x}\AgdaSpace{}%
\AgdaBound{x₁}\AgdaSpace{}%
\AgdaBound{x₂}\AgdaSpace{}%
\AgdaSymbol{|}\AgdaSpace{}%
\AgdaInductiveConstructor{true}\AgdaSpace{}%
\AgdaSymbol{|}\AgdaSpace{}%
\AgdaInductiveConstructor{true}\AgdaSpace{}%
\AgdaSymbol{|}\AgdaSpace{}%
\AgdaInductiveConstructor{false}\AgdaSpace{}%
\AgdaSymbol{|}\AgdaSpace{}%
\AgdaInductiveConstructor{false}\AgdaSpace{}%
\AgdaKeyword{with}%
\>[119]\AgdaSymbol{(}\AgdaFunction{isSigned}%
\>[130]\AgdaBound{msg₁}\AgdaSpace{}%
\AgdaBound{sig3}\AgdaSpace{}%
\AgdaBound{pbk5}\AgdaSymbol{)}\<%
\\
\>[0]\AgdaFunction{lemmaHoareTripleStackGeAux'22}\AgdaSpace{}%
\AgdaBound{msg₁}\AgdaSpace{}%
\AgdaBound{pbk1}\AgdaSpace{}%
\AgdaBound{pbk2}\AgdaSpace{}%
\AgdaBound{pbk3}\AgdaSpace{}%
\AgdaBound{pbk4}\AgdaSpace{}%
\AgdaBound{pbk5}\AgdaSpace{}%
\AgdaBound{sig1}\AgdaSpace{}%
\AgdaBound{sig2}\AgdaSpace{}%
\AgdaBound{sig3}\AgdaSpace{}%
\AgdaBound{x}\AgdaSpace{}%
\AgdaBound{x₁}\AgdaSpace{}%
\AgdaBound{x₂}\AgdaSpace{}%
\AgdaSymbol{|}\AgdaSpace{}%
\AgdaInductiveConstructor{true}\AgdaSpace{}%
\AgdaSymbol{|}\AgdaSpace{}%
\AgdaInductiveConstructor{true}\AgdaSpace{}%
\AgdaSymbol{|}\AgdaSpace{}%
\AgdaInductiveConstructor{false}\AgdaSpace{}%
\AgdaSymbol{|}\AgdaSpace{}%
\AgdaInductiveConstructor{false}\AgdaSpace{}%
\AgdaSymbol{|}\AgdaSpace{}%
\AgdaInductiveConstructor{true}\AgdaSpace{}%
\AgdaSymbol{=}\AgdaSpace{}%
\AgdaInductiveConstructor{tt}\<%
\\
\>[0]\AgdaFunction{lemmaHoareTripleStackGeAux'22}\AgdaSpace{}%
\AgdaBound{msg₁}\AgdaSpace{}%
\AgdaBound{pbk1}\AgdaSpace{}%
\AgdaBound{pbk2}\AgdaSpace{}%
\AgdaBound{pbk3}\AgdaSpace{}%
\AgdaBound{pbk4}\AgdaSpace{}%
\AgdaBound{pbk5}\AgdaSpace{}%
\AgdaBound{sig1}\AgdaSpace{}%
\AgdaBound{sig2}\AgdaSpace{}%
\AgdaBound{sig3}\AgdaSpace{}%
\AgdaBound{x}\AgdaSpace{}%
\AgdaBound{x₁}\AgdaSpace{}%
\AgdaBound{x₂}\AgdaSpace{}%
\AgdaSymbol{|}\AgdaSpace{}%
\AgdaInductiveConstructor{true}\AgdaSpace{}%
\AgdaSymbol{|}\AgdaSpace{}%
\AgdaInductiveConstructor{true}\AgdaSpace{}%
\AgdaSymbol{|}\AgdaSpace{}%
\AgdaInductiveConstructor{false}\AgdaSpace{}%
\AgdaSymbol{|}\AgdaSpace{}%
\AgdaInductiveConstructor{true}\AgdaSpace{}%
\AgdaSymbol{=}\AgdaSpace{}%
\AgdaInductiveConstructor{tt}\<%
\\
\>[0]\AgdaFunction{lemmaHoareTripleStackGeAux'22}\AgdaSpace{}%
\AgdaBound{msg₁}\AgdaSpace{}%
\AgdaBound{pbk1}\AgdaSpace{}%
\AgdaBound{pbk2}\AgdaSpace{}%
\AgdaBound{pbk3}\AgdaSpace{}%
\AgdaBound{pbk4}\AgdaSpace{}%
\AgdaBound{pbk5}\AgdaSpace{}%
\AgdaBound{sig1}\AgdaSpace{}%
\AgdaBound{sig2}\AgdaSpace{}%
\AgdaBound{sig3}\AgdaSpace{}%
\AgdaBound{x}\AgdaSpace{}%
\AgdaBound{x₁}\AgdaSpace{}%
\AgdaBound{x₂}\AgdaSpace{}%
\AgdaSymbol{|}\AgdaSpace{}%
\AgdaInductiveConstructor{true}\AgdaSpace{}%
\AgdaSymbol{|}\AgdaSpace{}%
\AgdaInductiveConstructor{true}\AgdaSpace{}%
\AgdaSymbol{|}\AgdaSpace{}%
\AgdaInductiveConstructor{true}\AgdaSpace{}%
\AgdaSymbol{=}\AgdaSpace{}%
\AgdaInductiveConstructor{tt}\<%
\\
\\[\AgdaEmptyExtraSkip]%
\\[\AgdaEmptyExtraSkip]%
\\[\AgdaEmptyExtraSkip]%
\>[0]\AgdaFunction{lemmaHoareTripleStackGeAux'23}%
\>[5804I]\AgdaSymbol{:}\AgdaSpace{}%
\AgdaSymbol{(}\AgdaBound{msg}\AgdaSpace{}%
\AgdaSymbol{:}\AgdaSpace{}%
\AgdaDatatype{Msg}\AgdaSymbol{)}\<%
\\
\>[5804I][@{}l@{\AgdaIndent{0}}]%
\>[31]\AgdaSymbol{(}\AgdaBound{pbk1}\AgdaSpace{}%
\AgdaBound{pbk2}\AgdaSpace{}%
\AgdaBound{pbk3}\AgdaSpace{}%
\AgdaBound{pbk4}\AgdaSpace{}%
\AgdaBound{pbk5}\AgdaSpace{}%
\AgdaBound{sig1}\AgdaSpace{}%
\AgdaBound{sig2}\AgdaSpace{}%
\AgdaBound{sig3}\AgdaSpace{}%
\AgdaSymbol{:}\AgdaSpace{}%
\AgdaDatatype{ℕ}\AgdaSymbol{)}\<%
\\
\>[31]\AgdaSymbol{→}\AgdaSpace{}%
\AgdaFunction{True}\AgdaSpace{}%
\AgdaSymbol{(}\AgdaFunction{isSigned}%
\>[49]\AgdaBound{msg}\AgdaSpace{}%
\AgdaBound{sig3}\AgdaSpace{}%
\AgdaBound{pbk5}\AgdaSymbol{)}\<%
\\
\>[.][@{}l@{}]\<[5804I]%
\>[30]\AgdaSymbol{→}%
\>[33]\AgdaFunction{True}\AgdaSpace{}%
\AgdaSymbol{(}\AgdaFunction{isSigned}%
\>[49]\AgdaBound{msg}\AgdaSpace{}%
\AgdaBound{sig2}\AgdaSpace{}%
\AgdaBound{pbk4}\AgdaSymbol{)}\<%
\\
\>[30]\AgdaSymbol{→}%
\>[33]\AgdaFunction{True}\AgdaSpace{}%
\AgdaSymbol{(}\AgdaFunction{isSigned}%
\>[49]\AgdaBound{msg}\AgdaSpace{}%
\AgdaBound{sig1}\AgdaSpace{}%
\AgdaBound{pbk3}\AgdaSymbol{)}\<%
\\
\>[30][@{}l@{\AgdaIndent{0}}]%
\>[31]\AgdaSymbol{→}\AgdaSpace{}%
\AgdaFunction{True}\AgdaSpace{}%
\AgdaSymbol{(}\AgdaFunction{compareSigsMultiSig}\AgdaSpace{}%
\AgdaBound{msg}\AgdaSpace{}%
\AgdaSymbol{(}\AgdaSpace{}%
\AgdaBound{sig1}\AgdaSpace{}%
\AgdaOperator{\AgdaInductiveConstructor{∷}}\AgdaSpace{}%
\AgdaBound{sig2}\AgdaSpace{}%
\AgdaOperator{\AgdaInductiveConstructor{∷}}\AgdaSpace{}%
\AgdaBound{sig3}\AgdaSpace{}%
\AgdaOperator{\AgdaInductiveConstructor{∷}}\AgdaSpace{}%
\AgdaInductiveConstructor{[]}\AgdaSymbol{)}\AgdaSpace{}%
\AgdaSymbol{(}\AgdaBound{pbk1}\AgdaSpace{}%
\AgdaOperator{\AgdaInductiveConstructor{∷}}\AgdaSpace{}%
\AgdaBound{pbk2}\AgdaSpace{}%
\AgdaOperator{\AgdaInductiveConstructor{∷}}\AgdaSpace{}%
\AgdaBound{pbk3}\AgdaSpace{}%
\AgdaOperator{\AgdaInductiveConstructor{∷}}\AgdaSpace{}%
\AgdaBound{pbk4}\AgdaSpace{}%
\AgdaOperator{\AgdaInductiveConstructor{∷}}\AgdaSpace{}%
\AgdaBound{pbk5}\AgdaSpace{}%
\AgdaOperator{\AgdaInductiveConstructor{∷}}\AgdaSpace{}%
\AgdaInductiveConstructor{[]}\AgdaSymbol{))}\<%
\\
\>[0]\AgdaFunction{lemmaHoareTripleStackGeAux'23}\AgdaSpace{}%
\AgdaBound{msg₁}\AgdaSpace{}%
\AgdaBound{pbk1}\AgdaSpace{}%
\AgdaBound{pbk2}\AgdaSpace{}%
\AgdaBound{pbk3}\AgdaSpace{}%
\AgdaBound{pbk4}\AgdaSpace{}%
\AgdaBound{pbk5}\AgdaSpace{}%
\AgdaBound{sig1}\AgdaSpace{}%
\AgdaBound{sig2}\AgdaSpace{}%
\AgdaBound{sig3}\AgdaSpace{}%
\AgdaBound{x}\AgdaSpace{}%
\AgdaBound{x₁}\AgdaSpace{}%
\AgdaBound{x₂}\AgdaSpace{}%
\AgdaKeyword{with}\AgdaSpace{}%
\AgdaSymbol{(}\AgdaFunction{isSigned}%
\>[99]\AgdaBound{msg₁}\AgdaSpace{}%
\AgdaBound{sig1}\AgdaSpace{}%
\AgdaBound{pbk1}\AgdaSymbol{)}\<%
\\
\>[0]\AgdaFunction{lemmaHoareTripleStackGeAux'23}\AgdaSpace{}%
\AgdaBound{msg₁}\AgdaSpace{}%
\AgdaBound{pbk1}\AgdaSpace{}%
\AgdaBound{pbk2}\AgdaSpace{}%
\AgdaBound{pbk3}\AgdaSpace{}%
\AgdaBound{pbk4}\AgdaSpace{}%
\AgdaBound{pbk5}\AgdaSpace{}%
\AgdaBound{sig1}\AgdaSpace{}%
\AgdaBound{sig2}\AgdaSpace{}%
\AgdaBound{sig3}\AgdaSpace{}%
\AgdaBound{x}\AgdaSpace{}%
\AgdaBound{x₁}\AgdaSpace{}%
\AgdaBound{x₂}\AgdaSpace{}%
\AgdaSymbol{|}\AgdaSpace{}%
\AgdaInductiveConstructor{false}\AgdaSpace{}%
\AgdaKeyword{with}\AgdaSpace{}%
\AgdaSymbol{(}\AgdaFunction{isSigned}%
\>[107]\AgdaBound{msg₁}\AgdaSpace{}%
\AgdaBound{sig1}\AgdaSpace{}%
\AgdaBound{pbk2}\AgdaSymbol{)}\<%
\\
\>[0]\AgdaFunction{lemmaHoareTripleStackGeAux'23}\AgdaSpace{}%
\AgdaBound{msg₁}\AgdaSpace{}%
\AgdaBound{pbk1}\AgdaSpace{}%
\AgdaBound{pbk2}\AgdaSpace{}%
\AgdaBound{pbk3}\AgdaSpace{}%
\AgdaBound{pbk4}\AgdaSpace{}%
\AgdaBound{pbk5}\AgdaSpace{}%
\AgdaBound{sig1}\AgdaSpace{}%
\AgdaBound{sig2}\AgdaSpace{}%
\AgdaBound{sig3}\AgdaSpace{}%
\AgdaBound{x}\AgdaSpace{}%
\AgdaBound{x₁}\AgdaSpace{}%
\AgdaBound{x₂}\AgdaSpace{}%
\AgdaSymbol{|}\AgdaSpace{}%
\AgdaInductiveConstructor{false}\AgdaSpace{}%
\AgdaSymbol{|}\AgdaSpace{}%
\AgdaInductiveConstructor{false}\AgdaSpace{}%
\AgdaKeyword{with}\AgdaSpace{}%
\AgdaSymbol{(}\AgdaFunction{isSigned}%
\>[115]\AgdaBound{msg₁}\AgdaSpace{}%
\AgdaBound{sig1}\AgdaSpace{}%
\AgdaBound{pbk3}\AgdaSymbol{)}\<%
\\
\>[0]\AgdaFunction{lemmaHoareTripleStackGeAux'23}\AgdaSpace{}%
\AgdaBound{msg₁}\AgdaSpace{}%
\AgdaBound{pbk1}\AgdaSpace{}%
\AgdaBound{pbk2}\AgdaSpace{}%
\AgdaBound{pbk3}\AgdaSpace{}%
\AgdaBound{pbk4}\AgdaSpace{}%
\AgdaBound{pbk5}\AgdaSpace{}%
\AgdaBound{sig1}\AgdaSpace{}%
\AgdaBound{sig2}\AgdaSpace{}%
\AgdaBound{sig3}\AgdaSpace{}%
\AgdaBound{x}\AgdaSpace{}%
\AgdaBound{x₁}\AgdaSpace{}%
\AgdaBound{x₂}\AgdaSpace{}%
\AgdaSymbol{|}\AgdaSpace{}%
\AgdaInductiveConstructor{false}\AgdaSpace{}%
\AgdaSymbol{|}\AgdaSpace{}%
\AgdaInductiveConstructor{false}\AgdaSpace{}%
\AgdaSymbol{|}\AgdaSpace{}%
\AgdaInductiveConstructor{true}\AgdaSpace{}%
\AgdaKeyword{with}\AgdaSpace{}%
\AgdaSymbol{(}\AgdaFunction{isSigned}%
\>[122]\AgdaBound{msg₁}\AgdaSpace{}%
\AgdaBound{sig2}\AgdaSpace{}%
\AgdaBound{pbk4}\AgdaSymbol{)}\<%
\\
\>[0]\AgdaFunction{lemmaHoareTripleStackGeAux'23}\AgdaSpace{}%
\AgdaBound{msg₁}\AgdaSpace{}%
\AgdaBound{pbk1}\AgdaSpace{}%
\AgdaBound{pbk2}\AgdaSpace{}%
\AgdaBound{pbk3}\AgdaSpace{}%
\AgdaBound{pbk4}\AgdaSpace{}%
\AgdaBound{pbk5}\AgdaSpace{}%
\AgdaBound{sig1}\AgdaSpace{}%
\AgdaBound{sig2}\AgdaSpace{}%
\AgdaBound{sig3}\AgdaSpace{}%
\AgdaBound{x}\AgdaSpace{}%
\AgdaBound{x₁}\AgdaSpace{}%
\AgdaBound{x₂}\AgdaSpace{}%
\AgdaSymbol{|}\AgdaSpace{}%
\AgdaInductiveConstructor{false}\AgdaSpace{}%
\AgdaSymbol{|}\AgdaSpace{}%
\AgdaInductiveConstructor{false}\AgdaSpace{}%
\AgdaSymbol{|}\AgdaSpace{}%
\AgdaInductiveConstructor{true}\AgdaSpace{}%
\AgdaSymbol{|}\AgdaSpace{}%
\AgdaInductiveConstructor{true}\AgdaSpace{}%
\AgdaKeyword{with}\AgdaSpace{}%
\AgdaSymbol{(}\AgdaFunction{isSigned}%
\>[129]\AgdaBound{msg₁}\AgdaSpace{}%
\AgdaBound{sig3}\AgdaSpace{}%
\AgdaBound{pbk5}\AgdaSymbol{)}\<%
\\
\>[0]\AgdaFunction{lemmaHoareTripleStackGeAux'23}\AgdaSpace{}%
\AgdaBound{msg₁}\AgdaSpace{}%
\AgdaBound{pbk1}\AgdaSpace{}%
\AgdaBound{pbk2}\AgdaSpace{}%
\AgdaBound{pbk3}\AgdaSpace{}%
\AgdaBound{pbk4}\AgdaSpace{}%
\AgdaBound{pbk5}\AgdaSpace{}%
\AgdaBound{sig1}\AgdaSpace{}%
\AgdaBound{sig2}\AgdaSpace{}%
\AgdaBound{sig3}\AgdaSpace{}%
\AgdaBound{x}\AgdaSpace{}%
\AgdaBound{x₁}\AgdaSpace{}%
\AgdaBound{x₂}\AgdaSpace{}%
\AgdaSymbol{|}\AgdaSpace{}%
\AgdaInductiveConstructor{false}\AgdaSpace{}%
\AgdaSymbol{|}\AgdaSpace{}%
\AgdaInductiveConstructor{false}\AgdaSpace{}%
\AgdaSymbol{|}\AgdaSpace{}%
\AgdaInductiveConstructor{true}\AgdaSpace{}%
\AgdaSymbol{|}\AgdaSpace{}%
\AgdaInductiveConstructor{true}\AgdaSpace{}%
\AgdaSymbol{|}\AgdaSpace{}%
\AgdaInductiveConstructor{true}\AgdaSpace{}%
\AgdaSymbol{=}\AgdaSpace{}%
\AgdaInductiveConstructor{tt}\<%
\\
\>[0]\AgdaFunction{lemmaHoareTripleStackGeAux'23}\AgdaSpace{}%
\AgdaBound{msg₁}\AgdaSpace{}%
\AgdaBound{pbk1}\AgdaSpace{}%
\AgdaBound{pbk2}\AgdaSpace{}%
\AgdaBound{pbk3}\AgdaSpace{}%
\AgdaBound{pbk4}\AgdaSpace{}%
\AgdaBound{pbk5}\AgdaSpace{}%
\AgdaBound{sig1}\AgdaSpace{}%
\AgdaBound{sig2}\AgdaSpace{}%
\AgdaBound{sig3}\AgdaSpace{}%
\AgdaBound{x}\AgdaSpace{}%
\AgdaBound{x₁}\AgdaSpace{}%
\AgdaBound{x₂}\AgdaSpace{}%
\AgdaSymbol{|}\AgdaSpace{}%
\AgdaInductiveConstructor{false}\AgdaSpace{}%
\AgdaSymbol{|}\AgdaSpace{}%
\AgdaInductiveConstructor{true}\AgdaSpace{}%
\AgdaKeyword{with}\AgdaSpace{}%
\AgdaSymbol{(}\AgdaFunction{isSigned}%
\>[114]\AgdaBound{msg₁}\AgdaSpace{}%
\AgdaBound{sig2}\AgdaSpace{}%
\AgdaBound{pbk3}\AgdaSymbol{)}\<%
\\
\>[0]\AgdaFunction{lemmaHoareTripleStackGeAux'23}\AgdaSpace{}%
\AgdaBound{msg₁}\AgdaSpace{}%
\AgdaBound{pbk1}\AgdaSpace{}%
\AgdaBound{pbk2}\AgdaSpace{}%
\AgdaBound{pbk3}\AgdaSpace{}%
\AgdaBound{pbk4}\AgdaSpace{}%
\AgdaBound{pbk5}\AgdaSpace{}%
\AgdaBound{sig1}\AgdaSpace{}%
\AgdaBound{sig2}\AgdaSpace{}%
\AgdaBound{sig3}\AgdaSpace{}%
\AgdaBound{x}\AgdaSpace{}%
\AgdaBound{x₁}\AgdaSpace{}%
\AgdaBound{x₂}\AgdaSpace{}%
\AgdaSymbol{|}\AgdaSpace{}%
\AgdaInductiveConstructor{false}\AgdaSpace{}%
\AgdaSymbol{|}\AgdaSpace{}%
\AgdaInductiveConstructor{true}\AgdaSpace{}%
\AgdaSymbol{|}\AgdaSpace{}%
\AgdaInductiveConstructor{false}\AgdaSpace{}%
\AgdaKeyword{with}\AgdaSpace{}%
\AgdaSymbol{(}\AgdaFunction{isSigned}%
\>[122]\AgdaBound{msg₁}\AgdaSpace{}%
\AgdaBound{sig2}\AgdaSpace{}%
\AgdaBound{pbk4}\AgdaSymbol{)}\<%
\\
\>[0]\AgdaFunction{lemmaHoareTripleStackGeAux'23}\AgdaSpace{}%
\AgdaBound{msg₁}\AgdaSpace{}%
\AgdaBound{pbk1}\AgdaSpace{}%
\AgdaBound{pbk2}\AgdaSpace{}%
\AgdaBound{pbk3}\AgdaSpace{}%
\AgdaBound{pbk4}\AgdaSpace{}%
\AgdaBound{pbk5}\AgdaSpace{}%
\AgdaBound{sig1}\AgdaSpace{}%
\AgdaBound{sig2}\AgdaSpace{}%
\AgdaBound{sig3}\AgdaSpace{}%
\AgdaBound{x}\AgdaSpace{}%
\AgdaBound{x₁}\AgdaSpace{}%
\AgdaBound{x₂}\AgdaSpace{}%
\AgdaSymbol{|}\AgdaSpace{}%
\AgdaInductiveConstructor{false}\AgdaSpace{}%
\AgdaSymbol{|}\AgdaSpace{}%
\AgdaInductiveConstructor{true}\AgdaSpace{}%
\AgdaSymbol{|}\AgdaSpace{}%
\AgdaInductiveConstructor{false}\AgdaSpace{}%
\AgdaSymbol{|}\AgdaSpace{}%
\AgdaInductiveConstructor{true}\AgdaSpace{}%
\AgdaKeyword{with}\AgdaSpace{}%
\AgdaSymbol{(}\AgdaFunction{isSigned}%
\>[129]\AgdaBound{msg₁}\AgdaSpace{}%
\AgdaBound{sig3}\AgdaSpace{}%
\AgdaBound{pbk5}\AgdaSymbol{)}\<%
\\
\>[0]\AgdaFunction{lemmaHoareTripleStackGeAux'23}\AgdaSpace{}%
\AgdaBound{msg₁}\AgdaSpace{}%
\AgdaBound{pbk1}\AgdaSpace{}%
\AgdaBound{pbk2}\AgdaSpace{}%
\AgdaBound{pbk3}\AgdaSpace{}%
\AgdaBound{pbk4}\AgdaSpace{}%
\AgdaBound{pbk5}\AgdaSpace{}%
\AgdaBound{sig1}\AgdaSpace{}%
\AgdaBound{sig2}\AgdaSpace{}%
\AgdaBound{sig3}\AgdaSpace{}%
\AgdaBound{x}\AgdaSpace{}%
\AgdaBound{x₁}\AgdaSpace{}%
\AgdaBound{x₂}\AgdaSpace{}%
\AgdaSymbol{|}\AgdaSpace{}%
\AgdaInductiveConstructor{false}\AgdaSpace{}%
\AgdaSymbol{|}\AgdaSpace{}%
\AgdaInductiveConstructor{true}\AgdaSpace{}%
\AgdaSymbol{|}\AgdaSpace{}%
\AgdaInductiveConstructor{false}\AgdaSpace{}%
\AgdaSymbol{|}\AgdaSpace{}%
\AgdaInductiveConstructor{true}\AgdaSpace{}%
\AgdaSymbol{|}\AgdaSpace{}%
\AgdaInductiveConstructor{true}\AgdaSpace{}%
\AgdaSymbol{=}\AgdaSpace{}%
\AgdaInductiveConstructor{tt}\<%
\\
\>[0]\AgdaFunction{lemmaHoareTripleStackGeAux'23}\AgdaSpace{}%
\AgdaBound{msg₁}\AgdaSpace{}%
\AgdaBound{pbk1}\AgdaSpace{}%
\AgdaBound{pbk2}\AgdaSpace{}%
\AgdaBound{pbk3}\AgdaSpace{}%
\AgdaBound{pbk4}\AgdaSpace{}%
\AgdaBound{pbk5}\AgdaSpace{}%
\AgdaBound{sig1}\AgdaSpace{}%
\AgdaBound{sig2}\AgdaSpace{}%
\AgdaBound{sig3}\AgdaSpace{}%
\AgdaBound{x}\AgdaSpace{}%
\AgdaBound{x₁}\AgdaSpace{}%
\AgdaBound{x₂}\AgdaSpace{}%
\AgdaSymbol{|}\AgdaSpace{}%
\AgdaInductiveConstructor{false}\AgdaSpace{}%
\AgdaSymbol{|}\AgdaSpace{}%
\AgdaInductiveConstructor{true}\AgdaSpace{}%
\AgdaSymbol{|}\AgdaSpace{}%
\AgdaInductiveConstructor{true}\AgdaSpace{}%
\AgdaKeyword{with}\AgdaSpace{}%
\AgdaSymbol{(}\AgdaFunction{isSigned}%
\>[121]\AgdaBound{msg₁}\AgdaSpace{}%
\AgdaBound{sig3}\AgdaSpace{}%
\AgdaBound{pbk4}\AgdaSymbol{)}\<%
\\
\>[0]\AgdaFunction{lemmaHoareTripleStackGeAux'23}\AgdaSpace{}%
\AgdaBound{msg₁}\AgdaSpace{}%
\AgdaBound{pbk1}\AgdaSpace{}%
\AgdaBound{pbk2}\AgdaSpace{}%
\AgdaBound{pbk3}\AgdaSpace{}%
\AgdaBound{pbk4}\AgdaSpace{}%
\AgdaBound{pbk5}\AgdaSpace{}%
\AgdaBound{sig1}\AgdaSpace{}%
\AgdaBound{sig2}\AgdaSpace{}%
\AgdaBound{sig3}\AgdaSpace{}%
\AgdaBound{x}\AgdaSpace{}%
\AgdaBound{x₁}\AgdaSpace{}%
\AgdaBound{x₂}\AgdaSpace{}%
\AgdaSymbol{|}\AgdaSpace{}%
\AgdaInductiveConstructor{false}\AgdaSpace{}%
\AgdaSymbol{|}\AgdaSpace{}%
\AgdaInductiveConstructor{true}\AgdaSpace{}%
\AgdaSymbol{|}\AgdaSpace{}%
\AgdaInductiveConstructor{true}\AgdaSpace{}%
\AgdaSymbol{|}\AgdaSpace{}%
\AgdaInductiveConstructor{false}\AgdaSpace{}%
\AgdaKeyword{with}\AgdaSpace{}%
\AgdaSymbol{(}\AgdaFunction{isSigned}%
\>[129]\AgdaBound{msg₁}\AgdaSpace{}%
\AgdaBound{sig3}\AgdaSpace{}%
\AgdaBound{pbk5}\AgdaSymbol{)}\<%
\\
\>[0]\AgdaFunction{lemmaHoareTripleStackGeAux'23}\AgdaSpace{}%
\AgdaBound{msg₁}\AgdaSpace{}%
\AgdaBound{pbk1}\AgdaSpace{}%
\AgdaBound{pbk2}\AgdaSpace{}%
\AgdaBound{pbk3}\AgdaSpace{}%
\AgdaBound{pbk4}\AgdaSpace{}%
\AgdaBound{pbk5}\AgdaSpace{}%
\AgdaBound{sig1}\AgdaSpace{}%
\AgdaBound{sig2}\AgdaSpace{}%
\AgdaBound{sig3}\AgdaSpace{}%
\AgdaBound{x}\AgdaSpace{}%
\AgdaBound{x₁}\AgdaSpace{}%
\AgdaBound{x₂}\AgdaSpace{}%
\AgdaSymbol{|}\AgdaSpace{}%
\AgdaInductiveConstructor{false}\AgdaSpace{}%
\AgdaSymbol{|}\AgdaSpace{}%
\AgdaInductiveConstructor{true}\AgdaSpace{}%
\AgdaSymbol{|}\AgdaSpace{}%
\AgdaInductiveConstructor{true}\AgdaSpace{}%
\AgdaSymbol{|}\AgdaSpace{}%
\AgdaInductiveConstructor{false}\AgdaSpace{}%
\AgdaSymbol{|}\AgdaSpace{}%
\AgdaInductiveConstructor{true}\AgdaSpace{}%
\AgdaSymbol{=}\AgdaSpace{}%
\AgdaInductiveConstructor{tt}\<%
\\
\>[0]\AgdaFunction{lemmaHoareTripleStackGeAux'23}\AgdaSpace{}%
\AgdaBound{msg₁}\AgdaSpace{}%
\AgdaBound{pbk1}\AgdaSpace{}%
\AgdaBound{pbk2}\AgdaSpace{}%
\AgdaBound{pbk3}\AgdaSpace{}%
\AgdaBound{pbk4}\AgdaSpace{}%
\AgdaBound{pbk5}\AgdaSpace{}%
\AgdaBound{sig1}\AgdaSpace{}%
\AgdaBound{sig2}\AgdaSpace{}%
\AgdaBound{sig3}\AgdaSpace{}%
\AgdaBound{x}\AgdaSpace{}%
\AgdaBound{x₁}\AgdaSpace{}%
\AgdaBound{x₂}\AgdaSpace{}%
\AgdaSymbol{|}\AgdaSpace{}%
\AgdaInductiveConstructor{false}\AgdaSpace{}%
\AgdaSymbol{|}\AgdaSpace{}%
\AgdaInductiveConstructor{true}\AgdaSpace{}%
\AgdaSymbol{|}\AgdaSpace{}%
\AgdaInductiveConstructor{true}\AgdaSpace{}%
\AgdaSymbol{|}\AgdaSpace{}%
\AgdaInductiveConstructor{true}\AgdaSpace{}%
\AgdaSymbol{=}\AgdaSpace{}%
\AgdaInductiveConstructor{tt}\<%
\\
\>[0]\AgdaFunction{lemmaHoareTripleStackGeAux'23}\AgdaSpace{}%
\AgdaBound{msg₁}\AgdaSpace{}%
\AgdaBound{pbk1}\AgdaSpace{}%
\AgdaBound{pbk2}\AgdaSpace{}%
\AgdaBound{pbk3}\AgdaSpace{}%
\AgdaBound{pbk4}\AgdaSpace{}%
\AgdaBound{pbk5}\AgdaSpace{}%
\AgdaBound{sig1}\AgdaSpace{}%
\AgdaBound{sig2}\AgdaSpace{}%
\AgdaBound{sig3}\AgdaSpace{}%
\AgdaBound{x}\AgdaSpace{}%
\AgdaBound{x₁}\AgdaSpace{}%
\AgdaBound{x₂}\AgdaSpace{}%
\AgdaSymbol{|}\AgdaSpace{}%
\AgdaInductiveConstructor{true}\AgdaSpace{}%
\AgdaKeyword{with}\AgdaSpace{}%
\AgdaSymbol{(}\AgdaFunction{isSigned}%
\>[106]\AgdaBound{msg₁}\AgdaSpace{}%
\AgdaBound{sig2}\AgdaSpace{}%
\AgdaBound{pbk2}\AgdaSymbol{)}\<%
\\
\>[0]\AgdaFunction{lemmaHoareTripleStackGeAux'23}\AgdaSpace{}%
\AgdaBound{msg₁}\AgdaSpace{}%
\AgdaBound{pbk1}\AgdaSpace{}%
\AgdaBound{pbk2}\AgdaSpace{}%
\AgdaBound{pbk3}\AgdaSpace{}%
\AgdaBound{pbk4}\AgdaSpace{}%
\AgdaBound{pbk5}\AgdaSpace{}%
\AgdaBound{sig1}\AgdaSpace{}%
\AgdaBound{sig2}\AgdaSpace{}%
\AgdaBound{sig3}\AgdaSpace{}%
\AgdaBound{x}\AgdaSpace{}%
\AgdaBound{x₁}\AgdaSpace{}%
\AgdaBound{x₂}\AgdaSpace{}%
\AgdaSymbol{|}\AgdaSpace{}%
\AgdaInductiveConstructor{true}\AgdaSpace{}%
\AgdaSymbol{|}\AgdaSpace{}%
\AgdaInductiveConstructor{false}\AgdaSpace{}%
\AgdaKeyword{with}\AgdaSpace{}%
\AgdaSymbol{(}\AgdaFunction{isSigned}%
\>[114]\AgdaBound{msg₁}\AgdaSpace{}%
\AgdaBound{sig2}\AgdaSpace{}%
\AgdaBound{pbk3}\AgdaSymbol{)}\<%
\\
\>[0]\AgdaFunction{lemmaHoareTripleStackGeAux'23}\AgdaSpace{}%
\AgdaBound{msg₁}\AgdaSpace{}%
\AgdaBound{pbk1}\AgdaSpace{}%
\AgdaBound{pbk2}\AgdaSpace{}%
\AgdaBound{pbk3}\AgdaSpace{}%
\AgdaBound{pbk4}\AgdaSpace{}%
\AgdaBound{pbk5}\AgdaSpace{}%
\AgdaBound{sig1}\AgdaSpace{}%
\AgdaBound{sig2}\AgdaSpace{}%
\AgdaBound{sig3}\AgdaSpace{}%
\AgdaBound{x}\AgdaSpace{}%
\AgdaBound{x₁}\AgdaSpace{}%
\AgdaBound{x₂}\AgdaSpace{}%
\AgdaSymbol{|}\AgdaSpace{}%
\AgdaInductiveConstructor{true}\AgdaSpace{}%
\AgdaSymbol{|}\AgdaSpace{}%
\AgdaInductiveConstructor{false}\AgdaSpace{}%
\AgdaSymbol{|}\AgdaSpace{}%
\AgdaInductiveConstructor{false}\AgdaSpace{}%
\AgdaKeyword{with}\AgdaSpace{}%
\AgdaSymbol{(}\AgdaFunction{isSigned}%
\>[122]\AgdaBound{msg₁}\AgdaSpace{}%
\AgdaBound{sig2}\AgdaSpace{}%
\AgdaBound{pbk4}\AgdaSymbol{)}\<%
\\
\>[0]\AgdaFunction{lemmaHoareTripleStackGeAux'23}\AgdaSpace{}%
\AgdaBound{msg₁}\AgdaSpace{}%
\AgdaBound{pbk1}\AgdaSpace{}%
\AgdaBound{pbk2}\AgdaSpace{}%
\AgdaBound{pbk3}\AgdaSpace{}%
\AgdaBound{pbk4}\AgdaSpace{}%
\AgdaBound{pbk5}\AgdaSpace{}%
\AgdaBound{sig1}\AgdaSpace{}%
\AgdaBound{sig2}\AgdaSpace{}%
\AgdaBound{sig3}\AgdaSpace{}%
\AgdaBound{x}\AgdaSpace{}%
\AgdaBound{x₁}\AgdaSpace{}%
\AgdaBound{x₂}\AgdaSpace{}%
\AgdaSymbol{|}\AgdaSpace{}%
\AgdaInductiveConstructor{true}\AgdaSpace{}%
\AgdaSymbol{|}\AgdaSpace{}%
\AgdaInductiveConstructor{false}\AgdaSpace{}%
\AgdaSymbol{|}\AgdaSpace{}%
\AgdaInductiveConstructor{false}\AgdaSpace{}%
\AgdaSymbol{|}\AgdaSpace{}%
\AgdaInductiveConstructor{true}\AgdaSpace{}%
\AgdaKeyword{with}%
\>[119]\AgdaSymbol{(}\AgdaFunction{isSigned}%
\>[130]\AgdaBound{msg₁}\AgdaSpace{}%
\AgdaBound{sig3}\AgdaSpace{}%
\AgdaBound{pbk5}\AgdaSymbol{)}\<%
\\
\>[0]\AgdaFunction{lemmaHoareTripleStackGeAux'23}\AgdaSpace{}%
\AgdaBound{msg₁}\AgdaSpace{}%
\AgdaBound{pbk1}\AgdaSpace{}%
\AgdaBound{pbk2}\AgdaSpace{}%
\AgdaBound{pbk3}\AgdaSpace{}%
\AgdaBound{pbk4}\AgdaSpace{}%
\AgdaBound{pbk5}\AgdaSpace{}%
\AgdaBound{sig1}\AgdaSpace{}%
\AgdaBound{sig2}\AgdaSpace{}%
\AgdaBound{sig3}\AgdaSpace{}%
\AgdaBound{x}\AgdaSpace{}%
\AgdaBound{x₁}\AgdaSpace{}%
\AgdaBound{x₂}\AgdaSpace{}%
\AgdaSymbol{|}\AgdaSpace{}%
\AgdaInductiveConstructor{true}\AgdaSpace{}%
\AgdaSymbol{|}\AgdaSpace{}%
\AgdaInductiveConstructor{false}\AgdaSpace{}%
\AgdaSymbol{|}\AgdaSpace{}%
\AgdaInductiveConstructor{false}\AgdaSpace{}%
\AgdaSymbol{|}\AgdaSpace{}%
\AgdaInductiveConstructor{true}\AgdaSpace{}%
\AgdaSymbol{|}\AgdaSpace{}%
\AgdaInductiveConstructor{true}\AgdaSpace{}%
\AgdaSymbol{=}\AgdaSpace{}%
\AgdaInductiveConstructor{tt}\<%
\\
\>[0]\AgdaFunction{lemmaHoareTripleStackGeAux'23}\AgdaSpace{}%
\AgdaBound{msg₁}\AgdaSpace{}%
\AgdaBound{pbk1}\AgdaSpace{}%
\AgdaBound{pbk2}\AgdaSpace{}%
\AgdaBound{pbk3}\AgdaSpace{}%
\AgdaBound{pbk4}\AgdaSpace{}%
\AgdaBound{pbk5}\AgdaSpace{}%
\AgdaBound{sig1}\AgdaSpace{}%
\AgdaBound{sig2}\AgdaSpace{}%
\AgdaBound{sig3}\AgdaSpace{}%
\AgdaBound{x}\AgdaSpace{}%
\AgdaBound{x₁}\AgdaSpace{}%
\AgdaBound{x₂}\AgdaSpace{}%
\AgdaSymbol{|}\AgdaSpace{}%
\AgdaInductiveConstructor{true}\AgdaSpace{}%
\AgdaSymbol{|}\AgdaSpace{}%
\AgdaInductiveConstructor{false}\AgdaSpace{}%
\AgdaSymbol{|}\AgdaSpace{}%
\AgdaInductiveConstructor{true}\AgdaSpace{}%
\AgdaKeyword{with}\AgdaSpace{}%
\AgdaSymbol{(}\AgdaFunction{isSigned}%
\>[121]\AgdaBound{msg₁}\AgdaSpace{}%
\AgdaBound{sig3}\AgdaSpace{}%
\AgdaBound{pbk4}\AgdaSymbol{)}\<%
\\
\>[0]\AgdaFunction{lemmaHoareTripleStackGeAux'23}\AgdaSpace{}%
\AgdaBound{msg₁}\AgdaSpace{}%
\AgdaBound{pbk1}\AgdaSpace{}%
\AgdaBound{pbk2}\AgdaSpace{}%
\AgdaBound{pbk3}\AgdaSpace{}%
\AgdaBound{pbk4}\AgdaSpace{}%
\AgdaBound{pbk5}\AgdaSpace{}%
\AgdaBound{sig1}\AgdaSpace{}%
\AgdaBound{sig2}\AgdaSpace{}%
\AgdaBound{sig3}\AgdaSpace{}%
\AgdaBound{x}\AgdaSpace{}%
\AgdaBound{x₁}\AgdaSpace{}%
\AgdaBound{x₂}\AgdaSpace{}%
\AgdaSymbol{|}\AgdaSpace{}%
\AgdaInductiveConstructor{true}\AgdaSpace{}%
\AgdaSymbol{|}\AgdaSpace{}%
\AgdaInductiveConstructor{false}\AgdaSpace{}%
\AgdaSymbol{|}\AgdaSpace{}%
\AgdaInductiveConstructor{true}\AgdaSpace{}%
\AgdaSymbol{|}\AgdaSpace{}%
\AgdaInductiveConstructor{false}\AgdaSpace{}%
\AgdaKeyword{with}%
\>[119]\AgdaSymbol{(}\AgdaFunction{isSigned}%
\>[130]\AgdaBound{msg₁}\AgdaSpace{}%
\AgdaBound{sig3}\AgdaSpace{}%
\AgdaBound{pbk5}\AgdaSymbol{)}\<%
\\
\>[0]\AgdaFunction{lemmaHoareTripleStackGeAux'23}\AgdaSpace{}%
\AgdaBound{msg₁}\AgdaSpace{}%
\AgdaBound{pbk1}\AgdaSpace{}%
\AgdaBound{pbk2}\AgdaSpace{}%
\AgdaBound{pbk3}\AgdaSpace{}%
\AgdaBound{pbk4}\AgdaSpace{}%
\AgdaBound{pbk5}\AgdaSpace{}%
\AgdaBound{sig1}\AgdaSpace{}%
\AgdaBound{sig2}\AgdaSpace{}%
\AgdaBound{sig3}\AgdaSpace{}%
\AgdaBound{x}\AgdaSpace{}%
\AgdaBound{x₁}\AgdaSpace{}%
\AgdaBound{x₂}\AgdaSpace{}%
\AgdaSymbol{|}\AgdaSpace{}%
\AgdaInductiveConstructor{true}\AgdaSpace{}%
\AgdaSymbol{|}\AgdaSpace{}%
\AgdaInductiveConstructor{false}\AgdaSpace{}%
\AgdaSymbol{|}\AgdaSpace{}%
\AgdaInductiveConstructor{true}\AgdaSpace{}%
\AgdaSymbol{|}\AgdaSpace{}%
\AgdaInductiveConstructor{false}\AgdaSpace{}%
\AgdaSymbol{|}\AgdaSpace{}%
\AgdaInductiveConstructor{true}\AgdaSpace{}%
\AgdaSymbol{=}\AgdaSpace{}%
\AgdaInductiveConstructor{tt}\<%
\\
\>[0]\AgdaFunction{lemmaHoareTripleStackGeAux'23}\AgdaSpace{}%
\AgdaBound{msg₁}\AgdaSpace{}%
\AgdaBound{pbk1}\AgdaSpace{}%
\AgdaBound{pbk2}\AgdaSpace{}%
\AgdaBound{pbk3}\AgdaSpace{}%
\AgdaBound{pbk4}\AgdaSpace{}%
\AgdaBound{pbk5}\AgdaSpace{}%
\AgdaBound{sig1}\AgdaSpace{}%
\AgdaBound{sig2}\AgdaSpace{}%
\AgdaBound{sig3}\AgdaSpace{}%
\AgdaBound{x}\AgdaSpace{}%
\AgdaBound{x₁}\AgdaSpace{}%
\AgdaBound{x₂}\AgdaSpace{}%
\AgdaSymbol{|}\AgdaSpace{}%
\AgdaInductiveConstructor{true}\AgdaSpace{}%
\AgdaSymbol{|}\AgdaSpace{}%
\AgdaInductiveConstructor{false}\AgdaSpace{}%
\AgdaSymbol{|}\AgdaSpace{}%
\AgdaInductiveConstructor{true}\AgdaSpace{}%
\AgdaSymbol{|}\AgdaSpace{}%
\AgdaInductiveConstructor{true}\AgdaSpace{}%
\AgdaSymbol{=}\AgdaSpace{}%
\AgdaInductiveConstructor{tt}\<%
\\
\>[0]\AgdaFunction{lemmaHoareTripleStackGeAux'23}\AgdaSpace{}%
\AgdaBound{msg₁}\AgdaSpace{}%
\AgdaBound{pbk1}\AgdaSpace{}%
\AgdaBound{pbk2}\AgdaSpace{}%
\AgdaBound{pbk3}\AgdaSpace{}%
\AgdaBound{pbk4}\AgdaSpace{}%
\AgdaBound{pbk5}\AgdaSpace{}%
\AgdaBound{sig1}\AgdaSpace{}%
\AgdaBound{sig2}\AgdaSpace{}%
\AgdaBound{sig3}\AgdaSpace{}%
\AgdaBound{x}\AgdaSpace{}%
\AgdaBound{x₁}\AgdaSpace{}%
\AgdaBound{x₂}\AgdaSpace{}%
\AgdaSymbol{|}\AgdaSpace{}%
\AgdaInductiveConstructor{true}\AgdaSpace{}%
\AgdaSymbol{|}\AgdaSpace{}%
\AgdaInductiveConstructor{true}\AgdaSpace{}%
\AgdaKeyword{with}\AgdaSpace{}%
\AgdaSymbol{(}\AgdaFunction{isSigned}%
\>[113]\AgdaBound{msg₁}\AgdaSpace{}%
\AgdaBound{sig3}\AgdaSpace{}%
\AgdaBound{pbk3}\AgdaSymbol{)}\<%
\\
\>[0]\AgdaFunction{lemmaHoareTripleStackGeAux'23}\AgdaSpace{}%
\AgdaBound{msg₁}\AgdaSpace{}%
\AgdaBound{pbk1}\AgdaSpace{}%
\AgdaBound{pbk2}\AgdaSpace{}%
\AgdaBound{pbk3}\AgdaSpace{}%
\AgdaBound{pbk4}\AgdaSpace{}%
\AgdaBound{pbk5}\AgdaSpace{}%
\AgdaBound{sig1}\AgdaSpace{}%
\AgdaBound{sig2}\AgdaSpace{}%
\AgdaBound{sig3}\AgdaSpace{}%
\AgdaBound{x}\AgdaSpace{}%
\AgdaBound{x₁}\AgdaSpace{}%
\AgdaBound{x₂}\AgdaSpace{}%
\AgdaSymbol{|}\AgdaSpace{}%
\AgdaInductiveConstructor{true}\AgdaSpace{}%
\AgdaSymbol{|}\AgdaSpace{}%
\AgdaInductiveConstructor{true}\AgdaSpace{}%
\AgdaSymbol{|}\AgdaSpace{}%
\AgdaInductiveConstructor{false}\AgdaSpace{}%
\AgdaKeyword{with}\AgdaSpace{}%
\AgdaSymbol{(}\AgdaFunction{isSigned}%
\>[121]\AgdaBound{msg₁}\AgdaSpace{}%
\AgdaBound{sig3}\AgdaSpace{}%
\AgdaBound{pbk4}\AgdaSymbol{)}\<%
\\
\>[0]\AgdaFunction{lemmaHoareTripleStackGeAux'23}\AgdaSpace{}%
\AgdaBound{msg₁}\AgdaSpace{}%
\AgdaBound{pbk1}\AgdaSpace{}%
\AgdaBound{pbk2}\AgdaSpace{}%
\AgdaBound{pbk3}\AgdaSpace{}%
\AgdaBound{pbk4}\AgdaSpace{}%
\AgdaBound{pbk5}\AgdaSpace{}%
\AgdaBound{sig1}\AgdaSpace{}%
\AgdaBound{sig2}\AgdaSpace{}%
\AgdaBound{sig3}\AgdaSpace{}%
\AgdaBound{x}\AgdaSpace{}%
\AgdaBound{x₁}\AgdaSpace{}%
\AgdaBound{x₂}\AgdaSpace{}%
\AgdaSymbol{|}\AgdaSpace{}%
\AgdaInductiveConstructor{true}\AgdaSpace{}%
\AgdaSymbol{|}\AgdaSpace{}%
\AgdaInductiveConstructor{true}\AgdaSpace{}%
\AgdaSymbol{|}\AgdaSpace{}%
\AgdaInductiveConstructor{false}\AgdaSpace{}%
\AgdaSymbol{|}\AgdaSpace{}%
\AgdaInductiveConstructor{false}\AgdaSpace{}%
\AgdaKeyword{with}\AgdaSpace{}%
\AgdaSymbol{(}\AgdaFunction{isSigned}%
\>[129]\AgdaBound{msg₁}\AgdaSpace{}%
\AgdaBound{sig3}\AgdaSpace{}%
\AgdaBound{pbk5}\AgdaSymbol{)}\<%
\\
\>[0]\AgdaFunction{lemmaHoareTripleStackGeAux'23}\AgdaSpace{}%
\AgdaBound{msg₁}\AgdaSpace{}%
\AgdaBound{pbk1}\AgdaSpace{}%
\AgdaBound{pbk2}\AgdaSpace{}%
\AgdaBound{pbk3}\AgdaSpace{}%
\AgdaBound{pbk4}\AgdaSpace{}%
\AgdaBound{pbk5}\AgdaSpace{}%
\AgdaBound{sig1}\AgdaSpace{}%
\AgdaBound{sig2}\AgdaSpace{}%
\AgdaBound{sig3}\AgdaSpace{}%
\AgdaBound{x}\AgdaSpace{}%
\AgdaBound{x₁}\AgdaSpace{}%
\AgdaBound{x₂}\AgdaSpace{}%
\AgdaSymbol{|}\AgdaSpace{}%
\AgdaInductiveConstructor{true}\AgdaSpace{}%
\AgdaSymbol{|}\AgdaSpace{}%
\AgdaInductiveConstructor{true}\AgdaSpace{}%
\AgdaSymbol{|}\AgdaSpace{}%
\AgdaInductiveConstructor{false}\AgdaSpace{}%
\AgdaSymbol{|}\AgdaSpace{}%
\AgdaInductiveConstructor{false}\AgdaSpace{}%
\AgdaSymbol{|}\AgdaSpace{}%
\AgdaInductiveConstructor{true}\AgdaSpace{}%
\AgdaSymbol{=}\AgdaSpace{}%
\AgdaInductiveConstructor{tt}\<%
\\
\>[0]\AgdaFunction{lemmaHoareTripleStackGeAux'23}\AgdaSpace{}%
\AgdaBound{msg₁}\AgdaSpace{}%
\AgdaBound{pbk1}\AgdaSpace{}%
\AgdaBound{pbk2}\AgdaSpace{}%
\AgdaBound{pbk3}\AgdaSpace{}%
\AgdaBound{pbk4}\AgdaSpace{}%
\AgdaBound{pbk5}\AgdaSpace{}%
\AgdaBound{sig1}\AgdaSpace{}%
\AgdaBound{sig2}\AgdaSpace{}%
\AgdaBound{sig3}\AgdaSpace{}%
\AgdaBound{x}\AgdaSpace{}%
\AgdaBound{x₁}\AgdaSpace{}%
\AgdaBound{x₂}\AgdaSpace{}%
\AgdaSymbol{|}\AgdaSpace{}%
\AgdaInductiveConstructor{true}\AgdaSpace{}%
\AgdaSymbol{|}\AgdaSpace{}%
\AgdaInductiveConstructor{true}\AgdaSpace{}%
\AgdaSymbol{|}\AgdaSpace{}%
\AgdaInductiveConstructor{false}\AgdaSpace{}%
\AgdaSymbol{|}\AgdaSpace{}%
\AgdaInductiveConstructor{true}\AgdaSpace{}%
\AgdaSymbol{=}\AgdaSpace{}%
\AgdaInductiveConstructor{tt}\<%
\\
\>[0]\AgdaFunction{lemmaHoareTripleStackGeAux'23}\AgdaSpace{}%
\AgdaBound{msg₁}\AgdaSpace{}%
\AgdaBound{pbk1}\AgdaSpace{}%
\AgdaBound{pbk2}\AgdaSpace{}%
\AgdaBound{pbk3}\AgdaSpace{}%
\AgdaBound{pbk4}\AgdaSpace{}%
\AgdaBound{pbk5}\AgdaSpace{}%
\AgdaBound{sig1}\AgdaSpace{}%
\AgdaBound{sig2}\AgdaSpace{}%
\AgdaBound{sig3}\AgdaSpace{}%
\AgdaBound{x}\AgdaSpace{}%
\AgdaBound{x₁}\AgdaSpace{}%
\AgdaBound{x₂}\AgdaSpace{}%
\AgdaSymbol{|}\AgdaSpace{}%
\AgdaInductiveConstructor{true}\AgdaSpace{}%
\AgdaSymbol{|}\AgdaSpace{}%
\AgdaInductiveConstructor{true}\AgdaSpace{}%
\AgdaSymbol{|}\AgdaSpace{}%
\AgdaInductiveConstructor{true}\AgdaSpace{}%
\AgdaSymbol{=}\AgdaSpace{}%
\AgdaInductiveConstructor{tt}\<%
\\
\\[\AgdaEmptyExtraSkip]%
\\[\AgdaEmptyExtraSkip]%
\\[\AgdaEmptyExtraSkip]%
\\[\AgdaEmptyExtraSkip]%
\\[\AgdaEmptyExtraSkip]%
\>[0]\AgdaFunction{lemmaHoareTripleStackGeAux'Comb3-5}\AgdaSpace{}%
\AgdaSymbol{:}\AgdaSpace{}%
\AgdaSymbol{(}\AgdaBound{msg₁}\AgdaSpace{}%
\AgdaSymbol{:}\AgdaSpace{}%
\AgdaDatatype{Msg}\AgdaSymbol{)}\<%
\\
\>[0][@{}l@{\AgdaIndent{0}}]%
\>[9]\AgdaSymbol{(}\AgdaBound{pbk1}\AgdaSpace{}%
\AgdaBound{pbk2}\AgdaSpace{}%
\AgdaBound{pbk3}\AgdaSpace{}%
\AgdaBound{pbk4}\AgdaSpace{}%
\AgdaBound{pbk5}\AgdaSpace{}%
\AgdaBound{sig1}\AgdaSpace{}%
\AgdaBound{sig2}\AgdaSpace{}%
\AgdaBound{sig3}\AgdaSpace{}%
\AgdaSymbol{:}\AgdaSpace{}%
\AgdaDatatype{ℕ}\AgdaSymbol{)}\<%
\\
\>[9][@{}l@{\AgdaIndent{0}}]%
\>[10]\AgdaSymbol{→}\AgdaSpace{}%
\AgdaFunction{True}%
\>[6499I]\AgdaSymbol{(}\AgdaFunction{compareSigsMultiSigAux}\AgdaSpace{}%
\AgdaBound{msg₁}\AgdaSpace{}%
\AgdaSymbol{(}\AgdaBound{sig2}\AgdaSpace{}%
\AgdaOperator{\AgdaInductiveConstructor{∷}}\AgdaSpace{}%
\AgdaBound{sig3}\AgdaSpace{}%
\AgdaOperator{\AgdaInductiveConstructor{∷}}\AgdaSpace{}%
\AgdaInductiveConstructor{[]}\AgdaSymbol{)}\AgdaSpace{}%
\AgdaSymbol{(}\AgdaSpace{}%
\AgdaBound{pbk2}\AgdaSpace{}%
\AgdaOperator{\AgdaInductiveConstructor{∷}}\AgdaSpace{}%
\AgdaBound{pbk3}\AgdaSpace{}%
\AgdaOperator{\AgdaInductiveConstructor{∷}}\AgdaSpace{}%
\AgdaBound{pbk4}\AgdaSpace{}%
\AgdaOperator{\AgdaInductiveConstructor{∷}}\AgdaSpace{}%
\AgdaBound{pbk5}\AgdaSpace{}%
\AgdaOperator{\AgdaInductiveConstructor{∷}}\AgdaSpace{}%
\AgdaInductiveConstructor{[]}\AgdaSymbol{)}%
\>[100]\AgdaBound{sig1}\<%
\\
\>[.][@{}l@{}]\<[6499I]%
\>[17]\AgdaSymbol{(}\AgdaFunction{isSigned}%
\>[28]\AgdaBound{msg₁}\AgdaSpace{}%
\AgdaBound{sig1}\AgdaSpace{}%
\AgdaBound{pbk1}\AgdaSpace{}%
\AgdaSymbol{))}\<%
\\
\>[0][@{}l@{\AgdaIndent{0}}]%
\>[6]\AgdaSymbol{→}%
\>[9]\AgdaSymbol{((}\AgdaFunction{True}\AgdaSpace{}%
\AgdaSymbol{(}\AgdaFunction{isSigned}%
\>[27]\AgdaBound{msg₁}\AgdaSpace{}%
\AgdaBound{sig1}\AgdaSpace{}%
\AgdaBound{pbk1}\AgdaSymbol{)}\AgdaSpace{}%
\AgdaOperator{\AgdaRecord{∧}}\AgdaSpace{}%
\AgdaFunction{True}\AgdaSpace{}%
\AgdaSymbol{(}\AgdaFunction{isSigned}%
\>[61]\AgdaBound{msg₁}\AgdaSpace{}%
\AgdaBound{sig2}\AgdaSpace{}%
\AgdaBound{pbk2}\AgdaSymbol{))}\AgdaSpace{}%
\AgdaOperator{\AgdaRecord{∧}}\AgdaSpace{}%
\AgdaFunction{True}\AgdaSpace{}%
\AgdaSymbol{(}\AgdaFunction{isSigned}%
\>[96]\AgdaBound{msg₁}\AgdaSpace{}%
\AgdaBound{sig3}\AgdaSpace{}%
\AgdaBound{pbk3}\AgdaSymbol{))}\AgdaSpace{}%
\AgdaOperator{\AgdaDatatype{⊎}}\<%
\\
\>[9][@{}l@{\AgdaIndent{0}}]%
\>[10]\AgdaSymbol{((}\AgdaFunction{True}\AgdaSpace{}%
\AgdaSymbol{(}\AgdaFunction{isSigned}%
\>[28]\AgdaBound{msg₁}\AgdaSpace{}%
\AgdaBound{sig1}\AgdaSpace{}%
\AgdaBound{pbk1}\AgdaSymbol{)}\AgdaSpace{}%
\AgdaOperator{\AgdaRecord{∧}}\AgdaSpace{}%
\AgdaFunction{True}\AgdaSpace{}%
\AgdaSymbol{(}\AgdaFunction{isSigned}%
\>[62]\AgdaBound{msg₁}\AgdaSpace{}%
\AgdaBound{sig2}\AgdaSpace{}%
\AgdaBound{pbk2}\AgdaSymbol{))}\AgdaSpace{}%
\AgdaOperator{\AgdaRecord{∧}}\AgdaSpace{}%
\AgdaFunction{True}\AgdaSpace{}%
\AgdaSymbol{(}\AgdaFunction{isSigned}%
\>[97]\AgdaBound{msg₁}\AgdaSpace{}%
\AgdaBound{sig3}\AgdaSpace{}%
\AgdaBound{pbk4}\AgdaSymbol{))}\AgdaSpace{}%
\AgdaOperator{\AgdaDatatype{⊎}}\<%
\\
\>[10]\AgdaSymbol{((}\AgdaFunction{True}\AgdaSpace{}%
\AgdaSymbol{(}\AgdaFunction{isSigned}%
\>[28]\AgdaBound{msg₁}\AgdaSpace{}%
\AgdaBound{sig1}\AgdaSpace{}%
\AgdaBound{pbk1}\AgdaSymbol{)}\AgdaSpace{}%
\AgdaOperator{\AgdaRecord{∧}}\AgdaSpace{}%
\AgdaFunction{True}\AgdaSpace{}%
\AgdaSymbol{(}\AgdaFunction{isSigned}%
\>[62]\AgdaBound{msg₁}\AgdaSpace{}%
\AgdaBound{sig2}\AgdaSpace{}%
\AgdaBound{pbk2}\AgdaSymbol{))}\AgdaSpace{}%
\AgdaOperator{\AgdaRecord{∧}}\AgdaSpace{}%
\AgdaFunction{True}\AgdaSpace{}%
\AgdaSymbol{(}\AgdaFunction{isSigned}%
\>[97]\AgdaBound{msg₁}\AgdaSpace{}%
\AgdaBound{sig3}\AgdaSpace{}%
\AgdaBound{pbk5}\AgdaSymbol{))}\AgdaSpace{}%
\AgdaOperator{\AgdaDatatype{⊎}}\<%
\\
\>[10]\AgdaSymbol{((}\AgdaFunction{True}\AgdaSpace{}%
\AgdaSymbol{(}\AgdaFunction{isSigned}%
\>[28]\AgdaBound{msg₁}\AgdaSpace{}%
\AgdaBound{sig1}\AgdaSpace{}%
\AgdaBound{pbk1}\AgdaSymbol{)}\AgdaSpace{}%
\AgdaOperator{\AgdaRecord{∧}}\AgdaSpace{}%
\AgdaFunction{True}\AgdaSpace{}%
\AgdaSymbol{(}\AgdaFunction{isSigned}%
\>[62]\AgdaBound{msg₁}\AgdaSpace{}%
\AgdaBound{sig2}\AgdaSpace{}%
\AgdaBound{pbk3}\AgdaSymbol{))}\AgdaSpace{}%
\AgdaOperator{\AgdaRecord{∧}}\AgdaSpace{}%
\AgdaFunction{True}\AgdaSpace{}%
\AgdaSymbol{(}\AgdaFunction{isSigned}%
\>[97]\AgdaBound{msg₁}\AgdaSpace{}%
\AgdaBound{sig3}\AgdaSpace{}%
\AgdaBound{pbk4}\AgdaSymbol{))}\AgdaSpace{}%
\AgdaOperator{\AgdaDatatype{⊎}}\<%
\\
\>[10]\AgdaSymbol{((}\AgdaFunction{True}\AgdaSpace{}%
\AgdaSymbol{(}\AgdaFunction{isSigned}%
\>[28]\AgdaBound{msg₁}\AgdaSpace{}%
\AgdaBound{sig1}\AgdaSpace{}%
\AgdaBound{pbk1}\AgdaSymbol{)}\AgdaSpace{}%
\AgdaOperator{\AgdaRecord{∧}}\AgdaSpace{}%
\AgdaFunction{True}\AgdaSpace{}%
\AgdaSymbol{(}\AgdaFunction{isSigned}%
\>[62]\AgdaBound{msg₁}\AgdaSpace{}%
\AgdaBound{sig2}\AgdaSpace{}%
\AgdaBound{pbk3}\AgdaSymbol{))}\AgdaSpace{}%
\AgdaOperator{\AgdaRecord{∧}}\AgdaSpace{}%
\AgdaFunction{True}\AgdaSpace{}%
\AgdaSymbol{(}\AgdaFunction{isSigned}%
\>[97]\AgdaBound{msg₁}\AgdaSpace{}%
\AgdaBound{sig3}\AgdaSpace{}%
\AgdaBound{pbk5}\AgdaSymbol{))}\AgdaSpace{}%
\AgdaOperator{\AgdaDatatype{⊎}}\<%
\\
\>[10]\AgdaSymbol{((}\AgdaFunction{True}\AgdaSpace{}%
\AgdaSymbol{(}\AgdaFunction{isSigned}%
\>[28]\AgdaBound{msg₁}\AgdaSpace{}%
\AgdaBound{sig1}\AgdaSpace{}%
\AgdaBound{pbk1}\AgdaSymbol{)}\AgdaSpace{}%
\AgdaOperator{\AgdaRecord{∧}}\AgdaSpace{}%
\AgdaFunction{True}\AgdaSpace{}%
\AgdaSymbol{(}\AgdaFunction{isSigned}%
\>[62]\AgdaBound{msg₁}\AgdaSpace{}%
\AgdaBound{sig2}\AgdaSpace{}%
\AgdaBound{pbk4}\AgdaSymbol{))}\AgdaSpace{}%
\AgdaOperator{\AgdaRecord{∧}}\AgdaSpace{}%
\AgdaFunction{True}\AgdaSpace{}%
\AgdaSymbol{(}\AgdaFunction{isSigned}%
\>[97]\AgdaBound{msg₁}\AgdaSpace{}%
\AgdaBound{sig3}\AgdaSpace{}%
\AgdaBound{pbk5}\AgdaSymbol{))}\AgdaSpace{}%
\AgdaOperator{\AgdaDatatype{⊎}}\<%
\\
\>[10]\AgdaSymbol{((}\AgdaFunction{True}\AgdaSpace{}%
\AgdaSymbol{(}\AgdaFunction{isSigned}%
\>[28]\AgdaBound{msg₁}\AgdaSpace{}%
\AgdaBound{sig1}\AgdaSpace{}%
\AgdaBound{pbk2}\AgdaSymbol{)}\AgdaSpace{}%
\AgdaOperator{\AgdaRecord{∧}}\AgdaSpace{}%
\AgdaFunction{True}\AgdaSpace{}%
\AgdaSymbol{(}\AgdaFunction{isSigned}%
\>[62]\AgdaBound{msg₁}\AgdaSpace{}%
\AgdaBound{sig2}\AgdaSpace{}%
\AgdaBound{pbk3}\AgdaSymbol{))}\AgdaSpace{}%
\AgdaOperator{\AgdaRecord{∧}}\AgdaSpace{}%
\AgdaFunction{True}\AgdaSpace{}%
\AgdaSymbol{(}\AgdaFunction{isSigned}%
\>[97]\AgdaBound{msg₁}\AgdaSpace{}%
\AgdaBound{sig3}\AgdaSpace{}%
\AgdaBound{pbk4}\AgdaSymbol{))}\AgdaSpace{}%
\AgdaOperator{\AgdaDatatype{⊎}}\<%
\\
\>[10]\AgdaSymbol{((}\AgdaFunction{True}\AgdaSpace{}%
\AgdaSymbol{(}\AgdaFunction{isSigned}%
\>[28]\AgdaBound{msg₁}\AgdaSpace{}%
\AgdaBound{sig1}\AgdaSpace{}%
\AgdaBound{pbk2}\AgdaSymbol{)}\AgdaSpace{}%
\AgdaOperator{\AgdaRecord{∧}}\AgdaSpace{}%
\AgdaFunction{True}\AgdaSpace{}%
\AgdaSymbol{(}\AgdaFunction{isSigned}%
\>[62]\AgdaBound{msg₁}\AgdaSpace{}%
\AgdaBound{sig2}\AgdaSpace{}%
\AgdaBound{pbk3}\AgdaSymbol{))}\AgdaSpace{}%
\AgdaOperator{\AgdaRecord{∧}}\AgdaSpace{}%
\AgdaFunction{True}\AgdaSpace{}%
\AgdaSymbol{(}\AgdaFunction{isSigned}%
\>[97]\AgdaBound{msg₁}\AgdaSpace{}%
\AgdaBound{sig3}\AgdaSpace{}%
\AgdaBound{pbk5}\AgdaSymbol{))}\AgdaSpace{}%
\AgdaOperator{\AgdaDatatype{⊎}}\<%
\\
\>[10]\AgdaSymbol{((}\AgdaFunction{True}\AgdaSpace{}%
\AgdaSymbol{(}\AgdaFunction{isSigned}%
\>[28]\AgdaBound{msg₁}\AgdaSpace{}%
\AgdaBound{sig1}\AgdaSpace{}%
\AgdaBound{pbk2}\AgdaSymbol{)}\AgdaSpace{}%
\AgdaOperator{\AgdaRecord{∧}}\AgdaSpace{}%
\AgdaFunction{True}\AgdaSpace{}%
\AgdaSymbol{(}\AgdaFunction{isSigned}%
\>[62]\AgdaBound{msg₁}\AgdaSpace{}%
\AgdaBound{sig2}\AgdaSpace{}%
\AgdaBound{pbk4}\AgdaSymbol{))}\AgdaSpace{}%
\AgdaOperator{\AgdaRecord{∧}}\AgdaSpace{}%
\AgdaFunction{True}\AgdaSpace{}%
\AgdaSymbol{(}\AgdaFunction{isSigned}%
\>[97]\AgdaBound{msg₁}\AgdaSpace{}%
\AgdaBound{sig3}\AgdaSpace{}%
\AgdaBound{pbk5}\AgdaSymbol{))}\AgdaSpace{}%
\AgdaOperator{\AgdaDatatype{⊎}}\<%
\\
\>[10]\AgdaSymbol{((}\AgdaFunction{True}\AgdaSpace{}%
\AgdaSymbol{(}\AgdaFunction{isSigned}%
\>[28]\AgdaBound{msg₁}\AgdaSpace{}%
\AgdaBound{sig1}\AgdaSpace{}%
\AgdaBound{pbk3}\AgdaSymbol{)}\AgdaSpace{}%
\AgdaOperator{\AgdaRecord{∧}}\AgdaSpace{}%
\AgdaFunction{True}\AgdaSpace{}%
\AgdaSymbol{(}\AgdaFunction{isSigned}%
\>[62]\AgdaBound{msg₁}\AgdaSpace{}%
\AgdaBound{sig2}\AgdaSpace{}%
\AgdaBound{pbk4}\AgdaSymbol{))}\AgdaSpace{}%
\AgdaOperator{\AgdaRecord{∧}}\AgdaSpace{}%
\AgdaFunction{True}\AgdaSpace{}%
\AgdaSymbol{(}\AgdaFunction{isSigned}%
\>[97]\AgdaBound{msg₁}\AgdaSpace{}%
\AgdaBound{sig3}\AgdaSpace{}%
\AgdaBound{pbk5}\AgdaSymbol{))}\<%
\\
\\[\AgdaEmptyExtraSkip]%
\>[0]\AgdaFunction{lemmaHoareTripleStackGeAux'Comb3-5}\AgdaSpace{}%
\AgdaBound{msg₁}\AgdaSpace{}%
\AgdaBound{pbk1}\AgdaSpace{}%
\AgdaBound{pbk2}\AgdaSpace{}%
\AgdaBound{pbk3}\AgdaSpace{}%
\AgdaBound{pbk4}\AgdaSpace{}%
\AgdaBound{pbk5}\AgdaSpace{}%
\AgdaBound{sig1}\AgdaSpace{}%
\AgdaBound{sig2}\AgdaSpace{}%
\AgdaBound{sig3}\AgdaSpace{}%
\AgdaBound{x}\AgdaSpace{}%
\AgdaKeyword{with}%
\>[88]\AgdaSymbol{(}\AgdaFunction{isSigned}%
\>[99]\AgdaBound{msg₁}\AgdaSpace{}%
\AgdaBound{sig1}\AgdaSpace{}%
\AgdaBound{pbk1}\AgdaSymbol{)}\<%
\\
\>[0]\AgdaFunction{lemmaHoareTripleStackGeAux'Comb3-5}\AgdaSpace{}%
\AgdaBound{msg₁}\AgdaSpace{}%
\AgdaBound{pbk1}\AgdaSpace{}%
\AgdaBound{pbk2}\AgdaSpace{}%
\AgdaBound{pbk3}\AgdaSpace{}%
\AgdaBound{pbk4}\AgdaSpace{}%
\AgdaBound{pbk5}\AgdaSpace{}%
\AgdaBound{sig1}\AgdaSpace{}%
\AgdaBound{sig2}\AgdaSpace{}%
\AgdaBound{sig3}\AgdaSpace{}%
\AgdaBound{x}\AgdaSpace{}%
\AgdaSymbol{|}\AgdaSpace{}%
\AgdaInductiveConstructor{false}\AgdaSpace{}%
\AgdaKeyword{with}\AgdaSpace{}%
\AgdaSymbol{(}\AgdaFunction{isSigned}%
\>[106]\AgdaBound{msg₁}\AgdaSpace{}%
\AgdaBound{sig1}\AgdaSpace{}%
\AgdaBound{pbk2}\AgdaSymbol{)}\<%
\\
\>[0]\AgdaFunction{lemmaHoareTripleStackGeAux'Comb3-5}\AgdaSpace{}%
\AgdaBound{msg₁}\AgdaSpace{}%
\AgdaBound{pbk1}\AgdaSpace{}%
\AgdaBound{pbk2}\AgdaSpace{}%
\AgdaBound{pbk3}\AgdaSpace{}%
\AgdaBound{pbk4}\AgdaSpace{}%
\AgdaBound{pbk5}\AgdaSpace{}%
\AgdaBound{sig1}\AgdaSpace{}%
\AgdaBound{sig2}\AgdaSpace{}%
\AgdaBound{sig3}\AgdaSpace{}%
\AgdaBound{x}\AgdaSpace{}%
\AgdaSymbol{|}\AgdaSpace{}%
\AgdaInductiveConstructor{false}\AgdaSpace{}%
\AgdaSymbol{|}\AgdaSpace{}%
\AgdaInductiveConstructor{false}\AgdaSpace{}%
\AgdaKeyword{with}\AgdaSpace{}%
\AgdaSymbol{(}\AgdaFunction{isSigned}%
\>[114]\AgdaBound{msg₁}\AgdaSpace{}%
\AgdaBound{sig1}\AgdaSpace{}%
\AgdaBound{pbk3}\AgdaSymbol{)}\<%
\\
\>[0]\AgdaFunction{lemmaHoareTripleStackGeAux'Comb3-5}\AgdaSpace{}%
\AgdaBound{msg₁}\AgdaSpace{}%
\AgdaBound{pbk1}\AgdaSpace{}%
\AgdaBound{pbk2}\AgdaSpace{}%
\AgdaBound{pbk3}\AgdaSpace{}%
\AgdaBound{pbk4}\AgdaSpace{}%
\AgdaBound{pbk5}\AgdaSpace{}%
\AgdaBound{sig1}\AgdaSpace{}%
\AgdaBound{sig2}\AgdaSpace{}%
\AgdaBound{sig3}\AgdaSpace{}%
\AgdaBound{x}\AgdaSpace{}%
\AgdaSymbol{|}\AgdaSpace{}%
\AgdaInductiveConstructor{false}\AgdaSpace{}%
\AgdaSymbol{|}\AgdaSpace{}%
\AgdaInductiveConstructor{false}\AgdaSpace{}%
\AgdaSymbol{|}\AgdaSpace{}%
\AgdaInductiveConstructor{false}\AgdaSpace{}%
\AgdaKeyword{with}\AgdaSpace{}%
\AgdaSymbol{(}\AgdaFunction{isSigned}%
\>[122]\AgdaBound{msg₁}\AgdaSpace{}%
\AgdaBound{sig1}\AgdaSpace{}%
\AgdaBound{pbk4}\AgdaSymbol{)}\<%
\\
\>[0]\AgdaFunction{lemmaHoareTripleStackGeAux'Comb3-5}\AgdaSpace{}%
\AgdaBound{msg₁}\AgdaSpace{}%
\AgdaBound{pbk1}\AgdaSpace{}%
\AgdaBound{pbk2}\AgdaSpace{}%
\AgdaBound{pbk3}\AgdaSpace{}%
\AgdaBound{pbk4}\AgdaSpace{}%
\AgdaBound{pbk5}\AgdaSpace{}%
\AgdaBound{sig1}\AgdaSpace{}%
\AgdaBound{sig2}\AgdaSpace{}%
\AgdaBound{sig3}\AgdaSpace{}%
\AgdaBound{x}\AgdaSpace{}%
\AgdaSymbol{|}\AgdaSpace{}%
\AgdaInductiveConstructor{false}\AgdaSpace{}%
\AgdaSymbol{|}\AgdaSpace{}%
\AgdaInductiveConstructor{false}\AgdaSpace{}%
\AgdaSymbol{|}\AgdaSpace{}%
\AgdaInductiveConstructor{false}\AgdaSpace{}%
\AgdaSymbol{|}\AgdaSpace{}%
\AgdaInductiveConstructor{false}\AgdaSpace{}%
\AgdaKeyword{with}\AgdaSpace{}%
\AgdaSymbol{(}\AgdaFunction{isSigned}%
\>[130]\AgdaBound{msg₁}\AgdaSpace{}%
\AgdaBound{sig1}\AgdaSpace{}%
\AgdaBound{pbk5}\AgdaSymbol{)}\<%
\\
\>[0]\AgdaFunction{lemmaHoareTripleStackGeAux'Comb3-5}\AgdaSpace{}%
\AgdaBound{msg₁}\AgdaSpace{}%
\AgdaBound{pbk1}\AgdaSpace{}%
\AgdaBound{pbk2}\AgdaSpace{}%
\AgdaBound{pbk3}\AgdaSpace{}%
\AgdaBound{pbk4}\AgdaSpace{}%
\AgdaBound{pbk5}\AgdaSpace{}%
\AgdaBound{sig1}\AgdaSpace{}%
\AgdaBound{sig2}\AgdaSpace{}%
\AgdaBound{sig3}\AgdaSpace{}%
\AgdaSymbol{()}\AgdaSpace{}%
\AgdaSymbol{|}\AgdaSpace{}%
\AgdaInductiveConstructor{false}\AgdaSpace{}%
\AgdaSymbol{|}\AgdaSpace{}%
\AgdaInductiveConstructor{false}\AgdaSpace{}%
\AgdaSymbol{|}\AgdaSpace{}%
\AgdaInductiveConstructor{false}\AgdaSpace{}%
\AgdaSymbol{|}\AgdaSpace{}%
\AgdaInductiveConstructor{false}\AgdaSpace{}%
\AgdaSymbol{|}\AgdaSpace{}%
\AgdaInductiveConstructor{false}\<%
\\
\>[0]\AgdaFunction{lemmaHoareTripleStackGeAux'Comb3-5}\AgdaSpace{}%
\AgdaBound{msg₁}\AgdaSpace{}%
\AgdaBound{pbk1}\AgdaSpace{}%
\AgdaBound{pbk2}\AgdaSpace{}%
\AgdaBound{pbk3}\AgdaSpace{}%
\AgdaBound{pbk4}\AgdaSpace{}%
\AgdaBound{pbk5}\AgdaSpace{}%
\AgdaBound{sig1}\AgdaSpace{}%
\AgdaBound{sig2}\AgdaSpace{}%
\AgdaBound{sig3}\AgdaSpace{}%
\AgdaSymbol{()}\AgdaSpace{}%
\AgdaSymbol{|}\AgdaSpace{}%
\AgdaInductiveConstructor{false}\AgdaSpace{}%
\AgdaSymbol{|}\AgdaSpace{}%
\AgdaInductiveConstructor{false}\AgdaSpace{}%
\AgdaSymbol{|}\AgdaSpace{}%
\AgdaInductiveConstructor{false}\AgdaSpace{}%
\AgdaSymbol{|}\AgdaSpace{}%
\AgdaInductiveConstructor{false}\AgdaSpace{}%
\AgdaSymbol{|}\AgdaSpace{}%
\AgdaInductiveConstructor{true}\<%
\\
\>[0]\AgdaFunction{lemmaHoareTripleStackGeAux'Comb3-5}\AgdaSpace{}%
\AgdaBound{msg₁}\AgdaSpace{}%
\AgdaBound{pbk1}\AgdaSpace{}%
\AgdaBound{pbk2}\AgdaSpace{}%
\AgdaBound{pbk3}\AgdaSpace{}%
\AgdaBound{pbk4}\AgdaSpace{}%
\AgdaBound{pbk5}\AgdaSpace{}%
\AgdaBound{sig1}\AgdaSpace{}%
\AgdaBound{sig2}\AgdaSpace{}%
\AgdaBound{sig3}\AgdaSpace{}%
\AgdaBound{x}\AgdaSpace{}%
\AgdaSymbol{|}\AgdaSpace{}%
\AgdaInductiveConstructor{false}\AgdaSpace{}%
\AgdaSymbol{|}\AgdaSpace{}%
\AgdaInductiveConstructor{false}\AgdaSpace{}%
\AgdaSymbol{|}\AgdaSpace{}%
\AgdaInductiveConstructor{false}\AgdaSpace{}%
\AgdaSymbol{|}\AgdaSpace{}%
\AgdaInductiveConstructor{true}\AgdaSpace{}%
\AgdaKeyword{with}%
\>[119]\AgdaSymbol{(}\AgdaFunction{isSigned}%
\>[130]\AgdaBound{msg₁}\AgdaSpace{}%
\AgdaBound{sig2}\AgdaSpace{}%
\AgdaBound{pbk5}\AgdaSymbol{)}\<%
\\
\>[0]\AgdaFunction{lemmaHoareTripleStackGeAux'Comb3-5}\AgdaSpace{}%
\AgdaBound{msg₁}\AgdaSpace{}%
\AgdaBound{pbk1}\AgdaSpace{}%
\AgdaBound{pbk2}\AgdaSpace{}%
\AgdaBound{pbk3}\AgdaSpace{}%
\AgdaBound{pbk4}\AgdaSpace{}%
\AgdaBound{pbk5}\AgdaSpace{}%
\AgdaBound{sig1}\AgdaSpace{}%
\AgdaBound{sig2}\AgdaSpace{}%
\AgdaBound{sig3}\AgdaSpace{}%
\AgdaSymbol{()}\AgdaSpace{}%
\AgdaSymbol{|}\AgdaSpace{}%
\AgdaInductiveConstructor{false}\AgdaSpace{}%
\AgdaSymbol{|}\AgdaSpace{}%
\AgdaInductiveConstructor{false}\AgdaSpace{}%
\AgdaSymbol{|}\AgdaSpace{}%
\AgdaInductiveConstructor{false}\AgdaSpace{}%
\AgdaSymbol{|}\AgdaSpace{}%
\AgdaInductiveConstructor{true}\AgdaSpace{}%
\AgdaSymbol{|}\AgdaSpace{}%
\AgdaInductiveConstructor{false}\<%
\\
\>[0]\AgdaFunction{lemmaHoareTripleStackGeAux'Comb3-5}\AgdaSpace{}%
\AgdaBound{msg₁}\AgdaSpace{}%
\AgdaBound{pbk1}\AgdaSpace{}%
\AgdaBound{pbk2}\AgdaSpace{}%
\AgdaBound{pbk3}\AgdaSpace{}%
\AgdaBound{pbk4}\AgdaSpace{}%
\AgdaBound{pbk5}\AgdaSpace{}%
\AgdaBound{sig1}\AgdaSpace{}%
\AgdaBound{sig2}\AgdaSpace{}%
\AgdaBound{sig3}\AgdaSpace{}%
\AgdaSymbol{()}\AgdaSpace{}%
\AgdaSymbol{|}\AgdaSpace{}%
\AgdaInductiveConstructor{false}\AgdaSpace{}%
\AgdaSymbol{|}\AgdaSpace{}%
\AgdaInductiveConstructor{false}\AgdaSpace{}%
\AgdaSymbol{|}\AgdaSpace{}%
\AgdaInductiveConstructor{false}\AgdaSpace{}%
\AgdaSymbol{|}\AgdaSpace{}%
\AgdaInductiveConstructor{true}\AgdaSpace{}%
\AgdaSymbol{|}\AgdaSpace{}%
\AgdaInductiveConstructor{true}\<%
\\
\>[0]\AgdaFunction{lemmaHoareTripleStackGeAux'Comb3-5}\AgdaSpace{}%
\AgdaBound{msg₁}\AgdaSpace{}%
\AgdaBound{pbk1}\AgdaSpace{}%
\AgdaBound{pbk2}\AgdaSpace{}%
\AgdaBound{pbk3}\AgdaSpace{}%
\AgdaBound{pbk4}\AgdaSpace{}%
\AgdaBound{pbk5}\AgdaSpace{}%
\AgdaBound{sig1}\AgdaSpace{}%
\AgdaBound{sig2}\AgdaSpace{}%
\AgdaBound{sig3}\AgdaSpace{}%
\AgdaBound{x}\AgdaSpace{}%
\AgdaSymbol{|}\AgdaSpace{}%
\AgdaInductiveConstructor{false}\AgdaSpace{}%
\AgdaSymbol{|}\AgdaSpace{}%
\AgdaInductiveConstructor{false}\AgdaSpace{}%
\AgdaSymbol{|}\AgdaSpace{}%
\AgdaInductiveConstructor{true}\AgdaSpace{}%
\AgdaKeyword{with}%
\>[111]\AgdaSymbol{(}\AgdaFunction{isSigned}%
\>[122]\AgdaBound{msg₁}\AgdaSpace{}%
\AgdaBound{sig2}\AgdaSpace{}%
\AgdaBound{pbk4}\AgdaSymbol{)}\<%
\\
\>[0]\AgdaFunction{lemmaHoareTripleStackGeAux'Comb3-5}\AgdaSpace{}%
\AgdaBound{msg₁}\AgdaSpace{}%
\AgdaBound{pbk1}\AgdaSpace{}%
\AgdaBound{pbk2}\AgdaSpace{}%
\AgdaBound{pbk3}\AgdaSpace{}%
\AgdaBound{pbk4}\AgdaSpace{}%
\AgdaBound{pbk5}\AgdaSpace{}%
\AgdaBound{sig1}\AgdaSpace{}%
\AgdaBound{sig2}\AgdaSpace{}%
\AgdaBound{sig3}\AgdaSpace{}%
\AgdaBound{x}\AgdaSpace{}%
\AgdaSymbol{|}\AgdaSpace{}%
\AgdaInductiveConstructor{false}\AgdaSpace{}%
\AgdaSymbol{|}\AgdaSpace{}%
\AgdaInductiveConstructor{false}\AgdaSpace{}%
\AgdaSymbol{|}\AgdaSpace{}%
\AgdaInductiveConstructor{true}\AgdaSpace{}%
\AgdaSymbol{|}\AgdaSpace{}%
\AgdaInductiveConstructor{false}\AgdaSpace{}%
\AgdaKeyword{with}\AgdaSpace{}%
\AgdaSymbol{(}\AgdaFunction{isSigned}%
\>[129]\AgdaBound{msg₁}\AgdaSpace{}%
\AgdaBound{sig2}\AgdaSpace{}%
\AgdaBound{pbk5}\AgdaSymbol{)}\<%
\\
\>[0]\AgdaFunction{lemmaHoareTripleStackGeAux'Comb3-5}\AgdaSpace{}%
\AgdaBound{msg₁}\AgdaSpace{}%
\AgdaBound{pbk1}\AgdaSpace{}%
\AgdaBound{pbk2}\AgdaSpace{}%
\AgdaBound{pbk3}\AgdaSpace{}%
\AgdaBound{pbk4}\AgdaSpace{}%
\AgdaBound{pbk5}\AgdaSpace{}%
\AgdaBound{sig1}\AgdaSpace{}%
\AgdaBound{sig2}\AgdaSpace{}%
\AgdaBound{sig3}\AgdaSpace{}%
\AgdaSymbol{()}\AgdaSpace{}%
\AgdaSymbol{|}\AgdaSpace{}%
\AgdaInductiveConstructor{false}\AgdaSpace{}%
\AgdaSymbol{|}\AgdaSpace{}%
\AgdaInductiveConstructor{false}\AgdaSpace{}%
\AgdaSymbol{|}\AgdaSpace{}%
\AgdaInductiveConstructor{true}\AgdaSpace{}%
\AgdaSymbol{|}\AgdaSpace{}%
\AgdaInductiveConstructor{false}\AgdaSpace{}%
\AgdaSymbol{|}\AgdaSpace{}%
\AgdaInductiveConstructor{false}\<%
\\
\>[0]\AgdaFunction{lemmaHoareTripleStackGeAux'Comb3-5}\AgdaSpace{}%
\AgdaBound{msg₁}\AgdaSpace{}%
\AgdaBound{pbk1}\AgdaSpace{}%
\AgdaBound{pbk2}\AgdaSpace{}%
\AgdaBound{pbk3}\AgdaSpace{}%
\AgdaBound{pbk4}\AgdaSpace{}%
\AgdaBound{pbk5}\AgdaSpace{}%
\AgdaBound{sig1}\AgdaSpace{}%
\AgdaBound{sig2}\AgdaSpace{}%
\AgdaBound{sig3}\AgdaSpace{}%
\AgdaSymbol{()}\AgdaSpace{}%
\AgdaSymbol{|}\AgdaSpace{}%
\AgdaInductiveConstructor{false}\AgdaSpace{}%
\AgdaSymbol{|}\AgdaSpace{}%
\AgdaInductiveConstructor{false}\AgdaSpace{}%
\AgdaSymbol{|}\AgdaSpace{}%
\AgdaInductiveConstructor{true}\AgdaSpace{}%
\AgdaSymbol{|}\AgdaSpace{}%
\AgdaInductiveConstructor{false}\AgdaSpace{}%
\AgdaSymbol{|}\AgdaSpace{}%
\AgdaInductiveConstructor{true}\<%
\\
\>[0]\AgdaFunction{lemmaHoareTripleStackGeAux'Comb3-5}\AgdaSpace{}%
\AgdaBound{msg₁}\AgdaSpace{}%
\AgdaBound{pbk1}\AgdaSpace{}%
\AgdaBound{pbk2}\AgdaSpace{}%
\AgdaBound{pbk3}\AgdaSpace{}%
\AgdaBound{pbk4}\AgdaSpace{}%
\AgdaBound{pbk5}\AgdaSpace{}%
\AgdaBound{sig1}\AgdaSpace{}%
\AgdaBound{sig2}\AgdaSpace{}%
\AgdaBound{sig3}\AgdaSpace{}%
\AgdaBound{x}\AgdaSpace{}%
\AgdaSymbol{|}\AgdaSpace{}%
\AgdaInductiveConstructor{false}\AgdaSpace{}%
\AgdaSymbol{|}\AgdaSpace{}%
\AgdaInductiveConstructor{false}\AgdaSpace{}%
\AgdaSymbol{|}\AgdaSpace{}%
\AgdaInductiveConstructor{true}\AgdaSpace{}%
\AgdaSymbol{|}\AgdaSpace{}%
\AgdaInductiveConstructor{true}\AgdaSpace{}%
\AgdaKeyword{with}%
\>[118]\AgdaSymbol{(}\AgdaFunction{isSigned}%
\>[129]\AgdaBound{msg₁}\AgdaSpace{}%
\AgdaBound{sig3}\AgdaSpace{}%
\AgdaBound{pbk5}\AgdaSymbol{)}\<%
\\
\>[0]\AgdaFunction{lemmaHoareTripleStackGeAux'Comb3-5}\AgdaSpace{}%
\AgdaBound{msg₁}\AgdaSpace{}%
\AgdaBound{pbk1}\AgdaSpace{}%
\AgdaBound{pbk2}\AgdaSpace{}%
\AgdaBound{pbk3}\AgdaSpace{}%
\AgdaBound{pbk4}%
\>[6955I]\AgdaBound{pbk5}\AgdaSpace{}%
\AgdaBound{sig1}\AgdaSpace{}%
\AgdaBound{sig2}\AgdaSpace{}%
\AgdaBound{sig3}\AgdaSpace{}%
\AgdaBound{x}\AgdaSpace{}%
\AgdaSymbol{|}\AgdaSpace{}%
\AgdaInductiveConstructor{false}\AgdaSpace{}%
\AgdaSymbol{|}\AgdaSpace{}%
\AgdaInductiveConstructor{false}\AgdaSpace{}%
\AgdaSymbol{|}\AgdaSpace{}%
\AgdaInductiveConstructor{true}\AgdaSpace{}%
\AgdaSymbol{|}\AgdaSpace{}%
\AgdaInductiveConstructor{true}\AgdaSpace{}%
\AgdaSymbol{|}\AgdaSpace{}%
\AgdaInductiveConstructor{true}\<%
\\
\>[6955I][@{}l@{\AgdaIndent{0}}]%
\>[62]\AgdaSymbol{=}\AgdaSpace{}%
\AgdaInductiveConstructor{inj₂}\AgdaSpace{}%
\AgdaSymbol{(}\AgdaInductiveConstructor{inj₂}\AgdaSpace{}%
\AgdaSymbol{(}\AgdaInductiveConstructor{inj₂}\AgdaSpace{}%
\AgdaSymbol{(}\AgdaInductiveConstructor{inj₂}\AgdaSpace{}%
\AgdaSymbol{(}\AgdaInductiveConstructor{inj₂}\AgdaSpace{}%
\AgdaSymbol{(}\AgdaInductiveConstructor{inj₂}\AgdaSpace{}%
\AgdaSymbol{(}\AgdaInductiveConstructor{inj₂}\AgdaSpace{}%
\AgdaSymbol{(}\AgdaInductiveConstructor{inj₂}\AgdaSpace{}%
\AgdaSymbol{(}\AgdaInductiveConstructor{inj₂}\AgdaSpace{}%
\AgdaSymbol{(}\AgdaInductiveConstructor{conj}\AgdaSpace{}%
\AgdaSymbol{(}\AgdaInductiveConstructor{conj}\AgdaSpace{}%
\AgdaInductiveConstructor{tt}\AgdaSpace{}%
\AgdaInductiveConstructor{tt}\AgdaSymbol{)}\AgdaSpace{}%
\AgdaInductiveConstructor{tt}\AgdaSymbol{)))))))))}\<%
\\
\>[0]\AgdaFunction{lemmaHoareTripleStackGeAux'Comb3-5}\AgdaSpace{}%
\AgdaBound{msg₁}\AgdaSpace{}%
\AgdaBound{pbk1}\AgdaSpace{}%
\AgdaBound{pbk2}\AgdaSpace{}%
\AgdaBound{pbk3}\AgdaSpace{}%
\AgdaBound{pbk4}\AgdaSpace{}%
\AgdaBound{pbk5}\AgdaSpace{}%
\AgdaBound{sig1}\AgdaSpace{}%
\AgdaBound{sig2}\AgdaSpace{}%
\AgdaBound{sig3}\AgdaSpace{}%
\AgdaBound{x}\AgdaSpace{}%
\AgdaSymbol{|}\AgdaSpace{}%
\AgdaInductiveConstructor{false}\AgdaSpace{}%
\AgdaSymbol{|}\AgdaSpace{}%
\AgdaInductiveConstructor{true}\AgdaSpace{}%
\AgdaKeyword{with}\AgdaSpace{}%
\AgdaSymbol{(}\AgdaFunction{isSigned}%
\>[113]\AgdaBound{msg₁}\AgdaSpace{}%
\AgdaBound{sig2}\AgdaSpace{}%
\AgdaBound{pbk3}\AgdaSymbol{)}\<%
\\
\>[0]\AgdaFunction{lemmaHoareTripleStackGeAux'Comb3-5}\AgdaSpace{}%
\AgdaBound{msg₁}\AgdaSpace{}%
\AgdaBound{pbk1}\AgdaSpace{}%
\AgdaBound{pbk2}\AgdaSpace{}%
\AgdaBound{pbk3}\AgdaSpace{}%
\AgdaBound{pbk4}\AgdaSpace{}%
\AgdaBound{pbk5}\AgdaSpace{}%
\AgdaBound{sig1}\AgdaSpace{}%
\AgdaBound{sig2}\AgdaSpace{}%
\AgdaBound{sig3}\AgdaSpace{}%
\AgdaBound{x}\AgdaSpace{}%
\AgdaSymbol{|}\AgdaSpace{}%
\AgdaInductiveConstructor{false}\AgdaSpace{}%
\AgdaSymbol{|}\AgdaSpace{}%
\AgdaInductiveConstructor{true}\AgdaSpace{}%
\AgdaSymbol{|}\AgdaSpace{}%
\AgdaInductiveConstructor{false}\AgdaSpace{}%
\AgdaKeyword{with}%
\>[111]\AgdaSymbol{(}\AgdaFunction{isSigned}%
\>[122]\AgdaBound{msg₁}\AgdaSpace{}%
\AgdaBound{sig2}\AgdaSpace{}%
\AgdaBound{pbk4}\AgdaSymbol{)}\<%
\\
\>[0]\AgdaFunction{lemmaHoareTripleStackGeAux'Comb3-5}\AgdaSpace{}%
\AgdaBound{msg₁}\AgdaSpace{}%
\AgdaBound{pbk1}\AgdaSpace{}%
\AgdaBound{pbk2}\AgdaSpace{}%
\AgdaBound{pbk3}\AgdaSpace{}%
\AgdaBound{pbk4}\AgdaSpace{}%
\AgdaBound{pbk5}\AgdaSpace{}%
\AgdaBound{sig1}\AgdaSpace{}%
\AgdaBound{sig2}\AgdaSpace{}%
\AgdaBound{sig3}\AgdaSpace{}%
\AgdaBound{x}\AgdaSpace{}%
\AgdaSymbol{|}\AgdaSpace{}%
\AgdaInductiveConstructor{false}\AgdaSpace{}%
\AgdaSymbol{|}\AgdaSpace{}%
\AgdaInductiveConstructor{true}\AgdaSpace{}%
\AgdaSymbol{|}\AgdaSpace{}%
\AgdaInductiveConstructor{false}\AgdaSpace{}%
\AgdaSymbol{|}\AgdaSpace{}%
\AgdaInductiveConstructor{false}\AgdaSpace{}%
\AgdaKeyword{with}\AgdaSpace{}%
\AgdaSymbol{(}\AgdaFunction{isSigned}%
\>[129]\AgdaBound{msg₁}\AgdaSpace{}%
\AgdaBound{sig2}\AgdaSpace{}%
\AgdaBound{pbk5}\AgdaSymbol{)}\<%
\\
\>[0]\AgdaFunction{lemmaHoareTripleStackGeAux'Comb3-5}\AgdaSpace{}%
\AgdaBound{msg₁}\AgdaSpace{}%
\AgdaBound{pbk1}\AgdaSpace{}%
\AgdaBound{pbk2}\AgdaSpace{}%
\AgdaBound{pbk3}\AgdaSpace{}%
\AgdaBound{pbk4}\AgdaSpace{}%
\AgdaBound{pbk5}\AgdaSpace{}%
\AgdaBound{sig1}\AgdaSpace{}%
\AgdaBound{sig2}\AgdaSpace{}%
\AgdaBound{sig3}\AgdaSpace{}%
\AgdaSymbol{()}\AgdaSpace{}%
\AgdaSymbol{|}\AgdaSpace{}%
\AgdaInductiveConstructor{false}\AgdaSpace{}%
\AgdaSymbol{|}\AgdaSpace{}%
\AgdaInductiveConstructor{true}\AgdaSpace{}%
\AgdaSymbol{|}\AgdaSpace{}%
\AgdaInductiveConstructor{false}\AgdaSpace{}%
\AgdaSymbol{|}\AgdaSpace{}%
\AgdaInductiveConstructor{false}\AgdaSpace{}%
\AgdaSymbol{|}\AgdaSpace{}%
\AgdaInductiveConstructor{false}\<%
\\
\>[0]\AgdaFunction{lemmaHoareTripleStackGeAux'Comb3-5}\AgdaSpace{}%
\AgdaBound{msg₁}\AgdaSpace{}%
\AgdaBound{pbk1}\AgdaSpace{}%
\AgdaBound{pbk2}\AgdaSpace{}%
\AgdaBound{pbk3}\AgdaSpace{}%
\AgdaBound{pbk4}\AgdaSpace{}%
\AgdaBound{pbk5}\AgdaSpace{}%
\AgdaBound{sig1}\AgdaSpace{}%
\AgdaBound{sig2}\AgdaSpace{}%
\AgdaBound{sig3}\AgdaSpace{}%
\AgdaSymbol{()}\AgdaSpace{}%
\AgdaSymbol{|}\AgdaSpace{}%
\AgdaInductiveConstructor{false}\AgdaSpace{}%
\AgdaSymbol{|}\AgdaSpace{}%
\AgdaInductiveConstructor{true}\AgdaSpace{}%
\AgdaSymbol{|}\AgdaSpace{}%
\AgdaInductiveConstructor{false}\AgdaSpace{}%
\AgdaSymbol{|}\AgdaSpace{}%
\AgdaInductiveConstructor{false}\AgdaSpace{}%
\AgdaSymbol{|}\AgdaSpace{}%
\AgdaInductiveConstructor{true}\<%
\\
\>[0]\AgdaFunction{lemmaHoareTripleStackGeAux'Comb3-5}\AgdaSpace{}%
\AgdaBound{msg₁}\AgdaSpace{}%
\AgdaBound{pbk1}\AgdaSpace{}%
\AgdaBound{pbk2}\AgdaSpace{}%
\AgdaBound{pbk3}\AgdaSpace{}%
\AgdaBound{pbk4}\AgdaSpace{}%
\AgdaBound{pbk5}\AgdaSpace{}%
\AgdaBound{sig1}\AgdaSpace{}%
\AgdaBound{sig2}\AgdaSpace{}%
\AgdaBound{sig3}\AgdaSpace{}%
\AgdaBound{x}\AgdaSpace{}%
\AgdaSymbol{|}\AgdaSpace{}%
\AgdaInductiveConstructor{false}\AgdaSpace{}%
\AgdaSymbol{|}\AgdaSpace{}%
\AgdaInductiveConstructor{true}\AgdaSpace{}%
\AgdaSymbol{|}\AgdaSpace{}%
\AgdaInductiveConstructor{false}\AgdaSpace{}%
\AgdaSymbol{|}\AgdaSpace{}%
\AgdaInductiveConstructor{true}\AgdaSpace{}%
\AgdaKeyword{with}\AgdaSpace{}%
\AgdaSymbol{(}\AgdaFunction{isSigned}%
\>[128]\AgdaBound{msg₁}\AgdaSpace{}%
\AgdaBound{sig3}\AgdaSpace{}%
\AgdaBound{pbk5}\AgdaSymbol{)}\<%
\\
\>[0]\AgdaFunction{lemmaHoareTripleStackGeAux'Comb3-5}\AgdaSpace{}%
\AgdaBound{msg₁}\AgdaSpace{}%
\AgdaBound{pbk1}\AgdaSpace{}%
\AgdaBound{pbk2}\AgdaSpace{}%
\AgdaBound{pbk3}%
\>[7109I]\AgdaBound{pbk4}\AgdaSpace{}%
\AgdaBound{pbk5}\AgdaSpace{}%
\AgdaBound{sig1}\AgdaSpace{}%
\AgdaBound{sig2}\AgdaSpace{}%
\AgdaBound{sig3}\AgdaSpace{}%
\AgdaBound{x}\AgdaSpace{}%
\AgdaSymbol{|}\AgdaSpace{}%
\AgdaInductiveConstructor{false}\AgdaSpace{}%
\AgdaSymbol{|}\AgdaSpace{}%
\AgdaInductiveConstructor{true}\AgdaSpace{}%
\AgdaSymbol{|}\AgdaSpace{}%
\AgdaInductiveConstructor{false}\AgdaSpace{}%
\AgdaSymbol{|}\AgdaSpace{}%
\AgdaInductiveConstructor{true}\AgdaSpace{}%
\AgdaSymbol{|}\AgdaSpace{}%
\AgdaInductiveConstructor{true}\<%
\\
\>[7109I][@{}l@{\AgdaIndent{0}}]%
\>[56]\AgdaSymbol{=}%
\>[59]\AgdaInductiveConstructor{inj₂}\AgdaSpace{}%
\AgdaSymbol{(}\AgdaInductiveConstructor{inj₂}\AgdaSpace{}%
\AgdaSymbol{(}\AgdaInductiveConstructor{inj₂}\AgdaSpace{}%
\AgdaSymbol{(}\AgdaInductiveConstructor{inj₂}\AgdaSpace{}%
\AgdaSymbol{(}\AgdaInductiveConstructor{inj₂}\AgdaSpace{}%
\AgdaSymbol{(}\AgdaInductiveConstructor{inj₂}\AgdaSpace{}%
\AgdaSymbol{(}\AgdaInductiveConstructor{inj₂}\AgdaSpace{}%
\AgdaSymbol{(}\AgdaInductiveConstructor{inj₂}\AgdaSpace{}%
\AgdaSymbol{(}\AgdaInductiveConstructor{inj₁}\AgdaSpace{}%
\AgdaSymbol{(}\AgdaInductiveConstructor{conj}\AgdaSpace{}%
\AgdaSymbol{(}\AgdaInductiveConstructor{conj}\AgdaSpace{}%
\AgdaInductiveConstructor{tt}\AgdaSpace{}%
\AgdaInductiveConstructor{tt}\AgdaSymbol{)}\AgdaSpace{}%
\AgdaInductiveConstructor{tt}\AgdaSymbol{)))))))))}\<%
\\
\>[0]\AgdaFunction{lemmaHoareTripleStackGeAux'Comb3-5}\AgdaSpace{}%
\AgdaBound{msg₁}\AgdaSpace{}%
\AgdaBound{pbk1}\AgdaSpace{}%
\AgdaBound{pbk2}\AgdaSpace{}%
\AgdaBound{pbk3}\AgdaSpace{}%
\AgdaBound{pbk4}\AgdaSpace{}%
\AgdaBound{pbk5}\AgdaSpace{}%
\AgdaBound{sig1}\AgdaSpace{}%
\AgdaBound{sig2}\AgdaSpace{}%
\AgdaBound{sig3}\AgdaSpace{}%
\AgdaBound{x}\AgdaSpace{}%
\AgdaSymbol{|}\AgdaSpace{}%
\AgdaInductiveConstructor{false}\AgdaSpace{}%
\AgdaSymbol{|}\AgdaSpace{}%
\AgdaInductiveConstructor{true}\AgdaSpace{}%
\AgdaSymbol{|}\AgdaSpace{}%
\AgdaInductiveConstructor{true}\AgdaSpace{}%
\AgdaKeyword{with}\AgdaSpace{}%
\AgdaSymbol{(}\AgdaFunction{isSigned}%
\>[120]\AgdaBound{msg₁}\AgdaSpace{}%
\AgdaBound{sig3}\AgdaSpace{}%
\AgdaBound{pbk4}\AgdaSymbol{)}\<%
\\
\>[0]\AgdaFunction{lemmaHoareTripleStackGeAux'Comb3-5}\AgdaSpace{}%
\AgdaBound{msg₁}\AgdaSpace{}%
\AgdaBound{pbk1}\AgdaSpace{}%
\AgdaBound{pbk2}\AgdaSpace{}%
\AgdaBound{pbk3}\AgdaSpace{}%
\AgdaBound{pbk4}\AgdaSpace{}%
\AgdaBound{pbk5}\AgdaSpace{}%
\AgdaBound{sig1}\AgdaSpace{}%
\AgdaBound{sig2}\AgdaSpace{}%
\AgdaBound{sig3}\AgdaSpace{}%
\AgdaBound{x}\AgdaSpace{}%
\AgdaSymbol{|}\AgdaSpace{}%
\AgdaInductiveConstructor{false}\AgdaSpace{}%
\AgdaSymbol{|}\AgdaSpace{}%
\AgdaInductiveConstructor{true}\AgdaSpace{}%
\AgdaSymbol{|}\AgdaSpace{}%
\AgdaInductiveConstructor{true}\AgdaSpace{}%
\AgdaSymbol{|}\AgdaSpace{}%
\AgdaInductiveConstructor{false}\AgdaSpace{}%
\AgdaKeyword{with}\AgdaSpace{}%
\AgdaSymbol{(}\AgdaFunction{isSigned}%
\>[128]\AgdaBound{msg₁}\AgdaSpace{}%
\AgdaBound{sig3}\AgdaSpace{}%
\AgdaBound{pbk5}\AgdaSymbol{)}\<%
\\
\>[0]\AgdaFunction{lemmaHoareTripleStackGeAux'Comb3-5}\AgdaSpace{}%
\AgdaBound{msg₁}\AgdaSpace{}%
\AgdaBound{pbk1}\AgdaSpace{}%
\AgdaBound{pbk2}\AgdaSpace{}%
\AgdaBound{pbk3}%
\>[7184I]\AgdaBound{pbk4}\AgdaSpace{}%
\AgdaBound{pbk5}\AgdaSpace{}%
\AgdaBound{sig1}\AgdaSpace{}%
\AgdaBound{sig2}\AgdaSpace{}%
\AgdaBound{sig3}\AgdaSpace{}%
\AgdaBound{x}\AgdaSpace{}%
\AgdaSymbol{|}\AgdaSpace{}%
\AgdaInductiveConstructor{false}\AgdaSpace{}%
\AgdaSymbol{|}\AgdaSpace{}%
\AgdaInductiveConstructor{true}\AgdaSpace{}%
\AgdaSymbol{|}\AgdaSpace{}%
\AgdaInductiveConstructor{true}\AgdaSpace{}%
\AgdaSymbol{|}\AgdaSpace{}%
\AgdaInductiveConstructor{false}\AgdaSpace{}%
\AgdaSymbol{|}\AgdaSpace{}%
\AgdaInductiveConstructor{true}\<%
\\
\>[7184I][@{}l@{\AgdaIndent{0}}]%
\>[56]\AgdaSymbol{=}\AgdaSpace{}%
\AgdaInductiveConstructor{inj₂}\AgdaSpace{}%
\AgdaSymbol{(}\AgdaInductiveConstructor{inj₂}\AgdaSpace{}%
\AgdaSymbol{(}\AgdaInductiveConstructor{inj₂}\AgdaSpace{}%
\AgdaSymbol{(}\AgdaInductiveConstructor{inj₂}\AgdaSpace{}%
\AgdaSymbol{(}\AgdaInductiveConstructor{inj₂}\AgdaSpace{}%
\AgdaSymbol{(}\AgdaInductiveConstructor{inj₂}\AgdaSpace{}%
\AgdaSymbol{(}\AgdaInductiveConstructor{inj₂}\AgdaSpace{}%
\AgdaSymbol{(}\AgdaInductiveConstructor{inj₁}\AgdaSpace{}%
\AgdaSymbol{(}\AgdaInductiveConstructor{conj}\AgdaSpace{}%
\AgdaSymbol{(}\AgdaInductiveConstructor{conj}\AgdaSpace{}%
\AgdaInductiveConstructor{tt}\AgdaSpace{}%
\AgdaInductiveConstructor{tt}\AgdaSymbol{)}\AgdaSpace{}%
\AgdaInductiveConstructor{tt}\AgdaSymbol{))))))))}\<%
\\
\>[0]\AgdaFunction{lemmaHoareTripleStackGeAux'Comb3-5}\AgdaSpace{}%
\AgdaBound{msg₁}\AgdaSpace{}%
\AgdaBound{pbk1}\AgdaSpace{}%
\AgdaBound{pbk2}\AgdaSpace{}%
\AgdaBound{pbk3}%
\>[7217I]\AgdaBound{pbk4}\AgdaSpace{}%
\AgdaBound{pbk5}\AgdaSpace{}%
\AgdaBound{sig1}\AgdaSpace{}%
\AgdaBound{sig2}\AgdaSpace{}%
\AgdaBound{sig3}\AgdaSpace{}%
\AgdaBound{x}\AgdaSpace{}%
\AgdaSymbol{|}\AgdaSpace{}%
\AgdaInductiveConstructor{false}\AgdaSpace{}%
\AgdaSymbol{|}\AgdaSpace{}%
\AgdaInductiveConstructor{true}\AgdaSpace{}%
\AgdaSymbol{|}\AgdaSpace{}%
\AgdaInductiveConstructor{true}\AgdaSpace{}%
\AgdaSymbol{|}\AgdaSpace{}%
\AgdaInductiveConstructor{true}\<%
\\
\>[7217I][@{}l@{\AgdaIndent{0}}]%
\>[56]\AgdaSymbol{=}\AgdaSpace{}%
\AgdaInductiveConstructor{inj₂}\AgdaSpace{}%
\AgdaSymbol{(}\AgdaInductiveConstructor{inj₂}\AgdaSpace{}%
\AgdaSymbol{(}\AgdaInductiveConstructor{inj₂}\AgdaSpace{}%
\AgdaSymbol{(}\AgdaInductiveConstructor{inj₂}\AgdaSpace{}%
\AgdaSymbol{(}\AgdaInductiveConstructor{inj₂}\AgdaSpace{}%
\AgdaSymbol{(}\AgdaInductiveConstructor{inj₂}\AgdaSpace{}%
\AgdaSymbol{(}\AgdaInductiveConstructor{inj₁}\AgdaSpace{}%
\AgdaSymbol{(}\AgdaInductiveConstructor{conj}\AgdaSpace{}%
\AgdaSymbol{(}\AgdaInductiveConstructor{conj}\AgdaSpace{}%
\AgdaInductiveConstructor{tt}\AgdaSpace{}%
\AgdaInductiveConstructor{tt}\AgdaSymbol{)}\AgdaSpace{}%
\AgdaInductiveConstructor{tt}\AgdaSymbol{)))))))}\<%
\\
\>[0]\AgdaFunction{lemmaHoareTripleStackGeAux'Comb3-5}\AgdaSpace{}%
\AgdaBound{msg₁}\AgdaSpace{}%
\AgdaBound{pbk1}\AgdaSpace{}%
\AgdaBound{pbk2}\AgdaSpace{}%
\AgdaBound{pbk3}\AgdaSpace{}%
\AgdaBound{pbk4}\AgdaSpace{}%
\AgdaBound{pbk5}\AgdaSpace{}%
\AgdaBound{sig1}\AgdaSpace{}%
\AgdaBound{sig2}\AgdaSpace{}%
\AgdaBound{sig3}\AgdaSpace{}%
\AgdaBound{x}\AgdaSpace{}%
\AgdaSymbol{|}\AgdaSpace{}%
\AgdaInductiveConstructor{true}\AgdaSpace{}%
\AgdaKeyword{with}\AgdaSpace{}%
\AgdaSymbol{(}\AgdaFunction{isSigned}%
\>[105]\AgdaBound{msg₁}\AgdaSpace{}%
\AgdaBound{sig2}\AgdaSpace{}%
\AgdaBound{pbk2}\AgdaSymbol{)}\<%
\\
\>[0]\AgdaFunction{lemmaHoareTripleStackGeAux'Comb3-5}\AgdaSpace{}%
\AgdaBound{msg₁}\AgdaSpace{}%
\AgdaBound{pbk1}\AgdaSpace{}%
\AgdaBound{pbk2}\AgdaSpace{}%
\AgdaBound{pbk3}\AgdaSpace{}%
\AgdaBound{pbk4}\AgdaSpace{}%
\AgdaBound{pbk5}\AgdaSpace{}%
\AgdaBound{sig1}\AgdaSpace{}%
\AgdaBound{sig2}\AgdaSpace{}%
\AgdaBound{sig3}\AgdaSpace{}%
\AgdaBound{x}\AgdaSpace{}%
\AgdaSymbol{|}\AgdaSpace{}%
\AgdaInductiveConstructor{true}\AgdaSpace{}%
\AgdaSymbol{|}\AgdaSpace{}%
\AgdaInductiveConstructor{false}\AgdaSpace{}%
\AgdaKeyword{with}\AgdaSpace{}%
\AgdaSymbol{(}\AgdaFunction{isSigned}%
\>[113]\AgdaBound{msg₁}\AgdaSpace{}%
\AgdaBound{sig2}\AgdaSpace{}%
\AgdaBound{pbk3}\AgdaSymbol{)}\<%
\\
\>[0]\AgdaFunction{lemmaHoareTripleStackGeAux'Comb3-5}\AgdaSpace{}%
\AgdaBound{msg₁}\AgdaSpace{}%
\AgdaBound{pbk1}\AgdaSpace{}%
\AgdaBound{pbk2}\AgdaSpace{}%
\AgdaBound{pbk3}\AgdaSpace{}%
\AgdaBound{pbk4}\AgdaSpace{}%
\AgdaBound{pbk5}\AgdaSpace{}%
\AgdaBound{sig1}\AgdaSpace{}%
\AgdaBound{sig2}\AgdaSpace{}%
\AgdaBound{sig3}\AgdaSpace{}%
\AgdaBound{x}\AgdaSpace{}%
\AgdaSymbol{|}\AgdaSpace{}%
\AgdaInductiveConstructor{true}\AgdaSpace{}%
\AgdaSymbol{|}\AgdaSpace{}%
\AgdaInductiveConstructor{false}\AgdaSpace{}%
\AgdaSymbol{|}\AgdaSpace{}%
\AgdaInductiveConstructor{false}\AgdaSpace{}%
\AgdaKeyword{with}%
\>[111]\AgdaSymbol{(}\AgdaFunction{isSigned}%
\>[122]\AgdaBound{msg₁}\AgdaSpace{}%
\AgdaBound{sig2}\AgdaSpace{}%
\AgdaBound{pbk4}\AgdaSymbol{)}\<%
\\
\>[0]\AgdaFunction{lemmaHoareTripleStackGeAux'Comb3-5}\AgdaSpace{}%
\AgdaBound{msg₁}\AgdaSpace{}%
\AgdaBound{pbk1}\AgdaSpace{}%
\AgdaBound{pbk2}\AgdaSpace{}%
\AgdaBound{pbk3}\AgdaSpace{}%
\AgdaBound{pbk4}\AgdaSpace{}%
\AgdaBound{pbk5}\AgdaSpace{}%
\AgdaBound{sig1}\AgdaSpace{}%
\AgdaBound{sig2}\AgdaSpace{}%
\AgdaBound{sig3}\AgdaSpace{}%
\AgdaBound{x}\AgdaSpace{}%
\AgdaSymbol{|}\AgdaSpace{}%
\AgdaInductiveConstructor{true}\AgdaSpace{}%
\AgdaSymbol{|}\AgdaSpace{}%
\AgdaInductiveConstructor{false}\AgdaSpace{}%
\AgdaSymbol{|}\AgdaSpace{}%
\AgdaInductiveConstructor{false}\AgdaSpace{}%
\AgdaSymbol{|}\AgdaSpace{}%
\AgdaInductiveConstructor{false}\AgdaSpace{}%
\AgdaKeyword{with}%
\>[119]\AgdaSymbol{(}\AgdaFunction{isSigned}%
\>[130]\AgdaBound{msg₁}\AgdaSpace{}%
\AgdaBound{sig2}\AgdaSpace{}%
\AgdaBound{pbk5}\AgdaSymbol{)}\<%
\\
\>[0]\AgdaFunction{lemmaHoareTripleStackGeAux'Comb3-5}\AgdaSpace{}%
\AgdaBound{msg₁}\AgdaSpace{}%
\AgdaBound{pbk1}\AgdaSpace{}%
\AgdaBound{pbk2}\AgdaSpace{}%
\AgdaBound{pbk3}\AgdaSpace{}%
\AgdaBound{pbk4}\AgdaSpace{}%
\AgdaBound{pbk5}\AgdaSpace{}%
\AgdaBound{sig1}\AgdaSpace{}%
\AgdaBound{sig2}\AgdaSpace{}%
\AgdaBound{sig3}\AgdaSpace{}%
\AgdaSymbol{()}\AgdaSpace{}%
\AgdaSymbol{|}\AgdaSpace{}%
\AgdaInductiveConstructor{true}\AgdaSpace{}%
\AgdaSymbol{|}\AgdaSpace{}%
\AgdaInductiveConstructor{false}\AgdaSpace{}%
\AgdaSymbol{|}\AgdaSpace{}%
\AgdaInductiveConstructor{false}\AgdaSpace{}%
\AgdaSymbol{|}\AgdaSpace{}%
\AgdaInductiveConstructor{false}\AgdaSpace{}%
\AgdaSymbol{|}\AgdaSpace{}%
\AgdaInductiveConstructor{false}\<%
\\
\>[0]\AgdaFunction{lemmaHoareTripleStackGeAux'Comb3-5}\AgdaSpace{}%
\AgdaBound{msg₁}\AgdaSpace{}%
\AgdaBound{pbk1}\AgdaSpace{}%
\AgdaBound{pbk2}\AgdaSpace{}%
\AgdaBound{pbk3}\AgdaSpace{}%
\AgdaBound{pbk4}\AgdaSpace{}%
\AgdaBound{pbk5}\AgdaSpace{}%
\AgdaBound{sig1}\AgdaSpace{}%
\AgdaBound{sig2}\AgdaSpace{}%
\AgdaBound{sig3}\AgdaSpace{}%
\AgdaSymbol{()}\AgdaSpace{}%
\AgdaSymbol{|}\AgdaSpace{}%
\AgdaInductiveConstructor{true}\AgdaSpace{}%
\AgdaSymbol{|}\AgdaSpace{}%
\AgdaInductiveConstructor{false}\AgdaSpace{}%
\AgdaSymbol{|}\AgdaSpace{}%
\AgdaInductiveConstructor{false}\AgdaSpace{}%
\AgdaSymbol{|}\AgdaSpace{}%
\AgdaInductiveConstructor{false}\AgdaSpace{}%
\AgdaSymbol{|}\AgdaSpace{}%
\AgdaInductiveConstructor{true}\<%
\\
\>[0]\AgdaFunction{lemmaHoareTripleStackGeAux'Comb3-5}\AgdaSpace{}%
\AgdaBound{msg₁}\AgdaSpace{}%
\AgdaBound{pbk1}\AgdaSpace{}%
\AgdaBound{pbk2}\AgdaSpace{}%
\AgdaBound{pbk3}\AgdaSpace{}%
\AgdaBound{pbk4}\AgdaSpace{}%
\AgdaBound{pbk5}\AgdaSpace{}%
\AgdaBound{sig1}\AgdaSpace{}%
\AgdaBound{sig2}\AgdaSpace{}%
\AgdaBound{sig3}\AgdaSpace{}%
\AgdaBound{x}\AgdaSpace{}%
\AgdaSymbol{|}\AgdaSpace{}%
\AgdaInductiveConstructor{true}\AgdaSpace{}%
\AgdaSymbol{|}\AgdaSpace{}%
\AgdaInductiveConstructor{false}\AgdaSpace{}%
\AgdaSymbol{|}\AgdaSpace{}%
\AgdaInductiveConstructor{false}\AgdaSpace{}%
\AgdaSymbol{|}\AgdaSpace{}%
\AgdaInductiveConstructor{true}\AgdaSpace{}%
\AgdaKeyword{with}%
\>[118]\AgdaSymbol{(}\AgdaFunction{isSigned}%
\>[129]\AgdaBound{msg₁}\AgdaSpace{}%
\AgdaBound{sig3}\AgdaSpace{}%
\AgdaBound{pbk5}\AgdaSymbol{)}\<%
\\
\>[0]\AgdaFunction{lemmaHoareTripleStackGeAux'Comb3-5}\AgdaSpace{}%
\AgdaBound{msg₁}\AgdaSpace{}%
\AgdaBound{pbk1}\AgdaSpace{}%
\AgdaBound{pbk2}\AgdaSpace{}%
\AgdaBound{pbk3}%
\>[7382I]\AgdaBound{pbk4}\AgdaSpace{}%
\AgdaBound{pbk5}\AgdaSpace{}%
\AgdaBound{sig1}\AgdaSpace{}%
\AgdaBound{sig2}\AgdaSpace{}%
\AgdaBound{sig3}\AgdaSpace{}%
\AgdaBound{x}\AgdaSpace{}%
\AgdaSymbol{|}\AgdaSpace{}%
\AgdaInductiveConstructor{true}\AgdaSpace{}%
\AgdaSymbol{|}\AgdaSpace{}%
\AgdaInductiveConstructor{false}\AgdaSpace{}%
\AgdaSymbol{|}\AgdaSpace{}%
\AgdaInductiveConstructor{false}\AgdaSpace{}%
\AgdaSymbol{|}\AgdaSpace{}%
\AgdaInductiveConstructor{true}\AgdaSpace{}%
\AgdaSymbol{|}\AgdaSpace{}%
\AgdaInductiveConstructor{true}\<%
\\
\>[7382I][@{}l@{\AgdaIndent{0}}]%
\>[56]\AgdaSymbol{=}\AgdaSpace{}%
\AgdaInductiveConstructor{inj₂}\AgdaSpace{}%
\AgdaSymbol{(}\AgdaInductiveConstructor{inj₂}\AgdaSpace{}%
\AgdaSymbol{(}\AgdaInductiveConstructor{inj₂}\AgdaSpace{}%
\AgdaSymbol{(}\AgdaInductiveConstructor{inj₂}\AgdaSpace{}%
\AgdaSymbol{(}\AgdaInductiveConstructor{inj₂}\AgdaSpace{}%
\AgdaSymbol{(}\AgdaInductiveConstructor{inj₁}\AgdaSpace{}%
\AgdaSymbol{(}\AgdaInductiveConstructor{conj}\AgdaSpace{}%
\AgdaSymbol{(}\AgdaInductiveConstructor{conj}\AgdaSpace{}%
\AgdaInductiveConstructor{tt}\AgdaSpace{}%
\AgdaInductiveConstructor{tt}\AgdaSymbol{)}\AgdaSpace{}%
\AgdaInductiveConstructor{tt}\AgdaSymbol{))))))}\<%
\\
\>[0]\AgdaFunction{lemmaHoareTripleStackGeAux'Comb3-5}\AgdaSpace{}%
\AgdaBound{msg₁}\AgdaSpace{}%
\AgdaBound{pbk1}\AgdaSpace{}%
\AgdaBound{pbk2}\AgdaSpace{}%
\AgdaBound{pbk3}\AgdaSpace{}%
\AgdaBound{pbk4}\AgdaSpace{}%
\AgdaBound{pbk5}\AgdaSpace{}%
\AgdaBound{sig1}\AgdaSpace{}%
\AgdaBound{sig2}\AgdaSpace{}%
\AgdaBound{sig3}\AgdaSpace{}%
\AgdaBound{x}\AgdaSpace{}%
\AgdaSymbol{|}\AgdaSpace{}%
\AgdaInductiveConstructor{true}\AgdaSpace{}%
\AgdaSymbol{|}\AgdaSpace{}%
\AgdaInductiveConstructor{false}\AgdaSpace{}%
\AgdaSymbol{|}\AgdaSpace{}%
\AgdaInductiveConstructor{true}\AgdaSpace{}%
\AgdaKeyword{with}%
\>[110]\AgdaSymbol{(}\AgdaFunction{isSigned}%
\>[121]\AgdaBound{msg₁}\AgdaSpace{}%
\AgdaBound{sig3}\AgdaSpace{}%
\AgdaBound{pbk4}\AgdaSymbol{)}\<%
\\
\>[0]\AgdaFunction{lemmaHoareTripleStackGeAux'Comb3-5}\AgdaSpace{}%
\AgdaBound{msg₁}\AgdaSpace{}%
\AgdaBound{pbk1}\AgdaSpace{}%
\AgdaBound{pbk2}\AgdaSpace{}%
\AgdaBound{pbk3}\AgdaSpace{}%
\AgdaBound{pbk4}\AgdaSpace{}%
\AgdaBound{pbk5}\AgdaSpace{}%
\AgdaBound{sig1}\AgdaSpace{}%
\AgdaBound{sig2}\AgdaSpace{}%
\AgdaBound{sig3}\AgdaSpace{}%
\AgdaBound{x}\AgdaSpace{}%
\AgdaSymbol{|}\AgdaSpace{}%
\AgdaInductiveConstructor{true}\AgdaSpace{}%
\AgdaSymbol{|}\AgdaSpace{}%
\AgdaInductiveConstructor{false}\AgdaSpace{}%
\AgdaSymbol{|}\AgdaSpace{}%
\AgdaInductiveConstructor{true}\AgdaSpace{}%
\AgdaSymbol{|}\AgdaSpace{}%
\AgdaInductiveConstructor{false}\AgdaSpace{}%
\AgdaKeyword{with}%
\>[118]\AgdaSymbol{(}\AgdaFunction{isSigned}%
\>[129]\AgdaBound{msg₁}\AgdaSpace{}%
\AgdaBound{sig3}\AgdaSpace{}%
\AgdaBound{pbk5}\AgdaSymbol{)}\<%
\\
\>[0]\AgdaFunction{lemmaHoareTripleStackGeAux'Comb3-5}\AgdaSpace{}%
\AgdaBound{msg₁}\AgdaSpace{}%
\AgdaBound{pbk1}\AgdaSpace{}%
\AgdaBound{pbk2}\AgdaSpace{}%
\AgdaBound{pbk3}%
\>[7453I]\AgdaBound{pbk4}\AgdaSpace{}%
\AgdaBound{pbk5}\AgdaSpace{}%
\AgdaBound{sig1}\AgdaSpace{}%
\AgdaBound{sig2}\AgdaSpace{}%
\AgdaBound{sig3}\AgdaSpace{}%
\AgdaBound{x}\AgdaSpace{}%
\AgdaSymbol{|}\AgdaSpace{}%
\AgdaInductiveConstructor{true}\AgdaSpace{}%
\AgdaSymbol{|}\AgdaSpace{}%
\AgdaInductiveConstructor{false}\AgdaSpace{}%
\AgdaSymbol{|}\AgdaSpace{}%
\AgdaInductiveConstructor{true}\AgdaSpace{}%
\AgdaSymbol{|}\AgdaSpace{}%
\AgdaInductiveConstructor{false}\AgdaSpace{}%
\AgdaSymbol{|}\AgdaSpace{}%
\AgdaInductiveConstructor{true}\<%
\\
\>[7453I][@{}l@{\AgdaIndent{0}}]%
\>[56]\AgdaSymbol{=}\AgdaSpace{}%
\AgdaInductiveConstructor{inj₂}\AgdaSpace{}%
\AgdaSymbol{(}\AgdaInductiveConstructor{inj₂}\AgdaSpace{}%
\AgdaSymbol{(}\AgdaInductiveConstructor{inj₂}\AgdaSpace{}%
\AgdaSymbol{(}\AgdaInductiveConstructor{inj₂}\AgdaSpace{}%
\AgdaSymbol{(}\AgdaInductiveConstructor{inj₁}\AgdaSpace{}%
\AgdaSymbol{(}\AgdaInductiveConstructor{conj}\AgdaSpace{}%
\AgdaSymbol{(}\AgdaInductiveConstructor{conj}\AgdaSpace{}%
\AgdaInductiveConstructor{tt}\AgdaSpace{}%
\AgdaInductiveConstructor{tt}\AgdaSymbol{)}\AgdaSpace{}%
\AgdaInductiveConstructor{tt}\AgdaSymbol{)))))}\<%
\\
\>[0]\AgdaFunction{lemmaHoareTripleStackGeAux'Comb3-5}\AgdaSpace{}%
\AgdaBound{msg₁}\AgdaSpace{}%
\AgdaBound{pbk1}\AgdaSpace{}%
\AgdaBound{pbk2}\AgdaSpace{}%
\AgdaBound{pbk3}%
\>[7483I]\AgdaBound{pbk4}\AgdaSpace{}%
\AgdaBound{pbk5}\AgdaSpace{}%
\AgdaBound{sig1}\AgdaSpace{}%
\AgdaBound{sig2}\AgdaSpace{}%
\AgdaBound{sig3}\AgdaSpace{}%
\AgdaBound{x}\AgdaSpace{}%
\AgdaSymbol{|}\AgdaSpace{}%
\AgdaInductiveConstructor{true}\AgdaSpace{}%
\AgdaSymbol{|}\AgdaSpace{}%
\AgdaInductiveConstructor{false}\AgdaSpace{}%
\AgdaSymbol{|}\AgdaSpace{}%
\AgdaInductiveConstructor{true}\AgdaSpace{}%
\AgdaSymbol{|}\AgdaSpace{}%
\AgdaInductiveConstructor{true}\<%
\\
\>[7483I][@{}l@{\AgdaIndent{0}}]%
\>[56]\AgdaSymbol{=}\AgdaSpace{}%
\AgdaInductiveConstructor{inj₂}\AgdaSpace{}%
\AgdaSymbol{(}\AgdaInductiveConstructor{inj₂}\AgdaSpace{}%
\AgdaSymbol{(}\AgdaInductiveConstructor{inj₂}\AgdaSpace{}%
\AgdaSymbol{(}\AgdaInductiveConstructor{inj₁}\AgdaSpace{}%
\AgdaSymbol{(}\AgdaInductiveConstructor{conj}\AgdaSpace{}%
\AgdaSymbol{(}\AgdaInductiveConstructor{conj}\AgdaSpace{}%
\AgdaInductiveConstructor{tt}\AgdaSpace{}%
\AgdaInductiveConstructor{tt}\AgdaSymbol{)}\AgdaSpace{}%
\AgdaInductiveConstructor{tt}\AgdaSymbol{))))}\<%
\\
\>[0]\AgdaFunction{lemmaHoareTripleStackGeAux'Comb3-5}\AgdaSpace{}%
\AgdaBound{msg₁}\AgdaSpace{}%
\AgdaBound{pbk1}\AgdaSpace{}%
\AgdaBound{pbk2}\AgdaSpace{}%
\AgdaBound{pbk3}\AgdaSpace{}%
\AgdaBound{pbk4}\AgdaSpace{}%
\AgdaBound{pbk5}\AgdaSpace{}%
\AgdaBound{sig1}\AgdaSpace{}%
\AgdaBound{sig2}\AgdaSpace{}%
\AgdaBound{sig3}\AgdaSpace{}%
\AgdaBound{x}\AgdaSpace{}%
\AgdaSymbol{|}\AgdaSpace{}%
\AgdaInductiveConstructor{true}\AgdaSpace{}%
\AgdaSymbol{|}\AgdaSpace{}%
\AgdaInductiveConstructor{true}\AgdaSpace{}%
\AgdaKeyword{with}%
\>[102]\AgdaSymbol{(}\AgdaFunction{isSigned}%
\>[113]\AgdaBound{msg₁}\AgdaSpace{}%
\AgdaBound{sig3}\AgdaSpace{}%
\AgdaBound{pbk3}\AgdaSymbol{)}\<%
\\
\>[0]\AgdaFunction{lemmaHoareTripleStackGeAux'Comb3-5}\AgdaSpace{}%
\AgdaBound{msg₁}\AgdaSpace{}%
\AgdaBound{pbk1}\AgdaSpace{}%
\AgdaBound{pbk2}\AgdaSpace{}%
\AgdaBound{pbk3}\AgdaSpace{}%
\AgdaBound{pbk4}\AgdaSpace{}%
\AgdaBound{pbk5}\AgdaSpace{}%
\AgdaBound{sig1}\AgdaSpace{}%
\AgdaBound{sig2}\AgdaSpace{}%
\AgdaBound{sig3}\AgdaSpace{}%
\AgdaBound{x}\AgdaSpace{}%
\AgdaSymbol{|}\AgdaSpace{}%
\AgdaInductiveConstructor{true}\AgdaSpace{}%
\AgdaSymbol{|}\AgdaSpace{}%
\AgdaInductiveConstructor{true}\AgdaSpace{}%
\AgdaSymbol{|}\AgdaSpace{}%
\AgdaInductiveConstructor{false}\AgdaSpace{}%
\AgdaKeyword{with}\AgdaSpace{}%
\AgdaSymbol{(}\AgdaFunction{isSigned}%
\>[120]\AgdaBound{msg₁}\AgdaSpace{}%
\AgdaBound{sig3}\AgdaSpace{}%
\AgdaBound{pbk4}\AgdaSymbol{)}\<%
\\
\>[0]\AgdaFunction{lemmaHoareTripleStackGeAux'Comb3-5}\AgdaSpace{}%
\AgdaBound{msg₁}\AgdaSpace{}%
\AgdaBound{pbk1}\AgdaSpace{}%
\AgdaBound{pbk2}\AgdaSpace{}%
\AgdaBound{pbk3}\AgdaSpace{}%
\AgdaBound{pbk4}\AgdaSpace{}%
\AgdaBound{pbk5}\AgdaSpace{}%
\AgdaBound{sig1}\AgdaSpace{}%
\AgdaBound{sig2}\AgdaSpace{}%
\AgdaBound{sig3}\AgdaSpace{}%
\AgdaBound{x}\AgdaSpace{}%
\AgdaSymbol{|}\AgdaSpace{}%
\AgdaInductiveConstructor{true}\AgdaSpace{}%
\AgdaSymbol{|}\AgdaSpace{}%
\AgdaInductiveConstructor{true}\AgdaSpace{}%
\AgdaSymbol{|}\AgdaSpace{}%
\AgdaInductiveConstructor{false}\AgdaSpace{}%
\AgdaSymbol{|}\AgdaSpace{}%
\AgdaInductiveConstructor{false}\AgdaSpace{}%
\AgdaKeyword{with}\AgdaSpace{}%
\AgdaSymbol{(}\AgdaFunction{isSigned}%
\>[128]\AgdaBound{msg₁}\AgdaSpace{}%
\AgdaBound{sig3}\AgdaSpace{}%
\AgdaBound{pbk5}\AgdaSymbol{)}\<%
\\
\>[0]\AgdaFunction{lemmaHoareTripleStackGeAux'Comb3-5}\AgdaSpace{}%
\AgdaBound{msg₁}\AgdaSpace{}%
\AgdaBound{pbk1}\AgdaSpace{}%
\AgdaBound{pbk2}\AgdaSpace{}%
\AgdaBound{pbk3}%
\>[7569I]\AgdaBound{pbk4}\AgdaSpace{}%
\AgdaBound{pbk5}\AgdaSpace{}%
\AgdaBound{sig1}\AgdaSpace{}%
\AgdaBound{sig2}\AgdaSpace{}%
\AgdaBound{sig3}\AgdaSpace{}%
\AgdaBound{x}\AgdaSpace{}%
\AgdaSymbol{|}\AgdaSpace{}%
\AgdaInductiveConstructor{true}\AgdaSpace{}%
\AgdaSymbol{|}\AgdaSpace{}%
\AgdaInductiveConstructor{true}\AgdaSpace{}%
\AgdaSymbol{|}\AgdaSpace{}%
\AgdaInductiveConstructor{false}\AgdaSpace{}%
\AgdaSymbol{|}\AgdaSpace{}%
\AgdaInductiveConstructor{false}\AgdaSpace{}%
\AgdaSymbol{|}\AgdaSpace{}%
\AgdaInductiveConstructor{true}\<%
\\
\>[7569I][@{}l@{\AgdaIndent{0}}]%
\>[56]\AgdaSymbol{=}\AgdaSpace{}%
\AgdaInductiveConstructor{inj₂}\AgdaSpace{}%
\AgdaSymbol{(}\AgdaInductiveConstructor{inj₂}\AgdaSpace{}%
\AgdaSymbol{(}\AgdaInductiveConstructor{inj₁}\AgdaSpace{}%
\AgdaSymbol{(}\AgdaInductiveConstructor{conj}\AgdaSpace{}%
\AgdaSymbol{(}\AgdaInductiveConstructor{conj}\AgdaSpace{}%
\AgdaInductiveConstructor{tt}\AgdaSpace{}%
\AgdaInductiveConstructor{tt}\AgdaSymbol{)}\AgdaSpace{}%
\AgdaInductiveConstructor{tt}\AgdaSymbol{)))}\<%
\\
\>[0]\AgdaFunction{lemmaHoareTripleStackGeAux'Comb3-5}\AgdaSpace{}%
\AgdaBound{msg₁}\AgdaSpace{}%
\AgdaBound{pbk1}\AgdaSpace{}%
\AgdaBound{pbk2}\AgdaSpace{}%
\AgdaBound{pbk3}\AgdaSpace{}%
\AgdaBound{pbk4}\AgdaSpace{}%
\AgdaBound{pbk5}\AgdaSpace{}%
\AgdaBound{sig1}\AgdaSpace{}%
\AgdaBound{sig2}\AgdaSpace{}%
\AgdaBound{sig3}\AgdaSpace{}%
\AgdaBound{x}\AgdaSpace{}%
\AgdaSymbol{|}\AgdaSpace{}%
\AgdaInductiveConstructor{true}\AgdaSpace{}%
\AgdaSymbol{|}\AgdaSpace{}%
\AgdaInductiveConstructor{true}\AgdaSpace{}%
\AgdaSymbol{|}\AgdaSpace{}%
\AgdaInductiveConstructor{false}\AgdaSpace{}%
\AgdaSymbol{|}\AgdaSpace{}%
\AgdaInductiveConstructor{true}\AgdaSpace{}%
\AgdaSymbol{=}\AgdaSpace{}%
\AgdaInductiveConstructor{inj₂}\AgdaSpace{}%
\AgdaSymbol{(}\AgdaInductiveConstructor{inj₁}\AgdaSpace{}%
\AgdaSymbol{(}\AgdaInductiveConstructor{conj}\AgdaSpace{}%
\AgdaSymbol{(}\AgdaInductiveConstructor{conj}\AgdaSpace{}%
\AgdaInductiveConstructor{tt}\AgdaSpace{}%
\AgdaInductiveConstructor{tt}\AgdaSymbol{)}\AgdaSpace{}%
\AgdaInductiveConstructor{tt}\AgdaSymbol{))}\<%
\\
\>[0]\AgdaFunction{lemmaHoareTripleStackGeAux'Comb3-5}\AgdaSpace{}%
\AgdaBound{msg₁}\AgdaSpace{}%
\AgdaBound{pbk1}\AgdaSpace{}%
\AgdaBound{pbk2}\AgdaSpace{}%
\AgdaBound{pbk3}\AgdaSpace{}%
\AgdaBound{pbk4}\AgdaSpace{}%
\AgdaBound{pbk5}\AgdaSpace{}%
\AgdaBound{sig1}\AgdaSpace{}%
\AgdaBound{sig2}\AgdaSpace{}%
\AgdaBound{sig3}\AgdaSpace{}%
\AgdaBound{x}\AgdaSpace{}%
\AgdaSymbol{|}\AgdaSpace{}%
\AgdaInductiveConstructor{true}\AgdaSpace{}%
\AgdaSymbol{|}\AgdaSpace{}%
\AgdaInductiveConstructor{true}\AgdaSpace{}%
\AgdaSymbol{|}\AgdaSpace{}%
\AgdaInductiveConstructor{true}\AgdaSpace{}%
\AgdaSymbol{=}\AgdaSpace{}%
\AgdaInductiveConstructor{inj₁}\AgdaSpace{}%
\AgdaSymbol{(}\AgdaInductiveConstructor{conj}\AgdaSpace{}%
\AgdaSymbol{(}\AgdaInductiveConstructor{conj}\AgdaSpace{}%
\AgdaInductiveConstructor{tt}\AgdaSpace{}%
\AgdaInductiveConstructor{tt}\AgdaSymbol{)}\AgdaSpace{}%
\AgdaInductiveConstructor{tt}\AgdaSymbol{)}\<%
\\
\>[0]\<%
\end{code}
} 

\newcommand{\multisigoppushlistofpublickey}{
\begin{code}%
\>[0]\AgdaFunction{opPushList}\AgdaSpace{}%
\AgdaSymbol{:}\AgdaSpace{}%
\AgdaSymbol{(}\AgdaBound{pbkList}\AgdaSpace{}%
\AgdaSymbol{:}\AgdaSpace{}%
\AgdaDatatype{List}\AgdaSpace{}%
\AgdaDatatype{ℕ}\AgdaSymbol{)}\AgdaSpace{}%
\AgdaSymbol{→}\AgdaSpace{}%
\AgdaFunction{BitcoinScriptBasic}\<%
\\
\>[0]\AgdaFunction{opPushList}\AgdaSpace{}%
\AgdaInductiveConstructor{[]}\AgdaSpace{}%
\AgdaSymbol{=}\AgdaSpace{}%
\AgdaInductiveConstructor{[]}\<%
\\
\>[0]\AgdaFunction{opPushList}\AgdaSpace{}%
\AgdaSymbol{(}\AgdaBound{pbk₁}\AgdaSpace{}%
\AgdaOperator{\AgdaInductiveConstructor{∷}}\AgdaSpace{}%
\AgdaBound{pbkList}\AgdaSymbol{)}\AgdaSpace{}%
\AgdaSymbol{=}\AgdaSpace{}%
\AgdaInductiveConstructor{opPush}\AgdaSpace{}%
\AgdaBound{pbk₁}\AgdaSpace{}%
\AgdaOperator{\AgdaInductiveConstructor{∷}}\AgdaSpace{}%
\AgdaFunction{opPushList}\AgdaSpace{}%
\AgdaBound{pbkList}\<%
\end{code}
} 

\AgdaHide{
\begin{code}%
\>[0]\<%
\\
\>[0]\AgdaComment{--\ The\ multisig\ script\ m\ out\ of\ (length\ pbkList)}\<%
\\
\>[0]\AgdaComment{--\ where\ pbkList\ is\ a\ list\ of\ public\ keys.}\<%
\\
\>[0]\<%
\end{code}
} 

\newcommand{\multisigmultiSigscriptmoutoflengthpbkList}{
\begin{code}%
\>[0]\AgdaFunction{multiSigScriptm-nᵇ}%
\>[7661I]\AgdaSymbol{:}\AgdaSpace{}%
\AgdaSymbol{(}\AgdaBound{m}\AgdaSpace{}%
\AgdaSymbol{:}\AgdaSpace{}%
\AgdaDatatype{ℕ}\AgdaSymbol{)(}\AgdaBound{pbkList}\AgdaSpace{}%
\AgdaSymbol{:}\AgdaSpace{}%
\AgdaDatatype{List}\AgdaSpace{}%
\AgdaDatatype{ℕ}\AgdaSymbol{)(}\AgdaBound{m<n}\AgdaSpace{}%
\AgdaSymbol{:}\AgdaSpace{}%
\AgdaBound{m}\AgdaSpace{}%
\AgdaOperator{\AgdaFunction{<}}\AgdaSpace{}%
\AgdaFunction{length}\AgdaSpace{}%
\AgdaBound{pbkList}\AgdaSymbol{)}\<%
\\
\>[7661I][@{}l@{\AgdaIndent{0}}]%
\>[20]\AgdaSymbol{→}\AgdaSpace{}%
\AgdaFunction{BitcoinScriptBasic}\<%
\\
\>[0]\AgdaFunction{multiSigScriptm-nᵇ}\AgdaSpace{}%
\AgdaBound{m}\AgdaSpace{}%
\AgdaBound{pbkList}\AgdaSpace{}%
\AgdaBound{m<n}\AgdaSpace{}%
\AgdaSymbol{=}\<%
\\
\>[0][@{}l@{\AgdaIndent{0}}]%
\>[3]\AgdaInductiveConstructor{opPush}\AgdaSpace{}%
\AgdaBound{m}%
\>[13]\AgdaOperator{\AgdaInductiveConstructor{∷}}\AgdaSpace{}%
\AgdaSymbol{(}\AgdaFunction{opPushList}\AgdaSpace{}%
\AgdaBound{pbkList}\AgdaSpace{}%
\AgdaOperator{\AgdaFunction{\ensuremath{+\!\!+}}}\AgdaSpace{}%
\AgdaSymbol{(}\AgdaInductiveConstructor{opPush}\AgdaSpace{}%
\AgdaSymbol{(}\AgdaFunction{length}\AgdaSpace{}%
\AgdaBound{pbkList}\AgdaSymbol{)}\AgdaSpace{}%
\AgdaOperator{\AgdaInductiveConstructor{∷}}\AgdaSpace{}%
\AgdaOperator{\AgdaFunction{[}}%
\>[68]\AgdaInductiveConstructor{opMultiSig}\AgdaSpace{}%
\AgdaOperator{\AgdaFunction{]}}\AgdaSymbol{))}\<%
\end{code}
} 

\AgdaHide{
\begin{code}%
\>[0]\<%
\\
\\[\AgdaEmptyExtraSkip]%
\>[0]\<%
\end{code}
} 

\newcommand{\multisigcomplexmultisig}{
\begin{code}%
\>[0]\AgdaFunction{multiSigScript2-4ᵇ}\AgdaSpace{}%
\AgdaSymbol{:}\AgdaSpace{}%
\AgdaSymbol{(}\AgdaBound{pbk₁}\AgdaSpace{}%
\AgdaBound{pbk₂}\AgdaSpace{}%
\AgdaBound{pbk₃}\AgdaSpace{}%
\AgdaBound{pbk₄}\AgdaSpace{}%
\AgdaSymbol{:}%
\>[45]\AgdaDatatype{ℕ}\AgdaSymbol{)}\AgdaSpace{}%
\AgdaSymbol{→}\AgdaSpace{}%
\AgdaFunction{BitcoinScriptBasic}\<%
\\
\>[0]\AgdaFunction{multiSigScript2-4ᵇ}\AgdaSpace{}%
\AgdaBound{pbk₁}\AgdaSpace{}%
\AgdaBound{pbk₂}\AgdaSpace{}%
\AgdaBound{pbk₃}\AgdaSpace{}%
\AgdaBound{pbk₄}\AgdaSpace{}%
\AgdaSymbol{=}\<%
\\
\>[0][@{}l@{\AgdaIndent{0}}]%
\>[3]\AgdaFunction{multiSigScriptm-nᵇ}\AgdaSpace{}%
\AgdaNumber{2}%
\>[25]\AgdaSymbol{(}\AgdaBound{pbk₁}\AgdaSpace{}%
\AgdaOperator{\AgdaInductiveConstructor{∷}}\AgdaSpace{}%
\AgdaBound{pbk₂}\AgdaSpace{}%
\AgdaOperator{\AgdaInductiveConstructor{∷}}\AgdaSpace{}%
\AgdaBound{pbk₃}\AgdaSpace{}%
\AgdaOperator{\AgdaInductiveConstructor{∷}}\AgdaSpace{}%
\AgdaOperator{\AgdaFunction{[}}\AgdaSpace{}%
\AgdaBound{pbk₄}\AgdaSpace{}%
\AgdaOperator{\AgdaFunction{]}}\AgdaSymbol{)}\AgdaSpace{}%
\AgdaSymbol{(}\AgdaInductiveConstructor{s≤s}\AgdaSpace{}%
\AgdaSymbol{(}\AgdaInductiveConstructor{s≤s}\AgdaSpace{}%
\AgdaSymbol{(}\AgdaInductiveConstructor{s≤s}\AgdaSpace{}%
\AgdaInductiveConstructor{z≤n}\AgdaSymbol{)))}\<%
\end{code}
} 

\AgdaHide{
\begin{code}%
\>[0]\<%
\\
\>[0]\AgdaComment{\{-}\<%
\\
\>[0]\AgdaComment{-@BEGIN@complexmultisig}\<%
\\
\>[0]\AgdaComment{multiSigScript2-4ᵇ\ :\ (pbk₁\ pbk₂\ pbk₃\ pbk₄\ :\ \ ℕ)\ →\ BitcoinScriptBasic}\<%
\\
\>[0]\AgdaComment{multiSigScript2-4ᵇ\ \ pbk₁\ pbk₂\ pbk₃\ pbk₄\ =\ \ (opPush\ 2)\ ∷\ (opPush\ pbk₁)\ ∷\ \ (opPush\ pbk₂)\ ∷}\<%
\\
\>[0]\AgdaComment{\ \ \ \ \ \ \ \ \ \ \ \ \ \ \ \ \ \ \ \ (opPush\ pbk₃)\ ∷\ \ (opPush\ pbk₄)\ ∷\ (opPush\ 4)\ ∷\ \ [\ \ opMultiSig\ ]}\<%
\\
\>[0]\AgdaComment{-@END}\<%
\\
\>[0]\AgdaComment{-\}}\<%
\\
\\[\AgdaEmptyExtraSkip]%
\>[0]\AgdaFunction{multiSigScript-3-5-b}\AgdaSpace{}%
\AgdaSymbol{:}\AgdaSpace{}%
\AgdaSymbol{(}\AgdaBound{pbk1}\AgdaSpace{}%
\AgdaBound{pbk2}\AgdaSpace{}%
\AgdaBound{pbk3}\AgdaSpace{}%
\AgdaBound{pbk4}\AgdaSpace{}%
\AgdaBound{pbk5}\AgdaSpace{}%
\AgdaSymbol{:}%
\>[52]\AgdaDatatype{ℕ}\AgdaSymbol{)}\AgdaSpace{}%
\AgdaSymbol{→}\AgdaSpace{}%
\AgdaFunction{BitcoinScriptBasic}\<%
\\
\>[0]\AgdaFunction{multiSigScript-3-5-b}\AgdaSpace{}%
\AgdaBound{pbk1}\AgdaSpace{}%
\AgdaBound{pbk2}\AgdaSpace{}%
\AgdaBound{pbk3}\AgdaSpace{}%
\AgdaBound{pbk4}\AgdaSpace{}%
\AgdaBound{pbk5}\AgdaSpace{}%
\AgdaSymbol{=}\<%
\\
\>[0][@{}l@{\AgdaIndent{0}}]%
\>[6]\AgdaSymbol{(}\AgdaInductiveConstructor{opPush}\AgdaSpace{}%
\AgdaNumber{3}\AgdaSymbol{)}\AgdaSpace{}%
\AgdaOperator{\AgdaInductiveConstructor{∷}}\AgdaSpace{}%
\AgdaSymbol{(}\AgdaInductiveConstructor{opPush}\AgdaSpace{}%
\AgdaBound{pbk1}\AgdaSymbol{)}\AgdaSpace{}%
\AgdaOperator{\AgdaInductiveConstructor{∷}}%
\>[36]\AgdaSymbol{(}\AgdaInductiveConstructor{opPush}\AgdaSpace{}%
\AgdaBound{pbk2}\AgdaSymbol{)}\AgdaSpace{}%
\AgdaOperator{\AgdaInductiveConstructor{∷}}%
\>[53]\AgdaSymbol{(}\AgdaInductiveConstructor{opPush}\AgdaSpace{}%
\AgdaBound{pbk3}\AgdaSymbol{)}\AgdaSpace{}%
\AgdaOperator{\AgdaInductiveConstructor{∷}}%
\>[70]\AgdaSymbol{(}\AgdaInductiveConstructor{opPush}\AgdaSpace{}%
\AgdaBound{pbk4}\AgdaSymbol{)}\AgdaSpace{}%
\AgdaOperator{\AgdaInductiveConstructor{∷}}%
\>[87]\AgdaSymbol{(}\AgdaInductiveConstructor{opPush}\AgdaSpace{}%
\AgdaBound{pbk5}\AgdaSymbol{)}\AgdaSpace{}%
\AgdaOperator{\AgdaInductiveConstructor{∷}}\AgdaSpace{}%
\AgdaSymbol{(}\AgdaInductiveConstructor{opPush}\AgdaSpace{}%
\AgdaNumber{5}\AgdaSymbol{)}\AgdaSpace{}%
\AgdaOperator{\AgdaInductiveConstructor{∷}}\AgdaSpace{}%
\AgdaInductiveConstructor{opMultiSig}\AgdaSpace{}%
\AgdaOperator{\AgdaInductiveConstructor{∷}}\AgdaSpace{}%
\AgdaInductiveConstructor{[]}\<%
\\
\\[\AgdaEmptyExtraSkip]%
\>[0]\AgdaComment{--\textbackslash{}multisigcheckTimeScript}\<%
\end{code}
} 

\newcommand{\multisigcheckTimeScript}{
\begin{code}%
\>[0]\AgdaFunction{checkTimeScriptᵇ}\AgdaSpace{}%
\AgdaSymbol{:}\AgdaSpace{}%
\AgdaSymbol{(}\AgdaBound{time₁}\AgdaSpace{}%
\AgdaSymbol{:}\AgdaSpace{}%
\AgdaFunction{Time}\AgdaSymbol{)}\AgdaSpace{}%
\AgdaSymbol{→}\AgdaSpace{}%
\AgdaFunction{BitcoinScriptBasic}\<%
\\
\>[0]\AgdaFunction{checkTimeScriptᵇ}\AgdaSpace{}%
\AgdaBound{time₁}\AgdaSpace{}%
\AgdaSymbol{=}\AgdaSpace{}%
\AgdaSymbol{(}\AgdaInductiveConstructor{opPush}\AgdaSpace{}%
\AgdaBound{time₁}\AgdaSymbol{)}\AgdaSpace{}%
\AgdaOperator{\AgdaInductiveConstructor{∷}}\AgdaSpace{}%
\AgdaInductiveConstructor{opCHECKLOCKTIMEVERIFY}\AgdaSpace{}%
\AgdaOperator{\AgdaInductiveConstructor{∷}}\AgdaSpace{}%
\AgdaOperator{\AgdaFunction{[}}\AgdaSpace{}%
\AgdaInductiveConstructor{opDrop}%
\>[76]\AgdaOperator{\AgdaFunction{]}}\<%
\end{code}
} 

\AgdaHide{
\begin{code}%
\>[0]\<%
\\
\\[\AgdaEmptyExtraSkip]%
\\[\AgdaEmptyExtraSkip]%
\\[\AgdaEmptyExtraSkip]%
\\[\AgdaEmptyExtraSkip]%
\\[\AgdaEmptyExtraSkip]%
\\[\AgdaEmptyExtraSkip]%
\>[0]\AgdaComment{\{-\ Next\ steps\ -\}}\<%
\\
\\[\AgdaEmptyExtraSkip]%
\\[\AgdaEmptyExtraSkip]%
\\[\AgdaEmptyExtraSkip]%
\>[0]\AgdaComment{--new}\<%
\\
\\[\AgdaEmptyExtraSkip]%
\>[0]\AgdaFunction{lemmaHoareTripleStackGeAux'5}%
\>[7763I]\AgdaSymbol{:}%
\>[7764I]\AgdaSymbol{(}\AgdaBound{msg}\AgdaSpace{}%
\AgdaSymbol{:}\AgdaSpace{}%
\AgdaDatatype{Msg}\AgdaSymbol{)}\<%
\\
\>[.][@{}l@{}]\<[7764I]%
\>[31]\AgdaSymbol{(}\AgdaBound{pbk1}\AgdaSpace{}%
\AgdaBound{pbk2}\AgdaSpace{}%
\AgdaBound{pbk3}\AgdaSpace{}%
\AgdaBound{sig1}\AgdaSpace{}%
\AgdaBound{sig3}\AgdaSpace{}%
\AgdaSymbol{:}\AgdaSpace{}%
\AgdaDatatype{ℕ}\AgdaSymbol{)}\<%
\\
\>[31]\AgdaSymbol{→}\AgdaSpace{}%
\AgdaFunction{True}\AgdaSpace{}%
\AgdaSymbol{(}\AgdaFunction{isSigned}%
\>[49]\AgdaBound{msg}\AgdaSpace{}%
\AgdaBound{sig1}\AgdaSpace{}%
\AgdaBound{pbk1}\AgdaSymbol{)}\<%
\\
\>[7763I][@{}l@{\AgdaIndent{0}}]%
\>[30]\AgdaSymbol{→}%
\>[33]\AgdaFunction{True}\AgdaSpace{}%
\AgdaSymbol{(}\AgdaFunction{isSigned}%
\>[49]\AgdaBound{msg}\AgdaSpace{}%
\AgdaBound{sig3}\AgdaSpace{}%
\AgdaBound{pbk3}\AgdaSymbol{)}\<%
\\
\>[30][@{}l@{\AgdaIndent{0}}]%
\>[31]\AgdaSymbol{→}\AgdaSpace{}%
\AgdaFunction{True}\AgdaSpace{}%
\AgdaSymbol{(}\AgdaFunction{compareSigsMultiSig}\AgdaSpace{}%
\AgdaBound{msg}\AgdaSpace{}%
\AgdaSymbol{(}\AgdaSpace{}%
\AgdaBound{sig1}\AgdaSpace{}%
\AgdaOperator{\AgdaInductiveConstructor{∷}}\AgdaSpace{}%
\AgdaBound{sig3}\AgdaSpace{}%
\AgdaOperator{\AgdaInductiveConstructor{∷}}\AgdaSpace{}%
\AgdaInductiveConstructor{[]}\AgdaSymbol{)}\AgdaSpace{}%
\AgdaSymbol{(}\AgdaBound{pbk1}\AgdaSpace{}%
\AgdaOperator{\AgdaInductiveConstructor{∷}}\AgdaSpace{}%
\AgdaBound{pbk2}\AgdaSpace{}%
\AgdaOperator{\AgdaInductiveConstructor{∷}}\AgdaSpace{}%
\AgdaBound{pbk3}\AgdaSpace{}%
\AgdaOperator{\AgdaInductiveConstructor{∷}}\AgdaSpace{}%
\AgdaInductiveConstructor{[]}\AgdaSymbol{))}\<%
\\
\>[0]\AgdaFunction{lemmaHoareTripleStackGeAux'5}\AgdaSpace{}%
\AgdaBound{msg}\AgdaSpace{}%
\AgdaBound{pbk1}\AgdaSpace{}%
\AgdaBound{pbk2}\AgdaSpace{}%
\AgdaBound{pbk3}\AgdaSpace{}%
\AgdaBound{sig1}\AgdaSpace{}%
\AgdaBound{sig3}\AgdaSpace{}%
\AgdaBound{x}\AgdaSpace{}%
\AgdaBound{x₁}%
\>[64]\AgdaKeyword{with}\AgdaSpace{}%
\AgdaSymbol{(}\AgdaFunction{isSigned}%
\>[80]\AgdaBound{msg}\AgdaSpace{}%
\AgdaBound{sig1}\AgdaSpace{}%
\AgdaBound{pbk1}\AgdaSymbol{)}\<%
\\
\>[0]\AgdaFunction{lemmaHoareTripleStackGeAux'5}\AgdaSpace{}%
\AgdaBound{msg}\AgdaSpace{}%
\AgdaBound{pbk1}\AgdaSpace{}%
\AgdaBound{pbk2}\AgdaSpace{}%
\AgdaBound{pbk3}\AgdaSpace{}%
\AgdaBound{sig1}\AgdaSpace{}%
\AgdaBound{sig3}\AgdaSpace{}%
\AgdaBound{x}\AgdaSpace{}%
\AgdaBound{x₁}\AgdaSpace{}%
\AgdaSymbol{|}\AgdaSpace{}%
\AgdaInductiveConstructor{true}\AgdaSpace{}%
\AgdaKeyword{with}\AgdaSpace{}%
\AgdaSymbol{(}\AgdaFunction{isSigned}%
\>[86]\AgdaBound{msg}\AgdaSpace{}%
\AgdaBound{sig3}\AgdaSpace{}%
\AgdaBound{pbk2}\AgdaSymbol{)}\<%
\\
\>[0]\AgdaFunction{lemmaHoareTripleStackGeAux'5}\AgdaSpace{}%
\AgdaBound{msg}\AgdaSpace{}%
\AgdaBound{pbk1}\AgdaSpace{}%
\AgdaBound{pbk2}\AgdaSpace{}%
\AgdaBound{pbk3}\AgdaSpace{}%
\AgdaBound{sig1}\AgdaSpace{}%
\AgdaBound{sig3}\AgdaSpace{}%
\AgdaBound{x}\AgdaSpace{}%
\AgdaBound{x₁}\AgdaSpace{}%
\AgdaSymbol{|}\AgdaSpace{}%
\AgdaInductiveConstructor{true}\AgdaSpace{}%
\AgdaSymbol{|}\AgdaSpace{}%
\AgdaInductiveConstructor{false}\AgdaSpace{}%
\AgdaKeyword{with}\AgdaSpace{}%
\AgdaSymbol{(}\AgdaFunction{isSigned}%
\>[94]\AgdaBound{msg}\AgdaSpace{}%
\AgdaBound{sig3}\AgdaSpace{}%
\AgdaBound{pbk3}\AgdaSymbol{)}\<%
\\
\>[0]\AgdaFunction{lemmaHoareTripleStackGeAux'5}\AgdaSpace{}%
\AgdaBound{msg}\AgdaSpace{}%
\AgdaBound{pbk1}\AgdaSpace{}%
\AgdaBound{pbk2}\AgdaSpace{}%
\AgdaBound{pbk3}\AgdaSpace{}%
\AgdaBound{sig1}\AgdaSpace{}%
\AgdaBound{sig3}\AgdaSpace{}%
\AgdaBound{x}\AgdaSpace{}%
\AgdaBound{x₁}\AgdaSpace{}%
\AgdaSymbol{|}\AgdaSpace{}%
\AgdaInductiveConstructor{true}\AgdaSpace{}%
\AgdaSymbol{|}\AgdaSpace{}%
\AgdaInductiveConstructor{false}\AgdaSpace{}%
\AgdaSymbol{|}\AgdaSpace{}%
\AgdaInductiveConstructor{true}\AgdaSpace{}%
\AgdaSymbol{=}\AgdaSpace{}%
\AgdaInductiveConstructor{tt}\<%
\\
\>[0]\AgdaFunction{lemmaHoareTripleStackGeAux'5}\AgdaSpace{}%
\AgdaBound{msg}\AgdaSpace{}%
\AgdaBound{pbk1}\AgdaSpace{}%
\AgdaBound{pbk2}\AgdaSpace{}%
\AgdaBound{pbk3}\AgdaSpace{}%
\AgdaBound{sig1}\AgdaSpace{}%
\AgdaBound{sig3}\AgdaSpace{}%
\AgdaBound{x}\AgdaSpace{}%
\AgdaBound{x₁}\AgdaSpace{}%
\AgdaSymbol{|}\AgdaSpace{}%
\AgdaInductiveConstructor{true}\AgdaSpace{}%
\AgdaSymbol{|}\AgdaSpace{}%
\AgdaInductiveConstructor{true}\AgdaSpace{}%
\AgdaSymbol{=}\AgdaSpace{}%
\AgdaInductiveConstructor{tt}\<%
\\
\\[\AgdaEmptyExtraSkip]%
\\[\AgdaEmptyExtraSkip]%
\\[\AgdaEmptyExtraSkip]%
\\[\AgdaEmptyExtraSkip]%
\\[\AgdaEmptyExtraSkip]%
\\[\AgdaEmptyExtraSkip]%
\\[\AgdaEmptyExtraSkip]%
\>[0]\AgdaComment{--\textbackslash{}multisigtimeCheckPreCond}\<%
\end{code}
} 

\newcommand{\multisigtimeCheckPreCond}{
\begin{code}%
\>[0]\AgdaFunction{timeCheckPreCond}\AgdaSpace{}%
\AgdaSymbol{:}\AgdaSpace{}%
\AgdaSymbol{(}\AgdaBound{time₁}\AgdaSpace{}%
\AgdaSymbol{:}\AgdaSpace{}%
\AgdaFunction{Time}\AgdaSymbol{)}\AgdaSpace{}%
\AgdaSymbol{→}\AgdaSpace{}%
\AgdaFunction{StackPredicate}\<%
\\
\>[0]\AgdaFunction{timeCheckPreCond}\AgdaSpace{}%
\AgdaBound{time₁}\AgdaSpace{}%
\AgdaBound{time₂}\AgdaSpace{}%
\AgdaBound{msg}\AgdaSpace{}%
\AgdaBound{stack₁}\AgdaSpace{}%
\AgdaSymbol{=}\AgdaSpace{}%
\AgdaBound{time₁}\AgdaSpace{}%
\AgdaOperator{\AgdaFunction{≤}}\AgdaSpace{}%
\AgdaBound{time₂}\<%
\end{code}
} 

\AgdaHide{
\begin{code}%
\>[0]\<%
\end{code}
} 


\AgdaHide{
\begin{code}%
\>[0]\AgdaKeyword{module}\AgdaSpace{}%
\AgdaModule{stack}\AgdaSpace{}%
\AgdaKeyword{where}\<%
\\
\\[\AgdaEmptyExtraSkip]%
\>[0]\AgdaKeyword{open}\AgdaSpace{}%
\AgdaKeyword{import}\AgdaSpace{}%
\AgdaModule{Data.Nat}%
\>[22]\AgdaKeyword{hiding}\AgdaSpace{}%
\AgdaSymbol{(}\AgdaOperator{\AgdaDatatype{\AgdaUnderscore{}≤\AgdaUnderscore{}}}\AgdaSymbol{)}\<%
\\
\>[0]\AgdaKeyword{open}\AgdaSpace{}%
\AgdaKeyword{import}\AgdaSpace{}%
\AgdaModule{Data.List}\AgdaSpace{}%
\AgdaKeyword{hiding}\AgdaSpace{}%
\AgdaSymbol{(}\AgdaOperator{\AgdaFunction{\AgdaUnderscore{}\ensuremath{+\!\!+}\AgdaUnderscore{}}}\AgdaSymbol{)}\<%
\\
\>[0]\AgdaKeyword{open}\AgdaSpace{}%
\AgdaKeyword{import}\AgdaSpace{}%
\AgdaModule{Data.Unit}%
\>[23]\AgdaKeyword{hiding}\AgdaSpace{}%
\AgdaSymbol{(}\AgdaOperator{\AgdaRecord{\AgdaUnderscore{}≤\AgdaUnderscore{}}}\AgdaSymbol{)}\<%
\\
\>[0]\AgdaKeyword{open}\AgdaSpace{}%
\AgdaKeyword{import}\AgdaSpace{}%
\AgdaModule{Data.Empty}\<%
\\
\>[0]\AgdaKeyword{open}\AgdaSpace{}%
\AgdaKeyword{import}\AgdaSpace{}%
\AgdaModule{Data.Maybe}\<%
\\
\>[0]\AgdaKeyword{open}\AgdaSpace{}%
\AgdaKeyword{import}\AgdaSpace{}%
\AgdaModule{Data.Bool}%
\>[23]\AgdaKeyword{hiding}\AgdaSpace{}%
\AgdaSymbol{(}\AgdaOperator{\AgdaDatatype{\AgdaUnderscore{}≤\AgdaUnderscore{}}}\AgdaSpace{}%
\AgdaSymbol{;}\AgdaSpace{}%
\AgdaOperator{\AgdaFunction{if\AgdaUnderscore{}then\AgdaUnderscore{}else\AgdaUnderscore{}}}\AgdaSpace{}%
\AgdaSymbol{)}\AgdaSpace{}%
\AgdaKeyword{renaming}\AgdaSpace{}%
\AgdaSymbol{(}\AgdaOperator{\AgdaFunction{\AgdaUnderscore{}∧\AgdaUnderscore{}}}\AgdaSpace{}%
\AgdaSymbol{to}\AgdaSpace{}%
\AgdaOperator{\AgdaFunction{\AgdaUnderscore{}∧b\AgdaUnderscore{}}}\AgdaSpace{}%
\AgdaSymbol{;}\AgdaSpace{}%
\AgdaOperator{\AgdaFunction{\AgdaUnderscore{}∨\AgdaUnderscore{}}}\AgdaSpace{}%
\AgdaSymbol{to}\AgdaSpace{}%
\AgdaOperator{\AgdaFunction{\AgdaUnderscore{}∨b\AgdaUnderscore{}}}\AgdaSpace{}%
\AgdaSymbol{;}\AgdaSpace{}%
\AgdaFunction{T}\AgdaSpace{}%
\AgdaSymbol{to}\AgdaSpace{}%
\AgdaFunction{True}\AgdaSymbol{)}\<%
\\
\>[0]\AgdaKeyword{open}\AgdaSpace{}%
\AgdaKeyword{import}\AgdaSpace{}%
\AgdaModule{Data.Bool.Base}\AgdaSpace{}%
\AgdaKeyword{hiding}\AgdaSpace{}%
\AgdaSymbol{(}\AgdaOperator{\AgdaDatatype{\AgdaUnderscore{}≤\AgdaUnderscore{}}}\AgdaSpace{}%
\AgdaSymbol{;}\AgdaSpace{}%
\AgdaOperator{\AgdaFunction{if\AgdaUnderscore{}then\AgdaUnderscore{}else\AgdaUnderscore{}}}\AgdaSpace{}%
\AgdaSymbol{)}\AgdaSpace{}%
\AgdaKeyword{renaming}\AgdaSpace{}%
\AgdaSymbol{(}\AgdaOperator{\AgdaFunction{\AgdaUnderscore{}∧\AgdaUnderscore{}}}\AgdaSpace{}%
\AgdaSymbol{to}\AgdaSpace{}%
\AgdaOperator{\AgdaFunction{\AgdaUnderscore{}∧b\AgdaUnderscore{}}}\AgdaSpace{}%
\AgdaSymbol{;}\AgdaSpace{}%
\AgdaOperator{\AgdaFunction{\AgdaUnderscore{}∨\AgdaUnderscore{}}}\AgdaSpace{}%
\AgdaSymbol{to}\AgdaSpace{}%
\AgdaOperator{\AgdaFunction{\AgdaUnderscore{}∨b\AgdaUnderscore{}}}\AgdaSpace{}%
\AgdaSymbol{;}\AgdaSpace{}%
\AgdaFunction{T}\AgdaSpace{}%
\AgdaSymbol{to}\AgdaSpace{}%
\AgdaFunction{True}\AgdaSymbol{)}\<%
\\
\>[0]\AgdaKeyword{open}\AgdaSpace{}%
\AgdaKeyword{import}\AgdaSpace{}%
\AgdaModule{Data.Product}\AgdaSpace{}%
\AgdaKeyword{renaming}\AgdaSpace{}%
\AgdaSymbol{(}\AgdaOperator{\AgdaInductiveConstructor{\AgdaUnderscore{},\AgdaUnderscore{}}}\AgdaSpace{}%
\AgdaSymbol{to}\AgdaSpace{}%
\AgdaOperator{\AgdaInductiveConstructor{\AgdaUnderscore{},,\AgdaUnderscore{}}}\AgdaSpace{}%
\AgdaSymbol{)}\<%
\\
\>[0]\AgdaKeyword{open}\AgdaSpace{}%
\AgdaKeyword{import}\AgdaSpace{}%
\AgdaModule{Data.Nat.Base}\AgdaSpace{}%
\AgdaKeyword{hiding}\AgdaSpace{}%
\AgdaSymbol{(}\AgdaOperator{\AgdaDatatype{\AgdaUnderscore{}≤\AgdaUnderscore{}}}\AgdaSymbol{)}\<%
\\
\>[0]\AgdaKeyword{open}\AgdaSpace{}%
\AgdaKeyword{import}\AgdaSpace{}%
\AgdaModule{Data.List.NonEmpty}\AgdaSpace{}%
\AgdaKeyword{hiding}\AgdaSpace{}%
\AgdaSymbol{(}\AgdaField{head}\AgdaSymbol{)}\<%
\\
\\[\AgdaEmptyExtraSkip]%
\>[0]\AgdaKeyword{open}\AgdaSpace{}%
\AgdaKeyword{import}\AgdaSpace{}%
\AgdaModule{libraries.listLib}\<%
\\
\>[0]\AgdaKeyword{open}\AgdaSpace{}%
\AgdaKeyword{import}\AgdaSpace{}%
\AgdaModule{libraries.natLib}\<%
\\
\>[0]\AgdaKeyword{open}\AgdaSpace{}%
\AgdaKeyword{import}\AgdaSpace{}%
\AgdaModule{libraries.boolLib}\<%
\\
\\[\AgdaEmptyExtraSkip]%
\>[0]\AgdaKeyword{open}\AgdaSpace{}%
\AgdaKeyword{import}\AgdaSpace{}%
\AgdaModule{libraries.andLib}\<%
\\
\>[0]\AgdaKeyword{open}\AgdaSpace{}%
\AgdaKeyword{import}\AgdaSpace{}%
\AgdaModule{libraries.miscLib}\<%
\\
\>[0]\AgdaKeyword{open}\AgdaSpace{}%
\AgdaKeyword{import}\AgdaSpace{}%
\AgdaModule{libraries.maybeLib}\<%
\\
\\[\AgdaEmptyExtraSkip]%
\>[0]\AgdaKeyword{open}\AgdaSpace{}%
\AgdaKeyword{import}\AgdaSpace{}%
\AgdaModule{basicBitcoinDataType}\<%
\\
\>[0]\<%
\end{code}
} 

\newcommand{\stackexamplestackdef}{
\begin{code}%
\>[0]\AgdaFunction{Stack}\AgdaSpace{}%
\AgdaSymbol{:}\AgdaSpace{}%
\AgdaPrimitive{Set}\<%
\\
\>[0]\AgdaFunction{Stack}\AgdaSpace{}%
\AgdaSymbol{=}\AgdaSpace{}%
\AgdaDatatype{List}\AgdaSpace{}%
\AgdaDatatype{ℕ}\<%
\end{code}
} 

\AgdaHide{
\begin{code}%
\>[0]\<%
\\
\>[0]\AgdaFunction{stackHasSingletonTop}%
\>[22]\AgdaSymbol{:}\AgdaSpace{}%
\AgdaDatatype{ℕ}\AgdaSpace{}%
\AgdaSymbol{→}%
\>[30]\AgdaDatatype{Maybe}\AgdaSpace{}%
\AgdaFunction{Stack}\AgdaSpace{}%
\AgdaSymbol{→}%
\>[45]\AgdaDatatype{Bool}\<%
\\
\>[0]\AgdaFunction{stackHasSingletonTop}\AgdaSpace{}%
\AgdaBound{l}\AgdaSpace{}%
\AgdaInductiveConstructor{nothing}\AgdaSpace{}%
\AgdaSymbol{=}\AgdaSpace{}%
\AgdaInductiveConstructor{false}\<%
\\
\>[0]\AgdaFunction{stackHasSingletonTop}\AgdaSpace{}%
\AgdaBound{l}\AgdaSpace{}%
\AgdaSymbol{(}\AgdaInductiveConstructor{just}\AgdaSpace{}%
\AgdaInductiveConstructor{[]}\AgdaSymbol{)}\AgdaSpace{}%
\AgdaSymbol{=}\AgdaSpace{}%
\AgdaInductiveConstructor{false}\<%
\\
\>[0]\AgdaFunction{stackHasSingletonTop}\AgdaSpace{}%
\AgdaBound{l}\AgdaSpace{}%
\AgdaSymbol{(}\AgdaInductiveConstructor{just}\AgdaSpace{}%
\AgdaSymbol{(}\AgdaBound{z}\AgdaSpace{}%
\AgdaOperator{\AgdaInductiveConstructor{∷}}\AgdaSpace{}%
\AgdaBound{y}\AgdaSymbol{))}\AgdaSpace{}%
\AgdaSymbol{=}%
\>[41]\AgdaBound{l}%
\>[44]\AgdaOperator{\AgdaFunction{==b}}\AgdaSpace{}%
\AgdaBound{z}\<%
\\
\\[\AgdaEmptyExtraSkip]%
\\[\AgdaEmptyExtraSkip]%
\>[0]\AgdaFunction{stackHasTop}%
\>[13]\AgdaSymbol{:}%
\>[17]\AgdaDatatype{List}\AgdaSpace{}%
\AgdaDatatype{ℕ}\AgdaSpace{}%
\AgdaSymbol{→}%
\>[27]\AgdaDatatype{Maybe}%
\>[34]\AgdaFunction{Stack}\AgdaSpace{}%
\AgdaSymbol{→}\AgdaSpace{}%
\AgdaPrimitive{Set}\<%
\\
\>[0]\AgdaFunction{stackHasTop}\AgdaSpace{}%
\AgdaInductiveConstructor{[]}\AgdaSpace{}%
\AgdaBound{m}\AgdaSpace{}%
\AgdaSymbol{=}\AgdaSpace{}%
\AgdaRecord{⊤}\<%
\\
\>[0]\AgdaFunction{stackHasTop}\AgdaSpace{}%
\AgdaSymbol{(}\AgdaBound{y}\AgdaSpace{}%
\AgdaOperator{\AgdaInductiveConstructor{∷}}\AgdaSpace{}%
\AgdaBound{n}\AgdaSymbol{)}\AgdaSpace{}%
\AgdaBound{m}\AgdaSpace{}%
\AgdaSymbol{=}\AgdaSpace{}%
\AgdaFunction{True}\AgdaSymbol{(}\AgdaFunction{stackHasSingletonTop}\AgdaSpace{}%
\AgdaBound{y}\AgdaSpace{}%
\AgdaBound{m}\AgdaSymbol{)}\<%
\\
\\[\AgdaEmptyExtraSkip]%
\>[0]\AgdaFunction{stackAuxFunction}\AgdaSpace{}%
\AgdaSymbol{:}%
\>[20]\AgdaFunction{Stack}\AgdaSpace{}%
\AgdaSymbol{→}\AgdaSpace{}%
\AgdaDatatype{Bool}\AgdaSpace{}%
\AgdaSymbol{→}\AgdaSpace{}%
\AgdaDatatype{Maybe}\AgdaSpace{}%
\AgdaFunction{Stack}\<%
\\
\>[0]\AgdaFunction{stackAuxFunction}\AgdaSpace{}%
\AgdaBound{s}\AgdaSpace{}%
\AgdaBound{b}\AgdaSpace{}%
\AgdaSymbol{=}\AgdaSpace{}%
\AgdaInductiveConstructor{just}\AgdaSpace{}%
\AgdaSymbol{(}\AgdaFunction{boolToNat}\AgdaSpace{}%
\AgdaBound{b}\AgdaSpace{}%
\AgdaOperator{\AgdaInductiveConstructor{∷}}\AgdaSpace{}%
\AgdaBound{s}\AgdaSymbol{)}\<%
\\
\\[\AgdaEmptyExtraSkip]%
\\[\AgdaEmptyExtraSkip]%
\>[0]\AgdaComment{--\ Stack\ transformer}\<%
\\
\>[0]\AgdaFunction{StackTransformer}\AgdaSpace{}%
\AgdaSymbol{:}\AgdaSpace{}%
\AgdaPrimitive{Set}\<%
\\
\>[0]\AgdaFunction{StackTransformer}\AgdaSpace{}%
\AgdaSymbol{=}\AgdaSpace{}%
\AgdaFunction{Time}\AgdaSpace{}%
\AgdaSymbol{→}\AgdaSpace{}%
\AgdaDatatype{Msg}\AgdaSpace{}%
\AgdaSymbol{→}\AgdaSpace{}%
\AgdaFunction{Stack}\AgdaSpace{}%
\AgdaSymbol{→}\AgdaSpace{}%
\AgdaDatatype{Maybe}\AgdaSpace{}%
\AgdaFunction{Stack}\<%
\\
\\[\AgdaEmptyExtraSkip]%
\>[0]\AgdaComment{--\ function\ that\ checking\ if\ the\ stack\ is\ empty\ or\ the\ top\ element\ is\ false}\<%
\\
\>[0]\AgdaFunction{checkStackAux}\AgdaSpace{}%
\AgdaSymbol{:}\AgdaSpace{}%
\AgdaFunction{Stack}\AgdaSpace{}%
\AgdaSymbol{→}\AgdaSpace{}%
\AgdaDatatype{Bool}\<%
\\
\>[0]\AgdaFunction{checkStackAux}\AgdaSpace{}%
\AgdaInductiveConstructor{[]}%
\>[18]\AgdaSymbol{=}\AgdaSpace{}%
\AgdaInductiveConstructor{false}\<%
\\
\>[0]\AgdaFunction{checkStackAux}\AgdaSpace{}%
\AgdaSymbol{(}\AgdaInductiveConstructor{zero}\AgdaSpace{}%
\AgdaOperator{\AgdaInductiveConstructor{∷}}\AgdaSpace{}%
\AgdaBound{bitcoinStack₁}\AgdaSymbol{)}%
\>[38]\AgdaSymbol{=}\AgdaSpace{}%
\AgdaInductiveConstructor{false}\<%
\\
\>[0]\AgdaFunction{checkStackAux}\AgdaSpace{}%
\AgdaSymbol{(}\AgdaInductiveConstructor{suc}\AgdaSpace{}%
\AgdaBound{x}\AgdaSpace{}%
\AgdaOperator{\AgdaInductiveConstructor{∷}}\AgdaSpace{}%
\AgdaBound{bitcoinStack₁}\AgdaSymbol{)}%
\>[39]\AgdaSymbol{=}\AgdaSpace{}%
\AgdaInductiveConstructor{true}\<%
\\
\\[\AgdaEmptyExtraSkip]%
\>[0]\AgdaComment{--\ lifting\ the\ checkStackAux\ to\ Maybe\ StackIfStack\ data\ type}\<%
\\
\>[0]\AgdaFunction{checkStack}\AgdaSpace{}%
\AgdaSymbol{:}%
\>[14]\AgdaDatatype{Maybe}\AgdaSpace{}%
\AgdaFunction{Stack}%
\>[28]\AgdaSymbol{→}\AgdaSpace{}%
\AgdaDatatype{Bool}\<%
\\
\>[0]\AgdaFunction{checkStack}\AgdaSpace{}%
\AgdaInductiveConstructor{nothing}\AgdaSpace{}%
\AgdaSymbol{=}\AgdaSpace{}%
\AgdaInductiveConstructor{false}\<%
\\
\>[0]\AgdaFunction{checkStack}\AgdaSpace{}%
\AgdaSymbol{(}\AgdaInductiveConstructor{just}\AgdaSpace{}%
\AgdaBound{x}\AgdaSymbol{)}\AgdaSpace{}%
\AgdaSymbol{=}\AgdaSpace{}%
\AgdaFunction{checkStackAux}\AgdaSpace{}%
\AgdaBound{x}\<%
\end{code}
} 


\AgdaHide{
\begin{code}%
\>[0]\AgdaComment{--\textbackslash{}stackPredicate}\<%
\\
\>[0]\AgdaKeyword{module}\AgdaSpace{}%
\AgdaModule{stackPredicate}\AgdaSpace{}%
\AgdaKeyword{where}\<%
\\
\\[\AgdaEmptyExtraSkip]%
\>[0]\AgdaKeyword{open}\AgdaSpace{}%
\AgdaKeyword{import}\AgdaSpace{}%
\AgdaModule{Data.Nat}%
\>[22]\AgdaKeyword{hiding}\AgdaSpace{}%
\AgdaSymbol{(}\AgdaOperator{\AgdaDatatype{\AgdaUnderscore{}≤\AgdaUnderscore{}}}\AgdaSymbol{)}\<%
\\
\>[0]\AgdaKeyword{open}\AgdaSpace{}%
\AgdaKeyword{import}\AgdaSpace{}%
\AgdaModule{Data.List}\AgdaSpace{}%
\AgdaKeyword{hiding}\AgdaSpace{}%
\AgdaSymbol{(}\AgdaOperator{\AgdaFunction{\AgdaUnderscore{}\ensuremath{+\!\!+}\AgdaUnderscore{}}}\AgdaSymbol{)}\<%
\\
\>[0]\AgdaKeyword{open}\AgdaSpace{}%
\AgdaKeyword{import}\AgdaSpace{}%
\AgdaModule{Data.Unit}%
\>[23]\AgdaKeyword{hiding}\AgdaSpace{}%
\AgdaSymbol{(}\AgdaOperator{\AgdaRecord{\AgdaUnderscore{}≤\AgdaUnderscore{}}}\AgdaSymbol{)}\<%
\\
\>[0]\AgdaKeyword{open}\AgdaSpace{}%
\AgdaKeyword{import}\AgdaSpace{}%
\AgdaModule{Data.Empty}\<%
\\
\>[0]\AgdaKeyword{open}\AgdaSpace{}%
\AgdaKeyword{import}\AgdaSpace{}%
\AgdaModule{Data.Sum}\<%
\\
\>[0]\AgdaKeyword{open}\AgdaSpace{}%
\AgdaKeyword{import}\AgdaSpace{}%
\AgdaModule{Data.Maybe}\<%
\\
\>[0]\AgdaKeyword{open}\AgdaSpace{}%
\AgdaKeyword{import}\AgdaSpace{}%
\AgdaModule{Data.Bool}%
\>[23]\AgdaKeyword{hiding}\AgdaSpace{}%
\AgdaSymbol{(}\AgdaOperator{\AgdaDatatype{\AgdaUnderscore{}≤\AgdaUnderscore{}}}\AgdaSpace{}%
\AgdaSymbol{;}\AgdaSpace{}%
\AgdaOperator{\AgdaFunction{if\AgdaUnderscore{}then\AgdaUnderscore{}else\AgdaUnderscore{}}}\AgdaSpace{}%
\AgdaSymbol{)}\AgdaSpace{}%
\AgdaKeyword{renaming}\AgdaSpace{}%
\AgdaSymbol{(}\AgdaOperator{\AgdaFunction{\AgdaUnderscore{}∧\AgdaUnderscore{}}}\AgdaSpace{}%
\AgdaSymbol{to}\AgdaSpace{}%
\AgdaOperator{\AgdaFunction{\AgdaUnderscore{}∧b\AgdaUnderscore{}}}\AgdaSpace{}%
\AgdaSymbol{;}\AgdaSpace{}%
\AgdaOperator{\AgdaFunction{\AgdaUnderscore{}∨\AgdaUnderscore{}}}\AgdaSpace{}%
\AgdaSymbol{to}\AgdaSpace{}%
\AgdaOperator{\AgdaFunction{\AgdaUnderscore{}∨b\AgdaUnderscore{}}}\AgdaSpace{}%
\AgdaSymbol{;}\AgdaSpace{}%
\AgdaFunction{T}\AgdaSpace{}%
\AgdaSymbol{to}\AgdaSpace{}%
\AgdaFunction{True}\AgdaSymbol{)}\<%
\\
\>[0]\AgdaKeyword{open}\AgdaSpace{}%
\AgdaKeyword{import}\AgdaSpace{}%
\AgdaModule{Data.Bool.Base}\AgdaSpace{}%
\AgdaKeyword{hiding}\AgdaSpace{}%
\AgdaSymbol{(}\AgdaOperator{\AgdaDatatype{\AgdaUnderscore{}≤\AgdaUnderscore{}}}\AgdaSpace{}%
\AgdaSymbol{;}\AgdaSpace{}%
\AgdaOperator{\AgdaFunction{if\AgdaUnderscore{}then\AgdaUnderscore{}else\AgdaUnderscore{}}}\AgdaSpace{}%
\AgdaSymbol{)}\AgdaSpace{}%
\AgdaKeyword{renaming}\AgdaSpace{}%
\AgdaSymbol{(}\AgdaOperator{\AgdaFunction{\AgdaUnderscore{}∧\AgdaUnderscore{}}}\AgdaSpace{}%
\AgdaSymbol{to}\AgdaSpace{}%
\AgdaOperator{\AgdaFunction{\AgdaUnderscore{}∧b\AgdaUnderscore{}}}\AgdaSpace{}%
\AgdaSymbol{;}\AgdaSpace{}%
\AgdaOperator{\AgdaFunction{\AgdaUnderscore{}∨\AgdaUnderscore{}}}\AgdaSpace{}%
\AgdaSymbol{to}\AgdaSpace{}%
\AgdaOperator{\AgdaFunction{\AgdaUnderscore{}∨b\AgdaUnderscore{}}}\AgdaSpace{}%
\AgdaSymbol{;}\AgdaSpace{}%
\AgdaFunction{T}\AgdaSpace{}%
\AgdaSymbol{to}\AgdaSpace{}%
\AgdaFunction{True}\AgdaSymbol{)}\<%
\\
\>[0]\AgdaKeyword{open}\AgdaSpace{}%
\AgdaKeyword{import}\AgdaSpace{}%
\AgdaModule{Data.Product}\AgdaSpace{}%
\AgdaKeyword{renaming}\AgdaSpace{}%
\AgdaSymbol{(}\AgdaOperator{\AgdaInductiveConstructor{\AgdaUnderscore{},\AgdaUnderscore{}}}\AgdaSpace{}%
\AgdaSymbol{to}\AgdaSpace{}%
\AgdaOperator{\AgdaInductiveConstructor{\AgdaUnderscore{},,\AgdaUnderscore{}}}\AgdaSpace{}%
\AgdaSymbol{)}\<%
\\
\>[0]\AgdaKeyword{open}\AgdaSpace{}%
\AgdaKeyword{import}\AgdaSpace{}%
\AgdaModule{Data.Nat.Base}\AgdaSpace{}%
\AgdaKeyword{hiding}\AgdaSpace{}%
\AgdaSymbol{(}\AgdaOperator{\AgdaDatatype{\AgdaUnderscore{}≤\AgdaUnderscore{}}}\AgdaSymbol{)}\<%
\\
\>[0]\AgdaKeyword{open}\AgdaSpace{}%
\AgdaKeyword{import}\AgdaSpace{}%
\AgdaModule{Data.List.NonEmpty}\AgdaSpace{}%
\AgdaKeyword{hiding}\AgdaSpace{}%
\AgdaSymbol{(}\AgdaField{head}\AgdaSymbol{)}\<%
\\
\>[0]\AgdaKeyword{import}\AgdaSpace{}%
\AgdaModule{Relation.Binary.PropositionalEquality}\AgdaSpace{}%
\AgdaSymbol{as}\AgdaSpace{}%
\AgdaModule{Eq}\<%
\\
\>[0]\AgdaKeyword{open}\AgdaSpace{}%
\AgdaModule{Eq}\AgdaSpace{}%
\AgdaKeyword{using}\AgdaSpace{}%
\AgdaSymbol{(}\AgdaOperator{\AgdaDatatype{\AgdaUnderscore{}≡\AgdaUnderscore{}}}\AgdaSymbol{;}\AgdaSpace{}%
\AgdaInductiveConstructor{refl}\AgdaSymbol{;}\AgdaSpace{}%
\AgdaFunction{cong}\AgdaSymbol{;}\AgdaSpace{}%
\AgdaKeyword{module}\AgdaSpace{}%
\AgdaModule{≡-Reasoning}\AgdaSymbol{;}\AgdaSpace{}%
\AgdaFunction{sym}\AgdaSymbol{)}\<%
\\
\>[0]\AgdaKeyword{open}\AgdaSpace{}%
\AgdaModule{≡-Reasoning}\<%
\\
\>[0]\AgdaKeyword{open}\AgdaSpace{}%
\AgdaKeyword{import}\AgdaSpace{}%
\AgdaModule{Agda.Builtin.Equality}\<%
\\
\\[\AgdaEmptyExtraSkip]%
\>[0]\AgdaKeyword{open}\AgdaSpace{}%
\AgdaKeyword{import}\AgdaSpace{}%
\AgdaModule{libraries.listLib}\<%
\\
\>[0]\AgdaKeyword{open}\AgdaSpace{}%
\AgdaKeyword{import}\AgdaSpace{}%
\AgdaModule{libraries.natLib}\<%
\\
\>[0]\AgdaKeyword{open}\AgdaSpace{}%
\AgdaKeyword{import}\AgdaSpace{}%
\AgdaModule{libraries.boolLib}\<%
\\
\>[0]\AgdaKeyword{open}\AgdaSpace{}%
\AgdaKeyword{import}\AgdaSpace{}%
\AgdaModule{libraries.andLib}\<%
\\
\>[0]\AgdaKeyword{open}\AgdaSpace{}%
\AgdaKeyword{import}\AgdaSpace{}%
\AgdaModule{libraries.miscLib}\<%
\\
\>[0]\AgdaKeyword{open}\AgdaSpace{}%
\AgdaKeyword{import}\AgdaSpace{}%
\AgdaModule{libraries.maybeLib}\<%
\\
\\[\AgdaEmptyExtraSkip]%
\\[\AgdaEmptyExtraSkip]%
\>[0]\AgdaKeyword{open}\AgdaSpace{}%
\AgdaKeyword{import}\AgdaSpace{}%
\AgdaModule{stack}\<%
\\
\>[0]\AgdaKeyword{open}\AgdaSpace{}%
\AgdaKeyword{import}\AgdaSpace{}%
\AgdaModule{basicBitcoinDataType}\<%
\\
\\[\AgdaEmptyExtraSkip]%
\\[\AgdaEmptyExtraSkip]%
\>[0]\AgdaFunction{StackPredicate}\AgdaSpace{}%
\AgdaSymbol{:}\AgdaSpace{}%
\AgdaPrimitive{Set₁}\<%
\\
\>[0]\AgdaComment{--\textbackslash{}stackPredicateStackPred}\<%
\end{code}
} 

\newcommand{\stackPredicateStackPred}{
\begin{code}[inline]%
\>[0]\AgdaFunction{StackPredicate}\AgdaSpace{}%
\AgdaSymbol{=}\AgdaSpace{}%
\AgdaFunction{Time}\AgdaSpace{}%
\AgdaSymbol{→}\AgdaSpace{}%
\AgdaDatatype{Msg}\AgdaSpace{}%
\AgdaSymbol{→}\AgdaSpace{}%
\AgdaFunction{Stack}\AgdaSpace{}%
\AgdaSymbol{→}\AgdaSpace{}%
\AgdaPrimitive{Set}\<%
\end{code}
} 

\AgdaHide{
\begin{code}%
\>[0]\<%
\\
\>[0]\AgdaOperator{\AgdaFunction{\AgdaUnderscore{}⊎sp\AgdaUnderscore{}}}%
\>[8]\AgdaSymbol{:}\AgdaSpace{}%
\AgdaSymbol{(}\AgdaBound{φ}\AgdaSpace{}%
\AgdaBound{ψ}\AgdaSpace{}%
\AgdaSymbol{:}\AgdaSpace{}%
\AgdaFunction{StackPredicate}\AgdaSymbol{)}\AgdaSpace{}%
\AgdaSymbol{→}\AgdaSpace{}%
\AgdaFunction{StackPredicate}\<%
\\
\>[0]\AgdaSymbol{(}\AgdaBound{φ}%
\>[4]\AgdaOperator{\AgdaFunction{⊎sp}}%
\>[9]\AgdaBound{ψ}\AgdaSymbol{)}%
\>[13]\AgdaBound{t}\AgdaSpace{}%
\AgdaBound{m}\AgdaSpace{}%
\AgdaBound{st}\AgdaSpace{}%
\AgdaSymbol{=}\AgdaSpace{}%
\AgdaBound{φ}%
\>[25]\AgdaBound{t}\AgdaSpace{}%
\AgdaBound{m}\AgdaSpace{}%
\AgdaBound{st}%
\>[33]\AgdaOperator{\AgdaDatatype{⊎}}\AgdaSpace{}%
\AgdaBound{ψ}%
\>[38]\AgdaBound{t}\AgdaSpace{}%
\AgdaBound{m}\AgdaSpace{}%
\AgdaBound{st}\<%
\\
\\[\AgdaEmptyExtraSkip]%
\>[0]\AgdaOperator{\AgdaFunction{\AgdaUnderscore{}∧sp\AgdaUnderscore{}}}\AgdaSpace{}%
\AgdaSymbol{:}\AgdaSpace{}%
\AgdaSymbol{(}\AgdaSpace{}%
\AgdaBound{φ}\AgdaSpace{}%
\AgdaBound{ψ}%
\>[15]\AgdaSymbol{:}\AgdaSpace{}%
\AgdaFunction{StackPredicate}\AgdaSpace{}%
\AgdaSymbol{)}\AgdaSpace{}%
\AgdaSymbol{→}\AgdaSpace{}%
\AgdaFunction{StackPredicate}\<%
\\
\>[0]\AgdaSymbol{(}\AgdaBound{φ}\AgdaSpace{}%
\AgdaOperator{\AgdaFunction{∧sp}}\AgdaSpace{}%
\AgdaBound{ψ}\AgdaSpace{}%
\AgdaSymbol{)}\AgdaSpace{}%
\AgdaBound{t}\AgdaSpace{}%
\AgdaBound{m}\AgdaSpace{}%
\AgdaBound{s}%
\>[19]\AgdaSymbol{=}%
\>[22]\AgdaBound{φ}\AgdaSpace{}%
\AgdaBound{t}\AgdaSpace{}%
\AgdaBound{m}\AgdaSpace{}%
\AgdaBound{s}%
\>[31]\AgdaOperator{\AgdaRecord{∧}}%
\>[34]\AgdaBound{ψ}\AgdaSpace{}%
\AgdaBound{t}\AgdaSpace{}%
\AgdaBound{m}\AgdaSpace{}%
\AgdaBound{s}\<%
\\
\\[\AgdaEmptyExtraSkip]%
\\[\AgdaEmptyExtraSkip]%
\>[0]\AgdaFunction{truePredaux}\AgdaSpace{}%
\AgdaSymbol{:}\AgdaSpace{}%
\AgdaFunction{StackPredicate}\AgdaSpace{}%
\AgdaSymbol{→}\AgdaSpace{}%
\AgdaFunction{StackPredicate}\<%
\\
\>[0]\AgdaFunction{truePredaux}\AgdaSpace{}%
\AgdaBound{φ}\AgdaSpace{}%
\AgdaBound{time}\AgdaSpace{}%
\AgdaBound{msg}\AgdaSpace{}%
\AgdaInductiveConstructor{[]}\AgdaSpace{}%
\AgdaSymbol{=}\AgdaSpace{}%
\AgdaDatatype{⊥}\<%
\\
\>[0]\AgdaFunction{truePredaux}\AgdaSpace{}%
\AgdaBound{φ}\AgdaSpace{}%
\AgdaBound{time}\AgdaSpace{}%
\AgdaBound{msg}\AgdaSpace{}%
\AgdaSymbol{(}\AgdaInductiveConstructor{zero}\AgdaSpace{}%
\AgdaOperator{\AgdaInductiveConstructor{∷}}\AgdaSpace{}%
\AgdaBound{st}\AgdaSymbol{)}\AgdaSpace{}%
\AgdaSymbol{=}\AgdaSpace{}%
\AgdaDatatype{⊥}\<%
\\
\>[0]\AgdaFunction{truePredaux}\AgdaSpace{}%
\AgdaBound{φ}\AgdaSpace{}%
\AgdaBound{time}\AgdaSpace{}%
\AgdaBound{msg}\AgdaSpace{}%
\AgdaSymbol{(}\AgdaInductiveConstructor{suc}\AgdaSpace{}%
\AgdaBound{x}\AgdaSpace{}%
\AgdaOperator{\AgdaInductiveConstructor{∷}}\AgdaSpace{}%
\AgdaBound{st}\AgdaSymbol{)}\AgdaSpace{}%
\AgdaSymbol{=}\AgdaSpace{}%
\AgdaBound{φ}\AgdaSpace{}%
\AgdaBound{time}\AgdaSpace{}%
\AgdaBound{msg}%
\>[50]\AgdaBound{st}\<%
\\
\\[\AgdaEmptyExtraSkip]%
\>[0]\AgdaComment{--\ acceptingBasic\ :\ Time\ →\ Msg\ →\ Stack\ →\ Bool}\<%
\\
\>[0]\AgdaComment{--\ acceptingBasic\ time\ msg₁\ []\ =\ false}\<%
\\
\>[0]\AgdaComment{--\ acceptingBasic\ time\ msg₁\ (x\ ∷\ stack₁)\ =\ notFalse\ x}\<%
\\
\\[\AgdaEmptyExtraSkip]%
\>[0]\AgdaComment{--\textbackslash{}stackPredicateacceptState}\<%
\end{code}
} 

\newcommand{\stackPredicateacceptState}{
\begin{code}%
\>[0]\AgdaFunction{acceptStateˢ}\AgdaSpace{}%
\AgdaSymbol{:}\AgdaSpace{}%
\AgdaFunction{StackPredicate}\<%
\\
\>[0]\AgdaFunction{acceptStateˢ}%
\>[14]\AgdaBound{time}%
\>[20]\AgdaBound{msg₁}%
\>[26]\AgdaInductiveConstructor{[]}%
\>[40]\AgdaSymbol{=}%
\>[43]\AgdaDatatype{⊥}\<%
\\
\>[0]\AgdaFunction{acceptStateˢ}%
\>[14]\AgdaBound{time}%
\>[20]\AgdaBound{msg₁}%
\>[26]\AgdaSymbol{(}\AgdaBound{x}\AgdaSpace{}%
\AgdaOperator{\AgdaInductiveConstructor{∷}}\AgdaSpace{}%
\AgdaBound{stack₁}\AgdaSymbol{)}%
\>[40]\AgdaSymbol{=}%
\>[43]\AgdaFunction{NotFalse}\AgdaSpace{}%
\AgdaBound{x}\<%
\\
\>[0]\<%
\end{code}
} 

\AgdaHide{
\begin{code}%
\>[0]\<%
\end{code}
} 


\AgdaHide{
\begin{code}%
\>[0]\AgdaKeyword{open}\AgdaSpace{}%
\AgdaKeyword{import}\AgdaSpace{}%
\AgdaModule{basicBitcoinDataType}\<%
\\
\\[\AgdaEmptyExtraSkip]%
\>[0]\AgdaKeyword{module}%
\>[8]\AgdaModule{verificationStackScripts.stackVerificationP2PKH}\AgdaSpace{}%
\AgdaSymbol{(}\AgdaBound{param}\AgdaSpace{}%
\AgdaSymbol{:}\AgdaSpace{}%
\AgdaRecord{GlobalParameters}\AgdaSymbol{)}\AgdaSpace{}%
\AgdaKeyword{where}\<%
\\
\\[\AgdaEmptyExtraSkip]%
\>[0]\AgdaKeyword{open}\AgdaSpace{}%
\AgdaKeyword{import}\AgdaSpace{}%
\AgdaModule{libraries.listLib}\<%
\\
\>[0]\AgdaKeyword{open}\AgdaSpace{}%
\AgdaKeyword{import}\AgdaSpace{}%
\AgdaModule{Data.List.Base}\<%
\\
\>[0]\AgdaKeyword{open}\AgdaSpace{}%
\AgdaKeyword{import}\AgdaSpace{}%
\AgdaModule{libraries.natLib}\<%
\\
\>[0]\AgdaKeyword{open}\AgdaSpace{}%
\AgdaKeyword{import}\AgdaSpace{}%
\AgdaModule{Data.Nat}%
\>[22]\AgdaKeyword{renaming}\AgdaSpace{}%
\AgdaSymbol{(}\AgdaOperator{\AgdaDatatype{\AgdaUnderscore{}≤\AgdaUnderscore{}}}\AgdaSpace{}%
\AgdaSymbol{to}\AgdaSpace{}%
\AgdaOperator{\AgdaDatatype{\AgdaUnderscore{}≤'\AgdaUnderscore{}}}\AgdaSpace{}%
\AgdaSymbol{;}\AgdaSpace{}%
\AgdaOperator{\AgdaFunction{\AgdaUnderscore{}<\AgdaUnderscore{}}}\AgdaSpace{}%
\AgdaSymbol{to}\AgdaSpace{}%
\AgdaOperator{\AgdaFunction{\AgdaUnderscore{}<'\AgdaUnderscore{}}}\AgdaSymbol{)}\<%
\\
\>[0]\AgdaKeyword{open}\AgdaSpace{}%
\AgdaKeyword{import}\AgdaSpace{}%
\AgdaModule{Data.List}\AgdaSpace{}%
\AgdaKeyword{hiding}\AgdaSpace{}%
\AgdaSymbol{(}\AgdaOperator{\AgdaFunction{\AgdaUnderscore{}\ensuremath{+\!\!+}\AgdaUnderscore{}}}\AgdaSymbol{)}\<%
\\
\>[0]\AgdaKeyword{open}\AgdaSpace{}%
\AgdaKeyword{import}\AgdaSpace{}%
\AgdaModule{Data.Unit}%
\>[23]\AgdaKeyword{hiding}\AgdaSpace{}%
\AgdaSymbol{(}\AgdaOperator{\AgdaRecord{\AgdaUnderscore{}≤\AgdaUnderscore{}}}\AgdaSymbol{)}\<%
\\
\>[0]\AgdaKeyword{open}\AgdaSpace{}%
\AgdaKeyword{import}\AgdaSpace{}%
\AgdaModule{Data.Empty}\<%
\\
\>[0]\AgdaKeyword{open}\AgdaSpace{}%
\AgdaKeyword{import}\AgdaSpace{}%
\AgdaModule{Data.Bool}%
\>[23]\AgdaKeyword{hiding}\AgdaSpace{}%
\AgdaSymbol{(}\AgdaOperator{\AgdaDatatype{\AgdaUnderscore{}≤\AgdaUnderscore{}}}\AgdaSpace{}%
\AgdaSymbol{;}\AgdaSpace{}%
\AgdaOperator{\AgdaDatatype{\AgdaUnderscore{}<\AgdaUnderscore{}}}\AgdaSpace{}%
\AgdaSymbol{;}\AgdaSpace{}%
\AgdaOperator{\AgdaFunction{if\AgdaUnderscore{}then\AgdaUnderscore{}else\AgdaUnderscore{}}}\AgdaSpace{}%
\AgdaSymbol{)}\AgdaSpace{}%
\AgdaKeyword{renaming}\AgdaSpace{}%
\AgdaSymbol{(}\AgdaOperator{\AgdaFunction{\AgdaUnderscore{}∧\AgdaUnderscore{}}}\AgdaSpace{}%
\AgdaSymbol{to}\AgdaSpace{}%
\AgdaOperator{\AgdaFunction{\AgdaUnderscore{}∧b\AgdaUnderscore{}}}\AgdaSpace{}%
\AgdaSymbol{;}\AgdaSpace{}%
\AgdaOperator{\AgdaFunction{\AgdaUnderscore{}∨\AgdaUnderscore{}}}\AgdaSpace{}%
\AgdaSymbol{to}\AgdaSpace{}%
\AgdaOperator{\AgdaFunction{\AgdaUnderscore{}∨b\AgdaUnderscore{}}}\AgdaSpace{}%
\AgdaSymbol{;}\AgdaSpace{}%
\AgdaFunction{T}\AgdaSpace{}%
\AgdaSymbol{to}\AgdaSpace{}%
\AgdaFunction{True}\AgdaSymbol{)}\<%
\\
\>[0]\AgdaKeyword{open}\AgdaSpace{}%
\AgdaKeyword{import}\AgdaSpace{}%
\AgdaModule{Data.Product}\AgdaSpace{}%
\AgdaKeyword{renaming}\AgdaSpace{}%
\AgdaSymbol{(}\AgdaOperator{\AgdaInductiveConstructor{\AgdaUnderscore{},\AgdaUnderscore{}}}\AgdaSpace{}%
\AgdaSymbol{to}\AgdaSpace{}%
\AgdaOperator{\AgdaInductiveConstructor{\AgdaUnderscore{},,\AgdaUnderscore{}}}\AgdaSpace{}%
\AgdaSymbol{)}\<%
\\
\>[0]\AgdaKeyword{open}\AgdaSpace{}%
\AgdaKeyword{import}\AgdaSpace{}%
\AgdaModule{Data.Nat.Base}\AgdaSpace{}%
\AgdaKeyword{hiding}\AgdaSpace{}%
\AgdaSymbol{(}\AgdaOperator{\AgdaDatatype{\AgdaUnderscore{}≤\AgdaUnderscore{}}}\AgdaSpace{}%
\AgdaSymbol{;}\AgdaSpace{}%
\AgdaOperator{\AgdaFunction{\AgdaUnderscore{}<\AgdaUnderscore{}}}\AgdaSymbol{)}\<%
\\
\>[0]\AgdaKeyword{open}\AgdaSpace{}%
\AgdaKeyword{import}\AgdaSpace{}%
\AgdaModule{Data.List.NonEmpty}\AgdaSpace{}%
\AgdaKeyword{hiding}\AgdaSpace{}%
\AgdaSymbol{(}\AgdaField{head}\AgdaSymbol{;}\AgdaSpace{}%
\AgdaOperator{\AgdaFunction{[\AgdaUnderscore{}]}}\AgdaSymbol{)}\<%
\\
\>[0]\AgdaKeyword{open}\AgdaSpace{}%
\AgdaKeyword{import}\AgdaSpace{}%
\AgdaModule{Data.Nat}\AgdaSpace{}%
\AgdaKeyword{using}\AgdaSpace{}%
\AgdaSymbol{(}\AgdaDatatype{ℕ}\AgdaSymbol{;}\AgdaSpace{}%
\AgdaOperator{\AgdaPrimitive{\AgdaUnderscore{}+\AgdaUnderscore{}}}\AgdaSymbol{;}\AgdaSpace{}%
\AgdaOperator{\AgdaFunction{\AgdaUnderscore{}≥\AgdaUnderscore{}}}\AgdaSymbol{;}\AgdaSpace{}%
\AgdaOperator{\AgdaFunction{\AgdaUnderscore{}>\AgdaUnderscore{}}}\AgdaSymbol{;}\AgdaSpace{}%
\AgdaInductiveConstructor{zero}\AgdaSymbol{;}\AgdaSpace{}%
\AgdaInductiveConstructor{suc}\AgdaSymbol{;}\AgdaSpace{}%
\AgdaInductiveConstructor{s≤s}\AgdaSymbol{;}\AgdaSpace{}%
\AgdaInductiveConstructor{z≤n}\AgdaSymbol{)}\<%
\\
\>[0]\AgdaKeyword{import}\AgdaSpace{}%
\AgdaModule{Relation.Binary.PropositionalEquality}\AgdaSpace{}%
\AgdaSymbol{as}\AgdaSpace{}%
\AgdaModule{Eq}\<%
\\
\>[0]\AgdaKeyword{open}\AgdaSpace{}%
\AgdaModule{Eq}\AgdaSpace{}%
\AgdaKeyword{using}\AgdaSpace{}%
\AgdaSymbol{(}\AgdaOperator{\AgdaDatatype{\AgdaUnderscore{}≡\AgdaUnderscore{}}}\AgdaSymbol{;}\AgdaSpace{}%
\AgdaInductiveConstructor{refl}\AgdaSymbol{;}\AgdaSpace{}%
\AgdaFunction{cong}\AgdaSymbol{;}\AgdaSpace{}%
\AgdaKeyword{module}\AgdaSpace{}%
\AgdaModule{≡-Reasoning}\AgdaSymbol{;}\AgdaSpace{}%
\AgdaFunction{sym}\AgdaSymbol{)}\<%
\\
\>[0]\AgdaKeyword{open}\AgdaSpace{}%
\AgdaModule{≡-Reasoning}\<%
\\
\>[0]\AgdaKeyword{open}\AgdaSpace{}%
\AgdaKeyword{import}\AgdaSpace{}%
\AgdaModule{Agda.Builtin.Equality}\<%
\\
\\[\AgdaEmptyExtraSkip]%
\\[\AgdaEmptyExtraSkip]%
\>[0]\AgdaComment{--our\ libraries}\<%
\\
\>[0]\AgdaKeyword{open}\AgdaSpace{}%
\AgdaKeyword{import}\AgdaSpace{}%
\AgdaModule{libraries.andLib}\<%
\\
\>[0]\AgdaKeyword{open}\AgdaSpace{}%
\AgdaKeyword{import}\AgdaSpace{}%
\AgdaModule{libraries.miscLib}\<%
\\
\>[0]\AgdaKeyword{open}\AgdaSpace{}%
\AgdaKeyword{import}\AgdaSpace{}%
\AgdaModule{libraries.maybeLib}\<%
\\
\>[0]\AgdaKeyword{open}\AgdaSpace{}%
\AgdaKeyword{import}\AgdaSpace{}%
\AgdaModule{libraries.boolLib}\<%
\\
\\[\AgdaEmptyExtraSkip]%
\>[0]\AgdaKeyword{open}\AgdaSpace{}%
\AgdaKeyword{import}\AgdaSpace{}%
\AgdaModule{stack}\<%
\\
\>[0]\AgdaKeyword{open}\AgdaSpace{}%
\AgdaKeyword{import}\AgdaSpace{}%
\AgdaModule{stackPredicate}\<%
\\
\>[0]\AgdaKeyword{open}\AgdaSpace{}%
\AgdaKeyword{import}\AgdaSpace{}%
\AgdaModule{semanticBasicOperations}\AgdaSpace{}%
\AgdaBound{param}\<%
\\
\>[0]\AgdaKeyword{open}\AgdaSpace{}%
\AgdaKeyword{import}\AgdaSpace{}%
\AgdaModule{instructionBasic}\<%
\\
\>[0]\AgdaKeyword{open}\AgdaSpace{}%
\AgdaKeyword{import}\AgdaSpace{}%
\AgdaModule{verificationP2PKHbasic}\AgdaSpace{}%
\AgdaBound{param}\<%
\\
\\[\AgdaEmptyExtraSkip]%
\>[0]\AgdaKeyword{open}\AgdaSpace{}%
\AgdaKeyword{import}\AgdaSpace{}%
\AgdaModule{verificationStackScripts.stackHoareTriple}\AgdaSpace{}%
\AgdaBound{param}\<%
\\
\>[0]\AgdaKeyword{open}\AgdaSpace{}%
\AgdaKeyword{import}\AgdaSpace{}%
\AgdaModule{verificationStackScripts.stackState}\<%
\\
\>[0]\AgdaKeyword{open}\AgdaSpace{}%
\AgdaKeyword{import}\AgdaSpace{}%
\AgdaModule{verificationStackScripts.sPredicate}\<%
\\
\>[0]\AgdaKeyword{open}\AgdaSpace{}%
\AgdaKeyword{import}\AgdaSpace{}%
\AgdaModule{verificationStackScripts.semanticsStackInstructions}\AgdaSpace{}%
\AgdaBound{param}\<%
\\
\>[0]\AgdaKeyword{open}\AgdaSpace{}%
\AgdaKeyword{import}\AgdaSpace{}%
\AgdaModule{verificationStackScripts.stackVerificationLemmas}\AgdaSpace{}%
\AgdaBound{param}\<%
\\
\\[\AgdaEmptyExtraSkip]%
\\[\AgdaEmptyExtraSkip]%
\\[\AgdaEmptyExtraSkip]%
\>[0]\AgdaComment{--\ note\ accept\AgdaUnderscore{}0\ is\ the\ same\ as\ acceptState}\<%
\\
\>[0]\AgdaFunction{accept-0}\AgdaSpace{}%
\AgdaSymbol{:}\AgdaSpace{}%
\AgdaFunction{StackStatePred}\<%
\\
\>[0]\AgdaFunction{accept-0}\AgdaSpace{}%
\AgdaSymbol{=}\AgdaSpace{}%
\AgdaFunction{stackPred2SPred}\AgdaSpace{}%
\AgdaFunction{accept-0Basic}\<%
\\
\\[\AgdaEmptyExtraSkip]%
\\[\AgdaEmptyExtraSkip]%
\>[0]\AgdaFunction{accept₁}\AgdaSpace{}%
\AgdaSymbol{:}\AgdaSpace{}%
\AgdaFunction{StackStatePred}\<%
\\
\>[0]\AgdaFunction{accept₁}%
\>[9]\AgdaSymbol{=}\AgdaSpace{}%
\AgdaFunction{stackPred2SPred}\AgdaSpace{}%
\AgdaFunction{accept₁ˢ}\<%
\\
\\[\AgdaEmptyExtraSkip]%
\>[0]\AgdaFunction{accept₂}\AgdaSpace{}%
\AgdaSymbol{:}\AgdaSpace{}%
\AgdaFunction{StackStatePred}\<%
\\
\>[0]\AgdaFunction{accept₂}\AgdaSpace{}%
\AgdaSymbol{=}\AgdaSpace{}%
\AgdaFunction{stackPred2SPred}\AgdaSpace{}%
\AgdaFunction{accept₂ˢ}\<%
\\
\\[\AgdaEmptyExtraSkip]%
\>[0]\AgdaFunction{accept₃}\AgdaSpace{}%
\AgdaSymbol{:}\AgdaSpace{}%
\AgdaFunction{StackStatePred}\<%
\\
\>[0]\AgdaFunction{accept₃}\AgdaSpace{}%
\AgdaSymbol{=}\AgdaSpace{}%
\AgdaFunction{stackPred2SPred}\AgdaSpace{}%
\AgdaFunction{accept₃ˢ}\<%
\\
\\[\AgdaEmptyExtraSkip]%
\>[0]\AgdaComment{--\ checked\ needs\ to\ be\ pbkh\ not\ pbk}\<%
\\
\>[0]\AgdaFunction{accept₄}\AgdaSpace{}%
\AgdaSymbol{:}\AgdaSpace{}%
\AgdaDatatype{ℕ}\AgdaSpace{}%
\AgdaSymbol{→}\AgdaSpace{}%
\AgdaFunction{StackStatePred}\<%
\\
\>[0]\AgdaFunction{accept₄}\AgdaSpace{}%
\AgdaBound{pbkh}%
\>[14]\AgdaSymbol{=}\AgdaSpace{}%
\AgdaFunction{stackPred2SPred}\AgdaSpace{}%
\AgdaSymbol{(}\AgdaFunction{accept₄ˢ}\AgdaSpace{}%
\AgdaBound{pbkh}\AgdaSymbol{)}\<%
\\
\\[\AgdaEmptyExtraSkip]%
\>[0]\AgdaComment{--\ checked\ needs\ to\ be\ pbkh\ not\ pbk}\<%
\\
\>[0]\AgdaFunction{accept₅}\AgdaSpace{}%
\AgdaSymbol{:}\AgdaSpace{}%
\AgdaDatatype{ℕ}\AgdaSpace{}%
\AgdaSymbol{→}\AgdaSpace{}%
\AgdaFunction{StackStatePred}\<%
\\
\>[0]\AgdaFunction{accept₅}\AgdaSpace{}%
\AgdaBound{pbkh}%
\>[14]\AgdaSymbol{=}\AgdaSpace{}%
\AgdaFunction{stackPred2SPred}\AgdaSpace{}%
\AgdaSymbol{(}\AgdaFunction{accept₅ˢ}\AgdaSpace{}%
\AgdaBound{pbkh}\AgdaSymbol{)}\<%
\\
\\[\AgdaEmptyExtraSkip]%
\>[0]\AgdaComment{--\ accept-6\ :\ ℕ\ →\ StackStatePred}\<%
\\
\>[0]\AgdaComment{--\ accept-6\ pbkh\ \ =\ stackPred2SPred\ (wPreCondP2PKHˢ\ pbkh)}\<%
\\
\\[\AgdaEmptyExtraSkip]%
\>[0]\AgdaComment{--\ checked\ needs\ to\ be\ pbkh\ not\ pbk}\<%
\\
\>[0]\AgdaFunction{wPreCondP2PKH}\AgdaSpace{}%
\AgdaSymbol{:}\AgdaSpace{}%
\AgdaSymbol{(}\AgdaBound{pbkh}\AgdaSpace{}%
\AgdaSymbol{:}\AgdaSpace{}%
\AgdaDatatype{ℕ}\AgdaSymbol{)}%
\>[28]\AgdaSymbol{→}\AgdaSpace{}%
\AgdaFunction{StackStatePred}\<%
\\
\>[0]\AgdaFunction{wPreCondP2PKH}\AgdaSpace{}%
\AgdaBound{pbkh}%
\>[20]\AgdaSymbol{=}\AgdaSpace{}%
\AgdaFunction{stackPred2SPred}\AgdaSpace{}%
\AgdaSymbol{(}\AgdaFunction{wPreCondP2PKHˢ}\AgdaSpace{}%
\AgdaBound{pbkh}\AgdaSymbol{)}\<%
\\
\\[\AgdaEmptyExtraSkip]%
\>[0]\AgdaComment{--\ we\ use\ pbk\ and\ not\ pbkh\ because\ that\ is\ what\ is\ provided\ by\ the\ unlocking\ script}\<%
\\
\>[0]\AgdaFunction{correct-opCheckSig-to}\AgdaSpace{}%
\AgdaSymbol{:}\AgdaSpace{}%
\AgdaSymbol{(}\AgdaBound{s}\AgdaSpace{}%
\AgdaSymbol{:}\AgdaSpace{}%
\AgdaRecord{StackState}\AgdaSymbol{)}\AgdaSpace{}%
\AgdaSymbol{→}\AgdaSpace{}%
\AgdaFunction{accept₁}\AgdaSpace{}%
\AgdaBound{s}\AgdaSpace{}%
\AgdaSymbol{→}\AgdaSpace{}%
\AgdaSymbol{(}\AgdaFunction{accept-0}\AgdaSpace{}%
\AgdaOperator{\AgdaFunction{⁺}}\AgdaSymbol{)}\AgdaSpace{}%
\AgdaSymbol{(}\AgdaOperator{\AgdaFunction{⟦}}\AgdaSpace{}%
\AgdaInductiveConstructor{opCheckSig}\AgdaSpace{}%
\AgdaOperator{\AgdaFunction{⟧ₛ}}\AgdaSpace{}%
\AgdaBound{s}\AgdaSpace{}%
\AgdaSymbol{)}\<%
\\
\>[0]\AgdaFunction{correct-opCheckSig-to}\AgdaSpace{}%
\AgdaOperator{\AgdaInductiveConstructor{⟨}}\AgdaSpace{}%
\AgdaBound{time}\AgdaSpace{}%
\AgdaOperator{\AgdaInductiveConstructor{,}}\AgdaSpace{}%
\AgdaBound{msg₁}\AgdaSpace{}%
\AgdaOperator{\AgdaInductiveConstructor{,}}\AgdaSpace{}%
\AgdaBound{pbk}%
\>[43]\AgdaOperator{\AgdaInductiveConstructor{∷}}\AgdaSpace{}%
\AgdaBound{sig}\AgdaSpace{}%
\AgdaOperator{\AgdaInductiveConstructor{∷}}\AgdaSpace{}%
\AgdaBound{st}\AgdaSpace{}%
\AgdaOperator{\AgdaInductiveConstructor{⟩}}\AgdaSpace{}%
\AgdaBound{p}\AgdaSpace{}%
\AgdaSymbol{=}%
\>[61]\AgdaFunction{boolToNatNotFalseLemma}\AgdaSpace{}%
\AgdaSymbol{(}\AgdaFunction{isSigned}%
\>[95]\AgdaBound{msg₁}\AgdaSpace{}%
\AgdaBound{sig}\AgdaSpace{}%
\AgdaBound{pbk}\AgdaSymbol{)}\AgdaSpace{}%
\AgdaBound{p}\<%
\\
\\[\AgdaEmptyExtraSkip]%
\>[0]\AgdaComment{\{-\ old}\<%
\\
\>[0]\AgdaComment{correct-1-to\ :\ (s\ :\ StackState)\ →\ accept₁\ s\ →\ (accept-0\ ⁺)\ (⟦\ opCheckSig\ ⟧s\ s\ )}\<%
\\
\>[0]\AgdaComment{correct-1-to\ ⟨\ time\ ,\ msg₁\ ,\ pbk\ \ ∷\ sig\ ∷\ st\ ⟩\ p\ =\ \ boolToNatNotFalseLemma\ (isSigned\ \ msg₁\ sig\ pbk)\ p}\<%
\\
\>[0]\AgdaComment{-\}}\<%
\\
\\[\AgdaEmptyExtraSkip]%
\>[0]\AgdaFunction{correct-opCheckSig-from}\AgdaSpace{}%
\AgdaSymbol{:}\AgdaSpace{}%
\AgdaSymbol{(}\AgdaBound{s}\AgdaSpace{}%
\AgdaSymbol{:}\AgdaSpace{}%
\AgdaRecord{StackState}\AgdaSymbol{)}\AgdaSpace{}%
\AgdaSymbol{→}\AgdaSpace{}%
\AgdaSymbol{(}\AgdaFunction{accept-0}\AgdaSpace{}%
\AgdaOperator{\AgdaFunction{⁺}}\AgdaSymbol{)}\AgdaSpace{}%
\AgdaSymbol{(}\AgdaOperator{\AgdaFunction{⟦}}\AgdaSpace{}%
\AgdaInductiveConstructor{opCheckSig}\AgdaSpace{}%
\AgdaOperator{\AgdaFunction{⟧ₛ}}\AgdaSpace{}%
\AgdaBound{s}\AgdaSpace{}%
\AgdaSymbol{)}\AgdaSpace{}%
\AgdaSymbol{→}\AgdaSpace{}%
\AgdaFunction{accept₁}\AgdaSpace{}%
\AgdaBound{s}\<%
\\
\>[0]\AgdaFunction{correct-opCheckSig-from}\AgdaSpace{}%
\AgdaOperator{\AgdaInductiveConstructor{⟨}}\AgdaSpace{}%
\AgdaBound{time}\AgdaSpace{}%
\AgdaOperator{\AgdaInductiveConstructor{,}}\AgdaSpace{}%
\AgdaBound{msg₁}\AgdaSpace{}%
\AgdaOperator{\AgdaInductiveConstructor{,}}\AgdaSpace{}%
\AgdaBound{pbk}\AgdaSpace{}%
\AgdaOperator{\AgdaInductiveConstructor{∷}}\AgdaSpace{}%
\AgdaBound{sig}\AgdaSpace{}%
\AgdaOperator{\AgdaInductiveConstructor{∷}}\AgdaSpace{}%
\AgdaBound{stack₁}%
\>[60]\AgdaOperator{\AgdaInductiveConstructor{⟩}}\AgdaSpace{}%
\AgdaBound{p}\AgdaSpace{}%
\AgdaSymbol{=}\AgdaSpace{}%
\AgdaFunction{boolToNatNotFalseLemma2}\AgdaSpace{}%
\AgdaSymbol{(}\AgdaFunction{isSigned}%
\>[101]\AgdaBound{msg₁}\AgdaSpace{}%
\AgdaBound{sig}\AgdaSpace{}%
\AgdaBound{pbk}\AgdaSymbol{)}\AgdaSpace{}%
\AgdaBound{p}\<%
\\
\\[\AgdaEmptyExtraSkip]%
\>[0]\AgdaComment{\{-old}\<%
\\
\>[0]\AgdaComment{correct-1-from\ :\ (s\ :\ StackState)\ →\ (accept-0\ ⁺)\ (⟦\ opCheckSig\ ⟧s\ s\ )\ →\ accept₁\ s}\<%
\\
\>[0]\AgdaComment{correct-1-from\ ⟨\ time\ ,\ msg₁\ ,\ pbk\ ∷\ sig\ ∷\ stack₁\ \ ⟩\ p\ =\ boolToNatNotFalseLemma2\ (isSigned\ \ msg₁\ sig\ pbk)\ p}\<%
\\
\>[0]\AgdaComment{-\}}\<%
\end{code}
} 

\newcommand{\stackVerificationPtoPKHcorrectone}{
\begin{code}[inline]%
\>[0]\AgdaFunction{correct-opCheckSig}\AgdaSpace{}%
\AgdaSymbol{:}\AgdaSpace{}%
\AgdaOperator{\AgdaRecord{<}}\AgdaSpace{}%
\AgdaFunction{accept₁}\AgdaSpace{}%
\AgdaOperator{\AgdaRecord{>ⁱᶠᶠ}}\AgdaSpace{}%
\>[37]\AgdaSymbol{(}\AgdaOperator{\AgdaFunction{[}}\AgdaSpace{}%
\AgdaInductiveConstructor{opCheckSig}\AgdaSpace{}%
\AgdaOperator{\AgdaFunction{]}}\AgdaSymbol{)}\AgdaSpace{}%
\AgdaOperator{\AgdaRecord{<}}\AgdaSpace{}%
\AgdaFunction{acceptState}\AgdaSpace{}%
\AgdaOperator{\AgdaRecord{>}}\<%
\end{code}
} 

\AgdaHide{
\begin{code}%
\>[0]\AgdaFunction{correct-opCheckSig}\AgdaSpace{}%
\AgdaSymbol{.}\AgdaField{==>}\AgdaSpace{}%
\AgdaSymbol{=}\AgdaSpace{}%
\AgdaFunction{correct-opCheckSig-to}\<%
\\
\>[0]\AgdaFunction{correct-opCheckSig}\AgdaSpace{}%
\AgdaSymbol{.}\AgdaField{<==}\AgdaSpace{}%
\AgdaSymbol{=}\AgdaSpace{}%
\AgdaFunction{correct-opCheckSig-from}\<%
\\
\>[0]\AgdaComment{\{-old}\<%
\\
\>[0]\AgdaComment{-@BEGININLINE@correctone}\<%
\\
\>[0]\AgdaComment{correct-1\ :\ <\ accept₁\ >iff\ \ ([\ opCheckSig\ ])\ <\ acceptState\ >}\<%
\\
\>[0]\AgdaComment{-@END}\<%
\\
\>[0]\AgdaComment{correct-1\ .==>\ =\ correct-1-to}\<%
\\
\>[0]\AgdaComment{correct-1\ .<==\ =\ correct-1-from}\<%
\\
\>[0]\AgdaComment{-\}}\<%
\\
\\[\AgdaEmptyExtraSkip]%
\>[0]\AgdaFunction{correct-opVerify-to}\AgdaSpace{}%
\AgdaSymbol{:}\AgdaSpace{}%
\AgdaSymbol{(}\AgdaBound{s}\AgdaSpace{}%
\AgdaSymbol{:}\AgdaSpace{}%
\AgdaRecord{StackState}\AgdaSymbol{)}\AgdaSpace{}%
\AgdaSymbol{→}\AgdaSpace{}%
\AgdaFunction{accept₂}\AgdaSpace{}%
\AgdaBound{s}\AgdaSpace{}%
\AgdaSymbol{→}\AgdaSpace{}%
\AgdaSymbol{(}\AgdaFunction{accept₁}\AgdaSpace{}%
\AgdaOperator{\AgdaFunction{⁺}}\AgdaSymbol{)}\AgdaSpace{}%
\AgdaSymbol{(}\AgdaOperator{\AgdaFunction{⟦}}\AgdaSpace{}%
\AgdaInductiveConstructor{opVerify}\AgdaSpace{}%
\AgdaOperator{\AgdaFunction{⟧ₛ}}\AgdaSpace{}%
\AgdaBound{s}\AgdaSpace{}%
\AgdaSymbol{)}\<%
\\
\>[0]\AgdaFunction{correct-opVerify-to}\AgdaSpace{}%
\AgdaOperator{\AgdaInductiveConstructor{⟨}}\AgdaSpace{}%
\AgdaBound{time}\AgdaSpace{}%
\AgdaOperator{\AgdaInductiveConstructor{,}}\AgdaSpace{}%
\AgdaBound{msg₁}\AgdaSpace{}%
\AgdaOperator{\AgdaInductiveConstructor{,}}\AgdaSpace{}%
\AgdaInductiveConstructor{suc}\AgdaSpace{}%
\AgdaBound{x}\AgdaSpace{}%
\AgdaOperator{\AgdaInductiveConstructor{∷}}\AgdaSpace{}%
\AgdaBound{x₁}\AgdaSpace{}%
\AgdaOperator{\AgdaInductiveConstructor{∷}}\AgdaSpace{}%
\AgdaBound{x₂}\AgdaSpace{}%
\AgdaOperator{\AgdaInductiveConstructor{∷}}\AgdaSpace{}%
\AgdaBound{stack₁}\AgdaSpace{}%
\AgdaOperator{\AgdaInductiveConstructor{⟩}}\AgdaSpace{}%
\AgdaBound{p}\AgdaSpace{}%
\AgdaSymbol{=}\AgdaSpace{}%
\AgdaBound{p}\<%
\\
\\[\AgdaEmptyExtraSkip]%
\>[0]\AgdaFunction{correct-opVerify-from}\AgdaSpace{}%
\AgdaSymbol{:}\AgdaSpace{}%
\AgdaSymbol{(}\AgdaBound{s}\AgdaSpace{}%
\AgdaSymbol{:}\AgdaSpace{}%
\AgdaRecord{StackState}\AgdaSymbol{)}\AgdaSpace{}%
\AgdaSymbol{→}\AgdaSpace{}%
\AgdaSymbol{(}\AgdaFunction{accept₁}\AgdaSpace{}%
\AgdaOperator{\AgdaFunction{⁺}}\AgdaSymbol{)}\AgdaSpace{}%
\AgdaSymbol{(}\AgdaOperator{\AgdaFunction{⟦}}\AgdaSpace{}%
\AgdaInductiveConstructor{opVerify}\AgdaSpace{}%
\AgdaOperator{\AgdaFunction{⟧ₛ}}\AgdaSpace{}%
\AgdaBound{s}\AgdaSpace{}%
\AgdaSymbol{)}\AgdaSpace{}%
\AgdaSymbol{→}\AgdaSpace{}%
\AgdaFunction{accept₂}\AgdaSpace{}%
\AgdaBound{s}\<%
\\
\>[0]\AgdaFunction{correct-opVerify-from}\AgdaSpace{}%
\AgdaOperator{\AgdaInductiveConstructor{⟨}}\AgdaSpace{}%
\AgdaBound{time}\AgdaSpace{}%
\AgdaOperator{\AgdaInductiveConstructor{,}}\AgdaSpace{}%
\AgdaBound{msg₁}\AgdaSpace{}%
\AgdaOperator{\AgdaInductiveConstructor{,}}\AgdaSpace{}%
\AgdaInductiveConstructor{suc}\AgdaSpace{}%
\AgdaBound{x}\AgdaSpace{}%
\AgdaOperator{\AgdaInductiveConstructor{∷}}\AgdaSpace{}%
\AgdaBound{x₁}\AgdaSpace{}%
\AgdaOperator{\AgdaInductiveConstructor{∷}}\AgdaSpace{}%
\AgdaBound{x₂}\AgdaSpace{}%
\AgdaOperator{\AgdaInductiveConstructor{∷}}\AgdaSpace{}%
\AgdaBound{stack₁}\AgdaSpace{}%
\AgdaOperator{\AgdaInductiveConstructor{⟩}}\AgdaSpace{}%
\AgdaBound{p}\AgdaSpace{}%
\AgdaSymbol{=}\AgdaSpace{}%
\AgdaBound{p}\<%
\\
\>[0]\<%
\end{code}
} 

\newcommand{\stackVerificationPtoPKHcorrecttwo}{
\begin{code}[inline]%
\>[0]\AgdaFunction{correct-opVerify}\AgdaSpace{}%
\AgdaSymbol{:}\AgdaSpace{}%
\AgdaOperator{\AgdaRecord{<}}\AgdaSpace{}%
\AgdaFunction{accept₂}\AgdaSpace{}%
\AgdaOperator{\AgdaRecord{>ⁱᶠᶠ}}\AgdaSpace{}%
\>[35]\AgdaSymbol{(}\AgdaOperator{\AgdaFunction{[}}\AgdaSpace{}%
\AgdaInductiveConstructor{opVerify}\AgdaSpace{}%
\AgdaOperator{\AgdaFunction{]}}\AgdaSymbol{)}\AgdaSpace{}%
\AgdaOperator{\AgdaRecord{<}}\AgdaSpace{}%
\AgdaFunction{accept₁}\AgdaSpace{}%
\AgdaOperator{\AgdaRecord{>}}\<%
\end{code}
} 

\AgdaHide{
\begin{code}%
\>[0]\AgdaFunction{correct-opVerify}\AgdaSpace{}%
\AgdaSymbol{.}\AgdaField{==>}\AgdaSpace{}%
\AgdaSymbol{=}\AgdaSpace{}%
\AgdaFunction{correct-opVerify-to}\<%
\\
\>[0]\AgdaFunction{correct-opVerify}\AgdaSpace{}%
\AgdaSymbol{.}\AgdaField{<==}\AgdaSpace{}%
\AgdaSymbol{=}\AgdaSpace{}%
\AgdaFunction{correct-opVerify-from}\<%
\\
\\[\AgdaEmptyExtraSkip]%
\>[0]\AgdaComment{\{-old}\<%
\\
\>[0]\AgdaComment{correct-2-to\ :\ (s\ :\ StackState)\ →\ accept₂\ s\ →\ (accept₁\ ⁺)\ (⟦\ opVerify\ ⟧s\ s\ )}\<%
\\
\>[0]\AgdaComment{correct-2-to\ ⟨\ time\ ,\ msg₁\ ,\ suc\ x\ ∷\ x₁\ ∷\ x₂\ ∷\ stack₁\ ⟩\ p\ =\ p}\<%
\\
\>[0]\<%
\\
\>[0]\AgdaComment{correct-2-from\ :\ (s\ :\ StackState)\ →\ (accept₁\ ⁺)\ (⟦\ opVerify\ ⟧s\ s\ )\ →\ accept₂\ s}\<%
\\
\>[0]\AgdaComment{correct-2-from\ ⟨\ time\ ,\ msg₁\ ,\ suc\ x\ ∷\ x₁\ ∷\ x₂\ ∷\ stack₁\ ⟩\ p\ =\ p}\<%
\\
\>[0]\<%
\\
\>[0]\AgdaComment{-@BEGININLINE@correcttwo}\<%
\\
\>[0]\AgdaComment{correct-2\ :\ <\ accept₂\ >iff\ \ ([\ opVerify\ ])\ <\ accept₁\ >}\<%
\\
\>[0]\AgdaComment{-@END}\<%
\\
\>[0]\AgdaComment{correct-2\ .==>\ =\ correct-2-to}\<%
\\
\>[0]\AgdaComment{correct-2\ .<==\ =\ correct-2-from}\<%
\\
\>[0]\AgdaComment{-\}}\<%
\\
\\[\AgdaEmptyExtraSkip]%
\\[\AgdaEmptyExtraSkip]%
\>[0]\AgdaFunction{correct-opEqual-to}\AgdaSpace{}%
\AgdaSymbol{:}\AgdaSpace{}%
\AgdaSymbol{(}\AgdaBound{s}\AgdaSpace{}%
\AgdaSymbol{:}\AgdaSpace{}%
\AgdaRecord{StackState}\AgdaSymbol{)}\AgdaSpace{}%
\AgdaSymbol{→}\AgdaSpace{}%
\AgdaFunction{accept₃}\AgdaSpace{}%
\AgdaBound{s}\AgdaSpace{}%
\AgdaSymbol{→}\AgdaSpace{}%
\AgdaSymbol{(}\AgdaFunction{accept₂}\AgdaSpace{}%
\AgdaOperator{\AgdaFunction{⁺}}\AgdaSymbol{)}\AgdaSpace{}%
\AgdaSymbol{(}\AgdaOperator{\AgdaFunction{⟦}}\AgdaSpace{}%
\AgdaInductiveConstructor{opEqual}\AgdaSpace{}%
\AgdaOperator{\AgdaFunction{⟧ₛ}}\AgdaSpace{}%
\AgdaBound{s}\AgdaSpace{}%
\AgdaSymbol{)}\<%
\\
\>[0]\AgdaFunction{correct-opEqual-to}\AgdaSpace{}%
\AgdaOperator{\AgdaInductiveConstructor{⟨}}\AgdaSpace{}%
\AgdaBound{time}\AgdaSpace{}%
\AgdaOperator{\AgdaInductiveConstructor{,}}\AgdaSpace{}%
\AgdaBound{msg₁}\AgdaSpace{}%
\AgdaOperator{\AgdaInductiveConstructor{,}}\AgdaSpace{}%
\AgdaBound{pbk1}%
\>[41]\AgdaOperator{\AgdaInductiveConstructor{∷}}\AgdaSpace{}%
\AgdaDottedPattern{\AgdaSymbol{.}}\AgdaDottedPattern{\AgdaBound{pbk1}}\AgdaSpace{}%
\AgdaOperator{\AgdaInductiveConstructor{∷}}\AgdaSpace{}%
\AgdaBound{pbk2}\AgdaSpace{}%
\AgdaOperator{\AgdaInductiveConstructor{∷}}%
\>[59]\AgdaBound{sig}\AgdaSpace{}%
\AgdaOperator{\AgdaInductiveConstructor{∷}}\AgdaSpace{}%
\AgdaInductiveConstructor{[]}%
\>[69]\AgdaOperator{\AgdaInductiveConstructor{⟩}}\AgdaSpace{}%
\AgdaSymbol{(}\AgdaInductiveConstructor{conj}\AgdaSpace{}%
\AgdaInductiveConstructor{refl}\AgdaSpace{}%
\AgdaBound{checkSig}\AgdaSymbol{)}%
\>[93]\AgdaKeyword{rewrite}\AgdaSpace{}%
\AgdaSymbol{(}\AgdaSpace{}%
\AgdaFunction{lemmaCompareNat}\AgdaSpace{}%
\AgdaBound{pbk1}\AgdaSpace{}%
\AgdaSymbol{)}\AgdaSpace{}%
\AgdaSymbol{=}%
\>[129]\AgdaBound{checkSig}\<%
\\
\>[0]\AgdaFunction{correct-opEqual-to}\AgdaSpace{}%
\AgdaOperator{\AgdaInductiveConstructor{⟨}}\AgdaSpace{}%
\AgdaBound{time}\AgdaSpace{}%
\AgdaOperator{\AgdaInductiveConstructor{,}}\AgdaSpace{}%
\AgdaBound{msg₁}\AgdaSpace{}%
\AgdaOperator{\AgdaInductiveConstructor{,}}\AgdaSpace{}%
\AgdaBound{pbk1}\AgdaSpace{}%
\AgdaOperator{\AgdaInductiveConstructor{∷}}\AgdaSpace{}%
\AgdaDottedPattern{\AgdaSymbol{.}}\AgdaDottedPattern{\AgdaBound{pbk1}}\AgdaSpace{}%
\AgdaOperator{\AgdaInductiveConstructor{∷}}\AgdaSpace{}%
\AgdaBound{pbk2}%
\>[56]\AgdaOperator{\AgdaInductiveConstructor{∷}}\AgdaSpace{}%
\AgdaBound{sig}\AgdaSpace{}%
\AgdaOperator{\AgdaInductiveConstructor{∷}}\AgdaSpace{}%
\AgdaBound{x}\AgdaSpace{}%
\AgdaOperator{\AgdaInductiveConstructor{∷}}\AgdaSpace{}%
\AgdaBound{rest}%
\>[74]\AgdaOperator{\AgdaInductiveConstructor{⟩}}\AgdaSpace{}%
\AgdaSymbol{(}\AgdaInductiveConstructor{conj}\AgdaSpace{}%
\AgdaInductiveConstructor{refl}\AgdaSpace{}%
\AgdaBound{checkSig}\AgdaSymbol{)}\AgdaSpace{}%
\AgdaKeyword{rewrite}%
\>[106]\AgdaSymbol{(}\AgdaSpace{}%
\AgdaFunction{lemmaCompareNat}\AgdaSpace{}%
\AgdaBound{pbk1}\AgdaSpace{}%
\AgdaSymbol{)}\AgdaSpace{}%
\AgdaSymbol{=}\AgdaSpace{}%
\AgdaBound{checkSig}\<%
\\
\\[\AgdaEmptyExtraSkip]%
\>[0]\AgdaFunction{correct-opEqual-from}%
\>[22]\AgdaSymbol{:}\AgdaSpace{}%
\AgdaSymbol{(}\AgdaBound{s}\AgdaSpace{}%
\AgdaSymbol{:}\AgdaSpace{}%
\AgdaRecord{StackState}\AgdaSymbol{)}%
\>[42]\AgdaSymbol{→}\AgdaSpace{}%
\AgdaSymbol{(}\AgdaFunction{accept₂}\AgdaSpace{}%
\AgdaOperator{\AgdaFunction{⁺}}\AgdaSymbol{)}\AgdaSpace{}%
\AgdaSymbol{(}\AgdaOperator{\AgdaFunction{⟦}}\AgdaSpace{}%
\AgdaInductiveConstructor{opEqual}\AgdaSpace{}%
\AgdaOperator{\AgdaFunction{⟧ₛ}}\AgdaSpace{}%
\AgdaBound{s}\AgdaSpace{}%
\AgdaSymbol{)}\AgdaSpace{}%
\AgdaSymbol{→}\AgdaSpace{}%
\AgdaFunction{accept₃}\AgdaSpace{}%
\AgdaBound{s}\<%
\\
\>[0]\AgdaFunction{correct-opEqual-from}\AgdaSpace{}%
\AgdaOperator{\AgdaInductiveConstructor{⟨}}\AgdaSpace{}%
\AgdaBound{time}\AgdaSpace{}%
\AgdaOperator{\AgdaInductiveConstructor{,}}\AgdaSpace{}%
\AgdaBound{msg₁}\AgdaSpace{}%
\AgdaOperator{\AgdaInductiveConstructor{,}}\AgdaSpace{}%
\AgdaBound{x}\AgdaSpace{}%
\AgdaOperator{\AgdaInductiveConstructor{∷}}\AgdaSpace{}%
\AgdaBound{x₁}\AgdaSpace{}%
\AgdaOperator{\AgdaInductiveConstructor{∷}}\AgdaSpace{}%
\AgdaBound{pbk}\AgdaSpace{}%
\AgdaOperator{\AgdaInductiveConstructor{∷}}\AgdaSpace{}%
\AgdaBound{sig}\AgdaSpace{}%
\AgdaOperator{\AgdaInductiveConstructor{∷}}\AgdaSpace{}%
\AgdaBound{stack₁}%
\>[66]\AgdaOperator{\AgdaInductiveConstructor{⟩}}\AgdaSpace{}%
\AgdaBound{p}\AgdaSpace{}%
\AgdaKeyword{rewrite}\AgdaSpace{}%
\AgdaSymbol{(}\AgdaSpace{}%
\AgdaFunction{lemmaCorrect3From}\AgdaSpace{}%
\AgdaBound{x}\AgdaSpace{}%
\AgdaBound{x₁}\AgdaSpace{}%
\AgdaBound{pbk}\AgdaSpace{}%
\AgdaBound{sig}\AgdaSpace{}%
\AgdaBound{time}\AgdaSpace{}%
\AgdaBound{msg₁}\AgdaSpace{}%
\AgdaBound{p}\AgdaSpace{}%
\AgdaSymbol{)}\<%
\\
\>[0][@{}l@{\AgdaIndent{0}}]%
\>[2]\AgdaSymbol{=}%
\>[6]\AgdaKeyword{let}\<%
\\
\>[6][@{}l@{\AgdaIndent{0}}]%
\>[8]\AgdaBound{q}\AgdaSpace{}%
\AgdaSymbol{:}\AgdaSpace{}%
\AgdaFunction{True}\AgdaSpace{}%
\AgdaSymbol{(}\AgdaFunction{isSigned}%
\>[28]\AgdaBound{msg₁}\AgdaSpace{}%
\AgdaBound{sig}\AgdaSpace{}%
\AgdaBound{pbk}\AgdaSymbol{)}\<%
\\
\>[8]\AgdaBound{q}\AgdaSpace{}%
\AgdaSymbol{=}\AgdaSpace{}%
\AgdaFunction{correct3Aux2}\AgdaSpace{}%
\AgdaSymbol{(}\AgdaFunction{compareNaturals}\AgdaSpace{}%
\AgdaBound{x}\AgdaSpace{}%
\AgdaBound{x₁}\AgdaSymbol{)}\AgdaSpace{}%
\AgdaBound{pbk}\AgdaSpace{}%
\AgdaBound{sig}\AgdaSpace{}%
\AgdaBound{stack₁}\AgdaSpace{}%
\AgdaBound{time}\AgdaSpace{}%
\AgdaBound{msg₁}\AgdaSpace{}%
\AgdaBound{p}\<%
\\
\>[6]\AgdaKeyword{in}\AgdaSpace{}%
\AgdaSymbol{(}\AgdaInductiveConstructor{conj}\AgdaSpace{}%
\AgdaInductiveConstructor{refl}\AgdaSpace{}%
\AgdaBound{q}\AgdaSymbol{)}\<%
\\
\>[0]\<%
\end{code}
} 

\newcommand{\stackVerificationPtoPKHcorrectthree}{
\begin{code}[inline]%
\>[0]\AgdaFunction{correct-opEqual}\AgdaSpace{}%
\AgdaSymbol{:}\AgdaSpace{}%
\AgdaOperator{\AgdaRecord{<}}\AgdaSpace{}%
\AgdaFunction{accept₃}\AgdaSpace{}%
\AgdaOperator{\AgdaRecord{>ⁱᶠᶠ}}\AgdaSpace{}%
\>[34]\AgdaSymbol{(}\AgdaOperator{\AgdaFunction{[}}\AgdaSpace{}%
\AgdaInductiveConstructor{opEqual}\AgdaSpace{}%
\AgdaOperator{\AgdaFunction{]}}\AgdaSymbol{)}\AgdaSpace{}%
\AgdaOperator{\AgdaRecord{<}}\AgdaSpace{}%
\AgdaFunction{accept₂}\AgdaSpace{}%
\AgdaOperator{\AgdaRecord{>}}\<%
\end{code}
} 

\AgdaHide{
\begin{code}%
\>[0]\AgdaFunction{correct-opEqual}\AgdaSpace{}%
\AgdaSymbol{.}\AgdaField{==>}\AgdaSpace{}%
\AgdaSymbol{=}\AgdaSpace{}%
\AgdaFunction{correct-opEqual-to}\<%
\\
\>[0]\AgdaFunction{correct-opEqual}\AgdaSpace{}%
\AgdaSymbol{.}\AgdaField{<==}\AgdaSpace{}%
\AgdaSymbol{=}\AgdaSpace{}%
\AgdaFunction{correct-opEqual-from}\<%
\\
\\[\AgdaEmptyExtraSkip]%
\>[0]\AgdaComment{\{-old}\<%
\\
\>[0]\<%
\\
\>[0]\AgdaComment{correct-3-to\ :\ (s\ :\ StackState)\ →\ accept₃\ s\ →\ (accept₂\ ⁺)\ (⟦\ opEqual\ ⟧s\ s\ )}\<%
\\
\>[0]\AgdaComment{correct-3-to\ ⟨\ time\ ,\ msg₁\ ,\ pbk1\ \ ∷\ .pbk1\ ∷\ pbk2\ ∷\ \ sig\ ∷\ []\ \ ⟩\ (conj\ refl\ checkSig)\ \ rewrite\ (\ lemmaCompareNat\ pbk1\ )\ =\ \ checkSig}\<%
\\
\>[0]\AgdaComment{correct-3-to\ ⟨\ time\ ,\ msg₁\ ,\ pbk1\ ∷\ .pbk1\ ∷\ pbk2\ \ ∷\ sig\ ∷\ x\ ∷\ rest\ \ ⟩\ (conj\ refl\ checkSig)\ rewrite\ \ (\ lemmaCompareNat\ pbk1\ )\ =\ checkSig}\<%
\\
\>[0]\<%
\\
\>[0]\AgdaComment{correct-3-from\ \ :\ (s\ :\ StackState)\ \ →\ (accept₂\ ⁺)\ (⟦\ opEqual\ ⟧s\ s\ )\ →\ accept₃\ s}\<%
\\
\>[0]\AgdaComment{correct-3-from\ ⟨\ time\ ,\ msg₁\ ,\ x\ ∷\ x₁\ ∷\ pbk\ ∷\ sig\ ∷\ stack₁\ \ ⟩\ p\ rewrite\ (\ lemmaCorrect3From\ x\ x₁\ pbk\ sig\ time\ msg₁\ p\ )}\<%
\\
\>[0]\AgdaComment{\ \ =\ \ \ let}\<%
\\
\>[0]\AgdaComment{\ \ \ \ \ \ \ \ q\ :\ True\ (isSigned\ \ msg₁\ sig\ pbk)}\<%
\\
\>[0]\AgdaComment{\ \ \ \ \ \ \ \ q\ =\ correct3Aux2\ (compareNaturals\ x\ x₁)\ pbk\ sig\ stack₁\ time\ msg₁\ p}\<%
\\
\>[0]\AgdaComment{\ \ \ \ \ \ in\ (conj\ refl\ q)}\<%
\\
\>[0]\<%
\\
\>[0]\AgdaComment{-@BEGININLINE@correctthree}\<%
\\
\>[0]\AgdaComment{correct-3\ :\ <\ accept₃\ >iff\ \ ([\ opEqual\ ])\ <\ accept₂\ >}\<%
\\
\>[0]\AgdaComment{-@END}\<%
\\
\>[0]\AgdaComment{correct-3\ .==>\ =\ correct-3-to}\<%
\\
\>[0]\AgdaComment{correct-3\ .<==\ =\ correct-3-from}\<%
\\
\>[0]\AgdaComment{-\}}\<%
\\
\\[\AgdaEmptyExtraSkip]%
\>[0]\AgdaComment{--\ needs\ to\ be\ pbkh\ since\ opPush\ refers\ to\ it}\<%
\\
\>[0]\AgdaFunction{correct-opPush-to}\AgdaSpace{}%
\AgdaSymbol{:}\AgdaSpace{}%
\AgdaSymbol{(}\AgdaSpace{}%
\AgdaBound{pbkh}\AgdaSpace{}%
\AgdaSymbol{:}\AgdaSpace{}%
\AgdaDatatype{ℕ}\AgdaSpace{}%
\AgdaSymbol{)}\AgdaSpace{}%
\AgdaSymbol{→}%
\>[36]\AgdaSymbol{(}\AgdaBound{s}\AgdaSpace{}%
\AgdaSymbol{:}\AgdaSpace{}%
\AgdaRecord{StackState}\AgdaSymbol{)}\AgdaSpace{}%
\AgdaSymbol{→}\AgdaSpace{}%
\AgdaFunction{accept₄}\AgdaSpace{}%
\AgdaBound{pbkh}%
\>[69]\AgdaBound{s}\AgdaSpace{}%
\AgdaSymbol{→}\AgdaSpace{}%
\AgdaSymbol{(}\AgdaFunction{accept₃}\AgdaSpace{}%
\AgdaOperator{\AgdaFunction{⁺}}\AgdaSymbol{)}\AgdaSpace{}%
\AgdaSymbol{(}\AgdaOperator{\AgdaFunction{⟦}}\AgdaSpace{}%
\AgdaInductiveConstructor{opPush}\AgdaSpace{}%
\AgdaBound{pbkh}\AgdaSpace{}%
\AgdaOperator{\AgdaFunction{⟧ₛ}}\AgdaSpace{}%
\AgdaBound{s}\AgdaSpace{}%
\AgdaSymbol{)}\<%
\\
\>[0]\AgdaFunction{correct-opPush-to}\AgdaSpace{}%
\AgdaBound{pbkh}\AgdaSpace{}%
\AgdaOperator{\AgdaInductiveConstructor{⟨}}\AgdaSpace{}%
\AgdaBound{currentTime₁}\AgdaSpace{}%
\AgdaOperator{\AgdaInductiveConstructor{,}}\AgdaSpace{}%
\AgdaBound{msg₁}\AgdaSpace{}%
\AgdaOperator{\AgdaInductiveConstructor{,}}\AgdaSpace{}%
\AgdaDottedPattern{\AgdaSymbol{.}}\AgdaDottedPattern{\AgdaBound{pbkh}}\AgdaSpace{}%
\AgdaOperator{\AgdaInductiveConstructor{∷}}\AgdaSpace{}%
\AgdaBound{x₁}\AgdaSpace{}%
\AgdaOperator{\AgdaInductiveConstructor{∷}}\AgdaSpace{}%
\AgdaBound{x₂}\AgdaSpace{}%
\AgdaOperator{\AgdaInductiveConstructor{∷}}\AgdaSpace{}%
\AgdaBound{stack₁}\AgdaSpace{}%
\AgdaOperator{\AgdaInductiveConstructor{⟩}}\AgdaSpace{}%
\AgdaSymbol{(}\AgdaInductiveConstructor{conj}\AgdaSpace{}%
\AgdaInductiveConstructor{refl}\AgdaSpace{}%
\AgdaBound{and4}\AgdaSymbol{)}\AgdaSpace{}%
\AgdaSymbol{=}\AgdaSpace{}%
\AgdaInductiveConstructor{conj}\AgdaSpace{}%
\AgdaInductiveConstructor{refl}\AgdaSpace{}%
\AgdaBound{and4}\<%
\\
\\[\AgdaEmptyExtraSkip]%
\\[\AgdaEmptyExtraSkip]%
\>[0]\AgdaFunction{correct-opPush-from}\AgdaSpace{}%
\AgdaSymbol{:}\AgdaSpace{}%
\AgdaSymbol{(}\AgdaSpace{}%
\AgdaBound{pbkh}\AgdaSpace{}%
\AgdaSymbol{:}\AgdaSpace{}%
\AgdaDatatype{ℕ}\AgdaSpace{}%
\AgdaSymbol{)}\AgdaSpace{}%
\AgdaSymbol{→}%
\>[38]\AgdaSymbol{(}\AgdaBound{s}\AgdaSpace{}%
\AgdaSymbol{:}\AgdaSpace{}%
\AgdaRecord{StackState}\AgdaSymbol{)}\AgdaSpace{}%
\AgdaSymbol{→}\AgdaSpace{}%
\AgdaSymbol{(}\AgdaFunction{accept₃}\AgdaSpace{}%
\AgdaOperator{\AgdaFunction{⁺}}\AgdaSymbol{)}\AgdaSpace{}%
\AgdaSymbol{(}\AgdaOperator{\AgdaFunction{⟦}}\AgdaSpace{}%
\AgdaInductiveConstructor{opPush}\AgdaSpace{}%
\AgdaBound{pbkh}\AgdaSpace{}%
\AgdaOperator{\AgdaFunction{⟧ₛ}}\AgdaSpace{}%
\AgdaBound{s}\AgdaSpace{}%
\AgdaSymbol{)}\AgdaSpace{}%
\AgdaSymbol{→}\AgdaSpace{}%
\AgdaFunction{accept₄}\AgdaSpace{}%
\AgdaBound{pbkh}%
\>[107]\AgdaBound{s}\<%
\\
\>[0]\AgdaFunction{correct-opPush-from}\AgdaSpace{}%
\AgdaBound{pbkh}\AgdaSpace{}%
\AgdaOperator{\AgdaInductiveConstructor{⟨}}\AgdaSpace{}%
\AgdaBound{currentTime₁}\AgdaSpace{}%
\AgdaOperator{\AgdaInductiveConstructor{,}}\AgdaSpace{}%
\AgdaBound{msg₁}\AgdaSpace{}%
\AgdaOperator{\AgdaInductiveConstructor{,}}\AgdaSpace{}%
\AgdaDottedPattern{\AgdaSymbol{.}}\AgdaDottedPattern{\AgdaBound{pbkh}}\AgdaSpace{}%
\AgdaOperator{\AgdaInductiveConstructor{∷}}\AgdaSpace{}%
\AgdaBound{x₁}\AgdaSpace{}%
\AgdaOperator{\AgdaInductiveConstructor{∷}}\AgdaSpace{}%
\AgdaBound{x₂}\AgdaSpace{}%
\AgdaOperator{\AgdaInductiveConstructor{∷}}\AgdaSpace{}%
\AgdaBound{stack₁}%
\>[76]\AgdaOperator{\AgdaInductiveConstructor{⟩}}\AgdaSpace{}%
\AgdaSymbol{(}\AgdaInductiveConstructor{conj}\AgdaSpace{}%
\AgdaInductiveConstructor{refl}\AgdaSpace{}%
\AgdaBound{and4}\AgdaSymbol{)}\AgdaSpace{}%
\AgdaSymbol{=}\AgdaSpace{}%
\AgdaInductiveConstructor{conj}\AgdaSpace{}%
\AgdaInductiveConstructor{refl}\AgdaSpace{}%
\AgdaBound{and4}\<%
\\
\>[0]\<%
\end{code}
} 

\newcommand{\stackVerificationPtoPKHcorrectfour}{
\begin{code}[inline]%
\>[0]\AgdaFunction{correct-opPush}\AgdaSpace{}%
\AgdaSymbol{: (}%
\AgdaBound{pbkh}\AgdaSpace{}%
\AgdaSymbol{:}\AgdaSpace{}%
\AgdaDatatype{ℕ}%
\AgdaSymbol{)}\AgdaSpace{}%
\AgdaSymbol{→}\AgdaSpace{}%
\>[32]\AgdaOperator{\AgdaRecord{<}}\AgdaSpace{}%
\AgdaFunction{accept₄}\AgdaSpace{}%
\AgdaBound{pbkh}\AgdaSpace{}%
\AgdaOperator{\AgdaRecord{>ⁱᶠᶠ}}\AgdaSpace{}%
\>[53]\AgdaSymbol{(}\AgdaOperator{\AgdaFunction{[}}\AgdaSpace{}%
\AgdaInductiveConstructor{opPush}\AgdaSpace{}%
\AgdaBound{pbkh}\AgdaSpace{}%
\AgdaOperator{\AgdaFunction{]}}\AgdaSymbol{)}\AgdaSpace{}%
\AgdaOperator{\AgdaRecord{<}}\AgdaSpace{}%
\AgdaFunction{accept₃}\AgdaSpace{}%
\AgdaOperator{\AgdaRecord{>}}\<%
\end{code}
} 

\AgdaHide{
\begin{code}%
\>[0]\AgdaFunction{correct-opPush}\AgdaSpace{}%
\AgdaBound{pbkh}\AgdaSpace{}%
\AgdaSymbol{.}\AgdaField{==>}\AgdaSpace{}%
\AgdaSymbol{=}\AgdaSpace{}%
\AgdaFunction{correct-opPush-to}\AgdaSpace{}%
\AgdaBound{pbkh}\<%
\\
\>[0]\AgdaFunction{correct-opPush}\AgdaSpace{}%
\AgdaBound{pbkh}\AgdaSpace{}%
\AgdaSymbol{.}\AgdaField{<==}\AgdaSpace{}%
\AgdaSymbol{=}\AgdaSpace{}%
\AgdaFunction{correct-opPush-from}\AgdaSpace{}%
\AgdaBound{pbkh}\<%
\\
\\[\AgdaEmptyExtraSkip]%
\\[\AgdaEmptyExtraSkip]%
\>[0]\AgdaComment{\{-old}\<%
\\
\>[0]\<%
\\
\>[0]\<%
\\
\>[0]\AgdaComment{correct-4-to\ :\ (\ pbk\ :\ ℕ\ )\ →\ \ (s\ :\ StackState)\ →\ accept₄\ pbk\ \ s\ →\ (accept₃\ ⁺)\ (⟦\ opPush\ pbk\ ⟧s\ s\ )}\<%
\\
\>[0]\AgdaComment{correct-4-to\ pbk\ ⟨\ currentTime₁\ ,\ msg₁\ ,\ .pbk\ ∷\ x₁\ ∷\ x₂\ ∷\ stack₁\ ⟩\ (conj\ refl\ and4)\ =\ conj\ refl\ and4}\<%
\\
\>[0]\<%
\\
\>[0]\<%
\\
\>[0]\AgdaComment{correct-4-from\ :\ (\ pbk\ :\ ℕ\ )\ →\ \ (s\ :\ StackState)\ →\ (accept₃\ ⁺)\ (⟦\ opPush\ pbk\ ⟧s\ s\ )\ →\ accept₄\ pbk\ \ s}\<%
\\
\>[0]\AgdaComment{correct-4-from\ pbk\ ⟨\ currentTime₁\ ,\ msg₁\ ,\ .pbk\ ∷\ x₁\ ∷\ x₂\ ∷\ stack₁\ \ \ ⟩\ (conj\ refl\ and4)\ =\ conj\ refl\ and4}\<%
\\
\>[0]\<%
\\
\>[0]\AgdaComment{-@BEGININLINE@correctfour}\<%
\\
\>[0]\AgdaComment{correct-4\ :(\ pbk\ :\ ℕ\ )\ →\ \ <\ accept₄\ pbk\ >iff\ \ ([\ opPush\ pbk\ ])\ <\ accept₃\ >}\<%
\\
\>[0]\AgdaComment{-@END}\<%
\\
\>[0]\AgdaComment{correct-4\ pbk\ .==>\ =\ correct-4-to\ pbk}\<%
\\
\>[0]\AgdaComment{correct-4\ pbk\ .<==\ =\ correct-4-from\ pbk}\<%
\\
\>[0]\<%
\\
\>[0]\AgdaComment{-\}}\<%
\\
\\[\AgdaEmptyExtraSkip]%
\>[0]\AgdaComment{--\ needs\ to\ be\ pbkh\ since\ accept₄\ and\ accept₅\ refer\ to\ pbkh}\<%
\\
\>[0]\AgdaFunction{correct-opHash-to}\AgdaSpace{}%
\AgdaSymbol{:}\AgdaSpace{}%
\AgdaSymbol{(}\AgdaBound{pbkh}\AgdaSpace{}%
\AgdaSymbol{:}\AgdaSpace{}%
\AgdaDatatype{ℕ}\AgdaSpace{}%
\AgdaSymbol{)}\AgdaSpace{}%
\AgdaSymbol{→}\AgdaSpace{}%
\AgdaSymbol{(}\AgdaBound{s}\AgdaSpace{}%
\AgdaSymbol{:}\AgdaSpace{}%
\AgdaRecord{StackState}\AgdaSymbol{)}\AgdaSpace{}%
\AgdaSymbol{→}\AgdaSpace{}%
\AgdaFunction{accept₅}\AgdaSpace{}%
\AgdaBound{pbkh}\AgdaSpace{}%
\AgdaBound{s}\AgdaSpace{}%
\AgdaSymbol{→}%
\>[71]\AgdaSymbol{((}\AgdaSpace{}%
\AgdaFunction{accept₄}\AgdaSpace{}%
\AgdaBound{pbkh}\AgdaSpace{}%
\AgdaSymbol{)}\AgdaSpace{}%
\AgdaOperator{\AgdaFunction{⁺}}\AgdaSymbol{)}\AgdaSpace{}%
\AgdaSymbol{(}\AgdaOperator{\AgdaFunction{⟦}}\AgdaSpace{}%
\AgdaInductiveConstructor{opHash}\AgdaSpace{}%
\AgdaOperator{\AgdaFunction{⟧ₛ}}\AgdaSpace{}%
\AgdaBound{s}\AgdaSpace{}%
\AgdaSymbol{)}\<%
\\
\>[0]\AgdaFunction{correct-opHash-to}\AgdaSpace{}%
\AgdaBound{pbkh}\AgdaSpace{}%
\AgdaOperator{\AgdaInductiveConstructor{⟨}}\AgdaSpace{}%
\AgdaBound{time}\AgdaSpace{}%
\AgdaOperator{\AgdaInductiveConstructor{,}}\AgdaSpace{}%
\AgdaBound{msg₁}\AgdaSpace{}%
\AgdaOperator{\AgdaInductiveConstructor{,}}\AgdaSpace{}%
\AgdaBound{x}\AgdaSpace{}%
\AgdaOperator{\AgdaInductiveConstructor{∷}}\AgdaSpace{}%
\AgdaBound{x₁}\AgdaSpace{}%
\AgdaOperator{\AgdaInductiveConstructor{∷}}\AgdaSpace{}%
\AgdaBound{x₂}\AgdaSpace{}%
\AgdaOperator{\AgdaInductiveConstructor{∷}}\AgdaSpace{}%
\AgdaBound{stack₁}\AgdaSpace{}%
\AgdaOperator{\AgdaInductiveConstructor{⟩}}\AgdaSpace{}%
\AgdaSymbol{(}\AgdaInductiveConstructor{conj}\AgdaSpace{}%
\AgdaInductiveConstructor{refl}\AgdaSpace{}%
\AgdaBound{checkSig}\AgdaSymbol{)}\AgdaSpace{}%
\AgdaSymbol{=}\AgdaSpace{}%
\AgdaSymbol{(}\AgdaInductiveConstructor{conj}\AgdaSpace{}%
\AgdaInductiveConstructor{refl}\AgdaSpace{}%
\AgdaBound{checkSig}\AgdaSymbol{)}\<%
\\
\\[\AgdaEmptyExtraSkip]%
\>[0]\AgdaFunction{correct-opHash-from}\AgdaSpace{}%
\AgdaSymbol{:}\AgdaSpace{}%
\AgdaSymbol{(}\AgdaSpace{}%
\AgdaBound{pbkh}\AgdaSpace{}%
\AgdaSymbol{:}\AgdaSpace{}%
\AgdaDatatype{ℕ}\AgdaSpace{}%
\AgdaSymbol{)}\AgdaSpace{}%
\AgdaSymbol{→}%
\>[38]\AgdaSymbol{(}\AgdaBound{s}\AgdaSpace{}%
\AgdaSymbol{:}\AgdaSpace{}%
\AgdaRecord{StackState}\AgdaSymbol{)}%
\>[56]\AgdaSymbol{→}\AgdaSpace{}%
\AgdaSymbol{((}\AgdaSpace{}%
\AgdaFunction{accept₄}\AgdaSpace{}%
\AgdaBound{pbkh}\AgdaSymbol{)}\AgdaSpace{}%
\AgdaOperator{\AgdaFunction{⁺}}\AgdaSymbol{)}\AgdaSpace{}%
\AgdaSymbol{(}\AgdaOperator{\AgdaFunction{⟦}}\AgdaSpace{}%
\AgdaInductiveConstructor{opHash}\AgdaSpace{}%
\AgdaOperator{\AgdaFunction{⟧ₛ}}\AgdaSpace{}%
\AgdaBound{s}\AgdaSpace{}%
\AgdaSymbol{)}\AgdaSpace{}%
\AgdaSymbol{→}\AgdaSpace{}%
\AgdaFunction{accept₅}\AgdaSpace{}%
\AgdaBound{pbkh}%
\>[111]\AgdaBound{s}\<%
\\
\>[0]\AgdaFunction{correct-opHash-from}\AgdaSpace{}%
\AgdaDottedPattern{\AgdaSymbol{.(}}\AgdaDottedPattern{\AgdaFunction{hashFun}}\AgdaSpace{}%
\AgdaDottedPattern{\AgdaBound{x}}\AgdaDottedPattern{\AgdaSymbol{)}}\AgdaSpace{}%
\AgdaOperator{\AgdaInductiveConstructor{⟨}}\AgdaSpace{}%
\AgdaBound{time}\AgdaSpace{}%
\AgdaOperator{\AgdaInductiveConstructor{,}}\AgdaSpace{}%
\AgdaBound{msg₁}\AgdaSpace{}%
\AgdaOperator{\AgdaInductiveConstructor{,}}\AgdaSpace{}%
\AgdaBound{x}\AgdaSpace{}%
\AgdaOperator{\AgdaInductiveConstructor{∷}}\AgdaSpace{}%
\AgdaBound{x₁}\AgdaSpace{}%
\AgdaOperator{\AgdaInductiveConstructor{∷}}\AgdaSpace{}%
\AgdaBound{x₂}\AgdaSpace{}%
\AgdaOperator{\AgdaInductiveConstructor{∷}}\AgdaSpace{}%
\AgdaBound{stack₁}\AgdaSpace{}%
\AgdaOperator{\AgdaInductiveConstructor{⟩}}\AgdaSpace{}%
\AgdaSymbol{(}\AgdaInductiveConstructor{conj}\AgdaSpace{}%
\AgdaInductiveConstructor{refl}\AgdaSpace{}%
\AgdaBound{checkSig}\AgdaSymbol{)}\AgdaSpace{}%
\AgdaSymbol{=}\AgdaSpace{}%
\AgdaInductiveConstructor{conj}\AgdaSpace{}%
\AgdaInductiveConstructor{refl}\AgdaSpace{}%
\AgdaBound{checkSig}\<%
\\
\>[0]\<%
\end{code}
} 

\newcommand{\stackVerificationPtoPKHcorrectfive}{
\begin{code}[inline]%
\>[0]\AgdaFunction{correct-opHash}\AgdaSpace{}%
\AgdaSymbol{: (}%
\AgdaBound{pbkh}\AgdaSpace{}%
\AgdaSymbol{:}\AgdaSpace{}%
\AgdaDatatype{ℕ}%
\AgdaSymbol{)}\AgdaSpace{}%
\AgdaSymbol{→}\AgdaSpace{}%
\>[32]\AgdaOperator{\AgdaRecord{<}}\AgdaSpace{}%
\AgdaFunction{accept₅}\AgdaSpace{}%
\AgdaBound{pbkh}\AgdaSpace{}%
\AgdaOperator{\AgdaRecord{>ⁱᶠᶠ}}\AgdaSpace{}%
\>[53]\AgdaSymbol{(}\AgdaOperator{\AgdaFunction{[}}\AgdaSpace{}%
\AgdaInductiveConstructor{opHash}%
\>[64]\AgdaOperator{\AgdaFunction{]}}\AgdaSymbol{)}\AgdaSpace{}%
\AgdaOperator{\AgdaRecord{<}}\AgdaSpace{}%
\AgdaFunction{accept₄}\AgdaSpace{}%
\AgdaBound{pbkh}\AgdaSpace{}%
\AgdaOperator{\AgdaRecord{>}}\<%
\end{code}
} 

\AgdaHide{
\begin{code}%
\>[0]\AgdaFunction{correct-opHash}\AgdaSpace{}%
\AgdaBound{pbkh}\AgdaSpace{}%
\AgdaSymbol{.}\AgdaField{==>}\AgdaSpace{}%
\AgdaSymbol{=}\AgdaSpace{}%
\AgdaFunction{correct-opHash-to}\AgdaSpace{}%
\AgdaBound{pbkh}\<%
\\
\>[0]\AgdaFunction{correct-opHash}\AgdaSpace{}%
\AgdaBound{pbkh}\AgdaSpace{}%
\AgdaSymbol{.}\AgdaField{<==}\AgdaSpace{}%
\AgdaSymbol{=}\AgdaSpace{}%
\AgdaFunction{correct-opHash-from}\AgdaSpace{}%
\AgdaBound{pbkh}\<%
\\
\\[\AgdaEmptyExtraSkip]%
\\[\AgdaEmptyExtraSkip]%
\>[0]\AgdaComment{\{-old}\<%
\\
\>[0]\<%
\\
\>[0]\<%
\\
\>[0]\AgdaComment{correct-5-to\ :\ (pbk\ :\ ℕ\ )\ →\ (s\ :\ StackState)\ →\ accept₅\ pbk\ s\ →\ \ ((\ accept₄\ pbk\ )\ ⁺)\ (⟦\ opHash\ ⟧s\ s\ )}\<%
\\
\>[0]\AgdaComment{correct-5-to\ pbk\ ⟨\ time\ ,\ msg₁\ ,\ x\ ∷\ x₁\ ∷\ x₂\ ∷\ stack₁\ ⟩\ (conj\ refl\ checkSig)\ =\ (conj\ refl\ checkSig)}\<%
\\
\>[0]\<%
\\
\>[0]\AgdaComment{correct-5-from\ :\ (\ pbk\ :\ ℕ\ )\ →\ \ (s\ :\ StackState)\ \ →\ ((\ accept₄\ pbk)\ ⁺)\ (⟦\ opHash\ ⟧s\ s\ )\ →\ accept₅\ pbk\ \ s}\<%
\\
\>[0]\AgdaComment{correct-5-from\ .(hashFun\ x)\ ⟨\ time\ ,\ msg₁\ ,\ x\ ∷\ x₁\ ∷\ x₂\ ∷\ stack₁\ ⟩\ (conj\ refl\ checkSig)\ =\ conj\ refl\ checkSig}\<%
\\
\>[0]\<%
\\
\>[0]\AgdaComment{-@BEGININLINE@correctfive}\<%
\\
\>[0]\AgdaComment{correct-5\ :(\ pbk\ :\ ℕ\ )\ →\ \ <\ accept₅\ pbk\ >iff\ \ ([\ opHash\ \ ])\ <\ accept₄\ pbk\ >}\<%
\\
\>[0]\AgdaComment{-@END}\<%
\\
\>[0]\AgdaComment{correct-5\ pbk\ .==>\ =\ correct-5-to\ pbk}\<%
\\
\>[0]\AgdaComment{correct-5\ pbk\ .<==\ =\ correct-5-from\ pbk}\<%
\\
\>[0]\<%
\\
\>[0]\AgdaComment{-\}}\<%
\\
\>[0]\AgdaComment{--\ needs\ to\ be\ pbkh\ since\ accept₅\ refer\ to\ pbkh}\<%
\\
\>[0]\AgdaFunction{correct-opDup-to}\AgdaSpace{}%
\AgdaSymbol{:}\AgdaSpace{}%
\AgdaSymbol{(}\AgdaBound{pbkh}\AgdaSpace{}%
\AgdaSymbol{:}\AgdaSpace{}%
\AgdaDatatype{ℕ}\AgdaSpace{}%
\AgdaSymbol{)}\AgdaSpace{}%
\AgdaSymbol{→}\AgdaSpace{}%
\AgdaSymbol{(}\AgdaBound{s}\AgdaSpace{}%
\AgdaSymbol{:}\AgdaSpace{}%
\AgdaRecord{StackState}\AgdaSymbol{)}\AgdaSpace{}%
\AgdaSymbol{→}\AgdaSpace{}%
\AgdaFunction{wPreCondP2PKH}\AgdaSpace{}%
\AgdaBound{pbkh}\AgdaSpace{}%
\AgdaBound{s}\AgdaSpace{}%
\AgdaSymbol{→}%
\>[76]\AgdaSymbol{((}\AgdaSpace{}%
\AgdaFunction{accept₅}\AgdaSpace{}%
\AgdaBound{pbkh}\AgdaSpace{}%
\AgdaSymbol{)}\AgdaSpace{}%
\AgdaOperator{\AgdaFunction{⁺}}\AgdaSymbol{)}\AgdaSpace{}%
\AgdaSymbol{(}\AgdaOperator{\AgdaFunction{⟦}}\AgdaSpace{}%
\AgdaInductiveConstructor{opDup}\AgdaSpace{}%
\AgdaOperator{\AgdaFunction{⟧ₛ}}\AgdaSpace{}%
\AgdaBound{s}\AgdaSpace{}%
\AgdaSymbol{)}\<%
\\
\>[0]\AgdaFunction{correct-opDup-to}\AgdaSpace{}%
\AgdaBound{pbkh}\AgdaSpace{}%
\AgdaOperator{\AgdaInductiveConstructor{⟨}}\AgdaSpace{}%
\AgdaBound{time}\AgdaSpace{}%
\AgdaOperator{\AgdaInductiveConstructor{,}}\AgdaSpace{}%
\AgdaBound{msg₁}\AgdaSpace{}%
\AgdaOperator{\AgdaInductiveConstructor{,}}\AgdaSpace{}%
\AgdaBound{x}\AgdaSpace{}%
\AgdaOperator{\AgdaInductiveConstructor{∷}}\AgdaSpace{}%
\AgdaBound{x₁}\AgdaSpace{}%
\AgdaOperator{\AgdaInductiveConstructor{∷}}\AgdaSpace{}%
\AgdaInductiveConstructor{[]}\AgdaSpace{}%
\AgdaOperator{\AgdaInductiveConstructor{⟩}}\AgdaSpace{}%
\AgdaBound{p}\AgdaSpace{}%
\AgdaSymbol{=}\AgdaSpace{}%
\AgdaBound{p}\<%
\\
\>[0]\AgdaFunction{correct-opDup-to}\AgdaSpace{}%
\AgdaBound{pbkh}\AgdaSpace{}%
\AgdaOperator{\AgdaInductiveConstructor{⟨}}\AgdaSpace{}%
\AgdaBound{time}\AgdaSpace{}%
\AgdaOperator{\AgdaInductiveConstructor{,}}\AgdaSpace{}%
\AgdaBound{msg₁}\AgdaSpace{}%
\AgdaOperator{\AgdaInductiveConstructor{,}}\AgdaSpace{}%
\AgdaBound{x}\AgdaSpace{}%
\AgdaOperator{\AgdaInductiveConstructor{∷}}\AgdaSpace{}%
\AgdaBound{x₁}\AgdaSpace{}%
\AgdaOperator{\AgdaInductiveConstructor{∷}}\AgdaSpace{}%
\AgdaBound{x₂}\AgdaSpace{}%
\AgdaOperator{\AgdaInductiveConstructor{∷}}\AgdaSpace{}%
\AgdaBound{stack₁}\AgdaSpace{}%
\AgdaOperator{\AgdaInductiveConstructor{⟩}}\AgdaSpace{}%
\AgdaBound{p}\AgdaSpace{}%
\AgdaSymbol{=}\AgdaSpace{}%
\AgdaBound{p}\<%
\\
\\[\AgdaEmptyExtraSkip]%
\>[0]\AgdaFunction{correct-opDup-from}\AgdaSpace{}%
\AgdaSymbol{:}\AgdaSpace{}%
\AgdaSymbol{(}\AgdaSpace{}%
\AgdaBound{pbkh}\AgdaSpace{}%
\AgdaSymbol{:}\AgdaSpace{}%
\AgdaDatatype{ℕ}\AgdaSpace{}%
\AgdaSymbol{)}\AgdaSpace{}%
\AgdaSymbol{→}%
\>[37]\AgdaSymbol{(}\AgdaBound{s}\AgdaSpace{}%
\AgdaSymbol{:}\AgdaSpace{}%
\AgdaRecord{StackState}\AgdaSymbol{)}%
\>[55]\AgdaSymbol{→}\AgdaSpace{}%
\AgdaSymbol{((}\AgdaSpace{}%
\AgdaFunction{accept₅}\AgdaSpace{}%
\AgdaBound{pbkh}\AgdaSymbol{)}\AgdaSpace{}%
\AgdaOperator{\AgdaFunction{⁺}}\AgdaSymbol{)}\AgdaSpace{}%
\AgdaSymbol{(}\AgdaOperator{\AgdaFunction{⟦}}\AgdaSpace{}%
\AgdaInductiveConstructor{opDup}\AgdaSpace{}%
\AgdaOperator{\AgdaFunction{⟧ₛ}}\AgdaSpace{}%
\AgdaBound{s}\AgdaSpace{}%
\AgdaSymbol{)}\AgdaSpace{}%
\AgdaSymbol{→}\AgdaSpace{}%
\AgdaFunction{wPreCondP2PKH}\AgdaSpace{}%
\AgdaBound{pbkh}%
\>[115]\AgdaBound{s}\<%
\\
\>[0]\AgdaFunction{correct-opDup-from}\AgdaSpace{}%
\AgdaBound{pbkh}\AgdaSpace{}%
\AgdaOperator{\AgdaInductiveConstructor{⟨}}\AgdaSpace{}%
\AgdaBound{time}\AgdaSpace{}%
\AgdaOperator{\AgdaInductiveConstructor{,}}\AgdaSpace{}%
\AgdaBound{msg₁}\AgdaSpace{}%
\AgdaOperator{\AgdaInductiveConstructor{,}}\AgdaSpace{}%
\AgdaBound{x}\AgdaSpace{}%
\AgdaOperator{\AgdaInductiveConstructor{∷}}\AgdaSpace{}%
\AgdaBound{x₁}\AgdaSpace{}%
\AgdaOperator{\AgdaInductiveConstructor{∷}}\AgdaSpace{}%
\AgdaBound{stack₁}\AgdaSpace{}%
\AgdaOperator{\AgdaInductiveConstructor{⟩}}\AgdaSpace{}%
\AgdaBound{p}\AgdaSpace{}%
\AgdaSymbol{=}\AgdaSpace{}%
\AgdaBound{p}\<%
\\
\>[0]\<%
\end{code}
} 

\newcommand{\stackVerificationPtoPKHcorrectsix}{
\begin{code}[inline]%
\>[0]\AgdaFunction{correct-opDup}\AgdaSpace{}%
\AgdaSymbol{: (}%
\AgdaBound{pbkh}\AgdaSpace{}%
\AgdaSymbol{:}\AgdaSpace{}%
\AgdaDatatype{ℕ}%
\AgdaSymbol{)}\AgdaSpace{}%
\AgdaSymbol{→}\AgdaSpace{}%
\>[31]\AgdaOperator{\AgdaRecord{<}}\AgdaSpace{}%
\AgdaFunction{wPreCondP2PKH}\AgdaSpace{}%
\AgdaBound{pbkh}\AgdaSpace{}%
\AgdaOperator{\AgdaRecord{>ⁱᶠᶠ}}\AgdaSpace{}%
\>[58]\AgdaSymbol{(}\AgdaOperator{\AgdaFunction{[}}\AgdaSpace{}%
\AgdaInductiveConstructor{opDup}%
\>[68]\AgdaOperator{\AgdaFunction{]}}\AgdaSymbol{)}\AgdaSpace{}%
\AgdaOperator{\AgdaRecord{<}}\AgdaSpace{}%
\AgdaFunction{accept₅}\AgdaSpace{}%
\AgdaBound{pbkh}\AgdaSpace{}%
\AgdaOperator{\AgdaRecord{>}}\<%
\end{code}
} 

\AgdaHide{
\begin{code}%
\>[0]\AgdaFunction{correct-opDup}\AgdaSpace{}%
\AgdaBound{pbkh}\AgdaSpace{}%
\AgdaSymbol{.}\AgdaField{==>}\AgdaSpace{}%
\AgdaSymbol{=}\AgdaSpace{}%
\AgdaFunction{correct-opDup-to}\AgdaSpace{}%
\AgdaBound{pbkh}\<%
\\
\>[0]\AgdaFunction{correct-opDup}\AgdaSpace{}%
\AgdaBound{pbkh}\AgdaSpace{}%
\AgdaSymbol{.}\AgdaField{<==}\AgdaSpace{}%
\AgdaSymbol{=}\AgdaSpace{}%
\AgdaFunction{correct-opDup-from}\AgdaSpace{}%
\AgdaBound{pbkh}\<%
\\
\\[\AgdaEmptyExtraSkip]%
\>[0]\AgdaComment{\{-old}\<%
\\
\>[0]\AgdaComment{correct-6-to\ :\ (pbkh\ :\ ℕ\ )\ →\ (s\ :\ StackState)\ →\ wPreCondP2PKH\ pbkh\ s\ →\ \ ((\ accept₅\ pbkh\ )\ ⁺)\ (⟦\ opDup\ ⟧s\ s\ )}\<%
\\
\>[0]\AgdaComment{correct-6-to\ pbkh\ ⟨\ time\ ,\ msg₁\ ,\ x\ ∷\ x₁\ ∷\ []\ ⟩\ p\ =\ p}\<%
\\
\>[0]\AgdaComment{correct-6-to\ pbkh\ ⟨\ time\ ,\ msg₁\ ,\ x\ ∷\ x₁\ ∷\ x₂\ ∷\ stack₁\ ⟩\ p\ =\ p}\<%
\\
\>[0]\<%
\\
\>[0]\AgdaComment{correct-6-from\ :\ (\ pbkh\ :\ ℕ\ )\ →\ \ (s\ :\ StackState)\ \ →\ ((\ accept₅\ pbkh)\ ⁺)\ (⟦\ opDup\ ⟧s\ s\ )\ →\ wPreCondP2PKH\ pbkh\ \ s}\<%
\\
\>[0]\AgdaComment{correct-6-from\ pbkh\ ⟨\ time\ ,\ msg₁\ ,\ x\ ∷\ x₁\ ∷\ stack₁\ ⟩\ p\ =\ p}\<%
\\
\>[0]\<%
\\
\>[0]\AgdaComment{-@BEGININLINE@correctsix}\<%
\\
\>[0]\AgdaComment{correct-6\ :(\ pbk\ :\ ℕ\ )\ →\ \ <\ wPreCondP2PKH\ pbk\ >iff\ \ ([\ opDup\ \ ])\ <\ accept₅\ pbk\ >}\<%
\\
\>[0]\AgdaComment{-@END}\<%
\\
\>[0]\AgdaComment{correct-6\ pbk\ .==>\ =\ correct-6-to\ pbk}\<%
\\
\>[0]\AgdaComment{correct-6\ pbk\ .<==\ =\ correct-6-from\ pbk}\<%
\\
\>[0]\<%
\\
\>[0]\AgdaComment{-\}}\<%
\\
\\[\AgdaEmptyExtraSkip]%
\>[0]\AgdaComment{--\ P2PKH\ script\ refers\ to\ pbkh\ not\ pbk}\<%
\end{code}
} 

\newcommand{\stackVerificationPtoPKHscriptppkh}{
\begin{code}%
\>[0]\AgdaFunction{scriptP2PKHᵇ}\AgdaSpace{}%
\AgdaSymbol{:}\AgdaSpace{}%
\AgdaSymbol{(}\AgdaBound{pbkh}\AgdaSpace{}%
\AgdaSymbol{:}\AgdaSpace{}%
\AgdaDatatype{ℕ}\AgdaSymbol{)}\AgdaSpace{}%
\AgdaSymbol{→}\AgdaSpace{}%
\AgdaFunction{BitcoinScriptBasic}\<%
\\
\>[0]\AgdaFunction{scriptP2PKHᵇ}\AgdaSpace{}%
\AgdaBound{pbkh}\AgdaSpace{}%
\AgdaSymbol{=}\AgdaSpace{}%
\AgdaInductiveConstructor{opDup}\AgdaSpace{}%
\AgdaOperator{\AgdaInductiveConstructor{∷}}\AgdaSpace{}%
\AgdaInductiveConstructor{opHash}\AgdaSpace{}%
\AgdaOperator{\AgdaInductiveConstructor{∷}}\AgdaSpace{}%
\AgdaSymbol{(}\AgdaInductiveConstructor{opPush}\AgdaSpace{}%
\AgdaBound{pbkh}\AgdaSymbol{)}\AgdaSpace{}%
\AgdaOperator{\AgdaInductiveConstructor{∷}}\AgdaSpace{}%
\AgdaInductiveConstructor{opEqual}\AgdaSpace{}%
\AgdaOperator{\AgdaInductiveConstructor{∷}}\AgdaSpace{}%
\AgdaInductiveConstructor{opVerify}\AgdaSpace{}%
\AgdaOperator{\AgdaInductiveConstructor{∷}}\AgdaSpace{}%
\AgdaOperator{\AgdaFunction{[}}\AgdaSpace{}%
\AgdaInductiveConstructor{opCheckSig}\AgdaSpace{}%
\AgdaOperator{\AgdaFunction{]}}\<%
\end{code}
} 

\AgdaHide{
\begin{code}%
\>[0]\<%
\\
\>[0]\AgdaComment{\{-\ Reminder\ \ from\ stackPredicate.agda}\<%
\\
\>[0]\<%
\\
\>[0]\<%
\\
\>[0]\AgdaComment{--\ acceptingBasic\ means\ that\ the\ stack\ has\ at\ hight\ >=\ 1\ and\ top\ element\ >0}\<%
\\
\>[0]\AgdaComment{acceptingBasic\ :\ Time\ →\ Msg\ →\ Stack\ →\ Bool}\<%
\\
\>[0]\AgdaComment{acceptingBasic\ time\ msg₁\ []\ =\ false}\<%
\\
\>[0]\AgdaComment{acceptingBasic\ time\ msg₁\ (x\ ∷\ stack₁)\ =\ notFalse\ x}\<%
\\
\>[0]\<%
\\
\>[0]\<%
\\
\>[0]\AgdaComment{acceptStateˢ\ :\ StackPredicate}\<%
\\
\>[0]\AgdaComment{acceptStateˢ\ time\ msg₁\ s\ =\ acceptingBasic\ time\ msg₁\ s)}\<%
\\
\>[0]\<%
\\
\>[0]\<%
\\
\>[0]\AgdaComment{--\ wPreCondP2PKHˢ\ expresses:}\<%
\\
\>[0]\AgdaComment{--\ \ \ \ stack\ has\ height\ at\ least\ 2}\<%
\\
\>[0]\AgdaComment{--\ \ \ \ for\ the\ two\ top\ elements\ pbk\ and\ sig\ we\ have}\<%
\\
\>[0]\AgdaComment{--\ \ \ \ hashFun\ pbk\ ≡\ \ pbkh\ \ \ and\ \ \ isSigned\ \ m\ sig\ pbk}\<%
\\
\>[0]\AgdaComment{wPreCondP2PKHˢ\ :\ (pbkh\ :\ ℕ\ )\ →\ StackPredicate}\<%
\\
\>[0]\AgdaComment{wPreCondP2PKHˢ\ pbkh\ time\ m\ []\ =\ ⊥}\<%
\\
\>[0]\AgdaComment{wPreCondP2PKHˢ\ pbkh\ time\ m\ (x\ ∷\ [])\ =\ ⊥}\<%
\\
\>[0]\AgdaComment{wPreCondP2PKHˢ\ pbkh\ time\ m\ (\ pbK\ ∷\ sig\ ∷\ st)}\<%
\\
\>[0]\AgdaComment{\ \ \ \ =\ \ (hashFun\ pbk\ ≡\ \ pbkh\ )\ ∧\ \ True\ (\ isSigned\ \ m\ sig\ pbk\ )}\<%
\\
\>[0]\<%
\\
\>[0]\AgdaComment{-\}}\<%
\\
\\[\AgdaEmptyExtraSkip]%
\\[\AgdaEmptyExtraSkip]%
\>[0]\AgdaComment{--\ acceptState\ :\ StackStatePred}\<%
\\
\>[0]\AgdaComment{--\ acceptState\ =\ stackPred2SPred\ acceptStateˢ\ --\ accept-0Basic}\<%
\\
\\[\AgdaEmptyExtraSkip]%
\>[0]\AgdaComment{--\ acceptState\ =\ accept-0Basic}\<%
\\
\\[\AgdaEmptyExtraSkip]%
\>[0]\AgdaComment{--\ wPreCondP2PKH\ pbkh\ =\ accept-6\ pbkh}\<%
\\
\\[\AgdaEmptyExtraSkip]%
\>[0]\<%
\end{code}
} 

\newcommand{\stackVerificationPtoPKHmainthro}{
\begin{code}%
\>[0]\AgdaFunction{theoremP2PKH}\AgdaSpace{}%
\AgdaSymbol{:}%
\>[16]\AgdaSymbol{(}\AgdaBound{pbkh}\AgdaSpace{}%
\AgdaSymbol{:}\AgdaSpace{}%
\AgdaDatatype{ℕ}\AgdaSymbol{)}\AgdaSpace{}%
\AgdaSymbol{→}\AgdaSpace{}%
\AgdaOperator{\AgdaRecord{<}}\AgdaSpace{}%
\AgdaFunction{wPreCondP2PKH}\AgdaSpace{}%
\AgdaBound{pbkh}\AgdaSpace{}%
\AgdaOperator{\AgdaRecord{>ⁱᶠᶠ}}\AgdaSpace{}%
\AgdaFunction{scriptP2PKHᵇ}\AgdaSpace{}%
\AgdaBound{pbkh}\AgdaSpace{}%
\AgdaOperator{\AgdaRecord{<}}\AgdaSpace{}%
\AgdaFunction{acceptState}\AgdaSpace{}%
\AgdaOperator{\AgdaRecord{>}}\<%
\\
\>[0]\AgdaFunction{theoremP2PKH}\AgdaSpace{}%
\AgdaBound{pbkh}%
\>[19]\AgdaSymbol{=}%
\>[829I]\AgdaFunction{wPreCondP2PKH}\AgdaSpace{}%
\AgdaBound{pbkh}\AgdaSpace{}%
\AgdaFunction{<><>⟨}\AgdaSpace{}%
\AgdaOperator{\AgdaFunction{[}}\AgdaSpace{}%
\AgdaInductiveConstructor{opDup}\AgdaSpace{}%
\AgdaOperator{\AgdaFunction{]}}%
\>[61]\AgdaFunction{⟩⟨}%
\>[65]\AgdaFunction{correct-opDup}%
\>[80]\AgdaBound{pbkh}\AgdaSpace{}%
\AgdaFunction{⟩}\<%
\\
\>[.][@{}l@{}]\<[829I]%
\>[21]\AgdaFunction{accept₅}%
\>[30]\AgdaBound{pbkh}%
\>[36]\AgdaFunction{<><>⟨}%
\>[43]\AgdaOperator{\AgdaFunction{[}}%
\>[46]\AgdaInductiveConstructor{opHash}\AgdaSpace{}%
\AgdaOperator{\AgdaFunction{]}}%
\>[61]\AgdaFunction{⟩⟨}%
\>[65]\AgdaFunction{correct-opHash}\AgdaSpace{}%
\AgdaBound{pbkh}\AgdaSpace{}%
\AgdaFunction{⟩}\<%
\\
\>[21]\AgdaFunction{accept₄}%
\>[30]\AgdaBound{pbkh}%
\>[36]\AgdaFunction{<><>⟨}%
\>[43]\AgdaOperator{\AgdaFunction{[}}%
\>[46]\AgdaInductiveConstructor{opPush}\AgdaSpace{}%
\AgdaBound{pbkh}\AgdaSpace{}%
\AgdaOperator{\AgdaFunction{]}}%
\>[61]\AgdaFunction{⟩⟨}%
\>[65]\AgdaFunction{correct-opPush}\AgdaSpace{}%
\AgdaBound{pbkh}\AgdaSpace{}%
\AgdaFunction{⟩}\<%
\\
\>[21]\AgdaFunction{accept₃}%
\>[36]\AgdaFunction{<><>⟨}%
\>[43]\AgdaOperator{\AgdaFunction{[}}%
\>[46]\AgdaInductiveConstructor{opEqual}\AgdaSpace{}%
\AgdaOperator{\AgdaFunction{]}}%
\>[61]\AgdaFunction{⟩⟨}%
\>[65]\AgdaFunction{correct-opEqual}%
\>[85]\AgdaFunction{⟩}\<%
\\
\>[21]\AgdaFunction{accept₂}%
\>[36]\AgdaFunction{<><>⟨}%
\>[43]\AgdaOperator{\AgdaFunction{[}}%
\>[46]\AgdaInductiveConstructor{opVerify}\AgdaSpace{}%
\AgdaOperator{\AgdaFunction{]}}%
\>[61]\AgdaFunction{⟩⟨}%
\>[65]\AgdaFunction{correct-opVerify}%
\>[85]\AgdaFunction{⟩}\<%
\\
\>[21]\AgdaFunction{accept₁}%
\>[36]\AgdaFunction{<><>⟨}%
\>[43]\AgdaOperator{\AgdaFunction{[}}%
\>[46]\AgdaInductiveConstructor{opCheckSig}\AgdaSpace{}%
\AgdaOperator{\AgdaFunction{]}}%
\>[61]\AgdaFunction{⟩⟨}%
\>[65]\AgdaFunction{correct-opCheckSig}%
\>[85]\AgdaFunction{⟩ᵉ}\<%
\\
\>[21]\AgdaFunction{acceptState}%
\>[34]\AgdaOperator{\AgdaFunction{∎p}}\<%
\end{code}
} 

\AgdaHide{
\begin{code}%
\>[0]\<%
\\
\>[0]\AgdaComment{\{-old}\<%
\\
\>[0]\<%
\\
\>[0]\<%
\\
\>[0]\AgdaComment{-@BEGIN@mainthro}\<%
\\
\>[0]\AgdaComment{theoremP2PKH\ :\ (pbkh\ :\ ℕ)\ →\ <\ wPreCondP2PKH\ pbkh\ >iff\ scriptP2PKHᵇ\ pbkh\ <\ acceptState\ >}\<%
\\
\>[0]\AgdaComment{theoremP2PKH\ pbkh\ \ =\ wPreCondP2PKH\ pbkh\ <><>⟨\ [\ opDup\ ]\ \ \ ⟩⟨\ \ correct-6\ \ pbkh\ \ ⟩}\<%
\\
\>[0]\AgdaComment{\ \ \ \ \ \ \ \ \ \ \ \ \ \ \ \ \ \ \ \ \ accept₅\ \ pbkh\ \ <><>⟨\ \ [\ \ opHash\ ]\ \ \ \ \ \ \ ⟩⟨\ \ correct-5\ \ pbkh\ \ ⟩}\<%
\\
\>[0]\AgdaComment{\ \ \ \ \ \ \ \ \ \ \ \ \ \ \ \ \ \ \ \ \ accept₄\ \ pbkh\ \ <><>⟨\ \ [\ \ opPush\ pbkh\ ]\ \ ⟩⟨\ \ correct-4\ \ pbkh\ \ ⟩}\<%
\\
\>[0]\AgdaComment{\ \ \ \ \ \ \ \ \ \ \ \ \ \ \ \ \ \ \ \ \ accept₃\ \ \ \ \ \ \ \ <><>⟨\ \ [\ \ opEqual\ ]\ \ \ \ \ \ ⟩⟨\ \ correct-3\ \ ⟩}\<%
\\
\>[0]\AgdaComment{\ \ \ \ \ \ \ \ \ \ \ \ \ \ \ \ \ \ \ \ \ accept₂\ \ \ \ \ \ \ \ <><>⟨\ \ [\ \ opVerify\ ]\ \ \ \ \ ⟩⟨\ \ correct-2\ \ ⟩}\<%
\\
\>[0]\AgdaComment{\ \ \ \ \ \ \ \ \ \ \ \ \ \ \ \ \ \ \ \ \ accept₁\ \ \ \ \ \ \ \ <><>⟨\ \ [\ \ opCheckSig\ ]\ \ \ ⟩⟨\ \ correct-1\ \ ⟩e\ \ acceptState\ \ ∎p}\<%
\\
\>[0]\AgdaComment{-@END}\<%
\\
\>[0]\AgdaComment{-\}}\<%
\\
\\[\AgdaEmptyExtraSkip]%
\\[\AgdaEmptyExtraSkip]%
\>[0]\AgdaComment{\{-\ We\ have\ natural\ accepting\ condition\ expressing\ that\ the\ stack\ has\ height\ >=\ 1\ and\ top\ element\ >\ 0}\<%
\\
\>[0]\AgdaComment{\ \ \ We\ have\ a\ natural\ weakest\ precondition\ expressing\ \ blablabla}\<%
\\
\>[0]\AgdaComment{\ \ \ and\ a\ proof\ that\ it\ is\ the\ weakest\ precondtion\ for\ the\ scriptP2PKH\ w.r.t.\ the\ acceptState}\<%
\\
\>[0]\<%
\\
\>[0]\AgdaComment{\ \ \ NOTE:\ \ here\ we\ used\ single\ instructions.\ However\ opPush\ pbkh\ is\ already\ a\ composed\ instruction}\<%
\\
\>[0]\AgdaComment{\ \ \ \ \ \ \ \ \ \ and\ we\ could\ use\ whole\ program\ pieces\ instead\ of\ single\ instructions.}\<%
\\
\>[0]\<%
\\
\>[0]\AgdaComment{\ \ \ \ \ \ \ \ \ \ The\ complication\ is\ due\ to\ the\ fact\ that\ instructions\ can\ lead\ to\ failure,\ and\ therefore}\<%
\\
\>[0]\AgdaComment{\ \ \ \ \ \ \ \ \ \ \ \ \ the\ operational\ semantics\ of\ the\ resulting\ programs\ become\ quite\ long\ and\ complicated}\<%
\\
\>[0]\AgdaComment{\ \ \ \ \ \ \ \ \ \ \ \ \ because\ if\ we\ have\ \ \ instructions\ p1\ p2\ p3\ \ in\ Agda\ syntax\ we\ get}\<%
\\
\>[0]\AgdaComment{\ \ \ \ \ \ \ \ \ \ \ \ \ [[\ p1\ ::\ p2\ ::\ p3\ ::[]\ ]]s\ =}\<%
\\
\>[0]\AgdaComment{\ \ \ \ \ \ \ \ \ \ \ \ \ \ \ \ case\ [[\ p1\ ]]s\ \ of}\<%
\\
\>[0]\AgdaComment{\ \ \ \ \ \ \ \ \ \ \ \ \ \ \ \ \ \ \ nothing\ ->\ nothing}\<%
\\
\>[0]\AgdaComment{\ \ \ \ \ \ \ \ \ \ \ \ \ \ \ \ \ \ \ just\ s1\ ->\ case\ [[\ p2\ ]]\ s1\ of}\<%
\\
\>[0]\AgdaComment{\ \ \ \ \ \ \ \ \ \ \ \ \ \ \ \ \ \ \ \ \ \ \ \ \ \ \ \ \ \ \ \ nothing\ ->\ nothing}\<%
\\
\>[0]\AgdaComment{\ \ \ \ \ \ \ \ \ \ \ \ \ \ \ \ \ \ \ \ \ \ \ \ \ \ \ \ \ \ \ \ just\ s2\ ->\ [[\ p3\ ]]\ s2}\<%
\\
\>[0]\AgdaComment{\ \ \ \ \ \ \ \ \ \ \ \ So\ one\ gets\ a\ very\ involved\ case\ distinction\ which\ quickly\ becomes\ difficult\ to\ view.}\<%
\\
\>[0]\AgdaComment{\ \ \ \ \ \ \ \ \ \ \ \ By\ having\ intermediate\ conditions\ we\ make\ this\ manageable.}\<%
\\
\>[0]\<%
\\
\>[0]\<%
\\
\>[0]\<%
\\
\>[0]\AgdaComment{Not\ for\ paper:}\<%
\\
\>[0]\AgdaComment{Current\ agda\ we\ have}\<%
\\
\>[0]\<%
\\
\>[0]\AgdaComment{f\ \ zero\ \ \ =\ ??}\<%
\\
\>[0]\AgdaComment{f\ (suc\ y)\ =\ ??}\<%
\\
\>[0]\<%
\\
\>[0]\AgdaComment{in\ Agda1\ we\ had}\<%
\\
\>[0]\AgdaComment{f\ x\ =\ \ case\ x\ of}\<%
\\
\>[0]\AgdaComment{\ \ \ \ \ \ \ \ \ \ \ \ zero\ ->\ ??}\<%
\\
\>[0]\AgdaComment{\ \ \ \ \ \ \ \ \ \ \ \ suc\ y\ ->\ ??}\<%
\\
\>[0]\<%
\\
\>[0]\AgdaComment{-\}}\<%
\end{code}
} 


\AgdaHide{
\begin{code}%
\>[0]\AgdaKeyword{open}\AgdaSpace{}%
\AgdaKeyword{import}\AgdaSpace{}%
\AgdaModule{basicBitcoinDataType}\<%
\\
\\[\AgdaEmptyExtraSkip]%
\>[0]\AgdaKeyword{module}\AgdaSpace{}%
\AgdaModule{verificationStackScripts.semanticsStackInstructions}\AgdaSpace{}%
\AgdaSymbol{(}\AgdaBound{param}\AgdaSpace{}%
\AgdaSymbol{:}\AgdaSpace{}%
\AgdaRecord{GlobalParameters}\AgdaSymbol{)}\AgdaSpace{}%
\AgdaKeyword{where}\<%
\\
\\[\AgdaEmptyExtraSkip]%
\\[\AgdaEmptyExtraSkip]%
\\[\AgdaEmptyExtraSkip]%
\>[0]\AgdaKeyword{open}\AgdaSpace{}%
\AgdaKeyword{import}\AgdaSpace{}%
\AgdaModule{Data.Nat}%
\>[22]\AgdaKeyword{hiding}\AgdaSpace{}%
\AgdaSymbol{(}\AgdaOperator{\AgdaDatatype{\AgdaUnderscore{}≤\AgdaUnderscore{}}}\AgdaSymbol{)}\<%
\\
\>[0]\AgdaKeyword{open}\AgdaSpace{}%
\AgdaKeyword{import}\AgdaSpace{}%
\AgdaModule{Data.List}\AgdaSpace{}%
\AgdaKeyword{hiding}\AgdaSpace{}%
\AgdaSymbol{(}\AgdaOperator{\AgdaFunction{\AgdaUnderscore{}\ensuremath{+\!\!+}\AgdaUnderscore{}}}\AgdaSymbol{)}\<%
\\
\>[0]\AgdaKeyword{open}\AgdaSpace{}%
\AgdaKeyword{import}\AgdaSpace{}%
\AgdaModule{Data.Unit}%
\>[23]\AgdaKeyword{hiding}\AgdaSpace{}%
\AgdaSymbol{(}\AgdaOperator{\AgdaRecord{\AgdaUnderscore{}≤\AgdaUnderscore{}}}\AgdaSymbol{)}\<%
\\
\>[0]\AgdaKeyword{open}\AgdaSpace{}%
\AgdaKeyword{import}\AgdaSpace{}%
\AgdaModule{Data.Empty}\<%
\\
\>[0]\AgdaKeyword{open}\AgdaSpace{}%
\AgdaKeyword{import}\AgdaSpace{}%
\AgdaModule{Data.Bool}%
\>[23]\AgdaKeyword{hiding}\AgdaSpace{}%
\AgdaSymbol{(}\AgdaOperator{\AgdaDatatype{\AgdaUnderscore{}≤\AgdaUnderscore{}}}\AgdaSpace{}%
\AgdaSymbol{;}\AgdaSpace{}%
\AgdaOperator{\AgdaFunction{if\AgdaUnderscore{}then\AgdaUnderscore{}else\AgdaUnderscore{}}}\AgdaSpace{}%
\AgdaSymbol{)}\AgdaSpace{}%
\AgdaKeyword{renaming}\AgdaSpace{}%
\AgdaSymbol{(}\AgdaOperator{\AgdaFunction{\AgdaUnderscore{}∧\AgdaUnderscore{}}}\AgdaSpace{}%
\AgdaSymbol{to}\AgdaSpace{}%
\AgdaOperator{\AgdaFunction{\AgdaUnderscore{}∧b\AgdaUnderscore{}}}\AgdaSpace{}%
\AgdaSymbol{;}\AgdaSpace{}%
\AgdaOperator{\AgdaFunction{\AgdaUnderscore{}∨\AgdaUnderscore{}}}\AgdaSpace{}%
\AgdaSymbol{to}\AgdaSpace{}%
\AgdaOperator{\AgdaFunction{\AgdaUnderscore{}∨b\AgdaUnderscore{}}}\AgdaSpace{}%
\AgdaSymbol{;}\AgdaSpace{}%
\AgdaFunction{T}\AgdaSpace{}%
\AgdaSymbol{to}\AgdaSpace{}%
\AgdaFunction{True}\AgdaSymbol{)}\<%
\\
\>[0]\AgdaKeyword{open}\AgdaSpace{}%
\AgdaKeyword{import}\AgdaSpace{}%
\AgdaModule{Data.Product}\AgdaSpace{}%
\AgdaKeyword{renaming}\AgdaSpace{}%
\AgdaSymbol{(}\AgdaOperator{\AgdaInductiveConstructor{\AgdaUnderscore{},\AgdaUnderscore{}}}\AgdaSpace{}%
\AgdaSymbol{to}\AgdaSpace{}%
\AgdaOperator{\AgdaInductiveConstructor{\AgdaUnderscore{},,\AgdaUnderscore{}}}\AgdaSpace{}%
\AgdaSymbol{)}\<%
\\
\>[0]\AgdaKeyword{open}\AgdaSpace{}%
\AgdaKeyword{import}\AgdaSpace{}%
\AgdaModule{Data.Nat.Base}\AgdaSpace{}%
\AgdaKeyword{hiding}\AgdaSpace{}%
\AgdaSymbol{(}\AgdaOperator{\AgdaDatatype{\AgdaUnderscore{}≤\AgdaUnderscore{}}}\AgdaSymbol{)}\<%
\\
\>[0]\AgdaKeyword{open}\AgdaSpace{}%
\AgdaKeyword{import}\AgdaSpace{}%
\AgdaModule{Data.Maybe}\<%
\\
\>[0]\AgdaKeyword{import}\AgdaSpace{}%
\AgdaModule{Relation.Binary.PropositionalEquality}\AgdaSpace{}%
\AgdaSymbol{as}\AgdaSpace{}%
\AgdaModule{Eq}\<%
\\
\>[0]\AgdaKeyword{open}\AgdaSpace{}%
\AgdaModule{Eq}\AgdaSpace{}%
\AgdaKeyword{using}\AgdaSpace{}%
\AgdaSymbol{(}\AgdaOperator{\AgdaDatatype{\AgdaUnderscore{}≡\AgdaUnderscore{}}}\AgdaSymbol{;}\AgdaSpace{}%
\AgdaInductiveConstructor{refl}\AgdaSymbol{;}\AgdaSpace{}%
\AgdaFunction{cong}\AgdaSymbol{;}\AgdaSpace{}%
\AgdaKeyword{module}\AgdaSpace{}%
\AgdaModule{≡-Reasoning}\AgdaSymbol{;}\AgdaSpace{}%
\AgdaFunction{sym}\AgdaSymbol{)}\<%
\\
\>[0]\AgdaKeyword{open}\AgdaSpace{}%
\AgdaModule{≡-Reasoning}\<%
\\
\>[0]\AgdaKeyword{open}\AgdaSpace{}%
\AgdaKeyword{import}\AgdaSpace{}%
\AgdaModule{Agda.Builtin.Equality}\<%
\\
\\[\AgdaEmptyExtraSkip]%
\\[\AgdaEmptyExtraSkip]%
\>[0]\AgdaComment{--our\ libraries}\<%
\\
\>[0]\AgdaKeyword{open}\AgdaSpace{}%
\AgdaKeyword{import}\AgdaSpace{}%
\AgdaModule{libraries.listLib}\<%
\\
\>[0]\AgdaKeyword{open}\AgdaSpace{}%
\AgdaKeyword{import}\AgdaSpace{}%
\AgdaModule{libraries.natLib}\<%
\\
\>[0]\AgdaKeyword{open}\AgdaSpace{}%
\AgdaKeyword{import}\AgdaSpace{}%
\AgdaModule{libraries.boolLib}\<%
\\
\>[0]\AgdaKeyword{open}\AgdaSpace{}%
\AgdaKeyword{import}\AgdaSpace{}%
\AgdaModule{libraries.andLib}\<%
\\
\>[0]\AgdaKeyword{open}\AgdaSpace{}%
\AgdaKeyword{import}\AgdaSpace{}%
\AgdaModule{libraries.miscLib}\<%
\\
\>[0]\AgdaKeyword{open}\AgdaSpace{}%
\AgdaKeyword{import}\AgdaSpace{}%
\AgdaModule{libraries.maybeLib}\<%
\\
\\[\AgdaEmptyExtraSkip]%
\>[0]\AgdaKeyword{open}\AgdaSpace{}%
\AgdaKeyword{import}\AgdaSpace{}%
\AgdaModule{stack}\<%
\\
\>[0]\AgdaKeyword{open}\AgdaSpace{}%
\AgdaKeyword{import}\AgdaSpace{}%
\AgdaModule{semanticBasicOperations}\AgdaSpace{}%
\AgdaBound{param}\<%
\\
\>[0]\AgdaKeyword{open}\AgdaSpace{}%
\AgdaKeyword{import}\AgdaSpace{}%
\AgdaModule{instructionBasic}\<%
\\
\>[0]\AgdaKeyword{open}\AgdaSpace{}%
\AgdaKeyword{import}\AgdaSpace{}%
\AgdaModule{verificationStackScripts.stackState}\<%
\\
\>[0]\AgdaKeyword{open}\AgdaSpace{}%
\AgdaKeyword{import}\AgdaSpace{}%
\AgdaModule{verificationStackScripts.sPredicate}\<%
\\
\>[0]\AgdaKeyword{open}\AgdaSpace{}%
\AgdaKeyword{import}\AgdaSpace{}%
\AgdaModule{verificationStackScripts.stackSemanticsInstructionsBasic}\AgdaSpace{}%
\AgdaBound{param}\<%
\\
\\[\AgdaEmptyExtraSkip]%
\\[\AgdaEmptyExtraSkip]%
\>[0]\AgdaComment{--\ Note\ that\ ⟦\ p\ ⟧sˢ\ \ is\ defined\ in}\<%
\\
\>[0]\AgdaComment{--\ verificationStackScripts.stackSemanticsInstructionsBasic}\<%
\\
\>[0]\AgdaComment{--\ \textbackslash{}semanticsStackInstructionssemanticS}\<%
\\
\>[0]\<%
\end{code}
} 

\newcommand{\semanticsStackInstructionssemanticS}{
\begin{code}%
\>[0]\AgdaOperator{\AgdaFunction{⟦\AgdaUnderscore{}⟧ₛ}}\AgdaSpace{}%
\AgdaSymbol{:}\AgdaSpace{}%
\AgdaDatatype{InstructionBasic}\AgdaSpace{}%
\AgdaSymbol{→}\AgdaSpace{}%
\AgdaRecord{StackState}\AgdaSpace{}%
\AgdaSymbol{→}\AgdaSpace{}%
\AgdaDatatype{Maybe}\AgdaSpace{}%
\AgdaRecord{StackState}\<%
\\
\>[0]\AgdaOperator{\AgdaFunction{⟦}}\AgdaSpace{}%
\AgdaBound{op}\AgdaSpace{}%
\AgdaOperator{\AgdaFunction{⟧ₛ}}%
\>[10]\AgdaSymbol{=}\AgdaSpace{}%
\AgdaFunction{liftStackFun2StackState}\AgdaSpace{}%
\AgdaOperator{\AgdaFunction{⟦}}\AgdaSpace{}%
\AgdaBound{op}\AgdaSpace{}%
\AgdaOperator{\AgdaFunction{⟧ₛˢ}}\<%
\end{code}
} 

\AgdaHide{
\begin{code}%
\>[0]\<%
\\
\\[\AgdaEmptyExtraSkip]%
\>[0]\AgdaComment{\{-}\<%
\\
\>[0]\AgdaComment{⟦\AgdaUnderscore{}⟧s\ :\ InstructionBasic\ →\ StackState\ →\ Maybe\ StackState}\<%
\\
\>[0]\AgdaComment{⟦\ opEqual\ ⟧s\ \ \ \ \ \ \ \ =\ \ liftStackToStackStateTransformer'\ \ executeStackEquality}\<%
\\
\>[0]\AgdaComment{⟦\ opAdd\ ⟧s\ \ \ \ \ \ \ \ \ \ =\ \ liftStackToStackStateTransformer'\ \ executeStackAdd}\<%
\\
\>[0]\AgdaComment{⟦\ (opPush\ x)\ ⟧s\ \ \ \ \ =\ \ liftStackToStackStateTransformer'\ \ (executeStackPush\ x)}\<%
\\
\>[0]\AgdaComment{⟦\ opSub\ ⟧s\ \ \ \ \ \ \ \ \ \ =\ \ liftStackToStackStateTransformer'\ \ executeStackSub}\<%
\\
\>[0]\AgdaComment{⟦\ opVerify\ ⟧s\ \ \ \ \ \ \ =\ \ liftStackToStackStateTransformer'\ \ executeStackVerify}\<%
\\
\>[0]\AgdaComment{⟦\ opCheckSig\ ⟧s\ \ \ \ \ =\ \ liftMsgStackToStateTransformer'\ \ \ \ executeStackCheckSig}\<%
\\
\>[0]\AgdaComment{⟦\ opEqualVerify\ ⟧s\ \ =\ \ liftStackToStackStateTransformer'\ \ executeStackVerify}\<%
\\
\>[0]\AgdaComment{⟦\ opDup\ ⟧s\ \ \ \ \ \ \ \ \ \ =\ \ liftStackToStackStateTransformer'\ \ executeStackDup}\<%
\\
\>[0]\AgdaComment{⟦\ opDrop\ ⟧s\ \ \ \ \ \ \ \ \ =\ \ liftStackToStackStateTransformer'\ \ executeStackDrop}\<%
\\
\>[0]\AgdaComment{⟦\ opSwap\ ⟧s\ \ \ \ \ \ \ \ \ =\ \ liftStackToStackStateTransformer'\ \ executeStackSwap}\<%
\\
\>[0]\AgdaComment{⟦\ opCHECKLOCKTIMEVERIFY\ ⟧s\ =\ liftTimeStackToStateTransformer'\ executeOpCHECKLOCKTIMEVERIFY}\<%
\\
\>[0]\AgdaComment{⟦\ opCheckSig3\ \ ⟧s\ \ \ =\ \ liftMsgStackToStateTransformer'\ \ \ \ executeStackCheckSig3Aux}\<%
\\
\>[0]\AgdaComment{⟦\ opHash\ \ ⟧s\ \ \ \ \ \ \ \ =\ \ liftStackToStackStateTransformer'\ \ executeOpHash}\<%
\\
\>[0]\AgdaComment{⟦\ opMultiSig\ ⟧s\ \ \ \ \ =\ \ liftMsgStackToStateTransformer'\ \ \ \ executeMultiSig}\<%
\\
\>[0]\AgdaComment{-\}}\<%
\\
\\[\AgdaEmptyExtraSkip]%
\\[\AgdaEmptyExtraSkip]%
\>[0]\AgdaOperator{\AgdaFunction{⟦\AgdaUnderscore{}⟧ₛ⁺}}\AgdaSpace{}%
\AgdaSymbol{:}\AgdaSpace{}%
\AgdaDatatype{InstructionBasic}\AgdaSpace{}%
\AgdaSymbol{→}\AgdaSpace{}%
\AgdaDatatype{Maybe}\AgdaSpace{}%
\AgdaRecord{StackState}\AgdaSpace{}%
\AgdaSymbol{→}\AgdaSpace{}%
\AgdaDatatype{Maybe}\AgdaSpace{}%
\AgdaRecord{StackState}\<%
\\
\>[0]\AgdaOperator{\AgdaFunction{⟦}}\AgdaSpace{}%
\AgdaBound{op}\AgdaSpace{}%
\AgdaOperator{\AgdaFunction{⟧ₛ⁺}}\AgdaSpace{}%
\AgdaBound{t}\AgdaSpace{}%
\AgdaSymbol{=}\AgdaSpace{}%
\AgdaBound{t}\AgdaSpace{}%
\AgdaOperator{\AgdaFunction{\ensuremath{\mathbin{{>}\mkern-8.5mu{>}\mkern-2mu{=}}}}}\AgdaSpace{}%
\AgdaOperator{\AgdaFunction{⟦}}\AgdaSpace{}%
\AgdaBound{op}\AgdaSpace{}%
\AgdaOperator{\AgdaFunction{⟧ₛ}}\<%
\\
\>[0]\AgdaComment{--\ ⟦\ op\ ⟧s⁺\ nothing\ =\ nothing}\<%
\\
\>[0]\AgdaComment{--\ ⟦\ op\ ⟧s⁺\ (just\ s\ )\ =\ ⟦\ op\ ⟧s\ s}\<%
\\
\\[\AgdaEmptyExtraSkip]%
\\[\AgdaEmptyExtraSkip]%
\>[0]\AgdaOperator{\AgdaFunction{⟦\AgdaUnderscore{}⟧}}\AgdaSpace{}%
\AgdaSymbol{:}\AgdaSpace{}%
\AgdaFunction{BitcoinScriptBasic}\AgdaSpace{}%
\AgdaSymbol{→}\AgdaSpace{}%
\AgdaRecord{StackState}\AgdaSpace{}%
\AgdaSymbol{→}\AgdaSpace{}%
\AgdaDatatype{Maybe}\AgdaSpace{}%
\AgdaRecord{StackState}\<%
\\
\>[0]\AgdaOperator{\AgdaFunction{⟦}}\AgdaSpace{}%
\AgdaInductiveConstructor{[]}%
\>[6]\AgdaOperator{\AgdaFunction{⟧}}%
\>[9]\AgdaSymbol{=}\AgdaSpace{}%
\AgdaInductiveConstructor{just}\<%
\\
\>[0]\AgdaOperator{\AgdaFunction{⟦}}\AgdaSpace{}%
\AgdaBound{op}\AgdaSpace{}%
\AgdaOperator{\AgdaInductiveConstructor{∷}}\AgdaSpace{}%
\AgdaInductiveConstructor{[]}%
\>[11]\AgdaOperator{\AgdaFunction{⟧}}%
\>[14]\AgdaSymbol{=}\AgdaSpace{}%
\AgdaOperator{\AgdaFunction{⟦}}\AgdaSpace{}%
\AgdaBound{op}\AgdaSpace{}%
\AgdaOperator{\AgdaFunction{⟧ₛ}}\<%
\\
\>[0]\AgdaCatchallClause{\AgdaOperator{\AgdaFunction{⟦}}}\AgdaSpace{}%
\AgdaCatchallClause{\AgdaBound{op}}\AgdaSpace{}%
\AgdaCatchallClause{\AgdaOperator{\AgdaInductiveConstructor{∷}}}\AgdaSpace{}%
\AgdaCatchallClause{\AgdaBound{p}}%
\>[11]\AgdaCatchallClause{\AgdaOperator{\AgdaFunction{⟧}}}%
\>[14]\AgdaCatchallClause{\AgdaBound{s}}\AgdaSpace{}%
\AgdaSymbol{=}\AgdaSpace{}%
\AgdaOperator{\AgdaFunction{⟦}}\AgdaSpace{}%
\AgdaBound{op}\AgdaSpace{}%
\AgdaOperator{\AgdaFunction{⟧ₛ}}\AgdaSpace{}%
\AgdaBound{s}\AgdaSpace{}%
\AgdaOperator{\AgdaFunction{\ensuremath{\mathbin{{>}\mkern-8.5mu{>}\mkern-2mu{=}}}}}%
\>[33]\AgdaOperator{\AgdaFunction{⟦}}\AgdaSpace{}%
\AgdaBound{p}\AgdaSpace{}%
\AgdaOperator{\AgdaFunction{⟧}}\<%
\\
\\[\AgdaEmptyExtraSkip]%
\>[0]\AgdaOperator{\AgdaFunction{⟦\AgdaUnderscore{}⟧⁺}}\AgdaSpace{}%
\AgdaSymbol{:}\AgdaSpace{}%
\AgdaFunction{BitcoinScriptBasic}\AgdaSpace{}%
\AgdaSymbol{→}\AgdaSpace{}%
\AgdaDatatype{Maybe}\AgdaSpace{}%
\AgdaRecord{StackState}\AgdaSpace{}%
\AgdaSymbol{→}\AgdaSpace{}%
\AgdaDatatype{Maybe}\AgdaSpace{}%
\AgdaRecord{StackState}\<%
\\
\>[0]\AgdaOperator{\AgdaFunction{⟦}}\AgdaSpace{}%
\AgdaBound{p}\AgdaSpace{}%
\AgdaOperator{\AgdaFunction{⟧⁺}}\AgdaSpace{}%
\AgdaBound{s}\AgdaSpace{}%
\AgdaSymbol{=}\AgdaSpace{}%
\AgdaBound{s}\AgdaSpace{}%
\AgdaOperator{\AgdaFunction{\ensuremath{\mathbin{{>}\mkern-8.5mu{>}\mkern-2mu{=}}}}}\AgdaSpace{}%
\AgdaOperator{\AgdaFunction{⟦}}\AgdaSpace{}%
\AgdaBound{p}\AgdaSpace{}%
\AgdaOperator{\AgdaFunction{⟧}}\<%
\\
\\[\AgdaEmptyExtraSkip]%
\\[\AgdaEmptyExtraSkip]%
\>[0]\AgdaFunction{validStackAux}\AgdaSpace{}%
\AgdaSymbol{:}\AgdaSpace{}%
\AgdaSymbol{(}\AgdaBound{pbkh}\AgdaSpace{}%
\AgdaSymbol{:}\AgdaSpace{}%
\AgdaDatatype{ℕ}\AgdaSymbol{)}\AgdaSpace{}%
\AgdaSymbol{→}%
\>[30]\AgdaSymbol{(}\AgdaBound{msg}\AgdaSpace{}%
\AgdaSymbol{:}\AgdaSpace{}%
\AgdaDatatype{Msg}\AgdaSymbol{)}\AgdaSpace{}%
\AgdaSymbol{→}%
\>[45]\AgdaFunction{Stack}\AgdaSpace{}%
\AgdaSymbol{→}%
\>[54]\AgdaDatatype{Bool}\<%
\\
\>[0]\AgdaFunction{validStackAux}\AgdaSpace{}%
\AgdaBound{pkh}\AgdaSpace{}%
\AgdaBound{msg[]}%
\>[25]\AgdaInductiveConstructor{[]}%
\>[29]\AgdaSymbol{=}\AgdaSpace{}%
\AgdaInductiveConstructor{false}\<%
\\
\>[0]\AgdaFunction{validStackAux}\AgdaSpace{}%
\AgdaBound{pkh}\AgdaSpace{}%
\AgdaBound{msg}\AgdaSpace{}%
\AgdaSymbol{(}\AgdaBound{pbk}\AgdaSpace{}%
\AgdaOperator{\AgdaInductiveConstructor{∷}}\AgdaSpace{}%
\AgdaInductiveConstructor{[]}\AgdaSymbol{)}\AgdaSpace{}%
\AgdaSymbol{=}\AgdaSpace{}%
\AgdaInductiveConstructor{false}\<%
\\
\>[0]\AgdaFunction{validStackAux}\AgdaSpace{}%
\AgdaBound{pkh}\AgdaSpace{}%
\AgdaBound{msg}\AgdaSpace{}%
\AgdaSymbol{(}\AgdaBound{pbk}\AgdaSpace{}%
\AgdaOperator{\AgdaInductiveConstructor{∷}}\AgdaSpace{}%
\AgdaBound{sig}\AgdaSpace{}%
\AgdaOperator{\AgdaInductiveConstructor{∷}}\AgdaSpace{}%
\AgdaBound{s}\AgdaSymbol{)}\AgdaSpace{}%
\AgdaSymbol{=}\AgdaSpace{}%
\AgdaFunction{hashFun}\AgdaSpace{}%
\AgdaBound{pbk}%
\>[53]\AgdaOperator{\AgdaFunction{==b}}\AgdaSpace{}%
\AgdaBound{pkh}\AgdaSpace{}%
\AgdaOperator{\AgdaFunction{∧b}}\AgdaSpace{}%
\AgdaFunction{isSigned}%
\>[74]\AgdaBound{msg}\AgdaSpace{}%
\AgdaBound{sig}\AgdaSpace{}%
\AgdaBound{pbk}\<%
\\
\\[\AgdaEmptyExtraSkip]%
\>[0]\AgdaFunction{validStack}\AgdaSpace{}%
\AgdaSymbol{:}\AgdaSpace{}%
\AgdaSymbol{(}\AgdaBound{pkh}\AgdaSpace{}%
\AgdaSymbol{:}\AgdaSpace{}%
\AgdaDatatype{ℕ}\AgdaSymbol{)}\AgdaSpace{}%
\AgdaSymbol{→}%
\>[26]\AgdaFunction{BStackStatePred}\<%
\\
\>[0]\AgdaFunction{validStack}\AgdaSpace{}%
\AgdaBound{pkh}\AgdaSpace{}%
\AgdaOperator{\AgdaInductiveConstructor{⟨}}\AgdaSpace{}%
\AgdaBound{time}\AgdaSpace{}%
\AgdaOperator{\AgdaInductiveConstructor{,}}\AgdaSpace{}%
\AgdaBound{msg₁}\AgdaSpace{}%
\AgdaOperator{\AgdaInductiveConstructor{,}}\AgdaSpace{}%
\AgdaBound{stack₁}\AgdaSpace{}%
\AgdaOperator{\AgdaInductiveConstructor{⟩}}\AgdaSpace{}%
\AgdaSymbol{=}\AgdaSpace{}%
\AgdaFunction{validStackAux}%
\>[57]\AgdaBound{pkh}%
\>[63]\AgdaBound{msg₁}%
\>[69]\AgdaBound{stack₁}\<%
\end{code}
} 


\AgdaHide{
\begin{code}%
\>[0]\AgdaComment{--\textbackslash{}verificationMultiSigBasicSymbolicExecutionPaper}\<%
\\
\>[0]\AgdaKeyword{open}\AgdaSpace{}%
\AgdaKeyword{import}\AgdaSpace{}%
\AgdaModule{basicBitcoinDataType}\<%
\\
\\[\AgdaEmptyExtraSkip]%
\>[0]\AgdaKeyword{module}\AgdaSpace{}%
\AgdaModule{paperTypes2021PostProceed.verificationMultiSigBasicSymbolicExecutionPaper}\AgdaSpace{}%
\AgdaSymbol{(}\AgdaBound{param}\AgdaSpace{}%
\AgdaSymbol{:}\AgdaSpace{}%
\AgdaRecord{GlobalParameters}\AgdaSymbol{)}\AgdaSpace{}%
\AgdaKeyword{where}\<%
\\
\\[\AgdaEmptyExtraSkip]%
\\[\AgdaEmptyExtraSkip]%
\>[0]\AgdaKeyword{open}\AgdaSpace{}%
\AgdaKeyword{import}\AgdaSpace{}%
\AgdaModule{Data.List.Base}\AgdaSpace{}%
\AgdaKeyword{hiding}\AgdaSpace{}%
\AgdaSymbol{(}\AgdaOperator{\AgdaFunction{\AgdaUnderscore{}\ensuremath{+\!\!+}\AgdaUnderscore{}}}\AgdaSpace{}%
\AgdaSymbol{)}\<%
\\
\>[0]\AgdaKeyword{open}\AgdaSpace{}%
\AgdaKeyword{import}\AgdaSpace{}%
\AgdaModule{Data.Nat}%
\>[22]\AgdaKeyword{renaming}\AgdaSpace{}%
\AgdaSymbol{(}\AgdaOperator{\AgdaDatatype{\AgdaUnderscore{}≤\AgdaUnderscore{}}}\AgdaSpace{}%
\AgdaSymbol{to}\AgdaSpace{}%
\AgdaOperator{\AgdaDatatype{\AgdaUnderscore{}≤'\AgdaUnderscore{}}}\AgdaSpace{}%
\AgdaSymbol{;}\AgdaSpace{}%
\AgdaOperator{\AgdaFunction{\AgdaUnderscore{}<\AgdaUnderscore{}}}\AgdaSpace{}%
\AgdaSymbol{to}\AgdaSpace{}%
\AgdaOperator{\AgdaFunction{\AgdaUnderscore{}<'\AgdaUnderscore{}}}\AgdaSymbol{)}\<%
\\
\>[0]\AgdaKeyword{open}\AgdaSpace{}%
\AgdaKeyword{import}\AgdaSpace{}%
\AgdaModule{Data.List}\AgdaSpace{}%
\AgdaKeyword{hiding}\AgdaSpace{}%
\AgdaSymbol{(}\AgdaOperator{\AgdaFunction{\AgdaUnderscore{}\ensuremath{+\!\!+}\AgdaUnderscore{}}}%
\>[36]\AgdaSymbol{)}\<%
\\
\>[0]\AgdaKeyword{open}\AgdaSpace{}%
\AgdaKeyword{import}\AgdaSpace{}%
\AgdaModule{Data.Sum}\<%
\\
\>[0]\AgdaKeyword{open}\AgdaSpace{}%
\AgdaKeyword{import}\AgdaSpace{}%
\AgdaModule{Data.Unit}%
\>[23]\AgdaKeyword{hiding}\AgdaSpace{}%
\AgdaSymbol{(}\AgdaOperator{\AgdaRecord{\AgdaUnderscore{}≤\AgdaUnderscore{}}}\AgdaSymbol{)}\<%
\\
\>[0]\AgdaKeyword{open}\AgdaSpace{}%
\AgdaKeyword{import}\AgdaSpace{}%
\AgdaModule{Data.Empty}\<%
\\
\>[0]\AgdaKeyword{open}\AgdaSpace{}%
\AgdaKeyword{import}\AgdaSpace{}%
\AgdaModule{Data.Maybe}\<%
\\
\>[0]\AgdaKeyword{open}\AgdaSpace{}%
\AgdaKeyword{import}\AgdaSpace{}%
\AgdaModule{Data.Bool}%
\>[23]\AgdaKeyword{hiding}\AgdaSpace{}%
\AgdaSymbol{(}\AgdaOperator{\AgdaDatatype{\AgdaUnderscore{}≤\AgdaUnderscore{}}}\AgdaSpace{}%
\AgdaSymbol{;}\AgdaSpace{}%
\AgdaOperator{\AgdaDatatype{\AgdaUnderscore{}<\AgdaUnderscore{}}}\AgdaSpace{}%
\AgdaSymbol{;}\AgdaSpace{}%
\AgdaOperator{\AgdaFunction{if\AgdaUnderscore{}then\AgdaUnderscore{}else\AgdaUnderscore{}}}%
\>[58]\AgdaSymbol{)}%
\>[61]\AgdaKeyword{renaming}\AgdaSpace{}%
\AgdaSymbol{(}\AgdaOperator{\AgdaFunction{\AgdaUnderscore{}∧\AgdaUnderscore{}}}\AgdaSpace{}%
\AgdaSymbol{to}\AgdaSpace{}%
\AgdaOperator{\AgdaFunction{\AgdaUnderscore{}∧b\AgdaUnderscore{}}}\AgdaSpace{}%
\AgdaSymbol{;}\AgdaSpace{}%
\AgdaOperator{\AgdaFunction{\AgdaUnderscore{}∨\AgdaUnderscore{}}}\AgdaSpace{}%
\AgdaSymbol{to}\AgdaSpace{}%
\AgdaOperator{\AgdaFunction{\AgdaUnderscore{}∨b\AgdaUnderscore{}}}\AgdaSpace{}%
\AgdaSymbol{;}\AgdaSpace{}%
\AgdaFunction{T}\AgdaSpace{}%
\AgdaSymbol{to}\AgdaSpace{}%
\AgdaFunction{True}\AgdaSymbol{)}\<%
\\
\>[0]\AgdaKeyword{open}\AgdaSpace{}%
\AgdaKeyword{import}\AgdaSpace{}%
\AgdaModule{Data.Product}\AgdaSpace{}%
\AgdaKeyword{renaming}\AgdaSpace{}%
\AgdaSymbol{(}\AgdaOperator{\AgdaInductiveConstructor{\AgdaUnderscore{},\AgdaUnderscore{}}}\AgdaSpace{}%
\AgdaSymbol{to}\AgdaSpace{}%
\AgdaOperator{\AgdaInductiveConstructor{\AgdaUnderscore{},,\AgdaUnderscore{}}}\AgdaSpace{}%
\AgdaSymbol{)}\<%
\\
\>[0]\AgdaKeyword{open}\AgdaSpace{}%
\AgdaKeyword{import}\AgdaSpace{}%
\AgdaModule{Data.Nat.Base}\AgdaSpace{}%
\AgdaKeyword{hiding}\AgdaSpace{}%
\AgdaSymbol{(}\AgdaOperator{\AgdaDatatype{\AgdaUnderscore{}≤\AgdaUnderscore{}}}\AgdaSpace{}%
\AgdaSymbol{;}\AgdaSpace{}%
\AgdaOperator{\AgdaFunction{\AgdaUnderscore{}<\AgdaUnderscore{}}}\AgdaSymbol{)}\<%
\\
\>[0]\AgdaKeyword{open}\AgdaSpace{}%
\AgdaKeyword{import}\AgdaSpace{}%
\AgdaModule{Data.List.NonEmpty}\AgdaSpace{}%
\AgdaKeyword{hiding}\AgdaSpace{}%
\AgdaSymbol{(}\AgdaField{head}\AgdaSymbol{;}\AgdaSpace{}%
\AgdaOperator{\AgdaFunction{[\AgdaUnderscore{}]}}\AgdaSymbol{)}\<%
\\
\>[0]\AgdaKeyword{open}\AgdaSpace{}%
\AgdaKeyword{import}\AgdaSpace{}%
\AgdaModule{Data.Nat}\AgdaSpace{}%
\AgdaKeyword{using}\AgdaSpace{}%
\AgdaSymbol{(}\AgdaDatatype{ℕ}\AgdaSymbol{;}\AgdaSpace{}%
\AgdaOperator{\AgdaPrimitive{\AgdaUnderscore{}+\AgdaUnderscore{}}}\AgdaSymbol{;}\AgdaSpace{}%
\AgdaOperator{\AgdaFunction{\AgdaUnderscore{}≥\AgdaUnderscore{}}}\AgdaSymbol{;}\AgdaSpace{}%
\AgdaOperator{\AgdaFunction{\AgdaUnderscore{}>\AgdaUnderscore{}}}\AgdaSymbol{;}\AgdaSpace{}%
\AgdaInductiveConstructor{zero}\AgdaSymbol{;}\AgdaSpace{}%
\AgdaInductiveConstructor{suc}\AgdaSymbol{;}\AgdaSpace{}%
\AgdaInductiveConstructor{s≤s}\AgdaSymbol{;}\AgdaSpace{}%
\AgdaInductiveConstructor{z≤n}\AgdaSymbol{)}\<%
\\
\>[0]\AgdaKeyword{import}\AgdaSpace{}%
\AgdaModule{Relation.Binary.PropositionalEquality}\AgdaSpace{}%
\AgdaSymbol{as}\AgdaSpace{}%
\AgdaModule{Eq}\<%
\\
\>[0]\AgdaKeyword{open}\AgdaSpace{}%
\AgdaModule{Eq}\AgdaSpace{}%
\AgdaKeyword{using}\AgdaSpace{}%
\AgdaSymbol{(}\AgdaOperator{\AgdaDatatype{\AgdaUnderscore{}≡\AgdaUnderscore{}}}\AgdaSymbol{;}\AgdaSpace{}%
\AgdaInductiveConstructor{refl}\AgdaSymbol{;}\AgdaSpace{}%
\AgdaFunction{cong}\AgdaSymbol{;}\AgdaSpace{}%
\AgdaKeyword{module}\AgdaSpace{}%
\AgdaModule{≡-Reasoning}\AgdaSymbol{;}\AgdaSpace{}%
\AgdaFunction{sym}\AgdaSymbol{)}\<%
\\
\>[0]\AgdaKeyword{open}\AgdaSpace{}%
\AgdaModule{≡-Reasoning}\<%
\\
\>[0]\AgdaKeyword{open}\AgdaSpace{}%
\AgdaKeyword{import}\AgdaSpace{}%
\AgdaModule{Agda.Builtin.Equality}\<%
\\
\\[\AgdaEmptyExtraSkip]%
\\[\AgdaEmptyExtraSkip]%
\>[0]\AgdaComment{--our\ libraries}\<%
\\
\>[0]\AgdaKeyword{open}\AgdaSpace{}%
\AgdaKeyword{import}\AgdaSpace{}%
\AgdaModule{libraries.listLib}\<%
\\
\>[0]\AgdaKeyword{open}\AgdaSpace{}%
\AgdaKeyword{import}\AgdaSpace{}%
\AgdaModule{libraries.emptyLib}\<%
\\
\>[0]\AgdaKeyword{open}\AgdaSpace{}%
\AgdaKeyword{import}\AgdaSpace{}%
\AgdaModule{libraries.natLib}\<%
\\
\>[0]\AgdaKeyword{open}\AgdaSpace{}%
\AgdaKeyword{import}\AgdaSpace{}%
\AgdaModule{libraries.boolLib}\<%
\\
\>[0]\AgdaKeyword{open}\AgdaSpace{}%
\AgdaKeyword{import}\AgdaSpace{}%
\AgdaModule{libraries.equalityLib}\<%
\\
\>[0]\AgdaKeyword{open}\AgdaSpace{}%
\AgdaKeyword{import}\AgdaSpace{}%
\AgdaModule{libraries.andLib}\<%
\\
\>[0]\AgdaKeyword{open}\AgdaSpace{}%
\AgdaKeyword{import}\AgdaSpace{}%
\AgdaModule{libraries.miscLib}\<%
\\
\>[0]\AgdaKeyword{open}\AgdaSpace{}%
\AgdaKeyword{import}\AgdaSpace{}%
\AgdaModule{libraries.maybeLib}\<%
\\
\\[\AgdaEmptyExtraSkip]%
\>[0]\AgdaKeyword{open}\AgdaSpace{}%
\AgdaKeyword{import}\AgdaSpace{}%
\AgdaModule{stack}\<%
\\
\>[0]\AgdaKeyword{open}\AgdaSpace{}%
\AgdaKeyword{import}\AgdaSpace{}%
\AgdaModule{stackPredicate}\<%
\\
\>[0]\AgdaKeyword{open}\AgdaSpace{}%
\AgdaKeyword{import}\AgdaSpace{}%
\AgdaModule{semanticBasicOperations}\AgdaSpace{}%
\AgdaBound{param}\<%
\\
\>[0][@{}l@{\AgdaIndent{0}}]%
\>[3]\AgdaKeyword{renaming}\AgdaSpace{}%
\AgdaSymbol{(}\AgdaFunction{compareSigsMultiSigAux}\AgdaSpace{}%
\AgdaSymbol{to}\AgdaSpace{}%
\AgdaFunction{cmpMultiSigsAux}\AgdaSymbol{)}\AgdaSpace{}%
\AgdaComment{--cmpMultiSigsAux)}\<%
\\
\>[0]\AgdaKeyword{open}\AgdaSpace{}%
\AgdaKeyword{import}\AgdaSpace{}%
\AgdaModule{instructionBasic}\<%
\\
\>[0]\AgdaKeyword{open}\AgdaSpace{}%
\AgdaKeyword{import}\AgdaSpace{}%
\AgdaModule{verificationMultiSig}\AgdaSpace{}%
\AgdaBound{param}%
\>[40]\AgdaKeyword{hiding}\AgdaSpace{}%
\AgdaSymbol{(}\AgdaFunction{multiSigScript2-4ᵇ}\AgdaSymbol{)}\<%
\\
\>[0]\AgdaKeyword{open}\AgdaSpace{}%
\AgdaKeyword{import}\AgdaSpace{}%
\AgdaModule{verificationStackScripts.semanticsStackInstructions}\AgdaSpace{}%
\AgdaBound{param}\<%
\\
\>[0]\AgdaKeyword{open}\AgdaSpace{}%
\AgdaKeyword{import}\AgdaSpace{}%
\AgdaModule{verificationStackScripts.stackVerificationLemmas}\AgdaSpace{}%
\AgdaBound{param}\<%
\\
\>[0]\AgdaKeyword{open}\AgdaSpace{}%
\AgdaKeyword{import}\AgdaSpace{}%
\AgdaModule{verificationStackScripts.stackHoareTriple}\AgdaSpace{}%
\AgdaBound{param}\<%
\\
\>[0]\AgdaKeyword{open}\AgdaSpace{}%
\AgdaKeyword{import}\AgdaSpace{}%
\AgdaModule{verificationStackScripts.sPredicate}\<%
\\
\>[0]\AgdaKeyword{open}\AgdaSpace{}%
\AgdaKeyword{import}\AgdaSpace{}%
\AgdaModule{verificationStackScripts.hoareTripleStackBasic}\AgdaSpace{}%
\AgdaBound{param}\<%
\\
\>[0]\AgdaKeyword{open}\AgdaSpace{}%
\AgdaKeyword{import}\AgdaSpace{}%
\AgdaModule{verificationStackScripts.stackState}\<%
\\
\>[0]\AgdaKeyword{open}\AgdaSpace{}%
\AgdaKeyword{import}\AgdaSpace{}%
\AgdaModule{verificationStackScripts.stackSemanticsInstructionsBasic}\AgdaSpace{}%
\AgdaBound{param}\<%
\\
\>[0]\AgdaKeyword{open}\AgdaSpace{}%
\AgdaKeyword{import}\AgdaSpace{}%
\AgdaModule{verificationStackScripts.verificationMultiSigBasic}\AgdaSpace{}%
\AgdaBound{param}\<%
\\
\\[\AgdaEmptyExtraSkip]%
\>[0]\AgdaKeyword{private}\<%
\\
\>[0][@{}l@{\AgdaIndent{0}}]%
\>[2]\AgdaKeyword{postulate}\AgdaSpace{}%
\AgdaPostulate{pbk₁}\AgdaSpace{}%
\AgdaPostulate{pbk₂}\AgdaSpace{}%
\AgdaPostulate{pbk₃}\AgdaSpace{}%
\AgdaPostulate{pbk₄}\AgdaSpace{}%
\AgdaSymbol{:}%
\>[35]\AgdaDatatype{ℕ}\<%
\\
\>[2]\AgdaKeyword{postulate}\AgdaSpace{}%
\AgdaPostulate{time₁}\AgdaSpace{}%
\AgdaSymbol{:}\AgdaSpace{}%
\AgdaFunction{Time}\<%
\\
\>[2]\AgdaKeyword{postulate}\AgdaSpace{}%
\AgdaPostulate{msg₁}\AgdaSpace{}%
\AgdaSymbol{:}\AgdaSpace{}%
\AgdaDatatype{Msg}\<%
\\
\>[2]\AgdaKeyword{postulate}\AgdaSpace{}%
\AgdaPostulate{stack₁}\AgdaSpace{}%
\AgdaSymbol{:}\AgdaSpace{}%
\AgdaDatatype{List}\AgdaSpace{}%
\AgdaDatatype{ℕ}\<%
\\
\>[2]\AgdaKeyword{postulate}\AgdaSpace{}%
\AgdaPostulate{sig₂}\AgdaSpace{}%
\AgdaPostulate{sig₁}\AgdaSpace{}%
\AgdaPostulate{dummy}\AgdaSpace{}%
\AgdaSymbol{:}\AgdaSpace{}%
\AgdaDatatype{ℕ}\<%
\\
\>[0]\AgdaFunction{multiSigScript2-4ᵇ}\AgdaSpace{}%
\AgdaSymbol{:}\AgdaSpace{}%
\AgdaSymbol{(}\AgdaBound{pbk1}\AgdaSpace{}%
\AgdaBound{pbk2}\AgdaSpace{}%
\AgdaBound{pbk3}\AgdaSpace{}%
\AgdaBound{pbk4}\AgdaSpace{}%
\AgdaSymbol{:}%
\>[45]\AgdaDatatype{ℕ}\AgdaSymbol{)}\AgdaSpace{}%
\AgdaSymbol{→}\AgdaSpace{}%
\AgdaFunction{BitcoinScriptBasic}\<%
\\
\>[0]\AgdaComment{--\textbackslash{}verificationMultiSigBasicSymbolicExecutionPapercomplexmultisig}\<%
\end{code}
} 

\newcommand{\verificationMultiSigBasicSymbolicExecutionPapercomplexmultisig}{
\begin{code}%
\>[0]\AgdaFunction{multiSigScript2-4ᵇ}%
\>[20]\AgdaBound{pbk1}\AgdaSpace{}%
\AgdaBound{pbk2}\AgdaSpace{}%
\AgdaBound{pbk3}\AgdaSpace{}%
\AgdaBound{pbk4}\AgdaSpace{}%
\AgdaSymbol{=}\AgdaSpace{}%
\AgdaSymbol{(}\AgdaInductiveConstructor{opPush}\AgdaSpace{}%
\AgdaNumber{2}\AgdaSymbol{)}%
\>[57]\AgdaOperator{\AgdaInductiveConstructor{∷}}%
\>[60]\AgdaSymbol{(}\AgdaInductiveConstructor{opPush}\AgdaSpace{}%
\AgdaBound{pbk1}\AgdaSymbol{)}%
\>[75]\AgdaOperator{\AgdaInductiveConstructor{∷}}%
\>[78]\AgdaSymbol{(}\AgdaInductiveConstructor{opPush}\AgdaSpace{}%
\AgdaBound{pbk2}\AgdaSymbol{)}\AgdaSpace{}%
\AgdaOperator{\AgdaInductiveConstructor{∷}}\<%
\\
\>[0][@{}l@{\AgdaIndent{0}}]%
\>[2]\AgdaSymbol{(}\AgdaInductiveConstructor{opPush}\AgdaSpace{}%
\AgdaBound{pbk3}\AgdaSymbol{)}\AgdaSpace{}%
\AgdaOperator{\AgdaInductiveConstructor{∷}}%
\>[23]\AgdaSymbol{(}\AgdaInductiveConstructor{opPush}\AgdaSpace{}%
\AgdaBound{pbk4}\AgdaSymbol{)}%
\>[38]\AgdaOperator{\AgdaInductiveConstructor{∷}}%
\>[41]\AgdaSymbol{(}\AgdaInductiveConstructor{opPush}\AgdaSpace{}%
\AgdaNumber{4}\AgdaSymbol{)}%
\>[56]\AgdaOperator{\AgdaInductiveConstructor{∷}}%
\>[59]\AgdaOperator{\AgdaFunction{[}}\AgdaSpace{}%
\AgdaInductiveConstructor{opMultiSig}\AgdaSpace{}%
\AgdaOperator{\AgdaFunction{]}}\<%
\end{code}
} 

\AgdaHide{
\begin{code}%
\>[0]\<%
\\
\\[\AgdaEmptyExtraSkip]%
\>[0]\AgdaFunction{multisigScript-2-4-symbolic}\AgdaSpace{}%
\AgdaSymbol{=}\<%
\\
\>[0]\AgdaComment{--\textbackslash{}verificationMultiSigBasicSymbolicExecutionPaper}\<%
\end{code}
} 

\newcommand{\verificationMultiSigBasicSymbolicExecutionPapermultisigsymbolic}{
\begin{code}[inline]%
\>[0][@{}l@{\AgdaIndent{1}}]%
\>[8]\AgdaOperator{\AgdaFunction{⟦}}\AgdaSpace{}%
\AgdaFunction{multiSigScript2-4ᵇ}\AgdaSpace{}%
\AgdaPostulate{pbk₁}\AgdaSpace{}%
\AgdaPostulate{pbk₂}\AgdaSpace{}%
\AgdaPostulate{pbk₃}\AgdaSpace{}%
\AgdaPostulate{pbk₄}\AgdaSpace{}%
\AgdaOperator{\AgdaFunction{⟧ˢ}}\AgdaSpace{}%
\AgdaPostulate{time₁}\AgdaSpace{}%
\AgdaPostulate{msg₁}\AgdaSpace{}%
\AgdaPostulate{stack₁}\<%
\end{code}
} 

\AgdaHide{
\begin{code}%
\>[0]\AgdaComment{\{-\ evaluated\ we\ get}\<%
\\
\>[0]\<%
\\
\>[0]\AgdaComment{executeMultiSig3\ msg₁\ (pbk₁\ ∷\ pbk₂\ ∷\ pbk₃\ ∷\ [\ pbk₄\ ])\ 2\ stack₁\ []}\<%
\\
\>[0]\<%
\\
\>[0]\AgdaComment{-\}}\<%
\\
\\[\AgdaEmptyExtraSkip]%
\>[0]\AgdaFunction{test2}\AgdaSpace{}%
\AgdaSymbol{:}\AgdaSpace{}%
\AgdaDatatype{Maybe}\AgdaSpace{}%
\AgdaFunction{Stack}\<%
\\
\>[0]\AgdaFunction{test2}%
\>[206I]\AgdaSymbol{=}\<%
\end{code}
} 

\newcommand{\verificationMultiSigBasicSymbolicExecutionPaperresultmultisig}{
\begin{code}[inline]%
\>[.][@{}l@{}]\<[206I]%
\>[6]\AgdaFunction{executeMultiSig3}\AgdaSpace{}%
\AgdaPostulate{msg₁}\AgdaSpace{}%
\AgdaSymbol{(}\AgdaPostulate{pbk₁}\AgdaSpace{}%
\AgdaOperator{\AgdaInductiveConstructor{∷}}\AgdaSpace{}%
\AgdaPostulate{pbk₂}\AgdaSpace{}%
\AgdaOperator{\AgdaInductiveConstructor{∷}}\AgdaSpace{}%
\AgdaPostulate{pbk₃}\AgdaSpace{}%
\AgdaOperator{\AgdaInductiveConstructor{∷}}\AgdaSpace{}%
\AgdaOperator{\AgdaFunction{[}}\AgdaSpace{}%
\AgdaPostulate{pbk₄}\AgdaSpace{}%
\AgdaOperator{\AgdaFunction{]}}\AgdaSymbol{)}\AgdaSpace{}%
\AgdaNumber{2}\AgdaSpace{}%
\AgdaPostulate{stack₁}\AgdaSpace{}%
\AgdaInductiveConstructor{[]}\<%
\end{code}
} 

\AgdaHide{
\begin{code}%
\>[0]\<%
\\
\>[0]\AgdaComment{--\ now\ we\ try\ out\ stack₁\ =\ []}\<%
\\
\\[\AgdaEmptyExtraSkip]%
\>[0]\AgdaFunction{multisigScript-2-4-symbolic-empty}\AgdaSpace{}%
\AgdaSymbol{=}\AgdaSpace{}%
\AgdaOperator{\AgdaFunction{⟦}}\AgdaSpace{}%
\AgdaFunction{multiSigScript2-4ᵇ}\AgdaSpace{}%
\AgdaPostulate{pbk₁}\AgdaSpace{}%
\AgdaPostulate{pbk₂}\AgdaSpace{}%
\AgdaPostulate{pbk₃}\AgdaSpace{}%
\AgdaPostulate{pbk₄}\AgdaSpace{}%
\AgdaOperator{\AgdaFunction{⟧ˢ}}\AgdaSpace{}%
\AgdaPostulate{time₁}\AgdaSpace{}%
\AgdaPostulate{msg₁}\AgdaSpace{}%
\AgdaInductiveConstructor{[]}\<%
\\
\\[\AgdaEmptyExtraSkip]%
\>[0]\AgdaComment{\{-}\<%
\\
\>[0]\AgdaComment{result\ nothing}\<%
\\
\>[0]\AgdaComment{-\}}\<%
\\
\\[\AgdaEmptyExtraSkip]%
\>[0]\AgdaFunction{multisigScript-2-4-symbolic-1stackelement}\AgdaSpace{}%
\AgdaSymbol{=}\AgdaSpace{}%
\AgdaOperator{\AgdaFunction{⟦}}\AgdaSpace{}%
\AgdaFunction{multiSigScript2-4ᵇ}\AgdaSpace{}%
\AgdaPostulate{pbk₁}\AgdaSpace{}%
\AgdaPostulate{pbk₂}\AgdaSpace{}%
\AgdaPostulate{pbk₃}\AgdaSpace{}%
\AgdaPostulate{pbk₄}\AgdaSpace{}%
\AgdaOperator{\AgdaFunction{⟧ˢ}}\AgdaSpace{}%
\AgdaPostulate{time₁}\AgdaSpace{}%
\AgdaPostulate{msg₁}\AgdaSpace{}%
\AgdaOperator{\AgdaFunction{[}}\AgdaSpace{}%
\AgdaPostulate{sig₂}\AgdaSpace{}%
\AgdaOperator{\AgdaFunction{]}}\<%
\\
\\[\AgdaEmptyExtraSkip]%
\>[0]\AgdaComment{\{-}\<%
\\
\>[0]\AgdaComment{result\ nothing}\<%
\\
\>[0]\AgdaComment{-\}}\<%
\\
\\[\AgdaEmptyExtraSkip]%
\>[0]\AgdaFunction{multisigScript-2-4-symbolic-2stackelement}\AgdaSpace{}%
\AgdaSymbol{=}\AgdaSpace{}%
\AgdaOperator{\AgdaFunction{⟦}}\AgdaSpace{}%
\AgdaFunction{multiSigScript2-4ᵇ}\AgdaSpace{}%
\AgdaPostulate{pbk₁}\AgdaSpace{}%
\AgdaPostulate{pbk₂}\AgdaSpace{}%
\AgdaPostulate{pbk₃}\AgdaSpace{}%
\AgdaPostulate{pbk₄}\AgdaSpace{}%
\AgdaOperator{\AgdaFunction{⟧ˢ}}\AgdaSpace{}%
\AgdaPostulate{time₁}\AgdaSpace{}%
\AgdaPostulate{msg₁}\AgdaSpace{}%
\AgdaSymbol{(}\AgdaPostulate{sig₂}\AgdaSpace{}%
\AgdaOperator{\AgdaInductiveConstructor{∷}}\AgdaSpace{}%
\AgdaOperator{\AgdaFunction{[}}\AgdaSpace{}%
\AgdaPostulate{sig₁}\AgdaSpace{}%
\AgdaOperator{\AgdaFunction{]}}\AgdaSymbol{)}\<%
\\
\\[\AgdaEmptyExtraSkip]%
\>[0]\AgdaComment{\{-}\<%
\\
\>[0]\AgdaComment{result\ nothing}\<%
\\
\>[0]\AgdaComment{-\}}\<%
\\
\\[\AgdaEmptyExtraSkip]%
\>[0]\AgdaFunction{stackNeededFirstStepMultiSig}\AgdaSpace{}%
\AgdaSymbol{:}\AgdaSpace{}%
\AgdaSymbol{(}\AgdaBound{sig₂}\AgdaSpace{}%
\AgdaBound{sig₁}\AgdaSpace{}%
\AgdaBound{dummy}\AgdaSpace{}%
\AgdaSymbol{:}\AgdaSpace{}%
\AgdaDatatype{ℕ}\AgdaSymbol{)(}\AgdaBound{stack₁}\AgdaSpace{}%
\AgdaSymbol{:}\AgdaSpace{}%
\AgdaFunction{Stack}\AgdaSymbol{)}\AgdaSpace{}%
\AgdaSymbol{→}\AgdaSpace{}%
\AgdaFunction{Stack}\<%
\\
\>[0]\AgdaFunction{stackNeededFirstStepMultiSig}\AgdaSpace{}%
\AgdaBound{sig₂}\AgdaSpace{}%
\AgdaBound{sig₁}\AgdaSpace{}%
\AgdaBound{dummy}\AgdaSpace{}%
\AgdaBound{stack₁}\AgdaSpace{}%
\AgdaSymbol{=}\<%
\\
\>[0][@{}l@{\AgdaIndent{0}}]%
\>[2]\AgdaBound{sig₂}\AgdaSpace{}%
\AgdaOperator{\AgdaInductiveConstructor{∷}}\AgdaSpace{}%
\AgdaBound{sig₁}\AgdaSpace{}%
\AgdaOperator{\AgdaInductiveConstructor{∷}}\AgdaSpace{}%
\AgdaBound{dummy}\AgdaSpace{}%
\AgdaOperator{\AgdaInductiveConstructor{∷}}\AgdaSpace{}%
\AgdaBound{stack₁}\<%
\\
\\[\AgdaEmptyExtraSkip]%
\\[\AgdaEmptyExtraSkip]%
\>[0]\AgdaFunction{stackNeededFirstStepMultiSig'}\AgdaSpace{}%
\AgdaSymbol{:}\AgdaSpace{}%
\AgdaFunction{Stack}\<%
\\
\>[0]\AgdaFunction{stackNeededFirstStepMultiSig'}\AgdaSpace{}%
\AgdaSymbol{=}\<%
\\
\>[0]\AgdaComment{--\textbackslash{}verificationMultiSigBasicSymbolicExecutionPaperstackNeededFirstStepMultiSig}\<%
\end{code}
} 

\newcommand{\verificationMultiSigBasicSymbolicExecutionPaperstackNeededFirstStepMultiSig}{
\begin{code}[inline]%
\>[0][@{}l@{\AgdaIndent{1}}]%
\>[2]\AgdaPostulate{sig₂}\AgdaSpace{}%
\AgdaOperator{\AgdaInductiveConstructor{∷}}\AgdaSpace{}%
\AgdaPostulate{sig₁}\AgdaSpace{}%
\AgdaOperator{\AgdaInductiveConstructor{∷}}\AgdaSpace{}%
\AgdaPostulate{dummy}\AgdaSpace{}%
\AgdaOperator{\AgdaInductiveConstructor{∷}}\AgdaSpace{}%
\AgdaPostulate{stack₁}\<%
\end{code}
} 

\AgdaHide{
\begin{code}%
\>[0]\<%
\\
\\[\AgdaEmptyExtraSkip]%
\>[0]\AgdaFunction{multisigScript-2-4-symbolic-3stackelement}\AgdaSpace{}%
\AgdaSymbol{=}\<%
\end{code}
} 

\newcommand{\verificationMultiSigBasicSymbolicExecutionPaperstackhasthreeelement}{
\begin{code}[inline]%
\>[0][@{}l@{\AgdaIndent{1}}]%
\>[2]\AgdaOperator{\AgdaFunction{⟦}}\AgdaSpace{}%
\AgdaFunction{multiSigScript2-4ᵇ}\AgdaSpace{}%
\AgdaPostulate{pbk₁}\AgdaSpace{}%
\AgdaPostulate{pbk₂}\AgdaSpace{}%
\AgdaPostulate{pbk₃}\AgdaSpace{}%
\AgdaPostulate{pbk₄}\AgdaSpace{}%
\AgdaOperator{\AgdaFunction{⟧ˢ}}\AgdaSpace{}%
\AgdaPostulate{time₁}\AgdaSpace{}%
\AgdaPostulate{msg₁}\AgdaSpace{}%
\AgdaSymbol{(}\AgdaPostulate{sig₂}\AgdaSpace{}%
\AgdaOperator{\AgdaInductiveConstructor{∷}}\AgdaSpace{}%
\AgdaPostulate{sig₁}\AgdaSpace{}%
\AgdaOperator{\AgdaInductiveConstructor{∷}}\AgdaSpace{}%
\AgdaPostulate{dummy}\AgdaSpace{}%
\AgdaOperator{\AgdaInductiveConstructor{∷}}\AgdaSpace{}%
\AgdaPostulate{stack₁}\AgdaSymbol{)}\<%
\end{code}
} 

\AgdaHide{
\begin{code}%
\>[0]\AgdaComment{\{-}\<%
\\
\>[0]\AgdaComment{just}\<%
\\
\>[0]\AgdaComment{(boolToNat}\<%
\\
\>[0]\AgdaComment{\ (cmpMultiSigsAux\ msg₁\ [\ sig₂\ ]\ (pbk₂\ ∷\ pbk₃\ ∷\ [\ pbk₄\ ])\ sig₁\ \ (isSigned\ msg₁\ sig₁\ pbk₁))}\<%
\\
\>[0]\AgdaComment{\ ∷\ \ stack₁)}\<%
\\
\>[0]\AgdaComment{-\}}\<%
\\
\\[\AgdaEmptyExtraSkip]%
\>[0]\AgdaFunction{multisigScript-2-4-symbolic-3stackelementNormalised}\AgdaSpace{}%
\AgdaSymbol{:}\AgdaSpace{}%
\AgdaDatatype{Maybe}\AgdaSpace{}%
\AgdaFunction{Stack}\<%
\\
\>[0]\AgdaFunction{multisigScript-2-4-symbolic-3stackelementNormalised}\AgdaSpace{}%
\AgdaSymbol{=}\<%
\\
\>[0]\AgdaComment{--\textbackslash{}verificationMultiSigBasicSymbolicExecutionPaperstackhasthreeelementNormalised}\<%
\end{code}
} 

\newcommand{\verificationMultiSigBasicSymbolicExecutionPaperstackhasthreeelementNormalised}{
\begin{code}%
\>[0][@{}l@{\AgdaIndent{1}}]%
\>[2]\AgdaInductiveConstructor{just}%
\>[8]\AgdaSymbol{(}\AgdaFunction{boolToNat}\AgdaSpace{}%
\AgdaSymbol{(}\AgdaFunction{cmpMultiSigsAux}\AgdaSpace{}%
\AgdaPostulate{msg₁}\AgdaSpace{}%
\AgdaOperator{\AgdaFunction{[}}\AgdaSpace{}%
\AgdaPostulate{sig₂}\AgdaSpace{}%
\AgdaOperator{\AgdaFunction{]}}\AgdaSpace{}%
\AgdaSymbol{(}\AgdaPostulate{pbk₂}\AgdaSpace{}%
\AgdaOperator{\AgdaInductiveConstructor{∷}}\AgdaSpace{}%
\AgdaPostulate{pbk₃}\AgdaSpace{}%
\AgdaOperator{\AgdaInductiveConstructor{∷}}\AgdaSpace{}%
\AgdaOperator{\AgdaFunction{[}}\AgdaSpace{}%
\AgdaPostulate{pbk₄}\AgdaSpace{}%
\AgdaOperator{\AgdaFunction{]}}\AgdaSymbol{)}\AgdaSpace{}%
\AgdaPostulate{sig₁}\<%
\\
\>[8]\AgdaSymbol{(}\AgdaFunction{isSigned}\AgdaSpace{}%
\AgdaPostulate{msg₁}\AgdaSpace{}%
\AgdaPostulate{sig₁}\AgdaSpace{}%
\AgdaPostulate{pbk₁}\AgdaSymbol{))}\AgdaSpace{}%
\AgdaOperator{\AgdaInductiveConstructor{∷}}%
\>[38]\AgdaPostulate{stack₁}\AgdaSymbol{)}\<%
\end{code}
} 

\AgdaHide{
\begin{code}%
\>[0]\<%
\\
\\[\AgdaEmptyExtraSkip]%
\>[0]\AgdaComment{\{-}\<%
\\
\>[0]\AgdaComment{So\ the\ program\ succeeds\ (we\ obtain\ result\ just)\ and\ all\ we\ need\ to\ check\ is\ whether\ the\ top\ element\ is}\<%
\\
\>[0]\<%
\\
\>[0]\AgdaComment{(boolToNat}\<%
\\
\>[0]\AgdaComment{\ (cmpMultiSigsAux\ msg₁\ [\ sig₂\ ]\ (pbk₂\ ∷\ pbk₃\ ∷\ [\ pbk₄\ ])\ sig₁\ \ (isSigned\ msg₁\ sig₁\ pbk₁))}\<%
\\
\>[0]\<%
\\
\>[0]\AgdaComment{is\ >\ 0}\<%
\\
\>[0]\<%
\\
\>[0]\<%
\\
\>[0]\AgdaComment{which\ is\ the\ case\ if}\<%
\\
\>[0]\AgdaComment{(cmpMultiSigsAux\ msg₁\ [\ sig₂\ ]\ (pbk₂\ ∷\ pbk₃\ ∷\ [\ pbk₄\ ])\ sig₁\ \ (isSigned\ msg₁\ sig₁\ pbk₁))}\<%
\\
\>[0]\<%
\\
\>[0]\AgdaComment{is\ true}\<%
\\
\>[0]\<%
\\
\>[0]\<%
\\
\>[0]\AgdaComment{so\ we\ symbolically\ evaluate}\<%
\\
\>[0]\<%
\\
\>[0]\AgdaComment{cmpMultiSigsAux\ msg₁\ [\ sig₂\ ]\ (pbk₂\ ∷\ pbk₃\ ∷\ [\ pbk₄\ ])\ sig₁\ \ (isSigned\ msg₁\ sig₁\ pbk₁)}\<%
\\
\>[0]\<%
\\
\>[0]\AgdaComment{-\}}\<%
\\
\\[\AgdaEmptyExtraSkip]%
\>[0]\AgdaFunction{topElementMultisigScript-2-4-symbolic-3'}\AgdaSpace{}%
\AgdaSymbol{:}\AgdaSpace{}%
\AgdaDatatype{Bool}\<%
\\
\>[0]\AgdaFunction{topElementMultisigScript-2-4-symbolic-3'}\AgdaSpace{}%
\AgdaSymbol{=}\<%
\\
\>[0][@{}l@{\AgdaIndent{0}}]%
\>[3]\AgdaFunction{cmpMultiSigsAux}\AgdaSpace{}%
\AgdaPostulate{msg₁}\AgdaSpace{}%
\AgdaOperator{\AgdaFunction{[}}\AgdaSpace{}%
\AgdaPostulate{sig₂}\AgdaSpace{}%
\AgdaOperator{\AgdaFunction{]}}\AgdaSpace{}%
\AgdaSymbol{(}\AgdaPostulate{pbk₂}\AgdaSpace{}%
\AgdaOperator{\AgdaInductiveConstructor{∷}}\AgdaSpace{}%
\AgdaPostulate{pbk₃}\AgdaSpace{}%
\AgdaOperator{\AgdaInductiveConstructor{∷}}\AgdaSpace{}%
\AgdaOperator{\AgdaFunction{[}}\AgdaSpace{}%
\AgdaPostulate{pbk₄}\AgdaSpace{}%
\AgdaOperator{\AgdaFunction{]}}\AgdaSymbol{)}\AgdaSpace{}%
\AgdaPostulate{sig₁}%
\>[64]\AgdaSymbol{(}\AgdaBound{param}\AgdaSpace{}%
\AgdaSymbol{.}\AgdaField{signed}\AgdaSpace{}%
\AgdaPostulate{msg₁}\AgdaSpace{}%
\AgdaPostulate{sig₁}\AgdaSpace{}%
\AgdaPostulate{pbk₁}\AgdaSymbol{)}\<%
\\
\\[\AgdaEmptyExtraSkip]%
\>[0]\AgdaFunction{topElementMultisigScript-2-4-symbolic-3}\AgdaSpace{}%
\AgdaSymbol{:}\AgdaSpace{}%
\AgdaDatatype{Bool}\<%
\\
\>[0]\AgdaFunction{topElementMultisigScript-2-4-symbolic-3}\AgdaSpace{}%
\AgdaSymbol{=}\<%
\end{code}
} 

\newcommand{\verificationMultiSigBasicSymbolicExecutionPapercompareSigmulti}{
\begin{code}[inline]%
\>[0][@{}l@{\AgdaIndent{1}}]%
\>[3]\AgdaFunction{cmpMultiSigsAux}\AgdaSpace{}%
\AgdaPostulate{msg₁}\AgdaSpace{}%
\AgdaOperator{\AgdaFunction{[}}\AgdaSpace{}%
\AgdaPostulate{sig₂}\AgdaSpace{}%
\AgdaOperator{\AgdaFunction{]}}\AgdaSpace{}%
\AgdaSymbol{(}\AgdaPostulate{pbk₂}\AgdaSpace{}%
\AgdaOperator{\AgdaInductiveConstructor{∷}}\AgdaSpace{}%
\AgdaPostulate{pbk₃}\AgdaSpace{}%
\AgdaOperator{\AgdaInductiveConstructor{∷}}\AgdaSpace{}%
\AgdaOperator{\AgdaFunction{[}}\AgdaSpace{}%
\AgdaPostulate{pbk₄}\AgdaSpace{}%
\AgdaOperator{\AgdaFunction{]}}\AgdaSymbol{)}\AgdaSpace{}%
\AgdaPostulate{sig₁}%
\>[64]\AgdaSymbol{(}\AgdaFunction{isSigned}\AgdaSpace{}%
\AgdaPostulate{msg₁}\AgdaSpace{}%
\AgdaPostulate{sig₁}\AgdaSpace{}%
\AgdaPostulate{pbk₁}\AgdaSymbol{)}\<%
\end{code}
} 

\AgdaHide{
\begin{code}%
\>[0]\<%
\\
\>[0]\AgdaFunction{subExpTopElementMultisigScript-2-4-symbolic-3}\AgdaSpace{}%
\AgdaSymbol{:}\AgdaSpace{}%
\AgdaSymbol{(}\AgdaBound{msg₁}\AgdaSpace{}%
\AgdaSymbol{:}\AgdaSpace{}%
\AgdaDatatype{Msg}\AgdaSymbol{)(}\AgdaBound{sig₁}\AgdaSpace{}%
\AgdaBound{pbk₁}\AgdaSpace{}%
\AgdaSymbol{:}\AgdaSpace{}%
\AgdaDatatype{ℕ}\AgdaSymbol{)}\AgdaSpace{}%
\AgdaSymbol{→}\AgdaSpace{}%
\AgdaDatatype{Bool}\<%
\\
\>[0]\AgdaFunction{subExpTopElementMultisigScript-2-4-symbolic-3}\AgdaSpace{}%
\AgdaBound{msg₁}\AgdaSpace{}%
\AgdaBound{sig₁}\AgdaSpace{}%
\AgdaBound{pbk₁}\AgdaSpace{}%
\AgdaSymbol{=}\<%
\\
\>[0][@{}l@{\AgdaIndent{0}}]%
\>[1]\AgdaFunction{isSigned}\AgdaSpace{}%
\AgdaBound{msg₁}\AgdaSpace{}%
\AgdaBound{sig₁}\AgdaSpace{}%
\AgdaBound{pbk₁}\<%
\\
\\[\AgdaEmptyExtraSkip]%
\>[0]\AgdaFunction{subExpTopElementMultisigScript-2-4-symbolic-3'}\AgdaSpace{}%
\AgdaSymbol{:}\AgdaSpace{}%
\AgdaDatatype{Bool}\<%
\\
\>[0]\AgdaFunction{subExpTopElementMultisigScript-2-4-symbolic-3'}\AgdaSpace{}%
\AgdaSymbol{=}\<%
\\
\>[0]\AgdaComment{--\textbackslash{}verificationMultiSigBasicSymbolicExecutionPapersubexTopElementOne}\<%
\end{code}
} 

\newcommand{\verificationMultiSigBasicSymbolicExecutionPapersubexTopElementOne}{
\begin{code}[inline]%
\>[0][@{}l@{\AgdaIndent{1}}]%
\>[1]\AgdaFunction{isSigned}\AgdaSpace{}%
\AgdaPostulate{msg₁}\AgdaSpace{}%
\AgdaPostulate{sig₁}\AgdaSpace{}%
\AgdaPostulate{pbk₁}\<%
\end{code}
} 

\AgdaHide{
\begin{code}%
\>[0]\<%
\\
\\[\AgdaEmptyExtraSkip]%
\\[\AgdaEmptyExtraSkip]%
\>[0]\AgdaFunction{testEqual}\AgdaSpace{}%
\AgdaSymbol{:}\AgdaSpace{}%
\AgdaFunction{topElementMultisigScript-2-4-symbolic-3'}\AgdaSpace{}%
\AgdaOperator{\AgdaDatatype{≡}}\AgdaSpace{}%
\AgdaFunction{topElementMultisigScript-2-4-symbolic-3}\<%
\\
\>[0]\AgdaFunction{testEqual}\AgdaSpace{}%
\AgdaSymbol{=}\AgdaSpace{}%
\AgdaInductiveConstructor{refl}\<%
\\
\\[\AgdaEmptyExtraSkip]%
\>[0]\AgdaComment{\{-\ So\ we\ can\ always\ write}\<%
\\
\>[0]\<%
\\
\>[0]\AgdaComment{isSigned\ \ \ instead\ of\ \ \ \ \ param\ .signed}\<%
\\
\>[0]\<%
\\
\>[0]\AgdaComment{-\}}\<%
\\
\\[\AgdaEmptyExtraSkip]%
\\[\AgdaEmptyExtraSkip]%
\\[\AgdaEmptyExtraSkip]%
\\[\AgdaEmptyExtraSkip]%
\>[0]\AgdaComment{\{-}\<%
\\
\>[0]\AgdaComment{We\ now\ make\ a\ casedistinction\ on\ (isSigned\ msg₁\ sig₁\ pbk₁)}\<%
\\
\>[0]\<%
\\
\>[0]\<%
\\
\>[0]\AgdaComment{-\}}\<%
\\
\\[\AgdaEmptyExtraSkip]%
\\[\AgdaEmptyExtraSkip]%
\\[\AgdaEmptyExtraSkip]%
\\[\AgdaEmptyExtraSkip]%
\>[0]\AgdaFunction{multisigAuxStep1True}\AgdaSpace{}%
\AgdaSymbol{=}\AgdaSpace{}%
\AgdaFunction{cmpMultiSigsAux}\AgdaSpace{}%
\AgdaPostulate{msg₁}\AgdaSpace{}%
\AgdaOperator{\AgdaFunction{[}}\AgdaSpace{}%
\AgdaPostulate{sig₂}\AgdaSpace{}%
\AgdaOperator{\AgdaFunction{]}}\AgdaSpace{}%
\AgdaSymbol{(}\AgdaPostulate{pbk₂}\AgdaSpace{}%
\AgdaOperator{\AgdaInductiveConstructor{∷}}\AgdaSpace{}%
\AgdaPostulate{pbk₃}\AgdaSpace{}%
\AgdaOperator{\AgdaInductiveConstructor{∷}}\AgdaSpace{}%
\AgdaOperator{\AgdaFunction{[}}\AgdaSpace{}%
\AgdaPostulate{pbk₄}\AgdaSpace{}%
\AgdaOperator{\AgdaFunction{]}}\AgdaSymbol{)}\AgdaSpace{}%
\AgdaPostulate{sig₁}\AgdaSpace{}%
\AgdaInductiveConstructor{true}\<%
\\
\>[0]\AgdaComment{\{-}\<%
\\
\>[0]\AgdaComment{\ \ compareSigsMultiSigAux\ msg₁\ []\ (pbk₃\ ∷\ [\ pbk₄\ ])\ sig₂\ (isSigned\ msg₁\ sig₂\ pbk₂)}\<%
\\
\>[0]\AgdaComment{-\}}\<%
\\
\\[\AgdaEmptyExtraSkip]%
\>[0]\AgdaFunction{resultMultisigAuxStep1True}\AgdaSpace{}%
\AgdaSymbol{:}\AgdaSpace{}%
\AgdaDatatype{Bool}\<%
\\
\>[0]\AgdaFunction{resultMultisigAuxStep1True}\AgdaSpace{}%
\AgdaSymbol{=}\<%
\\
\>[0]\AgdaComment{--\textbackslash{}verificationMultiSigBasicSymbolicExecutionPaperresultStepOnetrue}\<%
\end{code}
} 

\newcommand{\verificationMultiSigBasicSymbolicExecutionPaperresultStepOnetrue}{
\begin{code}[inline]%
\>[0][@{}l@{\AgdaIndent{1}}]%
\>[2]\AgdaFunction{cmpMultiSigsAux}\AgdaSpace{}%
\AgdaPostulate{msg₁}\AgdaSpace{}%
\AgdaInductiveConstructor{[]}\AgdaSpace{}%
\AgdaSymbol{(}\AgdaPostulate{pbk₃}\AgdaSpace{}%
\AgdaOperator{\AgdaInductiveConstructor{∷}}\AgdaSpace{}%
\AgdaOperator{\AgdaFunction{[}}\AgdaSpace{}%
\AgdaPostulate{pbk₄}\AgdaSpace{}%
\AgdaOperator{\AgdaFunction{]}}\AgdaSymbol{)}\AgdaSpace{}%
\AgdaPostulate{sig₂}\AgdaSpace{}%
\AgdaSymbol{(}\AgdaFunction{isSigned}\AgdaSpace{}%
\AgdaPostulate{msg₁}\AgdaSpace{}%
\AgdaPostulate{sig₂}\AgdaSpace{}%
\AgdaPostulate{pbk₂}\AgdaSymbol{)}\<%
\end{code}
} 

\AgdaHide{
\begin{code}%
\>[0]\<%
\\
\>[0]\AgdaFunction{resultMultisigAuxStep1TrueSubExp}\AgdaSpace{}%
\AgdaSymbol{:}\AgdaSpace{}%
\AgdaDatatype{Bool}\<%
\\
\>[0]\AgdaFunction{resultMultisigAuxStep1TrueSubExp}\AgdaSpace{}%
\AgdaSymbol{=}\<%
\\
\>[0]\AgdaComment{--\textbackslash{}verificationMultiSigBasicSymbolicExecutionPaperresultStepOnetrueSubExp}\<%
\end{code}
} 

\newcommand{\verificationMultiSigBasicSymbolicExecutionPaperresultStepOnetrueSubExp}{
\begin{code}[inline]%
\>[0][@{}l@{\AgdaIndent{1}}]%
\>[2]\AgdaFunction{isSigned}\AgdaSpace{}%
\AgdaPostulate{msg₁}\AgdaSpace{}%
\AgdaPostulate{sig₂}\AgdaSpace{}%
\AgdaPostulate{pbk₂}\<%
\end{code}
} 

\AgdaHide{
\begin{code}%
\>[0]\<%
\\
\\[\AgdaEmptyExtraSkip]%
\>[0]\AgdaFunction{multisigAuxStep1TrueStep2True}\AgdaSpace{}%
\AgdaSymbol{=}\AgdaSpace{}%
\AgdaFunction{cmpMultiSigsAux}\AgdaSpace{}%
\AgdaPostulate{msg₁}\AgdaSpace{}%
\AgdaInductiveConstructor{[]}\AgdaSpace{}%
\AgdaSymbol{(}\AgdaPostulate{pbk₃}\AgdaSpace{}%
\AgdaOperator{\AgdaInductiveConstructor{∷}}\AgdaSpace{}%
\AgdaOperator{\AgdaFunction{[}}\AgdaSpace{}%
\AgdaPostulate{pbk₄}\AgdaSpace{}%
\AgdaOperator{\AgdaFunction{]}}\AgdaSymbol{)}\AgdaSpace{}%
\AgdaPostulate{sig₂}\AgdaSpace{}%
\AgdaInductiveConstructor{true}\<%
\\
\\[\AgdaEmptyExtraSkip]%
\>[0]\AgdaComment{\{-\ returns\ true\ -\}}\<%
\\
\\[\AgdaEmptyExtraSkip]%
\\[\AgdaEmptyExtraSkip]%
\>[0]\AgdaFunction{multisigAuxStep1TrueStep2False}\AgdaSpace{}%
\AgdaSymbol{=}\AgdaSpace{}%
\AgdaFunction{cmpMultiSigsAux}\AgdaSpace{}%
\AgdaPostulate{msg₁}\AgdaSpace{}%
\AgdaInductiveConstructor{[]}\AgdaSpace{}%
\AgdaSymbol{(}\AgdaPostulate{pbk₃}\AgdaSpace{}%
\AgdaOperator{\AgdaInductiveConstructor{∷}}\AgdaSpace{}%
\AgdaOperator{\AgdaFunction{[}}\AgdaSpace{}%
\AgdaPostulate{pbk₄}\AgdaSpace{}%
\AgdaOperator{\AgdaFunction{]}}\AgdaSymbol{)}\AgdaSpace{}%
\AgdaPostulate{sig₂}\AgdaSpace{}%
\AgdaInductiveConstructor{false}\<%
\\
\\[\AgdaEmptyExtraSkip]%
\>[0]\AgdaComment{\{-\ returns}\<%
\\
\>[0]\AgdaComment{\ \ \ compareSigsMultiSigAux\ msg₁\ []\ [\ pbk₄\ ]\ sig₂\ (isSigned\ msg₁\ sig₂\ pbk₃)}\<%
\\
\>[0]\AgdaComment{-\}}\<%
\\
\\[\AgdaEmptyExtraSkip]%
\>[0]\AgdaFunction{resultMultisigAuxStep1Step2False}\AgdaSpace{}%
\AgdaSymbol{:}\AgdaSpace{}%
\AgdaDatatype{Bool}\<%
\\
\>[0]\AgdaFunction{resultMultisigAuxStep1Step2False}\AgdaSpace{}%
\AgdaSymbol{=}\<%
\\
\>[0]\AgdaComment{--\textbackslash{}verificationMultiSigBasicSymbolicExecutionPaperresultStepOnetrueStepTwoFalse}\<%
\end{code}
} 

\newcommand{\verificationMultiSigBasicSymbolicExecutionPaperresultStepOnetrueStepTwoFalse}{
\begin{code}[inline]%
\>[0][@{}l@{\AgdaIndent{1}}]%
\>[1]\AgdaFunction{cmpMultiSigsAux}\AgdaSpace{}%
\AgdaPostulate{msg₁}\AgdaSpace{}%
\AgdaInductiveConstructor{[]}\AgdaSpace{}%
\AgdaOperator{\AgdaFunction{[}}\AgdaSpace{}%
\AgdaPostulate{pbk₄}\AgdaSpace{}%
\AgdaOperator{\AgdaFunction{]}}\AgdaSpace{}%
\AgdaPostulate{sig₂}\AgdaSpace{}%
\AgdaSymbol{(}\AgdaFunction{isSigned}\AgdaSpace{}%
\AgdaPostulate{msg₁}\AgdaSpace{}%
\AgdaPostulate{sig₂}\AgdaSpace{}%
\AgdaPostulate{pbk₃}\AgdaSymbol{)}\<%
\end{code}
} 

\AgdaHide{
\begin{code}%
\>[0]\<%
\\
\>[0]\AgdaFunction{resultMultisigAuxStep1Step2FalseCoreExp}\AgdaSpace{}%
\AgdaSymbol{:}\AgdaSpace{}%
\AgdaDatatype{Bool}\<%
\\
\>[0]\AgdaFunction{resultMultisigAuxStep1Step2FalseCoreExp}\AgdaSpace{}%
\AgdaSymbol{=}\<%
\\
\>[0]\AgdaComment{--\textbackslash{}verificationMultiSigBasicSymbolicExecutionPaperresultStepOnetrueStepTwoFalseCore}\<%
\end{code}
} 

\newcommand{\verificationMultiSigBasicSymbolicExecutionPaperresultStepOnetrueStepTwoFalseCore}{
\begin{code}[inline]%
\>[0][@{}l@{\AgdaIndent{1}}]%
\>[1]\AgdaFunction{isSigned}\AgdaSpace{}%
\AgdaPostulate{msg₁}\AgdaSpace{}%
\AgdaPostulate{sig₂}\AgdaSpace{}%
\AgdaPostulate{pbk₃}\<%
\end{code}
} 

\AgdaHide{
\begin{code}%
\>[0]\<%
\\
\>[0]\AgdaFunction{multisigAuxStep1TrueStep2FalseStep3True}\AgdaSpace{}%
\AgdaSymbol{=}\AgdaSpace{}%
\AgdaFunction{cmpMultiSigsAux}\AgdaSpace{}%
\AgdaPostulate{msg₁}\AgdaSpace{}%
\AgdaInductiveConstructor{[]}\AgdaSpace{}%
\AgdaOperator{\AgdaFunction{[}}\AgdaSpace{}%
\AgdaPostulate{pbk₄}\AgdaSpace{}%
\AgdaOperator{\AgdaFunction{]}}\AgdaSpace{}%
\AgdaPostulate{sig₂}\AgdaSpace{}%
\AgdaInductiveConstructor{true}\<%
\\
\\[\AgdaEmptyExtraSkip]%
\\[\AgdaEmptyExtraSkip]%
\>[0]\AgdaComment{\{-\ returns\ true\ -\}}\<%
\\
\\[\AgdaEmptyExtraSkip]%
\>[0]\AgdaFunction{multisigAuxStep1TrueStep2FalseStep3False}\AgdaSpace{}%
\AgdaSymbol{=}\AgdaSpace{}%
\AgdaFunction{cmpMultiSigsAux}\AgdaSpace{}%
\AgdaPostulate{msg₁}\AgdaSpace{}%
\AgdaInductiveConstructor{[]}\AgdaSpace{}%
\AgdaOperator{\AgdaFunction{[}}\AgdaSpace{}%
\AgdaPostulate{pbk₄}\AgdaSpace{}%
\AgdaOperator{\AgdaFunction{]}}\AgdaSpace{}%
\AgdaPostulate{sig₂}\AgdaSpace{}%
\AgdaInductiveConstructor{false}\<%
\\
\\[\AgdaEmptyExtraSkip]%
\>[0]\AgdaComment{\{-\ returns}\<%
\\
\>[0]\AgdaComment{\ \ \ \ cmpMultiSigsAux\ msg₁\ []\ []\ sig₂\ (isSigned\ msg₁\ sig₂\ pbk₄)}\<%
\\
\>[0]\AgdaComment{-\}}\<%
\\
\\[\AgdaEmptyExtraSkip]%
\\[\AgdaEmptyExtraSkip]%
\\[\AgdaEmptyExtraSkip]%
\>[0]\AgdaFunction{multisigAuxStep1TrueStep2FalseStep3FalseStep4True}\AgdaSpace{}%
\AgdaSymbol{=}\AgdaSpace{}%
\AgdaFunction{cmpMultiSigsAux}\AgdaSpace{}%
\AgdaPostulate{msg₁}\AgdaSpace{}%
\AgdaInductiveConstructor{[]}\AgdaSpace{}%
\AgdaInductiveConstructor{[]}\AgdaSpace{}%
\AgdaPostulate{sig₂}\AgdaSpace{}%
\AgdaInductiveConstructor{true}\<%
\\
\\[\AgdaEmptyExtraSkip]%
\>[0]\AgdaComment{\{-\ returns\ true\ -\}}\<%
\\
\\[\AgdaEmptyExtraSkip]%
\>[0]\AgdaFunction{multisigAuxStep1TrueStep2FalseStep3FalseStep4False}\AgdaSpace{}%
\AgdaSymbol{=}\AgdaSpace{}%
\AgdaFunction{cmpMultiSigsAux}\AgdaSpace{}%
\AgdaPostulate{msg₁}\AgdaSpace{}%
\AgdaInductiveConstructor{[]}\AgdaSpace{}%
\AgdaInductiveConstructor{[]}\AgdaSpace{}%
\AgdaPostulate{sig₂}\AgdaSpace{}%
\AgdaInductiveConstructor{false}\<%
\\
\\[\AgdaEmptyExtraSkip]%
\>[0]\AgdaComment{\{-\ returns\ false\ -\}}\<%
\\
\\[\AgdaEmptyExtraSkip]%
\>[0]\AgdaFunction{multisigAuxStep1False}\AgdaSpace{}%
\AgdaSymbol{=}\AgdaSpace{}%
\AgdaFunction{cmpMultiSigsAux}\AgdaSpace{}%
\AgdaPostulate{msg₁}\AgdaSpace{}%
\AgdaOperator{\AgdaFunction{[}}\AgdaSpace{}%
\AgdaPostulate{sig₂}\AgdaSpace{}%
\AgdaOperator{\AgdaFunction{]}}\AgdaSpace{}%
\AgdaSymbol{(}\AgdaPostulate{pbk₂}\AgdaSpace{}%
\AgdaOperator{\AgdaInductiveConstructor{∷}}\AgdaSpace{}%
\AgdaPostulate{pbk₃}\AgdaSpace{}%
\AgdaOperator{\AgdaInductiveConstructor{∷}}\AgdaSpace{}%
\AgdaOperator{\AgdaFunction{[}}\AgdaSpace{}%
\AgdaPostulate{pbk₄}\AgdaSpace{}%
\AgdaOperator{\AgdaFunction{]}}\AgdaSymbol{)}\AgdaSpace{}%
\AgdaPostulate{sig₁}\AgdaSpace{}%
\AgdaInductiveConstructor{false}\<%
\\
\\[\AgdaEmptyExtraSkip]%
\>[0]\AgdaComment{\{-\ returns}\<%
\\
\>[0]\<%
\\
\>[0]\AgdaComment{\ \ \ cmpMultiSigsAux\ msg₁\ [\ sig₂\ ]\ (pbk₃\ ∷\ [\ pbk₄\ ])\ sig₁\ (isSigned\ msg₁\ sig₁\ pbk₂)}\<%
\\
\>[0]\<%
\\
\>[0]\AgdaComment{-\}}\<%
\\
\\[\AgdaEmptyExtraSkip]%
\>[0]\AgdaFunction{multisigAuxStep1FalseStep2True}%
\>[32]\AgdaSymbol{=}\AgdaSpace{}%
\AgdaFunction{cmpMultiSigsAux}\AgdaSpace{}%
\AgdaPostulate{msg₁}\AgdaSpace{}%
\AgdaOperator{\AgdaFunction{[}}\AgdaSpace{}%
\AgdaPostulate{sig₂}\AgdaSpace{}%
\AgdaOperator{\AgdaFunction{]}}\AgdaSpace{}%
\AgdaSymbol{(}\AgdaPostulate{pbk₃}\AgdaSpace{}%
\AgdaOperator{\AgdaInductiveConstructor{∷}}\AgdaSpace{}%
\AgdaOperator{\AgdaFunction{[}}\AgdaSpace{}%
\AgdaPostulate{pbk₄}\AgdaSpace{}%
\AgdaOperator{\AgdaFunction{]}}\AgdaSymbol{)}\AgdaSpace{}%
\AgdaPostulate{sig₁}\AgdaSpace{}%
\AgdaInductiveConstructor{true}\<%
\\
\\[\AgdaEmptyExtraSkip]%
\>[0]\AgdaComment{\{-\ returns}\<%
\\
\>[0]\AgdaComment{\ \ \ \ \ cmpMultiSigsAux\ msg₁\ []\ [\ pbk₄\ ]\ sig₂\ (isSigned\ msg₁\ sig₂\ pbk₃)}\<%
\\
\>[0]\AgdaComment{-\}}\<%
\\
\\[\AgdaEmptyExtraSkip]%
\>[0]\AgdaFunction{multisigAuxStep1FalseStep2TrueStep3True}%
\>[41]\AgdaSymbol{=}\AgdaSpace{}%
\AgdaFunction{cmpMultiSigsAux}\AgdaSpace{}%
\AgdaPostulate{msg₁}\AgdaSpace{}%
\AgdaInductiveConstructor{[]}\AgdaSpace{}%
\AgdaOperator{\AgdaFunction{[}}\AgdaSpace{}%
\AgdaPostulate{pbk₄}\AgdaSpace{}%
\AgdaOperator{\AgdaFunction{]}}\AgdaSpace{}%
\AgdaPostulate{sig₂}\AgdaSpace{}%
\AgdaInductiveConstructor{true}\<%
\\
\\[\AgdaEmptyExtraSkip]%
\>[0]\AgdaComment{\{-\ returns\ true\ -\}}\<%
\\
\\[\AgdaEmptyExtraSkip]%
\>[0]\AgdaFunction{multisigAuxStep1FalseStep2TrueStep3False}%
\>[42]\AgdaSymbol{=}\AgdaSpace{}%
\AgdaFunction{cmpMultiSigsAux}\AgdaSpace{}%
\AgdaPostulate{msg₁}\AgdaSpace{}%
\AgdaInductiveConstructor{[]}\AgdaSpace{}%
\AgdaOperator{\AgdaFunction{[}}\AgdaSpace{}%
\AgdaPostulate{pbk₄}\AgdaSpace{}%
\AgdaOperator{\AgdaFunction{]}}\AgdaSpace{}%
\AgdaPostulate{sig₂}\AgdaSpace{}%
\AgdaInductiveConstructor{false}\<%
\\
\\[\AgdaEmptyExtraSkip]%
\>[0]\AgdaComment{\{-\ returns}\<%
\\
\>[0]\AgdaComment{\ \ \ cmpMultiSigsAux\ msg₁\ []\ []\ sig₂\ (isSigned\ msg₁\ sig₂\ pbk₄)}\<%
\\
\>[0]\AgdaComment{-\}}\<%
\\
\\[\AgdaEmptyExtraSkip]%
\>[0]\AgdaFunction{multisigAuxStep1FalseStep2TrueStep3FalseStep4True}%
\>[51]\AgdaSymbol{=}\AgdaSpace{}%
\AgdaFunction{cmpMultiSigsAux}\AgdaSpace{}%
\AgdaPostulate{msg₁}\AgdaSpace{}%
\AgdaInductiveConstructor{[]}\AgdaSpace{}%
\AgdaInductiveConstructor{[]}\AgdaSpace{}%
\AgdaPostulate{sig₂}\AgdaSpace{}%
\AgdaInductiveConstructor{true}\<%
\\
\\[\AgdaEmptyExtraSkip]%
\>[0]\AgdaComment{\{-\ returns\ true\ -\}}\<%
\\
\\[\AgdaEmptyExtraSkip]%
\>[0]\AgdaFunction{multisigAuxStep1FalseStep2TrueStep3FalseStepFalse}%
\>[51]\AgdaSymbol{=}\AgdaSpace{}%
\AgdaFunction{cmpMultiSigsAux}\AgdaSpace{}%
\AgdaPostulate{msg₁}\AgdaSpace{}%
\AgdaInductiveConstructor{[]}\AgdaSpace{}%
\AgdaInductiveConstructor{[]}\AgdaSpace{}%
\AgdaPostulate{sig₂}\AgdaSpace{}%
\AgdaInductiveConstructor{false}\<%
\\
\\[\AgdaEmptyExtraSkip]%
\>[0]\AgdaComment{\{-\ returns\ false\ -\}}\<%
\\
\\[\AgdaEmptyExtraSkip]%
\\[\AgdaEmptyExtraSkip]%
\>[0]\AgdaFunction{multisigAuxStep1FalseStep2False}%
\>[33]\AgdaSymbol{=}\AgdaSpace{}%
\AgdaFunction{cmpMultiSigsAux}\AgdaSpace{}%
\AgdaPostulate{msg₁}\AgdaSpace{}%
\AgdaOperator{\AgdaFunction{[}}\AgdaSpace{}%
\AgdaPostulate{sig₂}\AgdaSpace{}%
\AgdaOperator{\AgdaFunction{]}}\AgdaSpace{}%
\AgdaSymbol{(}\AgdaPostulate{pbk₃}\AgdaSpace{}%
\AgdaOperator{\AgdaInductiveConstructor{∷}}\AgdaSpace{}%
\AgdaOperator{\AgdaFunction{[}}\AgdaSpace{}%
\AgdaPostulate{pbk₄}\AgdaSpace{}%
\AgdaOperator{\AgdaFunction{]}}\AgdaSymbol{)}\AgdaSpace{}%
\AgdaPostulate{sig₁}\AgdaSpace{}%
\AgdaInductiveConstructor{false}\<%
\\
\\[\AgdaEmptyExtraSkip]%
\>[0]\AgdaComment{\{-returns}\<%
\\
\>[0]\<%
\\
\>[0]\AgdaComment{\ \ cmpMultiSigsAux\ msg₁\ [\ sig₂\ ]\ [\ pbk₄\ ]\ sig₁\ (isSigned\ msg₁\ sig₁\ pbk₃)}\<%
\\
\>[0]\AgdaComment{-\}}\<%
\\
\\[\AgdaEmptyExtraSkip]%
\>[0]\AgdaFunction{multisigAuxStep1FalseStep2FalseStep3True}%
\>[42]\AgdaSymbol{=}\AgdaSpace{}%
\AgdaFunction{cmpMultiSigsAux}\AgdaSpace{}%
\AgdaPostulate{msg₁}\AgdaSpace{}%
\AgdaOperator{\AgdaFunction{[}}\AgdaSpace{}%
\AgdaPostulate{sig₂}\AgdaSpace{}%
\AgdaOperator{\AgdaFunction{]}}\AgdaSpace{}%
\AgdaOperator{\AgdaFunction{[}}\AgdaSpace{}%
\AgdaPostulate{pbk₄}\AgdaSpace{}%
\AgdaOperator{\AgdaFunction{]}}\AgdaSpace{}%
\AgdaPostulate{sig₁}\AgdaSpace{}%
\AgdaInductiveConstructor{true}\<%
\\
\\[\AgdaEmptyExtraSkip]%
\>[0]\AgdaComment{\{-\ returns}\<%
\\
\>[0]\AgdaComment{\ \ \ cmpMultiSigsAux\ msg₁\ []\ []\ sig₂\ (isSigned\ msg₁\ sig₂\ pbk₄)}\<%
\\
\>[0]\AgdaComment{-\}}\<%
\\
\\[\AgdaEmptyExtraSkip]%
\>[0]\AgdaFunction{multisigAuxStep1FalseStep2FalseStep3TrueStep4True}%
\>[51]\AgdaSymbol{=}\AgdaSpace{}%
\AgdaFunction{cmpMultiSigsAux}\AgdaSpace{}%
\AgdaPostulate{msg₁}\AgdaSpace{}%
\AgdaInductiveConstructor{[]}\AgdaSpace{}%
\AgdaInductiveConstructor{[]}\AgdaSpace{}%
\AgdaPostulate{sig₂}\AgdaSpace{}%
\AgdaInductiveConstructor{true}\<%
\\
\\[\AgdaEmptyExtraSkip]%
\>[0]\AgdaComment{\{-\ returns\ true\ -\}}\<%
\\
\\[\AgdaEmptyExtraSkip]%
\>[0]\AgdaFunction{multisigAuxStep1FalseStep2FalseStep3TrueStep4False}%
\>[52]\AgdaSymbol{=}\AgdaSpace{}%
\AgdaFunction{cmpMultiSigsAux}\AgdaSpace{}%
\AgdaPostulate{msg₁}\AgdaSpace{}%
\AgdaInductiveConstructor{[]}\AgdaSpace{}%
\AgdaInductiveConstructor{[]}\AgdaSpace{}%
\AgdaPostulate{sig₂}\AgdaSpace{}%
\AgdaInductiveConstructor{false}\<%
\\
\\[\AgdaEmptyExtraSkip]%
\>[0]\AgdaComment{\{-\ returns\ false\ -\}}\<%
\\
\\[\AgdaEmptyExtraSkip]%
\\[\AgdaEmptyExtraSkip]%
\\[\AgdaEmptyExtraSkip]%
\\[\AgdaEmptyExtraSkip]%
\>[0]\AgdaFunction{multisigAuxStep1FalseStep2FalseStep3False}%
\>[43]\AgdaSymbol{=}\AgdaSpace{}%
\AgdaFunction{cmpMultiSigsAux}\AgdaSpace{}%
\AgdaPostulate{msg₁}\AgdaSpace{}%
\AgdaOperator{\AgdaFunction{[}}\AgdaSpace{}%
\AgdaPostulate{sig₂}\AgdaSpace{}%
\AgdaOperator{\AgdaFunction{]}}\AgdaSpace{}%
\AgdaOperator{\AgdaFunction{[}}\AgdaSpace{}%
\AgdaPostulate{pbk₄}\AgdaSpace{}%
\AgdaOperator{\AgdaFunction{]}}\AgdaSpace{}%
\AgdaPostulate{sig₁}\AgdaSpace{}%
\AgdaInductiveConstructor{false}\<%
\\
\\[\AgdaEmptyExtraSkip]%
\>[0]\AgdaComment{\{-\ returns}\<%
\\
\>[0]\AgdaComment{\ \ \ \ cmpMultiSigsAux\ msg₁\ [\ sig₂\ ]\ []\ sig₁\ (isSigned\ msg₁\ sig₁\ pbk₄)}\<%
\\
\>[0]\AgdaComment{-\}}\<%
\\
\\[\AgdaEmptyExtraSkip]%
\>[0]\AgdaFunction{multisigAuxStep1FalseStep2FalseStep3FalseStep4True}%
\>[52]\AgdaSymbol{=}\AgdaSpace{}%
\AgdaFunction{cmpMultiSigsAux}\AgdaSpace{}%
\AgdaPostulate{msg₁}\AgdaSpace{}%
\AgdaOperator{\AgdaFunction{[}}\AgdaSpace{}%
\AgdaPostulate{sig₂}\AgdaSpace{}%
\AgdaOperator{\AgdaFunction{]}}\AgdaSpace{}%
\AgdaInductiveConstructor{[]}\AgdaSpace{}%
\AgdaPostulate{sig₁}\AgdaSpace{}%
\AgdaInductiveConstructor{true}\<%
\\
\\[\AgdaEmptyExtraSkip]%
\>[0]\AgdaComment{\{-\ returns\ false\ -\}}\<%
\\
\\[\AgdaEmptyExtraSkip]%
\>[0]\AgdaFunction{multisigAuxStep1FalseStep2FalseStep3FalseStep4False}%
\>[53]\AgdaSymbol{=}\AgdaSpace{}%
\AgdaFunction{cmpMultiSigsAux}\AgdaSpace{}%
\AgdaPostulate{msg₁}\AgdaSpace{}%
\AgdaOperator{\AgdaFunction{[}}\AgdaSpace{}%
\AgdaPostulate{sig₂}\AgdaSpace{}%
\AgdaOperator{\AgdaFunction{]}}\AgdaSpace{}%
\AgdaInductiveConstructor{[]}\AgdaSpace{}%
\AgdaPostulate{sig₁}\AgdaSpace{}%
\AgdaInductiveConstructor{false}\<%
\\
\\[\AgdaEmptyExtraSkip]%
\>[0]\AgdaComment{\{-\ returns\ false\ -\}}\<%
\\
\\[\AgdaEmptyExtraSkip]%
\\[\AgdaEmptyExtraSkip]%
\>[0]\AgdaComment{\{-\ So\ we\ see\ that\ that}\<%
\\
\>[0]\<%
\\
\>[0]\AgdaComment{(cmpMultiSigsAux\ msg₁\ [\ sig₂\ ]\ (pbk₂\ ∷\ pbk₃\ ∷\ [\ pbk₄\ ])\ sig₁\ \ (isSigned\ msg₁\ sig₁\ pbk₁))}\<%
\\
\>[0]\<%
\\
\>[0]\<%
\\
\>[0]\AgdaComment{returns\ true\ iff}\<%
\\
\>[0]\<%
\\
\>[0]\AgdaComment{(isSigned\ msg₁\ sig₂\ pbk₂)\ \ \ and\ (isSigned\ msg₁\ sig₂\ pbk₂)}\<%
\\
\>[0]\AgdaComment{or}\<%
\\
\>[0]\AgdaComment{(isSigned\ msg₁\ sig₂\ pbk₂)\ \ \ and\ ¬\ (isSigned\ msg₁\ sig₂\ pbk₂)\ \ and\ (isSigned\ msg₁\ sig₂\ pbk₃)}\<%
\\
\>[0]\AgdaComment{or}\<%
\\
\>[0]\AgdaComment{(isSigned\ msg₁\ sig₂\ pbk₂)\ \ \ and\ ¬\ (isSigned\ msg₁\ sig₂\ pbk₂)\ \ and\ ¬\ (isSigned\ msg₁\ sig₂\ pbk₃)\ and\ (isSigned\ msg₁\ sig₂\ pbk₄)}\<%
\\
\>[0]\AgdaComment{or}\<%
\\
\>[0]\AgdaComment{¬\ (isSigned\ msg₁\ sig₂\ pbk₂)\ \ and\ (isSigned\ msg₁\ sig₁\ pbk₂)\ and\ (isSigned\ msg₁\ sig₂\ pbk₃)}\<%
\\
\>[0]\AgdaComment{or}\<%
\\
\>[0]\AgdaComment{¬\ (isSigned\ msg₁\ sig₂\ pbk₂)\ \ and\ (isSigned\ msg₁\ sig₁\ pbk₂)\ and\ ¬\ (isSigned\ msg₁\ sig₂\ pbk₃)\ and\ (isSigned\ msg₁\ sig₂\ pbk₄)}\<%
\\
\>[0]\AgdaComment{or}\<%
\\
\>[0]\AgdaComment{¬\ (isSigned\ msg₁\ sig₂\ pbk₂)\ \ and\ ¬\ (isSigned\ msg₁\ sig₁\ pbk₂)\ \ and\ (isSigned\ msg₁\ sig₁\ pbk₃)\ \ and\ (isSigned\ msg₁\ sig₂\ pbk₄)}\<%
\\
\>[0]\<%
\\
\>[0]\<%
\\
\>[0]\<%
\\
\>[0]\AgdaComment{we\ simplify\ it\ to:}\<%
\\
\>[0]\<%
\\
\>[0]\AgdaComment{(isSigned\ msg₁\ sig₂\ pbk₂)\ \ \ and\ (isSigned\ msg₁\ sig₂\ pbk₂)}\<%
\\
\>[0]\AgdaComment{or}\<%
\\
\>[0]\AgdaComment{(isSigned\ msg₁\ sig₂\ pbk₂)\ \ \ and\ (isSigned\ msg₁\ sig₂\ pbk₃)}\<%
\\
\>[0]\AgdaComment{or}\<%
\\
\>[0]\AgdaComment{(isSigned\ msg₁\ sig₂\ pbk₂)\ \ \ and\ \ (isSigned\ msg₁\ sig₂\ pbk₄)}\<%
\\
\>[0]\AgdaComment{or}\<%
\\
\>[0]\AgdaComment{(isSigned\ msg₁\ sig₁\ pbk₂)\ and\ (isSigned\ msg₁\ sig₂\ pbk₃)}\<%
\\
\>[0]\AgdaComment{or}\<%
\\
\>[0]\AgdaComment{(isSigned\ msg₁\ sig₁\ pbk₂)\ and\ (isSigned\ msg₁\ sig₂\ pbk₄)}\<%
\\
\>[0]\AgdaComment{or}\<%
\\
\>[0]\AgdaComment{(isSigned\ msg₁\ sig₁\ pbk₃)\ \ and\ (isSigned\ msg₁\ sig₂\ pbk₄)}\<%
\\
\>[0]\<%
\\
\>[0]\<%
\\
\>[0]\<%
\\
\>[0]\AgdaComment{so\ the\ full\ script\ is\ accepted\ if\ and\ only\ if\ the\ stack\ has\ hight\ at\ least\ 3\ and}\<%
\\
\>[0]\AgdaComment{if\ the\ top\ elements\ are\ sig₁\ sig₂\ \ dummy}\<%
\\
\>[0]\AgdaComment{then\ \ the\ above\ condition\ holds}\<%
\\
\>[0]\<%
\\
\>[0]\AgdaComment{so\ the\ weakest\ precondition\ is\ ...\ \ name\ for\ weakest\ precondition}\<%
\\
\>[0]\<%
\\
\>[0]\<%
\\
\>[0]\AgdaComment{-\}}\<%
\\
\\[\AgdaEmptyExtraSkip]%
\\[\AgdaEmptyExtraSkip]%
\\[\AgdaEmptyExtraSkip]%
\\[\AgdaEmptyExtraSkip]%
\>[0]\AgdaComment{\{-}\<%
\\
\>[0]\AgdaComment{test\ :\ Set}\<%
\\
\>[0]\AgdaComment{test\ =\ \{!!\}}\<%
\\
\>[0]\AgdaComment{-\}}\<%
\end{code}
} 


\AgdaHide{
\begin{code}%
\>[0]\AgdaComment{--\textbackslash{}verificationMultiSigBasic}\<%
\\
\>[0]\AgdaKeyword{open}\AgdaSpace{}%
\AgdaKeyword{import}\AgdaSpace{}%
\AgdaModule{basicBitcoinDataType}\<%
\\
\\[\AgdaEmptyExtraSkip]%
\>[0]\AgdaKeyword{module}\AgdaSpace{}%
\AgdaModule{verificationStackScripts.verificationMultiSigBasic}\AgdaSpace{}%
\AgdaSymbol{(}\AgdaBound{param}\AgdaSpace{}%
\AgdaSymbol{:}\AgdaSpace{}%
\AgdaRecord{GlobalParameters}\AgdaSymbol{)}\AgdaSpace{}%
\AgdaKeyword{where}\<%
\\
\\[\AgdaEmptyExtraSkip]%
\\[\AgdaEmptyExtraSkip]%
\>[0]\AgdaKeyword{open}\AgdaSpace{}%
\AgdaKeyword{import}\AgdaSpace{}%
\AgdaModule{Data.List.Base}\AgdaSpace{}%
\AgdaKeyword{hiding}\AgdaSpace{}%
\AgdaSymbol{(}\AgdaOperator{\AgdaFunction{\AgdaUnderscore{}\ensuremath{+\!\!+}\AgdaUnderscore{}}}\AgdaSpace{}%
\AgdaSymbol{)}\<%
\\
\>[0]\AgdaKeyword{open}\AgdaSpace{}%
\AgdaKeyword{import}\AgdaSpace{}%
\AgdaModule{Data.Nat}%
\>[22]\AgdaKeyword{renaming}\AgdaSpace{}%
\AgdaSymbol{(}\AgdaOperator{\AgdaDatatype{\AgdaUnderscore{}≤\AgdaUnderscore{}}}\AgdaSpace{}%
\AgdaSymbol{to}\AgdaSpace{}%
\AgdaOperator{\AgdaDatatype{\AgdaUnderscore{}≤'\AgdaUnderscore{}}}\AgdaSpace{}%
\AgdaSymbol{;}\AgdaSpace{}%
\AgdaOperator{\AgdaFunction{\AgdaUnderscore{}<\AgdaUnderscore{}}}\AgdaSpace{}%
\AgdaSymbol{to}\AgdaSpace{}%
\AgdaOperator{\AgdaFunction{\AgdaUnderscore{}<'\AgdaUnderscore{}}}\AgdaSymbol{)}\<%
\\
\>[0]\AgdaKeyword{open}\AgdaSpace{}%
\AgdaKeyword{import}\AgdaSpace{}%
\AgdaModule{Data.List}\AgdaSpace{}%
\AgdaKeyword{hiding}\AgdaSpace{}%
\AgdaSymbol{(}\AgdaOperator{\AgdaFunction{\AgdaUnderscore{}\ensuremath{+\!\!+}\AgdaUnderscore{}}}%
\>[36]\AgdaSymbol{)}\<%
\\
\>[0]\AgdaKeyword{open}\AgdaSpace{}%
\AgdaKeyword{import}\AgdaSpace{}%
\AgdaModule{Data.Sum}\<%
\\
\>[0]\AgdaKeyword{open}\AgdaSpace{}%
\AgdaKeyword{import}\AgdaSpace{}%
\AgdaModule{Data.Unit}%
\>[23]\AgdaKeyword{hiding}\AgdaSpace{}%
\AgdaSymbol{(}\AgdaOperator{\AgdaRecord{\AgdaUnderscore{}≤\AgdaUnderscore{}}}\AgdaSymbol{)}\<%
\\
\>[0]\AgdaKeyword{open}\AgdaSpace{}%
\AgdaKeyword{import}\AgdaSpace{}%
\AgdaModule{Data.Empty}\<%
\\
\>[0]\AgdaKeyword{open}\AgdaSpace{}%
\AgdaKeyword{import}\AgdaSpace{}%
\AgdaModule{Data.Maybe}\<%
\\
\>[0]\AgdaKeyword{open}\AgdaSpace{}%
\AgdaKeyword{import}\AgdaSpace{}%
\AgdaModule{Data.Bool}%
\>[23]\AgdaKeyword{hiding}\AgdaSpace{}%
\AgdaSymbol{(}\AgdaOperator{\AgdaDatatype{\AgdaUnderscore{}≤\AgdaUnderscore{}}}\AgdaSpace{}%
\AgdaSymbol{;}\AgdaSpace{}%
\AgdaOperator{\AgdaDatatype{\AgdaUnderscore{}<\AgdaUnderscore{}}}\AgdaSpace{}%
\AgdaSymbol{;}\AgdaSpace{}%
\AgdaOperator{\AgdaFunction{if\AgdaUnderscore{}then\AgdaUnderscore{}else\AgdaUnderscore{}}}%
\>[58]\AgdaSymbol{)}%
\>[61]\AgdaKeyword{renaming}\AgdaSpace{}%
\AgdaSymbol{(}\AgdaOperator{\AgdaFunction{\AgdaUnderscore{}∧\AgdaUnderscore{}}}\AgdaSpace{}%
\AgdaSymbol{to}\AgdaSpace{}%
\AgdaOperator{\AgdaFunction{\AgdaUnderscore{}∧b\AgdaUnderscore{}}}\AgdaSpace{}%
\AgdaSymbol{;}\AgdaSpace{}%
\AgdaOperator{\AgdaFunction{\AgdaUnderscore{}∨\AgdaUnderscore{}}}\AgdaSpace{}%
\AgdaSymbol{to}\AgdaSpace{}%
\AgdaOperator{\AgdaFunction{\AgdaUnderscore{}∨b\AgdaUnderscore{}}}\AgdaSpace{}%
\AgdaSymbol{;}\AgdaSpace{}%
\AgdaFunction{T}\AgdaSpace{}%
\AgdaSymbol{to}\AgdaSpace{}%
\AgdaFunction{True}\AgdaSymbol{)}\<%
\\
\>[0]\AgdaKeyword{open}\AgdaSpace{}%
\AgdaKeyword{import}\AgdaSpace{}%
\AgdaModule{Data.Product}\AgdaSpace{}%
\AgdaKeyword{renaming}\AgdaSpace{}%
\AgdaSymbol{(}\AgdaOperator{\AgdaInductiveConstructor{\AgdaUnderscore{},\AgdaUnderscore{}}}\AgdaSpace{}%
\AgdaSymbol{to}\AgdaSpace{}%
\AgdaOperator{\AgdaInductiveConstructor{\AgdaUnderscore{},,\AgdaUnderscore{}}}\AgdaSpace{}%
\AgdaSymbol{)}\<%
\\
\>[0]\AgdaKeyword{open}\AgdaSpace{}%
\AgdaKeyword{import}\AgdaSpace{}%
\AgdaModule{Data.Nat.Base}\AgdaSpace{}%
\AgdaKeyword{hiding}\AgdaSpace{}%
\AgdaSymbol{(}\AgdaOperator{\AgdaDatatype{\AgdaUnderscore{}≤\AgdaUnderscore{}}}\AgdaSpace{}%
\AgdaSymbol{;}\AgdaSpace{}%
\AgdaOperator{\AgdaFunction{\AgdaUnderscore{}<\AgdaUnderscore{}}}\AgdaSymbol{)}\<%
\\
\>[0]\AgdaKeyword{open}\AgdaSpace{}%
\AgdaKeyword{import}\AgdaSpace{}%
\AgdaModule{Data.List.NonEmpty}\AgdaSpace{}%
\AgdaKeyword{hiding}\AgdaSpace{}%
\AgdaSymbol{(}\AgdaField{head}\AgdaSymbol{;}\AgdaSpace{}%
\AgdaOperator{\AgdaFunction{[\AgdaUnderscore{}]}}\AgdaSymbol{)}\<%
\\
\>[0]\AgdaKeyword{open}\AgdaSpace{}%
\AgdaKeyword{import}\AgdaSpace{}%
\AgdaModule{Data.Nat}\AgdaSpace{}%
\AgdaKeyword{using}\AgdaSpace{}%
\AgdaSymbol{(}\AgdaDatatype{ℕ}\AgdaSymbol{;}\AgdaSpace{}%
\AgdaOperator{\AgdaPrimitive{\AgdaUnderscore{}+\AgdaUnderscore{}}}\AgdaSymbol{;}\AgdaSpace{}%
\AgdaOperator{\AgdaFunction{\AgdaUnderscore{}≥\AgdaUnderscore{}}}\AgdaSymbol{;}\AgdaSpace{}%
\AgdaOperator{\AgdaFunction{\AgdaUnderscore{}>\AgdaUnderscore{}}}\AgdaSymbol{;}\AgdaSpace{}%
\AgdaInductiveConstructor{zero}\AgdaSymbol{;}\AgdaSpace{}%
\AgdaInductiveConstructor{suc}\AgdaSymbol{;}\AgdaSpace{}%
\AgdaInductiveConstructor{s≤s}\AgdaSymbol{;}\AgdaSpace{}%
\AgdaInductiveConstructor{z≤n}\AgdaSymbol{)}\<%
\\
\>[0]\AgdaKeyword{import}\AgdaSpace{}%
\AgdaModule{Relation.Binary.PropositionalEquality}\AgdaSpace{}%
\AgdaSymbol{as}\AgdaSpace{}%
\AgdaModule{Eq}\<%
\\
\>[0]\AgdaKeyword{open}\AgdaSpace{}%
\AgdaModule{Eq}\AgdaSpace{}%
\AgdaKeyword{using}\AgdaSpace{}%
\AgdaSymbol{(}\AgdaOperator{\AgdaDatatype{\AgdaUnderscore{}≡\AgdaUnderscore{}}}\AgdaSymbol{;}\AgdaSpace{}%
\AgdaInductiveConstructor{refl}\AgdaSymbol{;}\AgdaSpace{}%
\AgdaFunction{cong}\AgdaSymbol{;}\AgdaSpace{}%
\AgdaKeyword{module}\AgdaSpace{}%
\AgdaModule{≡-Reasoning}\AgdaSymbol{;}\AgdaSpace{}%
\AgdaFunction{sym}\AgdaSymbol{)}\<%
\\
\>[0]\AgdaKeyword{open}\AgdaSpace{}%
\AgdaModule{≡-Reasoning}\<%
\\
\>[0]\AgdaKeyword{open}\AgdaSpace{}%
\AgdaKeyword{import}\AgdaSpace{}%
\AgdaModule{Agda.Builtin.Equality}\<%
\\
\\[\AgdaEmptyExtraSkip]%
\\[\AgdaEmptyExtraSkip]%
\>[0]\AgdaComment{--our\ libraries}\<%
\\
\>[0]\AgdaKeyword{open}\AgdaSpace{}%
\AgdaKeyword{import}\AgdaSpace{}%
\AgdaModule{libraries.listLib}\<%
\\
\>[0]\AgdaKeyword{open}\AgdaSpace{}%
\AgdaKeyword{import}\AgdaSpace{}%
\AgdaModule{libraries.emptyLib}\<%
\\
\>[0]\AgdaKeyword{open}\AgdaSpace{}%
\AgdaKeyword{import}\AgdaSpace{}%
\AgdaModule{libraries.natLib}\<%
\\
\>[0]\AgdaKeyword{open}\AgdaSpace{}%
\AgdaKeyword{import}\AgdaSpace{}%
\AgdaModule{libraries.boolLib}\<%
\\
\>[0]\AgdaKeyword{open}\AgdaSpace{}%
\AgdaKeyword{import}\AgdaSpace{}%
\AgdaModule{libraries.equalityLib}\<%
\\
\>[0]\AgdaKeyword{open}\AgdaSpace{}%
\AgdaKeyword{import}\AgdaSpace{}%
\AgdaModule{libraries.andLib}\<%
\\
\>[0]\AgdaKeyword{open}\AgdaSpace{}%
\AgdaKeyword{import}\AgdaSpace{}%
\AgdaModule{libraries.miscLib}\<%
\\
\>[0]\AgdaKeyword{open}\AgdaSpace{}%
\AgdaKeyword{import}\AgdaSpace{}%
\AgdaModule{libraries.maybeLib}\<%
\\
\\[\AgdaEmptyExtraSkip]%
\>[0]\AgdaKeyword{open}\AgdaSpace{}%
\AgdaKeyword{import}\AgdaSpace{}%
\AgdaModule{stack}\<%
\\
\>[0]\AgdaKeyword{open}\AgdaSpace{}%
\AgdaKeyword{import}\AgdaSpace{}%
\AgdaModule{stackPredicate}\<%
\\
\>[0]\AgdaKeyword{open}\AgdaSpace{}%
\AgdaKeyword{import}\AgdaSpace{}%
\AgdaModule{semanticBasicOperations}\AgdaSpace{}%
\AgdaBound{param}\<%
\\
\>[0]\AgdaKeyword{open}\AgdaSpace{}%
\AgdaKeyword{import}\AgdaSpace{}%
\AgdaModule{instructionBasic}\<%
\\
\>[0]\AgdaKeyword{open}\AgdaSpace{}%
\AgdaKeyword{import}\AgdaSpace{}%
\AgdaModule{verificationMultiSig}\AgdaSpace{}%
\AgdaBound{param}\<%
\\
\>[0]\AgdaKeyword{open}\AgdaSpace{}%
\AgdaKeyword{import}\AgdaSpace{}%
\AgdaModule{hoareTripleStack}\AgdaSpace{}%
\AgdaBound{param}\<%
\\
\>[0]\AgdaKeyword{open}\AgdaSpace{}%
\AgdaKeyword{import}\AgdaSpace{}%
\AgdaModule{verificationStackScripts.semanticsStackInstructions}\AgdaSpace{}%
\AgdaBound{param}\<%
\\
\>[0]\AgdaKeyword{open}\AgdaSpace{}%
\AgdaKeyword{import}\AgdaSpace{}%
\AgdaModule{verificationStackScripts.stackVerificationLemmas}\AgdaSpace{}%
\AgdaBound{param}\<%
\\
\>[0]\AgdaKeyword{open}\AgdaSpace{}%
\AgdaKeyword{import}\AgdaSpace{}%
\AgdaModule{verificationStackScripts.stackHoareTriple}\AgdaSpace{}%
\AgdaBound{param}\<%
\\
\>[0]\AgdaKeyword{open}\AgdaSpace{}%
\AgdaKeyword{import}\AgdaSpace{}%
\AgdaModule{verificationStackScripts.sPredicate}\<%
\\
\>[0]\AgdaKeyword{open}\AgdaSpace{}%
\AgdaKeyword{import}\AgdaSpace{}%
\AgdaModule{verificationStackScripts.hoareTripleStackBasic}\AgdaSpace{}%
\AgdaBound{param}\<%
\\
\>[0]\AgdaKeyword{open}\AgdaSpace{}%
\AgdaKeyword{import}\AgdaSpace{}%
\AgdaModule{verificationStackScripts.stackState}\<%
\\
\>[0]\AgdaKeyword{open}\AgdaSpace{}%
\AgdaKeyword{import}\AgdaSpace{}%
\AgdaModule{verificationStackScripts.stackSemanticsInstructionsBasic}\AgdaSpace{}%
\AgdaBound{param}\<%
\\
\>[0]\AgdaKeyword{open}\AgdaSpace{}%
\AgdaKeyword{import}\AgdaSpace{}%
\AgdaModule{verificationStackScripts.stackVerificationLemmasPart2}\AgdaSpace{}%
\AgdaBound{param}\<%
\\
\>[0]\AgdaKeyword{open}\AgdaSpace{}%
\AgdaKeyword{import}\AgdaSpace{}%
\AgdaModule{verificationStackScripts.stackVerificationP2PKH}\AgdaSpace{}%
\AgdaBound{param}\<%
\\
\\[\AgdaEmptyExtraSkip]%
\>[0]\AgdaFunction{mainLemmaCorrectnessMultiSig-2-4}\AgdaSpace{}%
\AgdaSymbol{:}\AgdaSpace{}%
\AgdaSymbol{(}\AgdaBound{msg₁}\AgdaSpace{}%
\AgdaSymbol{:}\AgdaSpace{}%
\AgdaDatatype{Msg}\AgdaSymbol{)(}\AgdaBound{pbk1}\AgdaSpace{}%
\AgdaBound{pbk2}\AgdaSpace{}%
\AgdaBound{pbk3}\AgdaSpace{}%
\AgdaBound{pbk4}%
\>[69]\AgdaSymbol{:}\AgdaSpace{}%
\AgdaDatatype{ℕ}\AgdaSymbol{)}\AgdaSpace{}%
\AgdaSymbol{→}\<%
\\
\>[0][@{}l@{\AgdaIndent{0}}]%
\>[19]\AgdaOperator{\AgdaFunction{<}}\AgdaSpace{}%
\AgdaFunction{weakestPreCondMultiSig-2-4ˢ}\AgdaSpace{}%
\AgdaBound{pbk1}\AgdaSpace{}%
\AgdaBound{pbk2}\AgdaSpace{}%
\AgdaBound{pbk3}\AgdaSpace{}%
\AgdaBound{pbk4}\AgdaSpace{}%
\AgdaOperator{\AgdaFunction{>stackb}}\<%
\\
\>[19][@{}l@{\AgdaIndent{0}}]%
\>[20]\AgdaFunction{multiSigScript2-4ᵇ}\AgdaSpace{}%
\AgdaBound{pbk1}\AgdaSpace{}%
\AgdaBound{pbk2}\AgdaSpace{}%
\AgdaBound{pbk3}\AgdaSpace{}%
\AgdaBound{pbk4}\<%
\\
\>[0][@{}l@{\AgdaIndent{0}}]%
\>[18]\AgdaOperator{\AgdaFunction{<}}\AgdaSpace{}%
\AgdaFunction{acceptStateˢ}\AgdaSpace{}%
\AgdaOperator{\AgdaFunction{>}}\<%
\\
\>[0]\AgdaFunction{mainLemmaCorrectnessMultiSig-2-4}\AgdaSpace{}%
\AgdaBound{msg₁}\AgdaSpace{}%
\AgdaBound{pbk1}\AgdaSpace{}%
\AgdaBound{pbk2}\AgdaSpace{}%
\AgdaBound{pbk3}\AgdaSpace{}%
\AgdaBound{pbk4}\AgdaSpace{}%
\AgdaSymbol{.}\AgdaField{==>stg}\AgdaSpace{}%
\AgdaBound{time}\AgdaSpace{}%
\AgdaBound{msg₂}\AgdaSpace{}%
\AgdaSymbol{(}\AgdaBound{sig2}\AgdaSpace{}%
\AgdaOperator{\AgdaInductiveConstructor{∷}}\AgdaSpace{}%
\AgdaBound{sig1}\AgdaSpace{}%
\AgdaOperator{\AgdaInductiveConstructor{∷}}\AgdaSpace{}%
\AgdaBound{dummy}\AgdaSpace{}%
\AgdaOperator{\AgdaInductiveConstructor{∷}}\AgdaSpace{}%
\AgdaBound{stack}\AgdaSymbol{)}\<%
\\
\>[0][@{}l@{\AgdaIndent{0}}]%
\>[2]\AgdaSymbol{(}\AgdaInductiveConstructor{inj₁}\AgdaSpace{}%
\AgdaSymbol{(}\AgdaInductiveConstructor{conj}\AgdaSpace{}%
\AgdaBound{and3}\AgdaSpace{}%
\AgdaBound{and4}\AgdaSymbol{))}\AgdaSpace{}%
\AgdaSymbol{=}\<%
\\
\>[2][@{}l@{\AgdaIndent{0}}]%
\>[3]\AgdaFunction{boolToNatNotFalseLemma}\AgdaSpace{}%
\AgdaSymbol{(}\AgdaFunction{compareSigsMultiSigAux}\AgdaSpace{}%
\AgdaBound{msg₂}\AgdaSpace{}%
\AgdaSymbol{(}\AgdaBound{sig2}\AgdaSpace{}%
\AgdaOperator{\AgdaInductiveConstructor{∷}}\AgdaSpace{}%
\AgdaInductiveConstructor{[]}\AgdaSymbol{)}\AgdaSpace{}%
\AgdaSymbol{(}\AgdaBound{pbk2}\AgdaSpace{}%
\AgdaOperator{\AgdaInductiveConstructor{∷}}\AgdaSpace{}%
\AgdaBound{pbk3}\AgdaSpace{}%
\AgdaOperator{\AgdaInductiveConstructor{∷}}\AgdaSpace{}%
\AgdaBound{pbk4}\AgdaSpace{}%
\AgdaOperator{\AgdaInductiveConstructor{∷}}\AgdaSpace{}%
\AgdaInductiveConstructor{[]}\AgdaSymbol{)}\AgdaSpace{}%
\AgdaBound{sig1}\<%
\\
\>[3]\AgdaSymbol{(}\AgdaFunction{isSigned}%
\>[14]\AgdaBound{msg₂}\AgdaSpace{}%
\AgdaBound{sig1}\AgdaSpace{}%
\AgdaBound{pbk1}\AgdaSymbol{))}\<%
\\
\>[2]\AgdaSymbol{(}\AgdaFunction{lemmaHoareTripleStackGeAux'7}\AgdaSpace{}%
\AgdaBound{msg₂}\AgdaSpace{}%
\AgdaBound{pbk1}\AgdaSpace{}%
\AgdaBound{pbk2}\AgdaSpace{}%
\AgdaBound{pbk3}\AgdaSpace{}%
\AgdaBound{pbk4}\AgdaSpace{}%
\AgdaBound{sig1}\AgdaSpace{}%
\AgdaBound{sig2}\AgdaSpace{}%
\AgdaBound{and3}\AgdaSpace{}%
\AgdaBound{and4}\AgdaSymbol{)}\<%
\\
\>[0]\AgdaFunction{mainLemmaCorrectnessMultiSig-2-4}\AgdaSpace{}%
\AgdaBound{msg₁}\AgdaSpace{}%
\AgdaBound{pbk1}\AgdaSpace{}%
\AgdaBound{pbk2}\AgdaSpace{}%
\AgdaBound{pbk3}\AgdaSpace{}%
\AgdaBound{pbk4}\AgdaSpace{}%
\AgdaSymbol{.}\AgdaField{==>stg}\AgdaSpace{}%
\AgdaBound{time}\AgdaSpace{}%
\AgdaBound{msg₂}\AgdaSpace{}%
\AgdaSymbol{(}\AgdaBound{sig2}\AgdaSpace{}%
\AgdaOperator{\AgdaInductiveConstructor{∷}}\AgdaSpace{}%
\AgdaBound{sig1}\AgdaSpace{}%
\AgdaOperator{\AgdaInductiveConstructor{∷}}\AgdaSpace{}%
\AgdaBound{dummy}\AgdaSpace{}%
\AgdaOperator{\AgdaInductiveConstructor{∷}}\AgdaSpace{}%
\AgdaBound{stack}\AgdaSymbol{)}\<%
\\
\>[0][@{}l@{\AgdaIndent{0}}]%
\>[2]\AgdaSymbol{(}\AgdaInductiveConstructor{inj₂}\AgdaSpace{}%
\AgdaSymbol{(}\AgdaInductiveConstructor{inj₁}\AgdaSpace{}%
\AgdaSymbol{(}\AgdaInductiveConstructor{conj}\AgdaSpace{}%
\AgdaBound{and3}\AgdaSpace{}%
\AgdaBound{and4}\AgdaSymbol{)))}\AgdaSpace{}%
\AgdaSymbol{=}\<%
\\
\>[2][@{}l@{\AgdaIndent{0}}]%
\>[3]\AgdaFunction{boolToNatNotFalseLemma}\AgdaSpace{}%
\AgdaSymbol{(}\AgdaFunction{compareSigsMultiSigAux}\AgdaSpace{}%
\AgdaBound{msg₂}\AgdaSpace{}%
\AgdaSymbol{(}\AgdaBound{sig2}\AgdaSpace{}%
\AgdaOperator{\AgdaInductiveConstructor{∷}}\AgdaSpace{}%
\AgdaInductiveConstructor{[]}\AgdaSymbol{)}\AgdaSpace{}%
\AgdaSymbol{(}\AgdaBound{pbk2}\AgdaSpace{}%
\AgdaOperator{\AgdaInductiveConstructor{∷}}\AgdaSpace{}%
\AgdaBound{pbk3}\AgdaSpace{}%
\AgdaOperator{\AgdaInductiveConstructor{∷}}\AgdaSpace{}%
\AgdaBound{pbk4}\AgdaSpace{}%
\AgdaOperator{\AgdaInductiveConstructor{∷}}\AgdaSpace{}%
\AgdaInductiveConstructor{[]}\AgdaSymbol{)}\AgdaSpace{}%
\AgdaBound{sig1}\<%
\\
\>[3]\AgdaSymbol{(}\AgdaFunction{isSigned}%
\>[14]\AgdaBound{msg₂}\AgdaSpace{}%
\AgdaBound{sig1}\AgdaSpace{}%
\AgdaBound{pbk1}\AgdaSymbol{))}\<%
\\
\>[2]\AgdaSymbol{(}\AgdaFunction{lemmaHoareTripleStackGeAux'8}\AgdaSpace{}%
\AgdaBound{msg₂}\AgdaSpace{}%
\AgdaBound{pbk1}\AgdaSpace{}%
\AgdaBound{pbk2}\AgdaSpace{}%
\AgdaBound{pbk3}\AgdaSpace{}%
\AgdaBound{pbk4}\AgdaSpace{}%
\AgdaBound{sig1}\AgdaSpace{}%
\AgdaBound{sig2}\AgdaSpace{}%
\AgdaBound{and3}\AgdaSpace{}%
\AgdaBound{and4}\AgdaSymbol{)}\<%
\\
\>[0]\AgdaFunction{mainLemmaCorrectnessMultiSig-2-4}\AgdaSpace{}%
\AgdaBound{msg₁}\AgdaSpace{}%
\AgdaBound{pbk1}\AgdaSpace{}%
\AgdaBound{pbk2}\AgdaSpace{}%
\AgdaBound{pbk3}\AgdaSpace{}%
\AgdaBound{pbk4}\AgdaSpace{}%
\AgdaSymbol{.}\AgdaField{==>stg}\AgdaSpace{}%
\AgdaBound{time}\AgdaSpace{}%
\AgdaBound{msg₂}\AgdaSpace{}%
\AgdaSymbol{(}\AgdaBound{sig2}\AgdaSpace{}%
\AgdaOperator{\AgdaInductiveConstructor{∷}}\AgdaSpace{}%
\AgdaBound{sig1}\AgdaSpace{}%
\AgdaOperator{\AgdaInductiveConstructor{∷}}\AgdaSpace{}%
\AgdaBound{dummy}\AgdaSpace{}%
\AgdaOperator{\AgdaInductiveConstructor{∷}}\AgdaSpace{}%
\AgdaBound{stack}\AgdaSymbol{)}\<%
\\
\>[0][@{}l@{\AgdaIndent{0}}]%
\>[2]\AgdaSymbol{(}\AgdaInductiveConstructor{inj₂}\AgdaSpace{}%
\AgdaSymbol{(}\AgdaInductiveConstructor{inj₂}\AgdaSpace{}%
\AgdaSymbol{(}\AgdaInductiveConstructor{inj₁}\AgdaSpace{}%
\AgdaSymbol{(}\AgdaInductiveConstructor{conj}\AgdaSpace{}%
\AgdaBound{and3}\AgdaSpace{}%
\AgdaBound{and4}\AgdaSymbol{))))}\AgdaSpace{}%
\AgdaSymbol{=}\<%
\\
\>[2][@{}l@{\AgdaIndent{0}}]%
\>[3]\AgdaFunction{boolToNatNotFalseLemma}\AgdaSpace{}%
\AgdaSymbol{(}\AgdaFunction{compareSigsMultiSigAux}\AgdaSpace{}%
\AgdaBound{msg₂}\AgdaSpace{}%
\AgdaSymbol{(}\AgdaBound{sig2}\AgdaSpace{}%
\AgdaOperator{\AgdaInductiveConstructor{∷}}\AgdaSpace{}%
\AgdaInductiveConstructor{[]}\AgdaSymbol{)}\AgdaSpace{}%
\AgdaSymbol{(}\AgdaBound{pbk2}\AgdaSpace{}%
\AgdaOperator{\AgdaInductiveConstructor{∷}}\AgdaSpace{}%
\AgdaBound{pbk3}\AgdaSpace{}%
\AgdaOperator{\AgdaInductiveConstructor{∷}}\AgdaSpace{}%
\AgdaBound{pbk4}\AgdaSpace{}%
\AgdaOperator{\AgdaInductiveConstructor{∷}}\AgdaSpace{}%
\AgdaInductiveConstructor{[]}\AgdaSymbol{)}\AgdaSpace{}%
\AgdaBound{sig1}\<%
\\
\>[3]\AgdaSymbol{(}\AgdaFunction{isSigned}%
\>[14]\AgdaBound{msg₂}\AgdaSpace{}%
\AgdaBound{sig1}\AgdaSpace{}%
\AgdaBound{pbk1}\AgdaSymbol{))}\<%
\\
\>[3]\AgdaSymbol{(}\AgdaFunction{lemmaHoareTripleStackGeAux'9}\AgdaSpace{}%
\AgdaBound{msg₂}\AgdaSpace{}%
\AgdaBound{pbk1}\AgdaSpace{}%
\AgdaBound{pbk2}\AgdaSpace{}%
\AgdaBound{pbk3}\AgdaSpace{}%
\AgdaBound{pbk4}\AgdaSpace{}%
\AgdaBound{sig1}\AgdaSpace{}%
\AgdaBound{sig2}\AgdaSpace{}%
\AgdaBound{and3}\AgdaSpace{}%
\AgdaBound{and4}\AgdaSymbol{)}\<%
\\
\>[0]\AgdaFunction{mainLemmaCorrectnessMultiSig-2-4}\AgdaSpace{}%
\AgdaBound{msg₁}\AgdaSpace{}%
\AgdaBound{pbk1}\AgdaSpace{}%
\AgdaBound{pbk2}\AgdaSpace{}%
\AgdaBound{pbk3}\AgdaSpace{}%
\AgdaBound{pbk4}\AgdaSpace{}%
\AgdaSymbol{.}\AgdaField{==>stg}\AgdaSpace{}%
\AgdaBound{time}\AgdaSpace{}%
\AgdaBound{msg₂}\AgdaSpace{}%
\AgdaSymbol{(}\AgdaBound{sig2}\AgdaSpace{}%
\AgdaOperator{\AgdaInductiveConstructor{∷}}\AgdaSpace{}%
\AgdaBound{sig1}\AgdaSpace{}%
\AgdaOperator{\AgdaInductiveConstructor{∷}}\AgdaSpace{}%
\AgdaBound{dummy}\AgdaSpace{}%
\AgdaOperator{\AgdaInductiveConstructor{∷}}\AgdaSpace{}%
\AgdaBound{stack}\AgdaSymbol{)}\<%
\\
\>[0][@{}l@{\AgdaIndent{0}}]%
\>[2]\AgdaSymbol{(}\AgdaInductiveConstructor{inj₂}\AgdaSpace{}%
\AgdaSymbol{(}\AgdaInductiveConstructor{inj₂}\AgdaSpace{}%
\AgdaSymbol{(}\AgdaInductiveConstructor{inj₂}\AgdaSpace{}%
\AgdaSymbol{(}\AgdaInductiveConstructor{inj₁}\AgdaSpace{}%
\AgdaSymbol{(}\AgdaInductiveConstructor{conj}\AgdaSpace{}%
\AgdaBound{and3}\AgdaSpace{}%
\AgdaBound{and4}\AgdaSymbol{)))))}\AgdaSpace{}%
\AgdaSymbol{=}\<%
\\
\>[2][@{}l@{\AgdaIndent{0}}]%
\>[3]\AgdaFunction{boolToNatNotFalseLemma}\AgdaSpace{}%
\AgdaSymbol{(}\AgdaFunction{compareSigsMultiSigAux}\AgdaSpace{}%
\AgdaBound{msg₂}\AgdaSpace{}%
\AgdaSymbol{(}\AgdaBound{sig2}\AgdaSpace{}%
\AgdaOperator{\AgdaInductiveConstructor{∷}}\AgdaSpace{}%
\AgdaInductiveConstructor{[]}\AgdaSymbol{)}\AgdaSpace{}%
\AgdaSymbol{(}\AgdaBound{pbk2}\AgdaSpace{}%
\AgdaOperator{\AgdaInductiveConstructor{∷}}\AgdaSpace{}%
\AgdaBound{pbk3}\AgdaSpace{}%
\AgdaOperator{\AgdaInductiveConstructor{∷}}\AgdaSpace{}%
\AgdaBound{pbk4}\AgdaSpace{}%
\AgdaOperator{\AgdaInductiveConstructor{∷}}\AgdaSpace{}%
\AgdaInductiveConstructor{[]}\AgdaSymbol{)}\AgdaSpace{}%
\AgdaBound{sig1}\<%
\\
\>[3]\AgdaSymbol{(}\AgdaFunction{isSigned}%
\>[14]\AgdaBound{msg₂}\AgdaSpace{}%
\AgdaBound{sig1}\AgdaSpace{}%
\AgdaBound{pbk1}\AgdaSymbol{))}\<%
\\
\>[3]\AgdaSymbol{(}\AgdaFunction{lemmaHoareTripleStackGeAux'10}\AgdaSpace{}%
\AgdaBound{msg₂}\AgdaSpace{}%
\AgdaBound{pbk1}\AgdaSpace{}%
\AgdaBound{pbk2}\AgdaSpace{}%
\AgdaBound{pbk3}\AgdaSpace{}%
\AgdaBound{pbk4}\AgdaSpace{}%
\AgdaBound{sig1}\AgdaSpace{}%
\AgdaBound{sig2}\AgdaSpace{}%
\AgdaBound{and3}\AgdaSpace{}%
\AgdaBound{and4}\AgdaSymbol{)}\<%
\\
\>[0]\AgdaFunction{mainLemmaCorrectnessMultiSig-2-4}\AgdaSpace{}%
\AgdaBound{msg₁}\AgdaSpace{}%
\AgdaBound{pbk1}\AgdaSpace{}%
\AgdaBound{pbk2}\AgdaSpace{}%
\AgdaBound{pbk3}\AgdaSpace{}%
\AgdaBound{pbk4}\AgdaSpace{}%
\AgdaSymbol{.}\AgdaField{==>stg}\AgdaSpace{}%
\AgdaBound{time}\AgdaSpace{}%
\AgdaBound{msg₂}\AgdaSpace{}%
\AgdaSymbol{(}\AgdaBound{sig2}\AgdaSpace{}%
\AgdaOperator{\AgdaInductiveConstructor{∷}}\AgdaSpace{}%
\AgdaBound{sig1}\AgdaSpace{}%
\AgdaOperator{\AgdaInductiveConstructor{∷}}\AgdaSpace{}%
\AgdaBound{dummy}\AgdaSpace{}%
\AgdaOperator{\AgdaInductiveConstructor{∷}}\AgdaSpace{}%
\AgdaBound{stack}\AgdaSymbol{)}\<%
\\
\>[0][@{}l@{\AgdaIndent{0}}]%
\>[2]\AgdaSymbol{(}\AgdaInductiveConstructor{inj₂}\AgdaSpace{}%
\AgdaSymbol{(}\AgdaInductiveConstructor{inj₂}\AgdaSpace{}%
\AgdaSymbol{(}\AgdaInductiveConstructor{inj₂}\AgdaSpace{}%
\AgdaSymbol{(}\AgdaInductiveConstructor{inj₂}\AgdaSpace{}%
\AgdaSymbol{(}\AgdaInductiveConstructor{inj₁}\AgdaSpace{}%
\AgdaSymbol{(}\AgdaInductiveConstructor{conj}\AgdaSpace{}%
\AgdaBound{and3}\AgdaSpace{}%
\AgdaBound{and4}\AgdaSymbol{))))))}\AgdaSpace{}%
\AgdaSymbol{=}\<%
\\
\>[2][@{}l@{\AgdaIndent{0}}]%
\>[3]\AgdaFunction{boolToNatNotFalseLemma}\AgdaSpace{}%
\AgdaSymbol{(}\AgdaFunction{compareSigsMultiSigAux}\AgdaSpace{}%
\AgdaBound{msg₂}\AgdaSpace{}%
\AgdaSymbol{(}\AgdaBound{sig2}\AgdaSpace{}%
\AgdaOperator{\AgdaInductiveConstructor{∷}}\AgdaSpace{}%
\AgdaInductiveConstructor{[]}\AgdaSymbol{)}\AgdaSpace{}%
\AgdaSymbol{(}\AgdaBound{pbk2}\AgdaSpace{}%
\AgdaOperator{\AgdaInductiveConstructor{∷}}\AgdaSpace{}%
\AgdaBound{pbk3}\AgdaSpace{}%
\AgdaOperator{\AgdaInductiveConstructor{∷}}\AgdaSpace{}%
\AgdaBound{pbk4}\AgdaSpace{}%
\AgdaOperator{\AgdaInductiveConstructor{∷}}\AgdaSpace{}%
\AgdaInductiveConstructor{[]}\AgdaSymbol{)}\AgdaSpace{}%
\AgdaBound{sig1}\<%
\\
\>[3]\AgdaSymbol{(}\AgdaFunction{isSigned}%
\>[14]\AgdaBound{msg₂}\AgdaSpace{}%
\AgdaBound{sig1}\AgdaSpace{}%
\AgdaBound{pbk1}\AgdaSymbol{))}\<%
\\
\>[3]\AgdaSymbol{(}\AgdaFunction{lemmaHoareTripleStackGeAux'11}\AgdaSpace{}%
\AgdaBound{msg₂}\AgdaSpace{}%
\AgdaBound{pbk1}\AgdaSpace{}%
\AgdaBound{pbk2}\AgdaSpace{}%
\AgdaBound{pbk3}\AgdaSpace{}%
\AgdaBound{pbk4}\AgdaSpace{}%
\AgdaBound{sig1}\AgdaSpace{}%
\AgdaBound{sig2}\AgdaSpace{}%
\AgdaBound{and3}\AgdaSpace{}%
\AgdaBound{and4}\AgdaSymbol{)}\<%
\\
\>[0]\AgdaFunction{mainLemmaCorrectnessMultiSig-2-4}\AgdaSpace{}%
\AgdaBound{msg₁}\AgdaSpace{}%
\AgdaBound{pbk1}\AgdaSpace{}%
\AgdaBound{pbk2}\AgdaSpace{}%
\AgdaBound{pbk3}\AgdaSpace{}%
\AgdaBound{pbk4}\AgdaSpace{}%
\AgdaSymbol{.}\AgdaField{==>stg}\AgdaSpace{}%
\AgdaBound{time}\AgdaSpace{}%
\AgdaBound{msg₂}\AgdaSpace{}%
\AgdaSymbol{(}\AgdaBound{sig2}\AgdaSpace{}%
\AgdaOperator{\AgdaInductiveConstructor{∷}}\AgdaSpace{}%
\AgdaBound{sig1}\AgdaSpace{}%
\AgdaOperator{\AgdaInductiveConstructor{∷}}\AgdaSpace{}%
\AgdaBound{dummy}\AgdaSpace{}%
\AgdaOperator{\AgdaInductiveConstructor{∷}}\AgdaSpace{}%
\AgdaBound{stack}\AgdaSymbol{)}\<%
\\
\>[0][@{}l@{\AgdaIndent{0}}]%
\>[2]\AgdaSymbol{(}\AgdaInductiveConstructor{inj₂}\AgdaSpace{}%
\AgdaSymbol{(}\AgdaInductiveConstructor{inj₂}\AgdaSpace{}%
\AgdaSymbol{(}\AgdaInductiveConstructor{inj₂}\AgdaSpace{}%
\AgdaSymbol{(}\AgdaInductiveConstructor{inj₂}\AgdaSpace{}%
\AgdaSymbol{(}\AgdaInductiveConstructor{inj₂}\AgdaSpace{}%
\AgdaSymbol{(}\AgdaInductiveConstructor{conj}\AgdaSpace{}%
\AgdaBound{and3}\AgdaSpace{}%
\AgdaBound{and4}\AgdaSymbol{))))))}\AgdaSpace{}%
\AgdaSymbol{=}\<%
\\
\>[2][@{}l@{\AgdaIndent{0}}]%
\>[3]\AgdaFunction{boolToNatNotFalseLemma}\AgdaSpace{}%
\AgdaSymbol{(}\AgdaFunction{compareSigsMultiSigAux}\AgdaSpace{}%
\AgdaBound{msg₂}\AgdaSpace{}%
\AgdaSymbol{(}\AgdaBound{sig2}\AgdaSpace{}%
\AgdaOperator{\AgdaInductiveConstructor{∷}}\AgdaSpace{}%
\AgdaInductiveConstructor{[]}\AgdaSymbol{)}\AgdaSpace{}%
\AgdaSymbol{(}\AgdaBound{pbk2}\AgdaSpace{}%
\AgdaOperator{\AgdaInductiveConstructor{∷}}\AgdaSpace{}%
\AgdaBound{pbk3}\AgdaSpace{}%
\AgdaOperator{\AgdaInductiveConstructor{∷}}\AgdaSpace{}%
\AgdaBound{pbk4}\AgdaSpace{}%
\AgdaOperator{\AgdaInductiveConstructor{∷}}\AgdaSpace{}%
\AgdaInductiveConstructor{[]}\AgdaSymbol{)}\AgdaSpace{}%
\AgdaBound{sig1}\<%
\\
\>[3]\AgdaSymbol{(}\AgdaFunction{isSigned}%
\>[14]\AgdaBound{msg₂}\AgdaSpace{}%
\AgdaBound{sig1}\AgdaSpace{}%
\AgdaBound{pbk1}\AgdaSymbol{))}\<%
\\
\>[3]\AgdaSymbol{(}\AgdaFunction{lemmaHoareTripleStackGeAux'12}\AgdaSpace{}%
\AgdaBound{msg₂}\AgdaSpace{}%
\AgdaBound{pbk1}\AgdaSpace{}%
\AgdaBound{pbk2}\AgdaSpace{}%
\AgdaBound{pbk3}\AgdaSpace{}%
\AgdaBound{pbk4}\AgdaSpace{}%
\AgdaBound{sig1}\AgdaSpace{}%
\AgdaBound{sig2}\AgdaSpace{}%
\AgdaBound{and3}\AgdaSpace{}%
\AgdaBound{and4}\AgdaSymbol{)}\<%
\\
\>[0]\AgdaFunction{mainLemmaCorrectnessMultiSig-2-4}\AgdaSpace{}%
\AgdaBound{msg₁}\AgdaSpace{}%
\AgdaBound{pbk1}\AgdaSpace{}%
\AgdaBound{pbk2}\AgdaSpace{}%
\AgdaBound{pbk3}\AgdaSpace{}%
\AgdaBound{pbk4}\AgdaSpace{}%
\AgdaSymbol{.}\AgdaField{<==stg}\AgdaSpace{}%
\AgdaBound{time}\AgdaSpace{}%
\AgdaBound{msg₂}%
\>[77]\AgdaSymbol{(}\AgdaBound{sig2}\AgdaSpace{}%
\AgdaOperator{\AgdaInductiveConstructor{∷}}\AgdaSpace{}%
\AgdaBound{sig1}\AgdaSpace{}%
\AgdaOperator{\AgdaInductiveConstructor{∷}}\AgdaSpace{}%
\AgdaBound{dummy}\AgdaSpace{}%
\AgdaOperator{\AgdaInductiveConstructor{∷}}\AgdaSpace{}%
\AgdaBound{stack}\AgdaSymbol{)}\AgdaSpace{}%
\AgdaBound{x}\AgdaSpace{}%
\AgdaSymbol{=}\<%
\\
\>[0][@{}l@{\AgdaIndent{0}}]%
\>[2]\AgdaFunction{lemmaHoareTripleStackGeAux'Comb2-4}\AgdaSpace{}%
\AgdaBound{msg₂}\AgdaSpace{}%
\AgdaBound{pbk1}\AgdaSpace{}%
\AgdaBound{pbk2}\AgdaSpace{}%
\AgdaBound{pbk3}\AgdaSpace{}%
\AgdaBound{pbk4}\AgdaSpace{}%
\AgdaBound{sig1}\AgdaSpace{}%
\AgdaBound{sig2}\<%
\\
\>[2]\AgdaSymbol{(}\AgdaFunction{boolToNatNotFalseLemma2}\AgdaSpace{}%
\AgdaSymbol{(}\AgdaFunction{compareSigsMultiSigAux}\AgdaSpace{}%
\AgdaBound{msg₂}\AgdaSpace{}%
\AgdaSymbol{(}\AgdaBound{sig2}\AgdaSpace{}%
\AgdaOperator{\AgdaInductiveConstructor{∷}}\AgdaSpace{}%
\AgdaInductiveConstructor{[]}\AgdaSymbol{)}\AgdaSpace{}%
\AgdaSymbol{(}\AgdaBound{pbk2}\AgdaSpace{}%
\AgdaOperator{\AgdaInductiveConstructor{∷}}\AgdaSpace{}%
\AgdaBound{pbk3}\AgdaSpace{}%
\AgdaOperator{\AgdaInductiveConstructor{∷}}\AgdaSpace{}%
\AgdaBound{pbk4}\AgdaSpace{}%
\AgdaOperator{\AgdaInductiveConstructor{∷}}\AgdaSpace{}%
\AgdaInductiveConstructor{[]}\AgdaSymbol{)}\<%
\\
\>[2]\AgdaBound{sig1}\AgdaSpace{}%
\AgdaSymbol{(}\AgdaFunction{isSigned}%
\>[18]\AgdaBound{msg₂}\AgdaSpace{}%
\AgdaBound{sig1}\AgdaSpace{}%
\AgdaBound{pbk1}\AgdaSymbol{))}\AgdaSpace{}%
\AgdaBound{x}\AgdaSymbol{)}\<%
\\
\\[\AgdaEmptyExtraSkip]%
\\[\AgdaEmptyExtraSkip]%
\>[0]\AgdaFunction{weakestPreCondMultiSig-2-4}\AgdaSpace{}%
\AgdaSymbol{:}\AgdaSpace{}%
\AgdaSymbol{(}\AgdaBound{pbk1}\AgdaSpace{}%
\AgdaBound{pbk2}\AgdaSpace{}%
\AgdaBound{pbk3}\AgdaSpace{}%
\AgdaBound{pbk4}\AgdaSpace{}%
\AgdaSymbol{:}%
\>[53]\AgdaDatatype{ℕ}\AgdaSymbol{)→}\AgdaSpace{}%
\AgdaFunction{StackStatePred}\<%
\\
\>[0]\AgdaComment{--\textbackslash{}verificationMultiSigBasicweakestPreCondMultiSigNos}\<%
\end{code}
} 

\newcommand{\verificationMultiSigBasicweakestPreCondMultiSigNos}{
\begin{code}%
\>[0]\AgdaFunction{weakestPreCondMultiSig-2-4}\AgdaSpace{}%
\AgdaBound{pbk1}\AgdaSpace{}%
\AgdaBound{pbk2}\AgdaSpace{}%
\AgdaBound{pbk3}\AgdaSpace{}%
\AgdaBound{pbk4}\<%
\\
\>[0][@{}l@{\AgdaIndent{0}}]%
\>[5]\AgdaSymbol{=}\AgdaSpace{}%
\AgdaFunction{stackPred2SPred}\AgdaSpace{}%
\AgdaSymbol{(}\AgdaFunction{weakestPreCondMultiSig-2-4ˢ}\AgdaSpace{}%
\AgdaBound{pbk1}\AgdaSpace{}%
\AgdaBound{pbk2}\AgdaSpace{}%
\AgdaBound{pbk3}\AgdaSpace{}%
\AgdaBound{pbk4}\AgdaSymbol{)}\<%
\end{code}
} 

\AgdaHide{
\begin{code}%
\>[0]\<%
\\
\\[\AgdaEmptyExtraSkip]%
\\[\AgdaEmptyExtraSkip]%
\\[\AgdaEmptyExtraSkip]%
\>[0]\AgdaComment{--\ Main\ theorem\ for\ multisig-2-4}\<%
\\
\>[0]\AgdaComment{--\textbackslash{}verificationMultiSigBasictheoremCorrectnessMultiSigTwoFour}\<%
\end{code}
} 

\newcommand{\verificationMultiSigBasictheoremCorrectnessMultiSigTwoFour}{
\begin{code}%
\>[0]\AgdaFunction{theoremCorrectnessMultiSig-2-4}\AgdaSpace{}%
\AgdaSymbol{:}\AgdaSpace{}%
\AgdaSymbol{(}\AgdaBound{pbk1}\AgdaSpace{}%
\AgdaBound{pbk2}\AgdaSpace{}%
\AgdaBound{pbk3}\AgdaSpace{}%
\AgdaBound{pbk4}\AgdaSpace{}%
\AgdaSymbol{:}%
\>[57]\AgdaDatatype{ℕ}\AgdaSymbol{)}\<%
\\
\>[0][@{}l@{\AgdaIndent{0}}]%
\>[3]\AgdaSymbol{→}%
\>[6]\AgdaOperator{\AgdaRecord{<}}%
\>[9]\AgdaFunction{stackPred2SPred}\AgdaSpace{}%
\AgdaSymbol{(}\AgdaFunction{weakestPreCondMultiSig-2-4ˢ}\AgdaSpace{}%
\AgdaBound{pbk1}\AgdaSpace{}%
\AgdaBound{pbk2}\AgdaSpace{}%
\AgdaBound{pbk3}\AgdaSpace{}%
\AgdaBound{pbk4}\AgdaSymbol{)}\AgdaSpace{}%
\AgdaOperator{\AgdaRecord{>ⁱᶠᶠ}}\<%
\\
\>[6][@{}l@{\AgdaIndent{0}}]%
\>[7]\AgdaFunction{multiSigScript2-4ᵇ}\AgdaSpace{}%
\AgdaBound{pbk1}\AgdaSpace{}%
\AgdaBound{pbk2}\AgdaSpace{}%
\AgdaBound{pbk3}\AgdaSpace{}%
\AgdaBound{pbk4}\<%
\\
\>[7]\AgdaOperator{\AgdaRecord{<}}%
\>[10]\AgdaFunction{stackPred2SPred}\AgdaSpace{}%
\AgdaFunction{acceptStateˢ}%
\>[40]\AgdaOperator{\AgdaRecord{>}}\<%
\end{code}
} 

\AgdaHide{
\begin{code}%
\>[0]\AgdaFunction{theoremCorrectnessMultiSig-2-4}\AgdaSpace{}%
\AgdaBound{pbk1}\AgdaSpace{}%
\AgdaBound{pbk2}\AgdaSpace{}%
\AgdaBound{pbk3}\AgdaSpace{}%
\AgdaBound{pbk4}\<%
\\
\>[0][@{}l@{\AgdaIndent{0}}]%
\>[26]\AgdaSymbol{=}\AgdaSpace{}%
\AgdaFunction{hoareTripleStack2HoareTriple}\AgdaSpace{}%
\AgdaSymbol{(}\AgdaFunction{multiSigScript2-4ᵇ}\AgdaSpace{}%
\AgdaBound{pbk1}\AgdaSpace{}%
\AgdaBound{pbk2}\AgdaSpace{}%
\AgdaBound{pbk3}\AgdaSpace{}%
\AgdaBound{pbk4}\AgdaSymbol{)}\<%
\\
\>[26]\AgdaSymbol{(}\AgdaFunction{weakestPreCondMultiSig-2-4ˢ}\AgdaSpace{}%
\AgdaBound{pbk1}\AgdaSpace{}%
\AgdaBound{pbk2}\AgdaSpace{}%
\AgdaBound{pbk3}\AgdaSpace{}%
\AgdaBound{pbk4}\AgdaSpace{}%
\AgdaSymbol{)}\AgdaSpace{}%
\AgdaFunction{acceptStateˢ}\<%
\\
\>[26]\AgdaSymbol{(}\AgdaFunction{mainLemmaCorrectnessMultiSig-2-4}\AgdaSpace{}%
\AgdaSymbol{(}\AgdaInductiveConstructor{nat}\AgdaSpace{}%
\AgdaBound{pbk4}\AgdaSymbol{)}\AgdaSpace{}%
\AgdaBound{pbk1}\AgdaSpace{}%
\AgdaBound{pbk2}\AgdaSpace{}%
\AgdaBound{pbk3}\AgdaSpace{}%
\AgdaBound{pbk4}\AgdaSymbol{)}\<%
\end{code}
} 

\AgdaHide{
\begin{code}%
\>[0]\<%
\\
\\[\AgdaEmptyExtraSkip]%
\>[0]\AgdaComment{-------------------------------------------------------------}\<%
\\
\>[0]\AgdaComment{--\ \ Here\ we\ prove\ the\ correctness\ of\ a\ combined\ script}\<%
\\
\>[0]\AgdaComment{--\ \ namely\ the\ correctness\ of\ a\ combination\ of\ a\ check\ time\ script\ and\ themultisig\ script}\<%
\\
\>[0]\AgdaComment{--}\<%
\\
\>[0]\AgdaComment{--\ This\ demonstrates\ a\ combination\ of\ method1\ and\ method\ 2:}\<%
\\
\>[0]\AgdaComment{--\ \ we\ prove\ the\ correctness\ of\ the\ time\ script\ and\ the\ multisig\ script\ separately\ (the\ time\ script\ is\ trivial\ to\ verify)}\<%
\\
\>[0]\AgdaComment{--\ \ \ \ using\ \ method\ 2}\<%
\\
\>[0]\AgdaComment{--\ Then\ we\ prove\ the\ correctenss\ of\ the\ combined\ script\ using\ method1}\<%
\\
\>[0]\AgdaComment{--\ \ \ \ and\ this\ shows\ that\ we\ can\ make\ bigger\ jumps\ in\ method\ 1}\<%
\\
\>[0]\AgdaComment{-------------------------------}\<%
\\
\>[0]\AgdaComment{--\textbackslash{}verificationMultiSigBasictheoremCorrectnessTimeCheck}\<%
\end{code}
} 

\newcommand{\verificationMultiSigBasictheoremCorrectnessTimeCheck}{
\begin{code}%
\>[0]\AgdaFunction{theoremCorrectnessTimeCheck}\AgdaSpace{}%
\AgdaSymbol{:}\AgdaSpace{}%
\AgdaSymbol{(}\AgdaBound{φ}\AgdaSpace{}%
\AgdaSymbol{:}\AgdaSpace{}%
\AgdaFunction{StackPredicate}\AgdaSymbol{)(}\AgdaBound{time₁}\AgdaSpace{}%
\AgdaSymbol{:}\AgdaSpace{}%
\AgdaFunction{Time}\AgdaSymbol{)}\<%
\\
\>[0][@{}l@{\AgdaIndent{0}}]%
\>[3]\AgdaSymbol{→}%
\>[7]\AgdaOperator{\AgdaRecord{<}}%
\>[10]\AgdaFunction{stackPred2SPred}\AgdaSpace{}%
\AgdaSymbol{(}\AgdaFunction{timeCheckPreCond}\AgdaSpace{}%
\AgdaBound{time₁}\AgdaSpace{}%
\AgdaOperator{\AgdaFunction{∧sp}}\AgdaSpace{}%
\AgdaBound{φ}\AgdaSymbol{)}%
\>[59]\AgdaOperator{\AgdaRecord{>ⁱᶠᶠ}}\AgdaSpace{}%
\AgdaFunction{checkTimeScriptᵇ}\AgdaSpace{}%
\AgdaBound{time₁}\<%
\\
\>[7]\AgdaOperator{\AgdaRecord{<}}%
\>[10]\AgdaFunction{stackPred2SPred}\AgdaSpace{}%
\AgdaBound{φ}%
\>[30]\AgdaOperator{\AgdaRecord{>}}\<%
\end{code}
} 

\AgdaHide{
\begin{code}%
\>[0]\AgdaFunction{theoremCorrectnessTimeCheck}\AgdaSpace{}%
\AgdaBound{φ}\AgdaSpace{}%
\AgdaBound{time₁}\AgdaSpace{}%
\AgdaSymbol{.}\AgdaField{==>}\AgdaSpace{}%
\AgdaOperator{\AgdaInductiveConstructor{⟨}}\AgdaSpace{}%
\AgdaBound{currentTime₁}\AgdaSpace{}%
\AgdaOperator{\AgdaInductiveConstructor{,}}\AgdaSpace{}%
\AgdaBound{msg₁}\AgdaSpace{}%
\AgdaOperator{\AgdaInductiveConstructor{,}}\AgdaSpace{}%
\AgdaBound{stack₁}\AgdaSpace{}%
\AgdaOperator{\AgdaInductiveConstructor{⟩}}\AgdaSpace{}%
\AgdaSymbol{(}\AgdaInductiveConstructor{conj}\AgdaSpace{}%
\AgdaBound{and3}\AgdaSpace{}%
\AgdaBound{and4}\AgdaSymbol{)}\AgdaSpace{}%
\AgdaKeyword{with}\AgdaSpace{}%
\AgdaSymbol{(}\AgdaFunction{instructOpTime}\AgdaSpace{}%
\AgdaBound{currentTime₁}\AgdaSpace{}%
\AgdaBound{time₁}\AgdaSymbol{)}\<%
\\
\>[0]\AgdaFunction{theoremCorrectnessTimeCheck}\AgdaSpace{}%
\AgdaBound{φ}\AgdaSpace{}%
\AgdaBound{time₁}\AgdaSpace{}%
\AgdaSymbol{.}\AgdaField{==>}\AgdaSpace{}%
\AgdaOperator{\AgdaInductiveConstructor{⟨}}\AgdaSpace{}%
\AgdaBound{currentTime₁}\AgdaSpace{}%
\AgdaOperator{\AgdaInductiveConstructor{,}}\AgdaSpace{}%
\AgdaBound{msg₁}\AgdaSpace{}%
\AgdaOperator{\AgdaInductiveConstructor{,}}\AgdaSpace{}%
\AgdaBound{stack₁}\AgdaSpace{}%
\AgdaOperator{\AgdaInductiveConstructor{⟩}}\AgdaSpace{}%
\AgdaSymbol{(}\AgdaInductiveConstructor{conj}\AgdaSpace{}%
\AgdaBound{and3}\AgdaSpace{}%
\AgdaBound{and4}\AgdaSymbol{)}\AgdaSpace{}%
\AgdaSymbol{|}\AgdaSpace{}%
\AgdaInductiveConstructor{true}\AgdaSpace{}%
\AgdaSymbol{=}\AgdaSpace{}%
\AgdaBound{and4}\<%
\\
\>[0]\AgdaFunction{theoremCorrectnessTimeCheck}\AgdaSpace{}%
\AgdaBound{φ}\AgdaSpace{}%
\AgdaBound{time₁}\AgdaSpace{}%
\AgdaSymbol{.}\AgdaField{<==}\AgdaSpace{}%
\AgdaOperator{\AgdaInductiveConstructor{⟨}}\AgdaSpace{}%
\AgdaBound{currentTime₁}\AgdaSpace{}%
\AgdaOperator{\AgdaInductiveConstructor{,}}\AgdaSpace{}%
\AgdaBound{msg₁}\AgdaSpace{}%
\AgdaOperator{\AgdaInductiveConstructor{,}}\AgdaSpace{}%
\AgdaBound{stack₁}\AgdaSpace{}%
\AgdaOperator{\AgdaInductiveConstructor{⟩}}\AgdaSpace{}%
\AgdaBound{p}\AgdaSpace{}%
\AgdaKeyword{with}\AgdaSpace{}%
\AgdaSymbol{(}\AgdaFunction{instructOpTime}\AgdaSpace{}%
\AgdaBound{currentTime₁}\AgdaSpace{}%
\AgdaBound{time₁}\AgdaSymbol{)}\<%
\\
\>[0]\AgdaFunction{theoremCorrectnessTimeCheck}\AgdaSpace{}%
\AgdaBound{φ}\AgdaSpace{}%
\AgdaBound{time₁}\AgdaSpace{}%
\AgdaSymbol{.}\AgdaField{<==}\AgdaSpace{}%
\AgdaOperator{\AgdaInductiveConstructor{⟨}}\AgdaSpace{}%
\AgdaBound{currentTime₁}\AgdaSpace{}%
\AgdaOperator{\AgdaInductiveConstructor{,}}\AgdaSpace{}%
\AgdaBound{msg₁}\AgdaSpace{}%
\AgdaOperator{\AgdaInductiveConstructor{,}}\AgdaSpace{}%
\AgdaBound{stack₁}\AgdaSpace{}%
\AgdaOperator{\AgdaInductiveConstructor{⟩}}\AgdaSpace{}%
\AgdaBound{p}\AgdaSpace{}%
\AgdaSymbol{|}\AgdaSpace{}%
\AgdaInductiveConstructor{true}\AgdaSpace{}%
\AgdaSymbol{=}\AgdaSpace{}%
\AgdaInductiveConstructor{conj}\AgdaSpace{}%
\AgdaInductiveConstructor{tt}\AgdaSpace{}%
\AgdaBound{p}\<%
\\
\\[\AgdaEmptyExtraSkip]%
\\[\AgdaEmptyExtraSkip]%
\>[0]\AgdaComment{--\textbackslash{}verificationMultiSigBasictheoremCorrectnessCombinedMultiSigTimeCheck}\<%
\end{code}
} 

\newcommand{\verificationMultiSigBasictheoremCorrectnessCombinedMultiSigTimeCheck}{
\begin{code}%
\>[0]\AgdaFunction{theoremCorrectnessCombinedMultiSigTimeCheck}\AgdaSpace{}%
\AgdaSymbol{:}\AgdaSpace{}%
\AgdaSymbol{(}\AgdaBound{time₁}\AgdaSpace{}%
\AgdaSymbol{:}\AgdaSpace{}%
\AgdaFunction{Time}\AgdaSymbol{)}\AgdaSpace{}%
\AgdaSymbol{(}\AgdaBound{pbk1}\AgdaSpace{}%
\AgdaBound{pbk2}\AgdaSpace{}%
\AgdaBound{pbk3}\AgdaSpace{}%
\AgdaBound{pbk4}\AgdaSpace{}%
\AgdaSymbol{:}%
\>[85]\AgdaDatatype{ℕ}\AgdaSymbol{)}\<%
\\
\>[0][@{}l@{\AgdaIndent{0}}]%
\>[3]\AgdaSymbol{→}%
\>[7]\AgdaOperator{\AgdaRecord{<}}\AgdaSpace{}%
\AgdaFunction{stackPred2SPred}\AgdaSpace{}%
\AgdaSymbol{(}%
\>[28]\AgdaFunction{timeCheckPreCond}\AgdaSpace{}%
\AgdaBound{time₁}\AgdaSpace{}%
\AgdaOperator{\AgdaFunction{∧sp}}\<%
\\
\>[28][@{}l@{\AgdaIndent{0}}]%
\>[29]\AgdaFunction{weakestPreCondMultiSig-2-4ˢ}%
\>[58]\AgdaBound{pbk1}\AgdaSpace{}%
\AgdaBound{pbk2}\AgdaSpace{}%
\AgdaBound{pbk3}\AgdaSpace{}%
\AgdaBound{pbk4}\AgdaSymbol{)}\AgdaSpace{}%
\AgdaOperator{\AgdaRecord{>ⁱᶠᶠ}}\<%
\\
\>[7][@{}l@{\AgdaIndent{0}}]%
\>[8]\AgdaFunction{checkTimeScriptᵇ}\AgdaSpace{}%
\AgdaBound{time₁}\AgdaSpace{}%
\AgdaOperator{\AgdaFunction{\ensuremath{+\!\!+}}}\AgdaSpace{}%
\AgdaFunction{multiSigScript2-4ᵇ}\AgdaSpace{}%
\AgdaBound{pbk1}\AgdaSpace{}%
\AgdaBound{pbk2}\AgdaSpace{}%
\AgdaBound{pbk3}\AgdaSpace{}%
\AgdaBound{pbk4}\<%
\\
\>[8]\AgdaOperator{\AgdaRecord{<}}\AgdaSpace{}%
\AgdaFunction{acceptState}\AgdaSpace{}%
\AgdaOperator{\AgdaRecord{>}}\<%
\\
\>[0]\AgdaFunction{theoremCorrectnessCombinedMultiSigTimeCheck}\AgdaSpace{}%
\AgdaBound{time₁}\AgdaSpace{}%
\AgdaBound{pbk1}\AgdaSpace{}%
\AgdaBound{pbk2}\AgdaSpace{}%
\AgdaBound{pbk3}\AgdaSpace{}%
\AgdaBound{pbk4}\AgdaSpace{}%
\AgdaSymbol{=}\<%
\\
\>[0][@{}l@{\AgdaIndent{0}}]%
\>[2]\AgdaFunction{stackPred2SPred}\AgdaSpace{}%
\AgdaSymbol{(}\AgdaFunction{timeCheckPreCond}\AgdaSpace{}%
\AgdaBound{time₁}\AgdaSpace{}%
\AgdaOperator{\AgdaFunction{∧sp}}\<%
\\
\>[2][@{}l@{\AgdaIndent{0}}]%
\>[5]\AgdaFunction{weakestPreCondMultiSig-2-4ˢ}%
\>[34]\AgdaBound{pbk1}\AgdaSpace{}%
\AgdaBound{pbk2}\AgdaSpace{}%
\AgdaBound{pbk3}\AgdaSpace{}%
\AgdaBound{pbk4}\AgdaSymbol{)}\<%
\\
\>[5][@{}l@{\AgdaIndent{0}}]%
\>[11]\AgdaFunction{<><>⟨}%
\>[18]\AgdaFunction{checkTimeScriptᵇ}\AgdaSpace{}%
\AgdaBound{time₁}%
\>[42]\AgdaFunction{⟩⟨}%
\>[46]\AgdaFunction{theoremCorrectnessTimeCheck}\<%
\\
\>[18]\AgdaSymbol{(}\AgdaFunction{weakestPreCondMultiSig-2-4ˢ}\AgdaSpace{}%
\AgdaBound{pbk1}\AgdaSpace{}%
\AgdaBound{pbk2}\AgdaSpace{}%
\AgdaBound{pbk3}\AgdaSpace{}%
\AgdaBound{pbk4}\AgdaSymbol{)}\AgdaSpace{}%
\AgdaBound{time₁}%
\>[75]\AgdaFunction{⟩}\<%
\\
\>[2]\AgdaFunction{stackPred2SPred}\AgdaSpace{}%
\AgdaSymbol{(}\AgdaFunction{weakestPreCondMultiSig-2-4ˢ}\AgdaSpace{}%
\AgdaBound{pbk1}\AgdaSpace{}%
\AgdaBound{pbk2}\AgdaSpace{}%
\AgdaBound{pbk3}\AgdaSpace{}%
\AgdaBound{pbk4}\AgdaSymbol{)}\<%
\\
\>[2][@{}l@{\AgdaIndent{0}}]%
\>[11]\AgdaFunction{<><>⟨}%
\>[667I]\AgdaFunction{multiSigScript2-4ᵇ}\AgdaSpace{}%
\AgdaBound{pbk1}\AgdaSpace{}%
\AgdaBound{pbk2}\AgdaSpace{}%
\AgdaBound{pbk3}\AgdaSpace{}%
\AgdaBound{pbk4}\<%
\\
\>[.][@{}l@{}]\<[667I]%
\>[17]\AgdaFunction{⟩⟨}\AgdaSpace{}%
\AgdaFunction{theoremCorrectnessMultiSig-2-4}\AgdaSpace{}%
\AgdaBound{pbk1}\AgdaSpace{}%
\AgdaBound{pbk2}\AgdaSpace{}%
\AgdaBound{pbk3}\AgdaSpace{}%
\AgdaBound{pbk4}%
\>[73]\AgdaFunction{⟩ᵉ}\<%
\\
\>[2]\AgdaFunction{stackPred2SPred}\AgdaSpace{}%
\AgdaFunction{acceptStateˢ}\AgdaSpace{}%
\AgdaOperator{\AgdaFunction{∎p}}\<%
\end{code}
} 

\AgdaHide{
\begin{code}%
\>[0]\<%
\end{code}
} 


\AgdaHide{
\begin{code}%
\>[0]\AgdaComment{\{-\ OPTIONS\ --allow-unsolved-metas\ \#-\}}\<%
\\
\>[0]\AgdaKeyword{open}\AgdaSpace{}%
\AgdaKeyword{import}\AgdaSpace{}%
\AgdaModule{basicBitcoinDataType}\<%
\\
\\[\AgdaEmptyExtraSkip]%
\>[0]\AgdaKeyword{module}\AgdaSpace{}%
\AgdaModule{verificationP2PKHbasic}\AgdaSpace{}%
\AgdaSymbol{(}\AgdaBound{param}\AgdaSpace{}%
\AgdaSymbol{:}\AgdaSpace{}%
\AgdaRecord{GlobalParameters}\AgdaSymbol{)}\AgdaSpace{}%
\AgdaKeyword{where}\<%
\\
\\[\AgdaEmptyExtraSkip]%
\>[0]\AgdaKeyword{open}\AgdaSpace{}%
\AgdaKeyword{import}\AgdaSpace{}%
\AgdaModule{libraries.listLib}\<%
\\
\>[0]\AgdaKeyword{open}\AgdaSpace{}%
\AgdaKeyword{import}\AgdaSpace{}%
\AgdaModule{Data.List.Base}\<%
\\
\>[0]\AgdaKeyword{open}\AgdaSpace{}%
\AgdaKeyword{import}\AgdaSpace{}%
\AgdaModule{libraries.natLib}\<%
\\
\>[0]\AgdaKeyword{open}\AgdaSpace{}%
\AgdaKeyword{import}\AgdaSpace{}%
\AgdaModule{Data.Nat}%
\>[22]\AgdaKeyword{renaming}\AgdaSpace{}%
\AgdaSymbol{(}\AgdaOperator{\AgdaDatatype{\AgdaUnderscore{}≤\AgdaUnderscore{}}}\AgdaSpace{}%
\AgdaSymbol{to}\AgdaSpace{}%
\AgdaOperator{\AgdaDatatype{\AgdaUnderscore{}≤'\AgdaUnderscore{}}}\AgdaSpace{}%
\AgdaSymbol{;}\AgdaSpace{}%
\AgdaOperator{\AgdaFunction{\AgdaUnderscore{}<\AgdaUnderscore{}}}\AgdaSpace{}%
\AgdaSymbol{to}\AgdaSpace{}%
\AgdaOperator{\AgdaFunction{\AgdaUnderscore{}<'\AgdaUnderscore{}}}\AgdaSymbol{)}\<%
\\
\>[0]\AgdaKeyword{open}\AgdaSpace{}%
\AgdaKeyword{import}\AgdaSpace{}%
\AgdaModule{Data.List}\AgdaSpace{}%
\AgdaKeyword{hiding}\AgdaSpace{}%
\AgdaSymbol{(}\AgdaOperator{\AgdaFunction{\AgdaUnderscore{}\ensuremath{+\!\!+}\AgdaUnderscore{}}}\AgdaSymbol{)}\<%
\\
\>[0]\AgdaKeyword{open}\AgdaSpace{}%
\AgdaKeyword{import}\AgdaSpace{}%
\AgdaModule{Data.Unit}%
\>[23]\AgdaKeyword{hiding}\AgdaSpace{}%
\AgdaSymbol{(}\AgdaOperator{\AgdaRecord{\AgdaUnderscore{}≤\AgdaUnderscore{}}}\AgdaSymbol{)}\<%
\\
\>[0]\AgdaKeyword{open}\AgdaSpace{}%
\AgdaKeyword{import}\AgdaSpace{}%
\AgdaModule{Data.Empty}\<%
\\
\>[0]\AgdaKeyword{open}\AgdaSpace{}%
\AgdaKeyword{import}\AgdaSpace{}%
\AgdaModule{Data.Bool}%
\>[23]\AgdaKeyword{hiding}\AgdaSpace{}%
\AgdaSymbol{(}\AgdaOperator{\AgdaDatatype{\AgdaUnderscore{}≤\AgdaUnderscore{}}}\AgdaSpace{}%
\AgdaSymbol{;}\AgdaSpace{}%
\AgdaOperator{\AgdaDatatype{\AgdaUnderscore{}<\AgdaUnderscore{}}}\AgdaSpace{}%
\AgdaSymbol{;}\AgdaSpace{}%
\AgdaOperator{\AgdaFunction{if\AgdaUnderscore{}then\AgdaUnderscore{}else\AgdaUnderscore{}}}\AgdaSpace{}%
\AgdaSymbol{)}\AgdaSpace{}%
\AgdaKeyword{renaming}\AgdaSpace{}%
\AgdaSymbol{(}\AgdaOperator{\AgdaFunction{\AgdaUnderscore{}∧\AgdaUnderscore{}}}\AgdaSpace{}%
\AgdaSymbol{to}\AgdaSpace{}%
\AgdaOperator{\AgdaFunction{\AgdaUnderscore{}∧b\AgdaUnderscore{}}}\AgdaSpace{}%
\AgdaSymbol{;}\AgdaSpace{}%
\AgdaOperator{\AgdaFunction{\AgdaUnderscore{}∨\AgdaUnderscore{}}}\AgdaSpace{}%
\AgdaSymbol{to}\AgdaSpace{}%
\AgdaOperator{\AgdaFunction{\AgdaUnderscore{}∨b\AgdaUnderscore{}}}\AgdaSpace{}%
\AgdaSymbol{;}\AgdaSpace{}%
\AgdaFunction{T}\AgdaSpace{}%
\AgdaSymbol{to}\AgdaSpace{}%
\AgdaFunction{True}\AgdaSymbol{)}\<%
\\
\>[0]\AgdaKeyword{open}\AgdaSpace{}%
\AgdaKeyword{import}\AgdaSpace{}%
\AgdaModule{Data.Product}\AgdaSpace{}%
\AgdaKeyword{renaming}\AgdaSpace{}%
\AgdaSymbol{(}\AgdaOperator{\AgdaInductiveConstructor{\AgdaUnderscore{},\AgdaUnderscore{}}}\AgdaSpace{}%
\AgdaSymbol{to}\AgdaSpace{}%
\AgdaOperator{\AgdaInductiveConstructor{\AgdaUnderscore{},,\AgdaUnderscore{}}}\AgdaSpace{}%
\AgdaSymbol{)}\<%
\\
\>[0]\AgdaKeyword{open}\AgdaSpace{}%
\AgdaKeyword{import}\AgdaSpace{}%
\AgdaModule{Data.Nat.Base}\AgdaSpace{}%
\AgdaKeyword{hiding}\AgdaSpace{}%
\AgdaSymbol{(}\AgdaOperator{\AgdaDatatype{\AgdaUnderscore{}≤\AgdaUnderscore{}}}\AgdaSpace{}%
\AgdaSymbol{;}\AgdaSpace{}%
\AgdaOperator{\AgdaFunction{\AgdaUnderscore{}<\AgdaUnderscore{}}}\AgdaSymbol{)}\<%
\\
\>[0]\AgdaKeyword{open}\AgdaSpace{}%
\AgdaKeyword{import}\AgdaSpace{}%
\AgdaModule{Data.List.NonEmpty}\AgdaSpace{}%
\AgdaKeyword{hiding}\AgdaSpace{}%
\AgdaSymbol{(}\AgdaField{head}\AgdaSpace{}%
\AgdaSymbol{)}\<%
\\
\>[0]\AgdaKeyword{open}\AgdaSpace{}%
\AgdaKeyword{import}\AgdaSpace{}%
\AgdaModule{Data.Nat}\AgdaSpace{}%
\AgdaKeyword{using}\AgdaSpace{}%
\AgdaSymbol{(}\AgdaDatatype{ℕ}\AgdaSymbol{;}\AgdaSpace{}%
\AgdaOperator{\AgdaPrimitive{\AgdaUnderscore{}+\AgdaUnderscore{}}}\AgdaSymbol{;}\AgdaSpace{}%
\AgdaOperator{\AgdaFunction{\AgdaUnderscore{}≥\AgdaUnderscore{}}}\AgdaSymbol{;}\AgdaSpace{}%
\AgdaOperator{\AgdaFunction{\AgdaUnderscore{}>\AgdaUnderscore{}}}\AgdaSymbol{;}\AgdaSpace{}%
\AgdaInductiveConstructor{zero}\AgdaSymbol{;}\AgdaSpace{}%
\AgdaInductiveConstructor{suc}\AgdaSymbol{;}\AgdaSpace{}%
\AgdaInductiveConstructor{s≤s}\AgdaSymbol{;}\AgdaSpace{}%
\AgdaInductiveConstructor{z≤n}\AgdaSymbol{)}\<%
\\
\>[0]\AgdaKeyword{import}\AgdaSpace{}%
\AgdaModule{Relation.Binary.PropositionalEquality}\AgdaSpace{}%
\AgdaSymbol{as}\AgdaSpace{}%
\AgdaModule{Eq}\<%
\\
\>[0]\AgdaKeyword{open}\AgdaSpace{}%
\AgdaModule{Eq}\AgdaSpace{}%
\AgdaKeyword{using}\AgdaSpace{}%
\AgdaSymbol{(}\AgdaOperator{\AgdaDatatype{\AgdaUnderscore{}≡\AgdaUnderscore{}}}\AgdaSymbol{;}\AgdaSpace{}%
\AgdaInductiveConstructor{refl}\AgdaSymbol{;}\AgdaSpace{}%
\AgdaFunction{cong}\AgdaSymbol{;}\AgdaSpace{}%
\AgdaKeyword{module}\AgdaSpace{}%
\AgdaModule{≡-Reasoning}\AgdaSymbol{;}\AgdaSpace{}%
\AgdaFunction{sym}\AgdaSymbol{)}\<%
\\
\>[0]\AgdaKeyword{open}\AgdaSpace{}%
\AgdaModule{≡-Reasoning}\<%
\\
\>[0]\AgdaKeyword{open}\AgdaSpace{}%
\AgdaKeyword{import}\AgdaSpace{}%
\AgdaModule{Agda.Builtin.Equality}\<%
\\
\>[0]\AgdaComment{--open\ import\ Agda.Builtin.Equality.Rewrite}\<%
\\
\\[\AgdaEmptyExtraSkip]%
\>[0]\AgdaKeyword{open}\AgdaSpace{}%
\AgdaKeyword{import}\AgdaSpace{}%
\AgdaModule{libraries.andLib}\<%
\\
\>[0]\AgdaKeyword{open}\AgdaSpace{}%
\AgdaKeyword{import}\AgdaSpace{}%
\AgdaModule{libraries.miscLib}\<%
\\
\>[0]\AgdaKeyword{open}\AgdaSpace{}%
\AgdaKeyword{import}\AgdaSpace{}%
\AgdaModule{libraries.maybeLib}\<%
\\
\>[0]\AgdaKeyword{open}\AgdaSpace{}%
\AgdaKeyword{import}\AgdaSpace{}%
\AgdaModule{libraries.boolLib}\<%
\\
\\[\AgdaEmptyExtraSkip]%
\>[0]\AgdaKeyword{open}\AgdaSpace{}%
\AgdaKeyword{import}\AgdaSpace{}%
\AgdaModule{stack}\<%
\\
\>[0]\AgdaKeyword{open}\AgdaSpace{}%
\AgdaKeyword{import}\AgdaSpace{}%
\AgdaModule{stackPredicate}\<%
\\
\>[0]\AgdaKeyword{open}\AgdaSpace{}%
\AgdaKeyword{import}\AgdaSpace{}%
\AgdaModule{instruction}\<%
\\
\>[0]\AgdaKeyword{open}\AgdaSpace{}%
\AgdaKeyword{import}\AgdaSpace{}%
\AgdaModule{instructionBasic}\<%
\\
\>[0]\AgdaComment{--\ open\ import\ ledger\ param}\<%
\\
\>[0]\AgdaKeyword{open}\AgdaSpace{}%
\AgdaKeyword{import}\AgdaSpace{}%
\AgdaModule{semanticBasicOperations}\AgdaSpace{}%
\AgdaBound{param}\<%
\\
\\[\AgdaEmptyExtraSkip]%
\\[\AgdaEmptyExtraSkip]%
\>[0]\AgdaFunction{instruction-1}\AgdaSpace{}%
\AgdaSymbol{:}\AgdaSpace{}%
\AgdaDatatype{InstructionBasic}\<%
\\
\>[0]\AgdaFunction{instruction-1}\AgdaSpace{}%
\AgdaSymbol{=}\AgdaSpace{}%
\AgdaInductiveConstructor{opCheckSig}\<%
\\
\\[\AgdaEmptyExtraSkip]%
\>[0]\AgdaFunction{instruction-2}\AgdaSpace{}%
\AgdaSymbol{:}\AgdaSpace{}%
\AgdaDatatype{InstructionBasic}\<%
\\
\>[0]\AgdaFunction{instruction-2}\AgdaSpace{}%
\AgdaSymbol{=}\AgdaSpace{}%
\AgdaInductiveConstructor{opVerify}\<%
\\
\\[\AgdaEmptyExtraSkip]%
\>[0]\AgdaFunction{instruction-3}\AgdaSpace{}%
\AgdaSymbol{:}\AgdaSpace{}%
\AgdaDatatype{InstructionBasic}\<%
\\
\>[0]\AgdaFunction{instruction-3}\AgdaSpace{}%
\AgdaSymbol{=}\AgdaSpace{}%
\AgdaInductiveConstructor{opEqual}\<%
\\
\\[\AgdaEmptyExtraSkip]%
\>[0]\AgdaFunction{instruction-4}\AgdaSpace{}%
\AgdaSymbol{:}\AgdaSpace{}%
\AgdaDatatype{ℕ}\AgdaSpace{}%
\AgdaSymbol{→}%
\>[21]\AgdaDatatype{InstructionBasic}\<%
\\
\>[0]\AgdaFunction{instruction-4}\AgdaSpace{}%
\AgdaBound{pbkh}\AgdaSpace{}%
\AgdaSymbol{=}\AgdaSpace{}%
\AgdaInductiveConstructor{opPush}%
\>[29]\AgdaBound{pbkh}\<%
\\
\\[\AgdaEmptyExtraSkip]%
\>[0]\AgdaFunction{instruction-5}\AgdaSpace{}%
\AgdaSymbol{:}\AgdaSpace{}%
\AgdaDatatype{InstructionBasic}\<%
\\
\>[0]\AgdaFunction{instruction-5}\AgdaSpace{}%
\AgdaSymbol{=}\AgdaSpace{}%
\AgdaInductiveConstructor{opHash}\<%
\\
\\[\AgdaEmptyExtraSkip]%
\>[0]\AgdaFunction{instruction-6}\AgdaSpace{}%
\AgdaSymbol{:}\AgdaSpace{}%
\AgdaDatatype{InstructionBasic}\<%
\\
\>[0]\AgdaFunction{instruction-6}\AgdaSpace{}%
\AgdaSymbol{=}\AgdaSpace{}%
\AgdaInductiveConstructor{opDup}\<%
\\
\\[\AgdaEmptyExtraSkip]%
\>[0]\AgdaFunction{accept-0Basic}\AgdaSpace{}%
\AgdaSymbol{:}\AgdaSpace{}%
\AgdaFunction{StackPredicate}\<%
\\
\>[0]\AgdaFunction{accept-0Basic}\AgdaSpace{}%
\AgdaSymbol{=}\AgdaSpace{}%
\AgdaFunction{acceptStateˢ}\<%
\\
\>[0]\<%
\end{code}
} 

\newcommand{\verificationptopbasicposcontitionone}{
\begin{code}[inline]%
\>[0]\AgdaFunction{accept₁ˢ}\AgdaSpace{}%
\AgdaSymbol{:}\AgdaSpace{}%
\AgdaFunction{StackPredicate}\<%
\\
\>[0]\AgdaFunction{accept₁ˢ}\AgdaSpace{}%
\AgdaBound{time}\AgdaSpace{}%
\AgdaBound{m}\AgdaSpace{}%
\AgdaInductiveConstructor{[]}\AgdaSpace{}%
\AgdaSymbol{=}\AgdaSpace{}%
\AgdaDatatype{⊥}\<%
\\
\>[0]\AgdaFunction{accept₁ˢ}\AgdaSpace{}%
\AgdaBound{time}\AgdaSpace{}%
\AgdaBound{m}\AgdaSpace{}%
\AgdaSymbol{(}\AgdaBound{sig}\AgdaSpace{}%
\AgdaOperator{\AgdaInductiveConstructor{∷}}\AgdaSpace{}%
\AgdaInductiveConstructor{[]}\AgdaSymbol{)}\AgdaSpace{}%
\AgdaSymbol{=}\AgdaSpace{}%
\AgdaDatatype{⊥}\<%
\\
\>[0]\AgdaFunction{accept₁ˢ}\AgdaSpace{}%
\AgdaBound{time}\AgdaSpace{}%
\AgdaBound{m}\AgdaSpace{}%
\AgdaSymbol{(}\AgdaSpace{}%
\AgdaBound{pbk}%
\>[23]\AgdaOperator{\AgdaInductiveConstructor{∷}}\AgdaSpace{}%
\AgdaBound{sig}\AgdaSpace{}%
\AgdaOperator{\AgdaInductiveConstructor{∷}}\AgdaSpace{}%
\AgdaBound{st}\AgdaSymbol{)}\AgdaSpace{}%
\AgdaSymbol{=}\AgdaSpace{}%
\AgdaFunction{IsSigned}\AgdaSpace{}%
\AgdaBound{m}%
\>[49]\AgdaBound{sig}\AgdaSpace{}%
\AgdaBound{pbk}\<%
\end{code}
} 

\AgdaHide{
\begin{code}%
\>[0]\AgdaFunction{accept₂ˢCore}\AgdaSpace{}%
\AgdaSymbol{:}\AgdaSpace{}%
\AgdaFunction{Time}\AgdaSpace{}%
\AgdaSymbol{→}\AgdaSpace{}%
\AgdaDatatype{Msg}\AgdaSpace{}%
\AgdaSymbol{→}\AgdaSpace{}%
\AgdaDatatype{ℕ}\AgdaSpace{}%
\AgdaSymbol{→}\AgdaSpace{}%
\AgdaDatatype{ℕ}\AgdaSpace{}%
\AgdaSymbol{→}\AgdaSpace{}%
\AgdaDatatype{ℕ}\AgdaSpace{}%
\AgdaSymbol{→}\AgdaSpace{}%
\AgdaPrimitive{Set}\<%
\\
\>[0]\AgdaFunction{accept₂ˢCore}\AgdaSpace{}%
\AgdaBound{time}\AgdaSpace{}%
\AgdaBound{m}\AgdaSpace{}%
\AgdaInductiveConstructor{zero}\AgdaSpace{}%
\AgdaBound{pbk}\AgdaSpace{}%
\AgdaBound{sig}\AgdaSpace{}%
\AgdaSymbol{=}\AgdaSpace{}%
\AgdaDatatype{⊥}\<%
\\
\>[0]\AgdaFunction{accept₂ˢCore}\AgdaSpace{}%
\AgdaBound{time}\AgdaSpace{}%
\AgdaBound{m}\AgdaSpace{}%
\AgdaSymbol{(}\AgdaInductiveConstructor{suc}\AgdaSpace{}%
\AgdaBound{x}\AgdaSymbol{)}\AgdaSpace{}%
\AgdaBound{pbk}\AgdaSpace{}%
\AgdaBound{sig}\AgdaSpace{}%
\AgdaSymbol{=}%
\>[39]\AgdaFunction{IsSigned}%
\>[49]\AgdaBound{m}\AgdaSpace{}%
\AgdaBound{sig}\AgdaSpace{}%
\AgdaBound{pbk}\<%
\\
\>[0]\<%
\end{code}
} 

\newcommand{\verificationptopbasicacceptTwo}{
\begin{code}[inline]%
\>[0]\AgdaFunction{accept₂ˢ}\AgdaSpace{}%
\AgdaSymbol{:}\AgdaSpace{}%
\AgdaFunction{StackPredicate}\<%
\\
\>[0]\AgdaFunction{accept₂ˢ}\AgdaSpace{}%
\AgdaBound{time}\AgdaSpace{}%
\AgdaBound{m}\AgdaSpace{}%
\AgdaInductiveConstructor{[]}\AgdaSpace{}%
\AgdaSymbol{=}\AgdaSpace{}%
\AgdaDatatype{⊥}\<%
\\
\>[0]\AgdaFunction{accept₂ˢ}\AgdaSpace{}%
\AgdaBound{time}\AgdaSpace{}%
\AgdaBound{m}\AgdaSpace{}%
\AgdaSymbol{(}\AgdaBound{x}\AgdaSpace{}%
\AgdaOperator{\AgdaInductiveConstructor{∷}}\AgdaSpace{}%
\AgdaInductiveConstructor{[]}\AgdaSymbol{)}\AgdaSpace{}%
\AgdaSymbol{=}\AgdaSpace{}%
\AgdaDatatype{⊥}\<%
\\
\>[0]\AgdaFunction{accept₂ˢ}\AgdaSpace{}%
\AgdaBound{time}\AgdaSpace{}%
\AgdaBound{m}\AgdaSpace{}%
\AgdaSymbol{(}\AgdaBound{x}\AgdaSpace{}%
\AgdaOperator{\AgdaInductiveConstructor{∷}}\AgdaSpace{}%
\AgdaBound{x₁}\AgdaSpace{}%
\AgdaOperator{\AgdaInductiveConstructor{∷}}\AgdaSpace{}%
\AgdaInductiveConstructor{[]}\AgdaSymbol{)}\AgdaSpace{}%
\AgdaSymbol{=}\AgdaSpace{}%
\AgdaDatatype{⊥}\<%
\\
\>[0]\AgdaFunction{accept₂ˢ}\AgdaSpace{}%
\AgdaBound{time}\AgdaSpace{}%
\AgdaBound{m}\AgdaSpace{}%
\AgdaSymbol{(}\AgdaBound{x}\AgdaSpace{}%
\AgdaOperator{\AgdaInductiveConstructor{∷}}\AgdaSpace{}%
\AgdaBound{pbk}\AgdaSpace{}%
\AgdaOperator{\AgdaInductiveConstructor{∷}}\AgdaSpace{}%
\AgdaBound{sig}\AgdaSpace{}%
\AgdaOperator{\AgdaInductiveConstructor{∷}}\AgdaSpace{}%
\AgdaBound{rest}\AgdaSymbol{)}\AgdaSpace{}%
\AgdaSymbol{=}\AgdaSpace{}%
\AgdaFunction{accept₂ˢCore}\AgdaSpace{}%
\AgdaBound{time}\AgdaSpace{}%
\AgdaBound{m}\AgdaSpace{}%
\AgdaBound{x}\AgdaSpace{}%
\AgdaBound{pbk}\AgdaSpace{}%
\AgdaBound{sig}\<%
\end{code}
} 

\AgdaHide{
\begin{code}%
\>[0]\<%
\\
\>[0]\<%
\end{code}
} 

\newcommand{\verificationptopbasicacceptThree}{
\begin{code}[inline]%
\>[0]\AgdaFunction{accept₃ˢ}\AgdaSpace{}%
\AgdaSymbol{:}\AgdaSpace{}%
\AgdaFunction{StackPredicate}\<%
\\
\>[0]\AgdaFunction{accept₃ˢ}\AgdaSpace{}%
\AgdaBound{time}\AgdaSpace{}%
\AgdaBound{m}\AgdaSpace{}%
\AgdaInductiveConstructor{[]}\AgdaSpace{}%
\AgdaSymbol{=}\AgdaSpace{}%
\AgdaDatatype{⊥}\<%
\\
\>[0]\AgdaFunction{accept₃ˢ}\AgdaSpace{}%
\AgdaBound{time}\AgdaSpace{}%
\AgdaBound{m}\AgdaSpace{}%
\AgdaSymbol{(}\AgdaBound{x}\AgdaSpace{}%
\AgdaOperator{\AgdaInductiveConstructor{∷}}\AgdaSpace{}%
\AgdaInductiveConstructor{[]}\AgdaSymbol{)}\AgdaSpace{}%
\AgdaSymbol{=}\AgdaSpace{}%
\AgdaDatatype{⊥}\<%
\\
\>[0]\AgdaFunction{accept₃ˢ}\AgdaSpace{}%
\AgdaBound{time}\AgdaSpace{}%
\AgdaBound{m}\AgdaSpace{}%
\AgdaSymbol{(}\AgdaBound{x}\AgdaSpace{}%
\AgdaOperator{\AgdaInductiveConstructor{∷}}\AgdaSpace{}%
\AgdaBound{x₁}\AgdaSpace{}%
\AgdaOperator{\AgdaInductiveConstructor{∷}}\AgdaSpace{}%
\AgdaInductiveConstructor{[]}\AgdaSymbol{)}\AgdaSpace{}%
\AgdaSymbol{=}\AgdaSpace{}%
\AgdaDatatype{⊥}\<%
\\
\>[0]\AgdaFunction{accept₃ˢ}\AgdaSpace{}%
\AgdaBound{time}\AgdaSpace{}%
\AgdaBound{m}\AgdaSpace{}%
\AgdaSymbol{(}\AgdaBound{x}\AgdaSpace{}%
\AgdaOperator{\AgdaInductiveConstructor{∷}}\AgdaSpace{}%
\AgdaBound{x₁}\AgdaSpace{}%
\AgdaOperator{\AgdaInductiveConstructor{∷}}\AgdaSpace{}%
\AgdaBound{x2}\AgdaSpace{}%
\AgdaOperator{\AgdaInductiveConstructor{∷}}\AgdaSpace{}%
\AgdaInductiveConstructor{[]}\AgdaSymbol{)}\AgdaSpace{}%
\AgdaSymbol{=}\AgdaSpace{}%
\AgdaDatatype{⊥}\<%
\\
\>[0]\AgdaFunction{accept₃ˢ}\AgdaSpace{}%
\AgdaBound{time}\AgdaSpace{}%
\AgdaBound{m}\AgdaSpace{}%
\AgdaSymbol{(}\AgdaBound{pbkh2}\AgdaSpace{}%
\AgdaOperator{\AgdaInductiveConstructor{∷}}\AgdaSpace{}%
\AgdaBound{pbkh1}\AgdaSpace{}%
\AgdaOperator{\AgdaInductiveConstructor{∷}}\AgdaSpace{}%
\AgdaBound{pbk}\AgdaSpace{}%
\AgdaOperator{\AgdaInductiveConstructor{∷}}\AgdaSpace{}%
\AgdaBound{sig}\AgdaSpace{}%
\AgdaOperator{\AgdaInductiveConstructor{∷}}\AgdaSpace{}%
\AgdaBound{rest}\AgdaSymbol{)}\<%
\\
\>[0][@{}l@{\AgdaIndent{0}}]%
\>[5]\AgdaSymbol{=}%
\>[8]\AgdaSymbol{(}\AgdaBound{pbkh2}\AgdaSpace{}%
\AgdaOperator{\AgdaDatatype{≡}}%
\>[18]\AgdaBound{pbkh1}\AgdaSymbol{)}\AgdaSpace{}%
\AgdaOperator{\AgdaRecord{∧}}\AgdaSpace{}%
\AgdaFunction{IsSigned}%
\>[37]\AgdaBound{m}\AgdaSpace{}%
\AgdaBound{sig}\AgdaSpace{}%
\AgdaBound{pbk}\<%
\end{code}
} 

\AgdaHide{
\begin{code}%
\>[0]\<%
\\
\>[0]\<%
\end{code}
} 

\newcommand{\verificationptopbasicacceptFour}{
\begin{code}[inline]%
\>[0]\AgdaFunction{accept₄ˢ}\AgdaSpace{}%
\AgdaSymbol{:}\AgdaSpace{}%
\AgdaSymbol{(}\AgdaSpace{}%
\AgdaBound{pbkh1}\AgdaSpace{}%
\AgdaSymbol{:}\AgdaSpace{}%
\AgdaDatatype{ℕ}\AgdaSpace{}%
\AgdaSymbol{)}\AgdaSpace{}%
\AgdaSymbol{→}\AgdaSpace{}%
\AgdaFunction{StackPredicate}\<%
\\
\>[0]\AgdaFunction{accept₄ˢ}\AgdaSpace{}%
\AgdaBound{pbkh1}\AgdaSpace{}%
\AgdaBound{time}\AgdaSpace{}%
\AgdaBound{m}\AgdaSpace{}%
\AgdaInductiveConstructor{[]}\AgdaSpace{}%
\AgdaSymbol{=}\AgdaSpace{}%
\AgdaDatatype{⊥}\<%
\\
\>[0]\AgdaFunction{accept₄ˢ}\AgdaSpace{}%
\AgdaBound{pbkh1}\AgdaSpace{}%
\AgdaBound{time}\AgdaSpace{}%
\AgdaBound{m}\AgdaSpace{}%
\AgdaSymbol{(}\AgdaBound{x}\AgdaSpace{}%
\AgdaOperator{\AgdaInductiveConstructor{∷}}\AgdaSpace{}%
\AgdaInductiveConstructor{[]}\AgdaSymbol{)}\AgdaSpace{}%
\AgdaSymbol{=}\AgdaSpace{}%
\AgdaDatatype{⊥}\<%
\\
\>[0]\AgdaFunction{accept₄ˢ}\AgdaSpace{}%
\AgdaBound{pbkh1}\AgdaSpace{}%
\AgdaBound{time}\AgdaSpace{}%
\AgdaBound{m}\AgdaSpace{}%
\AgdaSymbol{(}\AgdaBound{x}\AgdaSpace{}%
\AgdaOperator{\AgdaInductiveConstructor{∷}}\AgdaSpace{}%
\AgdaBound{x1}%
\>[31]\AgdaOperator{\AgdaInductiveConstructor{∷}}\AgdaSpace{}%
\AgdaInductiveConstructor{[]}\AgdaSymbol{)}\AgdaSpace{}%
\AgdaSymbol{=}\AgdaSpace{}%
\AgdaDatatype{⊥}\<%
\\
\>[0]\AgdaFunction{accept₄ˢ}\AgdaSpace{}%
\AgdaBound{pbkh1}\AgdaSpace{}%
\AgdaBound{time}\AgdaSpace{}%
\AgdaBound{m}\AgdaSpace{}%
\AgdaSymbol{(}\AgdaSpace{}%
\AgdaBound{pbkh2}%
\>[32]\AgdaOperator{\AgdaInductiveConstructor{∷}}\AgdaSpace{}%
\AgdaBound{pbk}\AgdaSpace{}%
\AgdaOperator{\AgdaInductiveConstructor{∷}}\AgdaSpace{}%
\AgdaBound{sig}\AgdaSpace{}%
\AgdaOperator{\AgdaInductiveConstructor{∷}}\AgdaSpace{}%
\AgdaBound{st}\AgdaSymbol{)}\<%
\\
\>[0][@{}l@{\AgdaIndent{0}}]%
\>[8]\AgdaSymbol{=}\AgdaSpace{}%
\AgdaSymbol{(}\AgdaBound{pbkh2}\AgdaSpace{}%
\AgdaOperator{\AgdaDatatype{≡}}%
\>[20]\AgdaBound{pbkh1}\AgdaSymbol{)}\AgdaSpace{}%
\AgdaOperator{\AgdaRecord{∧}}%
\>[30]\AgdaFunction{IsSigned}%
\>[40]\AgdaBound{m}\AgdaSpace{}%
\AgdaBound{sig}\AgdaSpace{}%
\AgdaBound{pbk}\<%
\end{code}
} 

\AgdaHide{
\begin{code}%
\>[0]\<%
\\
\>[0]\<%
\end{code}
} 

\newcommand{\verificationptopbasicacceptFive}{
\begin{code}[inline]%
\>[0]\AgdaFunction{accept₅ˢ}\AgdaSpace{}%
\AgdaSymbol{:}\AgdaSpace{}%
\AgdaSymbol{(}\AgdaSpace{}%
\AgdaBound{pbkh1}\AgdaSpace{}%
\AgdaSymbol{:}\AgdaSpace{}%
\AgdaDatatype{ℕ}\AgdaSpace{}%
\AgdaSymbol{)}\AgdaSpace{}%
\AgdaSymbol{→}\AgdaSpace{}%
\AgdaFunction{StackPredicate}\<%
\\
\>[0]\AgdaFunction{accept₅ˢ}%
\>[10]\AgdaBound{pbkh1}\AgdaSpace{}%
\AgdaBound{time}\AgdaSpace{}%
\AgdaBound{m}\AgdaSpace{}%
\AgdaInductiveConstructor{[]}\AgdaSpace{}%
\AgdaSymbol{=}\AgdaSpace{}%
\AgdaDatatype{⊥}\<%
\\
\>[0]\AgdaFunction{accept₅ˢ}%
\>[10]\AgdaBound{pbkh1}\AgdaSpace{}%
\AgdaBound{time}\AgdaSpace{}%
\AgdaBound{m}\AgdaSpace{}%
\AgdaSymbol{(}\AgdaBound{x}\AgdaSpace{}%
\AgdaOperator{\AgdaInductiveConstructor{∷}}\AgdaSpace{}%
\AgdaInductiveConstructor{[]}\AgdaSymbol{)}\AgdaSpace{}%
\AgdaSymbol{=}\AgdaSpace{}%
\AgdaDatatype{⊥}\<%
\\
\>[0]\AgdaFunction{accept₅ˢ}%
\>[10]\AgdaBound{pbkh1}\AgdaSpace{}%
\AgdaBound{time}\AgdaSpace{}%
\AgdaBound{m}\AgdaSpace{}%
\AgdaSymbol{(}\AgdaBound{x}\AgdaSpace{}%
\AgdaOperator{\AgdaInductiveConstructor{∷}}\AgdaSpace{}%
\AgdaBound{x₁}\AgdaSpace{}%
\AgdaOperator{\AgdaInductiveConstructor{∷}}\AgdaSpace{}%
\AgdaInductiveConstructor{[]}\AgdaSymbol{)}\AgdaSpace{}%
\AgdaSymbol{=}\AgdaSpace{}%
\AgdaDatatype{⊥}\<%
\\
\>[0]\AgdaFunction{accept₅ˢ}%
\>[10]\AgdaBound{pbkh1}\AgdaSpace{}%
\AgdaBound{time}\AgdaSpace{}%
\AgdaBound{m}\AgdaSpace{}%
\AgdaSymbol{(}\AgdaSpace{}%
\AgdaBound{pbk1}%
\>[32]\AgdaOperator{\AgdaInductiveConstructor{∷}}\AgdaSpace{}%
\AgdaBound{pbk2}\AgdaSpace{}%
\AgdaOperator{\AgdaInductiveConstructor{∷}}\AgdaSpace{}%
\AgdaBound{sig}\AgdaSpace{}%
\AgdaOperator{\AgdaInductiveConstructor{∷}}\AgdaSpace{}%
\AgdaBound{st}\AgdaSymbol{)}\<%
\\
\>[0][@{}l@{\AgdaIndent{0}}]%
\>[7]\AgdaSymbol{=}%
\>[10]\AgdaSymbol{(}\AgdaFunction{hashFun}\AgdaSpace{}%
\AgdaBound{pbk1}\AgdaSpace{}%
\AgdaOperator{\AgdaDatatype{≡}}%
\>[27]\AgdaBound{pbkh1}\AgdaSymbol{)}\AgdaSpace{}%
\AgdaOperator{\AgdaRecord{∧}}%
\>[37]\AgdaFunction{IsSigned}%
\>[48]\AgdaBound{m}%
\>[51]\AgdaBound{sig}%
\>[56]\AgdaBound{pbk2}\<%
\end{code}
} 

\AgdaHide{
\begin{code}%
\>[0]\<%
\\
\>[0]\<%
\end{code}
} 

\newcommand{\verificationptopbasicweakestprecondition}{
\begin{code}%
\>[0]\AgdaFunction{wPreCondP2PKHˢ}\AgdaSpace{}%
\AgdaSymbol{:}\AgdaSpace{}%
\AgdaSymbol{(}\AgdaBound{pbkh}\AgdaSpace{}%
\AgdaSymbol{:}\AgdaSpace{}%
\AgdaDatatype{ℕ}\AgdaSpace{}%
\AgdaSymbol{)}\AgdaSpace{}%
\AgdaSymbol{→}\AgdaSpace{}%
\AgdaFunction{StackPredicate}\<%
\\
\>[0]\AgdaFunction{wPreCondP2PKHˢ}%
\>[16]\AgdaBound{pbkh}%
\>[22]\AgdaBound{time}%
\>[28]\AgdaBound{m}%
\>[31]\AgdaInductiveConstructor{[]}%
\>[53]\AgdaSymbol{=}%
\>[56]\AgdaDatatype{⊥}\<%
\\
\>[0]\AgdaFunction{wPreCondP2PKHˢ}%
\>[16]\AgdaBound{pbkh}%
\>[22]\AgdaBound{time}%
\>[28]\AgdaBound{m}%
\>[31]\AgdaSymbol{(}\AgdaBound{x}\AgdaSpace{}%
\AgdaOperator{\AgdaInductiveConstructor{∷}}\AgdaSpace{}%
\AgdaInductiveConstructor{[]}\AgdaSymbol{)}%
\>[53]\AgdaSymbol{=}%
\>[56]\AgdaDatatype{⊥}\<%
\\
\>[0]\AgdaFunction{wPreCondP2PKHˢ}%
\>[16]\AgdaBound{pbkh}%
\>[22]\AgdaBound{time}%
\>[28]\AgdaBound{m}%
\>[31]\AgdaSymbol{(}\AgdaSpace{}%
\AgdaBound{pbk}\AgdaSpace{}%
\AgdaOperator{\AgdaInductiveConstructor{∷}}\AgdaSpace{}%
\AgdaBound{sig}\AgdaSpace{}%
\AgdaOperator{\AgdaInductiveConstructor{∷}}\AgdaSpace{}%
\AgdaBound{st}\AgdaSymbol{)}%
\>[50]\AgdaSymbol{=}\<%
\\
\>[16]\AgdaSymbol{(}\AgdaFunction{hashFun}\AgdaSpace{}%
\AgdaBound{pbk}\AgdaSpace{}%
\AgdaOperator{\AgdaDatatype{≡}}%
\>[32]\AgdaBound{pbkh}\AgdaSpace{}%
\AgdaSymbol{)}\AgdaSpace{}%
\AgdaOperator{\AgdaRecord{∧}}\AgdaSpace{}%
\AgdaFunction{IsSigned}%
\>[51]\AgdaBound{m}\AgdaSpace{}%
\AgdaBound{sig}\AgdaSpace{}%
\AgdaBound{pbk}\<%
\end{code}
} 

\AgdaHide{
\begin{code}%
\>[0]\<%
\\
\\[\AgdaEmptyExtraSkip]%
\>[0]\AgdaFunction{correct3Aux1}\AgdaSpace{}%
\AgdaSymbol{:}%
\>[394I]\AgdaSymbol{(}\AgdaBound{x}\AgdaSpace{}%
\AgdaSymbol{:}\AgdaSpace{}%
\AgdaDatatype{ℕ}\AgdaSymbol{)(}\AgdaBound{rest}\AgdaSpace{}%
\AgdaSymbol{:}\AgdaSpace{}%
\AgdaDatatype{List}\AgdaSpace{}%
\AgdaDatatype{ℕ}\AgdaSymbol{)(}\AgdaBound{time}\AgdaSpace{}%
\AgdaSymbol{:}\AgdaSpace{}%
\AgdaFunction{Time}\AgdaSymbol{)(}\AgdaBound{msg}\AgdaSpace{}%
\AgdaSymbol{:}\AgdaSpace{}%
\AgdaDatatype{Msg}\AgdaSymbol{)}\AgdaSpace{}%
\AgdaSymbol{→}\AgdaSpace{}%
\AgdaFunction{accept₂ˢ}\AgdaSpace{}%
\AgdaBound{time}\AgdaSpace{}%
\AgdaBound{msg}\AgdaSpace{}%
\AgdaSymbol{(}\AgdaBound{x}\AgdaSpace{}%
\AgdaOperator{\AgdaInductiveConstructor{∷}}\AgdaSpace{}%
\AgdaBound{rest}\AgdaSymbol{)}\<%
\\
\>[.][@{}l@{}]\<[394I]%
\>[15]\AgdaSymbol{→}%
\>[18]\AgdaFunction{isTrueNat}\AgdaSpace{}%
\AgdaBound{x}\<%
\\
\>[0]\AgdaFunction{correct3Aux1}\AgdaSpace{}%
\AgdaInductiveConstructor{zero}\AgdaSpace{}%
\AgdaSymbol{(}\AgdaInductiveConstructor{zero}\AgdaSpace{}%
\AgdaOperator{\AgdaInductiveConstructor{∷}}\AgdaSpace{}%
\AgdaInductiveConstructor{[]}\AgdaSymbol{)}\AgdaSpace{}%
\AgdaBound{time}\AgdaSpace{}%
\AgdaBound{msg}\AgdaSpace{}%
\AgdaBound{accept}\AgdaSpace{}%
\AgdaSymbol{=}\AgdaSpace{}%
\AgdaBound{accept}\<%
\\
\>[0]\AgdaFunction{correct3Aux1}\AgdaSpace{}%
\AgdaInductiveConstructor{zero}\AgdaSpace{}%
\AgdaSymbol{(}\AgdaInductiveConstructor{zero}\AgdaSpace{}%
\AgdaOperator{\AgdaInductiveConstructor{∷}}\AgdaSpace{}%
\AgdaBound{x}\AgdaSpace{}%
\AgdaOperator{\AgdaInductiveConstructor{∷}}\AgdaSpace{}%
\AgdaBound{rest}\AgdaSymbol{)}\AgdaSpace{}%
\AgdaBound{time}\AgdaSpace{}%
\AgdaBound{msg}\AgdaSpace{}%
\AgdaBound{accept}\AgdaSpace{}%
\AgdaSymbol{=}\AgdaSpace{}%
\AgdaBound{accept}\<%
\\
\>[0]\AgdaFunction{correct3Aux1}\AgdaSpace{}%
\AgdaInductiveConstructor{zero}\AgdaSpace{}%
\AgdaSymbol{(}\AgdaInductiveConstructor{suc}\AgdaSpace{}%
\AgdaBound{x}\AgdaSpace{}%
\AgdaOperator{\AgdaInductiveConstructor{∷}}\AgdaSpace{}%
\AgdaInductiveConstructor{[]}\AgdaSymbol{)}\AgdaSpace{}%
\AgdaBound{time}\AgdaSpace{}%
\AgdaBound{msg}\AgdaSpace{}%
\AgdaBound{accept}\AgdaSpace{}%
\AgdaSymbol{=}\AgdaSpace{}%
\AgdaBound{accept}\<%
\\
\>[0]\AgdaFunction{correct3Aux1}\AgdaSpace{}%
\AgdaInductiveConstructor{zero}\AgdaSpace{}%
\AgdaSymbol{(}\AgdaInductiveConstructor{suc}\AgdaSpace{}%
\AgdaBound{x}\AgdaSpace{}%
\AgdaOperator{\AgdaInductiveConstructor{∷}}\AgdaSpace{}%
\AgdaBound{x₁}\AgdaSpace{}%
\AgdaOperator{\AgdaInductiveConstructor{∷}}\AgdaSpace{}%
\AgdaBound{rest}\AgdaSymbol{)}\AgdaSpace{}%
\AgdaBound{time}\AgdaSpace{}%
\AgdaBound{msg}\AgdaSpace{}%
\AgdaBound{accept}\AgdaSpace{}%
\AgdaSymbol{=}\AgdaSpace{}%
\AgdaBound{accept}\<%
\\
\>[0]\AgdaFunction{correct3Aux1}\AgdaSpace{}%
\AgdaSymbol{(}\AgdaInductiveConstructor{suc}\AgdaSpace{}%
\AgdaBound{x}\AgdaSymbol{)}\AgdaSpace{}%
\AgdaSymbol{(}\AgdaBound{x₁}\AgdaSpace{}%
\AgdaOperator{\AgdaInductiveConstructor{∷}}\AgdaSpace{}%
\AgdaBound{rest}\AgdaSymbol{)}\AgdaSpace{}%
\AgdaBound{time}\AgdaSpace{}%
\AgdaBound{msg}\AgdaSpace{}%
\AgdaBound{accept}\AgdaSpace{}%
\AgdaSymbol{=}\AgdaSpace{}%
\AgdaInductiveConstructor{tt}\<%
\\
\\[\AgdaEmptyExtraSkip]%
\\[\AgdaEmptyExtraSkip]%
\>[0]\AgdaFunction{correct3Aux2}\AgdaSpace{}%
\AgdaSymbol{:}%
\>[465I]\AgdaSymbol{(}\AgdaSpace{}%
\AgdaBound{x}\AgdaSpace{}%
\AgdaBound{pbk}\AgdaSpace{}%
\AgdaBound{sig}\AgdaSpace{}%
\AgdaSymbol{:}\AgdaSpace{}%
\AgdaDatatype{ℕ}\AgdaSpace{}%
\AgdaSymbol{)(}\AgdaSpace{}%
\AgdaBound{rest}\AgdaSpace{}%
\AgdaSymbol{:}\AgdaSpace{}%
\AgdaDatatype{List}\AgdaSpace{}%
\AgdaDatatype{ℕ}\AgdaSymbol{)(}\AgdaBound{time}\AgdaSpace{}%
\AgdaSymbol{:}\AgdaSpace{}%
\AgdaFunction{Time}\AgdaSymbol{)(}\AgdaBound{m}\AgdaSpace{}%
\AgdaSymbol{:}\AgdaSpace{}%
\AgdaDatatype{Msg}\AgdaSymbol{)}\<%
\\
\>[465I][@{}l@{\AgdaIndent{0}}]%
\>[16]\AgdaSymbol{→}\AgdaSpace{}%
\AgdaFunction{accept₂ˢ}\AgdaSpace{}%
\AgdaBound{time}\AgdaSpace{}%
\AgdaBound{m}\AgdaSpace{}%
\AgdaSymbol{(}\AgdaBound{x}\AgdaSpace{}%
\AgdaOperator{\AgdaInductiveConstructor{∷}}\AgdaSpace{}%
\AgdaBound{pbk}\AgdaSpace{}%
\AgdaOperator{\AgdaInductiveConstructor{∷}}\AgdaSpace{}%
\AgdaBound{sig}\AgdaSpace{}%
\AgdaOperator{\AgdaInductiveConstructor{∷}}\AgdaSpace{}%
\AgdaBound{rest}\AgdaSymbol{)}\<%
\\
\>[16]\AgdaSymbol{→}\AgdaSpace{}%
\AgdaFunction{IsSigned}%
\>[28]\AgdaBound{m}\AgdaSpace{}%
\AgdaBound{sig}\AgdaSpace{}%
\AgdaBound{pbk}\<%
\\
\>[0]\AgdaFunction{correct3Aux2}\AgdaSpace{}%
\AgdaSymbol{(}\AgdaInductiveConstructor{suc}\AgdaSpace{}%
\AgdaBound{x}\AgdaSymbol{)}\AgdaSpace{}%
\AgdaBound{pubkey}\AgdaSpace{}%
\AgdaBound{sig}\AgdaSpace{}%
\AgdaBound{rest}\AgdaSpace{}%
\AgdaBound{time}\AgdaSpace{}%
\AgdaBound{m}\AgdaSpace{}%
\AgdaBound{accept}\AgdaSpace{}%
\AgdaSymbol{=}\AgdaSpace{}%
\AgdaBound{accept}\<%
\\
\\[\AgdaEmptyExtraSkip]%
\\[\AgdaEmptyExtraSkip]%
\\[\AgdaEmptyExtraSkip]%
\>[0]\AgdaFunction{lemmaCorrect3From1}%
\>[503I]\AgdaSymbol{:}\AgdaSpace{}%
\AgdaSymbol{(}\AgdaBound{x}\AgdaSpace{}%
\AgdaBound{z}\AgdaSpace{}%
\AgdaBound{t}\AgdaSpace{}%
\AgdaSymbol{:}\AgdaSpace{}%
\AgdaDatatype{ℕ}\AgdaSymbol{)(}\AgdaBound{time}\AgdaSpace{}%
\AgdaSymbol{:}\AgdaSpace{}%
\AgdaFunction{Time}\AgdaSpace{}%
\AgdaSymbol{)(}\AgdaBound{m}\AgdaSpace{}%
\AgdaSymbol{:}\AgdaSpace{}%
\AgdaDatatype{Msg}\AgdaSymbol{)}\<%
\\
\>[503I][@{}l@{\AgdaIndent{0}}]%
\>[20]\AgdaSymbol{→}%
\>[23]\AgdaFunction{accept₂ˢCore}\AgdaSpace{}%
\AgdaBound{time}\AgdaSpace{}%
\AgdaBound{m}\AgdaSpace{}%
\AgdaBound{x}\AgdaSpace{}%
\AgdaBound{z}\AgdaSpace{}%
\AgdaBound{t}\AgdaSpace{}%
\AgdaSymbol{→}\AgdaSpace{}%
\AgdaFunction{isTrueNat}\AgdaSpace{}%
\AgdaBound{x}\<%
\\
\>[0]\AgdaFunction{lemmaCorrect3From1}\AgdaSpace{}%
\AgdaSymbol{(}\AgdaInductiveConstructor{suc}\AgdaSpace{}%
\AgdaBound{x}\AgdaSymbol{)}\AgdaSpace{}%
\AgdaBound{z}\AgdaSpace{}%
\AgdaBound{t}\AgdaSpace{}%
\AgdaBound{time}\AgdaSpace{}%
\AgdaBound{m}\AgdaSpace{}%
\AgdaBound{p}\AgdaSpace{}%
\AgdaSymbol{=}\AgdaSpace{}%
\AgdaInductiveConstructor{tt}\<%
\\
\\[\AgdaEmptyExtraSkip]%
\\[\AgdaEmptyExtraSkip]%
\>[0]\AgdaFunction{lemmaCorrect3From}\AgdaSpace{}%
\AgdaSymbol{:}%
\>[532I]\AgdaSymbol{(}\AgdaBound{x}\AgdaSpace{}%
\AgdaBound{y}\AgdaSpace{}%
\AgdaBound{z}\AgdaSpace{}%
\AgdaBound{t}\AgdaSpace{}%
\AgdaSymbol{:}\AgdaSpace{}%
\AgdaDatatype{ℕ}\AgdaSymbol{)(}\AgdaBound{time}\AgdaSpace{}%
\AgdaSymbol{:}\AgdaSpace{}%
\AgdaFunction{Time}\AgdaSymbol{)(}\AgdaBound{m}\AgdaSpace{}%
\AgdaSymbol{:}\AgdaSpace{}%
\AgdaDatatype{Msg}\AgdaSymbol{)}\<%
\\
\>[.][@{}l@{}]\<[532I]%
\>[20]\AgdaSymbol{→}%
\>[23]\AgdaFunction{accept₂ˢCore}\AgdaSpace{}%
\AgdaBound{time}\AgdaSpace{}%
\AgdaBound{m}\AgdaSpace{}%
\AgdaSymbol{(}\AgdaFunction{compareNaturals}\AgdaSpace{}%
\AgdaBound{x}\AgdaSpace{}%
\AgdaBound{y}\AgdaSymbol{)}\AgdaSpace{}%
\AgdaBound{z}\AgdaSpace{}%
\AgdaBound{t}\AgdaSpace{}%
\AgdaSymbol{→}\AgdaSpace{}%
\AgdaBound{x}\AgdaSpace{}%
\AgdaOperator{\AgdaDatatype{≡}}\AgdaSpace{}%
\AgdaBound{y}\<%
\\
\>[0]\AgdaFunction{lemmaCorrect3From}\AgdaSpace{}%
\AgdaBound{x}\AgdaSpace{}%
\AgdaBound{y}\AgdaSpace{}%
\AgdaBound{z}\AgdaSpace{}%
\AgdaBound{t}\AgdaSpace{}%
\AgdaBound{time}\AgdaSpace{}%
\AgdaBound{m}\AgdaSpace{}%
\AgdaBound{p}\AgdaSpace{}%
\AgdaSymbol{=}\AgdaSpace{}%
\AgdaFunction{compareNatToEq}\AgdaSpace{}%
\AgdaBound{x}%
\>[55]\AgdaBound{y}\AgdaSpace{}%
\AgdaSymbol{(}\AgdaFunction{lemmaCorrect3From1}\AgdaSpace{}%
\AgdaSymbol{(}\AgdaFunction{compareNaturals}\AgdaSpace{}%
\AgdaBound{x}\AgdaSpace{}%
\AgdaBound{y}\AgdaSymbol{)}\AgdaSpace{}%
\AgdaBound{z}\AgdaSpace{}%
\AgdaBound{t}\AgdaSpace{}%
\AgdaBound{time}\AgdaSpace{}%
\AgdaBound{m}\AgdaSpace{}%
\AgdaBound{p}\AgdaSymbol{)}\<%
\\
\\[\AgdaEmptyExtraSkip]%
\\[\AgdaEmptyExtraSkip]%
\>[0]\AgdaFunction{script-1-b}%
\>[12]\AgdaSymbol{:}\AgdaSpace{}%
\AgdaFunction{BitcoinScriptBasic}\<%
\\
\>[0]\AgdaFunction{script-1-b}\AgdaSpace{}%
\AgdaSymbol{=}\AgdaSpace{}%
\AgdaInductiveConstructor{opCheckSig}\AgdaSpace{}%
\AgdaOperator{\AgdaInductiveConstructor{∷}}\AgdaSpace{}%
\AgdaInductiveConstructor{[]}\<%
\\
\\[\AgdaEmptyExtraSkip]%
\>[0]\AgdaFunction{script-2-b}\AgdaSpace{}%
\AgdaSymbol{:}\AgdaSpace{}%
\AgdaFunction{BitcoinScriptBasic}\<%
\\
\>[0]\AgdaFunction{script-2-b}\AgdaSpace{}%
\AgdaSymbol{=}\AgdaSpace{}%
\AgdaInductiveConstructor{opVerify}\AgdaSpace{}%
\AgdaOperator{\AgdaInductiveConstructor{∷}}\AgdaSpace{}%
\AgdaFunction{script-1-b}\<%
\\
\\[\AgdaEmptyExtraSkip]%
\>[0]\AgdaFunction{script-3-b}\AgdaSpace{}%
\AgdaSymbol{:}\AgdaSpace{}%
\AgdaFunction{BitcoinScriptBasic}\<%
\\
\>[0]\AgdaFunction{script-3-b}\AgdaSpace{}%
\AgdaSymbol{=}\AgdaSpace{}%
\AgdaInductiveConstructor{opEqual}\AgdaSpace{}%
\AgdaOperator{\AgdaInductiveConstructor{∷}}\AgdaSpace{}%
\AgdaFunction{script-2-b}\<%
\\
\\[\AgdaEmptyExtraSkip]%
\>[0]\AgdaFunction{script-4-b}\AgdaSpace{}%
\AgdaSymbol{:}\AgdaSpace{}%
\AgdaDatatype{ℕ}\AgdaSpace{}%
\AgdaSymbol{→}\AgdaSpace{}%
\AgdaFunction{BitcoinScriptBasic}\<%
\\
\>[0]\AgdaFunction{script-4-b}\AgdaSpace{}%
\AgdaBound{pbkh}%
\>[17]\AgdaSymbol{=}%
\>[20]\AgdaInductiveConstructor{opPush}%
\>[28]\AgdaBound{pbkh}\AgdaSpace{}%
\AgdaOperator{\AgdaInductiveConstructor{∷}}\AgdaSpace{}%
\AgdaFunction{script-3-b}\<%
\\
\\[\AgdaEmptyExtraSkip]%
\>[0]\AgdaFunction{script-5-b}\AgdaSpace{}%
\AgdaSymbol{:}\AgdaSpace{}%
\AgdaDatatype{ℕ}\AgdaSpace{}%
\AgdaSymbol{→}%
\>[18]\AgdaFunction{BitcoinScriptBasic}\<%
\\
\>[0]\AgdaFunction{script-5-b}\AgdaSpace{}%
\AgdaBound{pbkh}\AgdaSpace{}%
\AgdaSymbol{=}\AgdaSpace{}%
\AgdaInductiveConstructor{opHash}%
\>[26]\AgdaOperator{\AgdaInductiveConstructor{∷}}\AgdaSpace{}%
\AgdaFunction{script-4-b}%
\>[40]\AgdaBound{pbkh}\<%
\\
\\[\AgdaEmptyExtraSkip]%
\>[0]\AgdaFunction{script-6-b}\AgdaSpace{}%
\AgdaSymbol{:}\AgdaSpace{}%
\AgdaDatatype{ℕ}\AgdaSpace{}%
\AgdaSymbol{→}\AgdaSpace{}%
\AgdaFunction{BitcoinScriptBasic}\<%
\\
\>[0]\AgdaFunction{script-6-b}\AgdaSpace{}%
\AgdaBound{pbkh}%
\>[17]\AgdaSymbol{=}\AgdaSpace{}%
\AgdaInductiveConstructor{opDup}\AgdaSpace{}%
\AgdaOperator{\AgdaInductiveConstructor{∷}}\AgdaSpace{}%
\AgdaFunction{script-5-b}\AgdaSpace{}%
\AgdaBound{pbkh}\<%
\\
\\[\AgdaEmptyExtraSkip]%
\\[\AgdaEmptyExtraSkip]%
\>[0]\AgdaComment{--new}\<%
\\
\>[0]\AgdaFunction{script-7-b}\AgdaSpace{}%
\AgdaSymbol{:}\AgdaSpace{}%
\AgdaDatatype{ℕ}\AgdaSpace{}%
\AgdaSymbol{→}\AgdaSpace{}%
\AgdaFunction{BitcoinScriptBasic}\<%
\\
\>[0]\AgdaFunction{script-7-b}\AgdaSpace{}%
\AgdaBound{pbkh}\AgdaSpace{}%
\AgdaSymbol{=}\AgdaSpace{}%
\AgdaInductiveConstructor{opMultiSig}\AgdaSpace{}%
\AgdaOperator{\AgdaInductiveConstructor{∷}}\AgdaSpace{}%
\AgdaFunction{script-6-b}\AgdaSpace{}%
\AgdaBound{pbkh}\<%
\\
\\[\AgdaEmptyExtraSkip]%
\>[0]\AgdaFunction{script-7'-b}\AgdaSpace{}%
\AgdaSymbol{:}\AgdaSpace{}%
\AgdaSymbol{(}\AgdaBound{pbkh}\AgdaSpace{}%
\AgdaBound{pbk1}\AgdaSpace{}%
\AgdaBound{pbk2}\AgdaSpace{}%
\AgdaSymbol{:}\AgdaSpace{}%
\AgdaDatatype{ℕ}\AgdaSymbol{)}\AgdaSpace{}%
\AgdaSymbol{→}\AgdaSpace{}%
\AgdaFunction{BitcoinScriptBasic}\<%
\\
\>[0]\AgdaFunction{script-7'-b}\AgdaSpace{}%
\AgdaBound{pbkh}\AgdaSpace{}%
\AgdaBound{pbk1}\AgdaSpace{}%
\AgdaBound{pbk2}\AgdaSpace{}%
\AgdaSymbol{=}\AgdaSpace{}%
\AgdaInductiveConstructor{opMultiSig}\AgdaSpace{}%
\AgdaOperator{\AgdaInductiveConstructor{∷}}\AgdaSpace{}%
\AgdaFunction{script-6-b}\AgdaSpace{}%
\AgdaBound{pbkh}\<%
\\
\\[\AgdaEmptyExtraSkip]%
\\[\AgdaEmptyExtraSkip]%
\>[0]\AgdaFunction{script-1}%
\>[10]\AgdaSymbol{:}\AgdaSpace{}%
\AgdaFunction{BitcoinScript}\<%
\\
\>[0]\AgdaFunction{script-1}%
\>[10]\AgdaSymbol{=}\AgdaSpace{}%
\AgdaFunction{basicBScript2BScript}\AgdaSpace{}%
\AgdaFunction{script-1-b}\<%
\\
\\[\AgdaEmptyExtraSkip]%
\>[0]\AgdaFunction{script-2}%
\>[10]\AgdaSymbol{:}\AgdaSpace{}%
\AgdaFunction{BitcoinScript}\<%
\\
\>[0]\AgdaFunction{script-2}%
\>[10]\AgdaSymbol{=}\AgdaSpace{}%
\AgdaFunction{basicBScript2BScript}\AgdaSpace{}%
\AgdaFunction{script-2-b}\<%
\\
\\[\AgdaEmptyExtraSkip]%
\>[0]\AgdaFunction{script-3}%
\>[10]\AgdaSymbol{:}\AgdaSpace{}%
\AgdaFunction{BitcoinScript}\<%
\\
\>[0]\AgdaFunction{script-3}%
\>[10]\AgdaSymbol{=}\AgdaSpace{}%
\AgdaFunction{basicBScript2BScript}\AgdaSpace{}%
\AgdaFunction{script-3-b}\<%
\\
\\[\AgdaEmptyExtraSkip]%
\>[0]\AgdaFunction{script-4}%
\>[10]\AgdaSymbol{:}\AgdaSpace{}%
\AgdaDatatype{ℕ}\AgdaSpace{}%
\AgdaSymbol{→}\AgdaSpace{}%
\AgdaFunction{BitcoinScript}\<%
\\
\>[0]\AgdaFunction{script-4}%
\>[10]\AgdaBound{pbk}\AgdaSpace{}%
\AgdaSymbol{=}\AgdaSpace{}%
\AgdaFunction{basicBScript2BScript}\AgdaSpace{}%
\AgdaSymbol{(}\AgdaFunction{script-4-b}\AgdaSpace{}%
\AgdaBound{pbk}\AgdaSymbol{)}\<%
\\
\\[\AgdaEmptyExtraSkip]%
\>[0]\AgdaFunction{script-5}%
\>[10]\AgdaSymbol{:}\AgdaSpace{}%
\AgdaDatatype{ℕ}\AgdaSpace{}%
\AgdaSymbol{→}\AgdaSpace{}%
\AgdaFunction{BitcoinScript}\<%
\\
\>[0]\AgdaFunction{script-5}%
\>[10]\AgdaBound{pbk}\AgdaSpace{}%
\AgdaSymbol{=}\AgdaSpace{}%
\AgdaFunction{basicBScript2BScript}\AgdaSpace{}%
\AgdaSymbol{(}\AgdaFunction{script-5-b}\AgdaSpace{}%
\AgdaBound{pbk}\AgdaSymbol{)}\<%
\\
\\[\AgdaEmptyExtraSkip]%
\>[0]\AgdaFunction{script-6}%
\>[10]\AgdaSymbol{:}\AgdaSpace{}%
\AgdaDatatype{ℕ}\AgdaSpace{}%
\AgdaSymbol{→}\AgdaSpace{}%
\AgdaFunction{BitcoinScript}\<%
\\
\>[0]\AgdaFunction{script-6}%
\>[10]\AgdaBound{pbk}\AgdaSpace{}%
\AgdaSymbol{=}\AgdaSpace{}%
\AgdaFunction{basicBScript2BScript}\AgdaSpace{}%
\AgdaSymbol{(}\AgdaFunction{script-6-b}\AgdaSpace{}%
\AgdaBound{pbk}\AgdaSymbol{)}\<%
\\
\\[\AgdaEmptyExtraSkip]%
\>[0]\AgdaFunction{script-7}%
\>[10]\AgdaSymbol{:}\AgdaSpace{}%
\AgdaDatatype{ℕ}\AgdaSpace{}%
\AgdaSymbol{→}\AgdaSpace{}%
\AgdaFunction{BitcoinScript}\<%
\\
\>[0]\AgdaFunction{script-7}%
\>[10]\AgdaBound{pbk}\AgdaSpace{}%
\AgdaSymbol{=}\AgdaSpace{}%
\AgdaFunction{basicBScript2BScript}\AgdaSpace{}%
\AgdaSymbol{(}\AgdaFunction{script-7-b}\AgdaSpace{}%
\AgdaBound{pbk}\AgdaSymbol{)}\<%
\\
\\[\AgdaEmptyExtraSkip]%
\\[\AgdaEmptyExtraSkip]%
\>[0]\AgdaFunction{script-7'}\AgdaSpace{}%
\AgdaSymbol{:}\AgdaSpace{}%
\AgdaSymbol{(}\AgdaBound{pbkh}\AgdaSpace{}%
\AgdaBound{pbk1}\AgdaSpace{}%
\AgdaBound{pbk2}\AgdaSpace{}%
\AgdaSymbol{:}\AgdaSpace{}%
\AgdaDatatype{ℕ}\AgdaSymbol{)}\AgdaSpace{}%
\AgdaSymbol{→}\AgdaSpace{}%
\AgdaFunction{BitcoinScript}\<%
\\
\>[0]\AgdaFunction{script-7'}%
\>[11]\AgdaBound{pbkh}\AgdaSpace{}%
\AgdaBound{pbk1}\AgdaSpace{}%
\AgdaBound{pbk2}\AgdaSpace{}%
\AgdaSymbol{=}\AgdaSpace{}%
\AgdaFunction{basicBScript2BScript}\AgdaSpace{}%
\AgdaSymbol{(}\AgdaFunction{script-7'-b}\AgdaSpace{}%
\AgdaBound{pbkh}\AgdaSpace{}%
\AgdaBound{pbk1}\AgdaSpace{}%
\AgdaBound{pbk2}\AgdaSymbol{)}\<%
\\
\\[\AgdaEmptyExtraSkip]%
\\[\AgdaEmptyExtraSkip]%
\\[\AgdaEmptyExtraSkip]%
\>[0]\AgdaFunction{instructionsBasic}\AgdaSpace{}%
\AgdaSymbol{:}\AgdaSpace{}%
\AgdaSymbol{(}\AgdaBound{pbkh}\AgdaSpace{}%
\AgdaSymbol{:}\AgdaSpace{}%
\AgdaDatatype{ℕ}\AgdaSymbol{)}\AgdaSpace{}%
\AgdaSymbol{(}\AgdaBound{n}\AgdaSpace{}%
\AgdaSymbol{:}\AgdaSpace{}%
\AgdaDatatype{ℕ}\AgdaSymbol{)}\AgdaSpace{}%
\AgdaSymbol{→}\AgdaSpace{}%
\AgdaBound{n}\AgdaSpace{}%
\AgdaOperator{\AgdaFunction{≤}}\AgdaSpace{}%
\AgdaNumber{5}\AgdaSpace{}%
\AgdaSymbol{→}\AgdaSpace{}%
\AgdaDatatype{InstructionBasic}\<%
\\
\>[0]\AgdaFunction{instructionsBasic}\AgdaSpace{}%
\AgdaBound{pbkh}\AgdaSpace{}%
\AgdaNumber{0}\AgdaSpace{}%
\AgdaSymbol{\AgdaUnderscore{}}\AgdaSpace{}%
\AgdaSymbol{=}\AgdaSpace{}%
\AgdaInductiveConstructor{opCheckSig}\<%
\\
\>[0]\AgdaFunction{instructionsBasic}\AgdaSpace{}%
\AgdaBound{pbkh}\AgdaSpace{}%
\AgdaNumber{1}\AgdaSpace{}%
\AgdaSymbol{\AgdaUnderscore{}}\AgdaSpace{}%
\AgdaSymbol{=}\AgdaSpace{}%
\AgdaInductiveConstructor{opVerify}\<%
\\
\>[0]\AgdaFunction{instructionsBasic}\AgdaSpace{}%
\AgdaBound{pbkh}\AgdaSpace{}%
\AgdaNumber{2}\AgdaSpace{}%
\AgdaSymbol{\AgdaUnderscore{}}\AgdaSpace{}%
\AgdaSymbol{=}\AgdaSpace{}%
\AgdaInductiveConstructor{opEqual}\<%
\\
\>[0]\AgdaFunction{instructionsBasic}\AgdaSpace{}%
\AgdaBound{pbkh}\AgdaSpace{}%
\AgdaNumber{3}\AgdaSpace{}%
\AgdaSymbol{\AgdaUnderscore{}}\AgdaSpace{}%
\AgdaSymbol{=}\AgdaSpace{}%
\AgdaInductiveConstructor{opPush}\AgdaSpace{}%
\AgdaBound{pbkh}\<%
\\
\>[0]\AgdaFunction{instructionsBasic}\AgdaSpace{}%
\AgdaBound{pbkh}\AgdaSpace{}%
\AgdaNumber{4}\AgdaSpace{}%
\AgdaSymbol{\AgdaUnderscore{}}\AgdaSpace{}%
\AgdaSymbol{=}\AgdaSpace{}%
\AgdaInductiveConstructor{opHash}\<%
\\
\>[0]\AgdaFunction{instructionsBasic}\AgdaSpace{}%
\AgdaBound{pbkh}\AgdaSpace{}%
\AgdaNumber{5}\AgdaSpace{}%
\AgdaSymbol{\AgdaUnderscore{}}\AgdaSpace{}%
\AgdaSymbol{=}\AgdaSpace{}%
\AgdaInductiveConstructor{opDup}\<%
\\
\\[\AgdaEmptyExtraSkip]%
\>[0]\AgdaFunction{scriptP2PKH}\AgdaSpace{}%
\AgdaSymbol{:}\AgdaSpace{}%
\AgdaSymbol{(}\AgdaBound{pbkh}\AgdaSpace{}%
\AgdaSymbol{:}\AgdaSpace{}%
\AgdaDatatype{ℕ}\AgdaSymbol{)}\AgdaSpace{}%
\AgdaSymbol{→}\AgdaSpace{}%
\AgdaFunction{BitcoinScript}\<%
\\
\>[0]\AgdaFunction{scriptP2PKH}\AgdaSpace{}%
\AgdaBound{pbkh}\AgdaSpace{}%
\AgdaSymbol{=}\AgdaSpace{}%
\AgdaInductiveConstructor{opDup}\AgdaSpace{}%
\AgdaOperator{\AgdaInductiveConstructor{∷}}\AgdaSpace{}%
\AgdaInductiveConstructor{opHash}\AgdaSpace{}%
\AgdaOperator{\AgdaInductiveConstructor{∷}}\AgdaSpace{}%
\AgdaSymbol{(}\AgdaInductiveConstructor{opPush}\AgdaSpace{}%
\AgdaBound{pbkh}\AgdaSymbol{)}\AgdaSpace{}%
\AgdaOperator{\AgdaInductiveConstructor{∷}}\AgdaSpace{}%
\AgdaInductiveConstructor{opEqual}\AgdaSpace{}%
\AgdaOperator{\AgdaInductiveConstructor{∷}}\AgdaSpace{}%
\AgdaInductiveConstructor{opVerify}\AgdaSpace{}%
\AgdaOperator{\AgdaInductiveConstructor{∷}}\AgdaSpace{}%
\AgdaInductiveConstructor{opCheckSig}\AgdaSpace{}%
\AgdaOperator{\AgdaInductiveConstructor{∷}}\AgdaSpace{}%
\AgdaInductiveConstructor{[]}\<%
\\
\\[\AgdaEmptyExtraSkip]%
\>[0]\AgdaFunction{weakestPreConditionP2PKHˢ}\AgdaSpace{}%
\AgdaSymbol{:}\AgdaSpace{}%
\AgdaSymbol{(}\AgdaBound{pbkh}\AgdaSpace{}%
\AgdaSymbol{:}\AgdaSpace{}%
\AgdaDatatype{ℕ}\AgdaSymbol{)}\AgdaSpace{}%
\AgdaSymbol{→}\AgdaSpace{}%
\AgdaFunction{StackPredicate}\<%
\\
\>[0]\AgdaFunction{weakestPreConditionP2PKHˢ}\AgdaSpace{}%
\AgdaSymbol{=}\AgdaSpace{}%
\AgdaFunction{wPreCondP2PKHˢ}\<%
\end{code}
} 
                          

\AgdaHide{
\begin{code}%
\>[0]\AgdaComment{--\textbackslash{}semanticBasicOperationsForTypeSetting}\<%
\\
\>[0]\AgdaKeyword{open}\AgdaSpace{}%
\AgdaKeyword{import}\AgdaSpace{}%
\AgdaModule{basicBitcoinDataType}\<%
\\
\\[\AgdaEmptyExtraSkip]%
\>[0]\AgdaKeyword{module}\AgdaSpace{}%
\AgdaModule{paperTypes2021PostProceed.semanticBasicOperationsForTypeSetting}\AgdaSpace{}%
\AgdaSymbol{(}\AgdaBound{param}\AgdaSpace{}%
\AgdaSymbol{:}\AgdaSpace{}%
\AgdaRecord{GlobalParameters}\AgdaSymbol{)}\AgdaSpace{}%
\AgdaKeyword{where}\<%
\\
\\[\AgdaEmptyExtraSkip]%
\>[0]\AgdaKeyword{open}\AgdaSpace{}%
\AgdaKeyword{import}\AgdaSpace{}%
\AgdaModule{Data.Nat}%
\>[22]\AgdaKeyword{hiding}\AgdaSpace{}%
\AgdaSymbol{(}\AgdaOperator{\AgdaDatatype{\AgdaUnderscore{}≤\AgdaUnderscore{}}}\AgdaSymbol{)}\<%
\\
\>[0]\AgdaKeyword{open}\AgdaSpace{}%
\AgdaKeyword{import}\AgdaSpace{}%
\AgdaModule{Data.List}\AgdaSpace{}%
\AgdaKeyword{hiding}\AgdaSpace{}%
\AgdaSymbol{(}\AgdaOperator{\AgdaFunction{\AgdaUnderscore{}\ensuremath{+\!\!+}\AgdaUnderscore{}}}\AgdaSymbol{)}\<%
\\
\>[0]\AgdaKeyword{open}\AgdaSpace{}%
\AgdaKeyword{import}\AgdaSpace{}%
\AgdaModule{Data.Unit}%
\>[23]\AgdaKeyword{hiding}\AgdaSpace{}%
\AgdaSymbol{(}\AgdaOperator{\AgdaRecord{\AgdaUnderscore{}≤\AgdaUnderscore{}}}\AgdaSymbol{)}\<%
\\
\>[0]\AgdaKeyword{open}\AgdaSpace{}%
\AgdaKeyword{import}\AgdaSpace{}%
\AgdaModule{Data.Empty}\<%
\\
\>[0]\AgdaKeyword{open}\AgdaSpace{}%
\AgdaKeyword{import}\AgdaSpace{}%
\AgdaModule{Data.Bool}%
\>[23]\AgdaKeyword{hiding}\AgdaSpace{}%
\AgdaSymbol{(}\AgdaOperator{\AgdaDatatype{\AgdaUnderscore{}≤\AgdaUnderscore{}}}\AgdaSpace{}%
\AgdaSymbol{;}\AgdaSpace{}%
\AgdaOperator{\AgdaFunction{if\AgdaUnderscore{}then\AgdaUnderscore{}else\AgdaUnderscore{}}}\AgdaSpace{}%
\AgdaSymbol{)}\AgdaSpace{}%
\AgdaKeyword{renaming}\AgdaSpace{}%
\AgdaSymbol{(}\AgdaOperator{\AgdaFunction{\AgdaUnderscore{}∧\AgdaUnderscore{}}}\AgdaSpace{}%
\AgdaSymbol{to}\AgdaSpace{}%
\AgdaOperator{\AgdaFunction{\AgdaUnderscore{}∧b\AgdaUnderscore{}}}\AgdaSpace{}%
\AgdaSymbol{;}\AgdaSpace{}%
\AgdaOperator{\AgdaFunction{\AgdaUnderscore{}∨\AgdaUnderscore{}}}\AgdaSpace{}%
\AgdaSymbol{to}\AgdaSpace{}%
\AgdaOperator{\AgdaFunction{\AgdaUnderscore{}∨b\AgdaUnderscore{}}}\AgdaSpace{}%
\AgdaSymbol{;}\AgdaSpace{}%
\AgdaFunction{T}\AgdaSpace{}%
\AgdaSymbol{to}\AgdaSpace{}%
\AgdaFunction{True}\AgdaSymbol{)}\<%
\\
\>[0]\AgdaKeyword{open}\AgdaSpace{}%
\AgdaKeyword{import}\AgdaSpace{}%
\AgdaModule{Data.Product}\AgdaSpace{}%
\AgdaKeyword{renaming}\AgdaSpace{}%
\AgdaSymbol{(}\AgdaOperator{\AgdaInductiveConstructor{\AgdaUnderscore{},\AgdaUnderscore{}}}\AgdaSpace{}%
\AgdaSymbol{to}\AgdaSpace{}%
\AgdaOperator{\AgdaInductiveConstructor{\AgdaUnderscore{},,\AgdaUnderscore{}}}\AgdaSpace{}%
\AgdaSymbol{)}\<%
\\
\>[0]\AgdaKeyword{open}\AgdaSpace{}%
\AgdaKeyword{import}\AgdaSpace{}%
\AgdaModule{Data.Nat.Base}\AgdaSpace{}%
\AgdaKeyword{hiding}\AgdaSpace{}%
\AgdaSymbol{(}\AgdaOperator{\AgdaDatatype{\AgdaUnderscore{}≤\AgdaUnderscore{}}}\AgdaSymbol{)}\<%
\\
\>[0]\AgdaComment{--\ open\ import\ Data.List.NonEmpty\ hiding\ (head)}\<%
\\
\>[0]\AgdaKeyword{open}\AgdaSpace{}%
\AgdaKeyword{import}\AgdaSpace{}%
\AgdaModule{Data.Maybe}\<%
\\
\\[\AgdaEmptyExtraSkip]%
\>[0]\AgdaKeyword{import}\AgdaSpace{}%
\AgdaModule{Relation.Binary.PropositionalEquality}\AgdaSpace{}%
\AgdaSymbol{as}\AgdaSpace{}%
\AgdaModule{Eq}\<%
\\
\>[0]\AgdaKeyword{open}\AgdaSpace{}%
\AgdaModule{Eq}\AgdaSpace{}%
\AgdaKeyword{using}\AgdaSpace{}%
\AgdaSymbol{(}\AgdaOperator{\AgdaDatatype{\AgdaUnderscore{}≡\AgdaUnderscore{}}}\AgdaSymbol{;}\AgdaSpace{}%
\AgdaInductiveConstructor{refl}\AgdaSymbol{;}\AgdaSpace{}%
\AgdaFunction{cong}\AgdaSymbol{;}\AgdaSpace{}%
\AgdaKeyword{module}\AgdaSpace{}%
\AgdaModule{≡-Reasoning}\AgdaSymbol{;}\AgdaSpace{}%
\AgdaFunction{sym}\AgdaSymbol{)}\<%
\\
\>[0]\AgdaKeyword{open}\AgdaSpace{}%
\AgdaModule{≡-Reasoning}\<%
\\
\>[0]\AgdaKeyword{open}\AgdaSpace{}%
\AgdaKeyword{import}\AgdaSpace{}%
\AgdaModule{Agda.Builtin.Equality}\<%
\\
\>[0]\AgdaComment{--open\ import\ Agda.Builtin.Equality.Rewrite}\<%
\\
\\[\AgdaEmptyExtraSkip]%
\\[\AgdaEmptyExtraSkip]%
\>[0]\AgdaKeyword{open}\AgdaSpace{}%
\AgdaKeyword{import}\AgdaSpace{}%
\AgdaModule{libraries.listLib}\<%
\\
\>[0]\AgdaKeyword{open}\AgdaSpace{}%
\AgdaKeyword{import}\AgdaSpace{}%
\AgdaModule{libraries.natLib}\<%
\\
\>[0]\AgdaKeyword{open}\AgdaSpace{}%
\AgdaKeyword{import}\AgdaSpace{}%
\AgdaModule{libraries.boolLib}\<%
\\
\>[0]\AgdaKeyword{open}\AgdaSpace{}%
\AgdaKeyword{import}\AgdaSpace{}%
\AgdaModule{libraries.andLib}\<%
\\
\>[0]\AgdaKeyword{open}\AgdaSpace{}%
\AgdaKeyword{import}\AgdaSpace{}%
\AgdaModule{libraries.miscLib}\<%
\\
\>[0]\AgdaKeyword{open}\AgdaSpace{}%
\AgdaKeyword{import}\AgdaSpace{}%
\AgdaModule{libraries.maybeLib}\<%
\\
\\[\AgdaEmptyExtraSkip]%
\>[0]\AgdaKeyword{open}\AgdaSpace{}%
\AgdaKeyword{import}\AgdaSpace{}%
\AgdaModule{stack}\<%
\\
\\[\AgdaEmptyExtraSkip]%
\>[0]\AgdaKeyword{open}\AgdaSpace{}%
\AgdaKeyword{import}\AgdaSpace{}%
\AgdaModule{semanticBasicOperations}\AgdaSpace{}%
\AgdaBound{param}\AgdaSpace{}%
\AgdaKeyword{hiding}\AgdaSpace{}%
\AgdaSymbol{(}\AgdaFunction{executeStackCheckSig3Aux}\AgdaSymbol{;}\AgdaFunction{compareSigsMultiSigAux}\AgdaSymbol{;}\AgdaFunction{compareSigsMultiSig}\AgdaSymbol{;}\AgdaFunction{executeMultiSig3}\AgdaSymbol{;}\AgdaFunction{executeMultiSig2}\AgdaSymbol{;}\AgdaFunction{executeMultiSig}\AgdaSymbol{)}\<%
\\
\\[\AgdaEmptyExtraSkip]%
\>[0]\AgdaFunction{executeStackCheckSig3Aux}\AgdaSpace{}%
\AgdaSymbol{:}\AgdaSpace{}%
\AgdaDatatype{Msg}\AgdaSpace{}%
\AgdaSymbol{→}%
\>[34]\AgdaFunction{Stack}\AgdaSpace{}%
\AgdaSymbol{→}\AgdaSpace{}%
\AgdaDatatype{Maybe}\AgdaSpace{}%
\AgdaFunction{Stack}\<%
\\
\>[0]\AgdaFunction{executeStackCheckSig3Aux}\AgdaSpace{}%
\AgdaBound{msg}\AgdaSpace{}%
\AgdaInductiveConstructor{[]}\AgdaSpace{}%
\AgdaSymbol{=}\AgdaSpace{}%
\AgdaInductiveConstructor{nothing}\<%
\\
\>[0]\AgdaFunction{executeStackCheckSig3Aux}\AgdaSpace{}%
\AgdaBound{mst}\AgdaSpace{}%
\AgdaSymbol{(}\AgdaBound{x}\AgdaSpace{}%
\AgdaOperator{\AgdaInductiveConstructor{∷}}\AgdaSpace{}%
\AgdaInductiveConstructor{[]}\AgdaSymbol{)}\AgdaSpace{}%
\AgdaSymbol{=}\AgdaSpace{}%
\AgdaInductiveConstructor{nothing}\<%
\\
\>[0]\AgdaFunction{executeStackCheckSig3Aux}\AgdaSpace{}%
\AgdaBound{msg}\AgdaSpace{}%
\AgdaSymbol{(}\AgdaBound{m}\AgdaSpace{}%
\AgdaOperator{\AgdaInductiveConstructor{∷}}\AgdaSpace{}%
\AgdaBound{k}\AgdaSpace{}%
\AgdaOperator{\AgdaInductiveConstructor{∷}}\AgdaSpace{}%
\AgdaInductiveConstructor{[]}\AgdaSymbol{)}\AgdaSpace{}%
\AgdaSymbol{=}%
\>[45]\AgdaInductiveConstructor{nothing}\<%
\\
\>[0]\AgdaFunction{executeStackCheckSig3Aux}\AgdaSpace{}%
\AgdaBound{msg}\AgdaSpace{}%
\AgdaSymbol{(}\AgdaBound{m}\AgdaSpace{}%
\AgdaOperator{\AgdaInductiveConstructor{∷}}\AgdaSpace{}%
\AgdaBound{k}\AgdaSpace{}%
\AgdaOperator{\AgdaInductiveConstructor{∷}}\AgdaSpace{}%
\AgdaBound{x}\AgdaSpace{}%
\AgdaOperator{\AgdaInductiveConstructor{∷}}\AgdaSpace{}%
\AgdaInductiveConstructor{[]}\AgdaSymbol{)}\AgdaSpace{}%
\AgdaSymbol{=}%
\>[49]\AgdaInductiveConstructor{nothing}\<%
\\
\>[0]\AgdaFunction{executeStackCheckSig3Aux}\AgdaSpace{}%
\AgdaBound{msg}\AgdaSpace{}%
\AgdaSymbol{(}\AgdaBound{m}\AgdaSpace{}%
\AgdaOperator{\AgdaInductiveConstructor{∷}}\AgdaSpace{}%
\AgdaBound{k}\AgdaSpace{}%
\AgdaOperator{\AgdaInductiveConstructor{∷}}\AgdaSpace{}%
\AgdaBound{x}\AgdaSpace{}%
\AgdaOperator{\AgdaInductiveConstructor{∷}}\AgdaSpace{}%
\AgdaBound{\,f\,}\AgdaSpace{}%
\AgdaOperator{\AgdaInductiveConstructor{∷}}\AgdaSpace{}%
\AgdaInductiveConstructor{[]}\AgdaSymbol{)}\AgdaSpace{}%
\AgdaSymbol{=}%
\>[53]\AgdaInductiveConstructor{nothing}\<%
\\
\>[0]\AgdaFunction{executeStackCheckSig3Aux}\AgdaSpace{}%
\AgdaBound{msg}\AgdaSpace{}%
\AgdaSymbol{(}\AgdaBound{m}\AgdaSpace{}%
\AgdaOperator{\AgdaInductiveConstructor{∷}}\AgdaSpace{}%
\AgdaBound{k}\AgdaSpace{}%
\AgdaOperator{\AgdaInductiveConstructor{∷}}\AgdaSpace{}%
\AgdaBound{x}\AgdaSpace{}%
\AgdaOperator{\AgdaInductiveConstructor{∷}}\AgdaSpace{}%
\AgdaBound{\,f\,}\AgdaSpace{}%
\AgdaOperator{\AgdaInductiveConstructor{∷}}\AgdaSpace{}%
\AgdaBound{l}\AgdaSpace{}%
\AgdaOperator{\AgdaInductiveConstructor{∷}}\AgdaSpace{}%
\AgdaInductiveConstructor{[]}\AgdaSymbol{)}\AgdaSpace{}%
\AgdaSymbol{=}%
\>[57]\AgdaInductiveConstructor{nothing}\<%
\\
\>[0]\AgdaFunction{executeStackCheckSig3Aux}\AgdaSpace{}%
\AgdaBound{msg}\AgdaSpace{}%
\AgdaSymbol{(}\AgdaBound{p1}\AgdaSpace{}%
\AgdaOperator{\AgdaInductiveConstructor{∷}}\AgdaSpace{}%
\AgdaBound{p2}\AgdaSpace{}%
\AgdaOperator{\AgdaInductiveConstructor{∷}}\AgdaSpace{}%
\AgdaBound{p3}\AgdaSpace{}%
\AgdaOperator{\AgdaInductiveConstructor{∷}}\AgdaSpace{}%
\AgdaBound{s1}\AgdaSpace{}%
\AgdaOperator{\AgdaInductiveConstructor{∷}}\AgdaSpace{}%
\AgdaBound{s2}\AgdaSpace{}%
\AgdaOperator{\AgdaInductiveConstructor{∷}}\AgdaSpace{}%
\AgdaBound{s3}\AgdaSpace{}%
\AgdaOperator{\AgdaInductiveConstructor{∷}}\AgdaSpace{}%
\AgdaBound{s}\AgdaSymbol{)}\AgdaSpace{}%
\AgdaSymbol{=}%
\>[66]\AgdaFunction{stackAuxFunction}\AgdaSpace{}%
\AgdaBound{s}\<%
\\
\>[0][@{}l@{\AgdaIndent{0}}]%
\>[7]\AgdaSymbol{((}\AgdaFunction{isSigned}\AgdaSpace{}%
\AgdaBound{msg}\AgdaSpace{}%
\AgdaBound{s1}\AgdaSpace{}%
\AgdaBound{p1}\AgdaSpace{}%
\AgdaSymbol{)}\AgdaSpace{}%
\AgdaOperator{\AgdaFunction{∧b}}\AgdaSpace{}%
\AgdaSymbol{(}\AgdaFunction{isSigned}\AgdaSpace{}%
\AgdaBound{msg}\AgdaSpace{}%
\AgdaBound{s2}\AgdaSpace{}%
\AgdaBound{p2}\AgdaSymbol{)}\AgdaSpace{}%
\AgdaOperator{\AgdaFunction{∧b}}\AgdaSpace{}%
\AgdaSymbol{(}\AgdaFunction{isSigned}\AgdaSpace{}%
\AgdaBound{msg}\AgdaSpace{}%
\AgdaBound{s3}\AgdaSpace{}%
\AgdaBound{p3}\AgdaSymbol{))}\<%
\\
\\[\AgdaEmptyExtraSkip]%
\\[\AgdaEmptyExtraSkip]%
\\[\AgdaEmptyExtraSkip]%
\>[0]\AgdaKeyword{mutual}\<%
\\
\>[0]\AgdaComment{--\textbackslash{}semanticBasicOperationsForTypeSettingcompareSigsMultiSig}\<%
\end{code}
} 

\newcommand{\semanticBasicOperationsForTypeSettingcompareSigsMultiSig}{
\begin{code}%
\>[0][@{}l@{\AgdaIndent{1}}]%
\>[1]\AgdaFunction{cmpMultiSigs}\AgdaSpace{}%
\AgdaSymbol{:}\AgdaSpace{}%
\AgdaSymbol{(}\AgdaBound{msg}\AgdaSpace{}%
\AgdaSymbol{:}\AgdaSpace{}%
\AgdaDatatype{Msg}\AgdaSymbol{)(}\AgdaBound{sigs}\AgdaSpace{}%
\AgdaBound{pbks}\AgdaSpace{}%
\AgdaSymbol{:}\AgdaSpace{}%
\AgdaDatatype{List}\AgdaSpace{}%
\AgdaDatatype{ℕ}\AgdaSymbol{)}%
\>[49]\AgdaSymbol{→}\AgdaSpace{}%
\AgdaDatatype{Bool}\<%
\\
\>[1]\AgdaFunction{cmpMultiSigs}\AgdaSpace{}%
\AgdaBound{msg}%
\>[19]\AgdaInductiveConstructor{[]}%
\>[36]\AgdaBound{pubkeys}%
\>[47]\AgdaSymbol{=}%
\>[50]\AgdaInductiveConstructor{true}\<%
\\
\>[1]\AgdaFunction{cmpMultiSigs}\AgdaSpace{}%
\AgdaBound{msg}%
\>[19]\AgdaSymbol{(}\AgdaBound{sig}\AgdaSpace{}%
\AgdaOperator{\AgdaInductiveConstructor{∷}}\AgdaSpace{}%
\AgdaBound{sigs}\AgdaSymbol{)}%
\>[33]\AgdaInductiveConstructor{[]}%
\>[47]\AgdaSymbol{=}%
\>[50]\AgdaInductiveConstructor{false}\<%
\\
\>[1]\AgdaFunction{cmpMultiSigs}\AgdaSpace{}%
\AgdaBound{msg}%
\>[19]\AgdaSymbol{(}\AgdaBound{sig}\AgdaSpace{}%
\AgdaOperator{\AgdaInductiveConstructor{∷}}\AgdaSpace{}%
\AgdaBound{sigs}\AgdaSymbol{)}%
\>[33]\AgdaSymbol{(}\AgdaBound{pbk}\AgdaSpace{}%
\AgdaOperator{\AgdaInductiveConstructor{∷}}\AgdaSpace{}%
\AgdaBound{pbks}\AgdaSymbol{)}%
\>[47]\AgdaSymbol{=}\AgdaSpace{}%
\AgdaFunction{cmpMultiSigsAux}\AgdaSpace{}%
\AgdaBound{msg}\AgdaSpace{}%
\AgdaBound{sigs}\AgdaSpace{}%
\AgdaBound{pbks}\AgdaSpace{}%
\AgdaBound{sig}\AgdaSpace{}%
\AgdaSymbol{(}\AgdaFunction{isSigned}\AgdaSpace{}%
\AgdaBound{msg}\AgdaSpace{}%
\AgdaBound{sig}\AgdaSpace{}%
\AgdaBound{pbk}\AgdaSymbol{)}\<%
\\
\\[\AgdaEmptyExtraSkip]%
\>[1]\AgdaFunction{cmpMultiSigsAux}\AgdaSpace{}%
\AgdaSymbol{:}\AgdaSpace{}%
\AgdaSymbol{(}\AgdaBound{msg}\AgdaSpace{}%
\AgdaSymbol{:}\AgdaSpace{}%
\AgdaDatatype{Msg}\AgdaSymbol{)(}\AgdaBound{sigs}\AgdaSpace{}%
\AgdaBound{pbks}\AgdaSpace{}%
\AgdaSymbol{:}\AgdaSpace{}%
\AgdaDatatype{List}\AgdaSpace{}%
\AgdaDatatype{ℕ}\AgdaSymbol{)(}\AgdaBound{sig}\AgdaSpace{}%
\AgdaSymbol{:}\AgdaSpace{}%
\AgdaDatatype{ℕ}\AgdaSymbol{)(}\AgdaBound{testRes}\AgdaSpace{}%
\AgdaSymbol{:}\AgdaSpace{}%
\AgdaDatatype{Bool}\AgdaSymbol{)}\AgdaSpace{}%
\AgdaSymbol{→}\AgdaSpace{}%
\AgdaDatatype{Bool}\<%
\\
\>[1]\AgdaFunction{cmpMultiSigsAux}%
\>[18]\AgdaBound{msg}%
\>[23]\AgdaBound{sigs}%
\>[29]\AgdaBound{pbks}%
\>[35]\AgdaBound{sig}%
\>[40]\AgdaInductiveConstructor{false}%
\>[47]\AgdaSymbol{=}%
\>[50]\AgdaFunction{cmpMultiSigs}%
\>[64]\AgdaBound{msg}%
\>[69]\AgdaSymbol{(}\AgdaBound{sig}\AgdaSpace{}%
\AgdaOperator{\AgdaInductiveConstructor{∷}}\AgdaSpace{}%
\AgdaBound{sigs}\AgdaSymbol{)}%
\>[83]\AgdaBound{pbks}\<%
\\
\>[1]\AgdaFunction{cmpMultiSigsAux}%
\>[18]\AgdaBound{msg}%
\>[23]\AgdaBound{sigs}%
\>[29]\AgdaBound{pbks}%
\>[35]\AgdaBound{sig}%
\>[40]\AgdaInductiveConstructor{true}%
\>[47]\AgdaSymbol{=}%
\>[50]\AgdaFunction{cmpMultiSigs}%
\>[64]\AgdaBound{msg}%
\>[69]\AgdaBound{sigs}%
\>[86]\AgdaBound{pbks}\<%
\end{code}
} 

\AgdaHide{
\begin{code}%
\>[0]\<%
\\
\>[0]\AgdaComment{--\textbackslash{}semanticBasicOperationsForTypeSettingexecuteMultiSigThree}\<%
\end{code}
} 

\newcommand{\semanticBasicOperationsForTypeSettingexecuteMultiSigThree}{
\begin{code}%
\>[0]\AgdaFunction{executeMultiSig3}\AgdaSpace{}%
\AgdaSymbol{:}\AgdaSpace{}%
\AgdaSymbol{(}\AgdaBound{msg}\AgdaSpace{}%
\AgdaSymbol{:}\AgdaSpace{}%
\AgdaDatatype{Msg}\AgdaSymbol{)(}\AgdaBound{pbks}\AgdaSpace{}%
\AgdaSymbol{:}\AgdaSpace{}%
\AgdaDatatype{List}\AgdaSpace{}%
\AgdaDatatype{ℕ}\AgdaSymbol{)(}\AgdaBound{numSigs}\AgdaSpace{}%
\AgdaSymbol{:}\AgdaSpace{}%
\AgdaDatatype{ℕ}\AgdaSymbol{)(}\AgdaBound{st}\AgdaSpace{}%
\AgdaSymbol{:}\AgdaSpace{}%
\AgdaFunction{Stack}\AgdaSymbol{)(}\AgdaBound{sigs}\AgdaSpace{}%
\AgdaSymbol{:}\AgdaSpace{}%
\AgdaDatatype{List}\AgdaSpace{}%
\AgdaDatatype{ℕ}\AgdaSymbol{)}\AgdaSpace{}%
\AgdaSymbol{→}\AgdaSpace{}%
\AgdaDatatype{Maybe}\AgdaSpace{}%
\AgdaFunction{Stack}\<%
\\
\>[0]\AgdaFunction{executeMultiSig3}%
\>[18]\AgdaBound{msg}%
\>[23]\AgdaBound{pbks}%
\>[29]\AgdaSymbol{\AgdaUnderscore{}}%
\>[43]\AgdaInductiveConstructor{[]}\AgdaSpace{}%
\AgdaBound{sigs}%
\>[65]\AgdaSymbol{=}%
\>[68]\AgdaInductiveConstructor{nothing}\<%
\\
\>[0]\AgdaFunction{executeMultiSig3}%
\>[18]\AgdaBound{msg}%
\>[23]\AgdaBound{pbks}%
\>[29]\AgdaInductiveConstructor{zero}%
\>[43]\AgdaSymbol{(}\AgdaBound{x}\AgdaSpace{}%
\AgdaOperator{\AgdaInductiveConstructor{∷}}\AgdaSpace{}%
\AgdaBound{restStack}\AgdaSymbol{)}\AgdaSpace{}%
\AgdaBound{sigs}\<%
\\
\>[18]\AgdaSymbol{=}%
\>[21]\AgdaInductiveConstructor{just}\AgdaSpace{}%
\AgdaSymbol{(}\AgdaFunction{boolToNat}\AgdaSpace{}%
\AgdaSymbol{(}\AgdaFunction{cmpMultiSigs}\AgdaSpace{}%
\AgdaBound{msg}\AgdaSpace{}%
\AgdaBound{sigs}\AgdaSpace{}%
\AgdaBound{pbks}\AgdaSymbol{)}\AgdaSpace{}%
\AgdaOperator{\AgdaInductiveConstructor{∷}}\AgdaSpace{}%
\AgdaBound{restStack}\AgdaSymbol{)}\<%
\\
\>[0]\AgdaFunction{executeMultiSig3}%
\>[18]\AgdaBound{msg}%
\>[23]\AgdaBound{pbks}%
\>[29]\AgdaSymbol{(}\AgdaInductiveConstructor{suc}\AgdaSpace{}%
\AgdaBound{numSigs}\AgdaSymbol{)}\AgdaSpace{}%
\AgdaSymbol{(}\AgdaBound{sig}\AgdaSpace{}%
\AgdaOperator{\AgdaInductiveConstructor{∷}}\AgdaSpace{}%
\AgdaBound{rest}\AgdaSymbol{)}\AgdaSpace{}%
\AgdaBound{sigs}\<%
\\
\>[18]\AgdaSymbol{=}%
\>[21]\AgdaFunction{executeMultiSig3}\AgdaSpace{}%
\AgdaBound{msg}\AgdaSpace{}%
\AgdaBound{pbks}\AgdaSpace{}%
\AgdaBound{numSigs}\AgdaSpace{}%
\AgdaBound{rest}\AgdaSpace{}%
\AgdaSymbol{(}\AgdaBound{sig}\AgdaSpace{}%
\AgdaOperator{\AgdaInductiveConstructor{∷}}\AgdaSpace{}%
\AgdaBound{sigs}\AgdaSymbol{)}\<%
\end{code}
} 

\AgdaHide{
\begin{code}%
\>[0]\<%
\\
\>[0]\AgdaFunction{coreExecuteMultiSigThree}\AgdaSpace{}%
\AgdaSymbol{:}\AgdaSpace{}%
\AgdaSymbol{(}\AgdaBound{msg}\AgdaSpace{}%
\AgdaSymbol{:}\AgdaSpace{}%
\AgdaDatatype{Msg}\AgdaSymbol{)(}\AgdaBound{sigs}\AgdaSpace{}%
\AgdaSymbol{:}\AgdaSpace{}%
\AgdaDatatype{List}\AgdaSpace{}%
\AgdaDatatype{ℕ}\AgdaSymbol{)(}\AgdaBound{pbks}\AgdaSpace{}%
\AgdaSymbol{:}\AgdaSpace{}%
\AgdaDatatype{List}\AgdaSpace{}%
\AgdaDatatype{ℕ}\AgdaSymbol{)(}\AgdaBound{restStack}\AgdaSpace{}%
\AgdaSymbol{:}\AgdaSpace{}%
\AgdaFunction{Stack}\AgdaSymbol{)}\AgdaSpace{}%
\AgdaSymbol{→}\AgdaSpace{}%
\AgdaDatatype{Maybe}\AgdaSpace{}%
\AgdaFunction{Stack}\<%
\\
\>[0]\AgdaFunction{coreExecuteMultiSigThree}\AgdaSpace{}%
\AgdaBound{msg}\AgdaSpace{}%
\AgdaBound{sigs}\AgdaSpace{}%
\AgdaBound{pbks}\AgdaSpace{}%
\AgdaBound{restStack}\AgdaSpace{}%
\AgdaSymbol{=}\<%
\\
\>[0]\AgdaComment{--\textbackslash{}semanticBasicOperationsForTypeSettingcoreExecuteMultiSigThree}\<%
\end{code}
} 

\newcommand{\semanticBasicOperationsForTypeSettingcoreExecuteMultiSigThree}{
\begin{code}[inline]%
\>[0][@{}l@{\AgdaIndent{1}}]%
\>[2]\AgdaInductiveConstructor{just}\AgdaSpace{}%
\AgdaSymbol{(}\AgdaFunction{boolToNat}\AgdaSpace{}%
\AgdaSymbol{(}\AgdaFunction{cmpMultiSigs}\AgdaSpace{}%
\AgdaBound{msg}\AgdaSpace{}%
\AgdaBound{sigs}\AgdaSpace{}%
\AgdaBound{pbks}\AgdaSymbol{)}\AgdaSpace{}%
\AgdaOperator{\AgdaInductiveConstructor{∷}}\AgdaSpace{}%
\AgdaBound{restStack}\AgdaSymbol{)}\<%
\end{code}
} 

\AgdaHide{
\begin{code}%
\>[0]\<%
\\
\>[0]\AgdaComment{--\textbackslash{}semanticBasicOperationsForTypeSettingexecuteMultiSigTwo}\<%
\end{code}
} 

\newcommand{\semanticBasicOperationsForTypeSettingexecuteMultiSigTwo}{
\begin{code}%
\>[0]\AgdaFunction{executeMultiSig2}\AgdaSpace{}%
\AgdaSymbol{:}\AgdaSpace{}%
\AgdaSymbol{(}\AgdaBound{msg}\AgdaSpace{}%
\AgdaSymbol{:}\AgdaSpace{}%
\AgdaDatatype{Msg}\AgdaSymbol{)(}\AgdaBound{numPbks}\AgdaSpace{}%
\AgdaSymbol{:}\AgdaSpace{}%
\AgdaDatatype{ℕ}\AgdaSymbol{)(}\AgdaBound{st}\AgdaSpace{}%
\AgdaSymbol{:}%
\>[51]\AgdaFunction{Stack}\AgdaSymbol{)(}\AgdaBound{pbks}\AgdaSpace{}%
\AgdaSymbol{:}\AgdaSpace{}%
\AgdaDatatype{List}\AgdaSpace{}%
\AgdaDatatype{ℕ}\AgdaSymbol{)}\AgdaSpace{}%
\AgdaSymbol{→}\AgdaSpace{}%
\AgdaDatatype{Maybe}\AgdaSpace{}%
\AgdaFunction{Stack}\<%
\\
\>[0]\AgdaFunction{executeMultiSig2}%
\>[18]\AgdaBound{msg}%
\>[23]\AgdaSymbol{\AgdaUnderscore{}}%
\>[32]\AgdaInductiveConstructor{[]}%
\>[50]\AgdaBound{pbks}%
\>[56]\AgdaSymbol{=}%
\>[59]\AgdaInductiveConstructor{nothing}\<%
\\
\>[0]\AgdaFunction{executeMultiSig2}%
\>[18]\AgdaBound{msg}%
\>[23]\AgdaInductiveConstructor{zero}%
\>[32]\AgdaSymbol{(}\AgdaBound{numSigs}\AgdaSpace{}%
\AgdaOperator{\AgdaInductiveConstructor{∷}}\AgdaSpace{}%
\AgdaBound{rest}\AgdaSymbol{)}%
\>[50]\AgdaBound{pbks}%
\>[56]\AgdaSymbol{=}%
\>[59]\AgdaFunction{executeMultiSig3}\AgdaSpace{}%
\AgdaBound{msg}\AgdaSpace{}%
\AgdaBound{pbks}\AgdaSpace{}%
\AgdaBound{numSigs}\AgdaSpace{}%
\AgdaBound{rest}\AgdaSpace{}%
\AgdaInductiveConstructor{[]}\<%
\\
\>[0]\AgdaFunction{executeMultiSig2}%
\>[18]\AgdaBound{msg}%
\>[23]\AgdaSymbol{(}\AgdaInductiveConstructor{suc}\AgdaSpace{}%
\AgdaBound{n}\AgdaSymbol{)}%
\>[32]\AgdaSymbol{(}\AgdaBound{pbk}\AgdaSpace{}%
\AgdaOperator{\AgdaInductiveConstructor{∷}}\AgdaSpace{}%
\AgdaBound{rest}\AgdaSymbol{)}%
\>[50]\AgdaBound{pbks}%
\>[56]\AgdaSymbol{=}%
\>[59]\AgdaFunction{executeMultiSig2}\AgdaSpace{}%
\AgdaBound{msg}\AgdaSpace{}%
\AgdaBound{n}\AgdaSpace{}%
\AgdaBound{rest}\AgdaSpace{}%
\AgdaSymbol{(}\AgdaBound{pbk}\AgdaSpace{}%
\AgdaOperator{\AgdaInductiveConstructor{∷}}\AgdaSpace{}%
\AgdaBound{pbks}\AgdaSymbol{)}\<%
\end{code}
} 

\AgdaHide{
\begin{code}%
\>[0]\AgdaComment{--open\ import\ addition10}\<%
\\
\\[\AgdaEmptyExtraSkip]%
\\[\AgdaEmptyExtraSkip]%
\>[0]\AgdaComment{--\ Add\ the\ notes\ to\ a\ file\ \ notes\ in\ notes}\<%
\\
\>[0]\AgdaComment{--\ next\ steps}\<%
\\
\>[0]\AgdaComment{--\ move\ executeMultiSig\ to\ semanticsInstructions.agda}\<%
\\
\>[0]\AgdaComment{--\ add\ opMultiSig\ to\ the\ instructions}\<%
\\
\>[0]\AgdaComment{--\ define\ ⟦\ OpMultiSig\ ⟧s\ =\ liftMsgStackToStateTransformerDepIfStack'\ executeMultiSig}\<%
\\
\>[0]\AgdaComment{--\ in\ semanticsInstructions}\<%
\\
\>[0]\AgdaComment{--\ define\ in\ addition4.agda}\<%
\\
\>[0]\AgdaComment{--\ \ \ ⟦\ OpMultiSig\ \ \ ⟧stacks\ time₁\ msg\ =\ \ executeMultiSig\ msg}\<%
\\
\\[\AgdaEmptyExtraSkip]%
\\[\AgdaEmptyExtraSkip]%
\>[0]\AgdaComment{--\ Try\ out\ evaluatings\ scripts\ as\ in\ scriptInterpreter.agda\ \ whether\ it\ works\ as\ intended}\<%
\\
\>[0]\AgdaComment{--\ \ \ write\ some\ multisignature\ script\ and\ a\ scriptsig\ \ and\ check\ whether\ they\ work\ correctly}\<%
\\
\\[\AgdaEmptyExtraSkip]%
\>[0]\AgdaComment{--\ Try\ out\ experiments\ like\ stackfunP2PKH\ \ \ in\ additionAntonExperiments.agda}\<%
\\
\>[0]\AgdaComment{--\ for}\<%
\\
\>[0]\AgdaComment{--\ ⟦\ (opPush\ 2)\ ∷\ (opPush\ pbk1)\ ∷\ \ (opPush\ pbk2)\ ∷\ \ (opPush\ pbk3)\ ∷\ \ (opPush\ 3)\ ∷\ \ opMultiSig\ ∷\ \ []\ \ \ \ \ ⟧stack\ time\ msg\ stack}\<%
\\
\>[0]\AgdaComment{--\ whether\ this\ is\ equal\ to}\<%
\\
\>[0]\AgdaComment{--\ executeMultiSig3\ msg\ (pbk1\ ∷\ pbk2\ ∷\ pbk3\ ∷\ [])\ 2\ stack\ []}\<%
\\
\>[0]\AgdaComment{--\ similarly\ as\ in\ lemma1\ by\ postulating\ \ \ pbk1,..,pbk3,time,msg,stack}\<%
\\
\\[\AgdaEmptyExtraSkip]%
\>[0]\AgdaComment{--\ if\ yes\ then\ the}\<%
\\
\>[0]\AgdaComment{--\ weakest\ precondtion\ for\ the\ multisig\ script\ would\ be:}\<%
\\
\>[0]\AgdaComment{--\ stack\ needs\ to\ have\ size\ at\ least\ 2\ and\ the\ for\ signatures\ sig1\ sig2\ on\ the\ stack\ we\ have}\<%
\\
\>[0]\AgdaComment{--\ cmpMultiSigs\ \ msg\ (sig1\ ∷\ [\ sig2\ ])\ (pbk1\ ∷\ pbk2\ ∷\ [\ pbk3\ ]\ )\ =\ true}\<%
\\
\\[\AgdaEmptyExtraSkip]%
\>[0]\AgdaComment{--\textbackslash{}semanticBasicOperationsForTypeSettingcheckpubk}\<%
\end{code}
} 

\newcommand{\semanticBasicOperationsForTypeSettingexecuteMultiSig}{
\begin{code}%
\>[0]\AgdaFunction{executeMultiSig}\AgdaSpace{}%
\AgdaSymbol{:}\AgdaSpace{}%
\AgdaDatatype{Msg}\AgdaSpace{}%
\AgdaSymbol{→}%
\>[25]\AgdaFunction{Stack}\AgdaSpace{}%
\AgdaSymbol{→}%
\>[34]\AgdaDatatype{Maybe}\AgdaSpace{}%
\AgdaFunction{Stack}\<%
\\
\>[0]\AgdaFunction{executeMultiSig}\AgdaSpace{}%
\AgdaBound{msg}\AgdaSpace{}%
\AgdaInductiveConstructor{[]}\AgdaSpace{}%
\AgdaSymbol{=}\AgdaSpace{}%
\AgdaInductiveConstructor{nothing}\<%
\\
\>[0]\AgdaFunction{executeMultiSig}\AgdaSpace{}%
\AgdaBound{msg}\AgdaSpace{}%
\AgdaSymbol{(}\AgdaBound{numberOfPbks}\AgdaSpace{}%
\AgdaOperator{\AgdaInductiveConstructor{∷}}\AgdaSpace{}%
\AgdaBound{st}\AgdaSymbol{)}\AgdaSpace{}%
\AgdaSymbol{=}\AgdaSpace{}%
\AgdaFunction{executeMultiSig2}\AgdaSpace{}%
\AgdaBound{msg}\AgdaSpace{}%
\AgdaBound{numberOfPbks}\AgdaSpace{}%
\AgdaBound{st}\AgdaSpace{}%
\AgdaInductiveConstructor{[]}\<%
\end{code}
} 

\AgdaHide{
\begin{code}%
\>[0]\<%
\end{code}
} 


\AgdaHide{
\begin{code}%
\>[0]\AgdaKeyword{open}\AgdaSpace{}%
\AgdaKeyword{import}\AgdaSpace{}%
\AgdaModule{basicBitcoinDataType}\<%
\\
\\[\AgdaEmptyExtraSkip]%
\>[0]\AgdaKeyword{module}\AgdaSpace{}%
\AgdaModule{verificationStackScripts.stackSemanticsInstructionsBasic}\AgdaSpace{}%
\AgdaSymbol{(}\AgdaBound{param}\AgdaSpace{}%
\AgdaSymbol{:}\AgdaSpace{}%
\AgdaRecord{GlobalParameters}\AgdaSymbol{)}%
\>[92]\AgdaKeyword{where}\<%
\\
\\[\AgdaEmptyExtraSkip]%
\>[0]\AgdaKeyword{open}\AgdaSpace{}%
\AgdaKeyword{import}\AgdaSpace{}%
\AgdaModule{Data.Nat}%
\>[22]\AgdaKeyword{hiding}\AgdaSpace{}%
\AgdaSymbol{(}\AgdaOperator{\AgdaDatatype{\AgdaUnderscore{}≤\AgdaUnderscore{}}}\AgdaSymbol{)}\<%
\\
\>[0]\AgdaKeyword{open}\AgdaSpace{}%
\AgdaKeyword{import}\AgdaSpace{}%
\AgdaModule{Data.List}\AgdaSpace{}%
\AgdaKeyword{hiding}\AgdaSpace{}%
\AgdaSymbol{(}\AgdaOperator{\AgdaFunction{\AgdaUnderscore{}\ensuremath{+\!\!+}\AgdaUnderscore{}}}\AgdaSymbol{)}\<%
\\
\>[0]\AgdaKeyword{open}\AgdaSpace{}%
\AgdaKeyword{import}\AgdaSpace{}%
\AgdaModule{Data.Unit}%
\>[23]\AgdaKeyword{hiding}\AgdaSpace{}%
\AgdaSymbol{(}\AgdaOperator{\AgdaRecord{\AgdaUnderscore{}≤\AgdaUnderscore{}}}\AgdaSymbol{)}\<%
\\
\>[0]\AgdaKeyword{open}\AgdaSpace{}%
\AgdaKeyword{import}\AgdaSpace{}%
\AgdaModule{Data.Empty}\<%
\\
\>[0]\AgdaKeyword{open}\AgdaSpace{}%
\AgdaKeyword{import}\AgdaSpace{}%
\AgdaModule{Data.Bool}%
\>[23]\AgdaKeyword{hiding}\AgdaSpace{}%
\AgdaSymbol{(}\AgdaOperator{\AgdaDatatype{\AgdaUnderscore{}≤\AgdaUnderscore{}}}\AgdaSpace{}%
\AgdaSymbol{;}\AgdaSpace{}%
\AgdaOperator{\AgdaFunction{if\AgdaUnderscore{}then\AgdaUnderscore{}else\AgdaUnderscore{}}}\AgdaSpace{}%
\AgdaSymbol{)}\AgdaSpace{}%
\AgdaKeyword{renaming}\AgdaSpace{}%
\AgdaSymbol{(}\AgdaOperator{\AgdaFunction{\AgdaUnderscore{}∧\AgdaUnderscore{}}}\AgdaSpace{}%
\AgdaSymbol{to}\AgdaSpace{}%
\AgdaOperator{\AgdaFunction{\AgdaUnderscore{}∧b\AgdaUnderscore{}}}\AgdaSpace{}%
\AgdaSymbol{;}\AgdaSpace{}%
\AgdaOperator{\AgdaFunction{\AgdaUnderscore{}∨\AgdaUnderscore{}}}\AgdaSpace{}%
\AgdaSymbol{to}\AgdaSpace{}%
\AgdaOperator{\AgdaFunction{\AgdaUnderscore{}∨b\AgdaUnderscore{}}}\AgdaSpace{}%
\AgdaSymbol{;}\AgdaSpace{}%
\AgdaFunction{T}\AgdaSpace{}%
\AgdaSymbol{to}\AgdaSpace{}%
\AgdaFunction{True}\AgdaSymbol{)}\<%
\\
\>[0]\AgdaKeyword{open}\AgdaSpace{}%
\AgdaKeyword{import}\AgdaSpace{}%
\AgdaModule{Data.Product}\AgdaSpace{}%
\AgdaKeyword{renaming}\AgdaSpace{}%
\AgdaSymbol{(}\AgdaOperator{\AgdaInductiveConstructor{\AgdaUnderscore{},\AgdaUnderscore{}}}\AgdaSpace{}%
\AgdaSymbol{to}\AgdaSpace{}%
\AgdaOperator{\AgdaInductiveConstructor{\AgdaUnderscore{},,\AgdaUnderscore{}}}\AgdaSpace{}%
\AgdaSymbol{)}\<%
\\
\>[0]\AgdaKeyword{open}\AgdaSpace{}%
\AgdaKeyword{import}\AgdaSpace{}%
\AgdaModule{Data.Nat.Base}\AgdaSpace{}%
\AgdaKeyword{hiding}\AgdaSpace{}%
\AgdaSymbol{(}\AgdaOperator{\AgdaDatatype{\AgdaUnderscore{}≤\AgdaUnderscore{}}}\AgdaSymbol{)}\<%
\\
\>[0]\AgdaKeyword{open}\AgdaSpace{}%
\AgdaKeyword{import}\AgdaSpace{}%
\AgdaModule{Data.Maybe}\<%
\\
\>[0]\AgdaKeyword{import}\AgdaSpace{}%
\AgdaModule{Relation.Binary.PropositionalEquality}\AgdaSpace{}%
\AgdaSymbol{as}\AgdaSpace{}%
\AgdaModule{Eq}\<%
\\
\>[0]\AgdaKeyword{open}\AgdaSpace{}%
\AgdaModule{Eq}\AgdaSpace{}%
\AgdaKeyword{using}\AgdaSpace{}%
\AgdaSymbol{(}\AgdaOperator{\AgdaDatatype{\AgdaUnderscore{}≡\AgdaUnderscore{}}}\AgdaSymbol{;}\AgdaSpace{}%
\AgdaInductiveConstructor{refl}\AgdaSymbol{;}\AgdaSpace{}%
\AgdaFunction{cong}\AgdaSymbol{;}\AgdaSpace{}%
\AgdaKeyword{module}\AgdaSpace{}%
\AgdaModule{≡-Reasoning}\AgdaSymbol{;}\AgdaSpace{}%
\AgdaFunction{sym}\AgdaSymbol{)}\<%
\\
\>[0]\AgdaKeyword{open}\AgdaSpace{}%
\AgdaModule{≡-Reasoning}\<%
\\
\>[0]\AgdaKeyword{open}\AgdaSpace{}%
\AgdaKeyword{import}\AgdaSpace{}%
\AgdaModule{Agda.Builtin.Equality}\<%
\\
\\[\AgdaEmptyExtraSkip]%
\\[\AgdaEmptyExtraSkip]%
\>[0]\AgdaComment{--our\ libraries}\<%
\\
\>[0]\AgdaKeyword{open}\AgdaSpace{}%
\AgdaKeyword{import}\AgdaSpace{}%
\AgdaModule{libraries.listLib}\<%
\\
\>[0]\AgdaKeyword{open}\AgdaSpace{}%
\AgdaKeyword{import}\AgdaSpace{}%
\AgdaModule{libraries.natLib}\<%
\\
\>[0]\AgdaKeyword{open}\AgdaSpace{}%
\AgdaKeyword{import}\AgdaSpace{}%
\AgdaModule{libraries.boolLib}\<%
\\
\>[0]\AgdaKeyword{open}\AgdaSpace{}%
\AgdaKeyword{import}\AgdaSpace{}%
\AgdaModule{libraries.andLib}\<%
\\
\>[0]\AgdaKeyword{open}\AgdaSpace{}%
\AgdaKeyword{import}\AgdaSpace{}%
\AgdaModule{libraries.miscLib}\<%
\\
\>[0]\AgdaKeyword{open}\AgdaSpace{}%
\AgdaKeyword{import}\AgdaSpace{}%
\AgdaModule{libraries.maybeLib}\<%
\\
\\[\AgdaEmptyExtraSkip]%
\\[\AgdaEmptyExtraSkip]%
\>[0]\AgdaKeyword{open}\AgdaSpace{}%
\AgdaKeyword{import}\AgdaSpace{}%
\AgdaModule{stack}\<%
\\
\>[0]\AgdaKeyword{open}\AgdaSpace{}%
\AgdaKeyword{import}\AgdaSpace{}%
\AgdaModule{semanticBasicOperations}\AgdaSpace{}%
\AgdaBound{param}\<%
\\
\>[0]\AgdaKeyword{open}\AgdaSpace{}%
\AgdaKeyword{import}\AgdaSpace{}%
\AgdaModule{instructionBasic}\<%
\\
\\[\AgdaEmptyExtraSkip]%
\>[0]\AgdaKeyword{open}\AgdaSpace{}%
\AgdaKeyword{import}\AgdaSpace{}%
\AgdaModule{verificationStackScripts.stackState}\<%
\\
\>[0]\AgdaKeyword{open}\AgdaSpace{}%
\AgdaKeyword{import}\AgdaSpace{}%
\AgdaModule{verificationStackScripts.sPredicate}\<%
\\
\>[0]\AgdaKeyword{open}\AgdaSpace{}%
\AgdaKeyword{import}\AgdaSpace{}%
\AgdaModule{verificationStackScripts.stackInstruction}\<%
\\
\\[\AgdaEmptyExtraSkip]%
\\[\AgdaEmptyExtraSkip]%
\\[\AgdaEmptyExtraSkip]%
\>[0]\AgdaComment{--\ The\ stack\ operation\ of\ type}\<%
\\
\>[0]\AgdaComment{--\ \ \ Time\ →\ Msg\ →\ \ Stack\ →\ Maybe\ Stack}\<%
\\
\>[0]\AgdaComment{--\ on\ which\ the\ semantics\ is\ based}\<%
\\
\>[0]\AgdaComment{--\ we\ have\ (see\ next\ lemma)\ for\ nonif\ instructions}\<%
\\
\>[0]\AgdaComment{--\ ⟦\ op\ ⟧s\ \ ≡\ stackTransform2StateTransform\ ⟦\ op\ ⟧sˢ\ \ ⟧ₛˢ}\<%
\\
\\[\AgdaEmptyExtraSkip]%
\>[0]\AgdaComment{--\ The\ stacks\ operations\ is\ only\ corret\ for\ non-if\ instructions}\<%
\\
\>[0]\AgdaComment{--\ original\ name\ was\ \ \ executeStackPartOfInstr}\<%
\\
\\[\AgdaEmptyExtraSkip]%
\>[0]\AgdaComment{--\ \textbackslash{}stackSemanticsInstructionsBasicsemanticsstype}\<%
\end{code}
} 

\newcommand{\stackSemanticsInstructionsBasicsemanticsstype}{
\begin{code}%
\>[0]\AgdaOperator{\AgdaFunction{⟦\AgdaUnderscore{}⟧ₛˢ}}\AgdaSpace{}%
\AgdaSymbol{:}\AgdaSpace{}%
\AgdaDatatype{InstructionBasic}\AgdaSpace{}%
\AgdaSymbol{→}%
\>[28]\AgdaFunction{Time}\AgdaSpace{}%
\AgdaSymbol{→}\AgdaSpace{}%
\AgdaDatatype{Msg}\AgdaSpace{}%
\AgdaSymbol{→}%
\>[42]\AgdaFunction{Stack}\AgdaSpace{}%
\AgdaSymbol{→}\AgdaSpace{}%
\AgdaDatatype{Maybe}\AgdaSpace{}%
\AgdaFunction{Stack}\<%
\\
\>[0]\AgdaOperator{\AgdaFunction{⟦}}\AgdaSpace{}%
\AgdaInductiveConstructor{opEqual}%
\>[11]\AgdaOperator{\AgdaFunction{⟧ₛˢ}}\AgdaSpace{}%
\AgdaBound{time₁}\AgdaSpace{}%
\AgdaBound{msg}\AgdaSpace{}%
\AgdaSymbol{=}\AgdaSpace{}%
\AgdaFunction{executeStackEquality}\<%
\end{code}
} 

\AgdaHide{
\begin{code}%
\>[0]\AgdaOperator{\AgdaFunction{⟦}}\AgdaSpace{}%
\AgdaInductiveConstructor{opAdd}%
\>[11]\AgdaOperator{\AgdaFunction{⟧ₛˢ}}\AgdaSpace{}%
\AgdaBound{time₁}\AgdaSpace{}%
\AgdaBound{msg}\AgdaSpace{}%
\AgdaSymbol{=}\AgdaSpace{}%
\AgdaFunction{executeStackAdd}\<%
\\
\>[0]\AgdaOperator{\AgdaFunction{⟦}}\AgdaSpace{}%
\AgdaInductiveConstructor{opPush}\AgdaSpace{}%
\AgdaBound{x}\AgdaSpace{}%
\AgdaOperator{\AgdaFunction{⟧ₛˢ}}\AgdaSpace{}%
\AgdaBound{time₁}\AgdaSpace{}%
\AgdaBound{msg}\AgdaSpace{}%
\AgdaSymbol{=}\AgdaSpace{}%
\AgdaFunction{executeStackPush}\AgdaSpace{}%
\AgdaBound{x}\<%
\\
\>[0]\AgdaOperator{\AgdaFunction{⟦}}\AgdaSpace{}%
\AgdaInductiveConstructor{opSub}%
\>[11]\AgdaOperator{\AgdaFunction{⟧ₛˢ}}\AgdaSpace{}%
\AgdaBound{time₁}\AgdaSpace{}%
\AgdaBound{msg}\AgdaSpace{}%
\AgdaSymbol{=}\AgdaSpace{}%
\AgdaFunction{executeStackSub}\<%
\\
\>[0]\AgdaOperator{\AgdaFunction{⟦}}\AgdaSpace{}%
\AgdaInductiveConstructor{opVerify}\AgdaSpace{}%
\AgdaOperator{\AgdaFunction{⟧ₛˢ}}\AgdaSpace{}%
\AgdaBound{time₁}\AgdaSpace{}%
\AgdaBound{msg}\AgdaSpace{}%
\AgdaSymbol{=}\AgdaSpace{}%
\AgdaFunction{executeStackVerify}\<%
\\
\>[0]\AgdaComment{--\ \textbackslash{}stackSemanticsInstructionsBasicsemanticss}\<%
\end{code}
} 

\newcommand{\stackSemanticsInstructionsBasicsemanticss}{
\begin{code}%
\>[0]\AgdaOperator{\AgdaFunction{⟦}}\AgdaSpace{}%
\AgdaInductiveConstructor{opCheckSig}\AgdaSpace{}%
\AgdaOperator{\AgdaFunction{⟧ₛˢ}}\AgdaSpace{}%
\AgdaBound{time₁}\AgdaSpace{}%
\AgdaBound{msg}\AgdaSpace{}%
\AgdaSymbol{=}\AgdaSpace{}%
\AgdaFunction{executeStackCheckSig}\AgdaSpace{}%
\AgdaBound{msg}\<%
\end{code}
} 

\AgdaHide{
\begin{code}%
\>[0]\AgdaOperator{\AgdaFunction{⟦}}\AgdaSpace{}%
\AgdaInductiveConstructor{opEqualVerify}\AgdaSpace{}%
\AgdaOperator{\AgdaFunction{⟧ₛˢ}}\AgdaSpace{}%
\AgdaBound{time₁}\AgdaSpace{}%
\AgdaBound{msg}\AgdaSpace{}%
\AgdaSymbol{=}\AgdaSpace{}%
\AgdaFunction{executeStackVerify}\<%
\\
\>[0]\AgdaOperator{\AgdaFunction{⟦}}\AgdaSpace{}%
\AgdaInductiveConstructor{opDup}%
\>[11]\AgdaOperator{\AgdaFunction{⟧ₛˢ}}\AgdaSpace{}%
\AgdaBound{time₁}\AgdaSpace{}%
\AgdaBound{msg}\AgdaSpace{}%
\AgdaSymbol{=}\AgdaSpace{}%
\AgdaFunction{executeStackDup}\<%
\\
\>[0]\AgdaOperator{\AgdaFunction{⟦}}\AgdaSpace{}%
\AgdaInductiveConstructor{opDrop}%
\>[11]\AgdaOperator{\AgdaFunction{⟧ₛˢ}}\AgdaSpace{}%
\AgdaBound{time₁}\AgdaSpace{}%
\AgdaBound{msg}\AgdaSpace{}%
\AgdaSymbol{=}\AgdaSpace{}%
\AgdaFunction{executeStackDrop}\<%
\\
\>[0]\AgdaOperator{\AgdaFunction{⟦}}\AgdaSpace{}%
\AgdaInductiveConstructor{opSwap}%
\>[11]\AgdaOperator{\AgdaFunction{⟧ₛˢ}}\AgdaSpace{}%
\AgdaBound{time₁}\AgdaSpace{}%
\AgdaBound{msg}\AgdaSpace{}%
\AgdaSymbol{=}\AgdaSpace{}%
\AgdaFunction{executeStackSwap}\<%
\\
\>[0]\AgdaOperator{\AgdaFunction{⟦}}\AgdaSpace{}%
\AgdaInductiveConstructor{opHash}%
\>[11]\AgdaOperator{\AgdaFunction{⟧ₛˢ}}\AgdaSpace{}%
\AgdaBound{time₁}\AgdaSpace{}%
\AgdaBound{msg}\AgdaSpace{}%
\AgdaSymbol{=}\AgdaSpace{}%
\AgdaFunction{executeOpHash}\<%
\\
\>[0]\AgdaOperator{\AgdaFunction{⟦}}\AgdaSpace{}%
\AgdaInductiveConstructor{opCHECKLOCKTIMEVERIFY}\AgdaSpace{}%
\AgdaOperator{\AgdaFunction{⟧ₛˢ}}\AgdaSpace{}%
\AgdaBound{time₁}\AgdaSpace{}%
\AgdaBound{msg}\AgdaSpace{}%
\AgdaSymbol{=}\AgdaSpace{}%
\AgdaFunction{executeOpCHECKLOCKTIMEVERIFY}\AgdaSpace{}%
\AgdaBound{time₁}\<%
\\
\>[0]\AgdaOperator{\AgdaFunction{⟦}}\AgdaSpace{}%
\AgdaInductiveConstructor{opCheckSig3}\AgdaSpace{}%
\AgdaOperator{\AgdaFunction{⟧ₛˢ}}\AgdaSpace{}%
\AgdaBound{time₁}\AgdaSpace{}%
\AgdaBound{msg}\AgdaSpace{}%
\AgdaSymbol{=}\AgdaSpace{}%
\AgdaFunction{executeStackCheckSig3Aux}\AgdaSpace{}%
\AgdaBound{msg}\<%
\\
\>[0]\AgdaOperator{\AgdaFunction{⟦}}\AgdaSpace{}%
\AgdaInductiveConstructor{opMultiSig}\AgdaSpace{}%
\AgdaOperator{\AgdaFunction{⟧ₛˢ}}%
\>[18]\AgdaBound{time₁}\AgdaSpace{}%
\AgdaBound{msg}\AgdaSpace{}%
\AgdaSymbol{=}%
\>[31]\AgdaFunction{executeMultiSig}\AgdaSpace{}%
\AgdaBound{msg}\<%
\\
\\[\AgdaEmptyExtraSkip]%
\>[0]\AgdaComment{--\ execute\ only\ the\ stack\ operations\ of\ a\ bitcoin\ script}\<%
\\
\>[0]\AgdaComment{--\ \ is\ correct\ only\ for\ non-if\ instructiosn}\<%
\\
\>[0]\AgdaOperator{\AgdaFunction{⟦\AgdaUnderscore{}⟧ˢ}}\AgdaSpace{}%
\AgdaSymbol{:}\AgdaSpace{}%
\AgdaSymbol{(}\AgdaBound{prog}\AgdaSpace{}%
\AgdaSymbol{:}\AgdaSpace{}%
\AgdaFunction{BitcoinScriptBasic}\AgdaSymbol{)(}\AgdaBound{time₁}\AgdaSpace{}%
\AgdaSymbol{:}\AgdaSpace{}%
\AgdaFunction{Time}\AgdaSymbol{)(}\AgdaBound{msg}\AgdaSpace{}%
\AgdaSymbol{:}\AgdaSpace{}%
\AgdaDatatype{Msg}\AgdaSymbol{)(}\AgdaBound{stack₁}\AgdaSpace{}%
\AgdaSymbol{:}\AgdaSpace{}%
\AgdaFunction{Stack}\AgdaSymbol{)}\AgdaSpace{}%
\AgdaSymbol{→}\AgdaSpace{}%
\AgdaDatatype{Maybe}\AgdaSpace{}%
\AgdaFunction{Stack}\<%
\\
\>[0]\AgdaOperator{\AgdaFunction{⟦}}\AgdaSpace{}%
\AgdaInductiveConstructor{[]}%
\>[12]\AgdaOperator{\AgdaFunction{⟧ˢ}}\AgdaSpace{}%
\AgdaBound{time₁}\AgdaSpace{}%
\AgdaBound{msg}\AgdaSpace{}%
\AgdaBound{stack₁}\AgdaSpace{}%
\AgdaSymbol{=}\AgdaSpace{}%
\AgdaInductiveConstructor{just}\AgdaSpace{}%
\AgdaBound{stack₁}\<%
\\
\>[0]\AgdaOperator{\AgdaFunction{⟦}}\AgdaSpace{}%
\AgdaBound{op}\AgdaSpace{}%
\AgdaOperator{\AgdaInductiveConstructor{∷}}\AgdaSpace{}%
\AgdaInductiveConstructor{[]}%
\>[12]\AgdaOperator{\AgdaFunction{⟧ˢ}}\AgdaSpace{}%
\AgdaBound{time₁}\AgdaSpace{}%
\AgdaBound{msg}\AgdaSpace{}%
\AgdaBound{stack₁}\AgdaSpace{}%
\AgdaSymbol{=}\AgdaSpace{}%
\AgdaOperator{\AgdaFunction{⟦}}\AgdaSpace{}%
\AgdaBound{op}\AgdaSpace{}%
\AgdaOperator{\AgdaFunction{⟧ₛˢ}}\AgdaSpace{}%
\AgdaBound{time₁}\AgdaSpace{}%
\AgdaBound{msg}\AgdaSpace{}%
\AgdaBound{stack₁}\<%
\\
\>[0]\AgdaCatchallClause{\AgdaOperator{\AgdaFunction{⟦}}}\AgdaSpace{}%
\AgdaCatchallClause{\AgdaBound{op}}\AgdaSpace{}%
\AgdaCatchallClause{\AgdaOperator{\AgdaInductiveConstructor{∷}}}\AgdaSpace{}%
\AgdaCatchallClause{\AgdaBound{prog}}\AgdaSpace{}%
\AgdaCatchallClause{\AgdaOperator{\AgdaFunction{⟧ˢ}}}\AgdaSpace{}%
\AgdaCatchallClause{\AgdaBound{time₁}}\AgdaSpace{}%
\AgdaCatchallClause{\AgdaBound{msg}}\AgdaSpace{}%
\AgdaCatchallClause{\AgdaBound{stack₁}}\AgdaSpace{}%
\AgdaSymbol{=}%
\>[35]\AgdaOperator{\AgdaFunction{⟦}}\AgdaSpace{}%
\AgdaBound{op}\AgdaSpace{}%
\AgdaOperator{\AgdaFunction{⟧ₛˢ}}\AgdaSpace{}%
\AgdaBound{time₁}\AgdaSpace{}%
\AgdaBound{msg}\AgdaSpace{}%
\AgdaBound{stack₁}\AgdaSpace{}%
\AgdaOperator{\AgdaFunction{\ensuremath{\mathbin{{>}\mkern-8.5mu{>}\mkern-2mu{=}}}}}\AgdaSpace{}%
\AgdaOperator{\AgdaFunction{⟦}}\AgdaSpace{}%
\AgdaBound{prog}\AgdaSpace{}%
\AgdaOperator{\AgdaFunction{⟧ˢ}}\AgdaSpace{}%
\AgdaBound{time₁}\AgdaSpace{}%
\AgdaBound{msg}\<%
\end{code}
} 

\maketitle

\begin{abstract}

This paper contributes to the verification of programs 
written in Bitcoin's smart contract language \Script{}
in the interactive theorem prover Agda. 
\ANNannotationDone{TODO: update abstract and title on easychair}%
%
\ANNannotationDone{according to review1, he said (in the text
it's occasionally written starting with a capital letter "Script".) (Done),
before was \emph{script}\\
AGS I have used a macro \Script{} when referring to Bitcoin Script as a
language.\\
Bitcoin wiki spells it as Script. I first was against the use of \Script{}
but 
feel now it is a good solution.}%
It focuses on the security property of access control for \Script{} programs 
that govern the distribution of Bitcoins.
It advocates that \emph{weakest preconditions} in the context of Hoare triples are
the appropriate notion for verifying access control.
It aims at obtaining human-readable descriptions of weakest preconditions
in order to close the validation gap between user requirements and
formal specification of smart contracts.

As examples for the proposed approach, 
the paper focuses on two standard \Script{} programs 
that govern the distribution of Bitcoins, 
\emph{Pay to Public Key Hash (P2PKH)} and \emph{Pay to Multisig (P2MS)}. 
\ANNannotationDone{AB: I think we never discuss P2MS in the paper. 
AGS: I have now introduced it in 6.2}%
The paper introduces an operational semantics of the \Script{} commands used in
P2PKH and P2MS, 
which is formalised in the Agda proof assistant and reasoned
about using Hoare triples. 
\ANNannotationDone{according to review2 (, which
we formalise in the Agda proof assistant and reason about using Hoare
triples) before was ,and formalises it in the Agda proof assistant using
Hoare triples}%
%
%
Two methodologies for obtaining human-readable descriptions of weakest preconditions are
discussed:
(1) a step-by-step approach, which works backwards instruction by instruction
through a script, 
sometimes grouping several instructions together; 
(2) symbolic execution of the code and translation into a
nested case distinction, which allows to read off weakest preconditions
as the disjunction of conjunctions of conditions along accepting paths.
A syntax for equational reasoning with Hoare Triples is defined 
in order to formalise those approaches in Agda. 

\ANNannotationinlineDone{AB: as we are not carrying out the correct by
construction approach, I would not mention it in the abstract.\\
It is mentioned as future work in the introduction.\\
AGS: agreed,so the following is omitted:\\
The approach allows to develop smart contracts which are correct
by construction by using weakest preconditions and
intermediate conditions as guidance.}%

\end{abstract}
\pagebreak  

\section{Introduction\ANNwhoIsDoingWhatDone{AB}}
%
\ANNannotationDone{Check all occurrences of pbk pbkh  whether they should be pbk or pbkh}
\ANNannotationDone{Suggestion: replace Smart contract in title by Bitcoin Script (using qualifiers like a weak form of smart contract would make the already long title too long.\\ 
One option would be Bitcoin's Contracts or Bitcoin Script Contracts.\\
Then change first sentence in abstract to something like:\\
This paper addresses the verification of Bitcoin Script which forms a simple form of smart contracts
in the sense of Szabo (reference needed). Then verification is carried out using the interactive theorem prover Agda. \\
Add a small discussion about history of smart contracts, see Wikipedia
\url{https://en.wikipedia.org/wiki/Smart_contract}
and \url{https://www.frontiersin.org/articles/10.3389/fbloc.2020.553671/full}
where it is clear: Szabo defined them before Bitcoin was introduced, Bitcoin fulfils the conditions of Szabo, the Ethereum white paper (to be cited) refers to Bitcoin \Script{} as a weak form of smart contract, and Wikipedia lists Bitcoin \Script{} as a smart contract language explicitly; however it is not Turing complete. Note as well that Bitcoin Script is a precursor of the EVM.}
\ANNannotationinlineDone{Main motivation for this work:\\
Security notion which is here relevant is access control\\
Access control is restriction to access for a resource.\\
- In case of cryptocurrencies the restriction is access to obtaining bitcoins.\\
- For access control, weakest precondition is the correct notion.\\
- a precondition says that if you fulfil the conditions you get access. But there
may be another way
- weakest precondition means: you get access if and only you fulfil the condition\\
- one can determine the weakest precondition easily:\\
weakest precondition(s) $<=>$ Accepting State($[[$ p $]]$(s) )\\
However we need to find a way which is understandable\\
Then the designer of a smart contract can decide whether the 
readable weakest precondition expresses the intention of the writers of the 
smart contract.\\
\mbox{\quad}\\
Three ways of obtaining a readable weakest precondition\\
1) by working oneself backwards instruction by instruction\\
In some cases it is easier to work oneself backwards several instructions.\\
- example will be multisig, where it is easier to express the weakest precondition
for the whole script.\\
\mbox{\quad}\\
2) One investigates the program in a symbolic way, and translates it into a
nested case distinction, where case distinctions are made on variables
(of type nat or stack) or on expressions formed from variables
by applying basic functions to it such as hashing function or checking for signature,
Then one can investigate the result and the weakest precondition is
the disjunction of a condition by fulfilling one of the successful branches.\\
\mbox{\quad}
3) Guessing directly the weakest precondition.\\
In all cases one needs to prove that we have the weakest precondition,
in case of 1) this follows by step wise operation, in 2) one proves that the program
is equivalent to the simplified program making the precondition and then
one can easily prove the weakest precondition. 3) would in many cases require
an automated theorem tool\\
\mbox{\quad}\\
Another contribution is that this is actually carried out in Agda for concrete
smart contracts, demonstrating the feasibility of this approach.\\
\mbox{\quad}\\
Important part is the use of Agda. Our approach is based on functional programming
and dependent type theory, which is different from other frameworks.\\
This is particularly important since the cryptocurrency Cardano will be /or is using a language 
related to Haskell for Smart contracts, and is listing on their home page
papers using Agda, so Agda is becoming important\\
\mbox{\quad}\\
Advantage of Agda of having proofs which re in checkable by hand as opposed to proofs
in other frameworks such as Coq, which use automatic proof search tools. 
Problem of automatic proof search that you need to trust those tools. Can be important
if one wants to put a lot of money on a single smart contract - trust could be higher
if we can actually verify the proof by hand.\\
\mbox{\quad}\\
Relation to correctness by construction. The use of weakest precondition with
explicit proofs could be a route of constructing verified smart contracts which are
correct by construction. You could start with an intended weakest precondition and postcondition
and then introduce some intermediate conditions and now put in the programs in between
rather than first writing a program and then verifying it.
That extends the approach as used in SPARK Ada where one uses Hoare logic without
weakest precondition to verify programs. \\
This has the advantage that the intermediate verification conditions serve as verified
comments - usually when we have comments and change the code, nothing enforces 
that the comments are updated. When we have verified conditions in the code
they need to be kept in sync. \\
Vision of a system which allows to create code while using conditions in side
and guaranteeing that they are weakest preconditions.
}
\ANNannotationinlineDone{Additional Motivation:\\
Whereas the usual approach to verifying smart contracts is to check for 
known problems in scripts, we want instead to create smart contracts which 
are correct from the start. This is achieved by using 
weakest preconditions, which express the correctness of the contract in such a
way that one directly sees that it is correct. So the script comes
with a precondition and it is correct from the start}
%
%


%
Bitcoin, the first cryptocurrency, was introduced in 2008 by Satoshi Nakamoto~\cite{nakamoto} 
to provide a public payment mechanism, the blockchain,
using pseudonymous keys in a peer-to-peer network with distributed control. 
\ANNannotationDone{Fahad regarding to review3: as a cryptocurrency
that uses in a peer-to-peer network.\\
AGS: He is right: it is neither private nor anonymous}%
%
%
Many other cryptocurrencies have been introduced since.

\ANNannotationDone{Arnold: The following paragraph discusses smart contracts.
I've tried to avoid any controversial elements, to make sure that we get
agreement from the 1st, but more importantly 3rd reviewer.}%
Bitcoin's blockchain provides a scripting system for transactions called \Script{}.
Lists of instructions in \Script{} are denoted \emph{Bitcoin scripts} or
simply \emph{scripts}.
The invention of Ethereum~\cite{Ethereum_WhitePaper} 
strengthened Bitcoin by
adding full (Turing complete) smart contracts to blockchains.
\ANNannotationDone{Anton: Arnold - please feel free to change this,
I'm not sure and this is more your domain.\\
I added ``full (Turing complete)'' because of comment by referee 1:\\
They were introduced as distributed applications first on Ethereum blockchain. --
     Isn't it the case that Bitcoin Script appeared before EVM and Ethereum?
}
In this context, smart contracts can be seen as programs that automatically
execute when transactions are performed on a blockchain.
Though not Turing complete as Ethereum~\cite{Ethereum_WhitePaper},
Bitcoin scripts can be viewed as a weak form of smart contracts, 
that provide important functionality, e.g.\ by governing the distribution of
the Bitcoin cryptocurrency.

\ANNannotationinlineDone{
Discussion on term smart contract\\
Several cryptocurrencies have been introduced such as
Ethereum~\cite{Ethereum_WhitePaper}.
%
%
With blockchain also a simple form of smart contracts 
in the original sense of
Szabo \cite{szabo:smartcontracts1994,szabo,szabo:IdeaOfSmartContracts} have
been
introduced~\cite{cuccuru:BeyondBitcoin} (we suggest that a better terminology 
would be to use ``blockchain contract'' for such kind of scripts).%
{\it footnote} {According
to \cite{viglotti:WhatMeanBySmartContracts}, smart contracts were introduced
1994 by Szabo \cite{szabo:smartcontracts1994}, where he wrote: 
``A smart contract is a computerized transaction protocol that executes the
terms of a contract. The general objectives of smart contract design are to
satisfy common contractual conditions (such as payment terms, liens,
confidentiality, and even enforcement), minimize exceptions both malicious
and accidental, and minimize the need for trusted intermediaries. Related
economic goals include lowering fraud loss, arbitration and enforcement
costs, and other transaction costs.''. In the white paper of Ethereum,
Buterin wrote \cite{Ethereum_WhitePaper}:
``Even without any extensions, the Bitcoin protocol actually does facilitate
a weak version of a concept of 'smart contracts'.''
Wikipedia \cite{Wiki:2022:SmartContract} summarises different definitions of
smart contracts as 
``A smart contract is a computer program or a transaction protocol which is
intended to automatically execute, control or document legally relevant
events and actions according to the terms of a contract or an agreement'',
and includes Bitcoin in its ``list of blockchain platforms supporting smart
contracts''. However, as reflected by the anonymous referees,
in the light of Ethereum's smart contracts 
it seems to becoming common practice to  have higher demands on the complexity of 
smart contract language to allow Turing completeness, the ability to call
other smart contracts, and because of this to allow
``fully automatic event-driven executions'' (from one anonymous referee),
as it is implemented in the Ethereum platform. Because
of this we believe that the notion of ``blockchain contract'' as opposed to ``smart
contract'' might be more in line with how the notion of smart contract has
evolved.}
%
%
\ANNannotationDone{according to review1: I removed ~\cite{luu}, before was
~\cite{szabo,luu}}%
Blockchain contracts are
programs in the Blockchain that are automatically executed when certain
conditions are fulfilled.
\ANNannotation{AGS: changed to ``blockchain contracts''. See as well
AnnotationDone by Fahad.}
\ANNannotationDone{Fahad: according to review1: the execution is triggered by an
external call, current blockchains do not allow fully automatic event-driven
executions. So I wrote the specific definition for smart contracts in
blockchain, before was (Smart contracts can be defined as programs that are
executed automatically when certain conditions are fulfilled).}
\ANNannotationDone{make it clarify here bitcoin is simple in order to
guarantee termination. Ethereum substantially extended it by allowing
jumps, therefore obtaining Turing completeness, and allowing to call other
programs, and solved the undecidability of the Turing halting problem by
requiring the account which is invoking a smart progress to pay a cost for
the execution of the program, with a cost called gas to be paid per
instruction.}
\ANNannotationDone{(Fahad regarding Dr. Anton notes and review3: (please check
my words)}  
The Bitcoin \Script{} language is not Turing-complete, which
means it cannot execute loops, jumps, or complicated control structures to
simplify and secure the transaction verification
process~\cite{bistarelli}.  In contrast, Ethereum is fully
Turing-complete~\cite{vujicic:ieee2018:etherandbitcoin}, which means it
supports all sorts of computations allowing jumps and calling other 
programs. In addition, it solved the undecidability of the Turing halting
problem by requiring the account to invoke smart progress to pay a cost
for the execution of the program, with a cost called gas to be paid per
instruction.
Blockchain contracts were introduced as distributed applications first on the
Bitcoin blockchain~\cite{Nawari:journalofbuildingengineering:blockchain:2019}
and then extended in Ethereum to full Smart Contracts. 
\ANNannotationDone{according to review1 I changed it to Bitcoin instead of
Ethereum and I added reference for it, before was They were introduced as
distributed applications first on Ethereum blockchain.}%
Bitcoin contracts can be written in \Script{} in Bitcoin~\cite{brakmic,
BartolettiMass}.
Full smart contracts can be written using a high-level language such as
Solidity or Vyper on Ethereum~\cite{Hildenbrandt}, 
however they are executed in low-level languages like
the EVM in Ethereum~\cite{Ethereum_WhitePaper},
which extends Bitcoin \Script{}, in particular by adding jumps, gas
limitation, the ability to call other contracts, and direct access to memory.
}

As smart contracts, including Bitcoin scripts, can control real world values 
and are immutable once deployed on the blockchain network, 
a method to demonstrate their security and correctness is 
needed~\cite{luu,wohrer}. 
According to~\cite{Almakhour, murray, Clarke}, there are two ways to verify
their correctness: 
\ANNannotationDone{before was accuracy, now correctness} %
\ANNannotationDone{regarding the review3 I replaced these sentences ((1) using
mathematical methods like formal verification, which utilise theorem proving,
model checking, 
automatically generated test cases, and similar techniques, 
and (2) testing.) please check it}%
(1) by using mathematical methods like
formal verification, which utilise theorem proving, model checking, and similar
techniques, and 
(2) by employing testing.
Theorem proving provides an extremely flexible verification method that can
be applied to various types of systems including smart contracts.
It can be done in interactive, automated, or hybrid mode.

%
\ANNannotationDone{Arnold: I did not check the next paragraph, Anton, can you
sign it off, please?}%
\ANNannotationDone{review1 change from (we are using ) to we use ), FA: Done}%
\ANNannotationDone{review2: In the next paragraph, you fail to mention that
your approach lies in of the *interactive* kind. FA : I added (in our
approach) to make it clear for a reader in our approach we use the
interactive.... }%
In our approach, we use the interactive theorem prover Agda \cite{agdaDoc}
for the verification of 
Bitcoin scripts.
Agda is designed to be both an interactive theorem prover and a dependently
typed programming language~\cite{Norell2009dependentlytyped}, therefore 
Agda allows us to 
define programs and reason about them in the same system.
This reduces the danger of producing errors when translating programs 
from a programming language to a theorem prover, and allows to execute
smart contracts in Agda directly.
Another advantage of Agda is to have proofs that are 
checkable
\ANNannotationDone{according to review3 we changed it to checkable}%
by hand. 
Other frameworks, such as Coq~\cite{Paulincoq}, use automatic proof search tools 
which usually do not 
provide proof certificates 
that could in principle be checked by hand.
Human checkable proof certificates
reduce the need to rely on a theorem prover being correct.
The latter is desirable because of potential attacks that exploit errors in theorem provers,
e.g.\ by creating a smart contract that contains a deliberate error 
together with a correctness proof that exploits the error in the theorem prover.%
\footnote{See the forum discussion on~\cite{StackExchange:onlinediscussion:2022}
for
a well documented list of incorrect protocols with false correctness proofs.
To hide backdoors using deliberately false correctness proofs is certainly 
conceivable.}
\ANNannotationDone{FA: Comment on footnote: I add the reference to our list of references }%
\ANNannotationDone{according to review1 and review2,
they asked why we choose Agda and the differences between Agda and Coq:}%
As a final point, there are some key distinctions between Agda and other theorem provers like Coq 
that 
suggest a different applicability of Agda. 
For example, Agda supports inductive-recursive types,
while Coq does not~\cite{Bove:Springer:Agdaoverivew:2009}. 
Agda also has a more flexible pattern matching system than Coq, 
including support for copattern matching~\cite{Bove:Springer:Agdaoverivew:2009}. 

\myparagraph{Main contributions.}
Our main contributions in this paper are:
\begin{itemize}
    \item We argue that weakest preconditions are the appropriate notion to verify
    access control for Bitcoin scripts.
    \item We propose to aim for human-readable descriptions of weakest
    preconditions
    to support judging whether the security property of access control is satisfied.
    \item We describe two methods for achieving human-readable descriptions
    of weakest preconditions: 
    a step-by-step approach, and a symbolic-execution-and-translation approach.
    \item We apply our proposed methodology to two standard Bitcoin scripts,
    providing fully formalised arguments in Agda.
\end{itemize}

In the following we explain our contributions in more detail.
The paper introduces an operational semantics of the \Script{} commands used in
\emph{Pay to Public Key Hash (P2PKH)} and \emph{Pay to Multisig (P2MS)}, 
two standard scripts that govern the distribution of Bitcoins. 
We define
\ANNannotationDone{review2, change this sentence "We
formalise their operational semantics in Agda using Hoare triples." to "We
define the operational semantics as stack operations....", FA: Done}%
the operational semantics as stack operations and
reason about the correctness of such operations using Hoare triples 
utilising pre- and postconditions.

\myparagraph{Weakest precondition for access control.}
Our verification focuses on the security property of \emph{access control}.
Access control is the restriction to access for a resource, 
which in our use case is access to cryptocurrencies like Bitcoin.
We advocate that, in the context of Hoare triples, \emph{weakest preconditions} 
are the appropriate notion to model access control:
A (general) precondition expresses that when it is satisfied, access is granted, 
but there may be other ways to gain access without satisfying the precondition.
The weakest precondition expresses that access is granted if and only if the
condition is satisfied.

\myparagraph{Human-readable descriptions.}
The weakest precondition can always be described in a direct way, 
for example as the set of states that after execution of the smart contract 
end in a state satisfying the given postcondition.
However, such a description is meaningless to humans who want to convince
themselves that the smart contract is 
secure, in the sense 
that they do not provide any further insights beyond the original 
smart contract.

It is known in software engineering, that failures of safety-critical systems
are often due to incomplete requirements or specifications rather than
coding errors.%
\footnote{For instance, \cite{martins:RequirementsEngineerging} writes:
``Almost all accidents with serious consequences in which
software was involved can be traced to requirements failures, and particularly to incomplete requirements.''}
The same applies to security related software.%
\footnote{The long list of protocols which were proven to be secure but had wrong proofs
\cite{cryptoStackExchange:SecureProtocolsRealWorldAttacks}
demonstrates that a proof of correctness is not sufficient. We assume that 
most of the examples had  correct proofs, but the statement shown
was not sufficient to guarantee security.} 
It is not sufficient to have
a proof of security of a protocol, if the statement does not express what is 
required. That the specification (here the formal statement of secure access control)
guarantees that the requirements are fulfilled (namely that it is impossible for a hacker
to access the resource, here the Bitcoin), needs to be checked by a human being,
who needs to be able to read  the specification and determine whether it
really is what is expressed by the requirements. 
%
%
\ANNannotationinlineDone{Anton: Added some stronger statements about the validation 
gap - I think this is important to convince the third referee of the importance of this paper.\\
AB: I would delete the sentence "This is the gap between 
validation and verification." as we do not explain what we mean by "validation".  We also do not explain what we mean by "verification", but that is fine, it can be assumed as implicite knowledge of the intended reader.  With validation I am not so sure.}%
Thus, the challenge is to obtain simple, human-readable descriptions of the
weakest precondition of a smart contract.
This would allow to close the validation gap between user requirements and
formal specification of smart contracts.


\myparagraph{Two methods for obtaining human-readable weakest preconditions.}
We discuss two methods for obtaining readable weakest preconditions:
The first, step-by-step approach,  
is obtained by working through the program backwards instruction by instruction.
In some cases it is easier to group several instructions together and deal
with them differently, 
as we will demonstrate with an example in Sect.~\ref{sec:combinedMultiSig}.
The second method, symbolic-execution-and-translation,
evaluates the program in a symbolic way, and translates it into a
nested case distinction.
The case distinctions are made on variables
(of type nat or stack) or on expressions formed from variables
by applying basic functions to them 
such as hashing or checking for signature.
From the resulting decision tree, the weakest precondition can be read off
as the disjunction of the conjunctions of the conditions that occur 
along branches that lead to a successful outcome.

For both methods, it is necessary to prove that the established weakest
precondition
is indeed the weakest precondition for the program under consideration.
For the first method, this follows by stepwise 
\ANNannotationDone{review2, change from (step wise) to stepwise, FA : Done}%
operation.
The second uses a proof that the original program is equivalent to the
transformed program from which the weakest precondition has been established,
or a direct proof which follows the case distinctions used in the symbolic
evaluation.


\myparagraph{Application of our proposed methodology.}
We demonstrate the feasibility of our approaches by carrying them out in Agda for
concrete smart contracts, including P2PKH and P2MS.

Our approach also provides opportunities for further
applications:
\ANNannotationDone{review2: change from (..opportunities for further
application) to (..opportunities for further applications), FA : Done}%
The usage of the weakest precondition with explicit proofs can be seen as a
method of building verified smart contracts that are \emph{correct by
construction}. 
Instead of constructing a program and then evaluating it, one can
start with the intended weakest precondition and postcondition, add
some intermediate conditions, and then develop the program between those
conditions. 
Such an approach would extend the SPARK Ada
framework~\cite{Adacore:SPARKAda:Homepage} 
\ANNannotationDone{review2: change from (would extend the SPARK Ada framework
of using Hoare logic without the weakest precondition to check programs) to
(..would extend the SPARK Ada framework to use Hoare logic (without the
weakest precondition) to check programs)}%
to use Hoare logic (without the weakest precondition) to check programs.

The remainder of this paper is organised as follows.
\ANNannotationDone{review2: change from (as follows:) to (as follows: in
Sect. 2 or as follows. In Sect. 2), FA: Done}%
In Sect.~\ref{sec:relatedwork}, we introduce related work on verification of
smart contracts.
Sect.~\ref{sec:verificationBitwithWP} introduces Bitcoin \Script{} and
defines its operational semantics.
In Sect.~\ref{sec:WPinHTforSecurity}, we specify the security of\ANNannotationDone{FA: change correctness to security }
Bitcoin \Script{} using Hoare logic and weakest preconditions.
We formalise these notions in Agda and 
introduce equational reasoning for Hoare triples to streamline our correctness
proofs.
%
%
\ANNannotationDone{we need
to change it regarding our new title: a new suggestion is 
( we specify the correctness of Bitcoin \Script{} using Hoare logic, 
weakest preconditions,
and equational reasoning for Hoare triples to streamline our correctness
proofs}%
\ANNannotationDone{review2: change this sentence (we also define a syntax for
equational reasoning with Hoare Triples to formalise those approaches in
Agda.) to (define a syntax for equational reasoning for Hoare triples to
streamline our correctness proofs.), FA: Done }%
Sect.~\ref{sec:proofusingstep} introduces our first,
step-by-step 
\ANNannotationDone{review2: change from (step by step) to (step-by-step), FA: Done}%
method of developing human-readable weakest preconditions and proving correctness of
%
\ANNannotationDone{before was Pay to Public Key Hash}%
P2PKH. 
In Sect.~\ref{sec:proofusingsymbloic}, we introduce our second method based
on symbolic execution and apply it to various examples.
In Sect.~\ref{sect:PracticalUsageOfAgda}, 
\ANNannotationDone{FA: I added section 7 }%
we explain how to practically use Agda to determine and prove
weakest preconditions using our library \cite{setzergithub}.
We conclude in Sect.~\ref{sec:conclusion}.\par

\myparagraph{Notations and git repository.} 
The formulas can be presented as full
Agda code, but often the formulas can also be presented in mathematical
style. In order to switch between Agda code and mathematical code easy, we
use the functional style for application (i.e.~writing $f\;a\;b\;c$ instead
of $f(a,b,c)$) and $x : A$ instead of $x \in A$. $s \;\AgdaDoubleColon\;l$
denotes prepending an element onto a list.
\ANNannotationDone{review2: change from (denotes the cons operation on lists)
to (denotes prepending an element onto a list), FA: Done}%
The original Agda definitions are also available~\cite{setzergithub}. Most display style
Agda code presented in this paper has been automatically extracted from the
Agda code, in some cases it was formatted by hand based on 
\LaTeX{} code generated by Agda to improve the presentation.

\ANNannotationinlineDone{
*****Old introduction*****

Cryptocurrency is an application of Blockchain technology. Bitcoin was
introduced in
2008 by Satoshi Nakamoto as a cryptocurrency that provides a private
anonymous payment mechanism in a peer-to-peer network
~\cite{Gencer}.
Since then, several cryptocurrencies
have been introduced such as Ethereum~\cite{nakamoto}. This is resulting a
new and emerging era
of decentralised and digital monetary where there is no need for any central
entity to manage and control users' transactions, such as central banks.
Another interesting technology that was re-emerged as a result of the new era
of Blockchain is the use of
smart contracts~\cite{szabo,luu}.
Smart contracts can be defined as programs automatically executed when
certain conditions are fulfilled.\ANNannotation{smart contracts definition:
Smart contracts are transactions that are defined through software and
executed automatically when conditions in the blockchains are
met. }
Smart contracts were introduced as a distributed application first on
Ethereum blockchain. They can be written using a high-level 
language such as Solidity or Vyper on Ethereum~\cite{Hildenbrandt}, and in
Bitcoin, they are written in Script~\cite{brakmic, BartolettiMass}.\par
As a result of the fact that smart contract is a gate to unlocking values and
are immutable
once they are deployed on the blockchain network, a method to evaluate the
security and correctness
of these programs are needed~\cite{luu, wohrer}. According
to~\cite{Almakhour, murray, Clarke},
there are two ways to verify its accuracy: (1) formal verification methods,
which are based on
theorem proving (mathematical methods), and (2) conducting test
cases. Theorem proving is an
extremely flexible verification method that can be applied to various types
of systems,
and it can be interactive, automated, or hybrid. The verification of software
programs such as
smart contract is important to ensure that they perform correctly without
interfering with other programs~\cite{Almakhour}.\par

In this paper, the Hoare triples was used to verify the correctness and
competence of
smart contract programs. Our approach is based on functional programming and
dependent type theory; 
therefore, we have chosen Agda as it is a suitable option to define our
programs and proofs.
Another advantage of Agda is having proofs that are checkable by hand. Other
frameworks,
such as Coq~\cite{Paulincoq}, use automatic proof search tools. The
verification of smart contracts needs to be carefully done; otherwise, its
vulnerabilities can be exposed,
and assets and money will be at risk. Therefore, verification should be
accomplished by hand.
An automated proof search framework would involve executing the verification
method through
automated proof search tools, which can sometimes lead to poor verification.\par
\ANNannotation{From Fahad for why we choose Agda: There are some key
distinctions between Agda and Coq that might suggest the wider applicability
of Agda. For example, Agda supports inductive-recursive types, while Coq does
not~\cite{Bove:Springer:Agdaoverivew:2009}. Agda also has a more flexible
pattern matching system than Coq and supports copattern
matching~\cite{Bove:Springer:Agdaoverivew:2009}.}
The usage of the weakest precondition with explicit proofs could be a
method of building verified smart contracts that are correct by
construction. Instead of 
constructing a program and then evaluating it, one may start with the
intended weakest
precondition and postcondition, then introduce some intermediate conditions,
and then put in
the program in between. This extends the SPARK
Ada~\cite{Adacore:SPARKAda:Homepage} approach of using Hoare logic
without the weakest precondition to check programs.\par
In this paper, we propose this approach because it is easy to understand and will
enable the smart contracts designer to decide whether the readable weakest
precondition
expresses the intention of the writers of the smart contract. Thus,
the principal purpose of this work is to introduce two methods for
obtaining a human-readable weakest precondition that can support smart
contracts developers in Bitcoin using Agda proof assistant with Hoare
triples. Nonetheless,
instead of checking for known flaws in scripts, which is the traditional
approach to smart contract verification, we want to construct smart contracts
that are correct from the start. This is accomplished by employing the
weakest preconditions, which explain the contract's validity in such a way
that it is
immediately obvious that it is correct. \par

The \textbf{remaining of this Paper is organised} as follows:
In Sect.~\ref{sec:relatedwork}, introduces related work that has been
addressed on the verification of smart contracts.
Sect.~\ref{sec:verificationBitwithWP}, verifies Bitcoin scripts using Weakest
Preconditions by giving a brief introduction to Bitcoin \Script, then
defining the operational semantics for Bitcoin \Script.
Sect.~\ref{sec:WPinHTforSecurity}, introduces Hoare logic and weakest
preconditions for security to adapt with the current framework and define a
syntax for equational reasoning with Hoare Triples to formalise those
approaches in Agda. 
Sect.~\ref{sec:proofusingstep}, introduces a way of proving the correctness
of Pay to Public Key Hash scripts (P2PKH) by working backwards step by step
through the script. We have designed this approach to look as a
human-readable weakest precondition.  
Sect.~\ref{sec:proofusingsymbloic}, explains another method that executes the
code symbolically and converts it to a nested case distinction, which enables
to read the weakest precondition.
Finally, Sect.~\ref{sec:conclusion} gives a short conclusion and further work.
\ANNannotationinline{AB: The punctuation in the previous paragraph has changed.  E.g.: `In Sect.~\ref{sec:relatedwork}, introduces...' the subject `we' is missing;  `Sect.~\ref{sec:WPinHTforSecurity}, introduces' - here `Sect.~\ref{sec:WPinHTforSecurity}' is the subject, so the comma has to go; etc.  It was correct on Friday.}

{\bf Notations and git repository.} The formulas can be presented as full
Agda code, but often the formulas can also be presented in mathematical
style. In order to switch between Agda code and mathematical code easy, we
use the functional style for application (i.e.~writing $f\;a\;b\;c$ instead
of $f(a,b,c)$) and $x : A$ instead of $x \in A$. $s \;\AgdaDoubleColon\;l$
denotes the cons operation on lists.
The original Agda definitions can be found in \cite{setzergithub}.
}

\section{Related Work\ANNwhoIsDoingWhatDone{AB}} \label{sec:relatedwork}
\ANNannotationinlineDone{%
The related work section should focus on smart contract verification\\
It should cover:\\
verification of script in bitcoin (around 3 papers)\\
verification of smart contracts in Ethereum (mention 2-3 most relevant)\\
formalisation of access control as a security property (related to smart contracts) - probably none\\
use of weakest precondition for verification\\
* in general (1-2 most important papers)  (maybe should just be mentioned in relevant section.)\\
* in verifying smart contracts\\
formalisations in Agda
}

In this section we describe research relevant to our approach.  
We start by discussing two papers introducing Hoare logic, predicate transformer semantics and weakest preconditions.
\ANNannotationDone{FA: I omit section reference}%
We then review papers that address verification of 
smart contracts,
\ANNannotationDone{FA: I omit reference}%
and Bitcoin scripts.
\ANNannotationDone{FA: I omit section reference}%
We present a number of approaches to use model-checking for the verification of smart contracts,
\ANNannotationDone{FA: I omit section reference}%
and finish with work with employs Agda in the verification of smart contracts.
\ANNannotationDone{add references to subsections}

\ANNannotationinlineDone{A comment on the tense in related work: we speak about results in published work in present tense, as those results are valid now.  
The only place where we would use past tense is when referring e.g.\ to experiments in such work, which have happened in the past.
}%
%

\myparagraph{Hoare Logic, Predicate Transformer Semantics and Weakest Preconditions.}
Hoare \cite{hoare:AxiomaticBasisForComputerProgramming:ACMcommunications:1969} defines a formal system using logical rules for reasoning about the correctness of computer programs.
It uses so-called Hoare triples which combine two predicates, a pre- and a postcondition, with a program to express that if the precondition holds for a state and the program executes successfully, then the postcondition holds for the resulting state.
Djikstra~\cite{Dijkstraguardedcommand:1975} introduces predicate transformer semantics that assigns to each statement in an imperative programming paradigm a corresponding total function between two predicates on the state space of the statement.
The predicate transformer defined by Djikstra applied to a postcondition returns the weakest precondition.

%
\ANNannotationDone{according to review2, You should cite Djikstra's "Guarded Command Language", which is the basis of most of the things you present in this paper,
preferably at the start of the Related Work section
}%
\ANNannotationDone{
AB: reviewer2 has requested us to refer to predicate transformer semantics, so we should mention that here as well.}%
%


\myparagraph{Verification of Smart Contracts.}
%
A number of authors have addressed the verification of smart contracts in Ethereum and similar platforms. 
\ANNannotationDone{removed "Ethereum" to fit scope of next subsection}%
Hirai~\cite{Hirai} used Isabelle/HOL theorem prover to validate Ethereum Virtual Machine (EVM) bytecode by developing a formal model for EVM using the Lem language~\cite{lem:ACM:2014}.
\ANNannotationDone{review2: ask to add the reference for Lem language, FA: Done}%
They use this model to prove invariants and safety properties of Ethereum smart contracts.
Amani et al.~\cite{Amani} extended Hirai's EVM formalisation in Isabelle/HOL by a sound program logic at bytecode level.  
To this end, they stored bytecode sequences into blocks of straight-line code, creating a program logic that could reason about these sequences. 
Ribeiro et al.~\cite{Ribeiro} developed an imperative language for a relevant subset of Solidity in the context of Ethereum, using a big-step semantics system. Additionally, they formalised smart contracts in Isabelle/HOL, extending the existing work. 
Their formalisation of semantics is based on Hoare logic and the weakest precondition
\ANNannotationDone{regarding to review1, he said change this weakest-precondition by weakest precondition (done) }%
calculus.
Their main contributions are proofs of soundness and relative completeness, as well as applications of their machinery to verify some smart contracts including modelling of smart contract vulnerabilities.
Bernardo et al.~\cite{Bernard:Tezossmartcontract2020:Springer}
\ANNannotationinlineDone{according to review1, he said add this reference to the related work ): Bernardo et al. Mi-Cho-Coq, FA: Done\\
AB: they do not fall under the topic of this section.  Change topic? - Done\\
Abstract: Tezos is a blockchain launched in June 2018. It is written in OCaml and supports smart contracts. Its smart contract language is called Michelson and it has been designed with formal verification in mind. In this article, we present Mi-Cho-Coq, a Coq framework for verifying the functional correctness of Michelson smart contracts.\\
Review1: Verification framework implemented in Coq and based on Hoare logic for smart contracts written in a stack-based language}%
present Mi-Cho-Coq, a Coq framework which has been used to formalise Tezos smart contracts written 
in the stack-based language Michelson. 
The framework is composed of a Michelson interpreter implemented in Coq, and the weakest precondition calculus to verify Michelson smart contracts' functional correctness.
O'Connor~\cite{OConnor:simplicity:acm2017}
\ANNannotationinlineDone{according to review1, he said add this reference to the related work ): Russell O'Connor. Simplicity: A New Language for Blockchains, FA: Done\\
AB: Same as previous, not verification of Ethereum.  Change topic? - Done\\
Abstract: 
Simplicity is a typed, combinator-based, functional language without loops and recursion, designed to be used for crypto-currencies and blockchain applications. It aims to improve upon existing crypto-currency languages, such as Bitcoin Script and Ethereum's EVM, while avoiding some of the problems they face. Simplicity comes with formal denotational semantics defined in Coq, a popular, general purpose software proof assistant. Simplicity also includes operational semantics that are defined with an abstract machine that we call the Bit Machine. The Bit Machine is used as a tool for measuring the computational space and time resources needed to evaluate Simplicity programs. Owing to its Turing incompleteness, Simplicity is amenable to static analysis that can be used to derive upper bounds on the computational resources needed, prior to execution. While Turing incomplete, Simplicity can express any finitary function, which we believe is enough to build useful ``smart contracts'' for blockchain applications.\\
Reviewer: a low-level language verified in Coq, designed to be an
improved Bitcoin Script.
}%
introduces Simplicity, a low-level, typed functional language, 
which is Turing incomplete.
The goal of Simplicity is to improve on existing blockchain-based languages, like Ethereum's EVM and Bitcoin \Script{}, while avoiding some of their issues. 
Simplicity is based on formal semantics and is specified in the Coq proof assistant. 
\ANNannotationinlineDone{AB: In relation to the next sentence, isn't any smart contract describing in some way a transaction? So what is the purpose of it?\\
To use Simplicity language to construct a smart contract, it must still be designed as a protocol with message exchanges and transactions.
}%
\ANNannotationinlineDone{AB: not sure where the following sentence is coming from, and why it is citing~\cite{bartoletti} which is discussed in the next paragraph.\\
It should be noted that these languages do not allow for the description of the whole contract but simply of the individual transactions utilised in the associated protocol~\cite{bartoletti}. Additionally, it is noteworthy that this language is not designed to describe the whole contract, therefore describing only the transactions used within the associated protocol~\cite{bartoletti}.
}%
Bhargavan et al.~\cite{Bhargavan} provided formalisations of EVM bytecode in F*, a functional programming language designed for program verification.  They defined a smart contract verification architecture that can compile Solidity contracts, and decompile EVM bytecode into F* using their shallow embedding, in order to express and analyse smart contracts.
%

\myparagraph{Verification of Bitcoin Scripts.} 
%
Klomp et al.~\cite{Klomp} proposed a symbolic verification theory, and  a tool to analyse and validate Bitcoin scripts, with a particular focus on characterising the conditions under which an output script, which controls the successful transfer of Bitcoins, will succeed.
Bartoletti et al.~\cite{bartoletti} presented BitML, a high-level domain-specific language for designing smart contracts in Bitcoin. 
They provided a compiler to convert smart contracts into Bitcoin transactions, and proved the correctness of their compiler w.r.t.\  a symbolic model for BitML and a computational model, which has been defined as well in~\cite{Atzei:spring2008:formalmodel}
\ANNannotationDone{regarding to review3, he said added the reference for this sentence "which has defined as well"  I did it}%
for Bitcoin. 
Setzer~\cite{Anton} developed models of the Bitcoin blockchain in the interactive theorem prover Agda.  
This work focuses on the formalisation of basic primitives in Agda as a basis for future work on verifying the protocols of cryptocurrencies and developing verified smart contracts.
\myparagraph{Verification of Smart Contracts using Model Checking.} 
A number of papers discuss tools for analysing and verifying smart contracts that utilise model checking.
Kalra et al.~\cite{kalra} developed a framework called ZEUS whose aim is to support automatic formal verification of smart contracts using abstract interpretation and symbolic model checking.
ZEUS starts from a high-level smart contract, and employs user assistance for capturing correctness and fairness requirements. 
The contract and policy specification are then transformed into an intermediate language with well defined execution semantics.
ZEUS then performs static analysis on the intermediate level and uses external SMT solvers to evaluate any verification properties discovered.
A main focus of the work is on reducing efficiently the state explosion problem inherent in any model checking approach.
Park et al.~\cite{Daejun} proposed a formal verification tool for EVM bytecode based on KEVM, a complete formal semantics of EVM bytecode developed in the K-framework.
To address performance challenges, they define EVM-specific abstractions and lemmas, which they then utilise to verify a number of concrete smart contracts.
Mavridou et al.~\cite{Mavridou} introduce the VeriSolid framework to support the verification of Ethereum smart contracts. 
VeriSolid is based on earlier work (FSolidM) which allows to graphically specify Ethereum smart contracts as transitions systems, and to generate Solidity code from those specification.
It uses model checking to verify smart contract models.
\ANNannotationinlineDone{Regarding Why3 system (email by Fahad)
The why3 system is a system which allows to write imperative programs in an ML-style syntax, add (normal) preconditions,
post- and intermediate conditions to it, generate verification conditions for Hoare triple which are then checked using various automated and interactive theorem tools (including Coq). SPARK Ada uses the why3 system to 
verify its verification conditions.\\
Add as well sentence about Oyente}%
Luu et al.~\cite{luu} provided operational semantics of a subset of Ethereum bytecode called EtherLite, which forms the bases of their symbolic execution tool Oyente for analysing Ethereum smart contracts.
Based on their tool they discovered a number of weaknesses in deployed smart contracts, including the DAO bug~\cite{DAObug:online:2016}.
%
Filli{\^a}tre et al.~\cite{Filliatre} introduced the Why3 system, which
allows writing imperative programs in WhyML, an ML dialect used for programming
and specification.
The system can add 
pre-, post- and intermediate conditions to it but does not make use of weakest precondition. 
Why3 can generate verification conditions for Hoare triple, which are checked using variously automated and interactive theorem provers.  
Why3 is used in SPARK Ada to verify its verification conditions.
%
%

%
\ANNannotationinlineDone{AB:ARBAC examples do not fit scope - delete?\\
AGS Omitted the following after discussion with Arnold:\\
Gofman et al.~\cite{Gofman} 
have presented a tool to analyse policies in both role-based access control (RBAC) and administrative role based access control (ARBAC). This tool can aid in analysing different properties such as the weakest precondition, flow of information, and accessibility.
%
Ferrara et al.~\cite{Ferrara} have used a Verifier of Access Control (VAC) tool to translate ARBAC rules into imperative programs and is used to verify security in programs.}
%
%

\myparagraph{Agda in the Verification of Blockchains.} 
Finally, Agda features in several papers discussing verification of blockchains. 
Chakravarty et al.~\cite{Chakravarty:ExtendedUTXOModel:2020}
\ANNannotationinlineDone{according to review2 
The literature review is slightly misinformed here. The Extended UTxO (EUTXO) model that is used by Cardano was introduced in (1)
and its main contribution is that "validation scripts" can now have more arguments, carry persistent state, etc... which brings at the same expressivity levels as Ethereum,
i.e. you can write any computation/validation on the chain.

(1) Chakravarty, Manuel MT, et al. "The extended UTXO model." International Conference on Financial Cryptography and Data Security. Springer, Cham, 2020.\\
now [16]

[12] is a further extension to EUTXO (EUTXOma, "ma" for "multi-asset") that adds support for "...custom assets and asset bundles of various forms.",
thereby generalising the native Ada currency of Cardano to possibly user-defined currencies with custom minting policies, which all reside on the ledger at the same time.\\
now [14]

[13] is a version of EUTXOma that is designed around a simplistic DSL that is not Turing-complete, for reasons of compatibility with earlier versions of Cardano that did not support EUTXO. Therefore, it relies on a plain UTxO ledger, as Bitcoin for that matter, hence its name "UTXOma".\\
now [15]

:  Chakravarty et al.~\cite{Chakravarty, Chakravarty1} extend existing UTXO accounting rules to custom assets and asset bundles of various form.  
Their approach avoids any form of global state for registering assets.
They provide a simple domain-specific language for forging policy scripts thus forgo the need for general-purpose smart contracts.
They formalise and reason about their models in Agda.}%
\ANNannotationinlineDone{FA: according to Arnold email, please check my words (With Extended UTXO (EUTXO)~\cite{Chakravarty:ExtendedUTXOModel:2020}, Bitcoin's UTXO paradigm was extended to enable more expressive validation scripts that implement large state machines and enforce invariants over whole transaction chains.
They created a new class of state machines based on Mealy machines called Constraint Emitting Machines (CEM). In addition to formalising CEMs, using Agda proof assistant, and demonstrating its conversion to EUTXO, they gave both systems a weak bisimulation.)}%
\ANNannotationinlineDone{
Abstract. Bitcoin and Ethereum, hosting the two currently most valuable and popular cryptocurrencies, use two rather different ledger models, known as the UTXO model and the account model, respectively. At the same time, these two public blockchains differ strongly in the expressiveness of the smart contracts that they support. This is no coincidence. Ethereum chose the account model explicitly to facilitate more expressive smart contracts. On the other hand, Bitcoin chose UTXO also for good reasons, including that its semantic model stays simple in a complex concurrent and distributed computing environment. This raises the question of whether it is possible to have expressive smart contracts, while keeping the semantic simplicity of the UTXO model.

In this paper, we answer this question affirmatively. We present Extended UTXO (EUTXO), an extension to Bitcoin’s UTXO model that supports a substantially more expressive form of validation scripts, including scripts that implement general state machines and enforce invariants across entire transaction chains.

To demonstrate the power of this model, we also introduce a form of state machines suitable for execution on a ledger, based on Mealy machines and called Constraint Emitting Machines (CEM). We formalise CEMs, show how to compile them to EUTXO, and show a weak bisimulation between the two systems. All of our work is formalised using the Agda proof assistant.
}%
introduce Extended UTXO (EUTXO), which extends Bitcoin's UTXO model to enable more expressive forms of validation scripts.
These scripts can express general state machines and reason about transaction chains:
The authors introduce a new class of state machines based on Mealy machines which they call Constraint Emitting Machines (CEM). 
In addition to formalising CEMs using Agda proof assistant, they demonstrate its conversion to EUTXO, and give a weak bisimulation between both systems.
In~\cite{Chakravarty1} Chakravarty et al.\ introduce a generalisation of the EUTXO ledger model using native tokens 
which they denote EUTXOma for EUTXO with multi-assets.
They provide a formalisation of the multi-asset EUTXO model in Agda.
Chakravarty et al.~\cite{Chakravarty} introduce a version of EUTXOma aligned to Bitcoin's UTXO model, hence denoted UTXOma.
They present a formal specification of the UTXOma ledger rules and formalise their model in Agda.
\ANNannotationinlineDone{AB: removed the following:\\
have discovered an alternative design based on Bitcoin-style UTXO ledgers by extending existing UTXO accounting rules to custom assets and asset bundles of various forms. 
Their approach avoids any form of global state for registering assets. 
They provide a basic domain-specific language for developing policy scripts, eliminating the need for general-purpose smart contracts. 
They present a formal specification of the UTXOma ledger rules and formalise their model in Agda.
Chakravarty et al.~\cite{Chakravarty1} investigate a generalisation of the Extended-UTXO ledger model using native tokens specified by the user based on the previous approach~\cite{Chakravarty}. 
They use transaction outputs to lock whole token bundles, including custom tokens whose forging is regulated by forging policy scripts rather than just coins of a single cryptocurrency. 
They explain how this results is a sophisticated ledger model that can accommodate many exciting use cases. 
They formalise this approach in the Agda proof assistant and prove that the custom token ledger is more expressive than the original Extended-UTXO.
}%
Chapman et al.~\cite{Chapmann} formalise System $F_{\omega\mu}$, which is polymorphic $\lambda$-calculus with higher-kinded and arbitrary recursive types, in Agda. 
System $F_{\omega\mu}$ corresponds
\ANNannotationDone{review2, change it from (correspondence to Plutus) to (corresponds to Plutus), FA : Done}%
to Plutus Core, which is the core of the smart contract language Plutus that features in the Cardano blockchain.
%
%
%
Melkonian~\cite{Melkonianformalizinginagda2019} introduces a formal Bitcoin transaction model 
to simulate transactions in the Bitcoin environment 
and to study their safety and correctness. 
The paper presents a formalisation of a process calculus for Bitcoin smart contracts, denoted BitML. 
The calculus 
can accept different types such as basic types, contracts, or small step semantics to outline a `certified compiler'~\cite{Melkoninan:MScThesis:FormalisingExtendedUTXO}.
\ANNannotationDone{according to review2, he asked to add this reference Melkonian~\cite{Melkonianformalizinginagda2019}}%
%


\section{Operational Semantics for Bitcoin Script} \label{sec:verificationBitwithWP}
We give a brief introduction of Bitcoin \Script{} in Subsec.~\ref{subsec:introb},
before defining its operational semantics in Subsec.~\ref{subsec:secops}.

\subsection{Introduction to Bitcoin Script\ANNwhoIsDoingWhatDone{AB}} \label{subsec:introb}
\ANNannotationinlineDone{This section describes the setup and architecture of 
bitcoin \Script:\\
The architecture is a stack machine inspired by forth.\\
we have basic operations, some of which can have quite a complex definition\\
Hint that there are OP\_IF etc. which are treated in a followup paper - in this paper
we ignore them\\
introduce the main examples multisig and P2PKH}
The scripting language for Bitcoin is stack-based,
\ANNannotationinlineDone{review2 said remove this citation ~\cite{bistarelli}
, a @stack-based@ is general elementary term that does not require a
reference, FA : Done}%
inspired
by the programming language Forth~\cite{Rather}, with the stack
being the only memory available.
Elements on the stack are byte vectors,
which we represent as natural numbers.
Values on the stack are also interpreted as truth values, 
any value ${>}0$ will be interpreted as true, and any other value as false.
\Script{} has its own set of commands called \emph{opcodes},
which manipulate the stack. 
They are similar to machine instructions, although
some instructions have a more complex behaviour.
The instructions of \Script{} are executed in sequence. In case of 
conditionals (which are not part of this paper)
the execution of instructions might be ignored
until the end of an if- or else-case has been reached, otherwise
the script is executed from left to right. Execution of instructions might
fail, in which case the execution of the script is aborted.
%
A full list of instructions with their meaning can be found in~\cite{wiki},
which is the defacto specification of \Script. 

\ANNannotationDone{according to review1, we do not need to write this sentence
(All opcodes considered in this paper have been formalised in Agda.
) because we use this sentence (We introduce a number of opcodes that are
relevant to this paper:) }%
The operational semantics of the opcodes can be found in the
source code \cite{setzergithub}.
%
%
%
%
%
We introduce here a number of opcodes that are relevant to this paper.
Execution of all opcodes fails, if there are not sufficiently many elements
on the stack to perform the operation in question.
\ANNannotationDone{to make it clear regarding to review1, I wrote this sentence
(We introduce a number of opcodes that are relevant to this paper, which  we
have been formalised in Agda:) instead of these sentences  (We introduce a
number of opcodes that are relevant to this :) and (All opcodes considered in
this paper have been formalised in Agda.) (note: if you agree I will replace
it)}%
\begin{itemize}
\item \texttt{OP\_DUP} duplicates the top element of the stack.
\item \texttt{OP\_HASH} takes the top item of the stack and replaces it with
its hash.
\item \texttt{OP\_EQUAL} pops the top two elements in the stack and checks
whether they are equal or not, pushing the Boolean result on the stack.
\ANNannotationDone{according to review3 : please add explicitly what happens to
the stack for OP$\_$EQUAL and
 OP$\_$CHECKSIG opcodes: e.g. are the top items removed?, before was (checks
 whether the top two elements in the stack are equal or not) }%
\item \texttt{OP\_VERIFY} invalidates the transaction if the top stack value
 is false. The top item on the stack will be removed.
\item \texttt{OP\_CHECKSIG} pops two elements from the
stack and checks whether
they form a correct pair of a signature and a
public key signing a serialised message obtained from the selected
input and all outputs of the transaction, and pushes
the Boolean result on the stack. 
\ANNannotationDone{according to review3 : please add explicitly what happens to
the stack for OP$\_$EQUAL and
 OP$\_$CHECKSIG opcodes: e.g. are the top items removed?, before was (hashes
 the entire transaction, and checks whether the top two items on the stack
 form a correct pair of a signature and a public key for this hash)}%
\item \texttt{OP\_CHECKLOCKTIMEVERIFY} fails if the time on the stack is
 greater than the current time.
\item \texttt{OP\_MULTISIG} is the multisig instruction, which will be 
discussed in detail in Sect.~\ref{subsec:multisig}.
\item There are a number of opcodes for
pushing byte vectors of differents lengths onto the stack.
We write \texttt{<number>} for the opcode together
with arguments pushing \texttt{number} onto the stack.
In Agda we will  have one instruction
$\AgdaOpPush\;n$ which pushes the number $n$ on the stack.
\end{itemize}%

Scripts can also contains control flow statements such as \texttt{OP\_IF}. 
The verification of scripts involving control statements is more involved and
will be considered in a follow-up paper.

In Bitcoin we consider the interplay between a locking script scriptPubKey
and  an unlocking script scriptSig.%
\footnote{We are using the terminology locking script and unlocking script
from \cite[Chapt~5]{antonopoulos}.}
The locking script is provided by the sender of a transaction to
lock the transaction, and the unlocking script is provided by the recipient
to unlock it. 
\ANNannotationinlineDone{AB: Isn't it exactly the other way round in the following two sentences: 
First unlocking (scriptSig) is executed, then locking (scriptPubKey)?  This is what we had in the submitted version:\\
First the unlocking script is executed.
If it terminates successfully, the stack returned is taken as starting point for the locking script. 
If the locking script terminates successfully and returns true on the stack, 
then the part of the transaction dealing with these particular two scripts succeeds.
}%
The unlocking script pushes the data required to unlock the transaction on the stack,
and the locking script then checks whether the stack contains the required
data. 
Therefore, the unlocking script is executed first, followed by the
locking script.%
%
\ANNannotationDone{according to review3: the notion of locking and unlocking
scripts may look strange for
someone not familiar with Bitcoin: some more explanations would be
welcome; we added this sentence The locking script acts like a contract that
can allow spending the money..... }%
%
\footnote{In the original version of Bitcoin both scripts
were concatenated and executed. However, because Bitcoin script has
non-local instructions
(e.g.~the conditionals \texttt{OP\_IF}, \texttt{OP\_ELSE}, \texttt{OP\_ENDIF}), 
when concatenating the two scripts
any non-local opcode occurring in the locking script (for instance as part of data) 
could be interpreted when running as the counterpart of a non-local opcode in
the locking script
and therefore result in an unintended execution of the unlocking script.
As a bug fix, in a later version of Bitcoin this was modified by having a break point in between the two, where 
only the stack is passed on. See Chapter 6, ``Separate execution of unlocking and locking scripts'' 
in~\cite[p.~136]{antonopoulos}. In this paper this problem doesn't occur because 
we don't consider non-local instructions. }
\ANNannotationinlineDone{before was \ footnote: regarding to REVIEW 2, he
said we need to move it inside the main text and we need to put it between
"(...)" I did it\\
AGS: I don't think he meant putting it into quotation marks}%
%


The main example in this paper is the pay-to-public-key-hash (P2PKH) script
consisting of the following locking and unlocking scripts:
\vspaceBeforeAgdaCode%
\begin{verbatim}
scriptPubKey: OP_DUP OP_HASH160 <pubKeyHash> OP_EQUAL OP_VERIFY OP_CHECKSIG
scriptSig:    <sig> <pubKey>
\end{verbatim}
\vspaceBeforeAgdaCode\par
The standard unlocking script \texttt{scriptSig} pushes
a signature \texttt{sig} and a public key \texttt{pubKey} onto the stack.
The locking script \texttt{scriptPubKey} checks whether \texttt{pubKey}
provided by the unlocking script
hashes to the provided \texttt{pubKeyHash}, and whether the signature is a
signature for the message signed by the public key. 
Full details will be discussed in Sect.~\ref{sec:proofusingstep}.

\subsection{Operational Semantics\ANNwhoIsDoingWhatDone{AB}} \label{subsec:secops}
\ANNannotationinlineDone{This section defines\\
$\semantic{p} : Stack \ar Maybe \;Stack$\\
first for opcodes and then for programs. \\
Discussion of monadic composition of programs, i.e. how to get from \\
$\semantic{p} : Stack \ar Maybe \;Stack$\\
and \\
$\semantic{q} : Stack \ar Maybe \;Stack$\\
to \\
$\semantic{p ++ q} : Stack \ar Maybe \;Stack$
}
Opcodes like \verb|OP_DUP| operate on the stack  defined in Agda as
a list of natural numbers $\AgdaStack$.
Opcodes like \verb|OP_CHECKSIG| check for signatures for the part of the transaction
which is to be signed -- what is to be signed is hard coded in Bitcoin. 
In order to abstract away from the precise format and the encoding,
we define a message type $\MsgAgda$ in Agda, which allows to represent 
messages such as those for the transaction to be signed,
and is to be instantiated with the concrete message to be signed.
Other opcodes like \verb|OP_CHECKLOCKTIMEVERIFY| refer to the current time,
for which we define a type $\TimeAgda$ in Agda.
Therefore, the operational semantics of opcodes depends on
$\TimeAgda \cross \MsgAgda \cross \AgdaStack$ which we define in Agda as the
record type $\StackStateAgda$.%
\footnote{The idea of packaging all components of the state into one product type,
which is then expanded into a more expanded state as more language constructs are added to the language, is
inspired by Peter Mosses' Modular SOS approach \cite{mosses:ModularStructuralOperationalSemantics:JLAP:2004}.
This approach was successful in creating a library of reusable components funcons for defining 
an executable operational semantics of language constructs, which require different sets of states.
One outcome was a ``component-based semantics for $\textsc{CAML LIGHT}$'' \cite{Churchill2015:ReusableComponents}.
}
 Note that $\TimeAgda$ and $\MsgAgda$ don't change
during the execution of a script.

The type of all opcodes is given as $\AgdaInstructionBasic$.%
\footnote{We are using in this paper
a sublanguage $\AgdaFunction{BitcoinScriptBasic}$
of Bitcoin, which doesn't contain conditionals, because they
require a more complex operational semantics and state
(see the discussion in the conclusion). 
We make the distinction between
the basic and full language in order to be compatible with
the planned follow up papers based on code under development,
which will extend the basic language.
We sometimes use notations such as $\AgdaFunction{ᵇ}$ to differentiate
between functions referring to the basic and full language.} 
Opcodes can also fail, for instance if there are not enough
elements on the stack as required by the operation. 
Hence, the operational semantics of an instruction 
\ANNannotationDone{according to review3: 1- variable p is used for both
instructions and script which is a bit confusing
2- same variable can be written with a different font if inside double square
brackets. So I used p (italic) for instruction and p (not italic for
script)\\
AGS: I adjusted it. When using symbolic evaluation we are postulating values,
therefore they become roman blue. I explained this}%
$op\;\AgdaColon\; \AgdaInstructionBasic$
is given as\\ 
$\AgdaSemanticsStackOp{\textit{op}}\;\AgdaColon\; \StackStateAgda\; \AgdaAr\; 
\AgdaMaybe\;\StackStateAgda$.%
\ANNannotationinlineDone{her we change p to \textit{p} according to review3
because we deal with instruction}%
\ANNannotationinlineDone{Shouldn't the two p's be with the same font?}%
\footnote{For the reader not familiar with the $\AgdaMaybe$
type, a set theoretic notation can be given as 
$\AgdaMaybe\;X := \{ \AgdaNothing \} \cup \{ \AgdaJust\;x \mid x : X\}$.
Here, $\AgdaNothing$ denotes undefined, and $\AgdaJust\;x$ denotes the
defined element $x$.
\AgdaMaybe{} forms a monad, with $\AgdaReturn := \AgdaJust :
A\; \AgdaAr\; \AgdaMaybe{}\;A$ and
the bind operation $(p\; \AgdaBind \;q : \AgdaMaybe{}\;B)$ for $p : \AgdaMaybe{}\;A$ 
and $q : A \; \AgdaAr \; \AgdaMaybe\;B$
defined by $(\AgdaNothing\;\AgdaBind\;q) = \AgdaNothing$ and
$(\AgdaJust\;a\;\AgdaBind\;q) = q\;a$.}
\ANNannotationDone{Arnold: do we need a reference for this obvious way? \\
Anton: I just defined it.}%

\ANNannotationinlineDone{it is not ideal to start details about message and
time here, might be better to have some initial part about the setup where
this is introduced. Some preliminary part in which we explain Msg, Time , and
that we have a stack (e.g. in ``Introduction to Bitcoin Script'' or directly
after it}%
The message and time never change, so $\AgdaSemanticsStackOp{\textit{p}}\;s$
\ANNannotationinlineDone{her we change p to \textit{p} according to review3
because we deal with instruction}%
will, if executed successfully, only change the stack part of $s$.
\ANNannotationinlineDone{What is `it' referring to?}%
As an example, we can define the semantics of the instruction $\AgdaOpEqual$.
\ANNannotationDone{AB: the underscore-placeholder in the following interpretations sometimes have different thickness.}%
We first define a simpler function \AgdaSemanticsStackOpssGeneric{},
which abstracts away the non-changing components $\TimeAgda$ and $\MsgAgda$:

\stackSemanticsInstructionsBasicsemanticsstype\par
The function
$\AgdaExecuteStackEquality\;\AgdaColon\;\StackAgda\;\AgdaAr\;\AgdaMaybe\;\StackAgda$
fails and returns $\AgdaNothing$
if the stack has height ${\leq} 1$, 
and otherwise compares the two top  numbers on the stack, replacing them by $1$ for
true in case they are equal, and by $0$ for false otherwise. 
\ANNannotationDone{according to review3, he said (definitions missing for
executeX for all opcodes X (except executeStackEquality that was informally
defined in section 3)}

\AgdaSemanticsStackOpssGeneric{} is then lifted to the semantics of the instructions  
\AgdaSemanticsStackOpGeneric{} using a generic function 
$\AgdaliftStackFunToStackState$:
\semanticsStackInstructionssemanticS

%
\ANNannotationDone{The above needs to be checked}%

As prerequisites for Sect~\ref{subsec:SymbolicExecutionExample P2PKH}, we 
define functions that define
the operational semantics of further Bitcoin instructions used in this paper:
$\AgdaExecuteStackDup\;\AgdaColon\;\StackAgda\;\AgdaAr\; \AgdaMaybe\;\StackAgda$
fails and returns \AgdaNothing{} if the stack is empty; otherwise, a duplicate of
the top element will be added onto the stack. 
The function
$\AgdaExecuteOpHash\;\AgdaColon\;\StackAgda\;\AgdaAr\;\AgdaMaybe\;\StackAgda$
fails and returns \AgdaNothing{} if the stack is empty; otherwise, the top element
is replaced by its
hash. $\AgdaexecuteStackVerify\;\AgdaColon\;\StackAgda\;\AgdaAr\;\AgdaMaybe\;\StackAgda$
fails and returns \AgdaNothing{} if the stack is empty or the top element is 0;
otherwise, it will remove the top element of the stack. 
$\AgdaExecuteStackCheckSig\;\AgdaColon\;\StackAgda \;\AgdaAr\;\AgdaMaybe\;\StackAgda$
fails and returns \AgdaNothing{} if the height of the stack ${\leq} 1$.
Otherwise it pops the two top elements  from the stack, and considers
them as a signature and public key.
It decides whether the message given by the argument
$msg : \MsgAgda$ is correctly signed by these data, and pushes the 
Boolean result on the stack.

\Script{} has instructions with more complex behaviour,
an example is the instruction \verb|OP_MULTISIG| which will be introduced in 
Sect.~\ref{subsec:multisig}. 
Some instructions depend on cryptographic functions for hashing and checking
signatures. 
We abstract away from their concrete definition and take
\ANNannotationDone{before was takes, done}%
them as parameters of the modules of the Agda code. 
This is not a problem in this paper, since the weakest preconditions only 
depend  on the results returned by these functions,
such as a check whether the part of the transaction to be signed is signed by a
signature corresponding to a given public key. 
%
\ANNannotationDone{Anton: I removed sentence`Weak
cryptographic functions make it easier...' since it is confusing (see annotationdone by Arnold)}
\ANNannotationDone{Arnold: I don't understand the last sentence`Weak
cryptographic functions make it easier...' Can we explain more?}
\ANNannotationDone{regarding to review2, he said we need to move from
footnote to the main text, done}%
%

General scripts are formalised in Agda as lists of
instructions, \AgdaBitcoinScriptBasic.
Let $p$ be a script.
We define $\AgdaSemantic{p}
: \StackStateAgda \; \AgdaAr \;\AgdaMaybe\;\StackStateAgda$ by monadic
composition,
that is 
\begin{itemize} 
\item $\AgdaSemantic{\AgdaEmptyList}:= \AgdaJust$, 
\item for an instruction $op$, script $q$ and 
$s : \StackStateAgda$ define
$\AgdaSemantic{op\; \AgdaDoubleColon\; q}\;s := 
\AgdaSemanticsStackOp{op}\; s \;\AgdaBind\;\AgdaSemantic{q}$.
\end{itemize} 
It follows that
$\all s \;\AgdaColon\; \StackStateAgda.\AgdaSemantic{p\; \AgdaDoubleAdd\;
q}\;s \;\AgdaEquiv\; \AgdaSemantic{p}\;s \;\AgdaBind\; \AgdaSemantic{q}$.\\
\ANNannotationDone{FA: regarding to review3, he said it should be: StackState
instead of Stack, So I changed \StackAgda to \StackStateAgda.
Anton: Correct}%
We lift as well $\AgdaSemantic{p}$ to $s : \AgdaMaybe\;\StackStateAgda$ by
defining $\AgdaSemanticPlus{\;p\;}\;s:= s \;\AgdaBind \; \AgdaSemantic{\;p\;}$.\\
Let\\ 
\sPredicatestackstatepred,\\
\stackPredicateStackPred, and \\
$\AgdastackPredtwoSPred\;\AgdaColon\;\AgdaStackPredicate\;\AgdaAr\;
\AgdaStackStatePred$ be the obvious lifting.







\section{Specifying Security of Bitcoin Scripts \ANNwhoIsDoingWhatDone{AGS / AB}}%
\ANNannotationDone{Should it be Bitcoin Scripts, as we are discussing correctness of programs, not a language.}
\ANNannotationinlineDone{new title: Specifying the Correctness of Bitcoin Script. We need to discuss it with Arnold\\[1ex]
Arnold: `Correctness' is very general, shall we go for `Specifying Security of Bitcoin Scripts'?\\[1ex]
Alternative titles\\
Specifying Correctness through Hoare Logic\\
How to Specify the Correctness of Bitcoin Script?\\
Reasoning about script correctness using Hoare logic\\
Denotational Semantics\\
Old title: Hoare Logic, Weakest Preconditions for Security, and Equational Reasoning}%
\label{sec:WPinHTforSecurity}

In this section we argue that weakest precondition in the context of Hoare logic 
are the appropriate notion to express security properties in Subsect.~\ref{subsec:WPforSecurity}.
We provide a formalisation of weakest preconditions in Agda in Subsect.~\ref{subsec:WPInAgda},
and discuss how weakest preconditions can be generated automatically in Subsect.~\ref{subsubsec:AutomaticallyGeneratedWP}, leading to the claim that we need human-readable descriptions of weakest preconditions.
To support our verification, we develop a library for equational reasoning  with Hoare triples in Subsect.~\ref{ourlibrary}\ANNannotationDone{FA: I add this sentence because we do not mention this section }.

\subsection{Weakest Precondition for Security}%
\label{subsec:WPforSecurity}

One widely used way to specify the correctness  of  imperative programs axiomatically
is Hoare logic~\cite{hoare:AxiomaticBasisForComputerProgramming:ACMcommunications:1969}.
Hoare logic is based on pre- and postconditions.
It works well for safety critical systems, 
where the set of inputs is controlled, and the aim is to guarantee a safe result.
An example of a commercial system for writing safety critical systems using Hoare logic is
SPARK 2014 \cite{Adacore:SPARKAda:Homepage}.

However, when dealing with security aspects, in particular access control, 
Hoare logic in general is not sufficient.
The issue is that for security it is necessary to guard against malicious entries to a program.
We argue that weakest preconditions in the context of Hoare logic 
is an appropriate notion to specify security properties.
A weakest precondition expresses that it is not only sufficient, but as well necessary
for the postcondition to hold after executing the program.

To explain our point, we specify the intended correctness of the locking script 
\verb|scriptPubKey| from Sect.~\ref{sec:verificationBitwithWP}. 
The intention, usually given by the user requirement, is that in order for a 
locking script to run successfully,
we need to provide a public key $pbk$ and a signature $sig$ such
that $pbk$ hashes to the value \verb|<pubKeyHash>| stored in the locking script, 
and that $sig$ validates the signed message using $pbk$.
The values $pbk$ and $sig$ need to be the top elements on the stack.
If we also fix their order and allow the stack to have 
arbitrary values otherwise,%
\footnote{Bitcoin scripts do not put any requirements on the stack
below the data required by the scripts.}
then we can express this condition 
as follows:\par 
\medskip
\begin{tabular}{@{}l@{}l}
\begin{tabular}{@{}l}
The two top elements of the stack are $pbk$ and $sig$,
$pbk$ hashes to \verb|<pubKeyHsh>|,\\
and $sig$ is a valid signature of the signed message w.r.t.~$pbk$.
\end{tabular}&(CondPBKH)
\end{tabular} \par
\medskip

We can define the specification of
the locking script \verb|scriptPubKey| as the property that
(CondPBKH) is the weakest precondition for the accepting postcondition.
We will show in Sect.~\ref{sec:proofusingstep} that (CondPBKH)
is indeed the weakest precondition of \verb|scriptPubKey|,
which verifies that \verb|scriptPubKey| fulfils the specification.

Let us now consider a faulty locking script instead of \verb|scriptPubKey|:\par 
\vspaceBeforeAgdaCode
\begin{verbatim}
scriptPubKeyFaulty: OP_DUP OP_HASH160 <pubKeyHash> OP_EQUAL 
\end{verbatim}
\vspaceBeforeAgdaCode
To see that it does not fulfil the specification given above, 
consider the weakest precondition for \verb|scriptPubKeyFaulty| for the accepting postcondition, 
which can be described by the following condition:\par
\medskip
\begin{tabular}{@{}lr}
The top element of the stack is $pbk$, and $pbk$ hashes to \verb|<pubKeyHsh>|. \mbox{\quad}&(CondPBKHfaulty)
\end{tabular}\par
\medskip
By inspection we see that (CondPBKHfaulty) is not equivalent to (CondPBKH), and therefore \verb|scriptPubKeyFaulty| doesn't  fulfil the specification. 
In fact we can identify states which satisfy (CondPBKHfaulty) but not (CondPBKH), 
e.g. a malicious attacker could just copy the public key of the sender onto the stack,
which violates the user requirements of a locking script. 

We observe that this example also demonstrates the inadequacy of general Hoare logic 
for the verification the security property of access control:
Using standard Hoare logic, we can prove that
(CondPBKH) is a precondition for the accepting postcondition for both \verb|scriptPubKey| and \verb|scriptPubKeyFaulty|.

As with all formal verification approaches,
there remains a gap between 
the user's intention expressed as requirements, 
and what is expressed as a formal specification. 
This gap cannot be filled in a  provably correct way, 
since requirements are a mental intention expressed in natural language.
However, the gap can be narrowed by 
expressing the specification in a 
human-readable format 
so that the validation
is as easy and clear as possible.
Here, validation means showing that the specification guarantees the
requirements, and is carried out by a human reader.

\subsection{Formalising Weakest Preconditions in Agda}%
\label{subsec:WPInAgda}

We now describe how weakest preconditions can be defined in Agda.
%
%
Let a precondition $\varphi$ and postcondition $\psi$ be given, 
both of type  \AgdaStackStatePred.
In order to accommodate $\Maybe$, we define a postfix operator $\AgdaUnderscore\AgdaLiftPlus$,
\ANNannotationDone{why are we lifting again?\\
Anton: We lift it to Maybe}%
to lift $\psi$ to $(\psi\; \AgdaLiftPlus) : \AgdaMaybe \;\StackStateAgda\;\AgdaAr\;
\AgdaSet$, defining $(\psi\;\AgdaLiftPlus)\; \AgdaNothing = \AgdaBot$
and $(\psi\;\AgdaLiftPlus) \circ \AgdaJust = \psi$.

A Hoare triple, consisting of a precondition, a program, and a postcondition, expresses that if the precondition is satisfied before execution of the program, then the postcondition holds after executing it.
We formalise Hoare triples as follows: 

%
%
\ANNannotationDone{this notes from review3, he asked to change $p$ to p because we use p as program not an instruction, so I changed it, before was {\;p\;}\\
Arnold: changed it to $p$ so that it is the same style as the second occurrence}%
%
\;\;\;\;\;$\Agdaleft\;\varphi\;\Agdaright\;p\;\Agdaleft\;\psi\;\Agdaright\;$:= $\all s$ $\in \StackStateAgda.\varphi(s) \;\AgdaAr\; (\psi\;\AgdaLiftPlus)\;(\AgdaSemantic{p}\;s)$

Weakest preconditions express that the precondition not only is sufficient, but as well necessary
for the postcondition to hold after executing the program:

%
\ANNannotationDone{FA FA : (Done and needs to be checked please): this notes from REVIEW 3 before was\[\AgdaHoareIff{\;\varphi\;}{\;p\;}{\;\psi\;} := \all s \in \StackStateAgda.\varphi(s) \;\iff\; (\psi\;\AgdaLiftPlus)\;(\AgdaSemantic{p}\;s)\]}%
%
\;\;\;\;\;$\Agdaleft\;\varphi\;\Agdarightiff
\;p\;
\Agdaleft\;\psi\;\Agdaright\;$:= $\all s$ $\in \StackStateAgda.\varphi(s) \;\iff\; (\psi\;\AgdaLiftPlus)\;(\AgdaSemantic{p}\;s)$
\ANNannotationDone{Arnold: shouldn't p's be of same font?}

Thus, for security the backwards direction of the equivalence in the previous formula is the important direction.

In Bitcoin we consider a locking script scriptPubKey
and  an unlocking script scriptSig, see Section~\ref{subsec:introb}.
Let us fix an unlocking script $unlock$ and a locking script $lock$.
Let $init$ be the
initial state consisting of an empty stack, and let $\AgdaaceeptCond$ be the accepting condition expressing that the stack is non empty with top element being not false, i.e.~${>}0$.
The combination of $unlock$ and $lock$ is accepted iff 
running $unlock$ on $init$ succeeds and running $lock$ on the resulting stack results in a state that satisfies  the accepting condition, 
i.e.~iff
$(\AgdaaceeptCond\;\AgdaLiftPlus)\;(\AgdaSemanticPlus{\;lock\;}\;(\AgdaSemantic{\;unlock\;}\;init))$.
Note that Bitcoin does not run the concatenation of the two scripts, as it did in its first version,
but runs 
first the unlocking scripts, and if it succeeds runs the locking script on the resulting stack.
%
Let $\varphi$ be the weakest precondition of $lock$,
i.e.~$\AgdaHoareIff{\;\varphi\;}{\;lock\;}{\;\AgdaaceeptCond\;}$.
Then the acceptance condition is equivalent to 
$(\varphi\AgdaLiftPlus)\;(\AgdaSemantic{\;unlock\;}\;init)$.
\ANNannotationDone{Anton: I clarified that we don't run the concatenation of scripts in Bitcoin but the
two scripts separately, in response to a comment by probably Arnold see ANNannotation}%
\ANNannotationDone{AB: Do we need a general statement supporting the above argument, something like:\\
Assume $<\phi>^{iff} p<\psi>$,
then $\psi^+([[p++q]] s)$ is equivalent to 
$\phi^+([[q]]s)$.
}%
Thus, $unlock$ succeeds 
iff running the unlocking script $unlock$ on the initial state $init$ produces a state fulfilling $\varphi$. 
Hence, by determining the weakest precondition for the 
locking script w.r.t.~the accepting condition we have obtained a characterisation of the set of unlocking scripts which unlock the locking script.
Note that we do not define inductively all successful unlocking scripts, since they could be arbitrary complex programs, but instead characterise them by the output they produce.

\subsection{Automatically Generated Weakest Preconditions}\label{subsubsec:AutomaticallyGeneratedWP}

\ANNannotationDone{needs to find its final place!. FA: I discussed it with Anton and the best place for it in 4.1 as subsubsection}

We start by giving a direct method for defining the weakest precondition for any Bitcoin script by describing the set of states that lead to a given final state.  
We then apply this general method to a toy example to demonstrate that the description obtained in this way is usually not helpful for a human to judge whether the script has the right properties, thus making the case that the task must be to find (equivalent) human-readable descriptions.

Weakest preconditions can be defined by the simple definition 
\vspaceBeforeAgdaCode%
\exampleGeneratedWeakPreCondweakestPreCond
\vspaceAfterAgdaCode\par
Consider a simple toy program which removes the top element from the stack three times:
\exampleGeneratedWeakPreCondtestprog\\
Its weakest precondition can be computed as\\
\exampleGeneratedWeakPreCondweakestPreCondTestProg\\
We obtain the following code (we slightly reformatted it to improve readability):
\vspaceBeforeAgdaCode%
\exampleGeneratedWeakPreCondweakestPreCondTestProgNormalised
\vspaceAfterAgdaCode\par
This condition is difficult to understand. 
The reason is that each instruction may cause the program to abort in case the stack is empty. 
The condition expresses: 
if the stack is empty then the condition is false. 
Otherwise, if after dropping the top element the stack is empty the condition is false. 
Otherwise, if after dropping again the top element the stack is empty the condition is false.
Otherwise the condition is true if after dropping again the top element the stack
is non empty and the top element is not false. 
The readable condition would express that the height of the stack is $\geq 4$ 
and the fourth element from the top is $>0$. 
In this simple example simplifying the condition would be easy, 
but when using different instructions the situation becomes more complicated.

What we did using our methods to avoid this problem was to create the weakest precondition by starting
from the end and improving it in each step, or by replacing the program by an easier program
(which in case of this example would return nothing if the stack has height $\leq 2$ and otherwise
returns the result of dropping the first three elements off the stack). 
An interesting project for future work 
would be to automate the steps we carried out manually, and obtain readable
weakest preconditions automatically.

\subsection{Equational Reasoning with Hoare Triples\ANNwhoIsDoingWhatDone{AB}}
\label{ourlibrary}
\ANNannotationinlineDone{We introduce the syntax introduced into Agda for
equational style reasoning  so that for instance what is currently
below the lemma can be shown by combining individual programs and proofs together.
}

To support the verification of Bitcoin scripts with Hoare triples and weakest preconditions in Agda, 
we have developed a library in Agda for equational reasoning with Hoare triples. 
The library is inspired by what is described in Wadler et al.~\cite{Wadler}. 
Let $p,q$ be scripts and $\phi,\phi',\psi,\psi': \PredicateAgda$.
If  we define
$\varphi\; \AgdaEquivPred \;\psi := \all s : \StackStateAgda.\varphi(s) \iff 
\psi(s)$,
\ANNannotationDoneQuestion{Anton: Added comment regarding :<=>  in the google docs document for referee 3}
\ANNannotationDone{write to 3rd referee and ref1:\\
In line 4 of 4.2 we replaced as suggested by the referee $:\Leftrightarrow$ by
$:=$. However, we want to stress that it was correct, it is common practice in logic when we define predicates to use the notation $\phi :<=> definition$.
$:=$ would be appropriate if the right side is exactly type theoretic code
or Agda code, but here it is just a logical formula which easily can be 
translated into a corresponding type theoretic formula (we used the logical 
formula to improve readability of the paper.}%
we can easily show 
\ANNannotationDone{FA (Done but need to be check it) :Regarding to review3: state briefly the type/nature of p,q and $\phi,\phi',\psi,\psi'$, I added (Note that: p and q are represented as programs, and $\phi,\phi',\psi,\psi'$ of types \PredicateAgda }%
\vspaceBeforeAgdaCode%
\[\begin{array}[t]{@{}lclcl} 
\AgdaHoareIff{\;\varphi\;}{\;\;p\;\;}{\;\psi\;} &\land &
\AgdaHoareIff{\;\psi\;}{\;\;q\;\;}{\;\rho\;}  &\AgdaAr&
\AgdaHoareIff{\;\varphi\;}{\;\;p\;\AgdaDoubleAdd\;q\;\;}{\;\rho\;}\\
\AgdaHoareIff{\;\varphi\;}{\;\;p\;\;}{\;\psi\;} &\land &\psi \;\AgdaEquivPred\;
\psi'\; &\AgdaAr& 
\AgdaHoareIff{\;\varphi\;}{\;\;p\;\;}{\;\psi'\;}\\
\varphi'\; \AgdaEquivPred \;\varphi
&\land & \AgdaHoareIff{\;\varphi\;}{\;\;p\;\;}{\;\psi\;}  &\AgdaAr& 
\AgdaHoareIff{\;\varphi'\;}{\;\;p\;\;}{\;\psi\;}
\end{array} \]
\vspaceBeforeAgdaCode\par
We demonstrate our syntax by an example, assuming (using Agda postulate)
programs $\AgdaFunction{prog1}$, $\AgdaFunction{prog2}$, 
$\AgdaFunction{prog3}$,
and proofs
\vspaceBeforeAgdaCode%
\demoEqualityReasoningproofs
\vspaceAfterAgdaCode\par
Then the proof for the Hoare triple for 
$\AgdaFunction{prog1}\; \AgdaDoubleAdd\; (\AgdaFunction{prog2} \;\AgdaDoubleAdd\; \AgdaFunction{prog3})$  
is given in Agda as follows:%
\footnote{In the last step we use 
$\Agdastepl$ instead of $\Agdasteplnoe$. This avoids concatenating the program with $\AgdaEmptyList$.
If we used $\Agdasteplnoe$, the theorem would prove the condition for program
$\AgdaPostulate{prog1}\AgdaSpace{}%
\AgdaOperator{\AgdaFunction{\ensuremath{+\!\!+}}}\AgdaSpace{}%
\AgdaSymbol{(}\AgdaPostulate{prog2}\AgdaSpace{}%
\AgdaOperator{\AgdaFunction{\ensuremath{+\!\!+}}}\AgdaSpace{}%
\AgdaSymbol{(}\AgdaPostulate{prog3}\AgdaSpace{}%
\AgdaOperator{\AgdaFunction{\ensuremath{+\!\!+}}}\AgdaSpace{}%
\AgdaEmptyList\AgdaSymbol{)} \AgdaSymbol{)} $, which is provably but not definitionally
equal to the original program, requiring an additional proof step.}
\vspaceBeforeAgdaCode%
\demoEqualityReasoninglemma 
\vspaceAfterAgdaCode\par
\ANNannotationDone{FA : (Done and needs to be checked please) review3 asked to explain $\Agdastepl$}%
%
%
%


\section{Proof of Correctness of the P2PKH script using the Step-by-Step Approach\ANNwhoIsDoingWhatDone{AB}} \label{sec:proofusingstep}
%
%
\ANNannotationinlineDone{We introduce how to get from the accepting condition 
instruction by instruction backwards to the weakest precondition.\\
We will just mention the condition, as far as possible in a mathematical
style (maybe give some Agda examples) as well, and leave the proofs
to the git repository.}
This section explains the usage of our approach by providing an example of
how to prove the correctness of the 
\ANNannotationDone{before was Pay to Public
Key Hash}%
P2PKH using step-by-step 
\ANNannotationinlineDone{before was step by
step}%
to obtain the weakest precondition. %
The 
\ANNannotationDone{before was Pay to Public Key Hash}%
P2PKH is the most used script in Bitcoin
transactions. The locking script, which depends on a public key hash, is
defined  as follows:
%
%
\ANNannotationDone{regarding to review2, he said we need to delete it superscript $ᵇ$ for basic indicates that we are
referring to the language of Bitcoin without opcodes for conditionals --
conditionals would require an extension of the state space.}%
\par 
\ANNannotationDone{AB: Should we add the type of scriptP@PKH as we do later for multiSigScript2-4 and checkTimeScript?}
\vspaceBeforeAgdaCode%
\stackVerificationPtoPKHscriptppkh
\vspaceAfterAgdaCode\par
In this section we develop a readable weakest precondition of the P2PKH
script and  prove its correctness by working backwards instruction by
instruction. 

Let \AgdaAcceptState{} be the accepting state where the stack is non-empty with top element  ${>}0$. 
We define intermediate conditions $\AgdaAcceptone$,
$\AgdaAccepttwo$, etc, 
the weakest precondition \ANNannotationDone{before was condition} $\AgdaWPreCondPtoPKH$,
and proofs $\AgdaCorrectopchecksig$,$\;\AgdaCorrectopverify$
\ANNannotationDone{FA: according
to review3, he needs to change the name to \AgdaCorrectopchecksig etc instead
of using \AgdaCorrectone so I added new macros for these corrects and I
changed the name for it}%
etc of corresponding
Hoare triples w.r.t.~the instructions of the Bitcoin script, working 
backwards starting from the last instruction $\AgdaOpCheckSig$:\\ 
%
\stackVerificationPtoPKHcorrectone\\
\stackVerificationPtoPKHcorrecttwo\\
\stackVerificationPtoPKHcorrectthree\\
\stackVerificationPtoPKHcorrectfour\\
\stackVerificationPtoPKHcorrectfive\\
\stackVerificationPtoPKHcorrectsix\\


The intermediate conditions can be read off from the operations.
We present them in mathematical notation below, using the following
conventions and abbreviations: 
\ANNannotationDone{I change only the order between t and m before was
$m: \AgdaMsg$, $t: \AgdaN$ denotes time, and now $t: \AgdaN$, $m: \AgdaMsg$}
$t: \AgdaN$ denotes time, $m: \AgdaMsg$, $st,st' : \AgdaStack$, $x : \AgdaN$;
for brevity we omit types after $\exists$ quantifiers. We use here and in the
remaining paper $ˢ$ for operations where the $\StackStateAgda$ argument has
been unfolded into its components.

\smallskip 
\ANNannotationDone{The mathematical formulas need to be checked carefully}\par
\ANNannotationDone{Should the ${}^s$ come after ${}_1$ as the unfolding is applied last?}
\noindent$\begin{array}[t]{@{}l@{\;}c@{\;}l@{}l@{}l} 
\AgdaAcceptStateSt\;t\;m\;st &\Leftrightarrow&
 \exists\; x,st'.&st\; \AgdaEquiv\; x\; \AgdaDoubleColon\;st' \;\land x > 0\\
\AgdaAcceptStone\;t\;m\;st &\Leftrightarrow&
 \exists\; pbk,sig, st'.&st\; \AgdaEquiv\; pbk\; \AgdaDoubleColon\;sig\; \AgdaDoubleColon\;st'
\\
&&\multicolumn{3}{l}{\land\; \AgdaIsSigned\;m\;sig\;pbk}\\
\AgdaAcceptSttwo\;t\;m\;st &\Leftrightarrow&
 \exists\; x,pbk,sig, st' .&st\; \AgdaEquiv\;x\; \AgdaDoubleColon\;pbk\; \AgdaDoubleColon\;sig\; \AgdaDoubleColon\;st' \\
&&\multicolumn{3}{l}{\land\;x > 0 \;\land\;\AgdaIsSigned\;m\;sig\;pbk}\\
\AgdaAcceptStthree\;t\;m\;st &\Leftrightarrow&
 \exists\; \pbkhtwo{},\pbkhone{},pbk,sig, st'
 .&st\; \AgdaEquiv\;\pbkhtwo{}\; \AgdaDoubleColon\;\pbkhone{}\; \AgdaDoubleColon\;pbk\; \AgdaDoubleColon\;sig\; \AgdaDoubleColon\;st'
\\ 
&&\multicolumn{3}{l}{\land\;\pbkhtwo{}\; \AgdaEquiv\; \pbkhone{} \land\; \AgdaIsSigned\;m\;sig\;pbk} \\
\AgdaAcceptStfour\;\pbkhone{}\;t\;m\;st &\Leftrightarrow&
 \exists\; \pbkhtwo{},pbk,sig,st'. &st\; \AgdaEquiv\;\pbkhtwo{}\;\AgdaDoubleColon\;pbk\; \AgdaDoubleColon\;sig\; 
\AgdaDoubleColon\;st'\\
&&\multicolumn{3}{l}{\land\; \pbkhtwo{}\; \AgdaEquiv\; \pbkhone{}\land\; \AgdaIsSigned\;m\;sig\;pbk} \\
\AgdaAcceptStfive\;\pbkhone{}\;t\;m\;st &\Leftrightarrow&
 \exists\; \pbkone{},pbk,sig, st'
 .&st\; \AgdaEquiv\;\pbkone{}\;\AgdaDoubleColon\;pbk\; \AgdaDoubleColon\;sig\; \AgdaDoubleColon\;st' \\
&&\multicolumn{3}{l}{\land\;\AgdaHashFun\; \pbkone{}\; \AgdaEquiv\; \pbkhone{} \land\; \AgdaIsSigned\;m\;sig\;pbk} \\
\multicolumn{3}{@{}l}{\AgdaWPreCondPtoPKHst\;\pbkhone{}\;t\;m\;st
\Leftrightarrow
 \exists\; pbk,sig, st'.}&st\; \AgdaEquiv\;pbk\; \AgdaDoubleColon\;sig\; \AgdaDoubleColon\;st'\\ 
&&\multicolumn{3}{l}{\land\;\AgdaHashFun\; pbk \;\AgdaEquiv\; \pbkhone\; \land\; \AgdaIsSigned\;m\;sig\;pbk} \\
\end{array} $ \par 
\ANNannotationinlineDone{Trying out an alternative, where  we omit the existential quantifiers:\\
In the following formulas the free variables on the right hand side are considered to be
existentially quantified (e.g. in the first formula
one needs to add $\exists x,st'.$)}
\ANNannotationinlineDone{$\begin{array}[t]{@{}l@{\;}c@{\;}l@{}l@{}l} 
\AgdaAcceptStateSt\;m\;t\;st &\Leftrightarrow&
st\; \AgdaEquiv\; x\; \AgdaDoubleColon\;st' \;\land x > 0\\
\AgdaAcceptStone\;m\;t\;st &\Leftrightarrow&
st\; \AgdaEquiv\; pbk\; \AgdaDoubleColon\;sig\; \AgdaDoubleColon\;st'\land \AgdaIsSigned\;m\;sig\;pbk\\
\AgdaAcceptSttwo\;m\;t\;st &\Leftrightarrow&
st\; \AgdaEquiv\;x\; \AgdaDoubleColon\;pbk\; \AgdaDoubleColon\;sig\; \AgdaDoubleColon\;st' \land\;x
> 0 \;\land\;\AgdaIsSigned\;m\;sig\;pbk\\
\AgdaAcceptStthree\;m\;t\;st &\Leftrightarrow&
st\; \AgdaEquiv\;\pbkhtwo{}\; \AgdaDoubleColon\;\pbkhone{}\; \AgdaDoubleColon\;pbk\; \AgdaDoubleColon\;sig\; \AgdaDoubleColon\;st' 
\land\;\pbkhtwo{}\; \AgdaEquiv\; \pbkhtwo{}\\ 
&&\land\; \AgdaIsSigned\;m\;sig\;pbk \\
\AgdaAcceptStfour\;\pbkhone{}\;m\;t\;st &\Leftrightarrow&
st\; \AgdaEquiv\;\pbkhtwo{}\;\AgdaDoubleColon\;pbk\; \AgdaDoubleColon\;sig\; \AgdaDoubleColon\;st'\land\; \pbkhtwo{}\; \AgdaEquiv\; 
\pbkhone{}\\
&&\land\; \AgdaIsSigned\;m\;sig\;pbk \\
\AgdaAcceptStfive\;\pbkhone{}\;m\;t\;st &\Leftrightarrow&
st\; \AgdaEquiv\;\pbkone{}\;\AgdaDoubleColon\;pbk\; \AgdaDoubleColon\;sig\; \AgdaDoubleColon\;st' \land\;\AgdaHashFun\; \pbkone{}\; 
\AgdaEquiv\; \pbkhone{}\\ 
&&\land \AgdaIsSigned\;m\;sig\;pb \\
\AgdaWPreCondPtoPKHst\;\pbkhone{}\;m\;t\;st
&\Leftrightarrow&
st\; \AgdaEquiv\;pbk\; \AgdaDoubleColon\;sig\; \AgdaDoubleColon\;st'
\land\;\AgdaHashFun\; pbk \;\AgdaEquiv\; \pbkhone\; \\
&&\land \AgdaIsSigned\;m\;sig\;pbk \\
\end{array} $
}
\smallskip 
\ANNannotationinlineDone{Arnold: observe that the order of arguments changes: above it is $m\ t$, below $time\ msg$. FA: I only change the formulas above because our code is correct and we do not need to change it}%
In Agda, these formulas are defined by case distinction on the stack.
As examples, the code for the 
accept condition (\AgdaAcceptState) and  the 
weakest precondition (\AgdaWPreCondPtoPKHst) is as follows: 
\vspacetwocm%
\stackPredicateacceptState 
\vspacethreecm \par
\vspacezeroonecm 
\verificationptopbasicweakestprecondition 
\vspacezerofourfivecm\par
Using our syntax for equational reasoning,
we can prove the weakest precondition for the P2PKH script as follows:
\vspaceBeforeAgdaCode%
\stackVerificationPtoPKHmainthro
\vspaceAfterAgdaCode\par

The locking script will be accepted if, after executing the code starting
with the stack returned by the unlocking script, the accept
condition \AgdaAcceptState{} is fulfilled. 
The verification  conditions and proofs were developed by working backwards starting from the last instruction 
and determining%
\ANNannotationDone{regarding review2, before was the weakest preconditions
(i.e.~conditions \AgdaAccepti{})} 
the weakest preconditions ``\AgdaAccepti{}''
w.r.t.~the end piece of the script starting with 
that instruction  and the accept condition as post-condition. 
The preconditions were obtained manually -- one could 
automate this by determining for each instruction depending on the post-condition a
corresponding pre-condition, where the challenge would be to simplify
the resulting pre-conditions in order to avoid a blowup in size.
We continued in this way until we reached the first instruction and obtained the weakest precondition for the locking script.
\AgdaFunction{theoremP2PKH} is using single
instructions in order to prove the correctness of P2PKH.
The proofs \AgdaCorrectopchecksig, \AgdaCorrectopverify,
\ANNannotationDone{I also changed it regarding review3 from \AgdaCorrectone, \AgdaCorrecttwo to \AgdaCorrectopchecksig and \AgdaCorrectopverify}%
etc are done by following the case
distinctions made in the corresponding verification conditions. The harder direction
is to prove that they are actually \emph{weakest} preconditions: 
Proving that the precondition implies the postcondition
after running the program, is easier since we are used to
mentally executing programs in forward direction.
Proving the opposite direction requires showing
that the only way, after running the program, to obtain 
the postcondition is to have the precondition fulfilled, 
which requires mentally reversing the execution of programs.
\ANNannotationDone{Anton: Made an attempt of explaining it, comment was:\\
according to review3 and review1, they ask to
explain it in more details (The harder direction...)}%

\ANNannotationinlineDone{Old text:}
\ANNannotationDone{To verify the correctness of scripts
in \AgdaFunction{scriptP2PKHbas} we have defined \AgdaFunction{theoremP2PKH}
where the correctness of this script is passed by using the Hoare logic with
the weakest precondition.   
\AgdaFunction{theoremP2PKH} verifies if the weakest precondition and the postcondition led to the accepting state. 
Since we are using the syntax \AgdaHoareTriple{$\Phi$}{p}{$\Psi$} which means
that if $\Phi$ holds, then $\Psi$ holds and vice versa.
In order to arrive at an accepting state, one should be left with a stack
that has its height of at least one element and the top element $>$ 0. 
Once all these conditions have been added to the stack and checked, the
script will be validated if it meets the weakest precondition (acceptCond)
and it restarts checking every condition by running the iteration backwards.Finally, \AgdaFunction{theoremP2PKH} is holding only if both iterations will
arrive in an accepting state.
However, when
pushing the public key hash on the stack, the instruction becomes
composed. 
}
\ANNannotationinlineDone{Add somewhere above:\\
The locking script will be accepted if after executing the code starting with
the stack returned by the unlocking script the acceptCond is fulfilled. 
We work backwards starting from the last instruction 
and determine the weakest preconditions (i.e.~conditions \AgdaAccepti{}) w.r.t.~the 
end piece of the program and the accept condition, 
until we have reached the first instruction and obtained the weakest precondition for the locking script.}
\ANNannotationDone{go above:\\ 
\AgdaFunction{theoremP2PKH} is using single
instructions in order to prove the correctness of P2PKH.}
\ANNannotationinlineDone{Should go to multiscript where 
actually there could be a short discussion that 
OP\_Multisig has a very complex precondition which is 
difficult to handle, but 
in combination with instructions before it becomes
easier
The precondition for the multiscript instruction
would need to express there is a number m on the
stack below it there are m public keys, then
there is anumber k on the stack
and below there are k signatures, and below there
is a dummy element and the public keys match
the signatures.\\
If we instead use the concrete script MULTISIG2-4
then the precondition is much simpler because
we know how many public keys we need to have.
This demonstrate that sometimes it is easier to work
with sequences of instructions rather than individual ones in the first approach.
}
\ANNannotationinlineDone{Does this need to go anywhere?\\
By noticing the latter, we understood that it is more efficient to
use whole programs than single instructions. The complication is due to the
fact that single instructions can lead to failure, and therefore the
operational semantics of the resulting programs become quite long and
complicated. In order to avoid long and complicated programs, we used
intermediate conditions, making the programs manageable.}

\section{Proof of Correctness using Symbolic Execution\ANNwhoIsDoingWhatDone{AGS- AB}} \label{sec:proofusingsymbloic}

\ANNannotationinlineDone{Referee 3 wants an explanation what problem to solve. 
We want to obtain readable weakest preconditions for bitcoin scripts, and develop a 
methodology for creating them. 
For the examples we are using, which are well studied examples, finding the correct weakest precondition
would not be so complex, although when finding them we had some surprises, because usually when 
thinking about the correctness one thinks about preconditions, not weakest preconditions.
We need to elaborate much more on this weakest precondition for access control. (We need a good explanation and maybe
Arnold can help with came). \\
The key question which is answered in this question is: what does it mean for a Script in Bitcoin to be correct, which is 
intermediate step towards answering the more difficult question towards what does it mean for a smart contract in 
Ethereum to be correct). 
(We need of course whenever we talk about Script say the weak form of smart contract as it occurs in Bitcoin, as formalised using Bitcoin Script,  and need to avoid using the word "Smart contract" and use "Bitcoin Script" instead).
The reader might think that it is that one looks for a precondition, but that doesn't guard against malicious 
interventions where one needs to bypass the intended precondition and still get access to the resources (Bitcoins).
In order to guard against malicious interventions we use instead weakest preconditions. 
Which means: in order to unlock it, the writer of the unlocking script needs to write a script which results in a 
Stack which fulfils the weakest precondition. 
Important: it is not possible to say inductively define all scripts which are accepted, because they can be arbitrary
programs. But we can define the condition any output of any unlocking script needs to be fulfilled.
So to unlock it in case of P2PKH you need to provide a script which produces at the end a stack containing the public key hashing to the given public key hash and a signature. Since the unlocker can execute the script the unlocker needs to have knowledge of this data.
The key point of the paper is: the correctness of a Bitcoin script can be specified by its weakest precondition.
The 2nd point is that these conditions need to be readable, because the writer of the locking script needs to be able to see that this precondition is what he or she intends to be meant. 
If we get an unreadable long formula, the writer of the locking script might misinterpret it and not see that this is not what he or she meant by it. 
So we want to develop a methodology which allows to determine such kind of readable weakest precondition. 
Of course, the examples we are using a well-understood scripts, so it is not so complicated to determine the weakest conditions, although when actually carrying them out we had some minor surprises in the intermediate conditions, where sometimes the weakest precondition wasn't what we expected. (would be nice to have an example).
}
\ANNannotationinlineDone{Describe symbolic execution.\\
We postulate variables for the arguments and look at the result of evaluating it.\\
Then we check where there is a case distinction, and make case distinctions
on that argument.\\
We determine in some cases the result directly and prove that it is equal.
in other cases we continue in the same way until we arrive at the final function.\\
We show that this function is for all arguments equal to the original function.\\
Now we determine a weakest pre condition by looking at the branches where it succeeds.\\
Then we obtain the weakest precondition
}

In this section, we will introduce a second method for obtaining readable
representations 
of weakest preconditions of Bitcoin scripts. 
This method is based on symbolic execution \cite{king:SymbolicExecutionProgramTesting} 
of the Bitcoin script,
and investigating the sequence of case distinctions carried out during the
execution. 
\ANNannotationDone{AB: I suggest to remove the following sentence:\\
By investigating each sequence, we are able to prove that the
script will run without errors, providing an efficient tool to verify
scripts.
}%
\ANNannotationDone{Anton: Added reference to classic article on symbolic execution. 
I'm not sure that this is what referee wants. Comment was:\\
according to review3: referencing about symbolic execution}%
We will consider three examples: 
The first will be the P2PKH script which we analysed already. 
We use it to explain the method 
and provide a second approach to determine and verify the already obtained
weakest precondition.
The second example will consider the multisig script which is a direct application 
of the \texttt{OP\_MULTISIG} instruction.
The third example will see an application of a combination of both methods.

\subsection{Example: P2PKH Script}
\label{subsec:SymbolicExecutionExample P2PKH}

When applying the symbolic evaluation method to the P2PKH script and
analysing the sequence of case distinctions carried out, we will see that
there will be exactly one path through the tree of case distinctions 
which results in an accepting condition. The conjunction of
the cases that determine this path will form the weakest
precondition. In examples  with more than one accepting path we would take
the disjunction of the conditions for each accepting path.%
\footnote{In our examples we got
only a few accepting paths, since concrete scripts in use are designed to 
deal with a small number of different scenarios for unlocking them,
so the majority of paths in the program are unsuccessful paths.
It could happen however that with more advanced
examples nested conditions result in an exponential
blowup of the number of cases -- if that occurs one would need to take an
approach where the nested case distinctions are preserved at least
partly and the resulting extracted formulas reflect those nested
case distinctions rather than flattening them out. This would avoid the
blowup in the size of the resulting weakest precondition.}
We will prove that the precondition is indeed the
weakest\ANNannotationinlineDone{before was the weakest precondition is indeed
a representation of the weakest precondition}
by developing an equivalent program \ptopkhFunctionDecoded{}
and showing that it fulfils the weakest precondition.

We start by declaring (using Agda's \AgdaKeyword{postulate})
symbolic values  
\AgdaPbkhPostulate, \AgdaMsgOnePostulate, \AgdaStackOnePostulate, 
\AgdaStackxOnePostulate, etc for the 
parameters (postulates are typeset in \AgdaFunction{blue}). This allows us to evaluate expressions up to $\AgdaexecuteStackVerify\;$
symbolically by using the normalisation procedure of Agda
\ANNannotationDone{FA: according to
review1 ("This allows us to evaluate expressions symbolically..." -- could
you clarify what
you use for evaluation? Is it Agda's normalisation?), I changed it the
sentence to This allows us to evaluate expressions symbolically by using the
normalisation in Agda. Before was This allows us to evaluate expressions
symbolically and.... In addition, according to review3 (say explicitly that
evaluation up to executeStackVerify has been done) (Done and needs to be checked please)} 
and to
determine the function 
\ptopkhFunctionDecoded{}. (In Sect.~\ref{sect:PracticalUsageOfAgda} we 
will elaborate how to do this practically in Agda).
Afterwards, we stop using those postulates (they were defined as \AgdaPrivate{})
and prove that the result of evaluating the P2PKH script for arbitrary parameters
is equivalent
to \ptopkhFunctionDecoded{}.\par 
%
When 
evaluating
\ANNannotationDone{AGS: Resolved - this was an Agda bug, resolved by
replacing two spaces by one in Agda\\
AB: Should there be a space between $[[..]]$stb and $\mathrm{time}_1$?}  
\ANNannotationDone{AB: Do we need to explain what $[[..]]$stb is?
AGS: I replaced it by $]]^s$, which uses now the same convention that ${}^s$
means that a stack predicate is unfolded. See explanation on top of page 9,
where this is introduced, I changed it from predicates to ``operations''. I'm
not sure whether this is enough.}
\stackVerificationPtoPKHsymbolicExecutiontestPtoP{} 
we obtain \par 
\vspaceBeforeAgdaCode%
\stackVerificationPtoPKHsymbolicExecutiontestPtoPKHscripttwoUsingMoreSpace
\vspaceAfterAgdaCode\par
We can write it equivalently using the do notation%
\footnote{The do notation is a widely used Haskell notation adapted to Agda,
which provides an alternative syntax for the same expression
making it appear as an imperative program if one reads $\Agadaleftarrow$ as
assignments. It demonstrates that we are consecutively executing the
instructions, with
the possibility of aborting in each step.}\par 
\vspaceBeforeAgdaCode
\stackVerificationPtoPKHsymbolicExecutiontestPtoPKHscripttwoUsingDo
\vspaceAfterAgdaCode\par
At this point further 
reduction is blocked by the first line of the previous expression, because 
$\AgdaFunction{executeStackDup}\;\AgdaPostulate{stack₁}$ makes a case
distinction on $\AgdaPostulate{stack₁}$.
Therefore, we introduce a symbolic case distinction on $\AgdaPostulate{stack₁}$:
\begin{itemize} 
\item 
\stackVerificationPtoPKHsymbolicExecutionEmpty{} evaluates to
$\AgdaNothing$.
\item 
\stackVerificationPtoPKHsymbolicExecutionnonestack{} evaluates to 
what in do notation can be written as\par 
\ANNannotationDone{FA: in the function below I
change it from stack₂ to stack₅ according to review3} 
\vspaceBeforeAgdaCode%
\stackVerificationPtoPKHsymbolicExecutionNonEmptyStackNormalFormUsingDo
\vspaceAfterAgdaCode\par
\end{itemize} 
Evaluation of the latter expression is blocked by the function
$\AgdaexecuteStackVerify$ which makes
a case distinction on the
expression \stackVerificationPtoPKHsymbolicExecutionNonEmptyStackNormalFormFirstPartZoomedIn.
We define\par
\ANNannotationDone{FA: in the function below I change it from stack₂ to
stack₅ according to review3} 
\vspaceBeforeAgdaCode%
\stackVerificationPtoPKHsymbolicExecutionabstractUsingVar
\vspaceAfterAgdaCode\par
\noindent hence \stackVerificationPtoPKHsymbolicExecutionnonestack{} evaluates to\\ 
\stackVerificationPtoPKHsymbolicExecutionnonEmptyStackNormalFormWithAbstractedFun{}.

Next we carry out a symbolic case distinction on the argument $cmp$ of 
\ANNannotationDone{AB: Should this be AbstrFun?}
\AgdaAbstrFun{}:
\begin{itemize} 
\item \stackVerificationPtoPKHsymbolicExecutionabstrFunZeroCase{} evaluates
to $\AgdaNothing$.
\item \stackVerificationPtoPKHsymbolicExecutionabstrFunSucCase{} evaluates
to \stackVerificationPtoPKHsymbolicExecutionabstrFunSucCaseNormal.  
\end{itemize} 

In order to normalise further, \AgdaExecuteStackCheckSig{} needs to make a
case distinction
on $\AgdaBound{stack₁}$, so we carry out a symbolic case distinction on that
argument:
\begin{itemize} 
\item \stackVerificationPtoPKHsymbolicExecutionabstrFunSucCaseEmpty{}
evaluates to \AgdaNothing.
\item \stackVerificationPtoPKHsymbolicExecutionabstrFunSucCaseNonEmpty{}
evaluates to \\
\stackVerificationPtoPKHsymbolicExecutionabstrFunSucCaseNonEmptyNormal
\end{itemize} 

We can now read off the weakest precondition. The only path which
ends up in a  $\AgdaJust$ result is when the stack is non empty
of the form $\AgdaPbkPostulate\;\AgdaDoubleColon\;\AgdaStackOne$, and\\
\stackVerificationPtoPKHsymbolicExecutionNonEmptyStackNormalFormFirstPartZoomedIn{} 
evaluates to $\AgdaSuc\;\AgdaXonePostulate$, i.e.\ it must be ${>}0$.
Furthermore, in this case $\AgdaStackOne$ needs to be itself non empty.
For $\AgdaStackOne = \AgdaSigOne\;\AgdaDoubleColon\;\AgdaStackTwo$,
the result returned is
\stackVerificationPtoPKHsymbolicExecutionabstrFunSucCaseNonEmptyNormal{}, 
which fulfils the accept condition if 
\stackVerificationPtoPKHsymbolicExecutionabstrFunSucCaseNonEmptySubTermOne\ $> 0$.
The latter is the case
if \stackVerificationPtoPKHsymbolicExecutionabstrFunSucCaseNonEmptySubTermTwo{}
is true.

Furthermore, $\AgdaCompareNaturals\;n\;m$ returns $1$ if $n$, $m$ are equal
otherwise $0$, so it is ${>}0$ if $n = m$. Therefore the P2PKH locking script
succeeds with an output stack fulfilling the acceptance condition,
if and only if the input stack has height at least two, and if it is 
$\AgdaPbkPostulate\;\AgdaDoubleColon\;\AgdaSigOne\;\AgdaDoubleColon\;\AgdaStackTwo$,
then $\AgdaPbkhPostulate$ is equal to $\AgdaHashFun\;\AgdaPbkPostulate $,  
and \stackVerificationPtoPKHsymbolicExecutionabstrFunSucCaseNonEmptySubTermTwo{}
is true. 
That is the same as the weakest precondition that we determined using the first approach.

In order to prove correctness, we first determine a more Agda style
formulation of the result of evaluation
of the P2PKH script, which we derive from the previous symbolic evaluation:
\vspaceBeforeAgdaCode%
\stackVerificationPtoPKHextractedProgramptopkhFunctionDecoded
\vspaceAfterAgdaCode\par
We prove that this function is equivalent to the result of evaluating the
P2PKH script. 
The proof is a simple case distinction following the cases
defining \ptopkhFunctionDecoded:
\vspaceBeforeAgdaCode%
\stackVerificationPtoPKHUsingEqualityOfProgramsptopkhFunctionDecodedcor
\vspaceAfterAgdaCode\par
We show that the extracted weakest precondition is a correct
for the extracted program:%
\footnote{\AgdaHoareGen{\_}{\_}{\_} is the generalisation of 
\AgdaHoareTripleIffNoSpace{\_}{\_}{\_} where Bitcoin scripts are replaced by 
Agda functions $\StackStateAgda \;\AgdaAr\;\Maybe\;\StackStateAgda$;
\AgdaHoareGens{\_}{\_}{\_} is the version, where the $\StackStateAgda$
is unfolded into its components.}%
\ANNannotationDoneQuestion{Anton: I put the footnote back in. The referee thought that
the notion $g^s$ is unnecessary, but that would require a substantial reworking
of the paper. The definition for $^g$ would still be necessary.\\
Fahad: according to review2, he said
unnecessary this footnote}%
\ANNannotationDone{footnote \AgdaHoareGen{\_}{\_}{\_} is the generalisation of 
\AgdaHoareTripleIffNoSpace{\_}{\_}{\_} where Bitcoin scripts are replaced by 
Agda functions $\StackStateAgda \;\AgdaAr\;\StackStateAgda$;
\AgdaHoareGens{\_}{\_}{\_} is the version, where the $\StackStateAgda$
is unfolded into its components.}%
\ANNannotationDone{AB: Should it be ${<>}g{<>}$ instead of ${<><>}g$ etc in
footnote?\\
AGS: you are right - corrected}%
\ANNannotationDone{Do we need to explain $<>$stgen?\\
AGS: made an attempt and changed as well the notation to something better}%
\vspaceBeforeAgdaCode%
\stackVerificationPtoPKHUsingEqualityOfProgramslemmaPTKHcoraux{}
\vspaceAfterAgdaCode\par
Afterwards, this is transferred into a proof of the weakest precondition for
the P2PKH script, 
using the equality proof from before:
\vspaceBeforeAgdaCode%
\stackVerificationPtoPKHUsingEqualityOfProgramstheoremPTPKHcor 
\vspaceAfterAgdaCode\par
Carrying out the symbolic execution was relatively easy, because Agda
supports evaluation of terms very well. It only becomes relatively long
in the Agda code \cite{setzergithub}
when documenting all the steps, which we did 
in order to explain how this is done in detail.
What matters is the resulting
program and a prove that it is equivalent, which was relatively short 
and easy. 
Maybe Agda's reflection mechanism \cite{agdadoc:Reflection}, once it is more fully developed, 
could be of help to find the successful branches of the program more easily.
To obtain a readable program rather than a machine generated program,
and therefore readable verification conditions, would however
require a lot of work, and probably require delegating some programming
tasks from Agda (in which tactics need to be written)
to its foreign language interface.



\subsection{Example: MultiSig Script (P2MS)} \label{subsec:multisig}
\ANNannotationinlineDone{Comments on Symbolic execution:\\
I looked with Fahad yesterday at the multisig script and we did a symbolic evaluation. It turns out that the script directly returns nothing if the stack is too small. Once it reaches a stack it results in a code which puts on the top of the stack the result of a function returning true or false.
We can symbolically evaluate that function and then see that it makes first a case distinction on the test whether  sig1 is a signature for pbk1. If yes, then it makes a case distinction on whether sig2 is a signature for
pbk2. If yes it succeeds, if not it checks whether sig2 is a signature for pbk3 etc.
Then one can directly read off the cases when it succeeds by symbolically evaluating the function.
From this one can read off the case distinction in a readable form.
I started and almost finished proving that the original function is equivalent to the one one gets, but realised that this is not even necessary. The resulting weakest condition is the one we used already, and one can easily
argue that now we prove directly that it is a weakest precondition (and that's done in the file we already have).
The symbolic evaluation is in the file verificationMultiSigBasicSymbolicExecution.agda}
\ANNannotationinlineDone{The above demonstrates that we can even do the symbolic evaluating with recursive functions, as long as the tree of case distinction is finite (which means roughly by koenig's lemma it is wellfounded and finitely branching.}
\ANNannotationinlineDone{The verificationMultiSigBasic.agda contains as well an example of a combined script,
namely the correctness of a combination of a check time script and themultisig script.\\
This demonstrates a combination of method1 and method 2:\\
$\ast$ we prove the correctness of the time script and the multisig script separately (the time script is trivial to verify)
using  method 2\\
$\ast$ Then we prove the correctenss of the combined script using method1
 and this shows that we can make bigger jumps in method 1
}

\ANNannotationinlineDone{The main complication is to actually define the
operational semantics of the mulstisig script, which needs to be elaborated.\\
\mbox{\quad}\\
The weakest precondition is easier to obtain for the whole multisignature script
then for just the multisig instruction.
The reason is that the weakest precondition for the multisig script itself
is very complex, whereas if we combine it with the pushes which come before it,
it is relatively easy.\\
It works in this case because the push always succeed, so when we evaluate
it the the whole script actually turns out to be equal to a subfunction
which is part of the definition of the multisig script.\\
then we can read of the weakest precondition easily}

The \texttt{OP\_MULTISIG} instruction is an instruction which has a more complex behaviour:
It assumes that the top elements of the stack are as follows:
\ANNannotationDone{Here we need to change k to m regarding this reference
http://dx.doi.org/10.1109/CVCBT.2018.00016. I did it}
\vspaceBeforeAgdaCode
\[\begin{array}[t]{@{}l} 
n \;\AgdaDoubleColon\;pbk_n\;\AgdaDoubleColon\;\cdots \;\AgdaDoubleColon\;pbk_2
\;\AgdaDoubleColon\;pbk_1\;\AgdaDoubleColon\;m\;\;\AgdaDoubleColon\;sig_m\;
\AgdaDoubleColon\;\cdots\;\AgdaDoubleColon\;sig_2
\;\AgdaDoubleColon\;sig_1\;\AgdaDoubleColon\;dummy\; 
\end{array} \] 
\vspaceAfterAgdaCode\par
\texttt{OP\_MULTISIG} checks whether the $m$
\ANNannotationDone{here I change it m instead of k}%
signatures are signatures corresponding to $m$
\ANNannotationDone{here I change it m instead of k}%
of the
$n$ public keys for the msg to be signed, where the matching public keys are
in the same order as the signatures. 
Observe that when pushed from a script, the public keys and signatures appear
in reverse order on the stack, 
as $pbk_1$ is pushed first onto the stack, etc. 
The $dummy$ element occurs due to a mistake in the Bitcoin protocol, which
has not been corrected as it
would require a hard
fork~\cite[p.~151-152]{antonopoulos}.
\ANNannotationDone{according to review2, he said added a 
reference for hard fork.}%

The operational semantics is given by a function $\AgdaExecuteMultiSig$,
which fetches the data from the stack as described before.
It fails if there are not enough elements on the stack and otherwise returns
\semanticBasicOperationsForTypeSettingcoreExecuteMultiSigThree, 
where $sigs$ and $pbks$ are the signatures and public keys fetched from the
stack in reverse order, 
and $restStack$ is the remainder of the stack.
%
%
The function \AgdacmpSigs{} compares whether signatures correspond
to public keys and is defined as follows:\par 
\vspaceBeforeAgdaCode%
\semanticBasicOperationsForTypeSettingcompareSigsMultiSig
\vspaceAfterAgdaCode\par 
We define now a generic multisig function.
\ANNannotationDone{according to review3, his note (I suspect that thehttps://www.overleaf.com/project/6242d67f42b2bc6d53ed57e5
generic Multisig M-N is not that more complicated to handle and explain. I
think it would add to the) 2 to k, So we define both generics for oppushlist
and multisig m out of length publickey}%
First we define \AgdaopPushList, which pushes 
a list of public keys on the stack:
\vspaceBeforeAgdaCode%
\multisigoppushlistofpublickey
\vspaceAfterAgdaCode\par
The $m$ out
of $n$ multi-signature script  P2MS
($n = \AgdaFunction{length}\;pbkList$) is defined as follows:
\vspaceBeforeAgdaCode%
\multisigmultiSigscriptmoutoflengthpbkList \par
The locking script
MultiSig script P2MS
applies \texttt{OP\_MULTISIG} to $m$ signatures and $n$%
\ANNannotationDone{review3, 4 to n} 
public keys. 
It pushes the number $m$ of required signatures, then
$n$ public keys, and then the number $n$ as the number of public keys, onto the stack, 
and executes \texttt{OP\_MULTISIG}.
If 
\ANNannotationDone{according to review3, he asked to added some info. about
unlocking script}
\texttt{OP\_MULTISIG} finds that the 
$m$ signatures are valid signature for the message to be signed
for $m$ out of the $n$ public keys in the same order as they
appear in the list of public keys,
then the script will be unlocked. 
As unlocking script one can use $\AgdaFunction{opPushList}$ applied to 
a list of $m$ appropriate signatures. 
In order to verify the script we will consider the concrete example of 
the 2-out-of-4 P2MS, for which we
obtain a very readable 
verification condition (the generic one becomes difficult to read). 
%
%
%
\ANNannotationDone{Add that the first approach is too defecault because 
difficult preconditions of multisig. Text from annotation at end of P2PKH.tex}

We will use the second approach of determining a readable form of the weakest
precondition and proving correctness by symbolic evaluation for 
the $2$ out of $4$ 
\AgdaFunction{multiSigScript2-4ᵇ}{}. 
The first approach is difficult to carry out since the instruction
\AgdaInductiveConstructor{opMultiSig} has a very complex precondition
that is difficult to handle -- 
it requires that the stack contains 
the number of public keys, then the public keys themselves,
then the number of signatures and the signatures, and a dummy element,
where the number of public keys and number of signatures can be arbitrary.
It is much easier to handle the full
\AgdaFunction{multiSigScript2-4ᵇ}{} script,
since, after the data has been inputted, the number of required signatures is known,
and the public keys are already provided by the script.

In order to demonstrate the first approach
we will instead, in Subsect.~\ref{sec:combinedMultiSig}, apply the 
step-by-step
\ANNannotationinlineDone{before was step by step} 
approach to a combined script, of which \AgdaFunction{multiSigScript2-4ᵇ}{} is one part.
This way we obtain a readable form of the weakest precondition and 
can then prove its correctness.
This will demonstrate that in some cases it is beneficial to interleave the
two processes, and apply the second method to sequences of instructions while
applying the first approach to the resulting sequences of instructions
instead of  single instructions.

\noindent We start the symbolic evaluation by computing the normal form of\par 
\verificationMultiSigBasicSymbolicExecutionPapermultisigsymbolic\\
and obtain \\
\verificationMultiSigBasicSymbolicExecutionPaperresultmultisig\\
Here, $\AgdaExecuteMultiSigThree$ is one of the auxiliary functions
\ANNannotationDone{AB: Should it be `executeMultiSig' with capital `S'
(twice)? (I can't find the definition of the two AdgaExecute macros to change
it myself)\\ FFA: I have updated to executeMultiSig3}
in the definition of $\AgdaExecuteMultiSig$.\\
That expression makes a case distinctions on $\AgdaStackOne$ and returns:
\begin{itemize}
\item $\AgdaNothing\;$ when the stack has height at most $2$ (obtained by evaluating it
symbolically for stacks of height $0$, $1$, $2$).
\item Otherwise, the stack has height $\geq 3$, and, if it is of the
form \verificationMultiSigBasicSymbolicExecutionPaperstackNeededFirstStepMultiSig{}, 
it reduces to \par 
\vspaceBeforeAgdaCode%
\verificationMultiSigBasicSymbolicExecutionPaperstackhasthreeelementNormalised
\vspaceAfterAgdaCode\par
\end{itemize} 
The script has terminated, because we obtain \AgdaJust{} as a result of the evaluation.
We now need to check whether the result fulfils the accept condition. 
For this the top element of the stack needs to be  ${>}0$, which is the case
if\\%
\ANNannotationDone{according to review3, I change cmpSigsAux to
cmpSigsMultiSigAux}
\verificationMultiSigBasicSymbolicExecutionPapercompareSigmulti\\
returns $\AgdaTrue$. 
Therefore, we perform symbolic case distinctions in the following way: 
\begin{itemize} 
\item In
case \verificationMultiSigBasicSymbolicExecutionPapersubexTopElementOne{}
evaluates to \AgdaTrue{}, i.e. if we replace that expression  by
\AgdaTrue{}, the reduction continues to\\
\verificationMultiSigBasicSymbolicExecutionPaperresultStepOnetrue, \\
which makes a case distinction on
\verificationMultiSigBasicSymbolicExecutionPaperresultStepOnetrueSubExp.
\begin{itemize} 
\item If that expression returns again \AgdaTrue, we obtain \AgdaTrue.
\item If it returns false, we obtain\\
\verificationMultiSigBasicSymbolicExecutionPaperresultStepOnetrueStepTwoFalse\\
which makes a case distinction
on \verificationMultiSigBasicSymbolicExecutionPaperresultStepOnetrueStepTwoFalseCore{}
\begin{itemize} 
\item In case of \AgdaTrue{}, we obtain \AgdaTrue. 
\item Otherwise the case distinctions continue, 
see the git repository \cite{setzergithub} for full details.\par 
\end{itemize} 
\end{itemize} 
\end{itemize} 
In total we see that we obtain \AgdaTrue{} iff one of the following cases holds:
\begin{itemize}
\item \AgdaFunction{(isSigned msg₁ sig₁
pbk₁)}\;\Agdaland\;\AgdaFunction{(isSigned msg₁ sig₂ pbk₂)}
\item \AgdaFunction{(isSigned msg₁ sig₁ pbk₁)}\;\Agdaland\;\AgdaFunction{¬
(isSigned msg₁ sig₂ pbk₂)}\;\Agdaland\;\AgdaFunction{(isSigned msg₁ sig₂ pbk₃)}
\item \AgdaFunction{(isSigned msg₁ sig₁ pbk₁)}\;\Agdaland\;\AgdaFunction{¬
(isSigned msg₁ sig₂ pbk₂)}\;\Agdaland\;\\
\AgdaFunction{¬ (isSigned msg₁ sig₂ pbk₃)}\;\Agdaland\;\AgdaFunction{(isSigned
msg₁ sig₂ pbk₄)}
\item $\ldots$ more cases.
\ANNannotationDone{before was or many cases.}
%
\end{itemize}

%
%
%
These cases can be simplified to an equivalent disjunction of the following cases:
\begin{itemize}
\item 
\AgdaFunction{(isSigned msg₁ sig₁ pbk₁)}\;\Agdaland\;\AgdaFunction{(isSigned
msg₁ sig₂ pbk₂)}
\item \AgdaFunction{(isSigned msg₁ sig₁
pbk₁)}\;\Agdaland\;\AgdaFunction{(isSigned msg₁ sig₂ pbk₃)}
\item \AgdaFunction{(isSigned msg₁ sig₁
pbk₁)}\;\Agdaland\;\AgdaFunction{(isSigned msg₁ sig₂ pbk₄)}
\item $\ldots$ more cases.
\end{itemize}
We obtain the following weakest precondition as a stack predicate:
\vspaceBeforeAgdaCode%
\multisigweakestPreCondTwoFour
\vspaceAfterAgdaCode\par
It expresses that the stack must have height at least
$3$, and if it is of the form
\verificationMultiSigBasicSymbolicExecutionPaperstackNeededFirstStepMultiSig{} 
then the signatures need to correspond to 2 out of the 4 public keys
in the same order as the public keys.\ANNannotationDone{according to reivew3, he said
(weakestPreConditionMultiSig only implicitly defined). So now we define the
weakest precondition}\ANNannotationDone{according to review3: his notes (1- stackPred2SPred
not defined
2- weakestPreConditionMultiSig only implicitly defined)}
Using the same case distinctions as they occurred in the symbolic
evaluation above we can now prove:
\vspaceBeforeAgdaCode%
\verificationMultiSigBasictheoremCorrectnessMultiSigTwoFour
\vspaceAfterAgdaCode\par
From the theorem above, we have obtained a readable weakest precondition by
symbolic execution, which will be used as a starting template for developing
a generic verification.%
\ANNannotationDone{In the follow-up paper, we aim togeneralise this theorem, therefor proving that our 
verification will work for
any arbitrary case. we move it to our conclusion}
The next step would be to generalise the verification conditions and
theorems to the generic case, however that would go beyond the scope of
the current paper.\par 
%
\ANNannotationDone{according to
Dr. Anton's note, I add this sentence}
\ANNannotationDone{Add sentences that we have now obtained by symbolic execution
a readable weakest precondition,
which can now be used as a template for the generic weakest precondition. We
leave it as future work to generalise
this theorem to a generic version. Note that the generic version would be
something which requires
an interactive  theorem proving tool such as Agda rather than an automated
theorem proving tool, at least with the
current state of technology.}
%


\subsection{Example: Combining the two Methods}\label{sec:combinedMultiSig}
In this subsection we show how to verify a combined script which consists of 
a simple script checking a certain amount of time has passed and the multisig script from the previous subsection. 
To determine a readable form of the weakest precondition and proving correctness 
we will combine both of our techniques: 
The weakest precondition for the multisig script has been determined by symbolic evaluation in the previous subsection. 
The weakest precondition for the simple time checking script 
will be obtained directly, as it is very simple. 
When we consider the combined scripts we will use the first method of moving backwards step-by-step\ANNannotationinlineDone{before was step by step}.
However, instead of using single instructions in each step, we now use several instructions as a single step.

We define the checktime script as follows:
\vspaceBeforeAgdaCode%
\multisigcheckTimeScript
\vspaceAfterAgdaCode\par
If we define 
\vspaceBeforeAgdaCode%
\multisigtimeCheckPreCond
\vspaceAfterAgdaCode\par
we can define its weakest precondition relative to a post condition
$\AgdaPhi$ only affecting the stack as in the following theorem:
\vspaceBeforeAgdaCode%
\verificationMultiSigBasictheoremCorrectnessTimeCheck
\vspaceAfterAgdaCode\par

Now we can determine the weakest precondition for the combined 
script and prove its correctness as follows:
\vspaceBeforeAgdaCode%
\verificationMultiSigBasictheoremCorrectnessCombinedMultiSigTimeCheck
\vspaceAfterAgdaCode\par
The weakest precondition states that the state time is $\geq  \AgdaTimeOne$,
and that the weakest precondition for the multisig script is fulfilled
(\AgdaAndsp{} forms the conjunction  of  the two  conditions).
For proving it we used a combination of both methods, the second method was used to determine
preconditions for the two parts of the scripts, and the first method, where we used whole 
scripts instead of basic instructions,  was used to determine the combined weakest precondition.

\ANNannotationinlineDone{AB: I would be nice to add a final sentence here, e.g. stating that the weakest precondition, or saying something more about the combined method\\
AGS: I have written something please check.\\
AB: Fine.  I think the first precondition expresses 'that the state time has not passed $\AgdaTimeOne$', however.\\
AGS: The instruction for time checking was defined wrongly by me and therefore the 
verification condition was the wrong  way round. Therefore the text is wrong as you noted it (it was correct with respect to the Agda  code).  I Have changed it and the Agda code. The checklock script is used to lock a transaction until the time  specified has passed.
See https://en.bitcoin.it/wiki/Script where it says:
Marks transaction as invalid if the top stack item is greater than the transaction's nLockTime field}

\section{Using Agda to Determine Readable Weakest Preconditions}
\label{sect:PracticalUsageOfAgda}
\ANNannotationDone{
-AB: OK\\
-AGS: What do you think about this title\\
-AB wrote: potentially other titles:\\
Using Agda to Describe Weakest Preconditions\\
Using our Agda Library}

Our library provides the operational semantics for (a subset of) Bitcoin \Script, and 
a framework for specifying and reasoning about weakest preconditions. 
The Agda user has to specify the script to be verified, 
and then consider suitable pieces of the 
specified script and provide weakest preconditions. 
Agda will then create goals, which are unimplemented holes in the code.
Agda will display the type of goals
and list of assumptions available for solving them, 
and provide considerable additional support for resolving those goals.
For instance, it allows to refine partial solutions provided by the user 
by applying it to sufficiently many new goals.
Agda will as well automatically create case distinctions (such as whether an element
of type \Maybe{} is \AgdaJust{} or \AgdaNothing{}). 
Agda can solve goals if the solution is unique and can be found in a direct way.
Agda's automated theorem proving support for finding solutions which are not unique
is not very strong due to the high complexity of the language. 

Agda Reflection \cite{agdadoc:Reflection} 
is an ongoing project 
which already now provides a considerable library for inspecting code inside a goal and computing solutions as Agda code. 
The aim is to provide something similar to  Coq's tactic language. 
In our code we frequently had to consider a nested case distinction for proving a goal, where most cases were solved because at one point one of the arguments became an element of the empty type. 
Automating this using Agda Reflection would make it much easier to use our library.

Finding a description of the weakest precondition has to be done manually at the moment.
We plan to create a library which computes such descriptions for instructions or small pieces of instructions.
Sometimes it is easier to provide weakest precondition for small pieces of code,  
for instance in case of the multisig instruction
the weakest precondition for the instruction itself is very complex, whereas
the weakest precondition for the P2MS script
is much easier to display. 
Defining and simplifying the weakest preconditions in the intermediate steps
has to be been done manually at the moment. 
Proofs have to be done manually in Agda, but they are relatively easy because of Agda's support for developing proofs. 
It would be desirable to have a more automated
support, where the user only needs to specify the verification conditions,
but proofs are carried out automatically. 
In general our impression is that for writing programs and specifying verification conditions Agda is very suitable: 
one obtains code which is very readable and close to standard mathematical notations. 
Where Agda is lacking is in providing support
for machine assisted proofs of the resulting conditions.  

Regarding the question, which of the two approaches to use 
(working backwards step-by-step or using symbolic evaluation), we 
have only some heuristics at the moment.
A good approach is that for pieces of code, where one
has an intuition what the underlying program written in Agda could be,
the symbolic evaluation is more suitable. 
For longer code, a good strategy is to cut the code into suitable pieces,
for which one can find a symbolic program and weakest preconditions, 
and then work oneself backwards using the first approach starting from the
acceptance condition. 
Note that symbolic execution can be done very fast: 
The user postulates variables for the arguments, applies the functions to be evaluated to those postulated arguments and then executes Agda's normalisation mechanism. 
Then the user needs to manually inspect the result to see which sub expression trigger the case distinction.
It would be nice project to develop a procedure which automates that process of symbolic execution -- 
this could be applicable to verification of other kinds of programs as well.

\section{Conclusion\ANNwhoIsDoingWhatDone{AB}} \label{sec:conclusion}
In this paper, we have implemented and tested two methods for developing
human-readable weakest preconditions and proving their correctness. 
These methods can help smart contract developers to fill the validation gap
between user requirements and formal specification. 
We have argued that weakest preconditions in Hoare logic is the correct
notion for specifying the security property of access control.
We have applied our approaches to P2PKH, P2MS, and a combination of P2MS with
a time lock.%
\ANNannotationDone{regarding review2, we use abbreviations such as
P2PKH and P2MS :before was Pay to Public Key Hash (P2PKH), Pay to
Multi-Signature (P2MS)}
The whole approach has been formalised in Agda~\cite{setzergithub}.

In future work, 
we will treat non-local instructions such
as \texttt{OP\_IF}, \texttt{OP\_ELSE}, and \texttt{OP\_ENDIF}, and will
formalise key instructions
to extend our approach to the whole of Bitcoin \Script{}. 
The difficulty is nesting of conditionals, and that 
Bitcoin scripts are not structured, and therefore some additional work
needs to be done to find the matching of 
if-then-else instructions.
In our approach we will use an expanded state space for dealing with those conditionals.
\ANNannotationDone{I changed it regarding review1, before other opcodes
such as conditionals like OP$\_$IF,}%
\ANNannotationDone{review1: * What are the
problems with $OP_IF$ and whether there could be issues with other
    instructions? Can the formalisation cover the whole Bitcoin Script in the
    future? (FA: Done)}%
Moreover, we plan to expand our library to support finding 
weakest preconditions for scripts having conditionals in a modular way.
Furthermore, we plan the make the process of script
verification more user-friendly by using a text parser that can record the
instructions used for verification.\par 
\ANNannotationDone{review1 :Do you plan to
embed the Script programs, e.g. by parsing the text representation or
extract by printing the Agda AST into text? How the users can start verifying their
 programs?}%
In addition, we aim to generalise the verification of P2MS to arbitrary
$m$ out of $n$ multiscripts, where the challenge is finding a suitable generic  human-readable
weakest precondition. 

Another route for future research is to develop our approach into a framework for
developing smart contracts that are correct by construction. One way to build such smart contracts is to use Hoare Type 
Theory~\cite{IMDEA:2015:HoareTypeTheory,Nanevski:ACMHeap-ManipulatingPrograms:2010}.%
%
\ANNannotationDone{review1: his note ("... developing smart contracts that
are correct by construction" -- maybe at this point something like Hoare Type
Theory (https://software.imdea.org/~a) could be useful.), FA
Done}
%
%





\newpage
\bibliographystyle{plainurl}
\bibliography{ref}
\end{document}